# Thirty Meter Telescope International Observatory Detailed Science Case

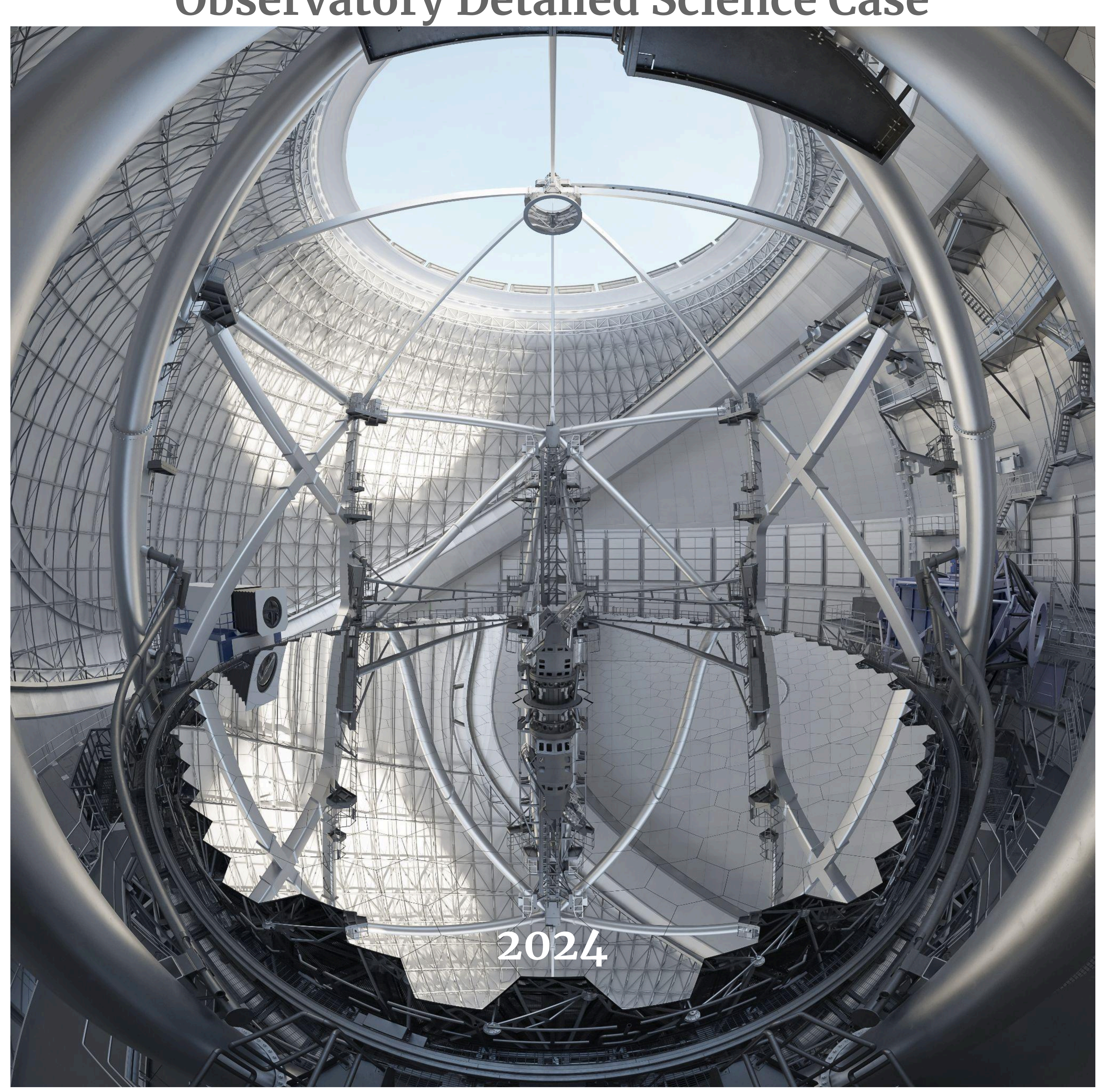

2024

# Prologue

Astronomy asks the big questions: where are we? How did we get here? Where are we going? The questions are thousands of years old, but today's answers started to be developed only a few hundred years ago with the advent of sharper tools for collecting light from the stars and the scientific approach to confronting ideas with data.

When Galileo first used a telescope to look at the sky, the question of whether celestial objects were made of different stuff than we know on Earth and whether the Earth went around the Sun were open for learned debate. But with the improved spatial resolution and better sensitivity of Galileo's instruments, combined with careful record-keeping and shrewd reasoning, the height of the moon's mountains, the phases of Venus as it circles the Sun, and the orbits of Jupiter's moons all became evidence that we live in a world where the Earth is a sample of the material of the universe as we orbit a star not so different from the more distant ones. We are not apart from the universe, but a part of the universe.

Caltech's 200-inch (5-meter) Hale telescope at Palomar Mountain was the marvel of its age when it came into operation in the 1950s. It was the culmination of 400 years of incremental progress in gathering more light. When closely coupled with our deeper understanding of gravitation and the quantum nature of the physical world, these telescopes led to a picture in which the atoms of Earth's elements had their origin in the nuclear furnaces that power the stars, the stars of the nighttime sky are themselves gathered into our spiral Milky Way Galaxy, and our galaxy and a hundred trillion other galaxies are distributed across a vast expanding universe that had its origin in a hot Big Bang about 14 billion years ago. The current state of the art in large telescopes came from brilliant technical insights by Jerry Nelson and his colleagues at the University of California. Nelson conceived of building a 10 meter telescope out of hexagonal tiles with sensors to detect each mirror's

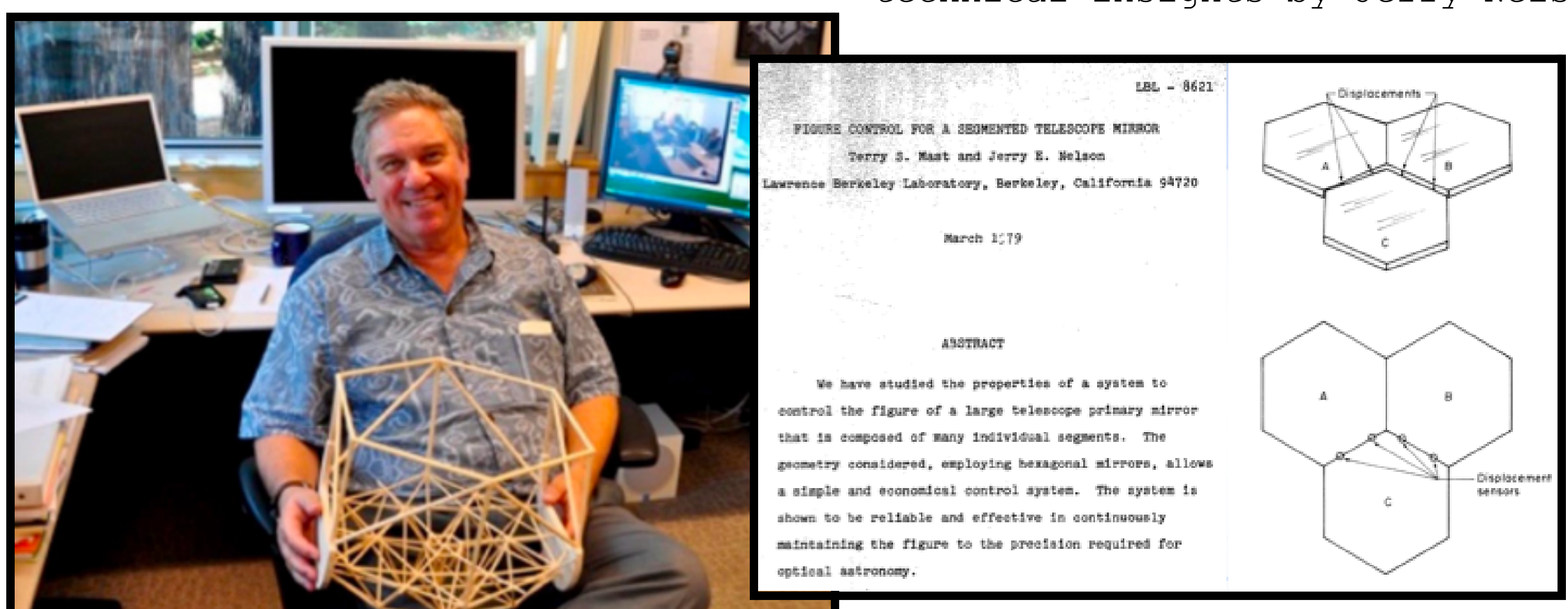

LBL - 8621

FIGURE CONTROL FOR A SEGMENTED TELESCOPE MIRROR

Terry S. Mast and Jerry E. Nelson

Lawrence Berkeley Laboratory, Berkeley, California 94720

March 1979

ABSTRACT

We have studied the properties of a system to control the figure of a large telescope primary mirror that is composed of many individual segments. The geometry considered, employing hexagonal mirrors, allows a simple and economical control system. The system is shown to be reliable and effective in continuously maintaining the figure to the precision required for optical astronomy.

position and computer controls to keep the segments aligned so that they function as a single mirror. This audacious design led to the Keck 10 meter in 1993 and succeeded so well a second 10 meter telescope was built at the same superb site on Maunakea, on the big island of Hawaiʻi in 1996.

A second technological thread, adaptive optics, compensates for the wiggles of the Earth's atmosphere so that telescopes with adaptive optics systems approach the theoretical limit of resolution that is set by the wavelength of light and the diameter of the telescope. Without adaptive optics, bigger telescopes just collect more light. With adaptive optics, bigger telescopes make sharper images and concentrate the light.

Together, these two advantages demonstrated by the Keck 10 meter make the prospect of yet larger telescopes irresistible. For many types of observations, their effectiveness scales as diameter to the fourth power. The Thirty Meter Telescope builds on the segmented mirror ideas of the Keck Telescopes, but has 3 times the aperture, 9 times the collecting power and, with adaptive optics, for some problems, 81 times better than today's best.

This document spells out what astronomers want to do with this tremendous increase in capability. Observations from the current generation of telescopes show that planets are common and we can look for signs of life. We know that black holes are an important part of the pageant of galaxy formation, but precisely how this happens is an enigma. Galaxies are bound by an invisible form of dark matter whose nature we do not know. The universe as a whole has its past and future bound up with the dark energy that is making the universe accelerate, but whose nature is also not known. As every chapter shows, mysteries abound. The way to make progress on these profound questions is by a more powerful telescope: the TMT.

TMT has been designed to function well for 50 years. The scientific questions of the future and the technology to answer those questions are not known today. But we can have confidence that there will be new forms of old questions and there will be new ways to investigate them. Where are we? How did we get here? Where are we going? This science case is just the beginning of a long adventure of discovery.

*Robert Kirshner*

TIO Executive Director

# DOCUMENT APPROVAL

**Author Release Note:**
See chapter 16 for the full list of contributors.

**Chief Editor:**

\Signature on file\

Warren Skidmore
TIO System Scientist

**Concurrence:**

\Signature on file\

David Andersen
TIO Project Scientist (acting)

\Signature on file\

Robert Kirshner
TIO Executive Director

**Approval:**

\Signature on file\

Gelys Trancho
TIO Project Systems Engineer

# Document Change Record

| Revision | Modifications | Release Approval | Date Released |
|---|---|---|---|
| CCR04 | The document was updated to include results from the JWST and other recent studies. Endorsed by the TIO SAC. | Docushare Routing #117263 | 15 October, 2024 |
| CCR03 | The document was updated to reflect changes from ASTRO 2020 and the 2022 updates. Endorsed by the TIO SAC. | DocuShare Routing #90291 | 22 September 2022 |
| REL02 | This version represents the public release of the  Detailed Science Case from 2015. Endorsed by the TIO SAC. | – | 29 April 2015 |
| REL01 | This was the first public release of the Detailed Science Case. Endorsed by the TIO SAC. | – | 8 Oct 2007 |

# 1 INTRODUCTION

This is the Detailed Science Case (DSC) for the Thirty Meter Telescope (TMT) International Observatory (TIO), a publicly available document, officially endorsed by the TIO Science Advisory Committee. This version builds upon the 2022 update of the original 2015 edition, incorporating significant changes in both TMT's instrumentation plans and key developments in the scientific landscape.

While certain aspects of the science cases have evolved since 2015, and new cases have been introduced, the observational capabilities needed to perform the studies outlined in this edition of the DSC are fully supported by the TIO design.

The original 2015 version was itself an update of the 2007 document, and was created with extensive contributions, expansions, and updates from members of the TIO International Science Development Teams. The 2007 document was shaped with input from the science instrument teams, the authors of the 2006 instrument feasibility studies, and a wide range of scientists from across the TMT collaboration. See chapter 17 for more details.

Throughout this document, the terms **TMT** and **TIO** are used interchangeably to refer to the observatory and to the telescope design.

## 1.1 PURPOSE

The Detailed Science Case is the highest-level statement of the Thirty Meter Telescope science cases. It provides examples of the kinds of exciting, groundbreaking science that will be enabled by TMT.

Wherever possible, synergies with other major new and upcoming facilities (e.g., the Atacama Large Millimeter Array, Rubin Observatory, and the James Webb Space Telescope) are discussed.

As appropriate, performance numbers (often conservative) are provided (e.g., sensitivities, integration times, spatial resolutions).

The DSC is not a requirements document. The Science Requirements Document (SRD, RD1) and the Operational Requirement Document (OpsRD, RD2) are the highest-level requirements document for TMT. Requirements flowdown from this document to the SRD and OpsRD, and the traceability between requirements at the system and subsystem levels and science cases is described in RD3.

This document is intended to provide a high-level overview of the TMT science case for the following audiences:

- Astronomers within the TMT partnership and international scientific colleagues
- Various astronomy strategic working groups and internal TMT groups
- Reviewers

## 1.2 ORGANIZATION OF CHAPTERS

This document is introduced in chapter 1. Chapter 2 provides the broader context for the Thirty Meter Telescope (TMT) Science Case, discussing the high-level *Big Questions* in astrophysics that drive the development of TMT. It also covers the synergies with other existing and future astronomical and scientific facilities and presents an overview of the observatory and instrument designs in response to these science cases.

The six motivating science cases, or Big Questions (see section 2.2), are further elaborated through nine science themes outlined in chapters 3 to 11. Each theme comprises multiple individual science cases, which in turn inspire specific observing programs. These programs lead to observing requirements, which are discussed separately in RD3. The nine science chapters were developed by the TMT International Science Development Teams (ISDTs), as described in section 16.1. The science chapters are as follows:

- **Chapter 3:** Fundamental Physics and Cosmology
- **Chapter 4:** Early Universe
- **Chapter 5:** Galaxy Formation and the Intergalactic Medium
- **Chapter 6:** Supermassive Black Holes
- **Chapter 7:** Exploration of the Milky Way and Nearby Galaxies
- **Chapter 8:** The Birth and Early Lives of Stars and Planets
- **Chapter 9:** Time Domain Science
- **Chapter 10:** Exoplanets
- **Chapter 11:** Our Solar System

**Chapter 12** explores new realms of discovery, explaining how the flexible capabilities of the TMT observatory will enable scientific investigations beyond current predictions.

**Chapter 13** summarizes the TMT science cases and instrument capabilities, linking the science themes, big questions, and observatory requirements.

The remaining chapters include:

- **Chapter 14:** Abbreviations and Acronyms
- **Chapter 15:** Index, including Tables of Contents, Figures, and Tables
- **Chapter 16:** Contributors
- **Chapter 17:** Acknowledgements

## 1.3 REFERENCE DOCUMENTS

RD1 Science Requirements Document, (TIO.PSC.DRD.05.001),

RD2 Operations Requirements Document, (TIO.OPS.DRD.07.002),

RD3 Science Cases Flowdown and Traceability (TIO.PSC.TEC.22.002)

RD4 "Pathways to Discovery in Astronomy and Astrophysics for the 2020s", 2021, National Academies of Sciences, Engineering, and Medicine

RD5 US-ELTP Common Project Description (TMT.PMO.MGT.20.004)

# 2 OVERVIEW

## 2.1 THE BIG PICTURE

Figure 2-1 encapsulates our current understanding of the history of our universe over the last 13.77 billion years from the first instants of cosmic inflation through the formation of the first stars and galaxies, all embedded in an accelerating framework of cosmic geometry.

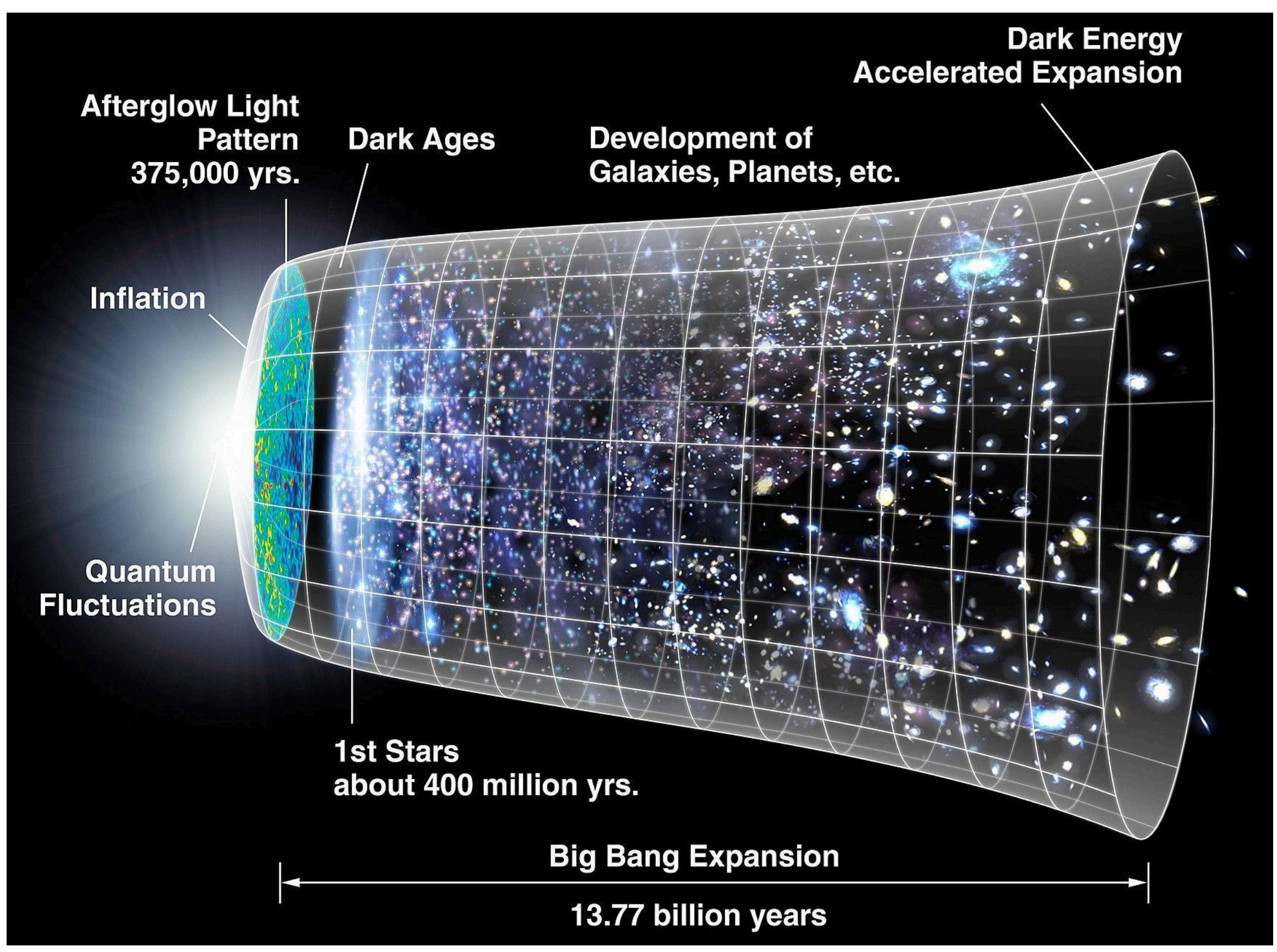

***Figure 2-1:*** *A schematic history of the universe. Time is represented from the Big Bang on the left, to our location in the present day on the right. The structure and state of the universe is illustrated as it evolves through cosmic time, with the first stars and galaxies depicted on the left with the advent of cosmic acceleration denoted by the flaring grid at the right. (Credit: NASA Science Team)*

Since the time of Galileo, astronomical observations with telescopes of ever-increasing power have provided the clues to build our understanding of the universe. But we know our understanding is limited: deep mysteries of the dark matter and the dark energy that dominate the universe remain unsolved. Though we know a great deal about the microscopic evolution of matter from the time of the hot Big Bang, through the nuclear furnaces of the stars and into the planets, stars and galaxies we see near us, there are many gaps in our understanding. For example, we do not know if life is rare or common on the planets that orbit other stars. The way to make progress across this sea of ignorance is to build sharper and more powerful tools to observe and infer what today we can only surmise. The Thirty Meter Telescope is part of that story: this chapter outlines how the design of the telescope and the instruments that will analyze the light it gathers are carefully crafted to match the most pressing scientific questions. We are confident that TMT will enable progress on these riddles and, if history is any guide, open new areas we have not yet imagined

Let's look at this figure in more detail and see what questions are lurking just behind it, starting from the left and working our way to the right:

1. What really is the nature and composition of the universe? What exactly is the 95% of the universe we euphemistically call *dark*?
2. When and how did the first galaxies form and how did they evolve?
3. What is the precise relationship between black holes and galaxy formation? Which came first?
4. How do stars and planets form?
5. What does the real population of exoplanets look like?
6. Is there life elsewhere in the universe?

Before we proceed more systematically through this list, take a moment to consider this last question - *is there life elsewhere in the universe?* That this is now a question for science and not philosophy, religion, or pseudoscience reflects the amazing time we are in. We now have the technical know-how and the political and social will to build facilities like TMT to address this question, shifting it from the realm of speculation and science fiction into the arena of evidence. TMT will help us understand the possibilities and possible abundance of life outside our solar system.

While figure 2-1 conveys a concise vision of how much we know, yet these 6 questions reveal how much we don't know starting with mysteries of the origin of the universe form and hierarchically cascading down to asking how life develops. These important questions remain with incomplete answers, but we have developed a good plan for advancing our understanding. By gathering the community and reaching a consensus, the US Astronomy 2020 Decadal Survey (Astro2020, RD4) articulated core science themes that will help organize our thoughts:

1. *Worlds and Suns in Context* — exploring the interconnected formation and evolution of exoplanets and their host stars, and possible origins and detection of life outside our own planet.

2. *New Messengers and New Physics* — exploring dark matter, dark energy, gravitational waves, multi-wavelength and multi-messenger astronomy; what is the universe made of and how can we learn about it? Exploring the dynamic universe to understand how things change and evolve on timescales from less than a second to billions of years.

3. *Cosmic Ecosystems* — exploring the vast range of temporal and spatial scales of the universe from the initial seeds of galaxies to the large scale structure revealed by galaxy surveys on the largest scales to the atomic-scale chemical change from the very first stars formed from the hydrogen and helium of from the Big Bang to the rich variety of the periodic table that we experience on Earth.

The Astro2020 report placed its highest priority for the coming decade to develop the next generation of large ground-based optical/infrared telescopes. The combination of scientific potential produced by diffraction-limited of the extremely large telescopes and technical readiness makes the TMT a compelling choice to explore these fields, answer these questions, and create a new set of unknowns for future innovators to resolve.

## 2.2 THE BIG QUESTIONS

TMT's extremely large primary mirror will enable observations from ultraviolet to mid-infrared wavelengths with up to 81 times the sensitivity of today's largest optical telescopes. State-of the-art adaptive optics systems will compensate for the blurring effects of Earth's atmosphere to deliver images at infrared wavelengths that are more than 12 times sharper than those of the famed Hubble Space Telescope, and 4 times sharper than those of the James Webb Space Telescope that marks today's state of the art.

This combination of sensitivity and image quality will allow TMT to probe deeper into the universe and to reveal details of cosmic objects that are invisible with today's observatories. TMT incorporates innovations in precision control that make it possible to acquire a new science target anywhere on the sky within a few minutes — a critical capability for rapid follow-up of gravitational wave sources and other cosmic explosions.

A common rule of thumb is that science takes significant steps forward when technology improves by an order of magnitude. With TMT, we have not one, but two orders of magnitude improvement to address the big questions introduced in section 2.1 and explored in more detail here.

***Q1. What is the nature and composition of the universe?***

The nature of dark matter and dark energy, the ingredients that dominate — by a factor of 20 to 1 — the composition of the universe, remains a complete mystery. The lower limit of the dark matter mass spectrum, for example, depends on the nature of the dark particle(s) (*warm* vs. *cold*). TMT will probe this limit down to levels at least ten times smaller than currently possible via diffraction-limited imaging of anomalies in strong gravitational lenses. Different dark energy models predict different rates of evolution for cosmic distances and structures: TMT will greatly expand the region of models that can be tested through deep spectroscopy of very distant supernovae (up to $z = 4$). The science cases in chapters 3, 4, 5, 6, 7, and 9 provide more details on how TMT will help us learn what our universe is made of.

***Q2. When did the first galaxies form and how did they evolve?***

Current telescopes do not have the sensitivity to probe the formation of early galaxies. Whereas existing 8 m class telescopes can provide detailed, spatially-resolved maps of morphology, chemistry, and kinematics for galaxies at $z < 1$, TMT will be able to provide comparable details out to $z = 5$–$6$. TMT will be able to detect the spectroscopic signatures of metal-free star formation, i.e., of the first stars forming in the first galaxies at redshifts well beyond 10. Closer to home, TMT will also be able to image spatially-resolved stellar populations in galaxies out to the distance of the Virgo Cluster to provide the first-ever archeological sample large enough to reveal the history of galaxy assembly. Closer still, TMT's large field of view and envisioned excellent ultraviolet throughput will enable powerful surveys of stellar abundances and ages in the halo of our own Milky Way over a volume nearly 100 times larger than previously possible and build a far more complete census of the smaller fragments that were hierarchically assembled into this halo. The science cases in chapters 3, 4, 5, 6, and 9 provide more details on how TMT will reveal the processes by which galaxies form and evolve.

***Q3. What is the relationship between black holes and galaxies?***

Black holes with masses as high as several billion times the mass of the Sun are now known to occupy the centers of galaxies. They seem intimately connected to galaxy formation and they warp space-time in fascinating ways, but we don't really understand their formation process(es) and the interplay between core black hole and galaxy formation. If mass is the link, then the black hole masses should track the galaxy masses across the entire Hubble sequence and across time, but current facilities lack the precision to measure this. TMT will be able to measure the black hole masses ten times smaller, and make dynamical measurements in galaxies more than 20 times farther than is currently possible. TMT will expand the number of measurable galaxy black hole masses by a factor of 1000. The science cases in chapters 4, 5, 6, and 9 provide more details on how TMT will help explore and disentangle the interplay between core black hole and galaxy formation.

***Q4. How do stars and planets form?***

The processes of stellar birth are closely intertwined with galaxy evolution and chemical enrichment in the universe that ultimately result in an astonishingly diverse range of planetary systems, some of which may harbor life. We know much about how stars are born, live, and die and yet, we cannot currently answer some very basic questions such as what determines a given star's mass, and we are far from understanding when, where, and how planets form. TMT's exquisite sensitivity, high spatial resolution, and efficient instrument suite will allow us to explore these questions from several different angles. TMT can measure the initial mass functions for distant young star clusters, not only in the Milky Way but also in other galaxies in a wide range of environments, providing one of the first building blocks in any complete model of stellar formation. Through mapping the velocity fields of protoplanetary disks, TMT will unveil the inner 10 AU of star forming regions, yielding detections of growing planets and revealing how water and organic molecules are distributed. The science cases in chapters 3, 4, 5, 7, 8, 10, and 11 provide more details on how TMT will uncover the processes involved in star and planet formation.

***Q5. What is the nature of extrasolar planets?***

Thanks to the Kepler space telescope, and a variety of other efforts, we now know that most stars harbor a planetary system, that Earth-like planets in habitable zones are quite common and that the easy-to-detect hot Jupiter-type planets are the exception rather than the rule, yet there still exist biases in our estimates of the population of exoplanets as a whole and we do not fully understand the formation and evolutionary processes that give rise to this variety of exoplanetary systems. Kepler was the tool we needed to move from a known handful to thousands of exoplanets. TMT will be the tool we need to start understanding the nature of these planets, able to explore the atmospheres of these planets beyond the closest and brightest hot Jupiters that existing 8 m class facilities can address and the very limited characterization of lava-ocean rocky planets possible with existing space missions, extending this work to greater distances and down toward Earth-like masses, including those in their respective habitable zones. The science cases in chapters 7, 8, 10, and 11 provide more details on how TMT will conduct its census of the nature and atmospheres of exoplanets.

***Q6. Is there life elsewhere in the universe?***

This question is at least as old as humanity itself: are we alone? Are conditions on terrestrial planets conducive to the development of life? Every star has a habitable zone in which a planet would have a surface temperature like that of Earth, implying the possibilities for Earth-like life elsewhere are quite compelling. If, as expected, exoplanetary systems have populations of small icy bodies like comets, it is possible that water and organic molecules could have been delivered to these planets by impacts with these bodies. TMT will not only explore the existence of such populations but will search for the biological signatures of life in exoplanet atmospheres. This exploration alone is worthy enough of this next generation facility, making the combined improvements in our understanding of science from the very beginning of the universe to the chances of life existing elsewhere a case too hard to ignore. The science cases in chapters 8, 10, and 11 demonstrate how TMT will support humanity in its quest for discovering the possibility of life beyond our fragile Earth.

## 2.3 THE OBSERVATORY

The answers to the Big Questions of section 2.2, will only come with advances in observing capabilities and the combination of multiwavelength studies from future ground and space-based facilities. TMT, along with a number of exciting new major observatories, provides for both these needs. The science cases to come in the later chapters demonstrate that although TMT is an essential tool to address these big questions, advancing our understanding will take a variety of other approaches. TMT will work to both complement — provide a new and unique way of addressing the question — and synergize — provide a combination more powerful than the sum of its parts — existing and future facilities.

Just as the unprecedented resolution of the Hubble Space Telescope was complemented by the greater light gathering power of the Keck, VLTs, Subaru, and Gemini telescopes, the space-based observatories of the future will need complementary observations at high spatial resolution and high spectral resolution that only large-aperture ground-based telescope can provide. Spatially resolved studies of small, faint objects such as the earliest galaxies or arcs of gravitationally lensed galaxies *require* the capabilities that TMT and its versatile instrumentation suite will provide.

Modern adaptive optics systems allow the largest optical-infrared telescopes on Earth to achieve higher spatial resolution than any telescope now in space. Adaptive optics compensates for atmospheric inhomogeneities to allow telescopes on Earth to reach their true diffraction limits set only by the laws of physics. Adaptive optics achieves an angular resolution that is proportional to the diameter of the telescope aperture: bigger telescopes make sharper images. TMT is designed to have high performance: high throughput, a clean aperture, and a rigid structure to realize the full potential of an extremely large aperture.

To achieve these huge gains in light collection and spatial resolution, TMT will be equipped with a suite of powerful adaptive optics facilities and science instruments. The selection of first light instruments from a suite of instrument concepts was made in December 2006 by the TMT Science Advisory Committee based on a careful consideration of the science programs accompanied by a vision of essential tools for pursuing science areas that are currently unforeseen. Consideration was also given to technical readiness, cost, and schedule. The initial selection of first light instruments consisted of a combined diffraction limited near-infrared integral field spectrograph and imager, a wide-field optical multi-object spectrograph and imager, and a multi-object slit-based near-infrared spectrograph. These choices were re-affirmed in 2011 at a project-wide science and instrumentation workshop in Victoria, BC.

However, the recent rise in importance of exoplanet science increased the need for TMT observatory to be able to characterize exoplanets changed in instrumentation priorities. In 2019, the Science Advisory Committee recommended a new first light instrument that uses the facility AO system to conduct wide ranging exoplanet science using a high resolution spectrograph with optional coronagraph. The capabilities of these instruments are summarized in the next subsections.

All instruments will be mounted and kept ready for operation. ready at all times (except scheduled maintenance, repair, or upgrades). The TMT can nimbly switch from one instrument to another during the night in less than 10 minutes..

### 2.3.1 The Facilities/Telescope

The Thirty Meter Telescope will be a 30 m ground-based telescope with a collecting area of 664.2 $m^2$. It will be able to support instruments that are sensitive through the atmospheric windows from 0.31 to 28 μm.

Advanced adaptive optics capabilities will allow highly sensitive, diffraction-limited observations from wavelengths shortward of 1 μm to the mid-infrared over most of the sky. A 20 arcmin diameter field of view at the telescope focal plane facilitates the deployment of wide-field, multi-object spectrographs. Figure 2-2 shows the major components of the planned TMT observatory and figure 2-3 shows the planned instrument layout on the Nasmyth platforms.

These capabilities will enable groundbreaking advances in a wide range of scientific areas, from the most distant reaches of the universe to our own solar system.

Powerful new facilities have often opened up unimagined areas of research and made unanticipated discoveries. We expect TMT will be no different: we look forward to a rich and diverse mix of both expected and unexpected scientific results (discussed further in chapter 12).

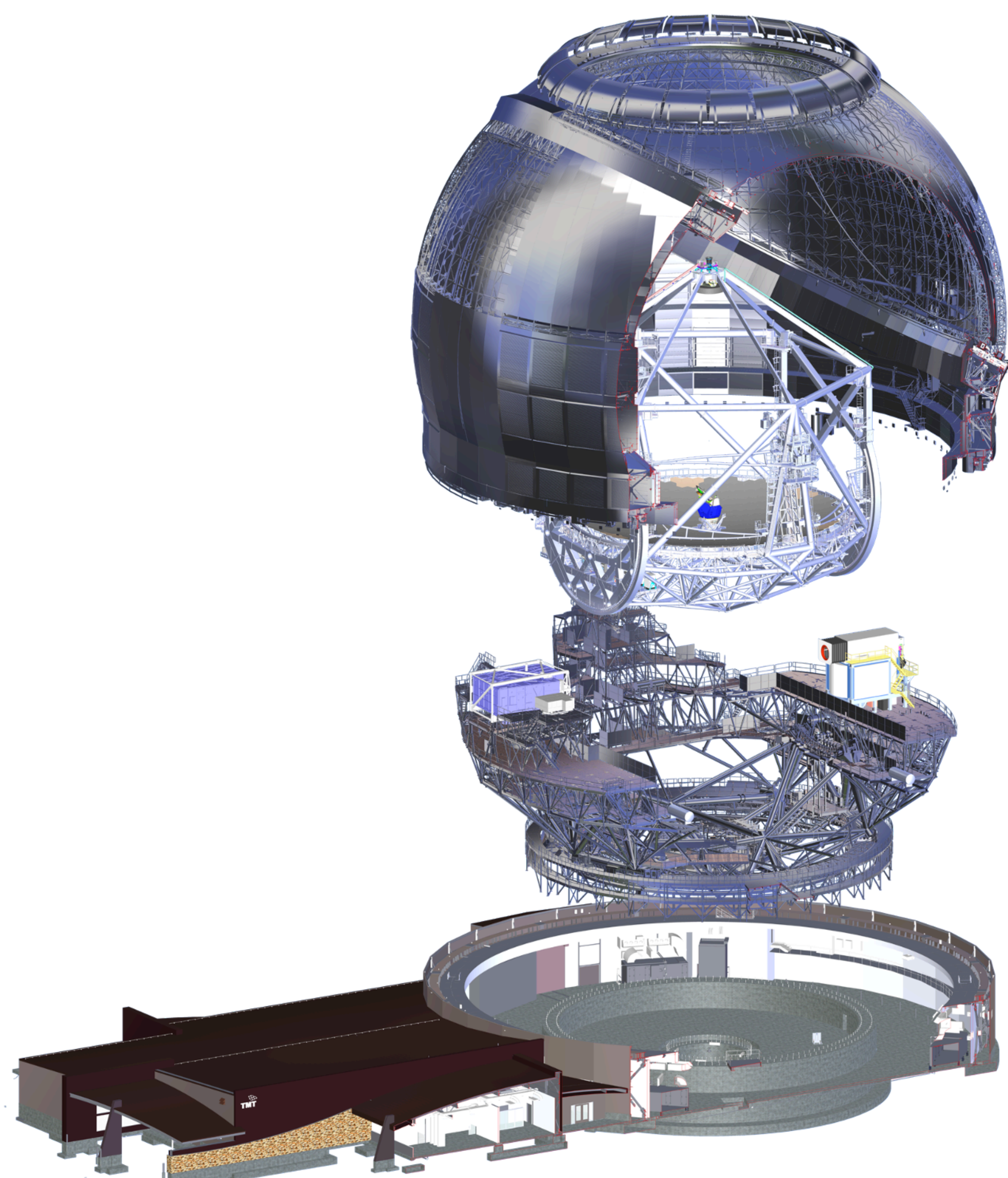

***Figure 2-2:*** *Exploded view of the TMT International Observatory. The summit facilities building and observing floor are shown in the lower segment. Above that is the telescope azimuth structure, including the first light instrument suite on the Nasmyth Platforms, with WFOS on the right and NFIRAOS and the Alignment and Phasing System on the left (see instrument descriptions in section 2.3.2). Above that are the telescope elevation structure, including the telescope optics (M1, M2 and M3) and at top, the enclosure.*

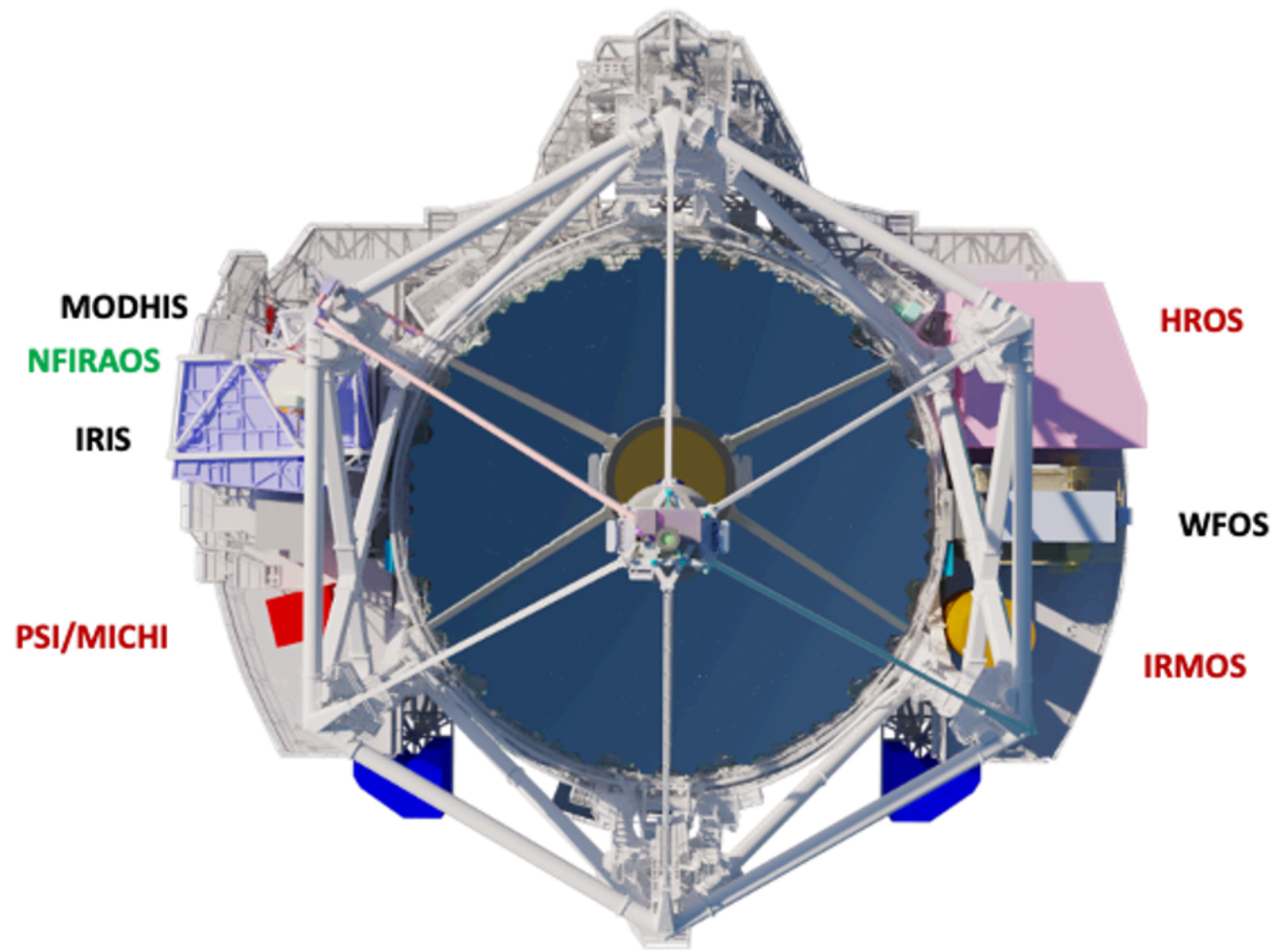

***Figure 2-3:*** *Top down view of TMT with its Nasmyth platforms. NFIRAOS feeding IRIS and MODHIS and WFOS will be built for TMT first light. There is room to accommodate first decade instruments as well including HROS, PSI, MICHI, and IRMOS (see instrument descriptions in section 2.3.2).*

### 2.3.2 First Light Suite Instruments and AO facilities

At first light, TMT plans to deploy a laser guide star supported, multi-conjugate adaptive optics (MCAO) system called **NFIRAOS (Narrow Field InfraRed Adaptive Optics System)**. NFIRAOS will provide diffraction-limited resolution ($\lambda/D$ = 0.0055 arcsec and 0.017 arcsec at 800 nm and 2.4 µm, respectively) and high Strehl ratios[1] (39% in J band at 1.25 µm, 58% in H band at 1.65 µm and 74% in K band at 2.2 µm, respectively) over a 34 arcsec science field-of-view and for 50% sky coverage at the Galactic poles. It will be possible to mount up to three (3) instruments on NFIRAOS.

One of these instruments will be **IRIS (InfraRed Imaging Spectrometer),** a combined high-resolution imager and integral-field unit (IFU) spectrometer. The direct imaging mode will provide a 34 x 34 arcsec field with 0.004 arcsec/pixel sampling. Both broadband and narrowband filters will be available. The IFU will provide spectral resolving powers of $R$ = 4000, 8000, and 10,000 over the entire J, H, and K bands (one band at a time) as well as a choice of coarse and fine plate scales for different science applications. The field-of-view of the IFU ranges from 2.25" x 4.4" (coarsest scale 0.050"/spaxel) to 0.512" x 0.512" (finest scale 0.004"/spaxel). The IRIS concept has a heritage with the Keck instruments OSIRIS and LIGER.

NFIRAOS will also feed **MODHIS (Multi-Objective Diffraction-limited High-resolution Infrared Spectrograph)**, a single object R > 100,000 fiber fed spectrograph, operating over a bandpass of 0.95 to 2.4 µm with instrumental stability yielding a precision radial velocity (PRV) of 30 cm/s (goal of 10 cm/s). In addition to traditional fiber fed spectroscopic capabilities that utilize the MCAO corrected image quality to allow spectroscopy of faint targets, MODHIS will include coronagraphic capabilities in front of fiber injection units (FIUs) that may be placed anywhere within a 4" diameter field of regard centered around the input optical axis. Exoplanet observations can reach an inner working angle of 2 λ/D in the high dispersion coronagraphic mode and can achieve an inner working angle of 0.25 λ/D in the special vortex fiber nulling mode. The MODHIS concept builds on experience with Palomar's PARVI and Keck's HISPEC.

[1] Strehl Ratio: When considering an unresolved point source, an adaptive optics system can be thought of as gathering a fraction of the light from the seeing disk and concentrating that fraction into an image with a diffraction-limited width, superimposed over the remaining uncorrected light. The Strehl Ratio is the ratio of the actual peak intensity divided by the theoretical diffraction-limited peak intensity for a perfect image.

The third first-light instrument is **WFOS (Wide-Field Optical Spectrometer)**. WFOS is a seeing-limited instrument that will cover a total spectral range from 0.31 to 1.0 µm using separate red and blue color channels. Two main spectral resolution modes will be available (R = 1500 and 3500 with 0.75" slits and R = 5000 or greater using narrower slits) with multiplexing slit configurations that use up to 500" total slit length over a 25 arcmin$^2$ field. The WFOS concept builds upon the heritage of such workhorse instruments as Keck's DEIMOS, MOSFIRE, and LRIS-2, and Magellan/IMACS.

Table 2-1 summarizes these instruments and their key capabilities.

***Table 2-1:*** *First Light instruments/AO Facilities*

| Instrument and Description | λ Range (µm) | Spectral Resolution | Modes | Field of View |
|---|---|---|---|---|
| **NFIRAOS**/Narrow Field Infrared Adaptive Optics System | 0.8 – 2.4 | N/A | NGSAO, LGS MCAO, SL Enhanced | LGS MCAO 2.0′ |
| **IRIS**/Diffraction-Limited NIR Imager and IFS | 0.84 – 2.4 | Z, Y, J, H, K, bandpass filters and multiple narrower band filters. 4,000 and 8,000 (some modes to 10,000) | NGSAO, LGS MCAO SL Enhanced | Imager: 34″ x 34″ @ 0.004″/pix<br>IFU with two slicing techniques<br>Lenslet: 0.512″ x 0.512″ @ 0.004″/spaxel<br>Slicer: 2.25″ x 4.4″ @ 0.050″/spaxel |
| **WFOS**/Wide Field Optical Spectrometer | 0.31 – 1.0 | 1,500 and 3,500 using 0.75″ slits. Goal of 5,000 currently achieved and higher R available with narrower slits. | Seeing-limited GLAO | 25 (8.3 x 3)-arcmin$^2$<br>500″ total slit length (up to 60 targets with 8″ slits)<br>Imaging: full field @ 0.05″/pixel |
| **MODHIS**/Multi-Objective Diffraction-Limited High-Resolution Infrared Spectrograph | 0.98 – 2.46 | > 100,000 with 30 cm/s (goal 10 cm/s) Doppler velocity precision | NGSAO, LGS MCAO | 4" diameter field of regard with positionable diffraction limited fiber bundle (target, sky, speckle, spare, calibration). 6"x6" imaging guider. |

### 2.3.3 First Decade Instruments and AO facilities

TMT has been designed to support a diverse set of instruments that will evolve over its 50-year lifetime, including instruments beyond those described here that take advantage of advances in technology. The first of the second-generation instruments is expected to be delivered two years after first light, and additional instruments should follow at the rate of one every 2.5 years thereafter.

Instrument priorities and requirements will be established by the TMT Board following SAC recommendations that include community, technical, and programmatic input. Future instrument concepts studied thus far are collectively referred to as the First Decade Instruments and are listed in table 2-2. They include:

**GLAO (Ground-Layer Adaptive Optics):** based on an adaptive secondary mirror and using the LGS facility, a GLAO system would provide corrections in the optical and near-IR that would support narrow slit widths of 0.25" and higher throughputs for WFOS and HROS.

**MIRAO (Mid-Infrared Adaptive Optics):** most likely based on a cooled deformable mirror downstream of the telescope optics, but possibly including an adaptive secondary mirror, this system would employ the LGS facility to obtain high Strehl ratios over a small field of view for mid-IR instruments.

**HROS (High-Resolution Optical Spectrometer):** a single-object, seeing-limited, high-resolution (*R*=50,000 for 1 arcsec slit) optical–UV echelle spectrograph in the heritage of HIRES at Keck, HDS at Subaru, and UVES at the VLT. Concepts considered all have a high stability of <10 cm/s/yr.

**IRMOS (Infrared Multi Object Spectrometer):** will have significant multiplexing capabilities. Scientific and technical trade studies will explore multi-object adaptive optics (MOAO), MCAO, multiple IFU, and slit based instrument options. Baselined for MOAO over a 5 arcmin field of view to feed up to 20 deployable IFUs, this future instrument expands on the diagnostic power of IRIS with high multiplexing capabilities.

**MIRES (Mid-Infrared Echelle Spectrometer):** is a diffraction-limited, high-resolution ($5000 < R < 100{,}000$) spectrometer and imager operating at 8–18 µm. It will employ a separate AO system optimized for the mid-infrared. A concept for a mid-IR high resolution spectrograph, imager and IFU instrument covering 3–14 µm called MICHI is being actively developed at this time.

**PFI (Planet Formation Instrument):** an extreme AO high contrast exoplanet imager with spectroscopic ($R \leq 100$) capability. The first version of the system will obtain contrasts of $10^6$ (goal: $10^7$) while the second version's requirement is $10^8$ (goal: $10^9$) in H-band for $R < 8$ mag. The next design phases benefit from the lessons learned from GPI at Gemini, SCExAO at Subaru, and SPHERE at the VLT. There is already a large overlap between the GPI and SCExAO development teams and those developing the Second Earth Imager for TMT (SEIT) and Planetary Systems Imager (PSI). PSI includes extensive spectroscopic capabilities to support detailed exoplanet characterization

***Table 2-2:*** *Baseline requirements for First Decade instruments. Included are two adaptive facilities (GLAO and MIRAO). Requirements are subject to change and instrument concepts not listed may be identified and developed.*

| Instrument and Description | λ Range (µm) | Spectral Resolution | Modes | Field of View |
|---|---|---|---|---|
| **GLAO**/Ground Layer Adaptive Optics (feeds WFOS and HROS) | 0.31–1.0 | N/A | GLAO | Large enough to cover WFOS |
| **MIRAO**/Mid-Infrared **Adaptive** Optics (feeds MICHI) | 4.5 – 28 | N/A | LGS MIRAO, high contrast | >10″ (1′ goal) |
| **PSI** PFI/Planet Formation Instrument | 0.6 – 5.3 | (fiber fed) High resolution R > 100K<br>(IFS) Medium resolution R > 5,000<br>(IFS) Low resolution R > 50 | ExAO | 2–5.3 µm only:<br>1.2″ x1.2″ (low resolution)<br>0.15″ x 0.15″ (medium resolution) |
| **MICHI** MIRES/ Mid-Infrared Echelle Spectrometer | 3.4 – 13.8 | Imager < 100, IFS 600–1,000, Spectrometer 120,000 | MIRAO | Imager: 28.1″ x 28.1″ @ 11 mas/pix N band<br>IFU: 0.175″ x 0.07″ (35 mas/spaxel) |
| **HROS**/High-Resolution Optical Spectrograph | 0.31 – 1 | Single Object: 100,000 & 50,000 (fibers)<br>40,000 & 20,000 (slits)<br>Multi-Object: 25,000 | Seeing-limited GLAO | > 10″ in diameter (single object mode)<br>10′–20′ diameter (multi-object mode) |
| **IRMOS**/IR Multi-Object Spectrograph | 0.8 – 2.5 | 2,000 – 10,000 | MOAO | > ten 3″ IFUs deployable within a 5′ diameter field |

TMT first light and first decade instrument capabilities are summarized in figure 2-4.

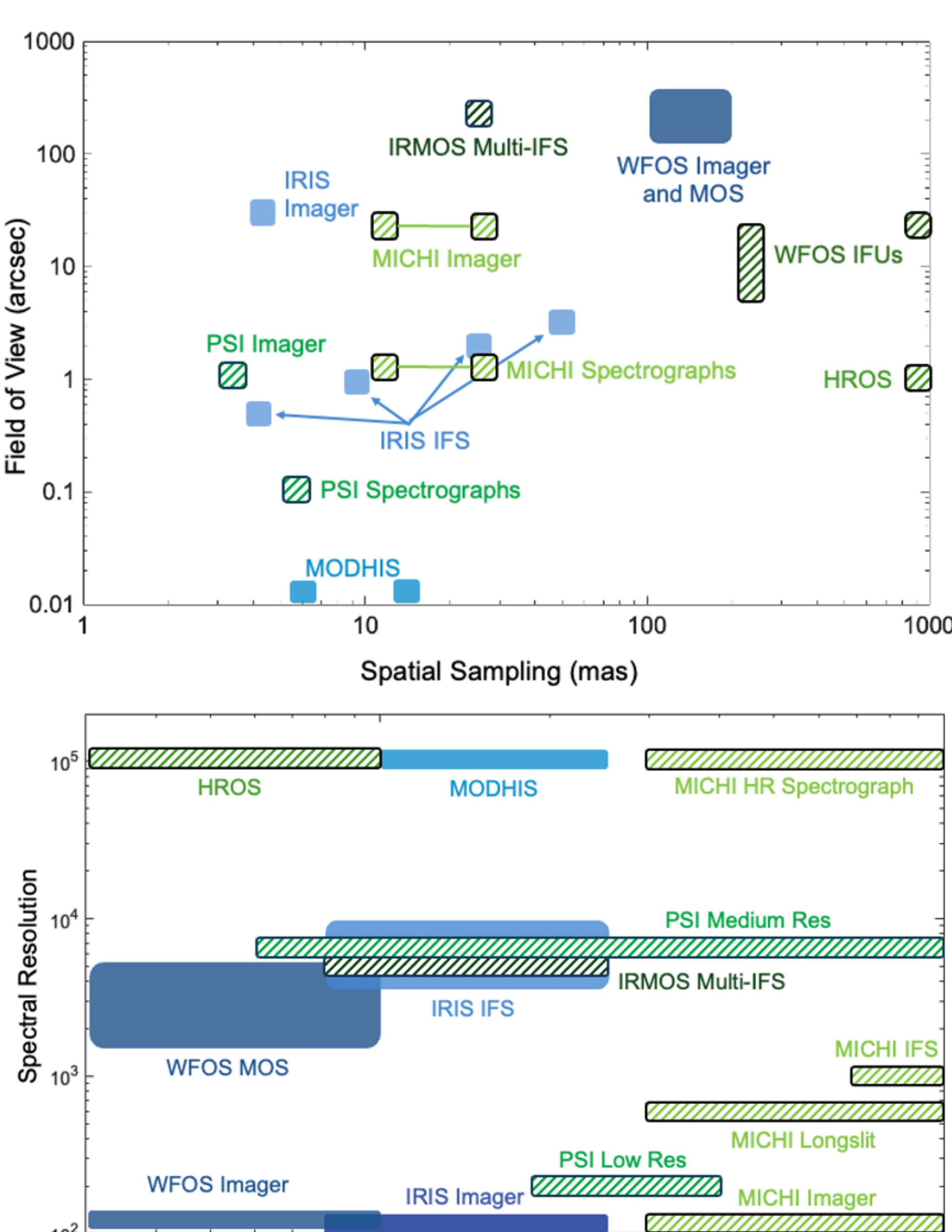

***Figure 2-4:*** *First light and first decade TMT instruments provide good coverage of the parameter space of instrumental field of view, spatial sampling, spectral resolution and spectral range. First light instruments are shown in solid boxes. Hatched boxes show parameters for proposed first decade instruments.*

## 2.4 NEW CAPABILITIES, NEW DISCOVERIES

Compared to existing telescopes, TMT will bring exceptional sensitivity and spatial resolution. The combination of these is transformational in terms of science that they will unlock. Figure 2-5 shows the level of sensitivity that TMT will provide for followup of the faintest targets detectable by major facilities or identified in a selection of major surveys.

In seeing-limited modes, sensitivity is dependent on the square of the primary mirror diameter ($D^2$), so the 14 times improvement in sensitivity compared to 8 m telescopes enables the detection of many more probes through the IGM thereby providing a much more detailed view of the structure of the universe on spatial scales more than a factor of 10 finer than presently possible (see section 5.3).

However, the true power of the 30 m aperture exploits adaptive optics that provide *diffraction-limited* corrected images. As listed in table 2-3, the sensitivity for point sources increases massively as $D^4$, but the advantage goes beyond this sensitivity improvement, particularly in fields where crowding is an issue. Finer spatial resolution allows more targets to be separately measured to much lower brightness limits, allowing the full range of members of dense star clusters to be measured, for example, covering the full stellar and sub-stellar mass range (see multiple cases in chapters 7 and 8).

***Table 2-3:*** *The basic scaling of the point source sensitivity with diameter and the advantage that TMT will have over the most common large telescopes today. The point source sensitivity is relevant for point-like sources, e.g., stars, exoplanets.* $\eta$*, S and D are telescope throughput, image Strehl ratio and telescope diameter respectively.*

| Point Source Sensitivity | Scaling | TMT gain compared to 8 m |
|---|---|---|
| Seeing-Limited | $\eta D^2$ | x 14 |
| MCAO | $\eta S^2 D^4$ | x 198 |
| ExAO* | $\eta S^2 D^4 /(1-S)$ | >>198 |

*Strehl ratio impact is inversely proportional to D for ExAO

The improved spatial resolution will support significant advances in the studies of exoplanet systems, population demographics, and exoplanet formation due to the much smaller inner working angle resulting from the smaller diffraction limit. This means that the inner portions of exoplanet systems and protoplanetary disks can be probed directly, pushing further into the regime of rocky planets in the habitable zones of their host stars (see chapters 8 and 10). Figure 2-6 shows how the seeing limited and diffraction-limited spatial resolution compares between some major facilities and to the equivalent physical scale for certain distances or redshifts.

Even in cases of resolved or extended targets, the ability to discern between different sources within a single system due to the improved spatial resolution is critical in many science cases, including measuring velocity dispersions of stars in the centers of galaxies, finding exoplanets, and characterizing their atmospheres, and separating AGN signatures from star formation regions, for example (see chapter 6). Table 2-4 specifies the spatial resolution advantage that TMT has compared to other facilities.

***Table 2-4:*** *The relative diffraction limited spatial resolution advantage of TMT over other facilities if observing at the same wavelengths. The linear value is simply the relative finer angular separations that targets can be distinguished from. The area value is the number of smaller spatial elements that a TMT AO fed instrument can supply within a single spatial element of the other facility.*

| **Diffraction-limited spatial resolution** | **TMT advantage relative to:** | | |
|---|---|---|---|
| | **8 m** | **JWST** | **HST** |
| Linear | x 3.75 | x 4.62 | x 12.5 |
| Area | x 14 | x 21 | x 156 |

We can clearly envision many amazing scientific opportunities that will be opened up by the leap in capabilities the TMT will bring forth, as evidenced by the vast range of science cases in this document.

Additionally, as mentioned in the preface and discussed in chapter 12, we do not truly yet know what scientific discoveries TMT will be used to make. However, by designing, building, and operating an observatory capable of tackling the broad range of ambitious science cases that we present here, we can be very confident of TMT's ability to produce groundbreaking results, in the science cases derived here, and in those yet to be imagined by the telescope's future users.

The ability to advance areas of science with TMT is illustrated in figures 2-5 and 2-6, which indicate the ability of the TMT observatory to do intensive follow up observations of targets identified in various imaging surveys.

Follow up observations can be spectroscopic or imaging to even greater depths and with much greater spatial resolution than the surveys. To gather spectroscopic follow up observations of any of the major imaging surveys requires the light gathering power and sensitivities of TMT and its 30m class counterparts.

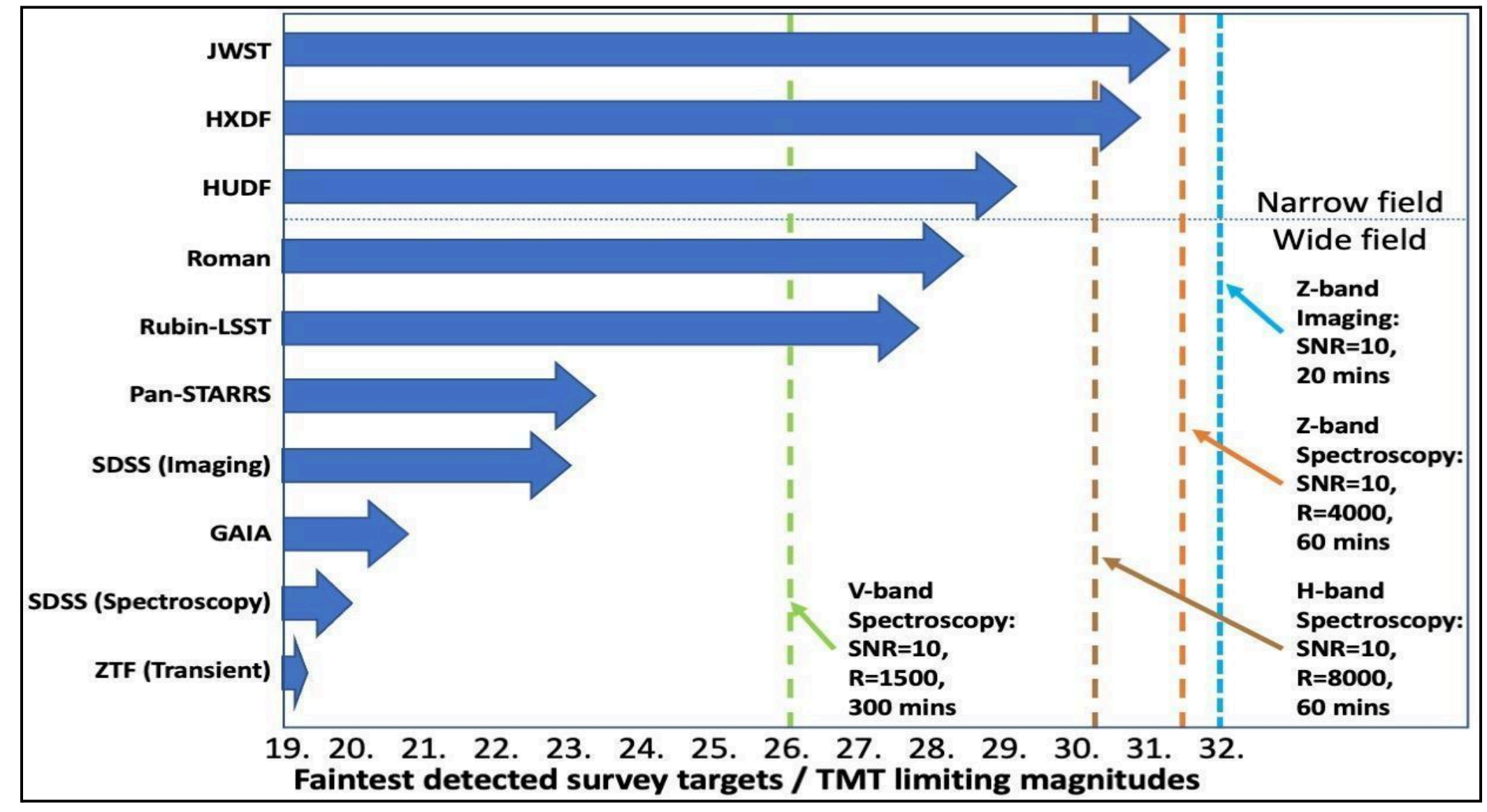

***Figure 2-5:*** *Arrows indicate the faintest targets that are reached by various imaging surveys. The deepest narrow field surveys have been conducted with Hubble and recently with JWST. All ground-based surveys and the GAIA survey cover optical wavelength ranges, Roman includes near-IR, HUDF and HXDF cover UV to near-IR, JWST survey is for 1 µm to 5 µm. Vertical lines indicate the approximate expected magnitudes that deliver a signal to noise of 10 in the specified modes.*

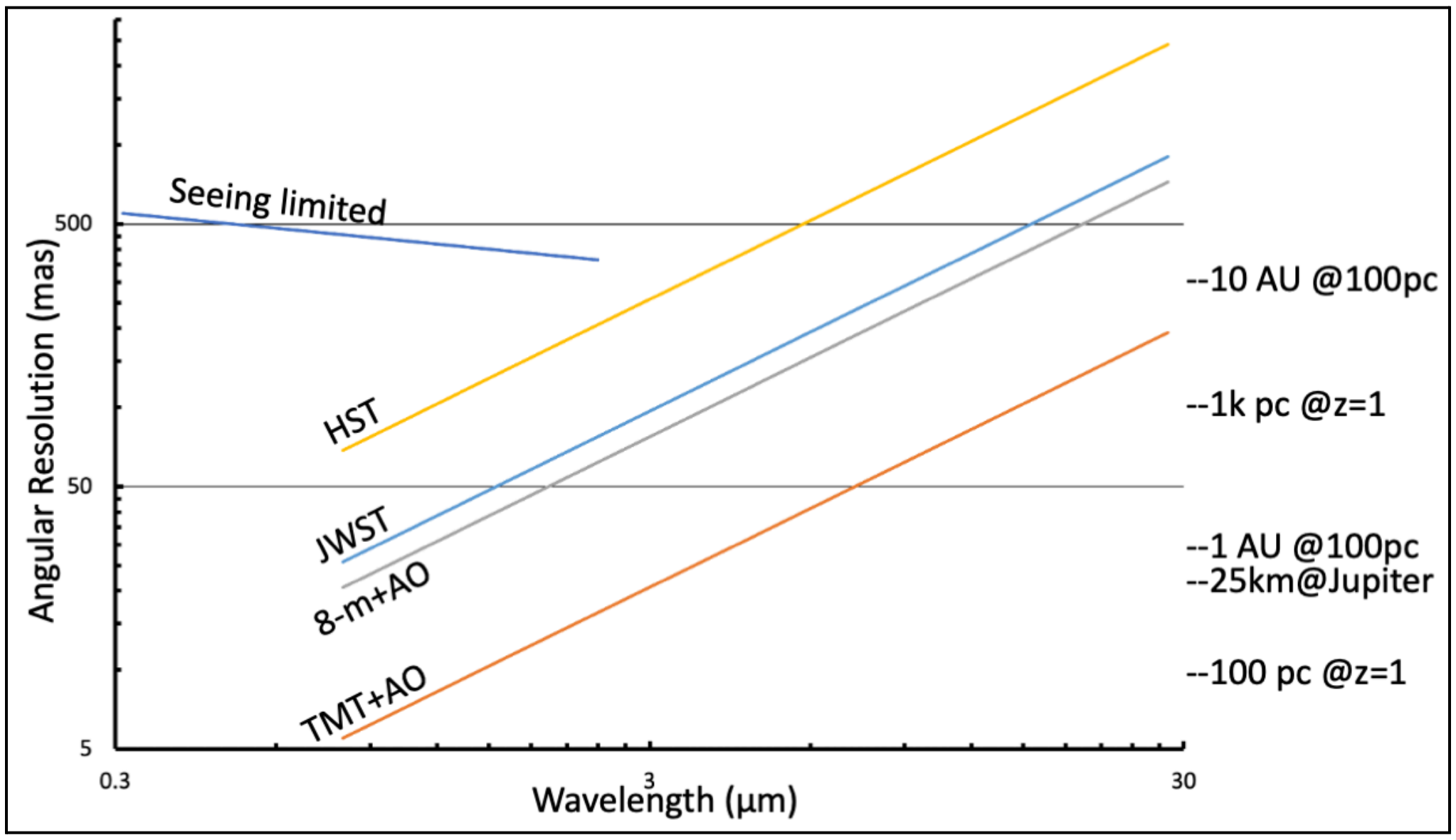

***Figure 2-6:*** *The angular resolution of TMT as a function of wavelength compared to other facilities. Seeing-limited spatial resolution is set by the median conditions at Maunakea. TMT's spatial resolution when using high-order correction adaptive optics surpasses all other existing major facilities. The angular scales corresponding to 1 AU & 10 AU and 100 pc & 1 Kpc at 100 pc and z=1 respectively, and 25 km at Jupiter are indicated on the right-hand side.*

## 2.5 Synergies and complementarity with current or coming new space and ground-based facilities

What *will* TMT bring to the table? The answer is multi-dimensional. In the first instance are TMT's instruments — selected with complementarity and synergies in mind. TMT's instruments need to be able to expand the range of objects other facilities can observe and provide new information with qualitatively different kinds of data on targets identified from other sources. TMT's instrument suite overlaps the broad characteristics of other optical and infrared facilities while also providing additional capability they do not have. As a result, the instruments highlighted in sections 2.3.2 and 2.3.3 provide a broad range of capabilities and observing modes.

Beyond this *overlap and expand* approach in instrument capability, TMT's operations model will add flexibility when compared to space observatories and geographic complementarity to many ground-based facilities. JWST's target of opportunity capability is very limited; TMT's design incorporates such targets and time domain science from the very start. TMT's adaptive queue mode, with a rich heritage from Gemini, allows flexible observation that optimizes the match between available and required conditions. Additionally, TMT's northern site can reach parts of the sky that overlap with Southern Hemisphere observatories like the Rubin Observatory, GMT and ESO's Extremely Large Telescope, and observe the northern sky the others cannot reach. Whether at Maunakea or in La Palma, TMT will be a significantly different longitude to expand the time a target that can be seen from Chile is visible.

Coordinated TMT observations with JWST could occur towards the end of JWST's extended mission lifetime and TMT will be a natural complement to the Habitable Worlds Observatory when it is launched. In addition, TMT will support the solar system probes that will study solar system planets, moons, and other small bodies in the next decades. With its exquisite angular resolution, TMT will be capable of studying solar system bodies at precisions that match those needed in these missions, accessing greater numbers of bodies than spacecraft can visit, while adding additional wavelengths and other instrument capabilities to those of the space missions themselves.

Natural synergies with JWST that depend on TMT's sensitivity and spatial resolution include studies of the structure and processes in early galaxies discovered by JWST that have different sizes, levels of chemical evolution and star formation **(section 4.2),** and high spatial resolution observations of star forming regions in low redshift galaxies **(section 7.8.7)**. Ground-based wide-angle surveys, in particular with the Rubin Observatory, will uncover galaxy clusters whose members will need detailed follow up in order to explore the influence of the intra-cluster environment on cluster members **(section 5.1.3)**. Due to matching spatial resolution obtained with TMT and ALMA, synergies include studies of different components and layers in protoplanetary disks **(section 8.6.1)**.

The high precision astrometric capability of TMT creates a valuable synergy with GAIA. For example, GAIA will eventually measure the proper motions of stars in many dwarf galaxies while TMT will provide higher precision measurements to a fainter limiting magnitude, for a small sample of dwarf galaxies **(section 3.1.1)**. Future TMT observations will also extend the reach of past surveys. Synergies with the original SDSS and SDSS-BOSS surveys, the ongoing DESI survey and forthcoming Subaru-PFS surveys will, for example, include TMT measurements of the cosmic structure to much greater depth and finer spatial/velocity detail in selected small fields **(section 5.2.1)**.

Multi-messenger and time domain astronomy will greatly benefit from TMT's light gathering power and spatial resolution to provide complementary observations in the near ultraviolet to infrared with those of other ground and space observatories **(chapter 9)**. TMT characterization of the optical counterparts of LIGO/VIRGO/KAGRA gravitational wave sources, for instance, will dramatically boost our understanding of gravitational mergers at lower mass regimes and the chemical enrichment of the universe with r-process heavy elements **(section 9.1)**.

Thanks to its rapid response ability TMT will also be able to choose targets from about 40% of the huge number of transient targets identified each night by the Rubin Observatory surveys, down to declinations around −30°. This overlap will facilitate the follow up with TMT of all forms of supernovae and periodic variable stars and tidal disruption events **(chapter 9)**. Supernovae follow up and periodic variable star characterization **(section 9.12)** will significantly improve the accuracy of the cosmic distance ladder and our understanding of cosmology in general. Follow up of tidal disruption events will advance our knowledge of the relation between supermassive black hole mass and host galaxy properties.

Space-based x-ray and $\gamma$-ray facilities will discover gamma ray bursts that can be subsequently observed with TMT in order to study the bursts themselves, or to use them as probes of their distant environments **(sections 9.6 and 9.7)**.

Other multi-messenger/time domain synergies with future ground facilities like ngVLA, SKA, and CTA will dramatically increase the potential of combined science programs between TMT and these observatories (e.g., **section 9.8)**.

NASA is pushing forward with the development of the flagship Habitable Worlds Observatory (HWO) that provides a broad set of observing capabilities that, in addition to seeking answer whether there is life on planets around other stars, will address questions concerning the evolution of large scale structure, galaxies and their contents over cosmic time through to characterizing objects in our solar system. HWO will have incredible sensitivity for biosignatures and evidence of biological processes. The aims for HWO are truly complementary to those of TMT (see table 2-6). The long lifetime of TMT and the ability of the TMT partnership to construct highly capable next generation instruments guarantees that the TMT will provide complementary capabilities that make significant contributions to science programs involving HWO. The inherent capabilities of TMT such as spatial resolution, flexible scheduling, and a wide range of observing modes, coupled with the development of extreme adaptive optics that perform well at shorter wavelengths mean TMT will bolster HWO's exoplanet characterization programs by providing context and critical cross checks of exoplanet studies and results.

Table 2-5 and 2-6 illustrates, in a compact manner, the science synergies between TMT and other ground and space-based astronomical facilities that are mentioned in the science chapters of this document.

### 2.5.1 US-ELTP

As part of the US-ELT Program (US-ELTP; https://useltp.org) TMT's association with the Giant Magellan Telescope and NOIRlab will provide the U.S. community with a set of extremely large telescopes (ELTs) capable of addressing all the key science priorities laid out in Astro2020, offering full sky coverage and a complementary suite of instruments. The longitudinal coverage offered by US-ELTP will enable optimal follow-up of transient events, increasing the time-base readily accessible for rapid follow-up and continuous coverage of rapidly changing physical phenomena over multiple wavelengths.

The location of the components of the US-ELTP in each of the Northern and Southern Hemispheres provides all-sky coverage and a significant area of overlap, allowing some observations to be optimized for both time availability (telescope and longitude) and the unique capabilities available at each facility.

NOIRLab partnership with the US-ELTP will enable the submission of single science proposals from eligible users that request the synergistic use of *multiple* NSF-funded facilities to carry out investigations across wavelength and messenger regimes, making programs like this easier to coordinate..

A full description on TMT and the US-ELTP synergies is provided in the US-ELTP Project Common Description (RD5)

***Table 2-5:*** *The science synergies between TMT and other ground based astronomical facilities. Double* ***d****ark* ***g****reen check boxes indicate science themes in which synergies with the listed facilities are specifically discussed in the science cases. Single* ***l****ight* ***g****reen check boxes indicate science themes where complementary observations are asserted in this document or specifically mentioned in science documents related to the listed facility.*

| Synergy | Ch#3 | Ch#4 | Ch#5 | Ch#6 | Ch#7 | Ch#8 | Ch#9 | Ch#10 | Ch#11 |
|---|---|---|---|---|---|---|---|---|---|
| | Fund. Physics and Cosmology | Early Universe | Galaxy Formation and the IGM | SMBH | MW and Nearby Galaxies | Birth & Early Lives of Stars & Planets | Time-Domain Science | Exoplanets | Our Solar System |
| ALMA | | ☑ | ☑ ☑ | ☑ ☑ | | ☑ ☑ | ☑ ☑ | ☑ ☑ | |
| CTA | | | ☑ | | | | ☑ ☑ | | |
| LIGO (-VIRGO-KAGRA) | | | | ☑ ☑ | | | ☑ ☑ | | |
| ngVLA | ☑ | ☑ | ☑ | ☑ | | ☑ | ☑ ☑ | | |
| Rubin Observatory | ☑ ☑ | ☑ ☑ | ☑ | ☑ ☑ | ☑ ☑ | | ☑ ☑ | | ☑ |
| SDSS (all surveys) | ☑ ☑ | ☑ ☑ | ☑ ☑ | ☑ ☑ | ☑ ☑ | ☑ | ☑ | | |
| SKA | ☑ ☑ | ☑ ☑ | | | | ☑ | ☑ ☑ | | |

***Table 2-6:*** *The science synergies between TMT space based astronomical facilities. Double dark green check boxes indicate science themes in which synergies with the listed facilities are specifically discussed in the science cases. Single* ***l****ight* ***g****reen check boxes indicate science themes where complementary observations are asserted in this document or specifically mentioned in science documents related to the listed facility.*

| Synergy | Ch#3 | Ch#4 | Ch#5 | Ch#6 | Ch#7 | Ch#8 | Ch#9 | Ch#10 | Ch#11 |
|---|---|---|---|---|---|---|---|---|---|
| | **Fund. Physics and Cosmology** | **Early Universe** | **Galaxy Formation and the IGM** | **SMBH** | **MW and Nearby Galaxies** | **Birth & Early Lives of Stars & Planets** | **Time-Domain Science** | **Exoplanets** | **Our Solar System** |
| **EUCLID** | ☑ ☑ | ☑ ☑ | | ☑ ☑ | | | ☑ ☑ | | |
| **GAIA** | ☑ ☑ | | | | ☑ ☑ | ☑ ☑ | ☑ ☑ | ☑ ☑ | |
| **JWST** | ☑ ☑ | ☑ ☑ | ☑ ☑ | ☑ ☑ | ☑ ☑ | ☑ ☑ | ☑ ☑ | ☑ ☑ | ☑ ☑ |
| **Roman Space Telescope** | ☑ ☑ | ☑ ☑ | ☑ | ☑ | ☑ | ☑ | ☑ ☑ | ☑ ☑ | |
| **HWO** | ☑ | ☑ | ☑ | ☑ | ☑ | ☑ | ☑ | ☑ ☑ | ☑ |
| **LYNX** | ☑ | | ☑ | ☑ | ☑ | ☑ | ☑ | | |
| **ORIGINS** | | ☑ | ☑ | ☑ | | ☑ | | | ☑ |
| **SPHEREx** | ☑ | ☑ | ☑ | | | ☑ | | | |

# 3. Fundamental physics and cosmology

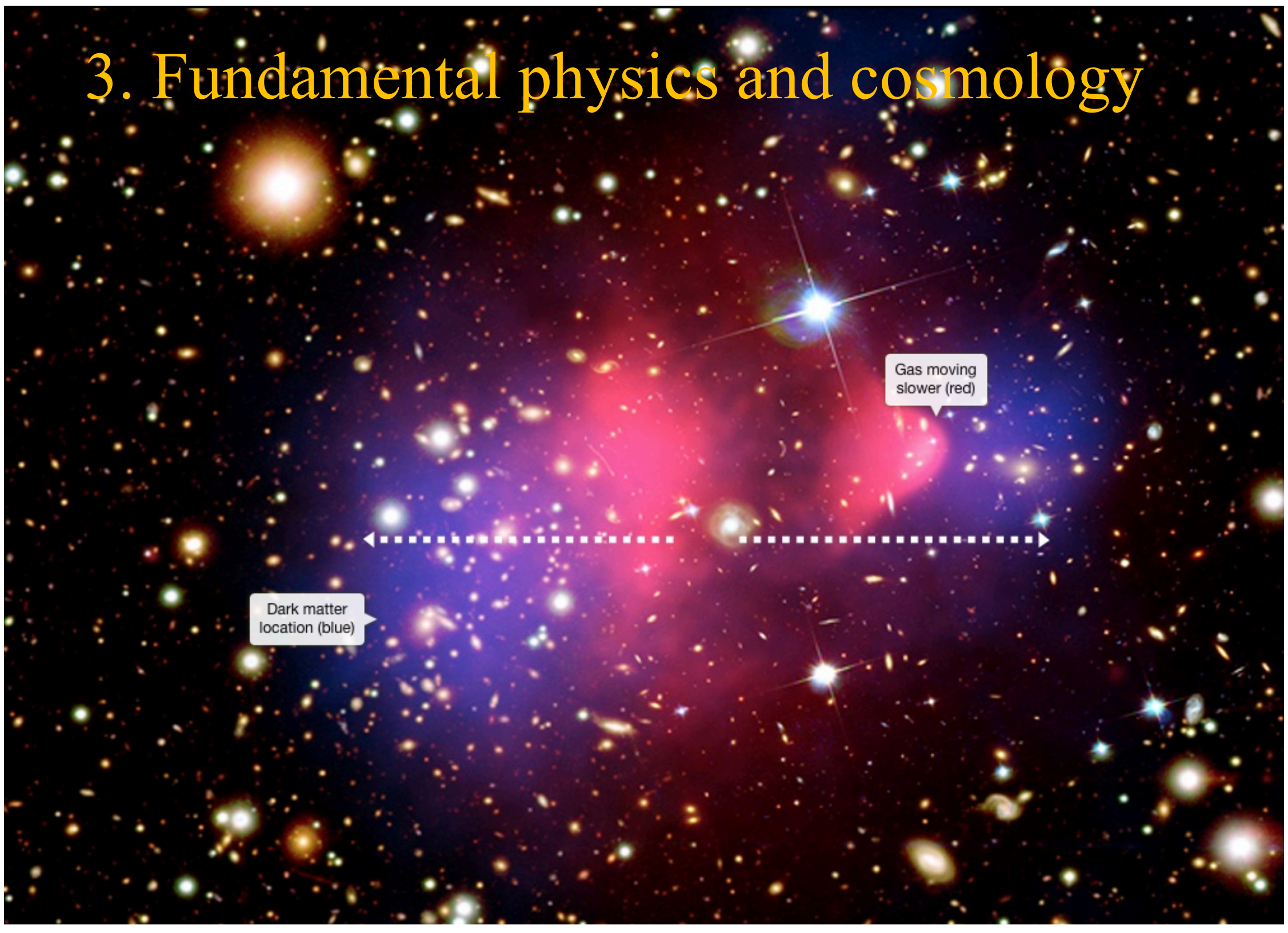

*Composite picture of the Bullet Cluster showing hot gas (pink), dark matter (blue) stellar light from galaxies. The difference in the distributions of baryonic matter (gas and stars) and dark matter arise due to the different levels of self-interaction between dark matter particles and baryonic matter (From -https://viewspace.org/)*

---

TMT will address the first big question from chapter 2, Q1-*What is the nature and composition of the universe?*, by exploring fundamental physics in the universe and testing models of dark matter, dark energy, modified gravity, and general relativity. These observations will also touch on two additional big questions addressing when and how the first galaxies, stars, and planets formed (Q2 and Q4). This chapter discusses a number of different areas of active research in fundamental physics and cosmology that are limited by current levels of available sensitivity and angular resolution. TMT observations will allow the precision necessary to fundamentally increase our knowledge of the fundamental components and building blocks of the universe.

The capabilities needed to advance this science fundamentally include TMT's immense light gathering power and fine angular resolution, supplemented by further AO improvements in spatial resolution. The envisioned observations require a range of spectral resolution from low resolution R~1000 to high resolution R~50,000. IFU capability is particularly important for dark matter substructure science as is MOS capability for lensing science. The neutron star science cases require a full range of wavelength coverage, from UV to IR.

---

**Contributors:** Marusa Bradac (UC Davis), James Bullock (UCI), Masashi Chiba (Tohoku University), Ian Dell'Antonio (Brown University), Jarah Evslin (IHEP), Christopher Fassnacht (UC Davis), Philip Lubin (UCSB), Surhud More (IUCAA), Julio Navarro (University of Victoria), Masamune Oguri (University of Tokyo), Joel Primack (UCSC), Anjan Ananda Sen (Jamia Millia Islamia Central University), Tommaso Treu (UC Los Angeles), David Tytler (UCSD), Gillian Wilson (UC Riverside), Renxin Xu (Peking University), Hongsheng Zhao (University of St. Andrews), Gongbo Zhao (NAOC)

---

# 3 FUNDAMENTAL PHYSICS AND COSMOLOGY

One of the most exciting areas of current scientific research lies at the interface between theoretical and high-energy physics and cosmology. Indeed, the discovery of dark matter and dark energy result from astrophysical observations of the universe. It is these kinds of careful quantitative cosmological and astrophysical measurements that offer the greatest promise for constraining fundamental physical theories. TMT will contribute in this area by providing highly accurate measurements of the most distant sources, large samples of objects probing the universe at intermediate distances, and precise studies of the dark matter in both nearby and distant objects.

## 3.1 THE NATURE OF DARK MATTER

There is now incontrovertible observational evidence that the main mass component of the universe is in the form of *dark matter* that interacts extremely weakly, if at all, with atoms and photons. In the standard paradigm, non-interacting dark matter controls the clustering of dark matter, making it a tractable computational problem in which the initial conditions and the evolution equations are known once the nature of dark matter has been specified. Simulations can then be compared to the observed galaxy clustering pattern to shed light onto the properties of dark matter. This has led to a scenario where dark matter takes the form of elementary particles that emerge from the early universe with negligible thermal velocities and a scale-free distribution of Gaussian density fluctuations. Coupled with a "dark energy" field that governs the late expansion of the universe, this *Cold Dark Matter* (ΛCDM) paradigm has now matured into a full theory without tunable parameters, from which detailed theoretical predictions are possible. Progress to reconcile ΛCDM predictions and observations is happening, for example, until the late 2010s, ΛCDM models predicted a number of low-mass dark matter halos that far exceeded the observed number of faint galaxies but more recent models (Du et al. 2024; Birrer 2021; Oh, Nierenberg, Gilman & Birrer 2024) predict numbers of satellite galaxies that are much more in line with observations (e.g., Gilman et al. 2020; Zelko et al. 2022; Keeley et al. 2024) (see section 3.1.2) . These predictions have been successfully verified on large scales (>1 Mpc), but the situation is far less clear on the scale of individual galaxies, where the ΛCDM model is challenged by a number of problems. The rotation curves of dwarf galaxies, for example, seem at face value inconsistent with the cuspy dark matter halos predicted by cosmological simulations (sections 3.1.1 and 7.6.2). Although some see these challenges as signaling the need to abandon the paradigm, others argue that these observations can be accommodated within ΛCDM once the effects of baryons on the scale of dwarf galaxies are properly included. This is a complex problem that is far from solved, and best approached by following multiple lines of enquiry. Despite its complexity, progress has been steady and breakthroughs seem within reach. TMT can play a key role in enabling these breakthroughs by placing definitive constraints on the mass profile of dwarf galaxies and enabling the detection of dark substructure in gravitationally lensed-systems.

### 3.1.1 Dwarf galaxy radial mass profiles

One of the most robust predictions of cosmological N-body simulations is that, in the absence of baryonic effects, CDM halos must have cuspy density profiles. Rotation curve studies of low surface brightness galaxies have hinted that the actual mass profile might be shallower than predicted by simulations, fostering a lively debate regarding the meaning of these results and their consistency with ΛCDM. One possibility is that baryons may transfer energy to the dark matter as dense star-forming regions dissolve, flattening the cusps and turning them into "cores" more easily reconciled with observations (Navarro et al. 1996; Governato et al. 2010; reviewed by Pontzen and Governato 2014) although this solution is not fully accepted (Papastergis et al. 2016). This mechanism, however, can only operate in galaxies where baryons are relatively important. The dwarf spheroidal galaxies of the local group (dSphs) have measured line-of-sight velocity dispersions that are much larger than would be inferred from their stellar content. It is believed that these are the most dark-matter-dominated galactic systems. Establishing the presence or absence of cusps/cores in dSphs is thus a direct test of one of the most distinctive predictions of the ΛCDM scenario. In addition to the cusp/core problem for the dSphs and the ultra-faint dwarfs (UFD), there is evidence for lower central densities and smaller overall numbers of the largest satellite galaxies of the Milky Way (MW). This is the "Too big to Fail" problem (Boylan-Kolchin, Bullock, Kaplinghat 2011), which, in part, may derive from the limits of prior observational material (Verbeke et al. 2017). By measuring the distribution of dark matter in these nearby satellites, TMT will make a direct test of the nature of dark matter, an important problem in cosmology.

The best constraints on dark matter profiles in these ultra-faint dwarf galaxies are presently those based on large samples of radial velocities, see figure 3-1. These analyses, however, suffer from the well-known degeneracies associated with the unknown stellar velocity anisotropy distribution (Binney & Tremaine 1988).

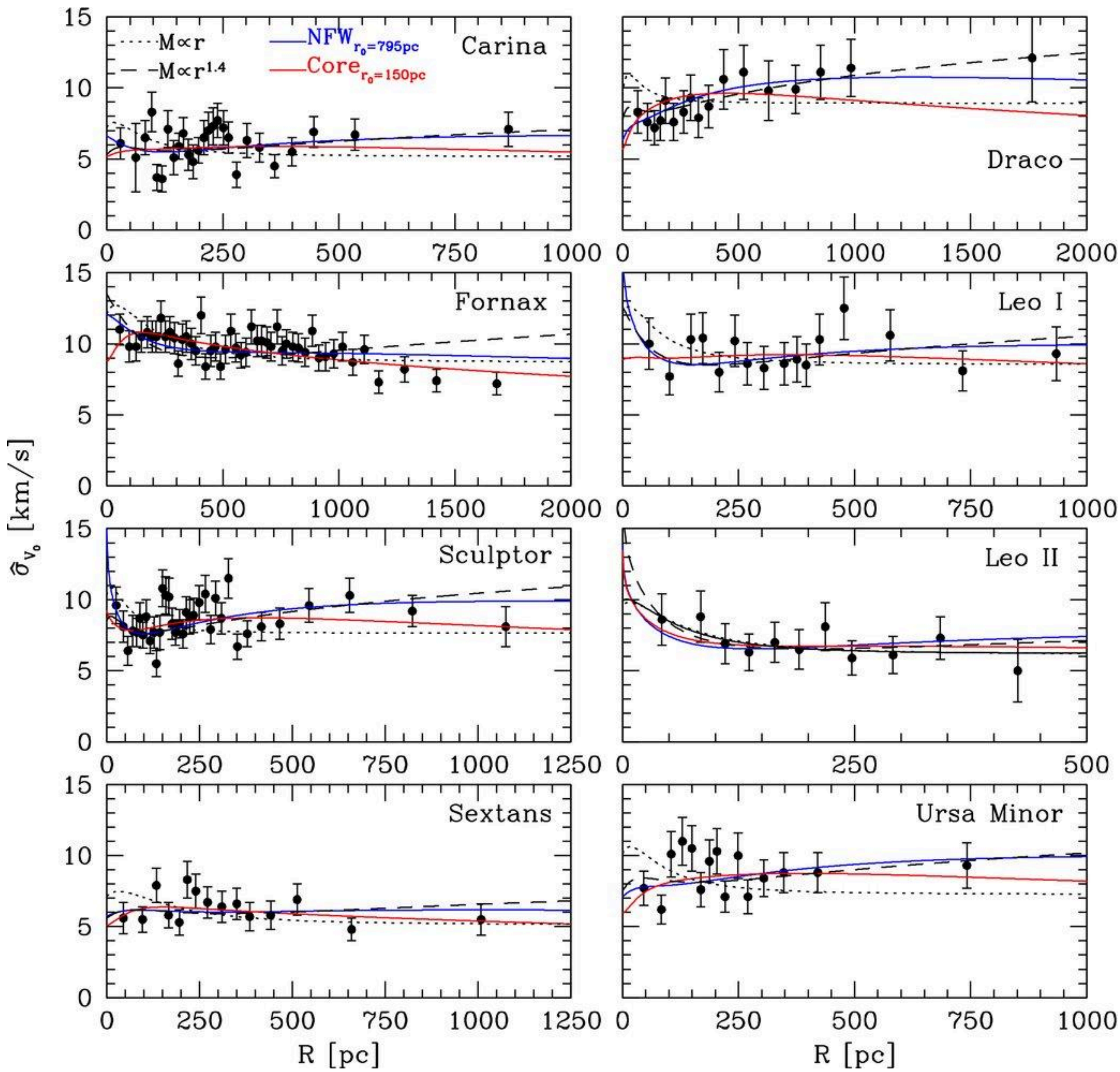

***Figure 3-1:*** *Measured line-of-sight velocity dispersion profiles in the Milky Way dSphs (Walker et al. 2009)*

If radial velocities alone are to be used, greatly increasing the sample of stars is essential to break the degeneracies. Figure 3-2 shows how the error on the logarithmic slope of the mass profile, γ, will be reduced as a function of the number of stars in a kinematic sample. Existing and imminent spectrographs such as Subaru's PFS will be able to efficiently measure radial velocities of hundreds of stars in these dwarfs. However, obtaining <5 km/s precision radial velocities for >10,000 stars will require reaching the main sequence in these dwarf galaxies. For Draco, that means g~23 (Strigari et al. 2007), which will be very challenging for telescopes in the pre-TMT era. Given the relatively small core radius of these dwarf galaxies (~6' for Draco), TMT/WFOS will be an efficient instrument, and should allow us to greatly improve the constraints on the innermost profile of these dwarf galaxy halos (see Hayashi & Chiba, 2012).

In addition to providing radial velocities for thousands of stars in multiple dwarf galaxies, TMT can also help to resolve the anisotropy degeneracies by providing proper motions (Massari et al. 2018). Strigari et al. (2007) have shown that if proper motion data could be obtained with ~ 5 km/s accuracy for just a few hundred stars per galaxy, the accuracy of current kinematic studies would be increased five fold. This level of accuracy is within the reach of TMT's IRIS spectro-imaging near-infrared camera with the NFIRAOS adaptive optics system. Relative astrometric measurements with the Keck AO system already reach the ~150 μas level. IRIS is designed to obtain an accuracy of ~30 μas at comparable signal to noise ratios. Thus, the relative positions of two measurements relative to a reference can be determined to ~40 mas per star. Over a three-year baseline this corresponds to a kinematic accuracy per star of ~7 km/s. Doubling the time baseline would allow IRIS to provide measurements of proper motion for stars brighter than K~18.5 with a kinematic accuracy better than 4 km/s.

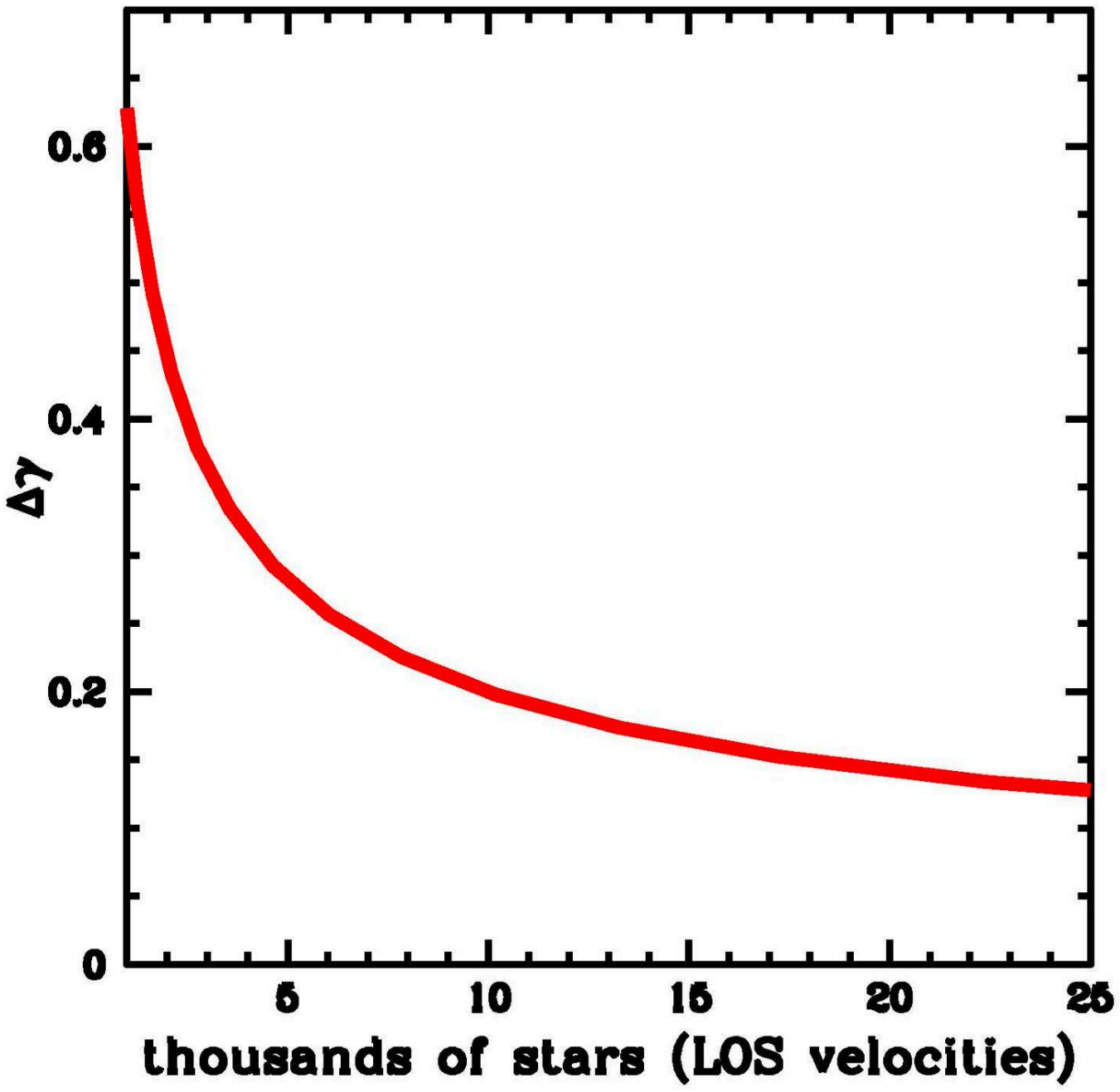

***Figure 3-2:*** *Statistical uncertainty on the logarithmic-slope of the density at the half-light radius (400 pc) of a Draco-like NFW dark matter density profile as a function of the number of stars observed. This is based on a Jeans analysis of the line-of-sight velocity dispersion, assuming 1km/s accuracy. The Fisher matrix yields its uncertainty in a fit to a 5-parameter family of profiles. In combination with the mass within the half-light radius, in the absence of baryonic effects this measurement provides a powerful test of ΛCDM vs alternative dark matter models. Figure credit: J. Bullock (UCI).*

Furthermore, for red giants brighter than V=18, corresponding to a G magnitude of 17.6, the GAIA mission is achieving a relative astrometric accuracy of ~80 μas in Data Release 4, based on their revised specifications and is expected to achieve better than 60 μas in DR5. The Sculptor dwarf contains more than 100 such stars; combining such a GAIA and TMT/IRIS position measurement separated by ~10 years would produce proper motions with sufficient accuracy to confidently distinguish cusp from core with a signal-to-noise ratio of four (Evslin 2015). A further TMT position measurement, several years later, could be used to measure velocity changes and thus identify binary systems with periods of years.

Such binaries can contaminate the velocity dispersions of ultra-faint dwarfs. The synergy with GAIA yields a competitive proper motion determination even in the case of the brighter but more distant Fornax dwarf, whose ecliptic latitude will provide it with more than twice the sky-averaged number of focal plane transits. TMT will be able to definitively distinguish power-law cusps from constant density cores in many of the Milky Way satellite galaxies.

Measuring accurate astrometry for stars covering the full tidal radius of the dwarf galaxies will not be easy, as it will require AO-assisted imaging with NFIRAOS-IRIS to get the required astrometric precision over a very large area.

The tidal radius of the Sculptor dwarf is nearly 75'.

Even with short (10 second) exposures (sufficient to reach K=22), fully mapping the proper motions across the entire tidal radius will take some 800 hours (40 nights per epoch for two epochs) of telescope time, assuming 100% overhead for the dithering. Most other dwarfs are smaller on the sky, and their motions can be mapped with <20 nights each of observing divided into two campaigns separated by multiple years. It will be important to observe both larger dSphs and UFDs, to control whether baryonic effects can affect the cuspiness of the mass profile (Onorbe et al. 2015, Cooke et al. 2022).

### 3.1.2 Dark Matter Substructure

Simulations of galaxy formation using ΛCDM under the standard model reveal that massive galaxy halos should be associated with numerous small subhalos (a.k.a. substructure). On these sub-galactic scales, however, while the previous discrepancy between the theory and observations of satellite galaxies (Boylan-Kolchin, Bullock, Kaplinghat 2011) appears to be almost reconciled (Tomozeiu, Mayer & Quinn 2016), there is still much to be done. Simulations predict that there should be significantly more substructures with masses between $4 \times 10^6$ and $4 \times 10^9$ solar masses than are observed Observations of the Milky Way, where the most complete investigations of substructure have been conducted, have found slightly fewer satellites than predicted.

This used to be known as the *missing satellites problem* on the low-mass end (e.g., Moore et al. 1999; Klypin et al. 1999) and a slight tension with excess satellites on the high-mass end of the mass function (e.g., Strigari et al. 2007) until, with improved modeling, the discrepancy has reduced but not completely.

Possible explanations for this discrepancy are (1) some type of astrophysical process (e.g., differences in star-formation efficiencies, incomplete modeling of dark matter halo formation and survival rates, e.g., Aguirre-Santaella 2023), (2) the Local Group is an outlier, (3) we have an incomplete understanding of the nature of dark matter (e.g.,Cerny et al., 2023), or (4) observational detections of substructures are significantly incomplete. To distinguish between these hypotheses requires a large sample of galaxies in which we can detect even non-luminous and ultra-faint substructures. The combination of gravitational lensing with high-resolution observations by the TMT will provide exactly such a sample and will do so via two independent techniques.

Figure 3-3 shows the Keck/NIRC2 imaging of the system, with the lensing galaxy masked out, and the enhanced convergence required to fit the data. The peak in the convergence map corresponds to the location of the substructure in this system.

**Flux ratio anomalies in lensed AGN.** The image configuration and predicted fluxes of lensed AGN are set by the smooth mass distribution in the main halo of the lensing galaxy. Deviations from the flux ratios predicted by the smooth mass models, i.e., flux ratio anomalies, are a sign of lumpiness in the mass distribution of the lensing galaxy (Richardson et al 2022). This lumpiness may be due to stars in the lens (microlensing), the dominant contaminant in searches for dark matter substructure. However, microlensing can be avoided if the angular size of the lensed object is larger than the lensing scale of the stars, which happens if the lensed emission is coming from a region larger than the broad line region in the background AGN. A promising approach utilizes the narrow-line region, which is large enough to avoid microlensing.

These observations require sensitive and high-resolution IFU observations with IRIS to obtain high-SNR spectroscopy of each of the lensed images that are typically separated by an arcsecond or less. This method was originally proposed by Moustakas & Metcalf (2003), and has seen a successful application by Nierenberg et al. (2014, 2017, 2020). The large gains in both sensitivity and angular resolution provided by TMT will allow this approach to be applied to a statistically significant sample of lens systems (Gilman et al. 2020). Alternatively, the dusty torus that is seen in the mid-IR can also provide a measurement of flux ratios that is immune from microlensing and dust. Current observations of lensed QSOs have found closely separated images (~0.3"), too faint and difficult to resolve with existing 8 m telescopes. JWST will be much more sensitive but its resolution will be significantly less than that of TMT. Diffraction-limited observations using TMT-IRIS will be essential to separate and distinguish these images, especially for the (expected) large number (100's) of small-separation lenses (<0.3") (Zelko, Nirenberg & Treu 2024).

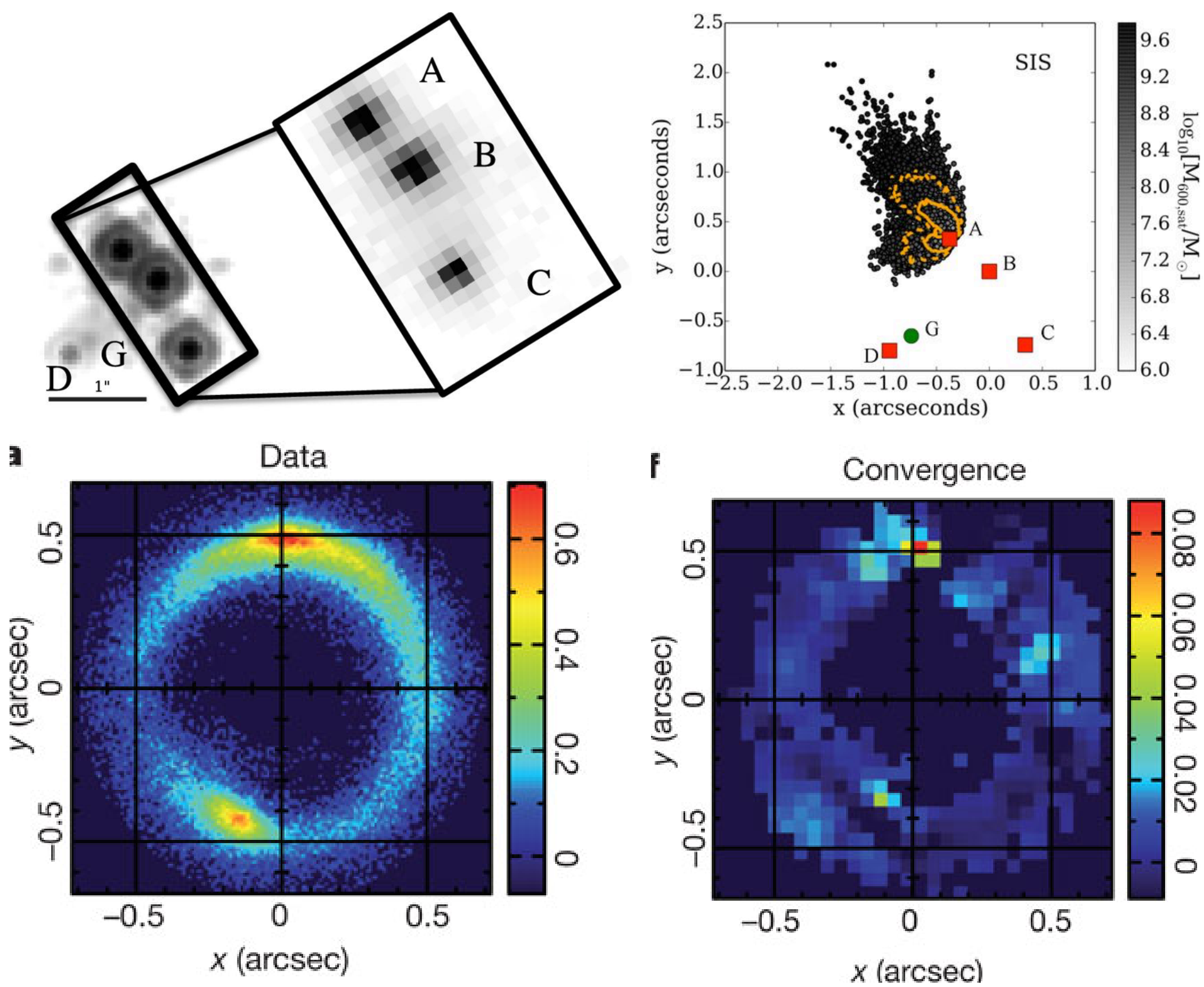

***Figure 3-3:*** *Demonstration of the detection of (dark) substructure via gravitational lensing and Keck adaptive optics imaging. (Top two panels) Substructure detection via flux ratio anomalies in the B1422+231 lens system. Figures show HST (small rectangle) and Keck/OSIRIS imaging (large rectangle) and the posterior probability for the location and mass of the substructure under the assumption of a single perturbing mass and a singular isothermal sphere profile. Figures from Nierenberg et al. (2014). (Bottom two panels) Substructure detection via gravitational imaging in the B1938+666 system. Figures from Vegetti et al. (2012).*

**Gravitational imaging.** Gravitational lensing plus sensitive high-resolution imaging provides a second path to quantifying the substructure mass function in distant galaxies. In the "gravitational imaging" technique (Koopmans 2005) the targeted systems are not lensed AGN, but rather systems in which a background galaxy is lensed into a long arc or a ring by a massive foreground galaxy. In these systems, the radius and overall appearance of the arc is set by the smooth mass distribution in the lensing galaxy. However, the presence of dark-matter substructure can perturb the shape of the ring on small angular scales. These small astrometric perturbations to the lensed emission allow the presence of substructures to be deduced, even with no detectable luminous component. An excellent demonstration of this technique with ground-based instrumentation resulted from NIR AO observations with Keck. A $10^8$ solar mass substructure in a z~0.9 galaxy was robustly detected in three independent data sets (Vegetti et al. 2012), and another detection of substructure with of a few of $10^9$ solar mass substructure has been detected with Hubble imaging (Vegetti, et al. 2010 ).

The substructure was detected at higher significance in the Keck AO data than in HST imaging of the same system, due to the higher angular resolution of the Keck observations, see figure 3-3. With its significant improvements compared to Keck in collecting area, three times better angular resolution than Keck and five times better than JWST, and number of systems accessible with AO, the TMT will move this technique into the regime where interesting constraints can be placed on the substructure mass functions and, furthermore, will allow these parameters to be determined as functions of galaxy mass, redshift, and other observables (Li et al. 2016).

**Combining flux ratio anomalies and gravitational imaging.** Recent developments in lensing techniques have shown that further progress can be made by combining the information from the flux ratios of the lensed quasars with that from the distortion of their host galaxy (Birrer 2021; Oh et al. 2024). This combined approach enables constraints that are more stringent than those obtained with each technique separately. It requires exquisite spatial resolution and control of the point spread function, and it is thus uniquely suited for TMT with the NFIRAOS-IRIS imager.

### 3.1.3 Dark Matter self-interaction cross-section

Many dark matter experiments are dedicated to searching for couplings of the dark matter particle to the Standard Model and can be classified into direct detection dark matter experiments (e.g., CDMS, LUX, etc), colliders (LHC) and annihilation experiments (FERMI). If, however, dark matter does not couple (or does so extremely weakly) to the Standard Model these searches will only allow us to rule out parts of a parameter space. Astrophysical experiments, however, can probe interactions between two dark matter particles via their effects on the structures and dynamics of dark-matter halos.

Interacting clusters have proven very fruitful for constraining the nature of dark matter (e.g., Andrade et al. 2022). In clusters with recent strong merger activity, the positions of dark matter halos and the main baryonic component may become separated if there is self-interaction between dark matter particles. The separation between the X-ray emitting gas and the dark matter provides a powerful upper limit on the dark matter self-interaction cross section, $\sigma_m$. Andrade et al. (2022) suggest that such measurements are consistent with no dark matter self-interaction. This measurement is controversial, and the significance depends on very accurate determination of the position of the weak lensing.

Early limits from studies of the Bullet Cluster (Randall et al. 2008), MACS J0025.4-1222 (Bradac et al. 2008), and DLSCLJ0916.2+2951 (Dawson et al. 2012) require $\sigma_m < 0.7$ cm$^2$g$^{-1}$. However, they leave a considerable room for improvement. The two main observational challenges to measure both gas-DM and galaxy-DM offsets are accurate measurements of the weak lensing centroid as well as accurate determination of cluster membership and hence galaxy centroid. Even more importantly, these studies all require detailed dynamical information of the system to be included in simulations, and only TMT-like light-gathering power can provide this for a large number of faint cluster members. Furthermore, because the weak lensing centroiding accuracy is a very strong function of the number density of lensed galaxies used, this measurement requires greater depth and resolution than HST or JWST will provide, meaning that TMT will be required if the issue of a dark matter self-interaction cross-section is to be resolved.

### 3.1.4 Baryonic power spectrum

The power spectrum of density fluctuations on small scales provides the strongest constraint on the fraction of hot and warm dark matter in the universe (Hu et al. 1998) as free streaming by this matter reduces the fluctuation power on small scales. Power spectrum data can also provide constraints on a wide variety of potential low-mass particles such as sterile neutrinos, axions and thermal relics (Narayan et al. 2000, Hannestad & Raffelt 2004, Viel et al. 2005, Iršič et al. 2024).

Studies of the distribution of matter can also tell us details about the early inflationary period of the universe. The power spectrum of density fluctuations is predicted by theory to have arisen from a "scale invariant" primordial spectrum that is a power law with a near unity slope. Small deviations from this unity slope are expected and are related to the degree by which the universe expanded during the period of inflation and hence to the physical properties of the fields that drove inflation. By measuring the distribution of matter over a wide range of scales, one can constrain both the slope and possible deviations with high precision. Measurement of the power spectrum on large scales is limited by "cosmic variance" (we have access to only one universe) meaning the most precise measurements will necessarily come from measures on small physical scales.

In the present-day universe, the memory of the initial conditions on small scales has been erased by the formation of complex structures such as galaxies. However, by studying the details of the distribution of diffuse hydrogen gas in the intergalactic medium at high redshift (figure 3-4), it is possible to measure the "primordial" power spectrum of all matter. This can be achieved by studying absorption line systems along the sightlines of many different sources.

Numerical simulations over the past decade (e.g., Viel et al. 2013) have further refined the power of IGM power spectrum measurements, allowing constraints to be determined even on scales that are somewhat non-linear.

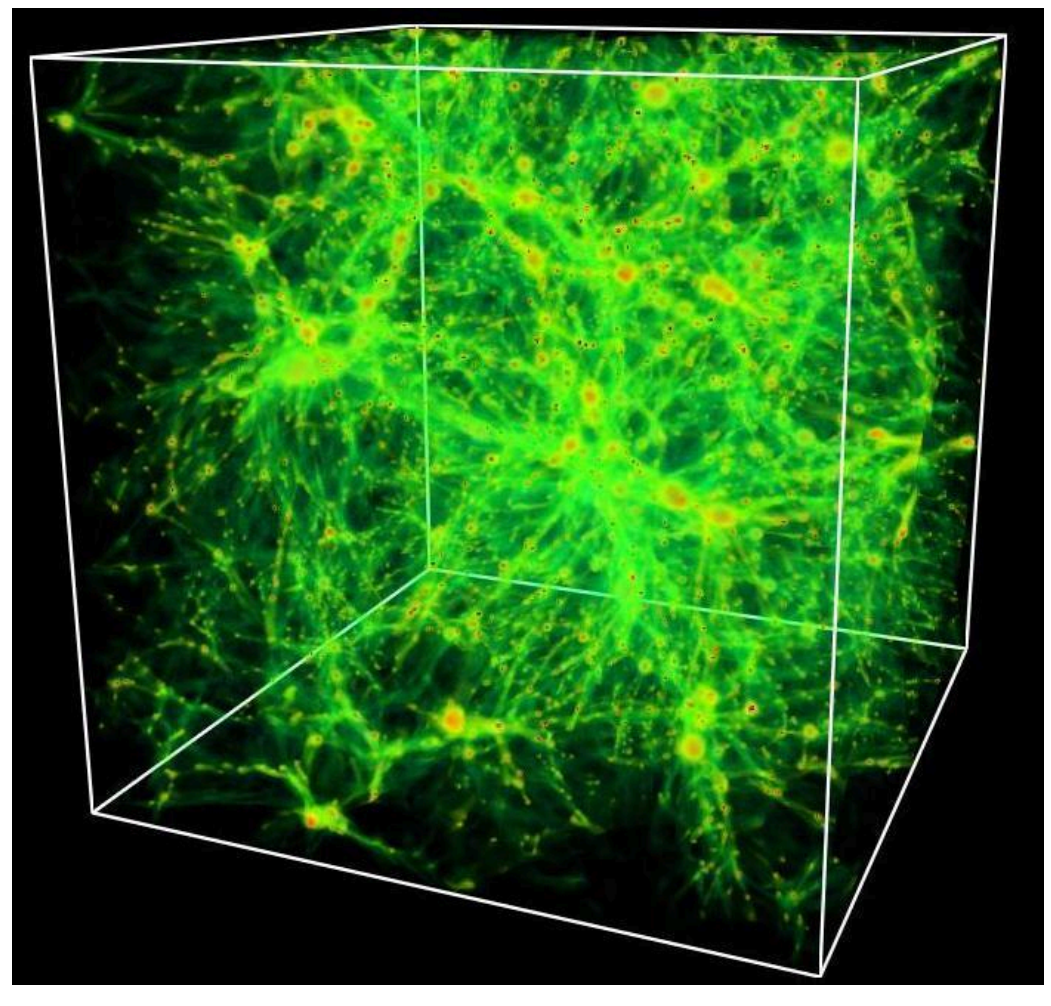

***Figure 3-4:*** *Gas tracing the dark matter structure at z ~ 3. TM probe multiple lines of sight using WFOS and background galax sources. This will provide a more complete understanding of the ba power spectrum and interaction of the IGM with baryonic mat galaxies. (Credit: R. Cen, Princeton U.)*

The technique of measuring the power spectrum from absorption line systems has been developed by the SDSS and BOSS (McDonald et al. 2006, Busca et al., 2013, Slozar et al. 2013) using quasar spectra at fairly low resolution (R ~ 2000) supplemented with smaller numbers of high-resolution spectra taken with Keck, Magellan and VLT (Viel et al. 2006, Becker et al. 2011). On large scales, the most sensitive sample to date is the one studied by the BOSS survey (Lee et al. 2013), which uses almost 55,000 quasars between $2.1<z<3.5$. Power spectrum measurements (combined with CMB constraints from WMAP and Planck) have placed upper mass limits of 0.23 eV for the sum of the masses of the standard model neutrinos (see Abazajian et al., 2015 for a discussion). Higher resolution and higher redshift observations (Viel et al. 2013) have placed the current lower limit of the warm dark matter mass $m_{wm}>3.3$ keV. The Dark Energy Spectroscopic Investigation (DESI) is increasing the sample of absorption line systems and the resolution on the power spectrum by a factor of >5 (Ravoux et al. 2023, Adame et al. 2024a). With TMT, the number of QSOs in the correct redshift range will increase by about an additional order of magnitude.

Absorption line studies using quasars are fundamentally limited by their angular density. Fainter but more abundant star forming Lyman-Break galaxies are being used in a limited way in the CLAMATO survey out to z~2-3 (Horowitz et al. 2022). However, with TMT it will be possible to use the fainter but much more abundant "normal" galaxies at even higher redshifts. By using galaxies as probes instead of quasars, TMT will allow for the first time the study of the three-dimensional distribution of diffuse hydrogen in the intergalactic medium, which is directly related to matter density, thereby increasing the precision of cosmological measurements by at least an order of magnitude over that possible with current telescopes. Here the distribution of matter is affected by complex physical processes, such as gas dynamics, star formation and feedback. TMT, using WFOS and IRMOS, will probe the distribution and composition of gas along the lines of sight to distant quasars and galaxies, providing unique high-quality data essential to an understanding of these processes.

## 3.2 COSMOLOGICAL PARAMETERS

### 3.2.1 Dark Energy

#### 3.2.1.1 Dark energy and modified gravity

Understanding the acceleration of the universe (Riess et al. 1998, Perlmutter et al. 1999, Eisenstein et al. 2005, Vikhlinin et al. 2009) is arguably the greatest challenge facing cosmology. In Einstein's theory of gravity, an accelerating expansion requires that the average pressure throughout space be negative. Cosmological fluids with negative pressure are called dark energy (DE). The simplest implementation of DE is a cosmological constant, which is incorporated in the concordance ΛCDM model. However, ΛCDM on its own provides no explanation for the smallness of the cosmological constant relative to most scales in the theory. Observations to date are consistent with a cosmological constant but also with several distinct dark energy models. The most common way to describe any dark energy candidate is through its equation of state parameter $w = p/\rho$, the ratio of its pressure (p) to its energy density (ρ). w=−1 is for an Einstein cosmological constant and for anything other than this, dark energy evolves with time.

Recent results from DESI (DESI collaboration; Calderon et al. 2024) suggest the exciting possibility that w may not be −1, thus making the case for pursuing this matter even more compelling.

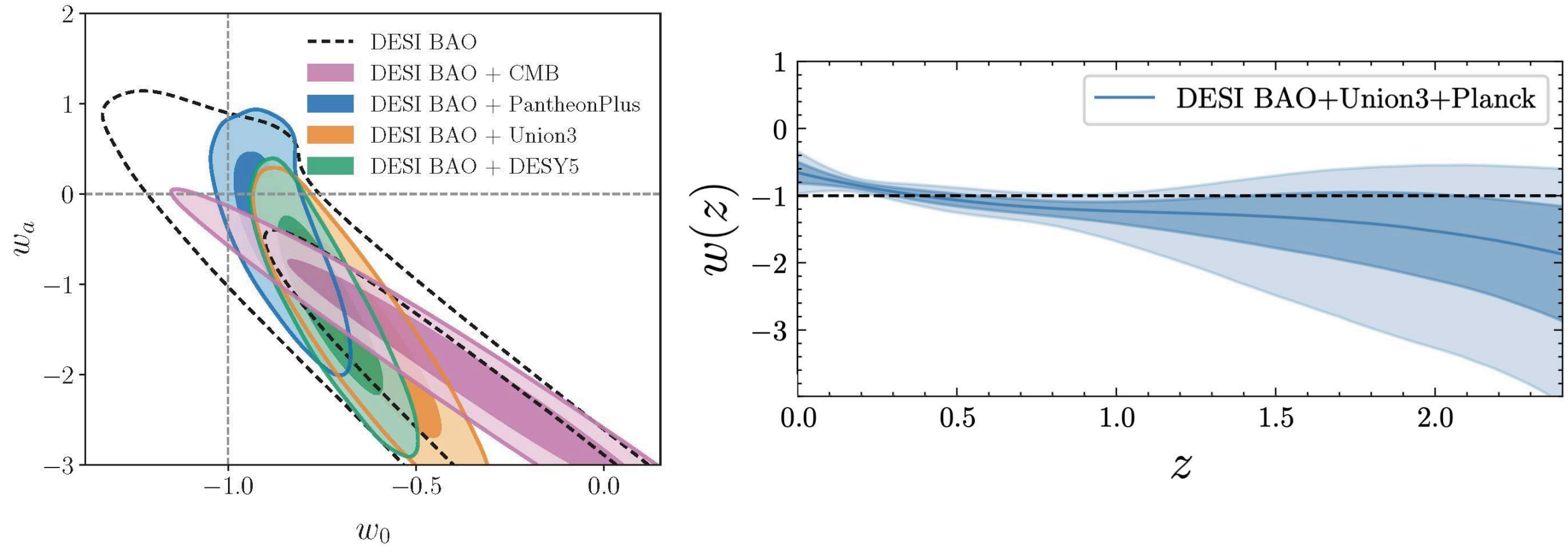

***Figure 3-5:*** *Left: Constraints in the w0–wa plane for the flat w0waCDM model, from different observations (figure 6 left from Adame et al. 2024b). Right: Dark energy reconstructions using the Chebyshev expansions of w(z) (top right of figure 1 of Calderon et al. 2024).*

Constraining w and its variation with time is one of the main goals for all present and future cosmological observations including TMT. For that, the CPL parametrization $w = w_0 + w_a(1 - a)$ by Chevalier & Polarski (2001) and Linder (2003) is the widely used parameterization for the equation of state. Present bounds on the evolution of w are shown in figure 3-5.

Another approach to accommodating the late time acceleration is to modify gravity at large cosmological scales without including exotic components in the energy budget of the universe (Dvali et al. 2000, Freese et al. 2002, De Felice & Tsujikawa 2010, Amendola & Tsujikawa 2010) and additional creative solutions are available (e.g., Anari et al. 2022). Often these models make predictions equivalent to those of standard dark energy models. TMT will differentiate and constrain these models through the observational tests described below. Additional tests of GR involving Super Massive Black Holes are described in section 6.1.1.

Measuring the properties of dark energy and testing alternate gravity theories is the goal of many experiments and observatories that will be operational at the same time as TMT (e.g., Rubin Observatory, Euclid, Roman). These missions will carry out large-scale surveys in order to exploit popular dark energy diagnostics such as baryonic acoustic oscillations or cosmic shear. TMT can be a competitive probe of dark energy and modified gravity by focusing on diagnostics that require its unique combination of sensitivity and angular resolution, like the ones described below (see also Jain et al. 2013). By measuring the anisotropy in $w_0$ and $w_a$ (e.g. with time delays), TMT may also be able to place competitive constraints on the speed of sound of the dark energy fluid.

Gravitational time delays between multiple images of lensed quasars and supernovae have emerged as powerful probes of dark energy (Wong et al 2020; Kelly et al 2023). A key requirement is the need to image the lensed host galaxy at sufficient resolution to chart the gravitational potential of the lens. In the next few decades, thousands of lensed quasars will be discovered and monitored by upcoming panoramic imaging surveys (Treu, Suyu & Marshall 2022), although identification is difficult without high spatial resolution AO observations, and high-resolution adaptive optics imaging will be used to identify and model selected systems and thereby provide absolute distances to <1% precision. Furthermore, optical or near-IR spectroscopy with resolutions of R>2500 are needed to measure the redshift of the lensed quasars and the redshift and internal kinematics of the deflector.

#### 3.2.1.2 Time-delay cosmography

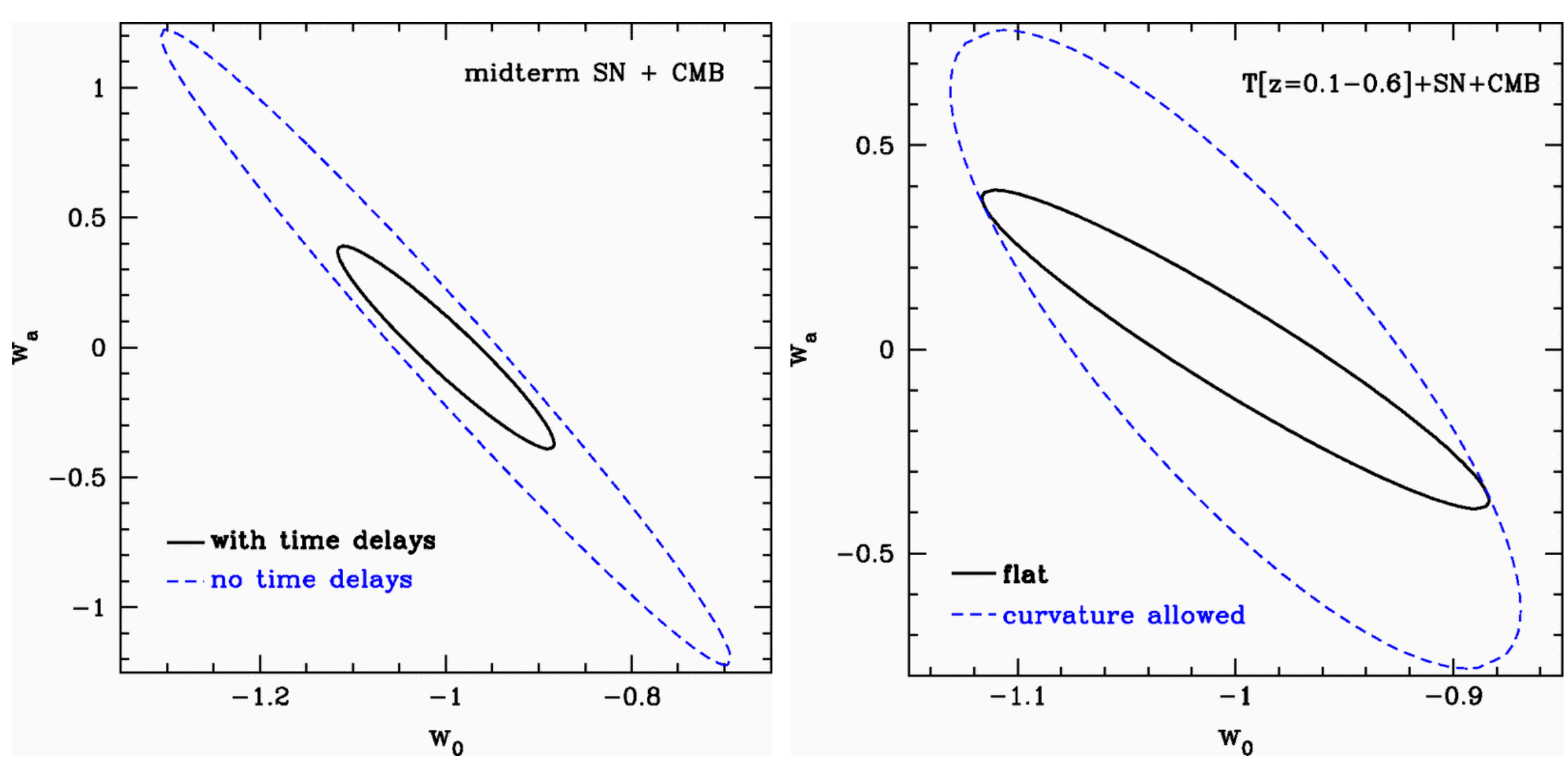

***Figure 3-6:*** *Forecasts for a time-delay experiment (Linder 2011) based on 150 time-delay distances. Top: 68% confidence level constraints on the dark energy equation of state parameters $w_0$ and $w_a$ using midterm supernova distances and CMB information with or without time delay measurements. Bottom: Similarly, 68% confidence level constraints using time delay, midterm supernovae, and CMB information, assuming spatial flatness or allowing curvature. In both panes the time delay probe demonstrates strong complementarity, tightening the area of uncertainty by a factor 4.8 and 20 respectively.*

The latter is vital to break the mass sheet degeneracy, the most fundamental limitation of the method (Birrer et al. 2020; Birrer & Treu 2021; Shajib et al 2023). The angular diameter of the gravitational lens systems is of order 3–4”, well matched to the capabilities of IRIS. The method is particularly powerful when combined with microwave background and Type Ia supernovae experiments, as it breaks many of the residual degeneracies (e.g., between curvature and *w*) as shown in figure 3-6. At the moment, this method is limited by the small number of known suitable systems. However, more than an order of magnitude increase in the number of known lenses with time delays is expected with Rubin Observatory (Oguri & Marshall 2010; Treu, Suyu & Marshall 2022).

Clearly, the cosmological parameters will be far more highly constrained with a larger number of systems (e.g., 1000; Shiralilou et al. 2020). Rubin Observatory will also find strongly lensed type-Ia supernovae, for which we can directly measure the magnification factor to further constrain the gravitational potential (Oguri & Kawano 2003; Suyu et al 2020), see also section 9.1.1. JWST observations of the multiply imaged supernova H0pe (Frye et al. 2024) clearly demonstrate the potential of using time delays measured photometrically and with time resolved spectroscopy. High resolution imaging and spectroscopy with TMT will transform these into high-precision cosmological probes.

#### 3.2.1.3 Redshift Drift

The evolution of the Hubble expansion causes the redshifts of cosmologically distant objects subject to the Hubble flow to change slowly with time (Sandage 1962), i.e. redshift drift. The ΛCDM model predicts a change in velocity of $\Delta v \sim 3$ cm $s^{-1}$ over a period of 10 years at $z \sim 3$. The Lyman-α forest absorption lines seen in the spectra of distant z=2–4 QSOs are ideal targets (Loeb 1998). Gathering R>50,000 optical spectra for multiple sightlines to QSOs over a 20 year timebase would allow the redshift drift to be measured.

However, this direct method could be modified to an effectively instantaneous method using multiply imaged QSOs with time delays of years caused by different sightlines (Helbig 2023). Instead of separating the redshift measurements by n years, the time delays of n years would be equivalent. Such observations would be similar, though with much higher spectral resolution than those described in section 3.1.2.

### 3.2.2 The Hubble Tension

Over the past decade, direct "late universe" measurements of the Hubble constant ($H_0$) with increasing accuracy and precision (Riess et al. 2022, Wong et al. 2020) have resulted in a growing discrepancy (in terms of statistical significance) with the expected value based on "early universe" observations of the Cosmic Microwave Background under ΛCDM (Planck Collaboration, 2020). This so-called *Hubble Tension* now exceeds 5σ for several combinations of early- and late-universe methods, leading to many intriguing possibilities of physics beyond the standard model (see review by Abdalla et al. 2022 and figure 3-7).

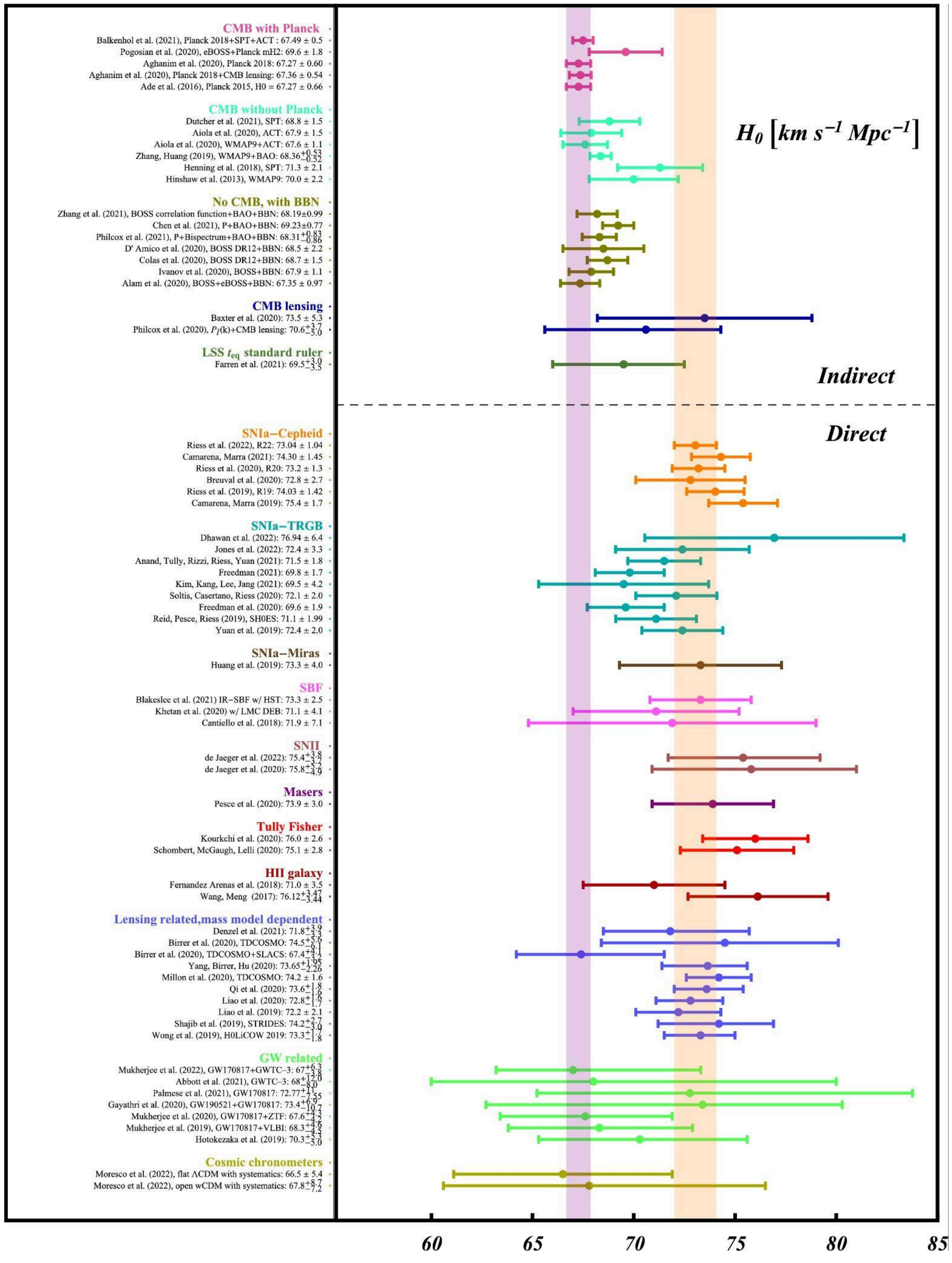

***Figure 3-7:*** *Compilation of early- and late-universe determinations of $H_0$ from Abdalla et al. (2022), highlighting the 5σ tension between the two epochs*

TMT will make significant contributions to this puzzling cosmological discrepancy in the post-JWST era by helping improve late-universe measurements of $H_0$, as described in section 9.12 and briefly summarized here. AO imaging will enable the discovery of Cepheids out to 100 Mpc, providing the required accurate near-infrared photometry to determine percent-level distances to their host galaxies. This will lead to the improved calibration of secondary distance indicators such as type Ia supernovae (SNe Ia).

Furthermore, spatially-resolved stellar kinematics of strong lenses using IFUs will help resolve the mass-sheet degeneracy that affects time-delay determinations of $H_0$ (Yıldırım et al. 2023; Shajib et al. 2023) using quasars and SNe Ia. Additionally, prompt imaging and spectroscopy of kilonovae with TMT will provide a complementary route to $H_0$ that is independent of the extragalactic distance ladder (Coughlin et al. 2020).

### 3.2.3 Cosmology from clusters of Galaxies

Clusters of galaxies are a very important laboratory for cosmology and fundamental physics, providing multiple probes into dark energy and its evolution, as well as the nature of dark matter and its interaction with baryons. Critical spectroscopic and astrometric measurements on lensed galaxies are only possible with the resolving power and sensitivity of an AO-fed thirty-meter class telescope.

#### 3.2.3.1 Strong and Weak-lensing tomography

The deflection of light due to gravitational lensing depends on the angular diameter distance ratios between the source, lens and observer. For the same lens, measuring the increase in lensing effect with source redshift provides a cosmographic measurement, whether the effect is weak lensing shear (using cross-correlation tomography (Jain & Taylor 2003), or the position of strong lensing arcs from background objects with different redshifts (Golse et al. 2002). In practice, the constraint increases very strongly with the number of arcs per cluster and the number of clusters (Jullo et al. 2010) as long as the large-scale structure along the line of sight is known (Dalal et al. 2004). A large number of multiple lensed objects have been discovered in the Hubble frontier fields (Jauzac et al. 2015), which can be combined with those discovered with IFU systems such as MUSE (Bergamini et al. 2023). Because many arcs are observed to have characteristic widths below 0.1", AO-based imaging is required to resolve and separate them from the background sky. At the same time, spectroscopic measurements required to securely establish arc redshifts and to identify structure along the line of sight require the aperture of TMT. For clusters with $z>0.5$, the entire strongly lensed region spans a <30" diameter region for all but the most massive clusters (Richard et al. 2014), well-matched to the planned 34"x34" FOV of the IRIS imager. Secure redshifts will be available through the IRIS IFU. Based on the HST WFC3 and JWST NIRCam imaging of the Frontier Fields, TMT will be able to observe up to hundreds of multiple images per cluster field (Jauzac et al. 2014; Bergamini et al. 2023) at a resolution (at 1μm) 5 times that of JWST's NIRCam. For higher redshift clusters, weak lensing tomography using the higher resolution of IRIS will allow detection and deblending of the background galaxies with higher efficiency than even JWST. In addition to tomography, this will enable mass profile measurement in the weak lensing regime, extending to scales >300 kpc.

#### 3.2.3.2 Masses for High Redshift Clusters

In hierarchical cluster formation scenarios, formation of clusters of galaxies occurs late in cosmic time. Therefore, detection of high mass systems at early redshifts poses a sensitive constraint on cosmological models and may point to primordial non-Gaussian forms or complex modifications of gravity. Because the mass function is so steep, the cosmological constraining power depends critically on the ability to measure the cluster mass. Clusters at $z>1.5$ are unlikely to be virialized, and masses derived from scaling relations from the X-ray or optical richness are unlikely to be well-calibrated at that distance. The two techniques that might have promise in determining the mass of high-z clusters both require the unique capability of a 30 m class aperture with both wide-field spectroscopic and high-resolution imaging capabilities. For weak lensing, the primary difficulties are measuring shapes for enough galaxies behind the clusters to adequately measure the shear and having sufficiently accurate redshifts to remove the huge foreground contribution (both to the galaxies and the shear). For galaxies at $z<1.5$, shear can be measured with HST or JWST (Kim et al. 2019; Schrabback et al. 2021), but for $z>>1$, higher spatial resolution integral field spectroscopy is required in order to enable deblending and separate identification of lensed sources. This requires a combination of deep optical and infrared spectroscopy (for infall measurements and foreground structure determination) and AO imaging (for shape measurements and redshift measurement of small background galaxies).

#### 3.2.3.3 Cluster abundance and mass function: Calibrating the Mass-Observable Relations

Cosmological constraints from clusters also derive from the evolution of the cluster mass function as a whole between $0<z<1.5$. Although TMT is not a survey machine and will not detect or measure masses for the majority of clusters for this measurement, it will play a critical role in many of the steps in the calibration of the mass measurements. For example, the contribution to the uncertainty in the calibration of the mass-temperature relation from X-ray observations that arises from uncertainty in assigning photometric redshifts for galaxies behind clusters of different redshifts is already comparable to the statistical errors (Finoguenov, Reiprich & Böhringer 2001). This uncertainty is expected to dominate the error in the era of large surveys such as eRosita, EUCLID, Roman and Rubin. In addition, measurement of the mass-observable relation at fixed redshift using weak lensing requires accurate measurements of shear. Defining a “golden set” of clusters with very accurate lensing measurements and especially with a dense and reliable set of redshift measurements for background galaxies will be key in deriving cosmological constraints from the evolution of the mass function. The angular sizes of the cluster cores are well-matched to the FOV of TMT’s AO system and tiling of observations can address limitations on the mass reconstruction due to mass-sheet degeneracy (Bradač, Lombardi & Schneider 2004).

### 3.2.4 Tests of general relativity

As described in chapter 6, *Supermassive Black Holes*, high precision determinations of the velocity and position of stars orbiting the supermassive black hole at the center of the Milky Way provide a stringent test of General Relativity (GR, e.g., Angelil & Saha 2011). Complementary tests can be performed on the scale of an entire galaxy using gravitational redshift measurements (Wojtak, Hansen & Hjorth 2011; Jimeno et al. 2015). Even for a face-on galaxy at a fixed distance, the apparent Doppler velocity is not constant due to GR effects. The light from the central part of the galaxy will be more redshifted compared to the outer part. This is because light from the inner part has to climb out of a deeper gravitational potential. There is also the Transverse Doppler Effect that makes the kinematically-hotter center be even more red-shifted. Both redshift effects are small (on the order of 0.1–1 km/s). By stacking high resolution spectra of many spatially resolved galaxies, however, the GR signal will be detectable by TMT (Zhao H., et al., 2013).

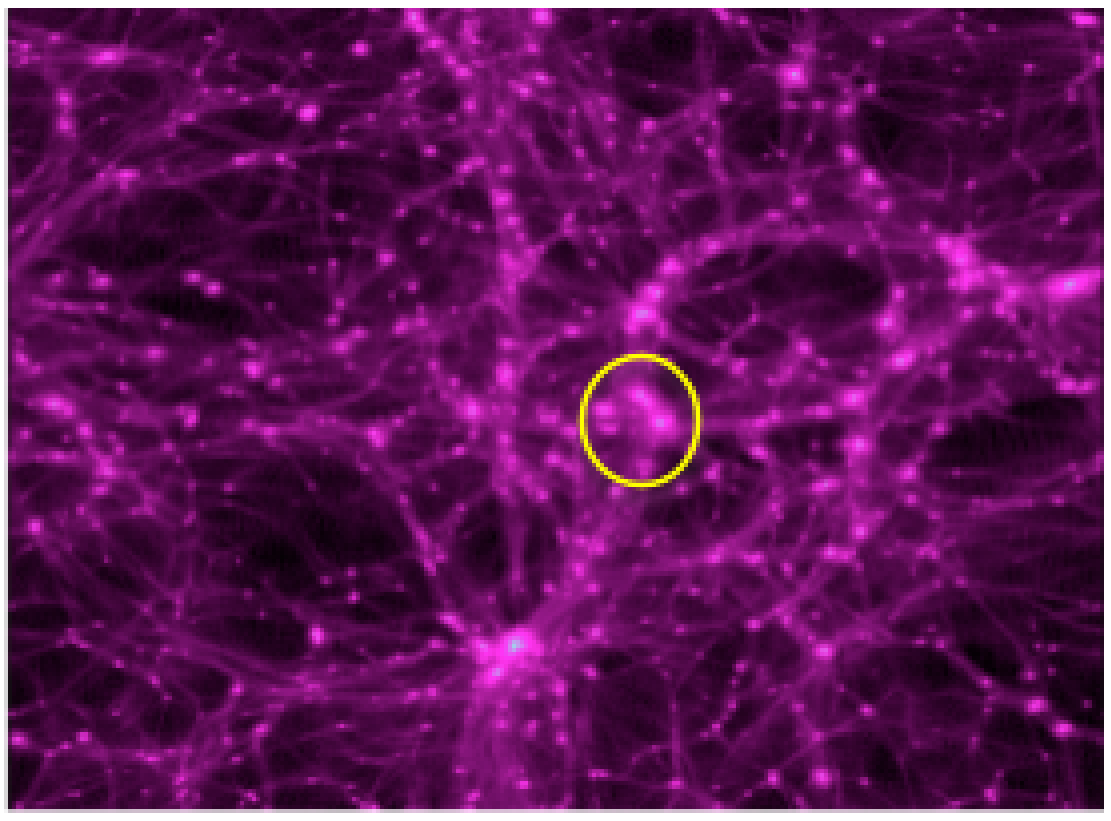

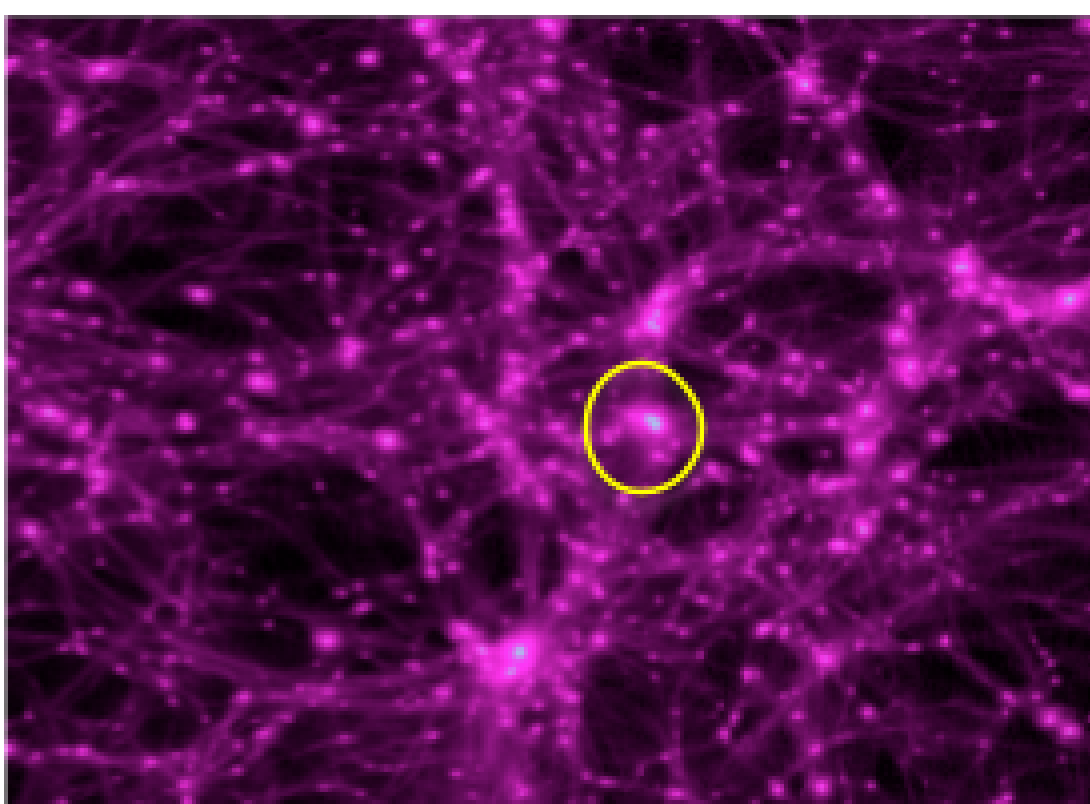

***Figure 3-8:*** *Snapshots at z=0 for N-body simulations of GR(Left) and f(R) gravity (right) . The eclipse highlights clear differences arisings between the predictions with the two forms of gravity.256^3 particles are simulated in a B=64 Mpc/h box. The Hu-Sawicki model (Hu & Zawicki 2007) is adopted for the f(R) simulation (Zhao, Li & Koyama 2011a),*

Alternatively, TMT can be used to test GR using the environmentally dependent screening effect. In principle, one can modify GR so that it mimics a cosmological background expansion history similar to that in the ΛCDM model (GR plus vacuum energy as dark energy). As shown in figure 3-8, this will automatically change the structure growth on all other scales, including the galactic, and even sub-galactic scales, which could violate the stringent solar system confirmation of GR. However, there are screening mechanisms (Khoury & Weltman 2004) in which the additional degrees of freedom (scalar fields) could be shielded to evade the solar system tests. The screening reduces the signal of modified gravity (MG) on (sub-) galactic scales, Therefore the high angular resolution and sensitivity of TMT makes it ideally suited to carry out these kinds of tests. For example, in modified gravity theories, the dynamical mass and lensing mass of the same object are in general different, unlike those in GR. Based on a high-resolution N-body simulation of a f(R) model (see figure 1 in Zhao, Li & Koyama 2011a), it is found that the fractional difference between dynamical and lensing masses (denoted by $D_M$) depends on the local environment (Zhao, et al. 2011b). On the other hand, tests using stellar distance indicators suggest that screening effects are small (Brax et al. 2021).

The lensing and dynamical masses of galaxies can be measured by TMT using strong lensing and velocity dispersion measurements (e.g., Schwab et al. 2010). The sample can then be subdivided by lensing mass and local environment indicators. Comparing the fractional mass difference $D_M$ of isolated low mass galaxies to those clustered massive galaxies can not only enhance the possible signal, but also minimize systematic effects.

In addition to the above tests, TMT can be potentially employed to observe the rich gravitational environment of the Galactic Center (GC) black hole to test black hole physics, GR effects, spacetime metrics and alternatives to GR (see section 6.1.1). The short period S-stars orbiting the GC black hole are strong probes for gravitational physics. Zhang et al. (2015) reported the possibility of constraining spin and mass of the black hole through orbital dynamics of six example S-stars in the nuclear star cluster of the black hole. It has been found that astrometric and spectroscopic precisions of 10 $\mu as$ and 1 km/sec respectively will be able to constrain spin of the black hole within less than two decades of monitoring of S-stars like S0-2 (15.92 years orbit with e = 0.89, semi major axis of around 970 AU and pericentre distance of 120 AU). Recently Qi, Shaughnessy and Brady (2021) reported that spin and quadrupole moment of the GC black hole will be measurable through tighter orbits of S4714 (12 years orbit with e = 0.98), S62 (9.9 years orbit with e = 0.976) and S4711 (7.6 years orbit with e =0.768). TMT can be employed for measurements of spin and quadrupole of the black hole through these stars which will enable us to test the Kerr metric hypothesis of astrophysical black holes and the no-hair theorem. Gravitational redshift of light emitted by S0-2 during its 2018 pericentre passage has been detected by the GRAVITY beam combiner on the VLT (Abuter et al. 2018, GRAVITY Collaboration). Schwarzschild pericentre shift of S0-2 has also been measured by the VLT+GRAVITY (Abuter et al. 2020, GRAVITY Collaboration). Astrometric and spectroscopic capabilities of the TMT will be better equipped to test these effects down to much more compact orbits below that of S0-2 thereby providing strong field tests of GR.

f(R) gravity theory corresponds to geometrical modifications to GR which are widely used to explain the dark universe (Capozziello 2002; Capozziello et al. 2003, 2007, 2014; Caroll et al. 2004; Starobinsky 2007) and early universe physics (Starobinsky 1980) without including exotic matter/energy components. f(R) gravity theory naturally predicts the existence of an extra gravitational degree of freedom known as scalaron. Kalita (2018) showed that the Yukawa correction to Newtonian gravity is a natural outcome of the weak field limit of f(R) gravity. It presents a Yukawa modification to Newtonian gravity and hence affects the pericentre shift of compact stellar orbits near the GC black hole. Earlier by analyzing the orbit of S0-2 Borka et al. (2016) reported the possibility of constraining the power law f(R) gravity. The very foundation of GR encoded in the principle of Local Position Invariance (LPI) has also been tested by using AO imaging data from Keck and radial velocity data of the VLT to constrain the range and coupling strength of Yukawa gravity through S0-2 and another S-star, S0-38 (19.2 years orbit with e =0.81) (Hees et al. 2017). The VLT with its three instruments NACO, SINFONI and GRAVITY has been able to measure dependency of hydrogen and helium absorption lines of the stellar absorption lines on gravitational potential as the star traverses through different regions of the gravitational potential well of the GC black hole (Amorim et al. 2019, GRAVITY Collaboration). After the detection of Schwarzschild pericentre shift of S0-2 De Martino et al. (2021) produced constraints on the parameters of f(R) modified gravity.

Competition between scalaron induced pericentre shift and GR pericentre shift of compact stellar orbits has been extensively studied in Kalita (2020, 2021). It has been found that scalarons with masses $10^{-22}$-$10^{-16}$eV show observable pericentre shift of compact stellar orbits. Paul, Kalita and Talukdar (2023) investigated screening of the scalarons and found that scalarons with $10^{-22}$eV are unscreened near the orbits of S0-2 (1000 AU) and S4716 (407 AU), and that scalarons with $10^{-19}$eV are unscreened at all orbits below S0-2 down to 45 AU. It has been observed that $10^{-16}$eV scalarons are screened in the orbital scale 45 – 1000 AU. Unscreened scalarons affect the GR pericentre shift. Scalarons with mass $10^{-16}$eV, which reproduce the angular scale of the GC black hole shadow can also show drastic departures from GR for orbits below 45 AU, were theoretically predicted by Kalita (2018) through UV and IR scales of gravity corrected vacuum fluctuations near the GC black hole. The prediction was made by an inverse relationship between scalaron mass and black hole mass (horizon). The Schwarzschild limit of the pericentre shift of this novel metric has been estimated and found that these scalarons restore GR down to 45 AU but show drastic departure from GR for orbits below 45 AU.

TMT will be employed to improve these constraints by resolving and tracking other S-stars. The discovery of more compact orbits below S0-2's orbit through TMT's capability will make it possible to improve the limit of the violation of the LPI. Lalremruati and Kalita (2021) showed that if the TMT achieves an astrometric capability in the range of (10-50)µas it will be able to measure Schwarzschild precession down to an orbit with a semi major axis of 500 AU (pericentre distance of around 50 AU with S0-2 like eccentricity). Therefore, TMT's measurements on pericentre shift in compact stellar orbits will help to constrain f(R) gravity (see also Paul, Kalita and Talukdar 2023). This will help in estimating the departure of f(R) scalaron gravity from GR in compact stellar orbits. Recent work has shown that scalarons with mass inversely related to black hole mass ensures asymptotic flatness of a modified Kerr metric near the black hole (Paul, Bhattacharjee and Kalita 2024). This metric, called the Kerr-scalaron metric with scalaron mass $10^{-17}$–$10^{-16}$eV, reproduces the angular scale of the GC black hole shadow. Therefore, one looks forward to TMT's state-of-the art capabilities to test scalaron gravity and constrain a cosmologically inspired and interesting alternative to GR in a new astrophysical laboratory.

## 3.3 PHYSICS OF EXTREME OBJECTS – NEUTRON STARS

Our knowledge of the strong and gravitational forces, among the four fundamental interactions conventionally recognized, is relatively incomplete. TMT can improve our understanding of both forces by constraining the equation of state of extremely dense matter and testing gravity theory with neutron star (NS) binary systems.

The measurement of NS masses provides clean and tight constraints on the state of dense matter at supranuclear density and has profound implications on the nature of the strong interaction. However, up to now, most accurate mass measurements are from observations of binary pulsars (Özel & Freire 2016). Gravitational Wave detections have generated a small number of well determined neutron star masses (Landry & Read 2021). Gravitational microlensing has been proposed as a possible way to measure the mass of isolated NSs (Schwarz & Seidel 2002), and it has been highlighted that microlensing events due to pulsars could be predicted based on pulsar proper motions obtained through radio observations (Dai et al., 2010). Future large radio telescopes, such as FAST and SKA, will discover thousands of new pulsars and determine their distances and proper motions; deeper optical surveys, such as those from the Roman Space Observatory and the Rubin Observatory, will provide billions of background sources. Therefore, it will become possible to predict potential microlensing events due to NSs, and further confirm and monitor them with radio telescopes (of order a few events per year can be expected from a dedicated monitoring campaign). With TMT's extremely high sensitivity and high-precision astrometry (~0.1 mas), we would be able to detect weak microlensing phenomena or provide precise measurements (see chapter 9), and then determine isolated NS masses (Tian & Mao 2012).

White dwarf-neutron star systems provide test-beds for gravity theories. By modeling the optical spectral lines of the white dwarf (WD) and high-precision radio pulsar timing, the masses of components in WD-NS binary systems can be determined independently of gravity theory, which makes WD-NS systems unique laboratories to test theories of gravity in strong-field regime (Freire et al. 2012; Antoniadis et al. 2013). The high spectral line sensitivity and resolution of TMT will greatly increase our sample of WD-NS systems and provide precise mass measurements of the binary components. By combining the derived masses with additional post-Keplerian parameters from radio timing, we will be able to probe various aspects of alternative gravity theories. Wide-orbit WD-NS systems are ideal systems to test strong equivalence principle, while tight-orbit WD-NS systems have proved to be extremely useful in probing the local Lorentz invariance of gravity (Shao & Wex, 2012).

In addition, the optical emissions from NSs carry important information about the underlying radiation processes in strong magnetic fields and hence the nature of NSs. Normally, the optical emission is weak, and only about 20 NSs have been observed in UV/optical/IR so far. However, it has been discovered that the optical/UV fluxes exceed extrapolations of X-ray blackbody emission (Tong et al., 2011; Wang et al., 2017) and thus, may reveal some new physical processes. To make breakthroughs in this field, it will be necessary to gather UV/optical spectroscopy of the very faint emission from NSs using large optical telescopes, such as TMT with WFOS.

## 3.4 VARIATION OF FUNDAMENTAL PHYSICAL CONSTANTS

In a large class of fundamental theories, the four-dimensional universe that we observe emerges from a higher-dimensional space time. In such theories, the fundamental physical constants, such as the fine structure constant and the electron-to-proton mass ratio, are expected to be time dependent (see Uzan 2007 for a review). Time variation of fundamental parameters is also generic in cosmological models of quintessence. It is possible to search for and constrain any such variations of the constants by accurate measurement of wavelengths of absorption lines seen in the spectra of quasars (Webb et al. 1999). Past observations have been inconclusive. More than a decade of study using 8 and 10 meter telescopes to constrain evolution of the fine structure constant has yielded mixed results (e.g., Murphy et al. 2003, Chand et al. 2004, Tzanavaris et al. 2005, Reinhold et al. 2006, Molaro et al. 2013). There has been recent tantalizing evidence for a variation in the electron-to-proton mass ratio (figure 3-9).

These observations have been limited both by signal to noise ratio, and by identification and control of systematic errors (Murphy et al. 2007). While both of these limitations need to be addressed, it is clear that the signal-to-noise ratio provided by the current generation of 8–10 m telescopes is at best marginal (Gonçalves et al., 2020). In the same amount of observing time, TMT with HROS would provide a three-fold improvement, providing a statistically definitive result. Combining TMT observations with results from other probes such as CMB power spectrum analysis (Ade et al. 2014), galaxy clusters (Galli 2013), or primordial nucleosynthesis (Coc et al. 2012), promises to set limits on the order of $10^{-4}$ on the variation of $\alpha$ , and of $10^{-3}$ on the variation of the electron mass over the past 13 billion years.

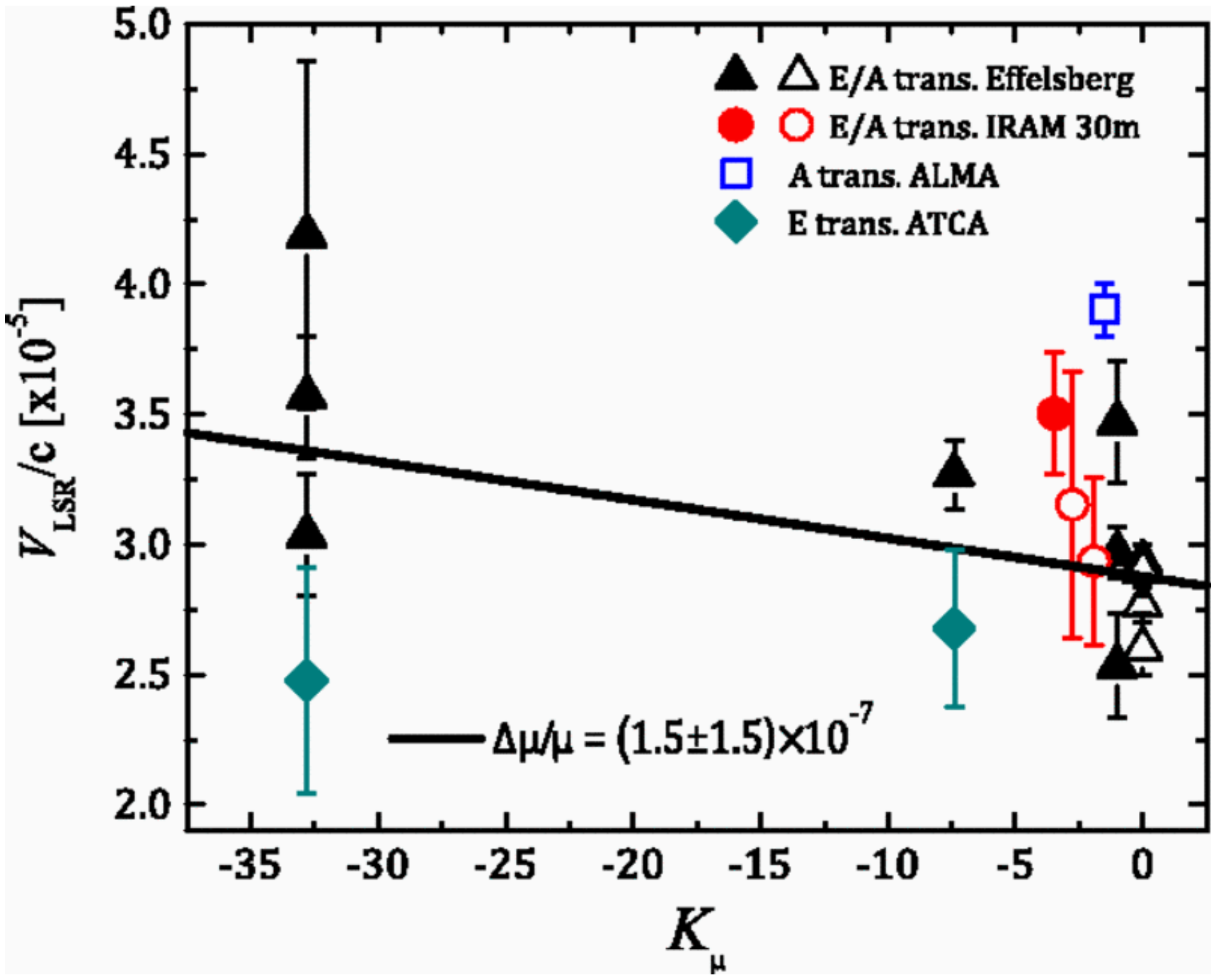

***Figure 3-9:*** *Measuring alpha. Wavelength residuals seen in QSO spectra vs. sensitivity coefficient. The slope indicates a variation of the ratio of the masses of the proton and electron (Bagdonaite et al. 2013a,b, 2014).*

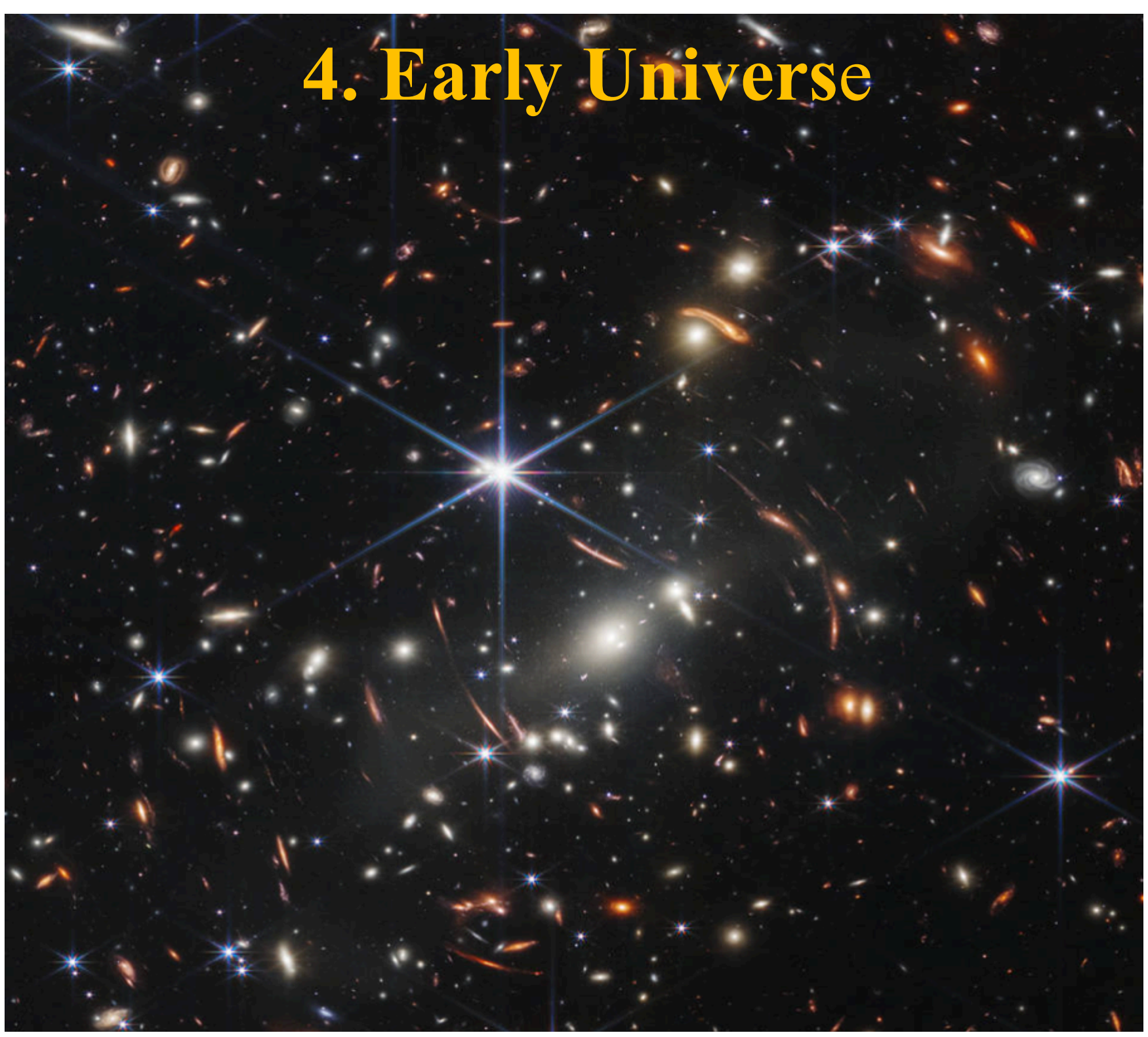

*NASA's James Webb Space Telescope has produced the deepest and sharpest infrared image of the distant universe to date. Known as Webb's First Deep Field, this image of galaxy cluster SMACS 0723 is overflowing with detail. Thousands of galaxies – including the faintest objects ever observed in the infrared – have appeared in Webb's view for the first time. This slice of the vast universe covers a patch of sky approximately the size of a grain of sand held at arm's length by someone on the ground. Image credits: NASA, ESA, CSA, and STScI*

---

TMT will address the second big question from chapter 2, Q2-*When did the first galaxies form and how did they evolve?,* by conducting a number of experiments aimed at looking deeper, further, earlier, and more precisely into the known universe than is currently possible. These observations will image distant, faint, galaxies; take spectra and spatially resolve early galaxies — including those early galaxies being discovered by the JWST; and use gravitational lenses to search for the most distant galaxies. Lensing-assisted observations will allow the measurement of star formation in early galaxies with rates of <1 $M_{\odot}$/yr, and the dissection of galaxies smaller than a few 100 pc. These observations will also contribute to additional big questions addressing when and how the first galaxies, stars, and planets formed and the role of black holes in early galaxy formation (questions 1, 3, and 4).

These observations require broad wavelength coverage from UV to IR and, in most cases, long exposures to reach the faintest targets. To go deeper and fainter, they also require the TMT's impressive sensitivity, exquisite angular resolution, multiplexing ability, and substantial field of view, taking the maximum advantage of the TMT's broad capabilities (see chapter 13 for a summary).

---

**Contributors:** Marusa Bradac (UC Davis), Scott Chapman (Dalhousie University), Ranga-Ram Chary (IPAC), Asantha Cooray (UC Irvine), Len Cowie (University of Hawaii), Ian Dell'Antonio (Brown University), Mark Dickinson (NOIRLab), Mike Pierce (University of Wyoming), Tommaso Treu (UCLA), Masanori Iye (NAOJ/TMT-J), Nobunari Kashikawa (NAOJ), Marie Lemoine-Busserolle (NOIRLab), Crystal Martin (UC Santa Barbara), Kimihiko Nakajima (NAOJ), Masami Ouchi (University of Tokyo), Sonali Sachdeva (Raman Research Institute), Toru Yamada (Tohoku University)

---

# 4 EARLY UNIVERSE

The expansion of the universe stretches the wavelengths of photons, causing spectral lines to shift to longer (redder) wavelengths. The redshift of a source is thus a measure of its distance, and the epoch at which the light was emitted. The frontier for spectroscopically confirmed galaxies is currently at a redshift of 13.2 (Curtis-Lake et al. 2023, Robertson et al. 2023) when the universe was roughly 500 million years old, and candidates have been identified as far back as z = 17 with JWST but have not yet been confirmed spectroscopically. Analysis of early JWST photometry and spectroscopy data has confirmed the presence of scores of galaxies (> 20) from redshift 10 to as high as 16, when the universe was less than 250 million years old. UV luminosity and stellar masses of a quarter of these sources, are so high ($M_{UV} < -19.5$ mag, $M_* \sim 10^{8-9}$ $M_\odot$), that several crucial modifications are required to the existing understanding of the formation of first sources.

The motivation for finding and studying the most distant galaxies is three-fold. Firstly, astronomers are inspired to undertake a census of the first galaxies seen in the first few hundred million years after the Big Bang. Studying the early universe provides a broad foundation that is fundamental to science and has driven astronomy to major discoveries in the past; it also excites great public interest. Secondly, and more fundamentally, hydrogen in intergalactic space is fully ionized by a redshift of about 5, whereas it was fully neutral soon after the time the microwave radiation emerged (the recombination epoch). At some point there was a landmark transition called *cosmic reionization*, akin to a phase transition in the intergalactic medium (IGM), when hydrogen became ionized over a period of time, the *Epoch of Reionization*. Most likely, the beginning of the epoch of reionization was closely related to the birth of the first galactic systems that released copious amounts of ionizing ultraviolet photons. Prior to the beginning of reionization, hydrogen was still in atomic form and the universe was devoid of any light emitting celestial sources — a period commonly termed the *Dark Ages*. Pinpointing when and how the Dark Ages ended and finding the earliest sources responsible for cosmic reionization is necessary to complete the story of galaxy evolution. Thirdly, the physical processes which accelerate or inhibit the cooling and collapse of hydrogen gas clouds into young galaxies at these early times provide the seeds from which later, more massive, galaxies such as our own Milky Way assembled, and perhaps supermassive black holes.

The motivation for finding and studying the most distant galaxies is the following:

1) Tracking the change in the properties of galaxies (i.e., their mass, size, star formation rate, etc.) with time, is crucial to pin down the physical processes that played a dominant role at each epoch. This exercise benefits from an increasingly clearer understanding of the first sources as they are the progenitors of the present-day galaxies.

2) Post the recombination epoch, the universe, devoid of any luminous sources, was filled with neutral Hydrogen. The first sources are thus responsible for putting an end to the Dark Ages and beginning the ionization of the neutral medium. To decipher the mechanism involved during the epoch of reionization and the time taken for this transition, we need an accurate census of the properties of the earliest sources.

## 4.1 EARLY GALAXIES AND COSMIC REIONIZATION

It is thought that the first luminous objects, forming from primordial hydrogen and helium, include giant short-lived stars, perhaps more than 200 times more massive than the Sun (Hirano et al. 2014). Their number and brightness is uncertain, but they are too faint to see with JWST or TMT without highly fortuitous lensing events, even if they lie in an accessible redshift range (redshift $z < 20$ for TMT; at higher redshift, Lyα shifts beyond the K-band window, and there will be no light detectable by ground-based optical/infrared telescopes). As gravity inexorably forces larger hydrogen clouds to cool and collapse, low mass galaxies will begin to shine. Calculations taking into account the transfer of UV radiation from these early galaxies through the IGM suggest that "bubbles" of ionized gas will develop around each prominent source. The bubbles will enlarge and eventually overlap, completing the process of cosmic reionization.

When might this have occurred? The patterns of temperature and polarization signals from microwave background photons detected by the WMAP and Planck satellites suggest that the free electrons responsible reside in ionized gas in the redshift range 5 to 20 with the midpoint of reionization at around redshift 8 (Planck Collaboration 2020). Thus, it is quite likely there will be an abundance of energetic, UV-luminous, star-forming sources in this interval that are responsible for this reionization. However, it is not yet known if reionization was a relatively quick or protracted event. Recent discoveries by the JWST of Lyman-alpha emission from early galaxies (Bunker et al. 2023) suggests that the epoch could have started as early as 200 Myr after the Big Bang, at redshifts as high as 20. The end of reionization appears to be constrained by current Lyman-alpha forest data at redshift < 5.5, about 1.1 Gyr after the Big Bang (Kulkarni et al. 2019a, Bosman et al. 2022). It also is not known whether the ionizing radiation comes from a highly abundant distribution of feeble sources, or a rarer population of brighter, more massive galaxies, though the former is favored (Bouwens et al. 2012; Matthee et al., 2022). Some cosmologists suspect there may also be contributions to the ionizing radiation from sources other than stars. This is because it is not yet understood how much ionizing radiation produced by stars escapes their host galaxy in order to propagate in the cosmological volume and cause reionization (Witten et al. 2023). Further, although contribution from black holes to reionization has so far been thought to be negligible (Kulkarni et al. 2019b), recent JWST discoveries of a surprisingly large number of faint black holes in the early universe (Harikane et al. 2023, Maiolino et al. 2023, Greene et al. 2023) suggests that these objects could also potentially drive, or at least significantly contribute to reionization. More exotic candidates for the sources of reionization have also been suggested, such as decaying particles (Hansen & Haiman 2003, Kasuya & Kawasaki 2004, Pierpaoli 2004). Only by locating and carefully studying the UV-emitting sources in this era can these important questions be answered.

Soon after the first sources emerge, the heated neutral hydrogen begins to glow more brightly than the microwave radiation. Powerful radio telescopes and interferometers such as the Square Kilometer Array (SKA) will trace the topology and growth of the bubbles around these sources by mapping selected regions of the sky using the redshifted 21 cm emission line of hydrogen. However, although low frequency radio surveys may improve our knowledge of when cosmic reionization occurred, they will not detect the sources responsible. Our physical understanding of the process will thus rely crucially on observations with JWST and TMT.

JWST should certainly be able to detect the brightest sources lying within each ionized bubble. However, TMT, with adaptive optics, will be able to detect and study objects an order of magnitude fainter. The gain will be particularly dramatic if, as expected, the earliest sources are physically small (Jaacks et al. 2019). TMT may also find the signatures of the much sought-after, chemically un-evolved, Population III sources expected at these early epochs. In general terms, therefore, we can expect the role of TMT to be one of providing a more detailed story of the properties and influence of earliest sources on the intergalactic medium, building on the basic progress made with radio surveys and JWST.

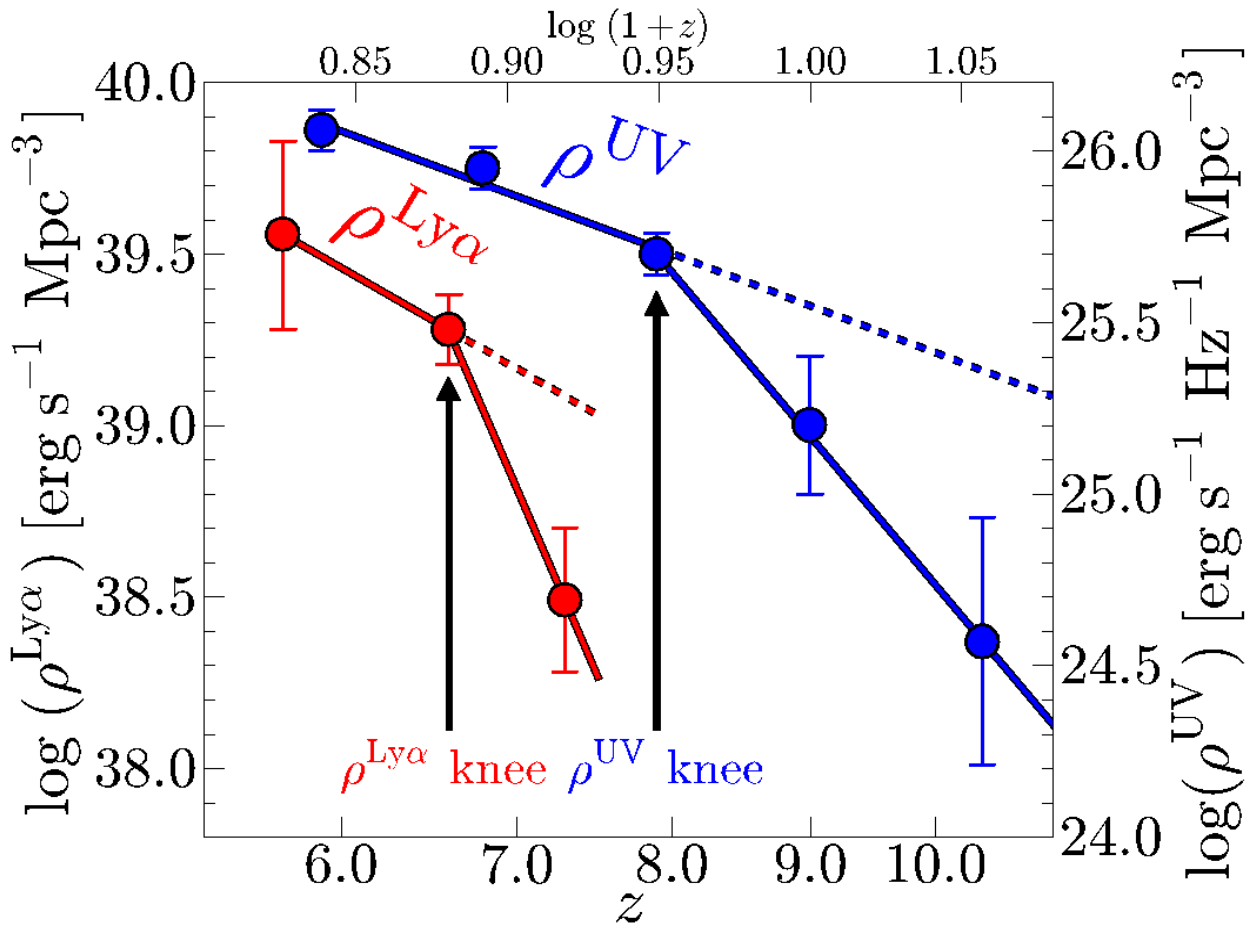

***Figure 4-1:*** *Evolution of Lyα and UV luminosity densities based on observations of Konno et al. (2014). The red and blue circles represent the Lyα and UV luminosity densities, respectively (labeled on the left and right vertical axes, respectively). Toward high-z, the decrease of Lyα luminosity density is clearly more rapid than that of UV at z~7, indicative of strong scattering given by neutral hydrogen gas in the IGM.*

The following subsections introduce several areas where TMT is likely to make a profound impact in its early years in addressing the physical processes in conjunction with the next generation facilities such as JWST and Rubin Observatory. Inevitably, given how little is known about the properties of the early stellar systems and their contributions to cosmic reionization, the quantitative details are more speculative than in other areas of this document. Flexibility in survey strategy will be crucial as more is learned about this uncharted era. Some of the performance uncertainties are discussed in terms of both the unknown size and abundance of star-forming sources.

### 4.1.1 Uncovering primordial stellar systems with TMT

The earliest galaxies will contain massive stars formed from primordial gas. As these stars evolve and eject processed material from their eventual supernovae, newly formed stars containing heavy elements will become more common. Simulations consistent with cosmological star formation rate density measurements using the HST suggest that primordial (so-called Population III) stars may start forming at redshifts of about 30 with the peak of Population III star formation at redshifts between 17 and 10 (Jaacks et al. 2019, Visbal et al. 2020). Rapid star formation has been confirmed with JWST observations in galaxies at redshifts 8.8, 10.2 and 11.4 (Heintz et al. 2024). The formation of metal-enriched Population II stars may proceed quite quickly, although this depends on how much gas is retained in the shallow gravitational potential wells of early galaxies; many of the early metals may be blown right out of galaxies by winds from massive stars and supernovae. Verifying the existence of chemically primordial galaxies and determining their redshift distribution would represent a major new constraint on the first stages of galaxy formation.

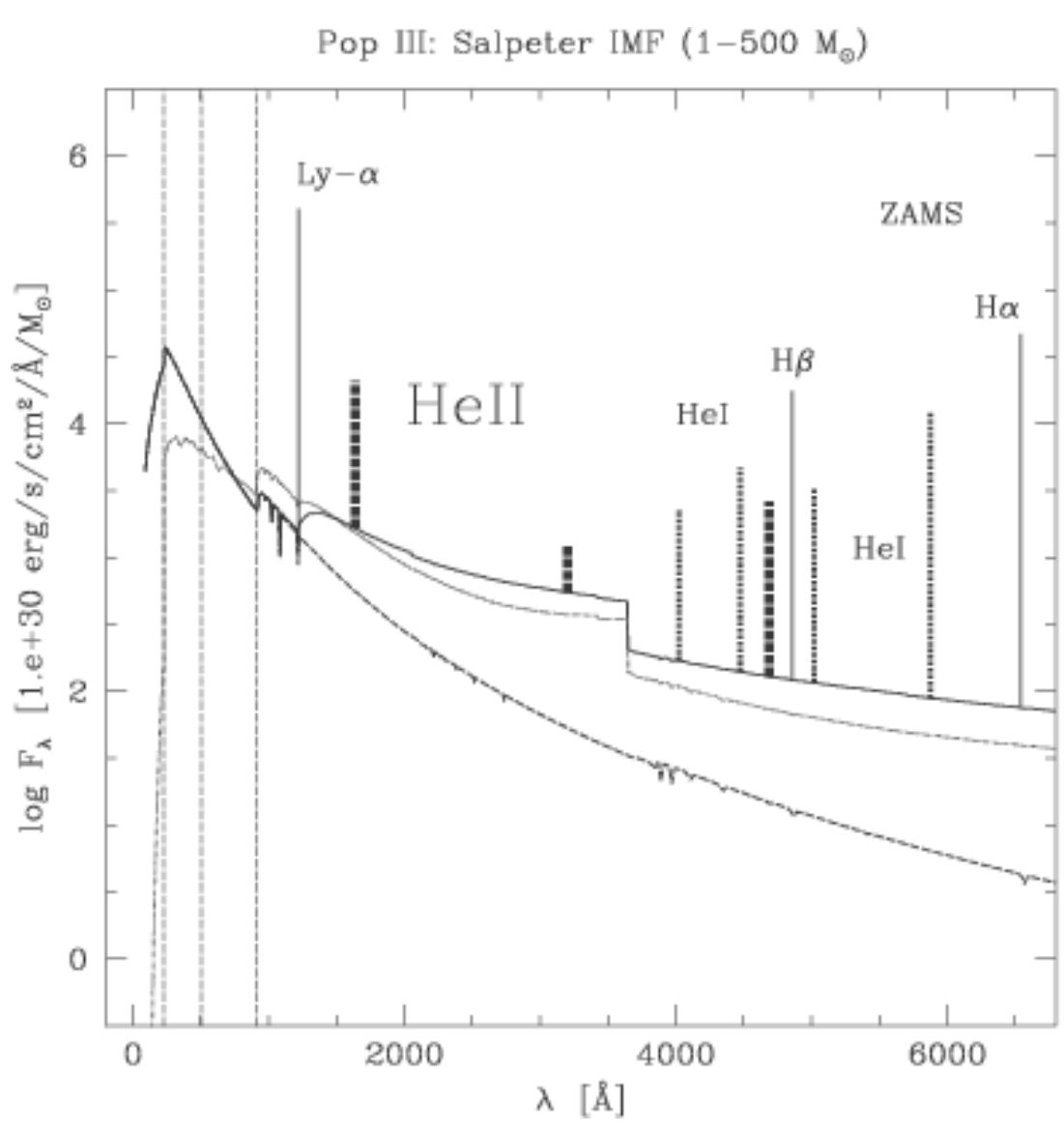

***Figure 4-2:*** *Predicted spectrum of a Pop III ZAMS burst (from Schaerer 2002) based on non-LTE model atmospheres including H and He recombination lines. The dashed line shows the pure stellar continuum (neglecting nebular emission). Note the prominent He II 1640 (thick dashed line) and the importance of nebular continuum emission. Simulations suggest that the He II line, which decays rapidly within 2 Myr, may be a valuable tracer of metal-free stellar populations.*

The hydrogen Lyα line at 1216 Å is a characteristic signature of photoionization by massive stars and is commonly observed in spectra of star-forming galaxies at $z < 6.5$. As much as 8% of the light of a metal-poor young galaxy can emerge in the form of Lyα, which has traditionally made it an effective tool for measuring galaxy redshifts. However, Lyα is a resonance line, and is easily scattered as the IGM becomes more neutral in the early universe. The presence and strength of Lyα will depend on the size and geometry of local ionization bubbles around young galaxies that may allow Lyα photons to escape. Observations suggest that the fraction of Lyα-emitting galaxies drops from z~6 to 7 (e.g., Pentericci et al. 2011, Ono et al. 2012, Schenker et al. 2012, 2014, Treu et al. 2013b; Mason et al. 2019). This evolution appears to steepen at z>7, in clear contrast with the evolution of UV continuum emission (Konno et al. 2014, figure 4-1). Thanks to the JWST, Lyα photons have now been detected out to redshift > 10 (Bunker et al. 2023).

For very young galaxies, the harder photo-ionizing spectrum from metal-poor stars should excite higher-ionization UV emission lines that are not ordinarily seen in more evolved systems. Observations of gravitationally lensed dwarf galaxies at redshifts 2 to 3, which may be close analogs to the galaxy population at $z > 7$ in terms of mass and metallicity, have detected moderately strong high-ionization metal lines such as CIII] 1909 Å, OIII] 1663 Å, and CIV 1549 Å, with equivalent widths as high as 15 Å (Stark et al. 2014). Indeed, such intense UV metal lines have been identified from luminous galaxies at $z>7$ with JWST/NIRSpec (e.g. Tang et al. 2023, Bunker et al. 2023, Zou et al. 2024). In principle, these UV metal lines can be detected out to $z\sim14$ with TMT, whose infrared spectrographs IRIS and IRMOS will be more sensitive than JWST/NIRSPEC to faint, narrow line emission in compact galaxies. These high-ionization metal lines are a promising tool for studying young star-forming galaxies in the early universe. This is due to evidence indicating a more highly ionized ISM condition, achieved by a harder ionizing radiation field at higher redshifts (e.g. Nakajima & Ouchi 2014, Steidel et al. 2016, Stark et al. 2017).

A related avenue for searching for the first stars is to use metal-line absorption in the spectra of background sources (Kulkarni et al. 2013, 2014). Foreground gas with chemical signatures produced by Population III stars can produce absorption lines that trace the properties of these stars. At high redshifts such lines are expected to not be affected by dust production or saturation, and as a result arguably provide the most direct method of inferring the presence of primordial stars. Indeed, the most high-quality current data may have started seeing signatures of unusual chemistry in such metal absorption lines (D'Odorico et al. 2024). However, this line of inquiry remains limited by telescope sensitivity (Becker et al. 2012). The superior sensitivity of the TMT will allow detection of unsaturated metal lines out to higher redshifts. It might then be possible to cross-correlate these inferences based on metal-line absorption with inferences from other probes.

The TMT could also directly detect individual Population III stars from high-redshifts via strong gravitational lensing. Indeed such detection of individual stars has already been reported at redshifts $z< 6$ using the JWST (Diego et al. 2023). With its near-IR capabilities, TMT could push this search to redshifts as high as 11, with much higher sensitivity than JWST (Zackrisson et al. 2024).

Spectroscopic measurements of early galaxies by the TMT can allow indirect inference of the formation of the first generation of stars via stellar population analysis. Using spectra of redshift-9 galaxies obtained from the HST, this technique has already produced constraints on star formation out to redshifts as high as 20 (Laporte et al. 2021).

For truly primordial galaxies, without metals, the last resort is observing the HeII emission line at 1640 Å (the equivalent of H$\alpha$, for ionized helium). Model atmosphere calculations by Schaerer (2002, figure 4-2) suggest this helium line could emit up to ~0.8% of the luminosity of a metal-poor young galaxy. Recent theoretical predictions clarify that HeII1640 emission, along with (non-detections of) other UV metal lines, provides reasonable diagnostics to identify such primordial galaxies (e.g. Nakajima & Maiolino 2022, Katz et al. 2023). Although the signal of HeII is weak, TMT would be able to see this line in galaxies up to redshift $z \sim 14$. Intriguingly, recent observations with JWST/NIRSpec-IFU found a tentative detection of strong HeII emission close to a luminous galaxy at $z=10.6$ (Maiolino et al. 2024), indicating primordial star formation in the vicinity of luminous objects. Deep follow-up observations with TMT/IRIS and IRMOS around high-redshift luminous objects, which will be identified by future surveys such as the Roman Space Telescope and Euclid, would be an important approach to identify the sources hosting the first generation of stars and examine their detailed properties. Assuming the sources are physically small, there will be a significant signal-to-noise advantage over JWST.

### 4.1.2 Detecting the sources of reionization

In the course of early galaxy formation, high-mass stars produce abundant UV photons that are believed to have been primarily responsible for cosmic reionization. Once reionization proceeds, the UV background radiation heats gas so as to suppress star formation in low-mass galaxies. Thus, early galaxy formation leads to reionization, initiating a "negative feedback loop" that actually delays the subsequent stellar mass build-up of the lowest mass systems. Thus, early galaxies and cosmic reionization would have a profound effect on the more recent history of galaxy formation.

At the end of the reionization era, at z ≈ 6, galaxies that are directly detected in deep Hubble Space Telescope observations (with MUV~<−18) appear to be sufficiently luminous and abundant to maintain the ionization of the IGM, provided that a substantial fraction (~30%) of the H-ionizing photons emitted by massive stars escapes the galaxies in which the stars are forming (the so-called "escape fraction"). However, at z ≈ 7–8 the directly-observed galaxy population appears to fall short (e.g., Finkelstein et al. 2012) unless a large population of galaxies below the current detection limit is present. Observations suggest that the slope of the galaxy UV luminosity function steepens significantly between z~4 and z~7–8, where its slope (at observed magnitudes) approaches that expected for the dark matter halo mass function (e.g., Bouwens et al. 2015). The validity of extrapolations of the luminosity function to unobserved faint magnitudes is currently untested.

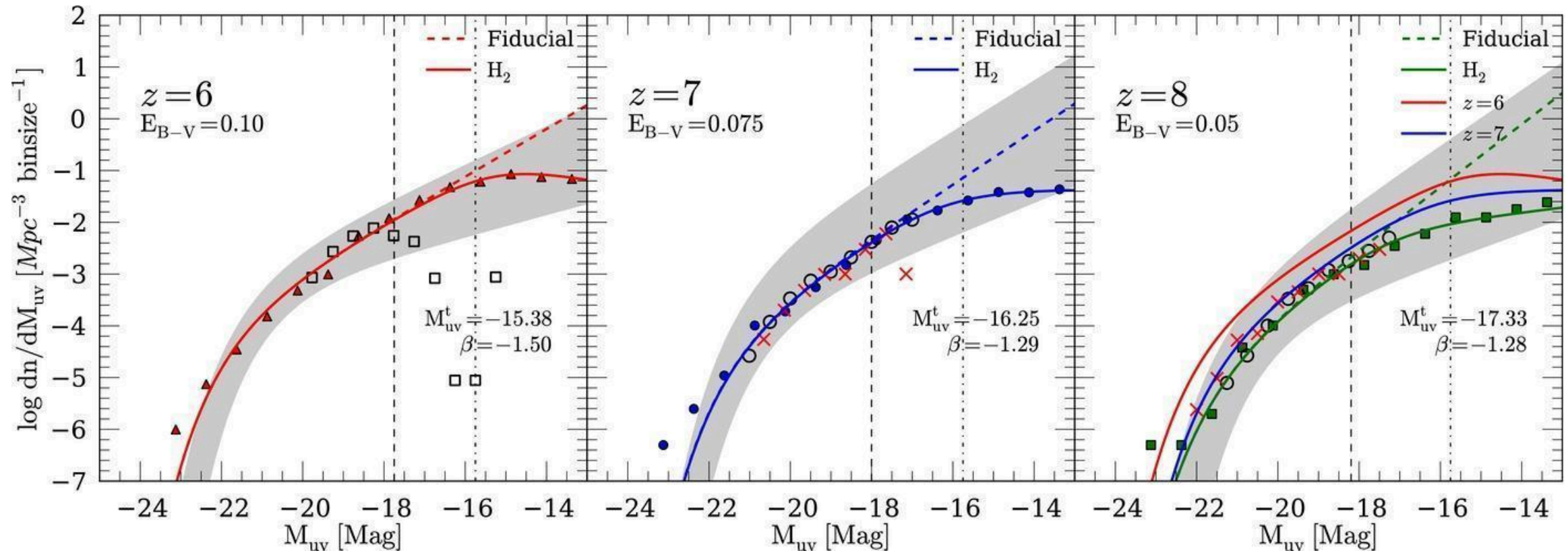

***Figure 4-3:*** *UV luminosity functions at z = 6 (left), 7 (center), and 8 (right), predicted by two sets of star-formation models with different assumptions about the dominant cooling processes (Jaacks et al. 2013). The vertical dashed lines show typical deep HST observational limits, and the vertical dash-dotted lines present the approximate observational limits of JWST and HFF-source plane. TMT will probe high-z sources at the observational limits of JWST+HFF and fainter.*

According to current theoretical models, the abundance of very faint galaxies is determined by the radiative cooling efficiency of gas within low-mass dark matter halos, which is extremely sensitive to the metallicity of the gas and (for truly zero-metallicity gas) by molecular hydrogen cooling (Jaacks et al. 2013, Boylan-Kolchin et al. 2014, figure 4-3). Moreover, the abundance of faint galaxies also depends on the effects of poorly understood feedback processes (such as supernovae) on subsequent star formation activity in the same regions. In short, the abundance of faint galaxies encodes the essential physics of star-formation and feedback in low mass dark matter halos.

Cosmic reionization models suggest that galaxies in the range $M_{UV} \sim -13$ to $-18$ produce more hydrogen ionizing photons than the luminous ($\sim M_{UV}^*$) galaxies at z > 7. However, the abundance of such faint galaxies is essentially unknown and unexplored even with the deepest data in the Hubble Ultra Deep Field (Ellis et al. 2013). The Hubble Frontier Fields (HFF) initiative exploited gravitational lensing to search for still fainter galaxies behind six massive galaxy clusters. reaching MUV~−14 (Ishigaki et al. 2018). Although this is an important step for exploring the high-z faint galaxy population, the HFF galaxy sample may be too small for robust statistical constraints on the faint end of the high-z luminosity function, and there may be lingering uncertainties on magnification factors that depend on mass models of the lensing clusters, and on the unknown size distribution of the source population. Moreover, the Hubble imaging data alone do not provide spectroscopic information needed to evaluate the metal abundance, ionization state, or velocity fields in this important early galaxy population.

JWST is providing new insights on these dwarf galaxies, but even at its resolution these galaxies are extremely compact (Yang et al. 2022) and therefore require TMT to fully resolve. An imaging campaign with TMT IRIS will unambiguously determine the abundance and morphology of faint galaxies at the epoch of reionization. Moreover, the combination of sensitivity and angular resolution of TMT/IRIS and IRMOS will permit detailed spectroscopic studies of metal enrichment and outflow processes of these key sources.

Apart from the discovery of the existence of galaxies that could potentially serve as sources of reionization, the TMT will also be an excellent telescope for measuring the escape fraction of ionizing photons from these galaxies. While direct measurements of the Lyman-continuum photons form galaxies deep inside the epoch of reionization will not be possible because of the large opacity of the neutral intergalactic medium, the high sensitivity of the TMT will still allow cross-correlation of galaxies and faint AGN in the epoch of reionization with faint Lyman-alpha transmission spikes in the spectra of background sources (Meyer et al. 2020).

TMT will also have interesting synergies with 21 cm experiments searching for sources of reionization. For instance, the 21 cm signal from the cosmic dawn is enabled by the Lyman-alpha photons from the very same stars that kickstart reionization. Indeed, the first reported detection of this signal already places very strong constraints on what a near-infrared telescope like the TMT should see (Mirocha and Furlanetto 2019, Mittal and Kulkarni 2022). In the next decade, such 21 cm measurements will be available from REACH (de Lera Acedo et al. 2022), NenuFAR, HERA, and SKA.TMT measurements of the luminosities of galaxies in the early stages of reionization, when combined with such 21 cm measurements, will shed light on when and how the source of reionization begin to form.

### 4.1.3 The process and history of reionization

Another major open question is the *history* and *process* of reionization, i.e., how the IGM ionization develops over time and is distributed in space. Although direct observations of the 21 cm line at high redshift offers the promise of mapping the neutral IGM at the epoch of reionization, this will not be accomplished with high fidelity and angular resolution during the decade prior to TMT first light. Alternatively, one can probe neutral hydrogen gas clouds along the line of sight to bright, high redshift background sources whose spectra show both Lyα and metal line absorption features. QSOs, GRBs, and perhaps even bright galaxies can serve as the source of background illumination for this purpose.

Recent work using VLT data has shown that the spatial fluctuations in the Lyman-alpha opacity of the IGM encodes information about the history of reionization (Kulkarni et al. 2019a, Bosman et al. 2022). TMT will be ideal to exploit this approach because on the one hand it will be able to measure the Lyman-alpha opacity on a greater number of sightlines because of the discovery of new quasars from surveys like LSST, Euclid, and Roman. On the other hand TMT will also be able to push the measurements to larger opacities, thanks to its superior sensitivity.

At redshifts $z > 6$, line blanketing in the Lyα forest removes nearly all the light shortward of Lyα in the object's rest frame. Very small amounts of residual flux occasionally penetrate at $z = 6.5$ (the so-called "dark gaps"), which can be analyzed statistically to infer the structure and neutral fraction of the intergalactic medium (Songaila & Cowie 2002; Paschos & Norman 2005; Gallerani et al. 2006). The number of quasars at $z > 6$ has grown more than fivefold in the last few years, but only a handful of bright quasars (mAB < 20) have been found at $z \sim 7$ and beyond, from the large-area near-IR surveys of UKIDSS and VISTA (e.g., Venemans et al. 2013). At the time of writing, the highest redshift quasar currently known, J0313-1806 has a redshift of 7.642 (Wang et al. 2021). The damped Lyα line absorption profile seen in this quasar suggests that the neutral fraction of IGM was fairly large, >10%, at $z \approx 7.1$, favoring late-epoch reionization (Mortlock et al. 2011, see figure 4-4). It is also possible that IGM ionization may decline at higher redshift as intergalactic gas becomes progressively neutral; searches may then have to focus on the strong lines of low-ionizing species like OI and CII. The abundance of high-ionization CIV absorbers decreases from $z{\sim}5$ to $z{\sim}6$ (Ryan-Weber et al. 2009; Becker et al. 2009), possibly indicating evolution in the ionizing background consistent with the expected progress of reionization. However, high-ionization species like CIV require relatively hard-spectrum sources of ionizing photons in close proximity. Low-ionization OI, CII, and SiII absorption lines have also been observed (Becker et al. 2011). OI is of particular interest because of its strong coupling with HI via the charge exchange reaction (so that OI/HI = O/H) but interpretation depends on the unknown abundance of O. Similar absorption line studies can be conducted using GRBs as background continuum sources. However, so far, there are only weak IGM constraints from observations of Lyα using GRBs at a redshift of 6.3 and below (Totani et al. 2014); progress at $z > 6$ has been limited by the lack of sufficiently bright GRBs (cf. the highest redshift GRB of $z=8.2$; Tanvir et al. 2009). With its large aperture and fast response time for target of opportunity observations, TMT will be able to obtain high-S/N spectra of faint GRBs out to very high redshifts. Nearly all of the information at these redshifts must come from Lyα damping profiles and metal lines longward of Lyα. At z > 7, all of these features lie at near-IR wavelengths — the domain of TMT/MODHIS.

As described in section 4.1.1, recent observations suggest that Lyα damping wing absorption due to a substantially neutral IGM suppresses Lyα emission from galaxies at z~8 and beyond. However, galaxies are also useful for probing the later progress of cosmic reionization history at z < 8, especially for investigating the topology of ionized bubbles in the late stages of reionization. Although neutral hydrogen in the IGM reduces the visibility of Lyα emission from distant sources, it also affects the line profile in a manner that is well understood (Miralda-Escude 1998). Accurately observed line profiles as a function of environment and redshift can thus constrain the neutrality of the IGM, although present studies cannot discern the subtle differences in Lyα damping wing profiles expected (Ouchi et al. 2010).

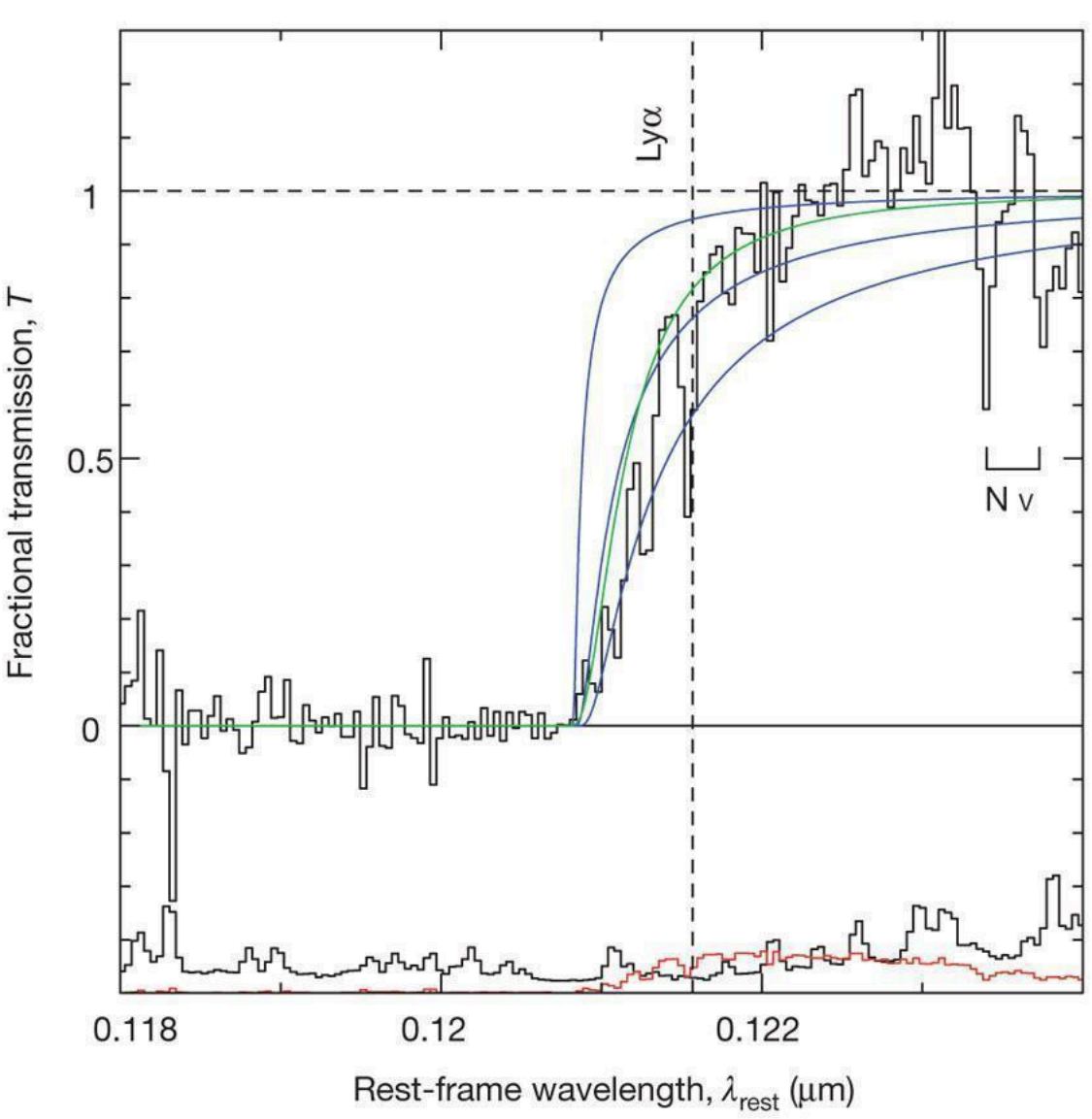

***Figure 4-4:*** *The transmission profile of the z=7.09 QSO ULAS J1120+0641 in the vicinity of Lyα at the emission redshift (black histogram; Mortlock et al. 2011). The transmission profile is formed by dividing the observed ULAS J1120+0641 spectrum by the SDSS QSO composite spectrum, to remove the typical intrinsic Lyα emission line. At the bottom of the panel, the random and systematic uncertainties are denoted with black and red histograms, respectively. The blue curves represent model Lya damping profiles from transmission through the IGM assuming neutral fractions of 10%, 50%, and 100% (top to bottom) and that the ionization front is located at a distance of 2.2 Mpc. The green curve indicates the absorption profile expected for neutral hydrogen column density of NHI=4x1020 cm-2 located 2.6 Mpc in front of the quasar.*

The measurements are challenging because of the need for high spectral resolution for individual sources, as well as the requirement to probe very faint, low-mass systems in low-density regions, where Lyα damping wing absorption would be prominent as a result of patchy reionization. Once a large number of high redshift Lyα line emitters are located, e.g., by the on-going deep narrowband survey of Subaru Hyper Suprime-Cam, more detailed TMT spectroscopic follow-up studies will become practical. As an example, an 8 hour observation with TMT/IRIS can achieve SNR = 22 per resolution element for an emitter up to $z \sim 8$ with a Lyα luminosity of $L >$ $5 \times 10^{41}$ erg $s^{-1}$. Such systems have a star formation rate of less than 10 $M_{\odot}$/yr (Le Delliou et al. 2006) and are more representative and likely more numerous than the more luminous emitters for which such observations have been attempted so far.

Recent JWST discoveries are suggestive of a potential significant contribution of faint black holes in the early universe (Harikane et al. 2023, Maiolino et al. 2023b, Greene et al. 2023). It is also likely that an even larger number of obscured black holes exists at these redshifts (Satyavolu et al. 2023). TMT could shed light on such objects by detecting Lyman-alpha halos around them. These are formed due to reprocessing of the UV radiation of these AGN by the surrounding gas. Presence of Lyman-alpha haloes can constrain their Lyman-alpha escape fraction and the concomitant contribution to reionization. JWST has shown the possibility of using damping wings in galaxy spectra to measure the neutral hydrogen fraction of the IGM (Umeda et al. 2023, Keating et al. 2023). With the TMT this can potentially allow one to map out the whole epoch of reionization, even out to redshifts where there are no AGN.

A novel direction to pursue with the TMT would be to measure the Lyman-alpha flux power spectrum within the proximity zones of AGN and bright galaxies in the epoch of reionization. At small scales this power spectrum has information of the thermal history of gas at earlier, even pre-reionization, epochs (Davies and Hennawi 2023). TMT will take this idea to smaller scales, deeper sensitivities, and higher redshifts. Any thermal history constraints obtained would be valuable to confront with 21 cm data and exotic heating mechanisms.

Discovery of Lyman-alpha emitting galaxies in the epoch of reionization will also allow constraints on the progress of reionization. The Lyman-alpha emission from such high-redshift galaxies is only visible if there is a large enough ionized region around the galaxies to enable the escape of Lyman-alpha photons. JWST has already shown that such Lyman-alpha emitters exist out to higher redshifts than expected (Bunker et al. 2023). TMT will have the capability of finding fainter LAEs out to higher redshifts, which will be able to push the measurements to the very early stages of reionization.

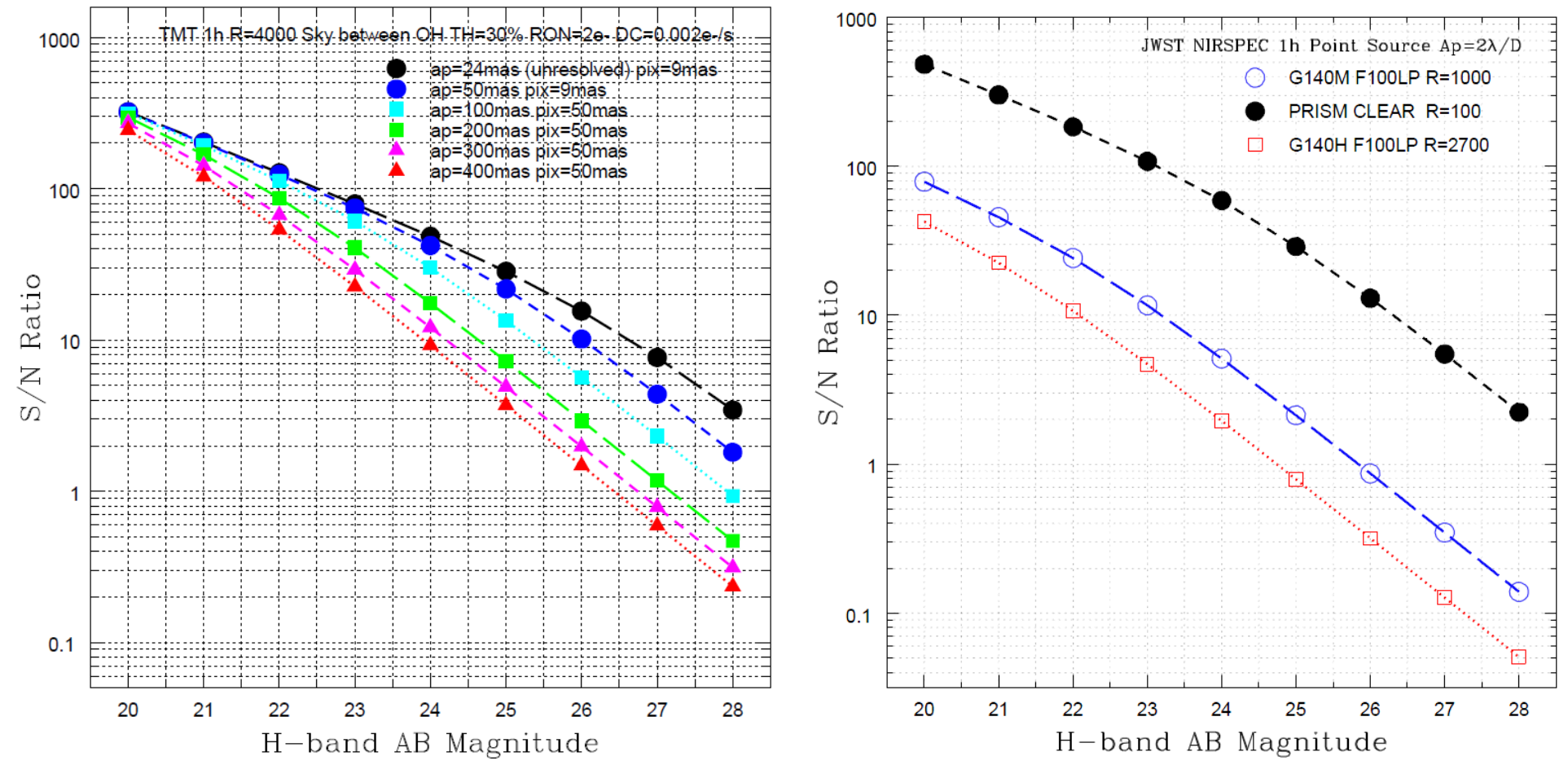

***Figure 4-5:*** *(Left) Expected S/N ratio for the case of various sizes of the spectroscopic aperture estimated assuming TMT IRIS. (Right) Expected S/N ratio for a spatially unresolved source with the different spectral resolution of JWST NIRSPEC obtained by NIRSPEC Exposure Time Calculator (https://jwst.etc.stsci.edu/workbook.html?wb_id=119420).*

## 4.2 ANGULAR SIZES AND THE SYNERGY WITH JWST AND FUTURE SPACE MISSIONS

Over the last 20 years there has been a vital synergy between HST and ground-based 8–10 meter class telescopes: resolved imaging from space, and deep spectroscopy from the ground. The relationship between JWST and TMT will be somewhat different, although just as important. Both TMT and JWST have impressive imaging and spectroscopic capabilities. JWST has the advantage of a reduced near-infrared background in space, and its sensitivity at wavelengths longer than 2.5 microns provides sensitivity to galaxies at redshifts beyond 20. However, TMT has a much larger collecting area than JWST, and superior angular resolution (a factor of 5 higher) with AO-fed instruments, offering greater sensitivity for physically small, faint galaxies. TMT can resolve and study finer structures in galaxies than is possible with JWST. Unlike HST, JWST offers sensitive multi-object spectroscopy with low to moderate spectral resolution (up to R = 2700 for JWST NIRSPEC), and at longer wavelengths, without atmospheric limitations. The higher resolution of TMT spectrometers (e.g., R ≈ 5000 for IRIS and IRMOS) will improve detection limits for narrow emission and absorption lines and offers velocity resolution that JWST cannot match.

JWST will have amassed a huge amount of observations before TMT first light. Based on their expected timetables, TMT will see first light well after the 5 year nominal lifetime of JWST operations, though there may be a chance of operational overlap if JWST can remain operational for 20+ years due to the highly fuel-efficient journey to L2. In any case, TMT observing programs will thus follow up galaxies detected in well-established JWST surveys, and study them in ways that JWST could not. TMT will be contemporaneous with future space missions that provide wide-field near-infrared survey capabilities, such as WFIRST (NASA) and WISH (JAXA). These will also supply an abundance of distant galaxy targets for TMT follow-up.

It is very important to know how small these early galaxies are. Oesch et al. (2010) studied the morphology and size of the galaxies at z=7–8 in the Hubble Ultra Deep Field (HUDF) and found that the galaxies with the UV absolute magnitude M(1600) = −18 to −21 (0.2L* to 3L* for z=7) are very compact, with the average size of 0.7±0.3 kpc (1kpc = 200 mas and 250 mas at z = 7 and 10, respectively, with the standard cosmology), but they are clearly resolved on the HST WFC3 image. Ono et al. (2013) also analyzed the images of the candidate z=8–12 galaxies in HUDF and found that their half-light radii are extremely small, 0.3–0.4 kpc. JWST confirms that galaxies at z>7 are very compact, even in the observed K-band (Yang et al. 2022), with sizes of order 100-200pc, barely resolved by JWST. These results suggest that galaxies at z>7 will require TMT+AO (~12 mas) which corresponds to 0.15–0.25 kpc (50 mas) and 0.04–0.06 kpc at z=7–15 to be fully resolved and characterized in detail.

As TMT observations in this regime will be background-limited, the sensitivity is a strong function of source angular size. Figure 4-5 shows the signal-to-noise ratio for spectroscopic detection of continuum emission from high-redshift galaxies observed with R= λ /Δλ =4000 (the spectral resolution of TMT's IRIS) as a function of apparent magnitude for various spectroscopic aperture sizes. For extremely compact sources that are unresolved or only marginally resolved, corresponding to objects smaller than ~1 kpc (200 mas at z=7) in their size, TMT will detect continuum emission with a S/N that is greater than can be achieved with JWST's NIRSPEC at R=1000 and 2700 and greater than 60% of the S/N achieved with NIRSPEC for R=100. Note that $H_{AB}$ =27 corresponds to 2.3$L_*$ at z=8, objects at the bright end of the luminosity distribution, and to 5.2$L_*$ at z=10 if the luminosity evolution from z=6 to 8 is extrapolated to z~10. If the galaxies are resolved to ~200 mas, which appears to be the case from the HUDF studies (figure 4-6), TMT will still be able to detect the continuum of $H_{AB}$ ~27mag source with S/N>10 in a 10hr integration with wavelength binning to R~400.

High-resolution imaging from JWST Cycle 1 programs reveal a high fraction of reionization-era Lyman-alpha emitters have close companions. Detection of Lyman-alpha when reionization was just beginning was a surprise, as Lyman alpha was predicted to become visible only at z < 8. Frequent mergers at very early stages generate bursty star formation histories. Short bursts of star formation emit more light per unit stellar mass than does a history with a constant star formation rate. If most galaxies have very bursty star formation histories, then the aforementioned tension suggested by the surprisingly high stellar masses reported at z > 8 could be greatly reduced, and possibly eliminated.

The high frequency of mergers among Lyman alpha emitters alone provides important new insight about the production and escape of ionizing radiation. Galactic winds powered by supernova feedback are expected to open up pathways through which ionizing radiation can escape from low mass galaxies (Martin et al, 2023). However, there is a well-known timing problem in young galaxies. A single instantaneous burst of star formation produces strong ionizing radiation for less than 10 Myr, after which all the stars hot enough to produce ionizing radiation have expired. Although some of these very massive stars likely produce supernova explosions, a great deal of uncertainty remains about the mass range where core collapse generates an explosion, especially at very low stellar metallicity. Even if all stars massive enough to produce ionizing radiation also produce supernovae, the number of stars (and explosions) is small compared to the number of core-collapse supernovae produced by the more numerous 8–20 solar mass stars.

The feedback timing problem for an instantaneous burst recognizes that this mechanical feedback follows the peak production of ionizing radiation in time. Hence, most of the ionizing radiation is generated before the holes in the interstellar medium, which are required for radiation leakage, could be generated. Mergers potentially solve this problem because they produce two bursts of star formation. First passage produces an initial burst, which could generate the holes. The second passage often produces a stronger burst. This scenario places the most efficient production of ionizing radiation in galaxies with open escape channels.

Figure 4-7 shows numerical simulation of merging dwarf galaxies at high redshift (Witten et al 2024). Radiative transfer modeling illustrates the direct and scattered Lyman-alpha emission. The higher angular resolution of TMT relative to JWST will play an important role in identifying and characterizing reionization-era mergers.

Imaging will identify multiple components at smaller angular separations. Spectroscopy with narrow slits or an IFU will directly measure the relative velocity of the merging galaxies thereby enabling dynamical modeling.

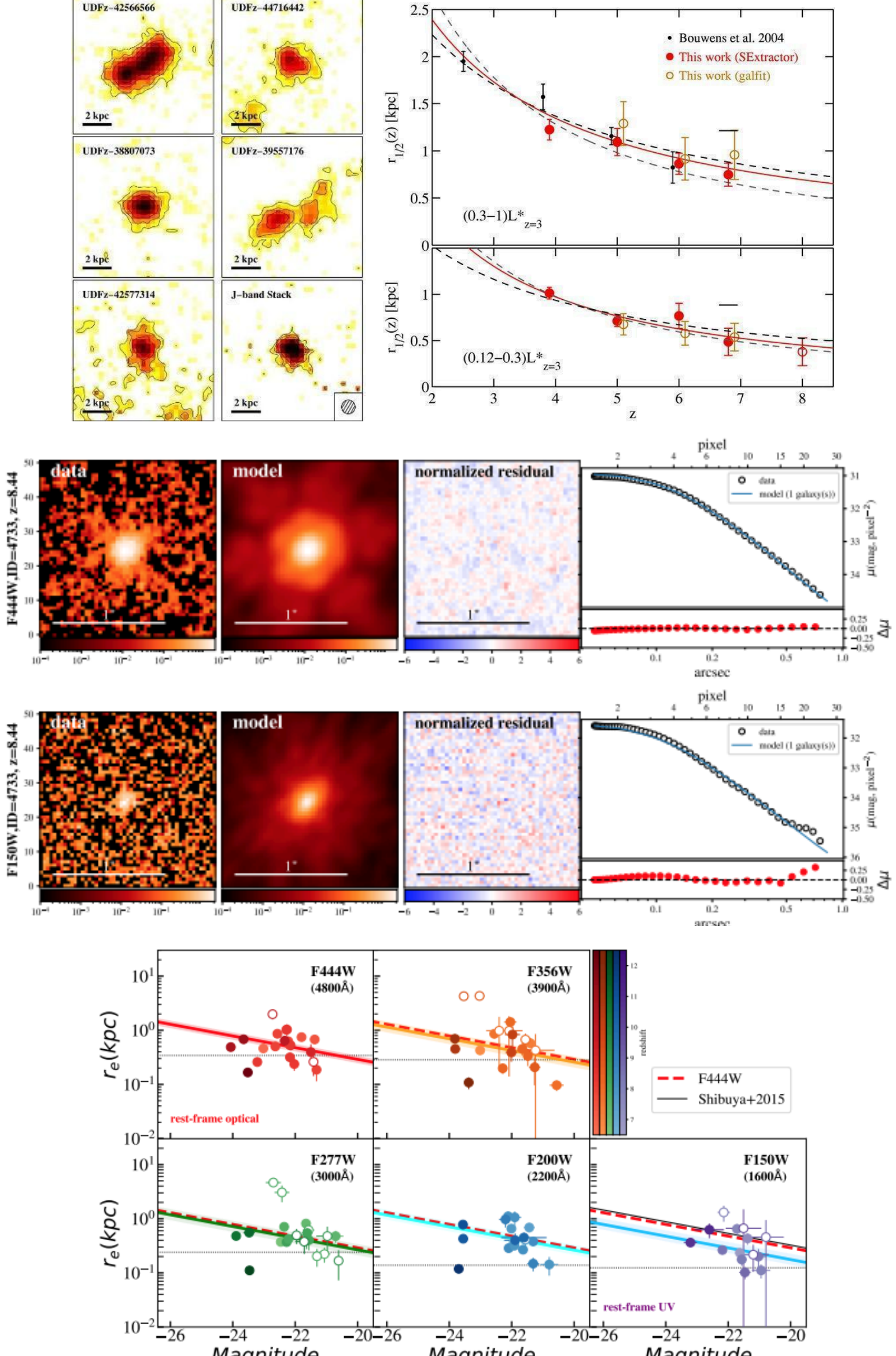

***Figure 4-6:*** *Upper Left: Morphology and size of galaxies at z=7–8 (Oesch et al. 2010). Upper Right: Size evolution of galaxies at 2 < z < 12 (Ono et al. 2013). Middle: Example of size measurements of a galaxy (ID=4733, z = 8.44) in F444W (rest-frame optical; upper) and F150W (UV; bottom) bands using Galight. In each panel, from left to right, the columns represent (1) observed data, (2) best-fit model image, (3) normalized residual map, and (4) surface brightness profiles in 1D and its residual (Yang et al. 2022). Bottom: Size-luminosity distribution of galaxies observed in 5 NIRCam bands from F444W to F150W band, corresponding to rest-frame optical (∼ 4800 ˚A) to UV (∼ 1600 ˚A) (Yang et al. 2022).*

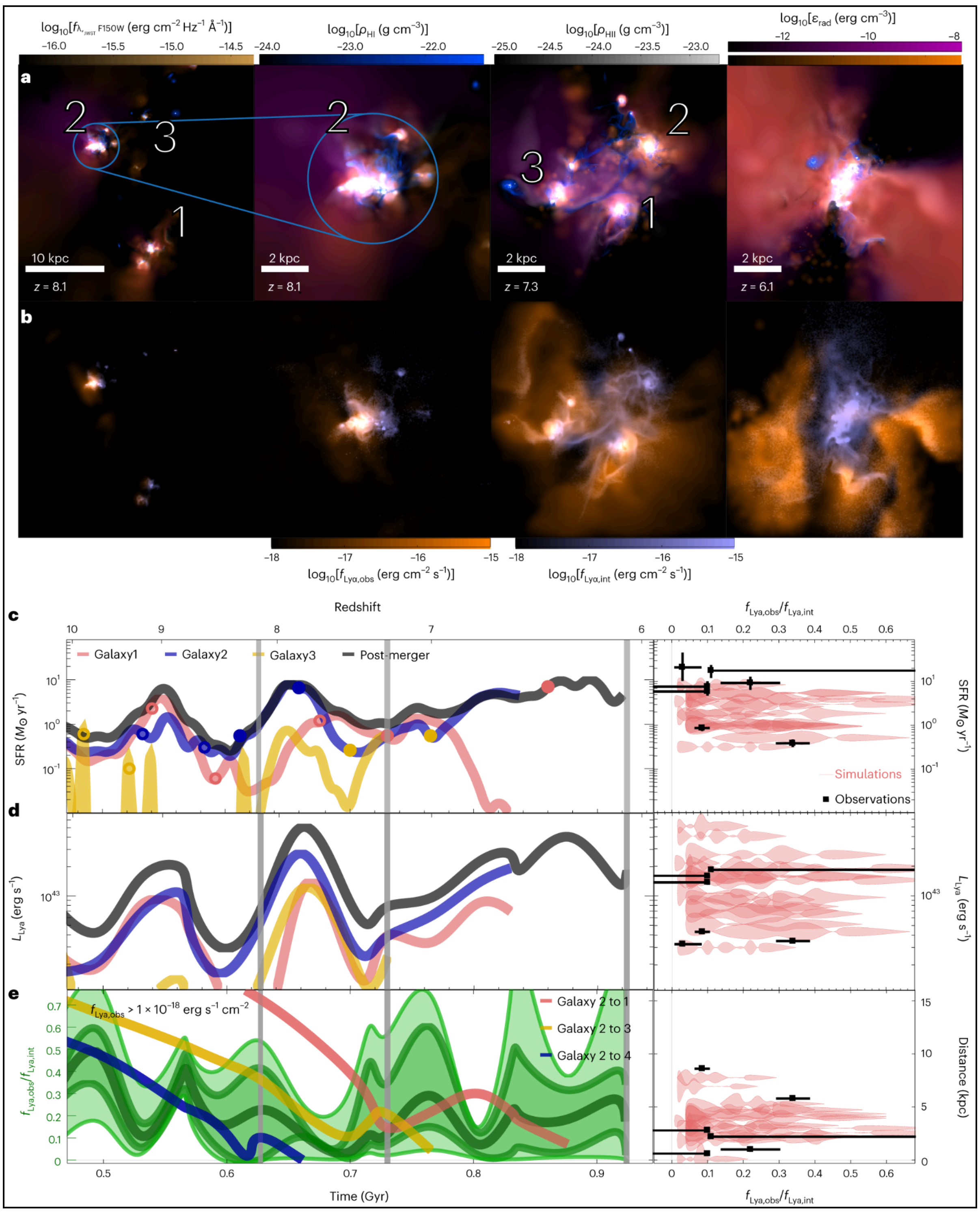

***Figure 4-7:*** *Figure 3 from Witten, et al., 2024, shows views on different spatial scales of a simulated high redshift galaxy merger system at different redshifts/times showing a) HI and HII densities, and b) intrinsic and scattered Lya flux. Row c shows the star formation rate as a function of time in the simulated merger. Row d is the Lya luminosity and e is the escape fraction.*

## 4.3 GRAVITATIONAL LENSING

Detecting galaxies at z > 6 is challenging, not only due to the large distance to these objects (which makes them faint) but also due to the fact that galaxies had few stars when they first formed, and therefore are intrinsically less massive and luminous than galaxies in the more mature universe. To overcome this problem, one can use galaxy clusters as cosmic telescopes that, similar to ordinary glass lenses, magnify objects behind them. These cosmic telescopes allow us to find and study a unique set of faint, high-redshift galaxies, achieving an order-of-magnitude gain in sensitivity over blank-field surveys, and due to the magnified apparent size, allowing us to probe an order of magnitude smaller physical scales. Several large campaigns have taken advantage of this, particularly CLASH (Postman et al. 2012) and the Hubble Frontier Fields (HFF; Lotz et al. 2017).

The results from these surveys have been extremely encouraging. Not only did they deliver a long-time record holder in redshift space (z=10.7, Coe et al. 2013), (see figure 4-8), these kinds of observations allow us to push the intrinsic luminosity limits further than the UDF and study properties of representative galaxies at z ~ 7 and beyond with new and future tools. For example, one can detect these galaxies with Spitzer and measure mature stellar populations as far as z=9.5 (Zheng et al. 2012, Bradac et al. 2014). TMT/IRIS imaging of clusters will discover and resolve star-forming regions in lensed objects more efficiently than even the HFF and CLASH (see section 3.1.7), and in some cases, JWST with TMT's superior resolution.

Targeting sub-L* LBGs spectroscopically at z ~ 7 are very difficult in blank fields due to their faintness. Indeed, the only spectroscopically detected sub-L* galaxies at z > 6.5 to date are a z=7.045 galaxy from Schenker et al. 2012 and a z=6.740 galaxy from Bradac et al. 2012. Both of these galaxies are lensed by a foreground cluster (A1703 and the Bullet Cluster, respectively), confirming the power of cosmic telescopes. Increasing the sample size is crucial, because with measurements of the equivalent-width (EW) distribution in Lyα Emitters (LAEs) we can distinguish between effects of ISM dust and neutral IGM and study the epoch of reionization (Treu et al. 2012, 2013a; Mason et al. 2019). We expect different EW distribution for sub-L* and >L* populations respectively if reionization is playing a role at these redshifts (Stark et al. 2010).

The main missing observational ingredient is a measurement of the EW distribution for both luminous and sub-L* galaxies at the redshifts of reionization. Cosmic telescopes allow us to probe the sub-L* regime particularly efficiently. However, despite the magnification, spectroscopic redshifts have been measured for only a handful of sources close to the reionization epoch. As discussed earlier (section 4.1.1), Lyα may be disappearing due to the increasing opacity of the intergalactic medium at z > 6. TMT detections for much weaker high-excitation emission lines like CIII] 1909 may be the most effective means to measure redshifts for these gravitationally lensed galaxies from the ground.

TMT will revolutionize the field and offer the well-matched capabilities for such of the above studies. In particular IRMOS, with its multiplexing capabilities will allow for an efficient follow up of galaxies behind cosmic telescopes. The one hour, 10-sigma TMT/IRMOS detection limit is $\sim 10^{41}$ ergs $s^{-1}$. With assistance from lensing we can however push these limits by an order of magnitude lower and study 0.1L* Lyman Break Galaxies spectroscopically. Such systems will have a star formation rate of much less than 1 solar mass/year. Deep JWST imaging from the PEARLS program [−1, −2] has revealed numerous high redshift lensed galaxies that would be ideal for spectroscopic follow-up with TMT/IRMOS and serve to illustrate the enormous power of JWST as a cosmic telescope and its synergism with TMT. TMT will give us an answer to whether these seeds of today's galaxies are responsible for reionizing the universe and what the topology of this process was. It will also allow us to explore galaxies and their stars in great detail from the time when the universe was only a few percent of its present age.

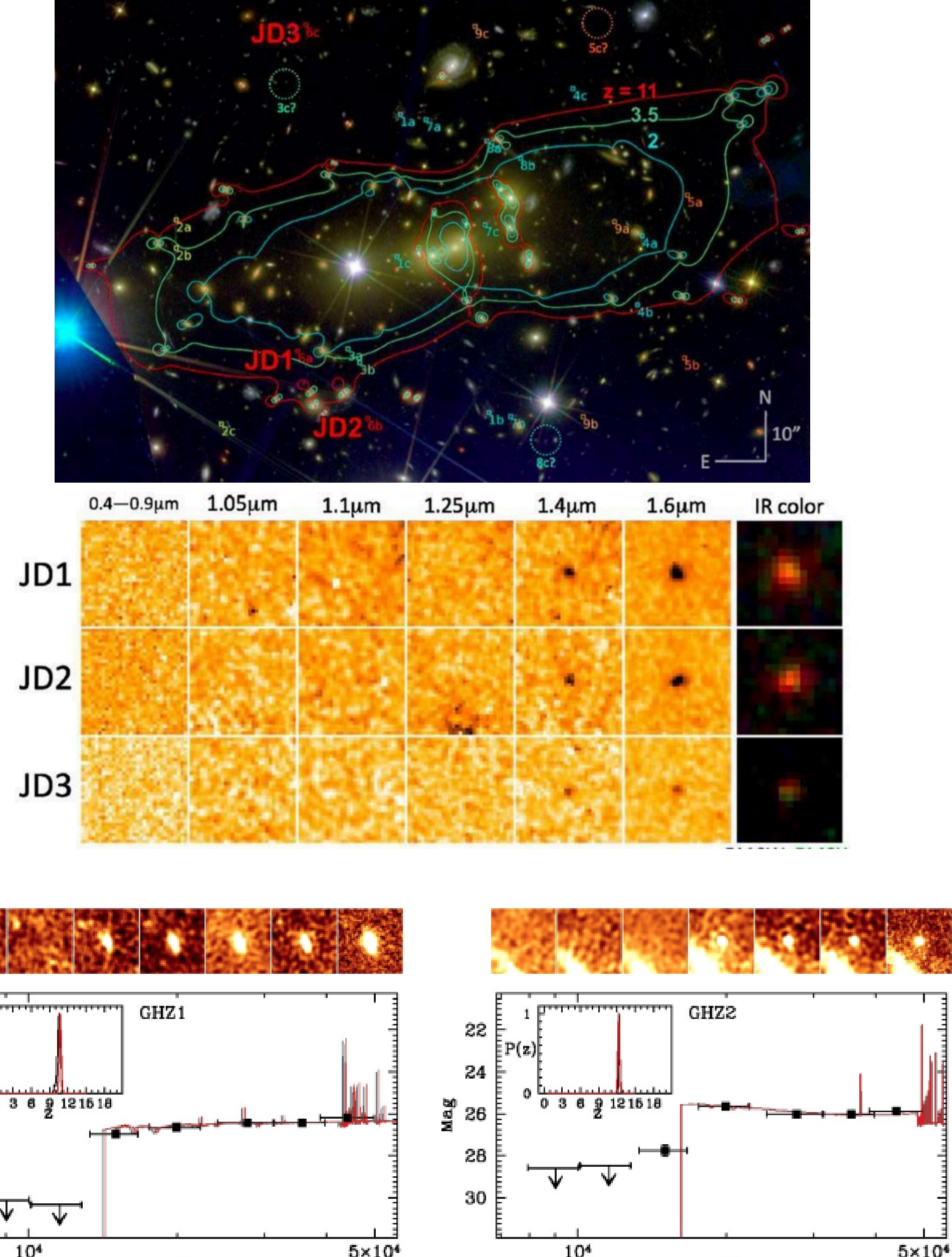

***Figure 4-8:*** *(Top) A massive galaxy cluster, MACS J0647.7+7015, acts as a powerful gravitational lens, producing three highly magnified images of a background galaxy estimated to be at a redshift z = 10.7. HST image of the cluster, from the CLASH survey, with the critical magnification lines at various redshifts (2, 3.5 and 11) indicated. The locations of the three images of the faint galaxy are marked JD1, JD2, JD3. (Middle) Images of the three lensed images from optical through near-infrared wavelengths. The galaxy "drops out" at wavelengths shorter than 1.4 microns, strongly suggesting a very high redshift. From Coe et al. 2013. (Bottom) The two high-quality bright high-redshift candidates from the GLASS-JWST NIRCAM field taken in parallel to NIRISS. Photometry and best-fit SEDs at the best-fit redshift are given in the main quadrant. Redshift probability distributions (PDF(z)) from zphot (grey) and EAzY (red) are shown in the inset. Thumbnails, from left to right, show the objects in the F090W, F115W, F150W, F200W, F277W, F356W and F444W bands.(Castellano et al. 2022).*

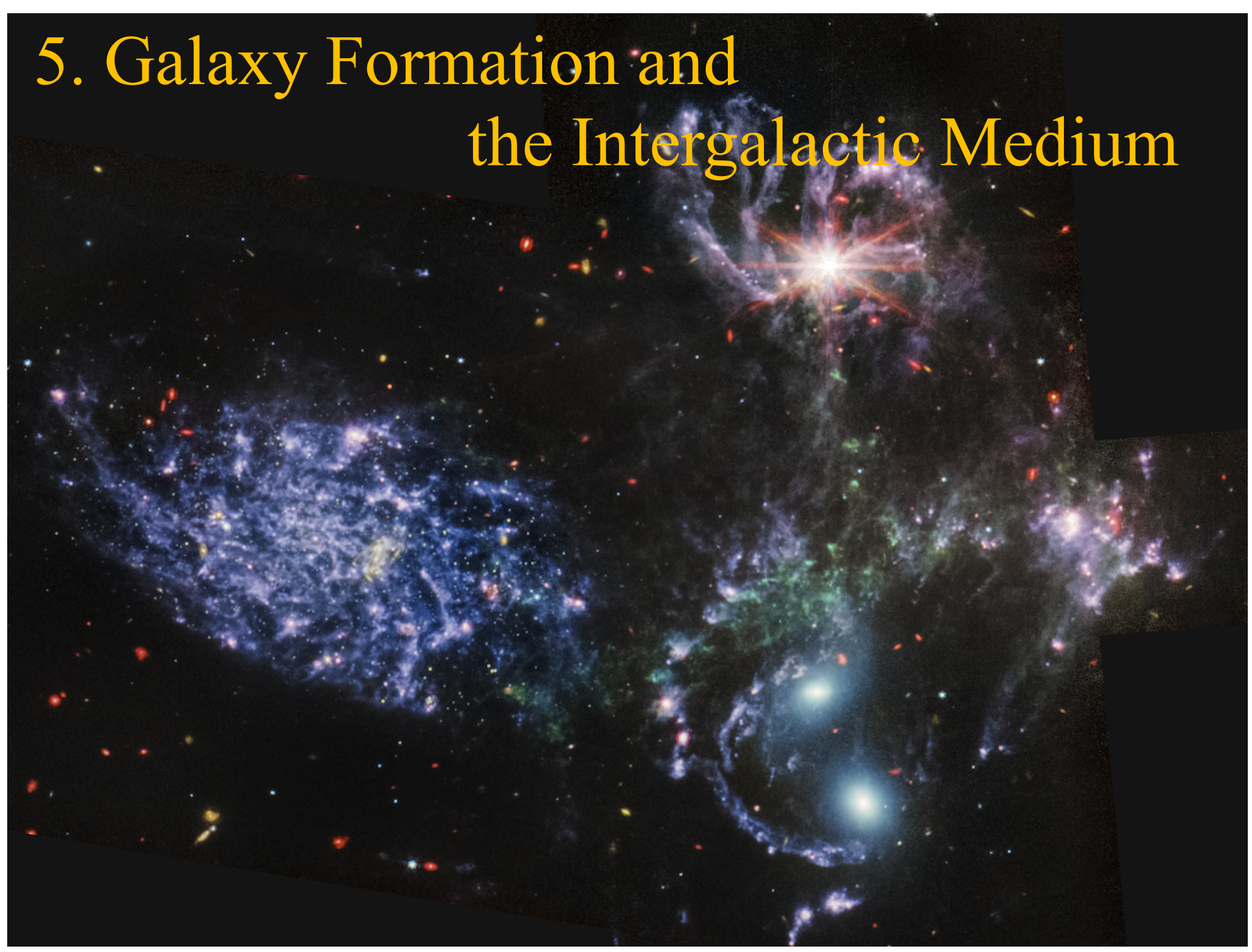

*With its powerful, mid-infrared vision, JWST-MIRI shows never-before-seen details of Stephan's Quintet, a visual grouping of five galaxies. MIRI pierced through dust-enshrouded regions to reveal huge shock waves and tidal tails, gas and stars stripped from the outer regions of the galaxies by interactions. It also unveiled hidden areas of star formation. The new information from MIRI provides invaluable insights into how galactic interactions may have driven galaxy evolution in the early universe. Image credits: NASA, ESA, CSA, and STScI*

---

To address the third big question in chapter 2, Q3-*What is the relationship between black holes and galaxies?*. TMT will conduct a variety of observations that will characterize both the host galaxies of supermassive blackholes (SMBHs) and measure the mass of the SMBHs themselves. TMT will study the chemical and kinematic environment in and around the host galaxy in context with the star formation rate and any activity from the SMBH itself. Such studies will also work to address additional big questions in Q1, Q2 and Q4.

To conduct this range of studies requires ultra-high sensitivity, optical and near-infrared, multi-object, medium and high-resolution spectroscopic capabilities to probe the chemistry and kinematics of the intergalactic and circum-galactic material, and to study more distant galaxies and determine the chemistry and amount of star formation taking place. AO corrected spatially resolved spectroscopy at the TMT diffraction limit is needed to determine the spatial distribution of star formation across galaxies separately from any AGN activity that can also drive outflows and alter star formation processes.

---

**Contributors**: Roberto Abraham (University of Toronto), Michael Cooper (UCI), Taotao Fang (Xiamen University), Roy Gal (University of Hawaii), Mauro Giavalisco (University of Massachusetts), Tadayuki Kodama (NAOJ), Mariko Kubo (Tohoku University), Marie Lemoine-Busserolle (NOIRLab), Jennifer Lotz (NOIRLab-Gemini), Crystal Martin (UC Santa Barbara), Michael Pierce (University of Wyoming), Jason X. Prochaska (UC Santa Cruz), Naveen Reddy (UC Riverside), Kartik Sheth (NRAO), R. Srianand (IUCAA), C.S. Steidel (Caltech), Masayuki Tanaka (NAOJ), Vivian U (UC Irvine), Shelley Wright (UC San Diego), Len Cowie (University of Hawaii)

---

# 5 GALAXY FORMATION AND THE INTERGALACTIC MEDIUM

Tremendous progress has been made over the last two decades in establishing a broad cosmological framework in which galaxies and large-scale structure develop hierarchically over time. However, there remain many unanswered questions about how the observable universe of galaxies is related to the growth of the underlying distribution of dark matter; most of this uncertainty relates to our poor understanding of the complex baryonic processes that must be included in any successful theory of galaxy formation: gas cooling, star formation, feedback, and merging. Understanding how these processes operate on galaxy scales is an inherently multi-wavelength observational problem that is limited at present by the sensitivity of our tools for extracting detailed physical information, for both stars and gas, as a function of cosmic time.

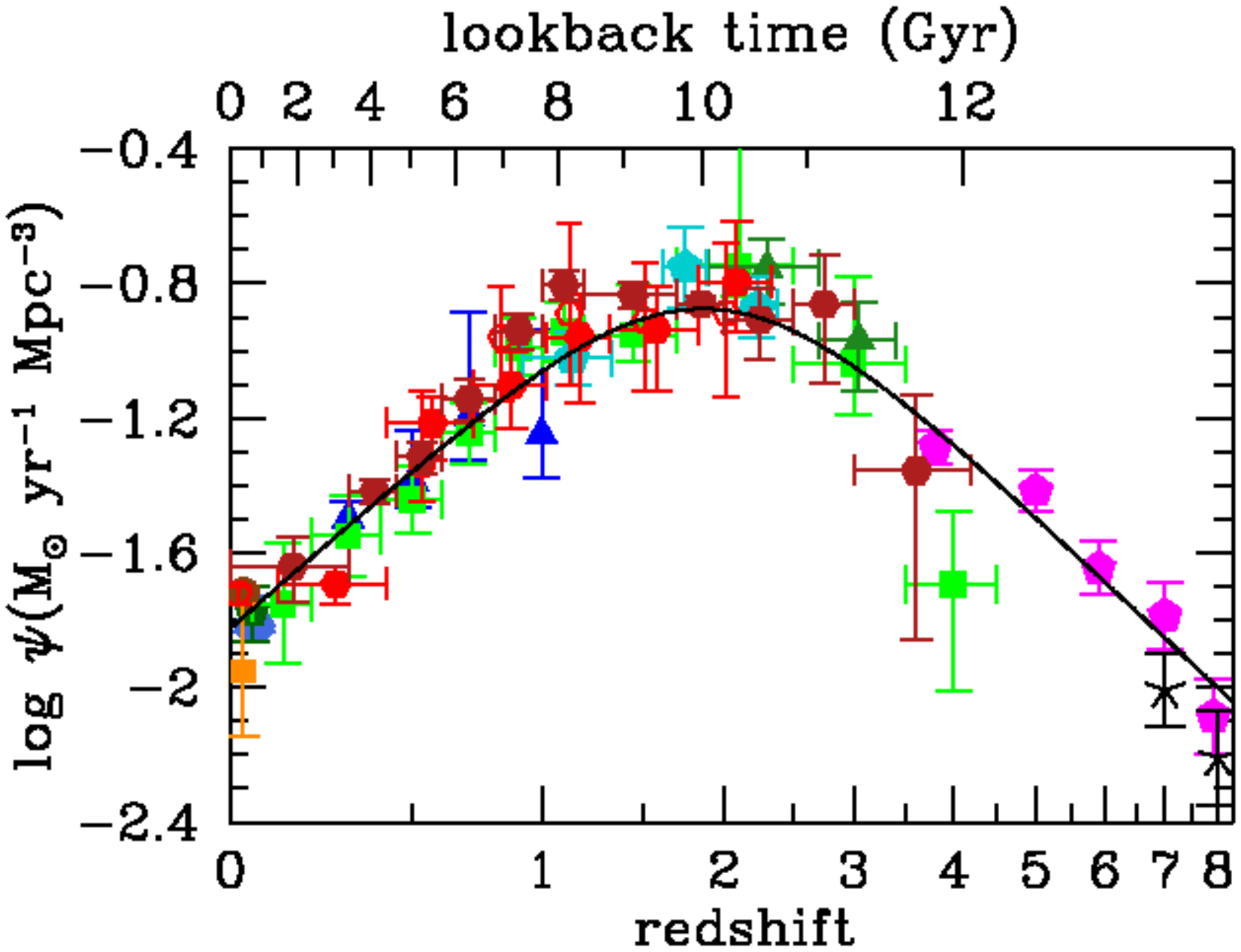

***Figure 5-1:*** *Current census of the integrated star formation rate per unit co-moving volume since redshift $z \sim 8$.. The star formation rate density (SFRD) is ~15 times higher at its peak near $z \sim 2$, when the universe was about 3 Gyr old, than at $z \sim 0$, and increases by a factor of ~25 from $z \sim 8$ to $z \sim 2$. Approximately 50% of the stars in the present day universe were formed over the ~2.5 billion year interval between $z \sim 1.5 – 3.5$. The history of black hole mass accretion follows a surprisingly similar evolution with redshift (from Madau & Dickinson 2014).*

Observations of galaxies beyond $z \sim 1$, selected using initial surveys at UV, optical, near-IR, and far-IR rest-frame wavelengths, have begun to establish in broad-brush what the universal history of star formation has been like over the last ~90% of the age of the universe. As figure 5-1 shows, the current census paints a picture in which the peak era for star formation in galaxies occurred in the distant past; indeed, while precise numbers remain somewhat controversial, there is now little doubt that star formation peaked at redshifts between $1.5<z<3$ and a large fraction of the stars observed in the present-day universe formed during a relatively brief but very intense period stretching between z~6 and z~1.5, corresponding to an interval of cosmic time over which the age of the universe was between 1 and 4 Gyr. JWST is now finding galaxies out to redshifts of 14 or more (Naidu et al 2022), well beyond the limits of figure 5-1. Qualitatively similar behavior with cosmic time to that shown in figure 5-1 is observed for the growth of supermassive black holes — a broad peak near $z \sim 1.5–3$ with a rapid ramp-up prior to $z \sim 3$ and a rapid decline after $z \sim 1$, hinting at a deep connection between between galaxy growth and black hole growth.

What is the interplay between star formation and AGN/supermassive black hole growth? Physical understanding of galaxy formation has been limited by the observational difficulties in studying the detailed physical processes acting on high redshift galaxies, due both to their faintness and their small angular sizes. Through the recent advent of IFU (integral field unit spectroscopy) and adaptive optics (AO) systems on ground-based 8–10 m class telescopes, however, we are just now beginning to witness the relevant processes *in situ* by resolving galaxies both spatially and kinematically; at present, however, such studies are limited to the brightest examples. TMT, with its huge light gathering power coupled with diffraction-limited angular resolution, will revolutionize our understanding of the structure of forming galaxies throughout the peak epoch of galaxy formation.

At all redshifts, most of the baryons in the universe actually reside *outside* of galaxies, in the intergalactic medium (IGM) and circum-galactic medium (CGM). The IGM/CGM is thus a vast reservoir of normal material acting as both a source of gas to fuel galaxy formation, and a sink of the enriched products of that process. The structure of the IGM/CGM is increasingly modified by the energetic processes occurring in forming galaxies, and thereby acts as a laboratory where the most sensitive measurements of the physics of the galaxy formation process may be made. For these reasons, understanding galaxy formation in the universe requires measurements of all material, both inside and outside of galaxies. Fortuitously, observational access to both galaxies and the IGM during this crucial epoch is excellent for large-aperture telescopes placed at the very best terrestrial sites.

TMT will thus play a fundamental role in providing the detailed physical measurements necessary to understand the development of present-day galaxies. TMT will provide rest-frame UV spectra of unprecedented sensitivity and depth: WFOS and HROS will together provide exquisite information on the physics of gas associated with forming galaxies and the IGM using the rich rest-frame far-UV transitions made observable from the ground by the redshift of the sources. IRIS and IRMOS, all taking advantage of the diffraction-limited capabilities of TMT in the near-IR, will provide spectra of extremely high sensitivity and spatial resolution, allowing the first detailed studies of galaxy structure, dynamics, and chemistry while they are in the process of forming the bulk of their stars. The near-IR observations with TMT will enable studies of distant galaxies, as they are growing, using the same well-calibrated rest-frame optical diagnostic spectral features of both stars and gas, that have been established by decades of research on nearby galaxies. Section 12 contains a summary mapping of the science cases discussed here and their dominant pressures on the design capabilities of the TMT and its instruments.

## 5.1 THE PEAK ERA OF GALAXY ASSEMBLY

### 5.1.1 TMT and galaxy formation

In hierarchical structure formation, gravitationally bound structures begin to collapse on small scales, and the largest bound objects are expected to be the last to form. This activity is driven by dark matter, and a universal peak in overall star-formation and black hole accretion, as observed, is not necessarily expected. Neither is the observation that the most massive known galaxies appear to finish forming their stars earliest, nor that typical galaxies at redshifts $z \sim 1$–3 appear to be forming stars at rates that are 10–100 times higher than for typical galaxies in the local universe. The symbiotic relationship between the formation of galaxy spheroids and central black holes also does not follow directly from a cosmological framework alone. The complexity of galaxy formation requires following the behavior of dark matter *and* the astrophysics of star formation, feedback of energy, accretion and expulsion of gas, and effects of the environment. Paradoxically, the former task is easier, even if we know very little of the properties or the precise nature of dark matter, because the physics of baryons are much more complex to model in a cosmological context. To make progress, observations of the baryonic component (both gas and stars) must inform theoretical developments with substantially better precision than is possible today.

Among the questions in galaxy formation for which our knowledge is acutely incomplete are:

- How does the distribution of dark matter relate to the luminous stars and gas that we see?
- Why does the growth of stellar mass and black hole mass in galaxies apparently proceed in an anti-hierarchical fashion — that is, the largest forming earliest? (Commonly referred to as *downsizing*).
- What is the physical mechanism through which galaxies at different epochs acquire their gas, and how much cross-talk and exchange is there between galaxies and the IGM/CGM?

- What internal or environmental effects modulate the rate at which galaxy formation can occur, and what ultimately shuts down ("quenches") star formation in massive galaxies?
- How do energetic processes—massive star formation, black hole accretion, supernovae — influence the formation and evolution of galaxies?
- What physics are responsible for the correlation between central supermassive black holes and the properties of their host galaxies (which is the "chicken" and which, the "egg")?
- What factors drive apparent differences between the low-redshift and high-redshift universe for galaxy scale star formation?
- How do the stellar birth rates as a function of stellar mass (the stellar initial mass function, or IMF) differ, and what are the ramifications of the differences?

Answers to these questions will come from detailed studies of galaxies and the intergalactic medium at high redshift. One needs to measure the distribution and metallicity of gas inside and outside of galaxies, the star formation rates, internal structure, stellar content, and dynamical states of galaxies, all as a function of environment and of cosmic time. The combination of WFOS, IRIS, and (later) IRMOS and HROS mounted on TMT provide a near ideal instrument suite for attacking the problem of structure formation in the distant universe. Across a wide wavelength range (0.3–2.4 µm), TMT complements or exceeds many of the spectroscopic capabilities of JWST while providing higher spatial resolution. Observations with JWST will provide complementary measures out to z-8 (Shaerer et al. 2022). Figure 5-2 shows the unparalleled power of IRMOS to make the required observations. In combination with ALMA, which tracks emission from dust and provides information on the molecular gas content and kinematics of galaxies at similar spatial resolution, TMT + AO will be a powerful tool for tracking the process of galaxy formation during the most active period in the history of the universe.

Broad wavelength coverage (0.3–2.4 µm) is key to probing the critical spectral diagnostics, from the far-UV stellar, interstellar, and intergalactic lines of H and metals, to the nebular emission and stellar absorption features in the rest-frame optical. Figure 5-3 shows examples of low-resolution spectra obtained using current 8 m class telescopes for some of the brightest star-forming galaxies at redshift $z \sim 2$, in order to illustrate some of the spectral features of interest. The planned TMT instrument suite (see section 2.3) will play complementary roles in addressing these fundamental issues:

- WFOS is optimized both for obtaining identification-quality spectra of the faintest galaxies and AGN throughout the dominant epoch of galaxy formation, and for high quality rest-frame UV spectra of large samples of galaxies and AGN that will provide diagnostics of gas and stars in galaxies, as well as the distribution, metallicity, and physical conditions of the IGM. Since most of the baryons are outside galaxies, 'fueling the fire' of star formation, connecting the physics of the IGM to the properties of galaxies is central to understanding how galaxies form and evolve.

- IRIS will dissect galaxies with velocity resolution of 30 km/s on physical scales as fine as 50–70 pc at any redshift in the range z=1–6.

- An IFU based IRMOS behind an MOAO system will combine the diagnostic power of IRIS with multiplexing capabilities, providing integral field spectroscopy in the near-IR (0.8–2.5 µm) for multiple objects distributed over a several arcmin field of view, allowing the study of key diagnostic nebular and stellar lines in the rest-frame optical (e.g., [OIII], Hβ, Mgβ, Hα, [NII]). IRMOS will obtain extremely sensitive multiplexed spectra in both emission lines and stellar continua in the observed frame near-IR, providing information on chemistry, kinematics, and stellar populations.

- HROS on TMT will have significantly greater sensitivity than the best current-generation high resolution spectroscopic capabilities, vastly increasing the number of lines of sight that can be used for the most detailed studies of the IGM and of individual galaxies and AGN.

Below, we highlight a few of the possible science avenues that can be explored using these capabilities

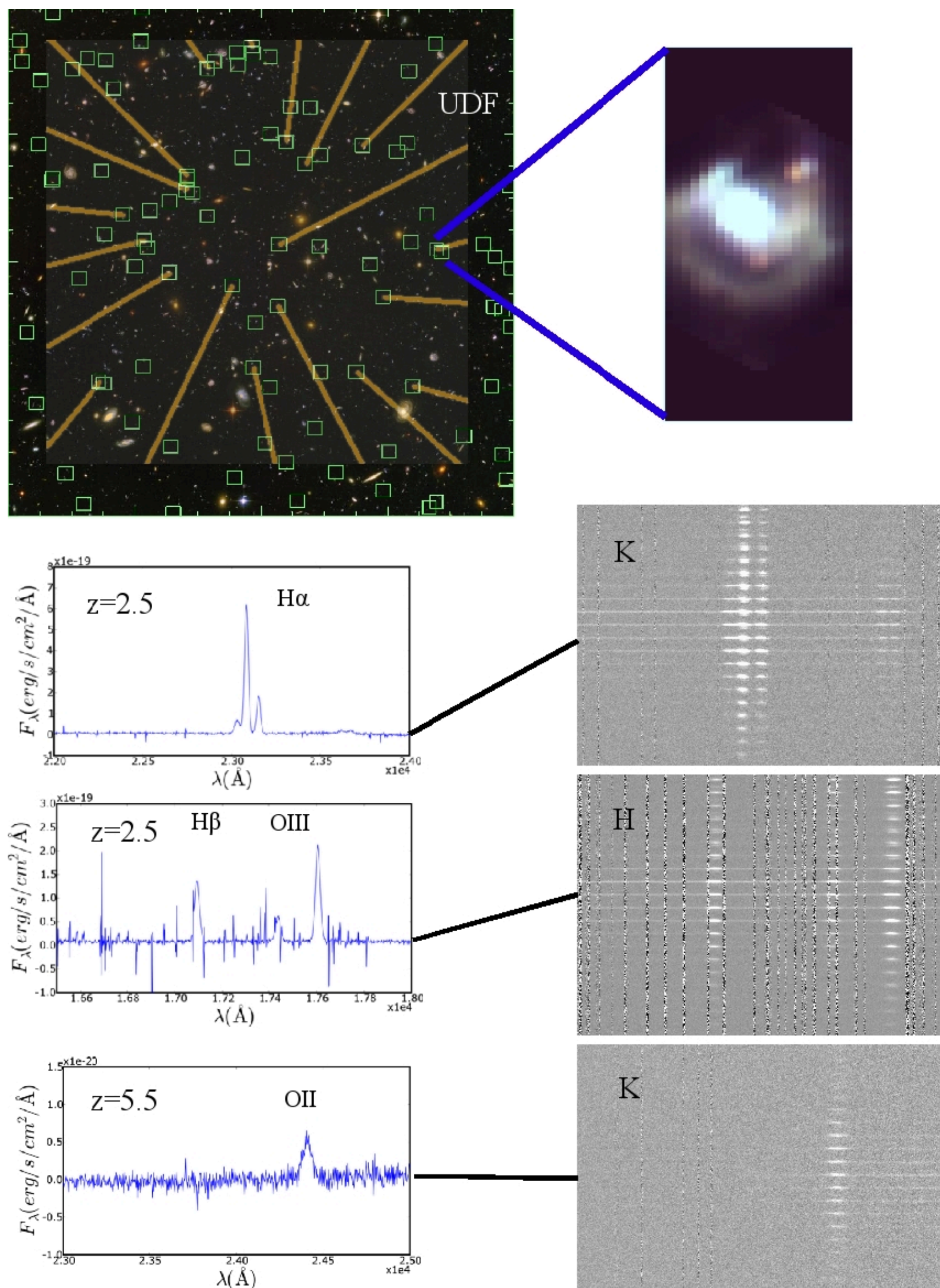

***Figure 5-2:*** *Simulated IRMOS observations of a typical Lyman-break galaxy at redshifts z = 2.5 and z = 5.5. (Top left) Image of the Hubble Ultra Deep Field (UDF). Lyman-break galaxy candidates at z = 2–6 are identified with green squares. MOS probes configured to cover the candidates are shown as yellow lines. Top right panel: simulated deep near-IR image of a typical Lyman-break galaxy as seen in a single IRMOS IFU. Middle panels: simulated 2D and 1D spectra showing characteristic emission lines of Lyman-break galaxies at z = 2.5 and 5. Courtesy: IRMOS-UF/HIA team.*

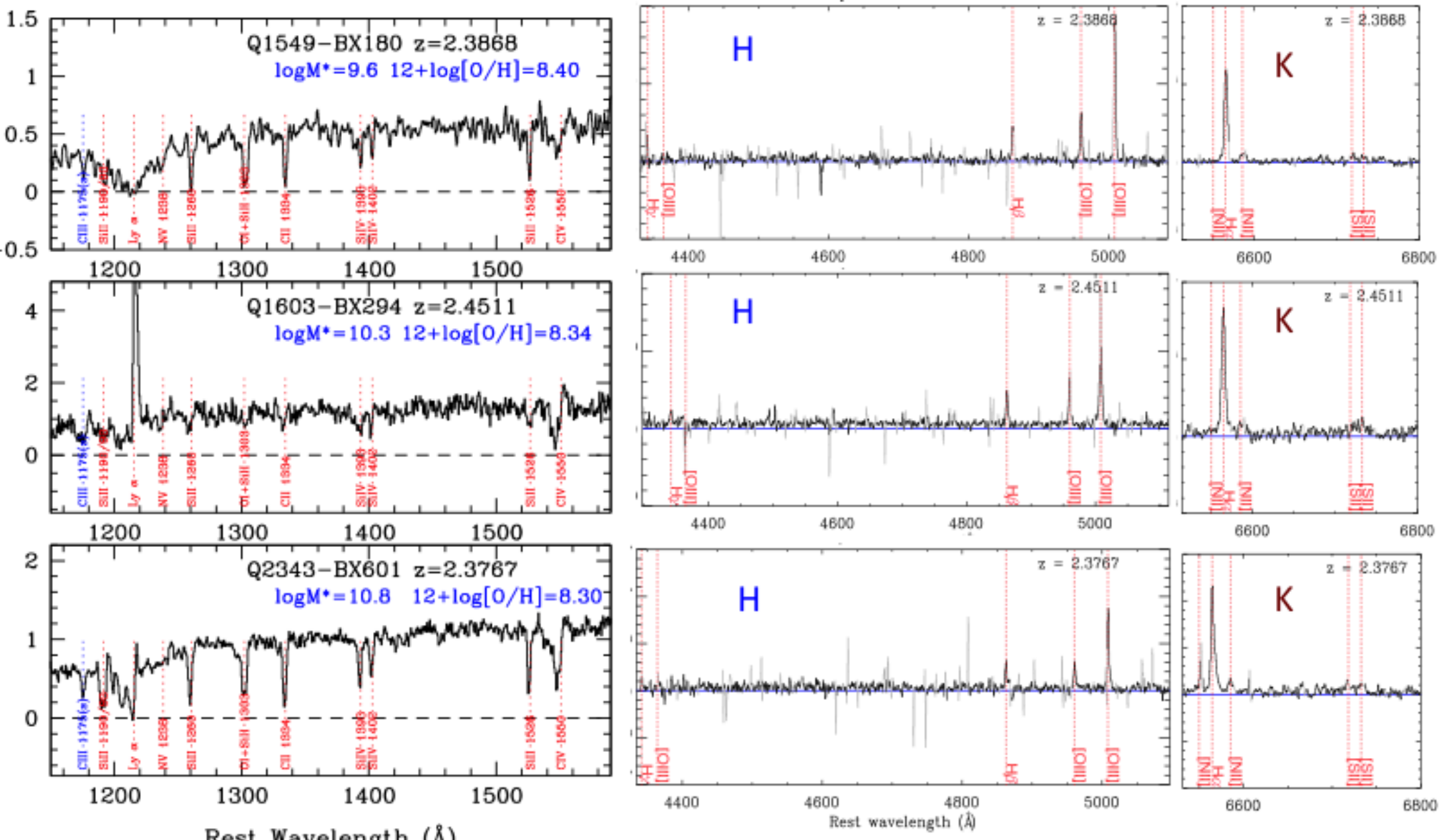

***Figure 5-3:*** *Example spectra of 3 z~2.4 galaxies; the left panels show the rest-frame far-UV spectra observed in the optical (0.31–0.6 µm), while the center and right panels show the rest-optical nebular emission lines observed in the H (1.6µm) and K (2.2 µm) atmospheric windows. The rest- far-UV spectra contain a wealth of interstellar absorption features, nebular emission lines, and stellar wind and photospheric features from massive stars. The rest-optical spectra are used for measuring kinematics, gas-phase metallicities, and ISM physical conditions. Credit: C. Steidel (CIT).*

### 5.1.2 How does the distribution of dark matter relate to the luminous stars and gas we see?

In the present day universe, the distribution of galaxy stellar masses (M*) described by the stellar mass function appears to have a different shape from that of the theoretically expected distribution of dark matter halo masses. Galaxies with small M* have a much flatter distribution (i.e., fewer galaxies with lower M*) than expected if galaxies formed with the same ratio of stellar-to-dark matter at all masses. At high M*, the stellar mass function drops exponentially with increasing halo mass, following the Schechter function (e.g., Cole et al. 2001). This behavior, where galaxy formation appears to be less efficient at low- and high-M*, is almost universally ascribed to “feedback”; the most prevalent qualitative explanation at present is that ionizing photons, energy, and momentum produced by massive star formation and supernovae are the dominant effects for lower-M* galaxies, while AGN (i.e., energy/momentum produced by accreting supermassive black holes) play the dominant role at the high-M* end of the galaxy mass function.

While the need for feedback is universally recognized for any viable galaxy formation model, it is fair to say that we do not yet understand how, when, and where it operates, nor which of the many physical processes dominates at a given time in a given galaxy. Making progress will require observing signatures over a range of galaxy masses that are sensitive to the details of all of the relevant physical processes within each galaxy, and over a range of cosmic times. These observations, at a minimum, must include measurement of AGN activity, massive star formation, stellar content, and gas-phase physical conditions both inside forming galaxies and in the surrounding IGM/CGM. TMT, using multiple instruments and techniques that each constitute a huge gain in sensitivity, spatial and spectral resolution over existing facilities, will play a central role in future science in this field. TMT’s extremely high spatial resolution behind AO, combined with the diagnostic power of spectroscopy (AO+IRMOS, IRIS, and MODHIS) in the near-IR, and unmatched complementary sensitivity for spectroscopic observations in the observed-frame optical (WFOS, HROS), will provide access to the relevant astrophysical diagnostics needed in the rest-frame UV (figure 5-3).

### 5.1.3 The Growth of Stars: Star-Formation Histories, Dust, and Chemical Evolution

Understanding galaxy evolution requires an understanding of the mechanisms by which galaxies accumulate their gas and form stars. The effect of these mechanisms (e.g., gas inflows, outflows, galaxy mergers) is imprinted on the star-formation histories of galaxies. From the theoretical side, hydrodynamical simulations suggest that galaxies living in relatively massive halos at intermediate redshifts accumulate most of their gas through relatively smooth accretion, rather than through major mergers, resulting in star-formation histories that can be well-approximated as smoothly varying functions, typically as exponentially declining or rising, or constant (e.g., Noeske et al. 2007a; Noeske et al. 2007b; Reddy et al. 2012). Clustering studies and cosmological simulations suggest that star-formation histories may be more stochastic at lower stellar masses ($\leq$109 $M_\odot$) and at higher redshifts (z>1.5). Outflows in faint galaxies may be enhanced by short bursts of star formation, and because fast-moving gas can more easily escape from less massive halos, low mass galaxies may be responsible for a majority of the globally-averaged metal enrichment of the IGM (Oppenheimer et al. 2012).

A promising method to deduce the level of stochasticity in star formation is to use probes that are sensitive to star formation on different timescales, such as UV continuum and Hα emission (Flores Velázquez et al., 2020). Measuring these tracers for very faint and low mass galaxies will become routine with TMT's WFOS and IRMOS instruments, and can be used for detailed studies of the "burstiness" of star formation in low mass galaxies. Adding information from WFOS and HROS on the CGM/IGM occupied by these objects will lead to a more complete picture of IGM chemical enrichment from the outflows of faint galaxies. Further constraints on stellar populations will come from modeling high-resolution spectra (e.g., from WFOS and HROS) of stellar emission/absorption features in the rest-UV, giving detailed information on the massive star populations and the impact of binarity and mass loss on the effective temperatures, ionizing parameters, Lyman continuum escape fractions, and recombination line strengths in HII regions. Similarly, rest-optical spectroscopy of the Balmer absorption lines, Ca HK, MgB, and Fe lines will yield constraints on star-formation history from both overall metal enrichment and measurement of metallicity ratios [α/Fe].

The interpretation of high-redshift galaxy observations depends critically on the amount of dust obscuration impacting these objects. Local prescriptions for dust corrections appear to work reasonably well on average for ~L* galaxies at z~2, but there are subsets of galaxies (e.g., the youngest galaxies, and the dustiest galaxies) where such relations break down (e.g., Reddy et al. 2010). At redshifts of z~5, there is evidence that the attenuation curve in typical Lyman-break galaxies may be more consistent with that found in present-day low metallicity galaxies, like the Small Magellanic Cloud (Shim et al. 2011; Capak et al 2015). WFOS, IRIS, and IRMOS will enable detailed studies of the shape of the attenuation curve, potentially for individual galaxies where the intrinsic spectrum revealed through dust-free lines of sight can be spatially disentangled from the dustier regions of the galaxies. The measurement of multiple Balmer line ratios (Hα/Hβ, Hβ/Hγ, Hγ/Hδ) will constrain the shape of the nebular reddening curve in individual galaxies. Together, the stellar and nebular reddening curves can be used to obtain more robust constraints on the stellar masses, ages, reddening, and star-formation rates of galaxies, yielding a better understanding of the impact of nebular line emission on stellar population modeling from photometry alone.

In addition to providing a better understanding of the effects of dust obscuration, TMT will enhance our understanding of gas-phase abundances (i.e., the metallicity) and dust formation, which can be used as a complementary probe of star-formation history and chemical evolution. Gas-phase metallicities measured with rest-optical emission lines, in concert with galaxy stellar masses and star-formation rates, provide important empirical constraints on the effects of galactic winds (e.g., Davé et al. 2012) on the chemical evolution of galaxies. IRIS and IRMOS will extend what is possible using current near-IR spectrometers on 8 m class telescopes, bringing into reach diagnostics of metallicity for galaxies as faint as L*/100 at redshifts 1.4<z<3.8. High spatial resolution emission line maps will reveal variations in kinematics, excitation mechanisms, stellar populations, and metallicities as a function of spatial location within galaxies—all essential ingredients to understanding galaxy formation and evolution.

### 5.1.4 The formation of passive galaxies and the birth of the Hubble Sequence

A fundamental property of galaxies that still eludes our full understanding is the Hubble Sequence, including the physical mechanisms that lead to the formation of early-type galaxies, namely the quenching of star formation. Gas removal and heating by feedback from AGN and/or star-formation have long been identified as likely mechanisms responsible for quenching, but the processes that terminate star formation on galaxy scales and prevent star formation from happening again over long time scales have not been firmly identified yet.

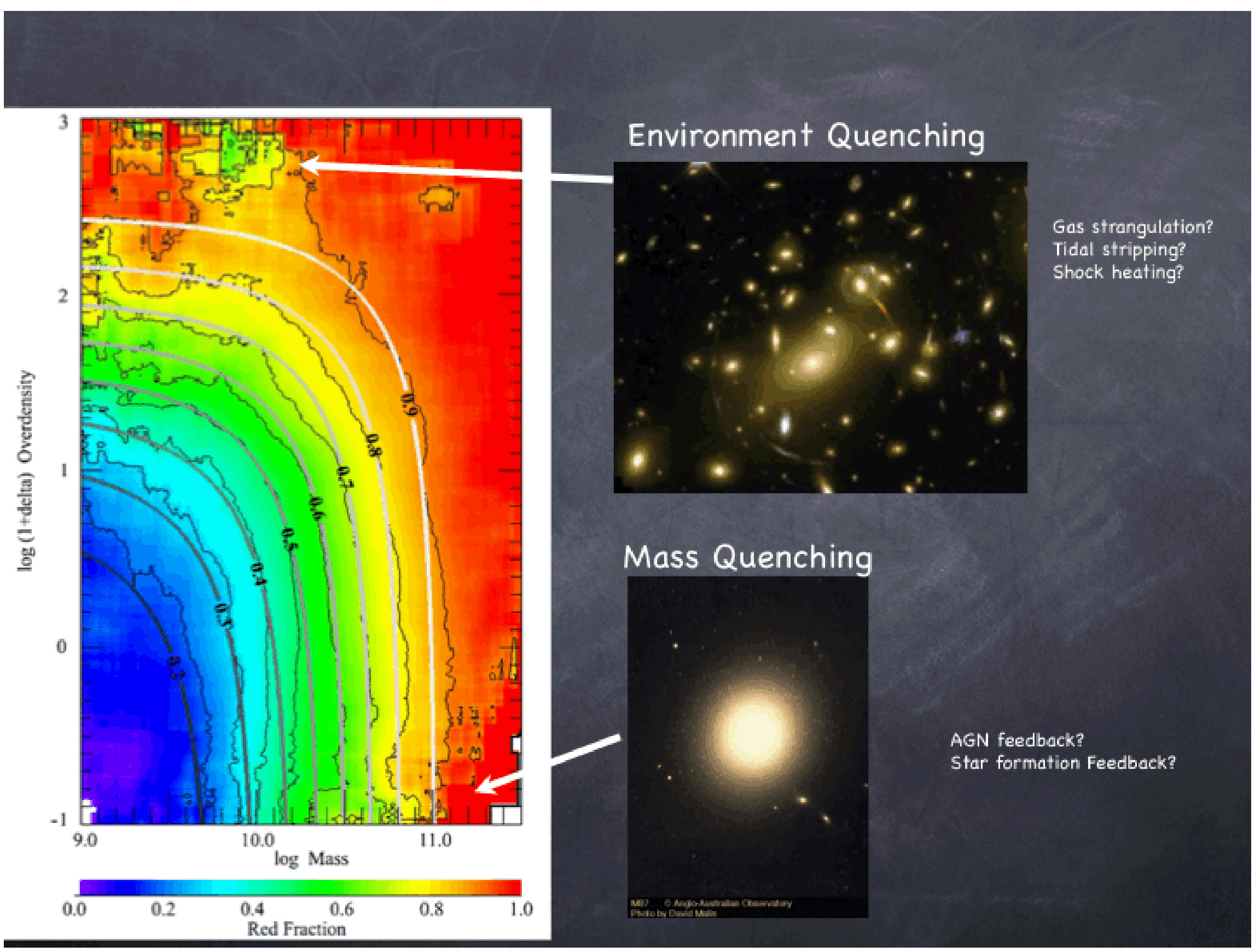

***Figure 5-4:*** *The fraction of red and passive galaxies, a proxy for the probability that a galaxy's star formation has been "quenched", as a function of local overdensity and stellar mass. Evidently, processes related both to the mass of individual galaxies and to their environment cause the termination of star formation, with the former being the primary driver and the latter contributing at the ~10% level (from Peng et al. 2012). Which stellar mass-dependent processes control the quenching of star formation activity remains unknown.*

Massive quiescent galaxies have been found at redshifts up to z~4 (e.g., Forrest et al. 2022; Tanaka et al. 2024; Valentino et al. 2023) and higher (z=4.65 to date) (e.g., Carnall et al. 2023) have been confirmed using 10-m class ground-based telescopes and JWST, respectively. The Small sky coverage of JWST means that relatively massive quenched galaxies must be abundant at high redshifts (e.g., Valentino et al. 2023). This is puzzling, because most cosmological simulations suggest that, at z >2, high mass galaxies can maintain high star formation rates over long timescales from continuous accretion of cool gas from the IGM (e.g., Kerez et al. 2005; Dekel et al. 2009), while at lower redshift both the accretion rate and the physical state of accreted gas changes at fixed stellar mass, leading to a dwindling of the gas supply available for star formation. While direct evidence that "quenching" of high mass galaxies must occur is plentiful, the processes physically responsible for it have not been clearly identified. At present, it seems that quenching depends on both galaxy mass and larger-scale environment (e.g., Peng et al. 2012) as suggested in figure 5-4. AGN are predicted to play key roles in mass quenching (Croton et al., 2006; Somerville et al., 2008) and several massive quiescent galaxies show strong evidence of AGN outflows (e.g., Kubo et al. 2022; Kocevski et al. 2023; Carnall et al. 2023).

At z>2 the population of galaxies in which star formation has ceased (i.e., they have become "passive'') is dominated by those having very compact morphologies and relatively large stellar masses (e.g., van der Wel et al. 2014; Ito et al. 2023), suggesting that these systems are the first to quench and pointing to very high density of gas and stars as conditions conducive to effective "mass" quenching (Bell et al. 2012). Indeed compact, massive post-starburst galaxies have been observed with outflow speeds in excesses of ~1000 km/sec (Diamond-Stanic et al. 2012), significantly higher than those of less compact galaxies of similar stellar mass. Very deep stacks of WFC3/HST images of compact passive galaxies at z~2 have found extremely steep light profiles (n>3) and no evidence of the more diffuse component expected in merger remnants (e.g., Hopkins et al. 2008; Wuyts et al. 2010), adding support to the notion that early massive galaxies formed more effectively via highly dissipative gas accretion than by hierarchical merging of stellar sub-systems (Hopkins et al. 2008; Williams et al. 2014) and possibly were subsequently quenched when infalling ICM gas met strong outflows from an active AGN at their core (Lacerda et al. 2020). Meanwhile, the latest large-volume cosmological numerical simulations have predicted that galaxy quenching does not immediately cause morphological transformations (Cortese et al. 2019; Park et al. 2022). Indeed, several disky quiescent galaxies at z>2 have been reported (e.g., Toft et al. 2017; Newman et al. 2018; Ito et al. 2023). Our understanding of the morphological evolution of galaxies at z>2 remains quite incomplete; even with the capability of JWST, $M_\star > 10^{11}\ M_\odot$ quiescent galaxies at z=3-4 are hardly resolved (Ito et al. 2023). Higher spatial resolution studies are needed to understand how quenching in galaxies has proceeded, and the limited samples afforded by strong lensing studies provide limited scope for this work.

To make progress in understanding what causes star-formation to cease in high-redshift galaxies, we need to explore the complex multi-dimensional manifold of galaxy properties and local environment. Diffraction limited near-IR images with TMT will detail the morphology of the individual galaxies at optical rest-frame wavelengths with better spatial resolution than JWST NIRCam. TMT spectroscopy at optical and near-IR wavelengths, where most of the relevant spectroscopic diagnostics of gas kinematics, stellar dynamics, star-formation, and AGN activity are found for z<4, will allow detailed studies of the stars and gas of individual galaxies and of their circumgalactic gas (via absorption features of suitable background sources, as discussed in section 5.2 below).

### 5.1.5 The Stellar Initial Mass Function, Early Black Holes and the Growth of Quasars

A variety of observational data suggest the possibility of environmental dependence of the shape of the stellar initial mass function in galaxies over cosmic time. The clues arise from observations of the rates of high-redshift gamma-ray bursts (e.g., Robertson et al. 2012, Chary et al. 2007), specific star-formation rates of high-redshift Lyman break galaxies (Shim et al. 2011, Davé et al. 2008) and the metal yields in Damped Lyman-alpha systems (Cooke et al. 2013, 2014). Observations of low metallicity dwarf galaxies in the Milky Way's neighborhood further argue in favor of an IMF that is "bottom-light" e.g., showing an underabundance of stars between 0.5–0.7 $M_\odot$ compared to a Salpeter IMF. Is there a metallicity dependence to the IMF? Does a top-heavy IMF in the first galaxies provide the intermediate mass seed black holes for the formation of quasars? These are key questions which require spectroscopic follow-up of both star-forming and AGN candidates in the distant universe identified through wide-field imaging surveys.

For example, spectroscopic measurements of the rest-frame far-UV spectra of LBGs offer direct constraints on the stellar birth rates at the "top end" of the stellar IMF, as well as the duty cycle of unusual and/or short-lived phases of high mass star formation (e.g., Pettini et al 2000; Shapley et al 2003; Jones et al 2012). The same spectra provide constraints on O-star metallicity, nebular abundances and excitation, and models of massive star main sequence evolution (see figure 5-5). The most important UV features, found in the rest-wavelength range $1000 < \lambda_0 < 2000$ Å, are observationally accessible (using WFOS and IRMOS) at all redshifts $1.5 < z < 7$.

Measurement of Fe-peak element abundance patterns in Damped Lyman Alpha QSO absorption line systems (DLAs; e.g., Cooke et al. 2011), have proved to be a powerful technique for tracing the nucleosynthetic yields of massive stars. At the present time, there is considerable degeneracy in the stellar population synthesis models such that the elemental yields can be fit very well by several models of intermediate mass stars (15–40 $M_\odot$). Precision measurements of the [C, O/Fe] abundances in metal poor systems will be a powerful technique to constrain the mass and metallicity of the stars responsible for enriching the DLA systems and potentially the stellar IMF at intermediate masses.

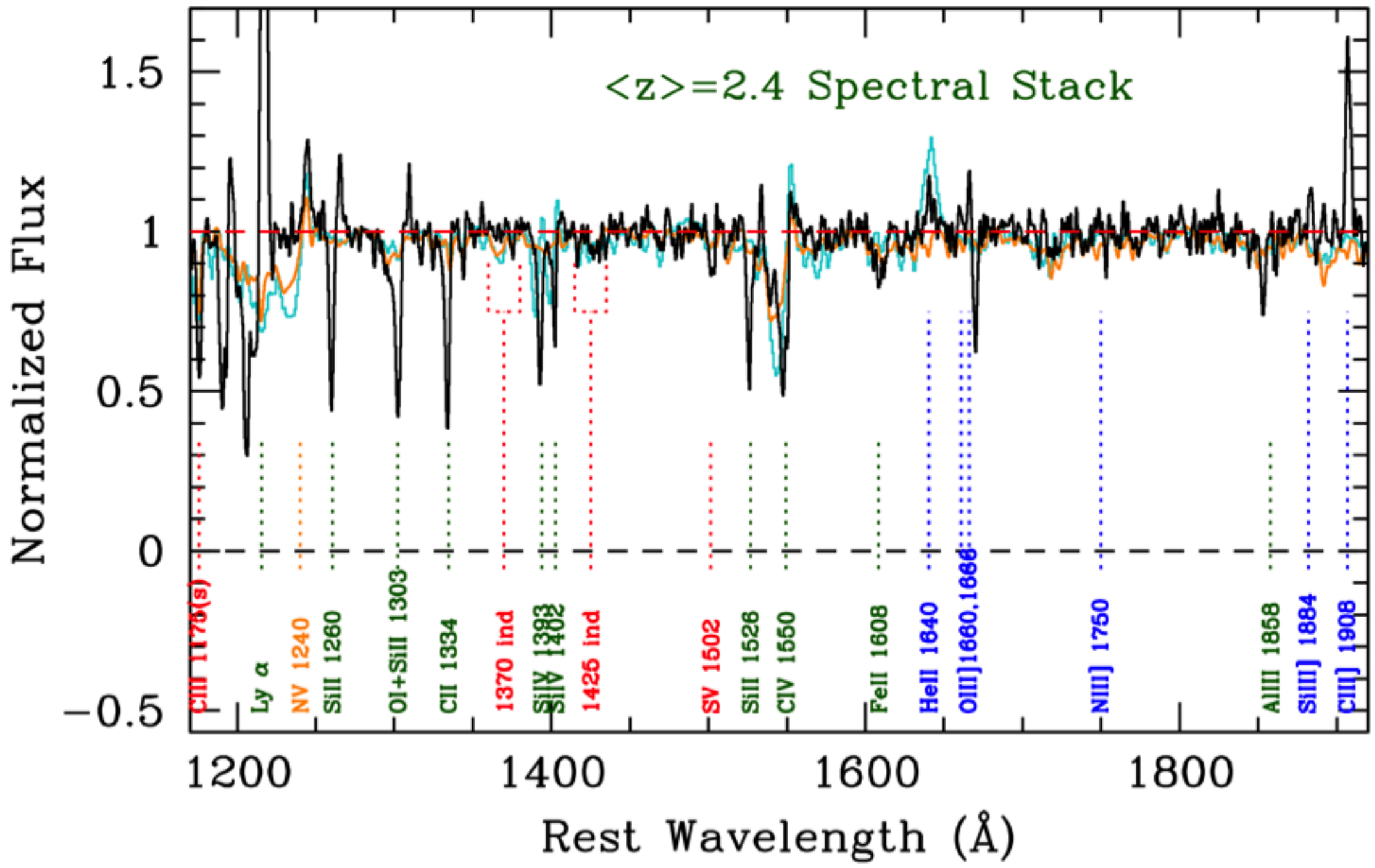

***Figure 5-5:*** *Rest-UV composite spectrum of star-forming galaxies at <z>~2.4, obtained with R~1500. The labeled features are color-coded according to their origin: red=stellar photospheric (O-stars); blue=nebular; green=primarily interstellar; orange: hot star wind lines. The CIV feature includes both a stellar P-Cygni emission/absorption as well as interstellar; HeII includes both nebular (narrow) and stellar (broad) components. Superposed on the observed spectra are population synthesis models, both assuming Salpeter IMF and 0.2 solar metal abundance: cyan: BPASS (Eldridge & Stanway 2009); light orange: Rix et al (2004).*

Finally, the presence of very bright QSOs at z>6 (see also section 4.1.3) has come as a significant surprise, since it implies very early formation of black hole masses as large as $10^{10}$ $M_{\odot}$. Do these QSOs harbor significant stellar populations in their host galaxies? Are the black holes being fed via "super-Eddington" accretion on to intermediate mass black holes? Are these intermediate mass black holes the end states of the earliest generations of massive stars? Measuring the metal abundances in QSO host galaxies, as well as studying the dynamics of the gas in these systems, will provide a better understanding of the formation of the most massive black holes at early times. High spatial resolution, IFU observations are key to separating QSO light from host galaxy light. Clever techniques to obtain high dynamic range in the observations will likely need to be pursued while spectroscopic measurements of the gas kinematics and outflows will provide a better handle on the accretion rates of these systems.

### 5.1.6 The Census of Baryons and the Baryon Cycle

Most baryonic matter is actually outside of galaxies. Intergalactic space is filled with a tenuous gas that provides "fuel" for forming galaxies and is a repository for material expelled by galaxies by energetic processes such as supernova explosions and accretion-powered AGN activity. The interplay between the galaxies and the intergalactic medium is of prime importance for understanding the history of normal matter in the universe, the process of galaxy formation, and the effects of the "feedback" of energy produced by forming galaxies. A number of studies have found the ubiquitous presence of galaxy-scale outflows at high redshift, and the direct correspondence between HI and metal absorption systems in the intergalactic medium (as probed by background quasars) and galaxies in the same volumes (e.g., Pettini et al. 2002, Adelberger et al. 2003, Steidel et al. 2010, Rudie et al. 2012, Hennawi & Prochaska 2013; Turner et al 2015). The imprint of such galaxy flows can be seen most readily in the rest-UV, where the naturally dark terrestrial background in the optical (defined for the present purposes as 0.31–1 µm) makes spectroscopy of faint objects feasible for at least the brightest examples with 8–10 m telescopes. As illustrated earlier in figure 5-3 the rest-UV spectra of galaxies contain a vast quantity of information because of the very large number of ground-state transitions of astrophysically abundant ions at these wavelengths, including hydrogen Lyα, CII, CIII, CIV, NI, NV, OI, OVI, SiII, SiIII, SiIV, FeII, AlII, AlIII, to name a small subset. The offsets between lines arising from the nebular regions (e.g., HeII in the rest-UV, or [OII]/[OIII] and Balmer lines in the rest-optical), and those arising from the ISM allow for direct measurements of galaxy-scale outflow velocities, along with a host of other physical information including the shape of the stellar initial mass function (IMF) at the high mass end, stellar chemical abundances, the chemistry of the interstellar gas, and the kinematics of gas motions within the galaxy. Currently, these details can be ascertained with spectral stacks or for individual bright lensed LBGs (e.g., cB58; Pettini et al. 2002).

***Table 5-1:*** *Predicted sensitivity for a slit-based version of IRMOS on TMT. Spectroscopic limits assume 0.01 e-/s dark current and effective read noise of 5e-/pixel, with background for spectral regions between OH lines, evaluated over a 3 pixel spectral resolution element and assuming a 0.5 arcsec extraction aperture.*

| **Passband** | **Spec. Sky Brightness, mag arc sec$^{-2}$ Vega (AB)** | **Continuum SNR = 10 ,1 hour, *R* = 3600 (0.23" slit) Vega (AB)** | **Line flux SNR = 10, 1 hour, *R* = 3600 (0.23" slit) (erg s$^{-1}$ cm$^{-2}$)** |
|---|---|---|---|
| Y (0.97 to 1.13µm) | 18.8 (19.4) | 23.4 (24.4) | 7.4 x $10^{-19}$ |
| J (1.15 to 1.35µm) | 18.2 (19.1) | 23.0 (24.1) | 6.4 x $10^{-19}$ |
| H (1.48 to 1.80µm) | 17.2 (18.6) | 22.7 (24.1) | 5.6 x $10^{-19}$ |
| K (1.95 to 2.25µm) | 16.6 (18.4) | 22.1 (23.9) | 4.8 x $10^{-19}$ |
| K (2.25 to 2.40µm) | 14.2 (16.0) | 20.9 (22.7) | 1.9 x $10^{-18}$ |

As shown in table 5-1, spectra with the SNR with sufficient diagnostic power will be within reach for relatively typical high-redshift ~L* galaxies using TMT/WFOS, while obtaining spectra of high quality but lower resolution will be possible for objects significantly fainter than L* even at $z \sim 6$ with modest ~1 hour integration times – the gain in sensitivity over current 8 m telescopes for spectroscopy over the 0.31–1 µm wavelength range will be a factor of at least 14. The sensitivity gain will bring intrinsically faint galaxies during the epoch of galaxy formation within spectroscopic reach for the first time, and will vastly increase the fraction of the high redshift populations for which extremely high quality spectra will be achievable, and where gas inflows and outflows can be measured in detail.

Using IRMOS, it will be practical to obtain nebular spectra of hundreds of high redshift objects within a survey region, where every spectrum will be of high enough quality to measure velocity dispersion, chemical abundance, star formation rate, stellar excitation, and characteristic electron density — and to evaluate whether or not active AGN are present — for galaxies at the limits of current *imaging* surveys.

Near-IR IFU spectroscopy will also be essential for finding evidence of outflows induced by AGN winds and jets, which are thought to play a crucial role in preventing runaway star formation in the most massive galaxies (Fabian 2012). Current 8 m class telescopes are only capable of finding such outflows in rare, highly-luminous AGN at z~0.5–2 where these feedback effects are most important (e.g., Cano-Diaz et al. 2012). TMT will be able to observe these outflows on smaller spatial scales, and in more typical, less luminous AGN. Together with ALMA observations of the corresponding molecular outflows, and sensitive VLA observations of the synchrotron emission from jets and outflows (even in traditionally "radio quiet" AGN), we can build up a complete picture of the nature and energetics of AGN feedback.

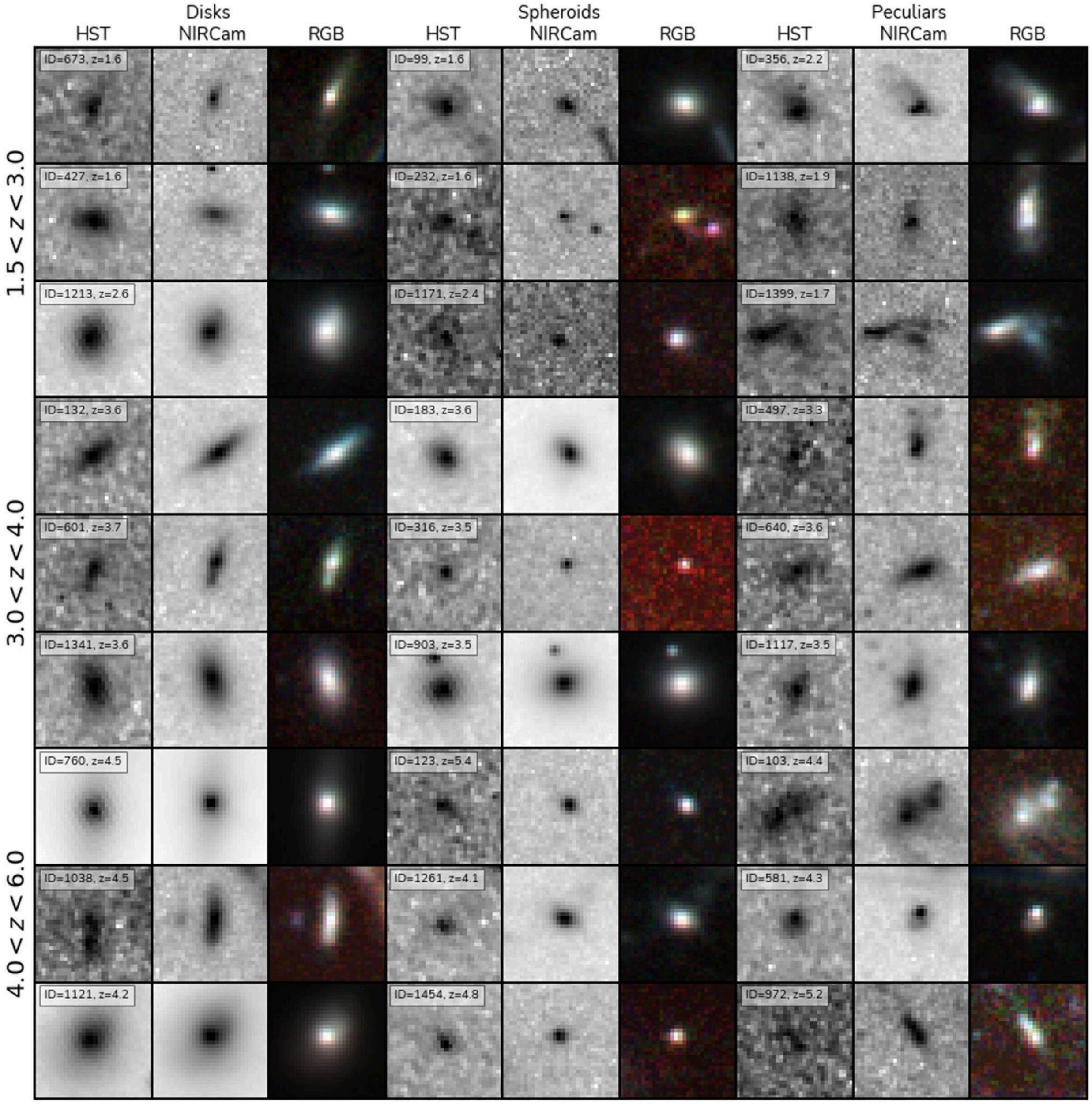

***Figure 5-6:*** *HST F160W and JWST rest-frame optical images of galaxies at a range of redshifts from Ferreira et al. (2022). Right hand column shows JWST NIRCam F277W, F356W, and F444W color images. Galaxies over this range of epochs have diverse morphologies, including highly compact objects, dust-enshrouded disks, and highly disturbed merger candidates. With IRIS and IRMOS in both imaging and IFU modes, it will be possible to observe systems of this nature with unprecedented angular resolution in the rest-frame visible.*

### 5.1.7 Spatial dissection of forming galaxies

A major breakthrough scientific application for the TMT will be the spatial dissection of galaxies during the peak epoch of galaxy formation. Observations of these galaxies will exploit both the light gathering power and the high angular and spectral resolution at near-IR wavelengths provided by TMT. Large samples of galaxies throughout the redshift range z = 1–6 are already known, and HST, JWST, and the current generation of 8–10 m telescopes will help us learn a great deal more in the coming decade. However, *spatially resolved* spectroscopy, allowing differences in chemistry, kinematics, and physical conditions to be mapped as a function of spatial position within the galaxies, is required to go beyond measurements of crude global properties, and thereby gain fresh understanding into the physics of galaxy formation.

Galaxies at z = 1–5 exhibit a variety of structures in the high spatial resolution images such as those recorded in the near-IR (H band) with the Hubble Space Telescope Wide Field Camera 3 (WFC3) (see figure 5-6). Massive galaxies with low star-formation rates typically appear compact and smooth at these times, with sizes significantly smaller than local early type galaxies. Star-forming galaxies have a wide diversity of spatial structures, from the blue compact analogues to the quenched compact galaxies, to dusty disk-dominated systems, to highly irregular, clumpy and/or diffuse systems. Interpreting the physics underlying these varied morphologies is particularly challenging in the early universe, when frequent galaxy mergers, high gas accretion rates, and dynamically unstable disks are predicted. A key question for TMT will be to understand how the structures of distant galaxies are connected to their assembly and star-formation histories.

To maintain the typical high star formation rates of ~100 $M_\odot$/yr observed in these galaxies without a major merger event requires rapid and semi-continuous replenishment of gas, which favors smooth accretion mechanisms such as through “cold flows” or rapid series of minor mergers (e.g., Dekel et al. 2009). Some numerical simulations predict that gas-rich heavy disks are fragmented into several clumps due to gravitational instabilities and that gas and stars in the clumps migrate to the center of galaxies, where they merge together and form a bulge (e.g., Hopkins et al. 2012). Despite efforts to constrain the origin and evolution of the z~2 massive disk galaxies by numerical simulations, these evolution scenarios have not been observationally confirmed yet. To put constraints on the process of galaxy assembly, it is essential to obtain the spatial distributions of stellar mass and ongoing star formation activities, since past/ongoing dynamical processes must be imprinted/visible in the stellar populations. Current studies with 8–10 m telescopes are limited to bright (thus massive, or at least highly star-forming) galaxies, and they need to be expanded to much wider domain of the SFR-M* diagram. Figure 5-7 shows how TMT’s NFIRAOS facility AO system feeding the IRIS IFS channel will provide the necessary sensitivity. The key question here is how the internal structures or physical processes depend on the location in the SFR-M* diagram, namely SFR (burstiness) and stellar mass. In particular, we can investigate whether the compactness of star forming regions is related to the mode of star formation (burstiness/dustiness), and whether smaller galaxies undergo different formation processes from massive galaxies related to the down-sizing effect (e.g., Cowie et al. 1996). Moreover, the internal properties should also be investigated as a function of environment (such as the number density of galaxies), as some environmental impacts must also be imprinted in the spatial distribution of stellar populations, such as a centralized starburst driven by ample gas inflow during galaxy-galaxy mergers, for example.

To understand the physics of galaxy assembly, it is essential to probe the velocity field of the gas and stars within galaxies. As in the local universe, sensitive measures of galaxy kinematics come from analysis of the narrow nebular emission lines such as those illustrated in figure 5-8. These lines, whose diagnostic power has been calibrated through decades of work in the local universe at rest-frame optical wavelengths, also provide information on chemistry, density, and excitation in the regions where stars are forming. Integral-field spectroscopy can be used to obtain spectral and spatial information over contiguous regions simultaneously. Initial attempts using AO-equipped 8 m class telescopes and IFU spectroscopy (e.g., Genzel et al. 2006; Law et al. 2007, 2009; Forster Schreiber et al. 2009; Wright et al. 2009; Lemoine-Busserolle & Lamareille 2010a, Lemoine-Busserolle et al. 2010b, Jones et al. 2010, 2013) to observe galaxies at z > 2 have shown the promise, and revealed the limitations, of this technique with current facilities (Price et al. 2021). Similarly, IFU observations of z > 3 galaxies with VLT/MUSE offer promise but are limited in spatial resolution (Mackenzie et al. 2019).

The James Webb Space Telescope (JWST) is significantly advancing our understanding of galaxy morphological evolution, revealing a rich diversity of galaxy shapes at high redshifts with its unparalleled infrared sensitivity and resolution. Studies have shown a higher fraction of disk galaxies in JWST images compared to those captured by the Hubble Space Telescope (HST), (Ferreira et al. 2022; Kartaltepe et al. 2023), with deep learning tools identifying faint, distant disks previously unnoticed by HST (Robertson et al. 2023, Huertas-Company et al. 2023; Tohill et al. 2024).

However, challenges remain in differentiating 3D galaxy shapes, such as prolate versus oblate geometries and the identification of triaxial (oval) disks. TMT's sensitive spectroscopy will disambiguate between these possibilities. The need for such observations is highlighted by recent research leveraging machine learning on JWST observations, which has identified many misclassifications of the morphologies of galaxies (Vega-Ferrero et al. 2024), and further studies have explored the 3D shapes of galaxies, suggesting a mixture of oblate and prolate geometries across different mass and redshift ranges (Nelson et al. 2023). Pandya et al. (2024) introduced a novel approach by modeling galaxies as triaxial ellipsoids, finding variations in shape that indicate both the presence of prolate geometries in younger dwarf galaxies and the evolution towards disk shapes over time.

This work hints at a complex evolution of galaxy structures from triaxial to more axisymmetric forms, contributing to our understanding of galaxy formation and morphology with implications for both prolate and oblate galaxy populations, including their size-mass relationships and morphological characteristics. TMT spectroscopy with IRIS and IRMOS will allow these ideas to be tested.

TMT with IRIS and IRMOS will represent sensitivity gains of a factor of between 10 and 100, with angular resolution gains of a factor of > 3–5, over current capabilities. The gains will vastly expand the range of targets that may be observed, making it feasible to construct the statistically representative samples needed to connect individual galaxies to the larger framework of cosmological evolution.

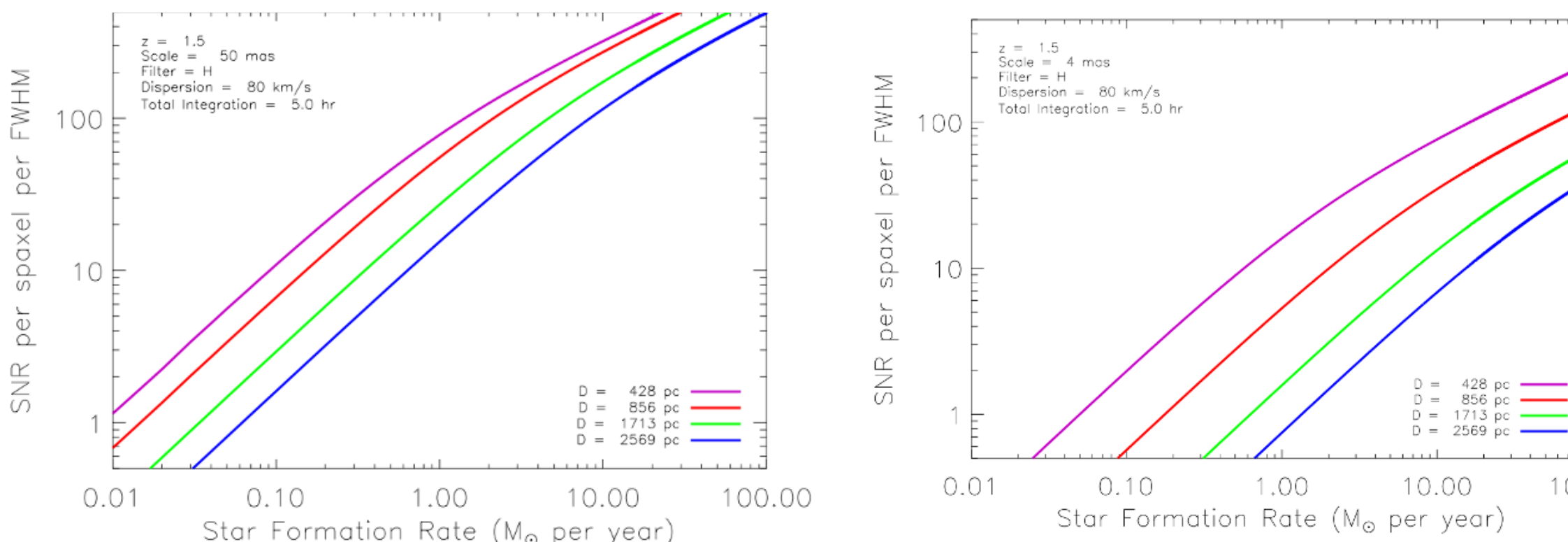

***Figure 5-7:*** *Estimated signal-to-noise ratio of the Hα emission line, per spatial element ("spaxel"), 0.050" (left) and 0.004" (right) per spectral resolution element, as a function of assumed SFR at z=1.5 (look back time 9.3 Gyr). The curves assume a range of scales over which the flux is distributed, from angular diameters of 0.05" to 0.30" and a fixed intrinsic velocity dispersion (80 km $s^{-1}$). A total integration time of 5 hours comprised of twenty 900 second exposures (H-band) was assumed (from Wright et al. 2010).*

With TMT instruments IRIS and IRMOS, the structure, velocity field, abundances, and excitation of forming galaxies may be examined at 50–100 pc resolution (8–15 mas, an order of magnitude higher than the highest resolution achieved by HST, and 3-5 times better than JWST at these wavelengths) throughout the redshift range of interest. At such scales, even individual giant HII regions and the largest rich star clusters within compact galaxies at high-redshift may be resolved and detected (see figure 5-9), providing a level of astrophysical detail currently only accessible in nearby, bright galaxies. With these extremely high-resolution velocity fields, the motions of a galaxy's stars and gas may be traced within its dark-matter dominated potential, allowing astronomers to track the emergence of rotation and dispersion-dominated galaxies at early times, as well as the physics of gas flows on sub-kpc scales. By studying the evolution of these processes over cosmic time, we can see how the local correlations between galaxy structure, kinematics, and star-formation came to be.

Together, the observations required to address these science cases place significant pressure on the design of the TMT and its instruments (see section 12), particularly aperture and optical design (signal to noise and angular resolution), AO (angular resolution), spectral resolution, wavelength coverage (UV to IR), multiplexing (MOS), and IFU capability.

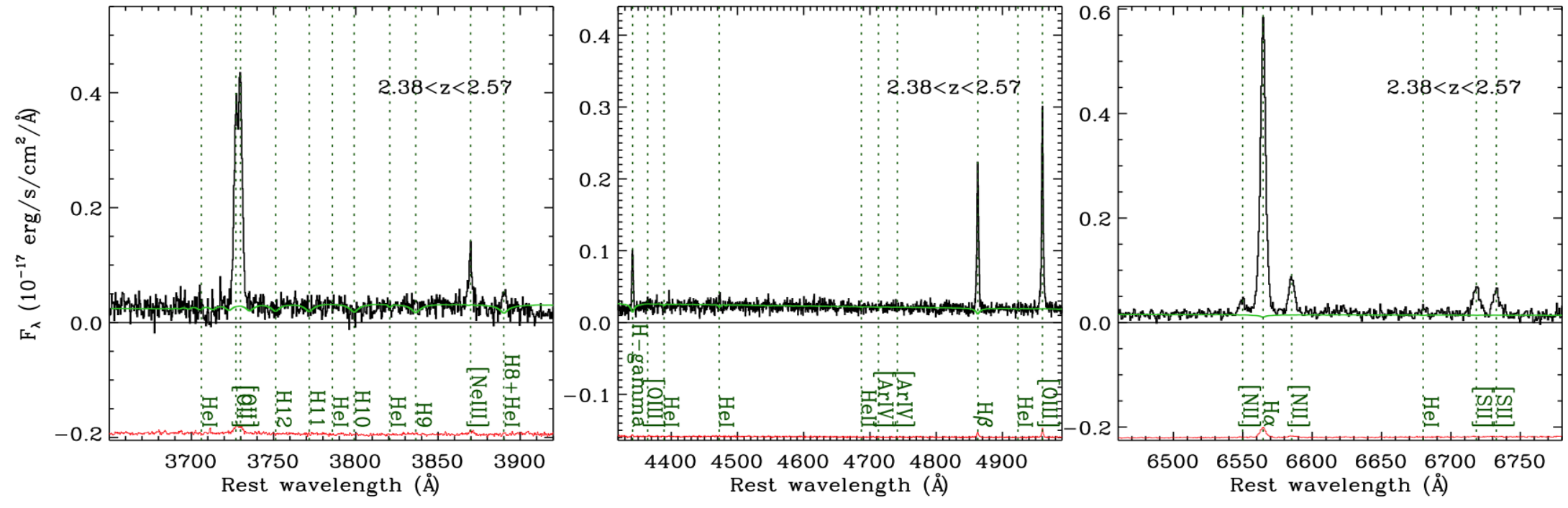

***Figure 5-8:*** *The composite spectra for a sample of 70 galaxies at <z> ~ 2.4, obtained in seeing-limited mode for whole galaxies using Keck/MOSFIRE. The line ratios indicate approximately 0.5 solar metallicity in the HII regions, and the ratio of the [OII] and [SII] lines indicate electron densities of ~220 cm*$^{-3}$*. Using IRIS/IRMOS, spectra of similar quality could be obtained for positions in individual galaxies at a spatial scale of 0.1–0.2 kpc at any redshift < 4. (A. Strom, C. Steidel CIT).*

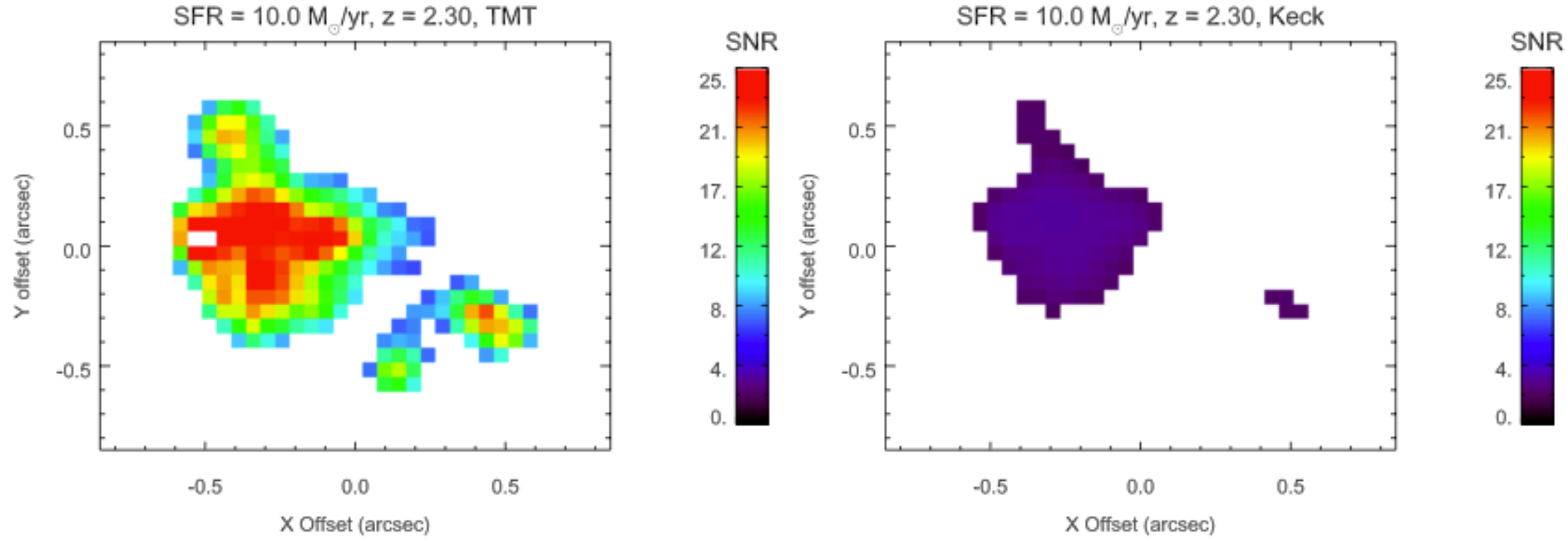

***Figure 5-9:*** *A simulated z=2.3 galaxy, with Hα redshifted into the K-band window and an integrated star formation rate of 10 M*$_\odot$ *yr*$^{-1}$ *(left). This is compared to Keck observations of the same simulated galaxy (right). Complex structures with multiple knots can be seen with IRIS and TMT, while only the brightest portion of the galaxy is barely detectable (SNR~3) with Keck.*

## 5.2 The Age of Maturity and Quiescence

### 5.2.1 Morphological and Kinematic Growth of Galaxies

More than half of the age of the universe is contained in the redshift range $z < 1.5$. During this vast time period of ~10 Gyr, the level of cosmic star-formation activity has declined by roughly an order of magnitude, and the abundance of bulge-dominated or elliptical galaxies has increased significantly. While the kinematic and morphological structure of present-day galaxies follows various well-known scaling relations, the establishment and evolution of these relations remains poorly understood. The capabilities of TMT, in particular the ability to constrain the kinematics and morphology of galaxies at $z < 1.5$, will allow the growth and development of galactic structures to be quantified at an unprecedented level of detail.

**The Build-Up of Galactic Disks:** Beyond $z > 1.5$ the majority of star forming galaxies have irregular or peculiar morphologies characterized by merging knots of star formation. In contrast, the most obvious feature of star forming galaxies today is the presence of a rotationally supported disk containing gas, dust, and active star formation. The stars span a range of metallicities and ages at any given radius, but HII region abundances reveal that gas-phase metallicities decrease with galactocentric radius (Zaritsky et al. 1990) suggesting that disks form from the inside out. The empirical data and semi-analytic simulations (Cole et al. 2000) suggest that the bulges of disk galaxies are comprised of old, metal-rich stars that form first though a series of major mergers followed by a more gradual and protracted formation of a disk. The luminosity of present-day spiral galaxies correlates strongly with their rotational velocity (Tully & Fisher 1977; Tully & Pierce 2000).

The disk surface brightness is also correlated with rotational velocity, those with higher surface brightness are found in galaxies with higher rotational velocities, and there is evidence for bimodality (McDonald et al. 2009). The shapes of galaxy rotation curves are also correlated with their rotational velocities (e.g., Persic, Salucci & Stel 1996). Together these correlations imply that the depths of the dark matter halos of disk galaxies drives their accretion of gas and their subsequent star formation history. TMT offers an opportunity to characterize the build up of galaxy disks. This includes using kinematics to trace the build up in mass and emission line diagnostics to trace the build up in metallicity.

Deep Keck spectroscopy of $z \sim 1$ disk galaxies has shown that the turnover in the rotation curves can be reached and the disk rotation velocity accurately measured. The Tully-Fisher relation is already in place for morphologically normal disk galaxies by $z \sim 1$ (Miller et al. 2011) but shows evidence of evolution at higher redshifts (e.g. Übler, et al. 2017; Straatman, et al. 2017). The large aperture and AO capabilities of TMT will allow the characterization of disk rotation curves at the epoch of peak galaxy assembly ($z \sim 1.5$) and higher redshifts. This will require Hα emission-line kinematics with spectral resolution R ~ 5000. Scaling the Miller et al. (2011) results to higher redshifts suggests that disk rotational velocities can be obtained with TMT and WFOS/IRMOS for moderate mass disk galaxies ($V_{circ} \sim 100$ km s$^{-1}$) at $z \sim 1.5$ in about 4 hours. With sufficiently deep near-infrared surface photometry from JWST the scaling relations, metallicities, and the assembly history of disks could be characterized for the first time.

**Growth and Assembly of Ellipticals:** The fundamental plane of elliptical galaxies (FP; Djorgovski & Davis 1987; Dressler et al. 1987) is thought to preserve a fossil record of their assembly history (Bender et al. 1992). This includes any differences between cluster and field samples but also differences between ellipticals within a given cluster. Data on nearby clusters suggest that the structural properties of the giant ellipticals ($L > L^*$) are significantly different from the lower luminosity systems ($L < L^*$ ; Pierce & Berrington 2014). Specifically, lower luminosity galaxies appear to comprise a dissipational sequence suggestive of gaseous (wet) mergers while the most luminous galaxies form a separate family suggestive of a dry merger history. This implies that the distribution of galaxies within the FP provides a measure of the relative role of wet vs. dry mergers for a given population.

However, high redshift studies of passive galaxies and their scaling relations have been limited to small samples of the most luminous galaxies because of the heroic efforts required to obtain diagnostic spectra of the stellar light (see figure 5-10). TMT WFOS and IRMOS will provide the increase in sensitivity necessary to extend such diagnostic spectroscopy to lower-luminosity passive galaxies, both within and outside of distant clusters.

Accurate measurements of stellar velocity dispersions and metallicity-sensitive line indices will require S/N ~ 50 spectroscopy at R ~ 3000. To sample 2–3 magnitudes below $L^*$ at z~2 will require exposure times of ~6 hours.

**Galactic Bars as Signposts of Disk Assembly:** As has already been noted, over half of the history of the universe is contained in the redshift range $z < 1.5$, over a look back time of about 10 Gyr. During this time period, star formation activity has declined and disk galaxies began to mature. Previous studies of the physics of galaxy disk stability have shown that as soon as a disk is sufficiently massive, dynamically cold and rotationally-supported, it should buckle and form a bar at its center. Stellar bars thus serve as important signposts for measuring maturity of disks. However, because disks can be stabilized against bar formation with other common structures, such as large central masses (Saha et al. 2018) and dark matter halos, we also probe those components as well. These types of studies are challenging due to the required sensitivity, spatial and spectral resolution, but are needed to improve out understanding of bar and disk galaxy formation. JWST imaging is showing that our full understanding of the process is limited because bars are seen at much higher redshift than previously expected (Costantin, et al. 2023; Guo, et al. 2023) and therefore dynamically stable disks are present in the universe at early times (Le Conte et al. 2024).

Using the *Hubble Space Telescope* and Keck, a precise measurements of the rate of assembly for $L^*$ and brighter disk galaxies on to the Hubble sequence were apparent prior to the latest HST and JWST measurements (see gray points in figure 5-11, Sheth et al. 2003, 2008; Cameron et al. 2010; Kraljic et al. 2012; Melvin et al. 2014). These studies have found that the decline in the bar fraction is not uniform across all galaxy types and the largest rate of evolution is seen in the low mass galaxies ($M^* < 10^{10.5}\ M_{\odot}$), while redder, bulge-dominated massive disks are already in place 7 Gyr ago.

Building on the the latest HST and JWST results, s combination of JWST and TMT measurements will allow us to extend this type of research in three important areas: disk-wide stellar velocity dispersion measurements for $L^*$ disks at $z < 1$, investigation of the bar fraction and disk stability for lower luminosity 1/10th $L^*$ disks at $z < 1$, and the study of the assembly of massive disks from $z \sim 1$–3 using bars as a signpost for disk maturity using rest-frame optical light.

The TMT will allow us to finally measure the stellar velocity dispersion across $L^*$ and brighter disks to $z \sim 1$. Analysis of the gas velocity dispersion from $0 < z < 1$ from the DEEP2 Keck/DEIMOS survey (Newman et al. 2013) showed that a small fraction of $L^*$ disk galaxies may be offset from the stellar Tully-Fisher relation (Kassin et al. 2007), but deeper observations have demonstrated that this is a spurious result (Miller et al. 2011). In any case, studies of galactic structures in these high-redshift disks have revealed that stellar bars are only present in disks that are on the Tully-Fisher relation (see figure 5-12; Sheth et al. 2012). In other words, bars form only in rotationally supported and dynamically cold disks, as in the nearby universe (Courteau et al. 2003), but not all rotationally supported and dynamically cold disks have a bar. TMT will, for the first time, allow us to probe the stellar dynamics of the disks at a range of redshifts to shed light on the role of the dark matter halo in determining the formation of structure in disk galaxies.

Another exciting capability of the TMT is to provide two-dimensional gas kinematics in individual galaxies at sub-kpc resolution to ascertain the gas inflows by stellar structures like bars and spiral arms. These studies will allow us to constrain how quickly the gas is being funneled to the galaxy centers and how that flow impacts central starburst and AGN activity and the build up of bulges. Sub-kpc resolution observations will also allow us to measure the gas outflows by starbursts and AGN activity.

The large collecting area of TMT will allow us to probe 2.5 magnitudes deeper than currently possible with Keck. As a result we will be able to trace in detail the dynamics of sub-$L^*$ disk populations. We can study the growth of structure and kinematics at $z < 1$ and connect this evolution ultimately to low mass, Magellanic disks in the nearby universe. A key question to be addressed is why the evolution of lower mass disk galaxies was delayed, and what physical process was responsible.

Finally the study of the evolution and assembly of disks at $z > 0.85$ has been hampered by lack of high resolution, high-sensitivity near-infrared dynamical observations that can probe rest-frame optical light. The time period between 7–11 Gyr of cosmic look back time is when disks are believed to have been assembled. What were the disk shapes and profiles of the first objects? What were the stellar and gas kinematics and how did they affect the build up of stars? Hα spectroscopy at sub-kpc resolution with TMT will allow us to measure the velocity fields and dispersions in the earliest disks. Imaging with JWST and internal dynamics probed by TMT will reveal the mechanisms driving the formation of the structure of these systems which can then be placed in context of our current understanding of $L^*$ disks at $z < 1$.

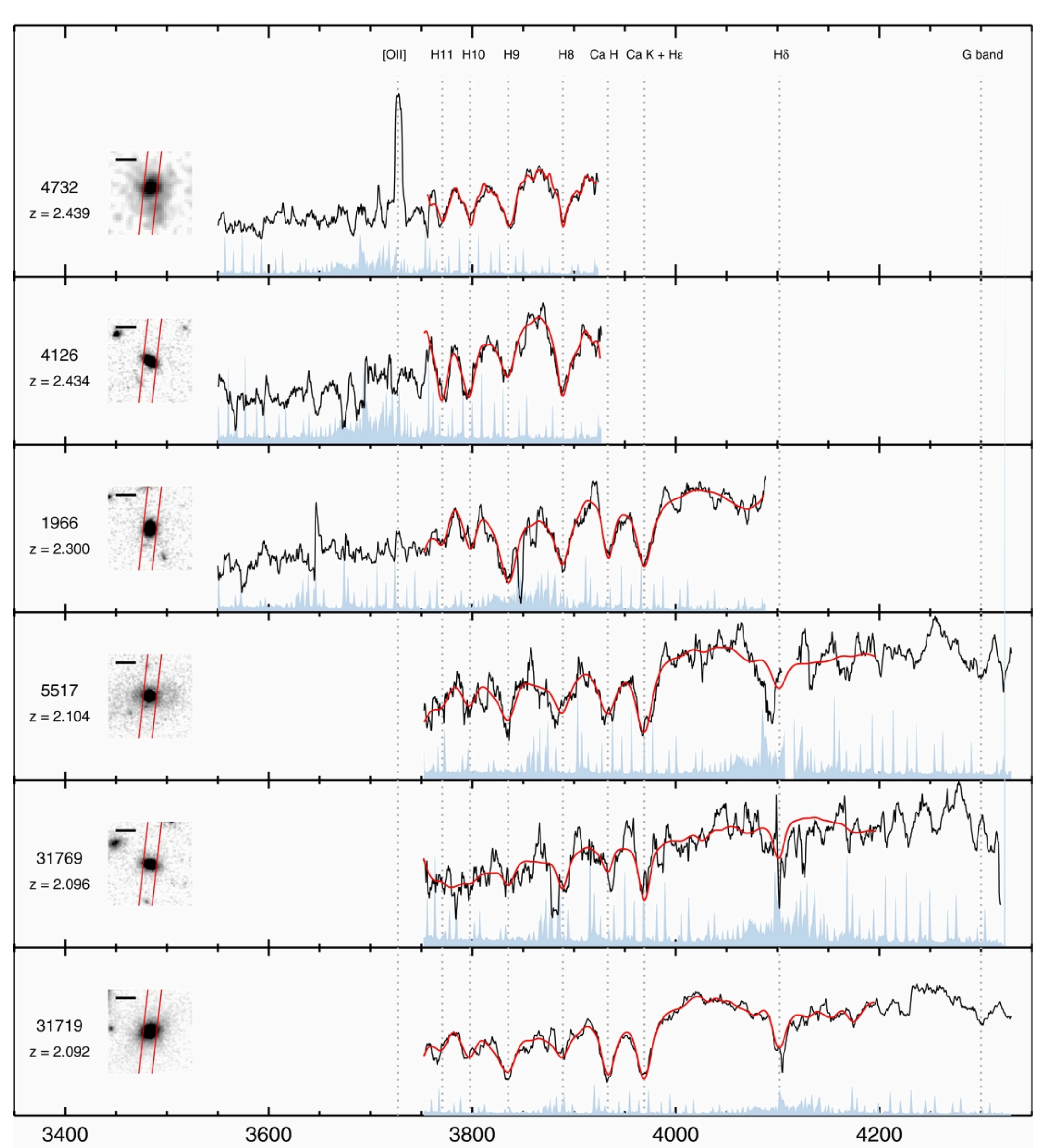

***Figure 5-10:*** *Spectra of massive early type galaxies at z>2 obtained using Keck/MOSFIRE with 8 hour integrations, demonstrating the ability to obtain high-quality continuum spectra of high redshift passive galaxies. IRMOS and WFOS will extend such capabilities to well below L* at similar redshifts. (from Belli et al 2014).*

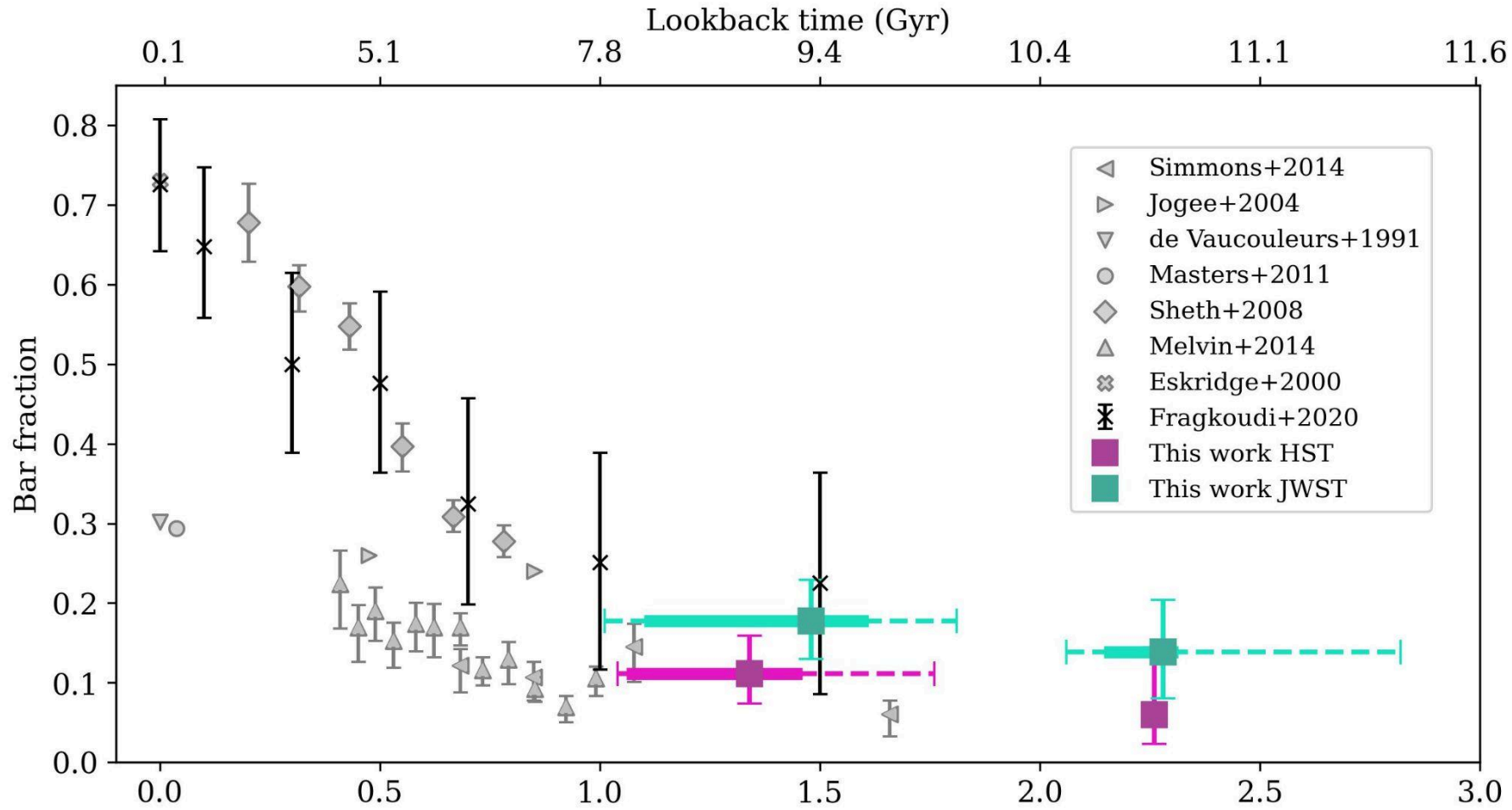

***Figure 5-11:*** *Figure 7 from Conte et al. (2024). The fraction of stellar bars in disc galaxies with redshift as determined in previous studies, compared to the most recent HST and JWST results. The gray triangles and diamonds are works earlier than 2020. The latest HST and JWST results demonstrate the departure from the apparent trend for redshifts of $z<\sim1.2$ and motivate higher spatial and spectral resolution studies with TMT.*

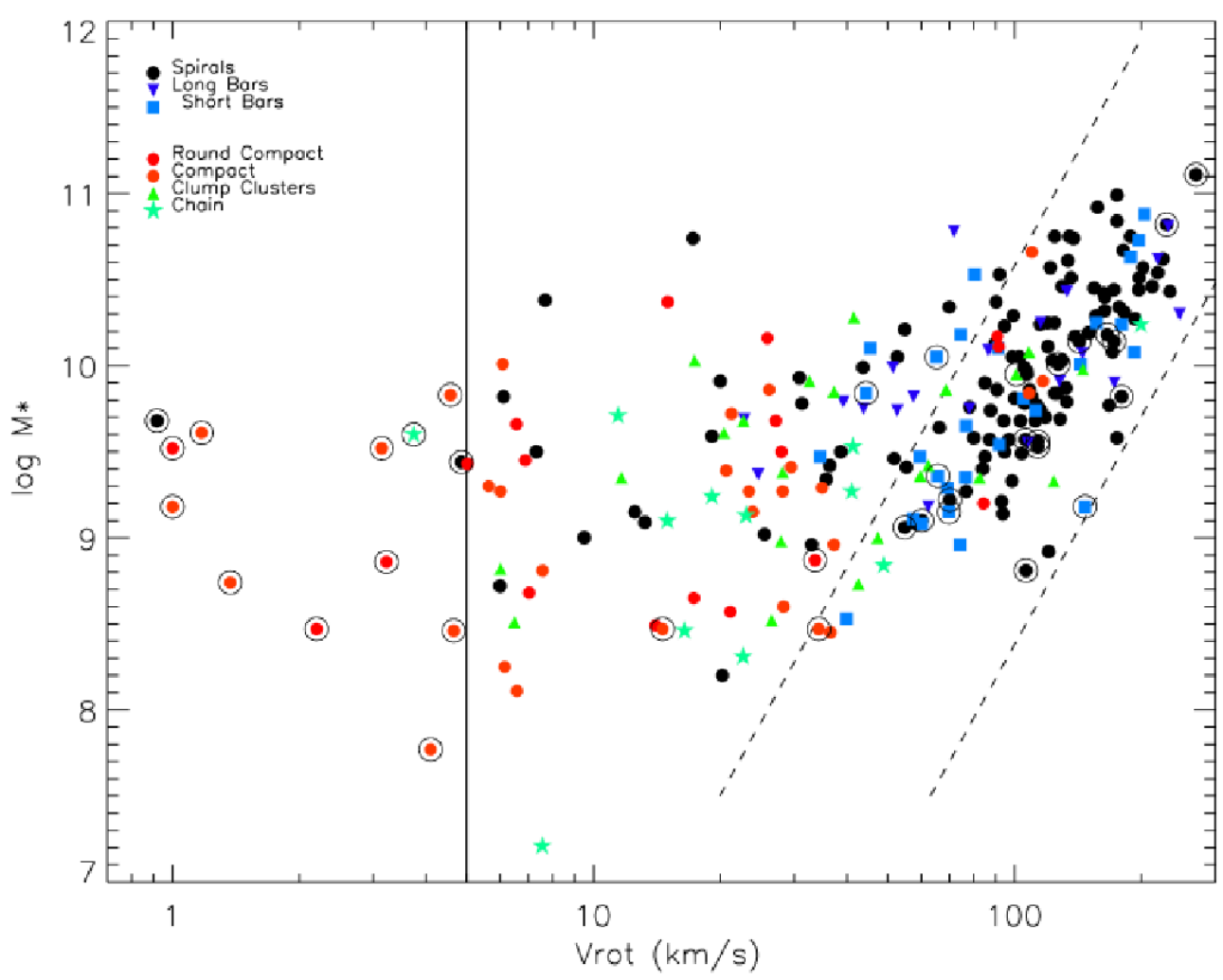

***Figure 5-12:*** *Stellar mass versus rotational velocity is shown for galaxies from DEEP2/AEGIS sample (Sheth et al., 2012). All black points are unbarred disks. Notice their scatter away from the TF which is indicated by the dashed lines. The blue points are barred spirals that tend to be on or near the TF.*

**The Impact of Galaxy Mergers:** The role of mergers in galaxy formation and evolution throughout cosmic time remains poorly understood despite substantial effort, using both HST and ground-based telescopes, to understand its importance. Studies of high-redshift galaxies, particularly since the peak of cosmic history at $z \sim 2$, diverge on the evolution of galaxy-pair fractions (e.g., Kartaltepe et al. 2007; Lotz et al. 2008) and the dominance of mergers in fueling nuclear starbursts and active galactic nuclei (AGN) activities (e.g., Pearson et al. 2019; Kocevski et al. 2012; Comerford et al. 2009). Settling these debates requires a thorough understanding of the detailed physical processes occurring in the extremely energetic and dusty environments within the nuclei of galaxy mergers.

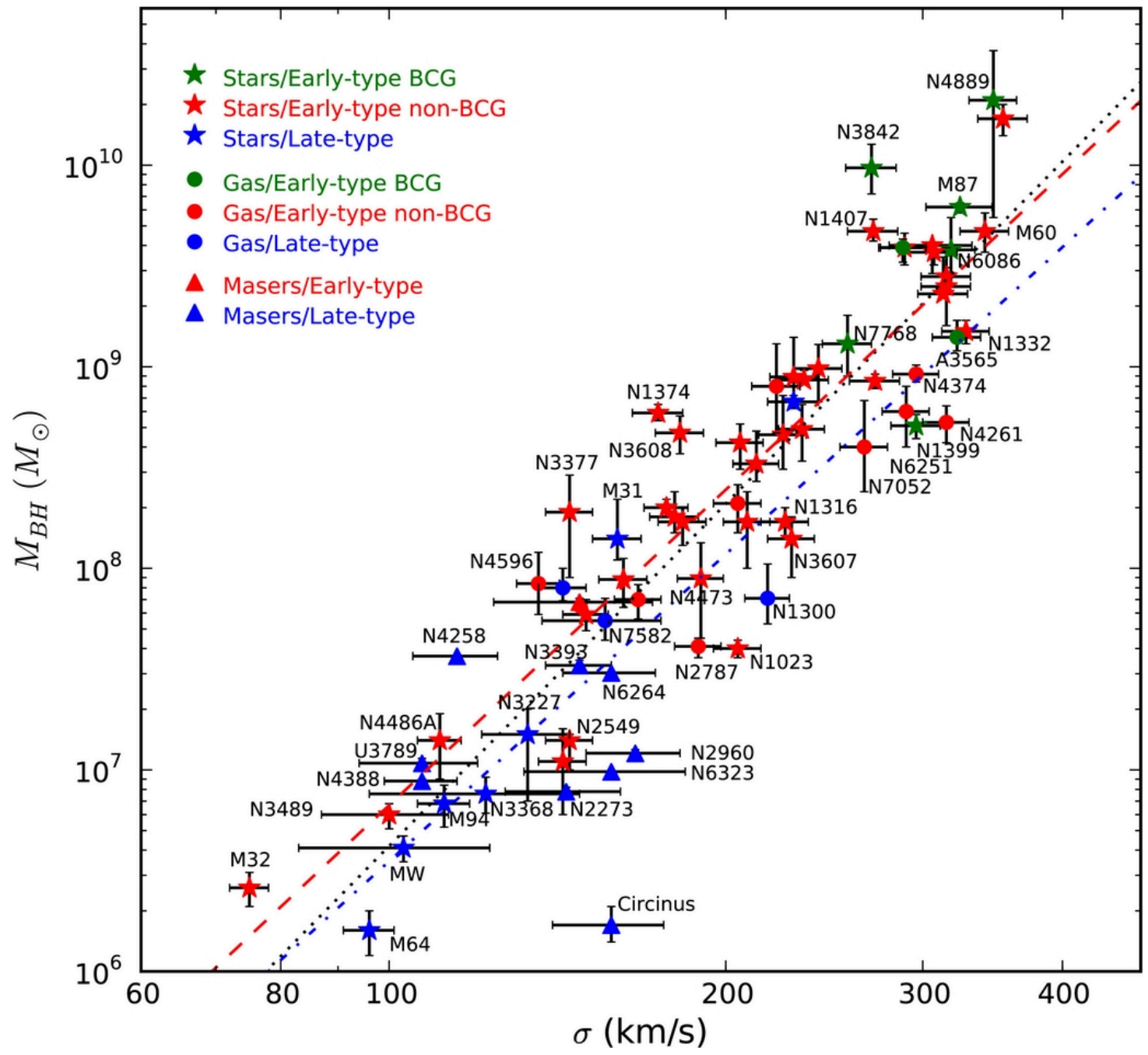

*Figure 5-13: A compilation of MBl measurements from McConnell and (2013). The black hole masses are mea using the dynamics of masers (trian stars (stars), or gas (circles). Error indicate 68% confidence intervals. The dotted line shows the best-fitting powe for the entire sample.*

Activity within the central few hundred parsecs of a galactic nucleus may govern the properties of its host. Indeed, the observed scaling relations of the masses and kinematic properties of galactic bulges with their central supermassive black holes (SMBH) are prime examples of such an effect: $M_{BH}$-$M_{bulge}$ (Kormendy & Richstone 1995; Magorrian et al. 1998), $M_{BH}$-$L_{bulge}$ (Marconi & Hunt 2003), and $M_{BH}$-$\sigma_\star$ (figure 5-13; Tremaine et al. 2002; Ferrarese & Merritt 2000; Gebhardt et al. 2000), see chapter 6. However, the possibility that the $M_{BH}$-$\sigma_\star$ relation may evolve with redshift (e.g., Zhang et al. 2012) has led to suggestions that the scaling relation depends most directly on stellar mass (Jahnke et al. 2009; Cisternas et al. 2011); in this case, the evolution of the $M_{BH}$-$\sigma_\star$ relation with redshift would reflect the changing fractions of mass in galaxy bulges versus disks. Cogent arguments have also been made that the growth of black holes and the stellar content of galaxies are both "self-regulated" via feedback from the central source (Marsden et al. 2020) and that ultimately it is galaxy dark matter halo mass which dictates the equilibrium state (Booth & Schaye 2010).

In any case, the coevolution of SMBHs and their host galaxies is one of the keys to understanding the driving force of galaxy evolution, as it has direct implications on the interplay between the central source and the surrounding interstellar medium. However, understanding this interaction has proven challenging because of observational difficulties. Detailed observations of the morphology and kinematics of nuclei in interacting galaxies is hard because of their dusty environments,and has effectively been impossible until the development of near-infrared integral-field spectrographs with adaptive optics (AO) (Davies et al. 2006; Nowak et al. 2008; McConnell et al. 2011; Gebhardt et al. 2011; Medling et al. 2011; U et al. 2013; Medling et al. 2014). TMT, with its nearly two orders of magnitude increase in sensitivity for diffraction-limited imaging/spectroscopy, will revolutionize our understanding of processes in galactic nuclei.

The mechanism through which black hole masses correlate with galaxy properties has often been associated with gas-rich galaxy mergers (e.g., Hopkins et al. 2006). Gravitational torques funnel the gas into their centers, triggering two phenomena: an intense burst of star formation to feed the bulge, and accretion of gas on to the black holes in the centers of each galaxy. It has been postulated that black hole growth can regulate this process through AGN feedback (Springel et al. 2005; Pearson et al. 2020) via massive winds that evacuate the gas from the galaxy on short timescales, cutting off star formation and future black hole growth. This sense of self-regulation is supported observationally by Kauffmann & Heckman (2009), who find that the Eddington ratio of a sample of AGNs depends on the supply of cold gas in the galaxy. Though the detailed mechanisms causing these correlations are still unconfirmed, star formation and black hole growth are fed by the same reservoir of inflowing gas; their growth histories are intertwined. It is likely that these two processes compete for fuel in a predictable fashion.

To understand this interplay, it is critical to examine systems during the period of maximum fueling of central starbursts and supermassive black holes. During a merger, does the black hole grow first, leaving the stars to slowly consume the remaining gas? Or is star formation quenched once the black hole reaches a bright quasar phase of extreme growth? Identifying a merged system consistent with black hole scaling relations would indicate the relative growth timescales and confirm whether the putative quasar-mode feedback occurs. Theoretical arguments have suggested that the starburst has preferential access to the available gas and that, therefore, a black hole would grow substantially only after star formation has quenched itself and the galaxy bulge is in place (Cen 2012); this scenario would predict that these mergers fall below black hole scaling relations.

Observationally, no clear verdict on the location of mergers along the $M_{BH}$-galaxy scaling relations has yet to be established, as different studies have presented apparently-conflicting evidence. Kormendy & Ho (2013) show five merging galaxies falling below the $M_{BH}$-$\sigma_\star$ and $M_{BH}$-$L_{K,\mathrm{bulge}}$ relations and suggest that black hole growth lags bulge formation in mergers; however, using Keck/OSIRIS integral-field spectrograph behind AO, Medling et al. (2014) find the black holes in a sample of gas-rich mergers at or above black hole scaling relations, i.e., significantly more massive than expected given their host galaxy properties. This is contrary to the predictions of quasar-mode feedback theories, in which black hole growth is delayed until the final stages of a galaxy merger, at which point energetic AGN feedback cuts off subsequent star formation.

The observations needed to support the science here will greatly benefit from the TMT's aperture (signal to noise and angular resolution), AO (angular resolution), multiplexing (MOS), broad wavelength range (UV to IR) and IFU capabilities, often using the entire package of these capabilities to make fundamental advances in these areas.

### 5.2.2 Feedback and the Physics of Galaxy Quenching

As has been noted in section 5.1.4, the mechanism quenching star formation in massive galaxies at $z>2$ is not understood. The possible relationship between quenching at high redshift and the more gradual quenching that is seen at lower redshifts is likewise not understood. Roughly half of the stellar mass in present-day galaxies was assembled since $z \sim 1$, over an era when the cosmic star formation rate steadily declined. This quenching was not predicted by theoretical models for the hierarchical growth of gravitationally bound structures, although our understanding of the relevant physical processes is certainly improving (Zheng et al. 2022). Understanding the growth of the baryonic component of galaxies requires following the accretion and ejection of gas from galaxies. This gas physics depends on the highly non-linear gas cooling rate and the feedback of energy, momentum, and heavy elements produced by star formation and active galactic nuclei (AGN), and these topics are all areas in which TMT will contribute.

Critical questions that can be investigated with TMT include the following:

- How do the properties of the circumgalactic gas (on scales from roughly the galactic radius to the virial radius) influence the rate of baryonic accretion onto the star-forming (Toomre $Q \sim 1$) galactic disk?
- How does the mass ejection rate from star-forming galaxies vary with galaxy mass and cosmic time?
- How does the coupling among the various feedback processes (stellar winds, supernovae, AGN) shape the efficiency of mass and metal ejection from galaxies?
- Which physical processes (supernova thermal energy, radiative momentum, cosmic rays, etc.) dominate the acceleration of galactic winds?
- Why is the characteristic stellar mass at which galaxies cease forming stars essentially independent of redshift?
- Why is star formation most efficient in galaxies similar in mass to the Milky Way?
- How does the environment affect the cycling of matter between galaxies and their surroundings?

Progress here requires more detailed studies of galaxies over the era of strongly evolving star formation activity from roughly $z \sim 1.5$ to the present. Multiple sight-lines to beacons behind the halos of individual galaxies will clarify the global structure and filling factor of circumgalactic gas. Spectral imaging of galaxies will resolve metallicity gradients within galaxies and provide sensitivity to filamentary structures connecting galaxies to their surroundings.

TMT instruments offer powerful capabilities for these types of studies. WFOS provides the blue sensitivity required to attack the halo absorption issues with observations of prominent resonance lines all the way down to $z \sim 0.1$; this science is not accessible with JWST. The IRIS instrument will generate spectral maps of star-forming structures within distant galaxies at higher resolution than JWST; and, combined with maps of molecular gas and dust obtained with ALMA, directly reveal which regions within galaxies form stars most efficiently. IRIS and IRMOS provide infrared spectroscopy of diagnostic emission lines that begin redshifting out of the optical bandpass at $z \gtrsim 0.5$, and complement JWST by offering higher spectral resolution.

**Absorption-line Measurements of Galaxy Outflows:** Adequate Spectral resolution (R ~ 5000) is required to obtain accurate measurements of the properties of gas outflows from galaxy spectra. Figure 5-14 shows examples of galaxy spectra at intermediate redshifts. The resonance absorption lines typically show a net blueshift relative to the galaxy, indicating outflowing gas along the sightlines to the star-forming regions. Each absorption trough, however, is a blend of interstellar absorption at the systemic velocity, blueshifted absorption from a galactic wind, and resonance emission (from both the near and far sides of the galactic halo). Optical spectroscopy with TMT will improve the resolution of these galaxy spectra by a factor of 9 at a comparable S/N ratio per resolution element. At similar resolution (60 km $s^{-1}$), previous studies of lower redshift galaxies not only separate these spectral components but also derive properties of the flow as a function of the velocity (Martin & Bouché 2009). TMT will do the same for the intermediate redshift sample. More will be said about these issues in section 5.3 below, on the intergalactic medium.

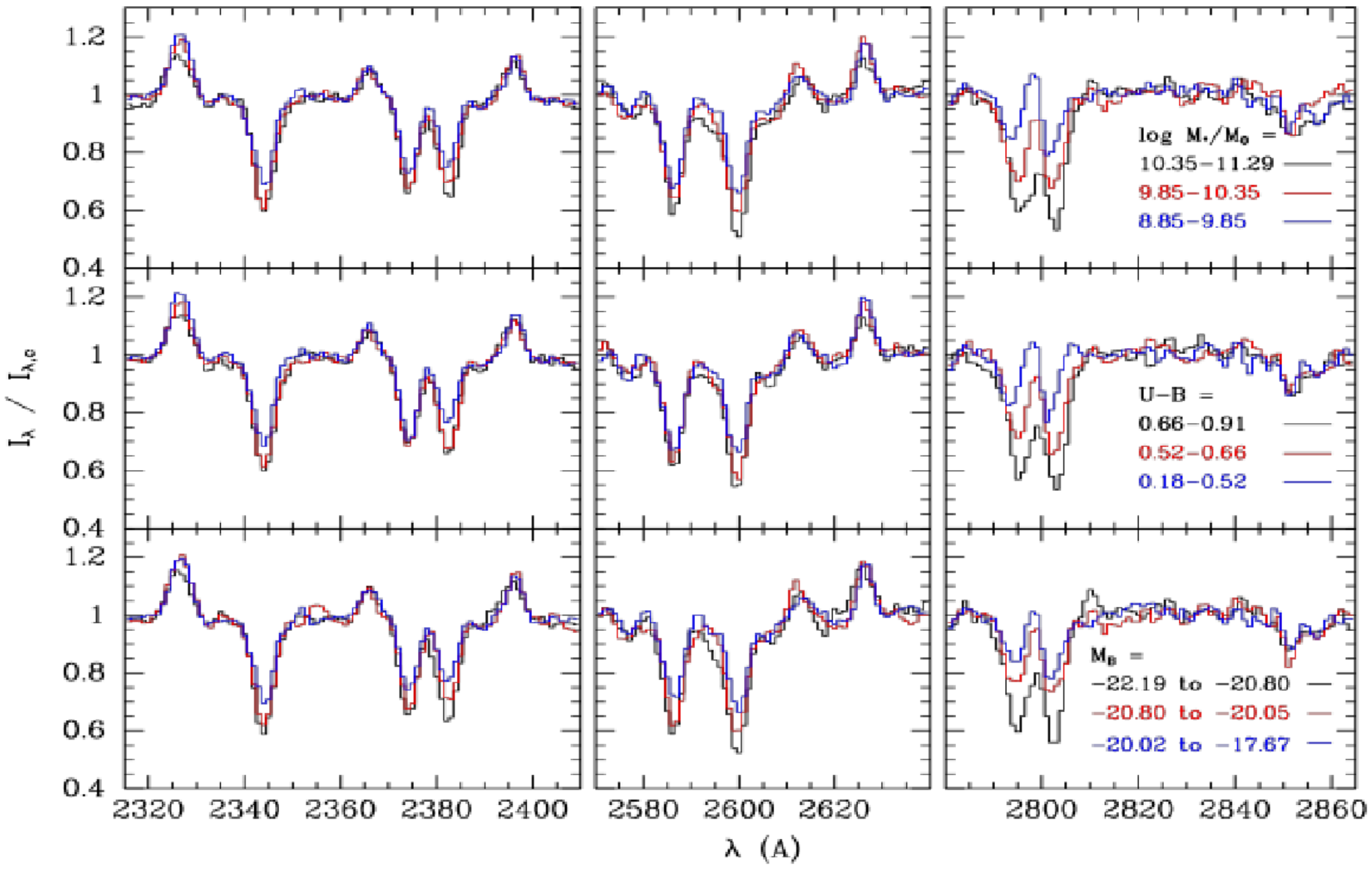

***Figure 5-14:*** *Comparison of the Fe II and Mg II line profiles in average spectra of galaxies with different properties. Top: toward lower mass galaxies, the Mg II absorption troughs become shallower, the Mg II emission becomes stronger, and the Fe II* emission (center) in λλ2612, 2626 increases. In the highest mass tertile, the Mg II doublet ratio at λ2796, 2803 (right) is not inverted, and Fe II λ2383 (left) has a higher equivalent width than Fe II λ2374, as it should in the absence of emission filling. Middle: we see similar trends with color because color is strongly correlated with stellar mass (redder galaxies are more massive on average). Bottom: the resonance absorption troughs become deeper in more luminous galaxies.*

### 5.2.3 The Influence of Local and Large-Scale Environment

It is well established that the physical properties of galaxies strongly correlate with the environment in the local universe, see section 7.8.6. In contrast to star-forming field galaxies, those in dense environments such as galaxy clusters are typically red and quiescent at low and intermediate redshift (Dressler 1980; Blanton et al. 2005; Cooper et al. 2006). Within these overdense environments, the galaxy population evolves strongly with redshift. Higher redshift systems have greater fractions of blue, star-forming, late-type galaxies (Butcher & Oemler 1984; Dressler et al. 1997; van Dokkum et al. 2001; Lubin et al. 2002). Recent studies of large-scale structures at $z \sim 1$ (Lubin et al. 2009) have found that many of the infalling galaxies experience significantly increased star-forming, starburst, and nuclear activity, and rapid quenching of star-formation (e.g., Dressler et al. 2004; Tran et al. 2003; De Lucia et al. 2004; Tanaka et al. 2005; Koyama et al. 2007; Best et al. 2003; Eastman et al. 2007; Kocevski et al. 2009). Altogether, these results imply that strong galaxy evolution occurs while galaxies assemble into clusters.

Several mechanisms had been proposed for these environmental effects. One class of mechanisms is cluster-related, such as ram-pressure stripping, galaxy harassment and strangulation (Gunn & Gott, 1972; Larson et al., 1980; Balogh & Morris, 2000). Other events such as galaxy interactions and galaxy mergers, which occur in both the field and in clusters, have also been suggested (Mihos & Hernquist, 1994; Barnes & Hernquist, 1992). We do not know which of the many possible physical mechanisms associated with the cluster environment are responsible for eventually suppressing this star formation and nuclear activity and transforming gaseous disk galaxies into passive spheroids. We *do* know that many of these mechanisms are associated *not* with the densest cluster regions, but rather with the infall regions and lower-density environments far from the cluster cores (Treu et al. 2003). We must therefore examine galaxy populations over the full range of environments to form a complete picture of galaxy evolution.

We observe strong differences in the high-mass end of the stellar mass function, even for clusters with similar dynamical (halo) masses and redshifts, implying different evolutionary histories. Thus both late quenching and merging are responsible for populating the bright end of the red sequence and that halo mass and mass specific processes play crucial roles in galaxy evolution (Lemaux et al. 2012, Wu & Zhang 2013). We also observe [OII] in many passive galaxies that is the result of nuclear emission from LINERs, not normal star formation (Lemaux et al. 2010). These observations, for instance, require a combination of rest-frame optical and IR spectroscopy, for which WFOS and IRIS are ideally suited. There is also an increasing fraction (10–15%) of post-starburst (K+A) galaxies than the < 3% in the distant field (Tran et al. 2004). These K+As are as massive as the passive (red) galaxies, implying a connection between the two populations (Wu & Zhang 2013). Their distribution seems to be intimately connected to their environments at z~1, indicating an important role for different processes in different environments.

Today, we require the use of the world's largest telescopes (Keck, Subaru, VLT, Gemini, etc.) to study even modest samples of the brighter galaxies in these environments. The challenge increases as we approach redshifts z>1, where many of the spectral features needed to measure galaxy metallicities (stellar and gas-phase), identify quenching events (post-starburst signatures), distinguish starbursts from AGN, and determine other essential galaxy properties, all shift into the near-infrared. These realities necessitate multi-object near-IR spectroscopy, which is only feasible on select facilities. But even there, large amounts of telescope time must be invested to observe modest samples at moderate resolution and S/N. Because much of the galaxy evolution at z~1–2 occurs in the lower mass (and lower luminosity) galaxies, TMT is especially well-suited to observe the most interesting targets. Workhorse instruments such as WFOS will enable optical spectroscopy of potentially interesting targets, identifying objects at z~1 for followup rest-frame optical spectroscopy using IRMOS. The most interesting candidates will be dissected using IRIS, whose extremely high spatial resolution will finally allow us to clearly see where AGN, starbursts, mergers and tidal interactions are at play. The two infrared spectrometers will probe the most important spectral features — [OIII], Hβ, Mg*b*, Hα, and [NII] — that can be used to distinguish the internal processes at work during this important epoch of galaxy transformation.

**Evolution of Satellite Galaxies:** models of galaxy formation and evolution have made significant progress in explaining the observed properties of massive galaxies as a function of cosmic time. Among other improvements, the inclusion of feedback from star formation and active galactic nuclei has aided greatly in matching the color and luminosity distributions of massive galaxies at z < 2 (e.g., Croton et al. 2006; Di Matteo et al. 2005). However, observational results have also illustrated a fundamental problem with the ability of these same models to predict the evolution of low-mass galaxies $M_\star \lesssim 10^8 M_\odot$ (Somerville et al. 2008; Weinmann et. al. 2011; Weinmannet al. 2012).

The shortcomings of modern models are perhaps best illustrated by studies of the Milky Way and its satellites. Beyond the well-known missing satellites problem (Moore et al. 1999; Klypin et al. 1999), additional work has shown that the most massive subhalos in simulated Milky Way-like systems are dramatically inconsistent with the dynamics of the brightest Milky Way satellites (otherwise known as the *Too Big To Fail* problem (Boylan-Kolchin et al. 2011, 2012). This result suggests a serious problem with our understanding of how low-mass galaxies populate dark matter halos within ΛCDM. In particular, the *Too Big To Fail* problem indicates a failure of subhalo abundance matching (SHAM) at low stellar masses (< $10^8$ $M_\odot$). Abundance matching or SHAM, a common technique for populating simulated dark matter distributions with galaxies, assumes a one-to-one relation between a galaxy's stellar mass and the mass of its parent dark matter (sub)halo (Behroozi et al. 2013; Moster et al. 2013). This simple empirical approach to modeling galaxy formation has yielded great success for massive galaxies, matching a wide range of clustering statistics as a function of cosmic time (e.g., Berrier et al. 2006; Conroy et al. 2006). The importance of this success is that it indicates that a galaxy's stellar mass is determined, to a very large extent, solely by the mass of its parent halo.

To robustly probe the dark matter content on these low-mass scales requires observations of a statistically meaningful sample of systems beyond the very local universe, ideally extending to higher redshift. Unfortunately, current kinematic constraints on the stellar mass-halo mass relation are barely able to reach the critical scales where breakdown in abundance matching arises ($M_\star$ < $10^8$ $M_\odot$; see figure 5-15). Moreover, the required observations are very challenging, requiring 5–10 hours of integration with Keck/DEIMOS and are limited by the spatial and spectral resolution afforded by current facilities. The sensitivity of TMT combined with the spectral capabilities of IRIS and WFOS will open up the ability to constrain the dynamics (and thus dark matter halo masses) of low-mass systems far beyond current limits and across a range of environments (including both field and satellite populations).

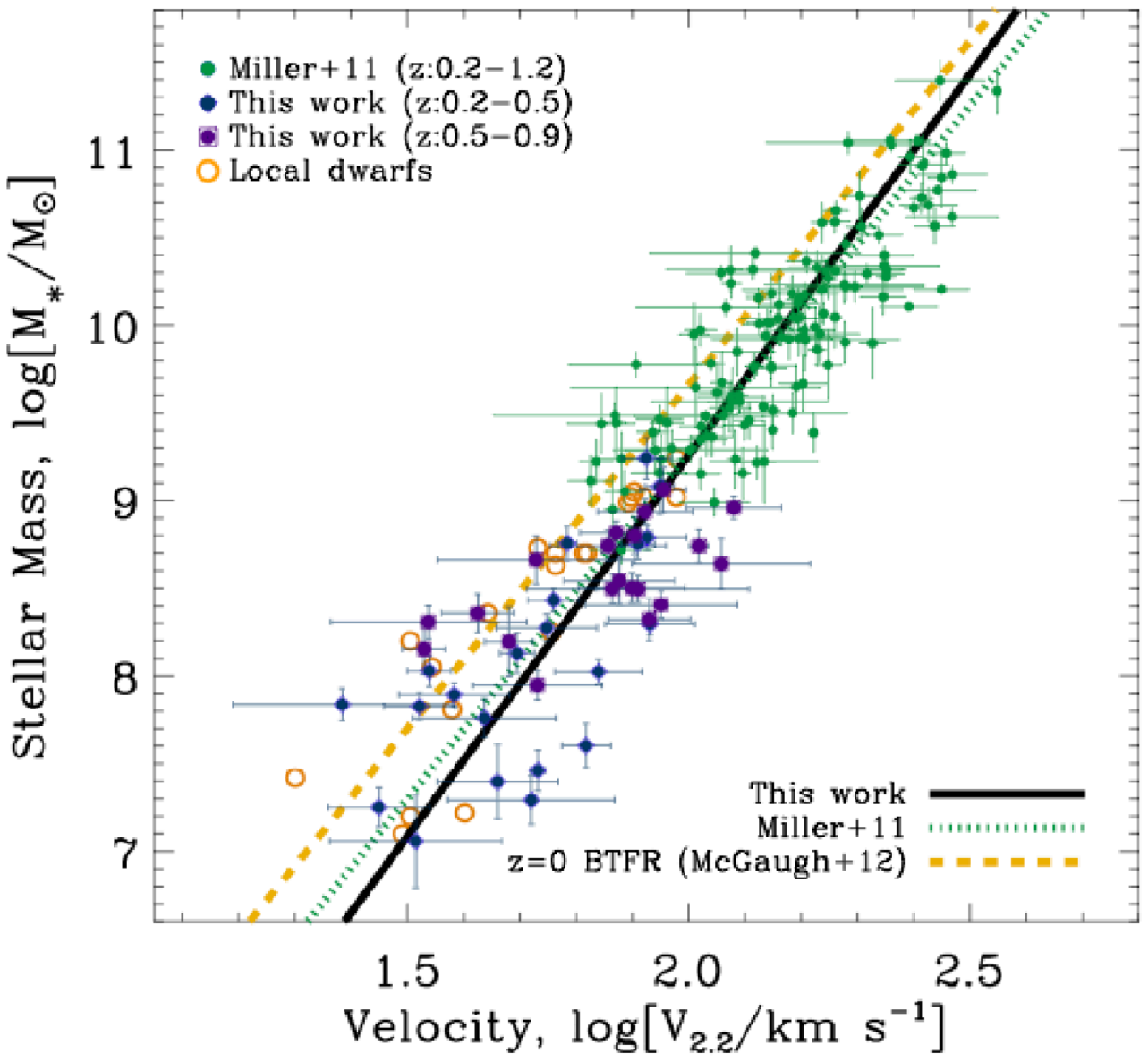

***Figure 5-15:*** *Fig. 5, Miller et al. 2014 - The Tully–Fisher (TF) relation for stellar mass for a sample of low mass galaxies in redshift intervals 0.2<z<0.5 (blue diamonds) and 0.5<z<0.9 (purple squares), and a higher mass sample over the full redshift range (green points). Golden rings show the Local Group dwarf galaxies in the same mass range. Lines represent best-fit TF relations, black line for dwarf mass, dotted green for intermediate redshift samples and yellow line for the local baryonic TF relation.*

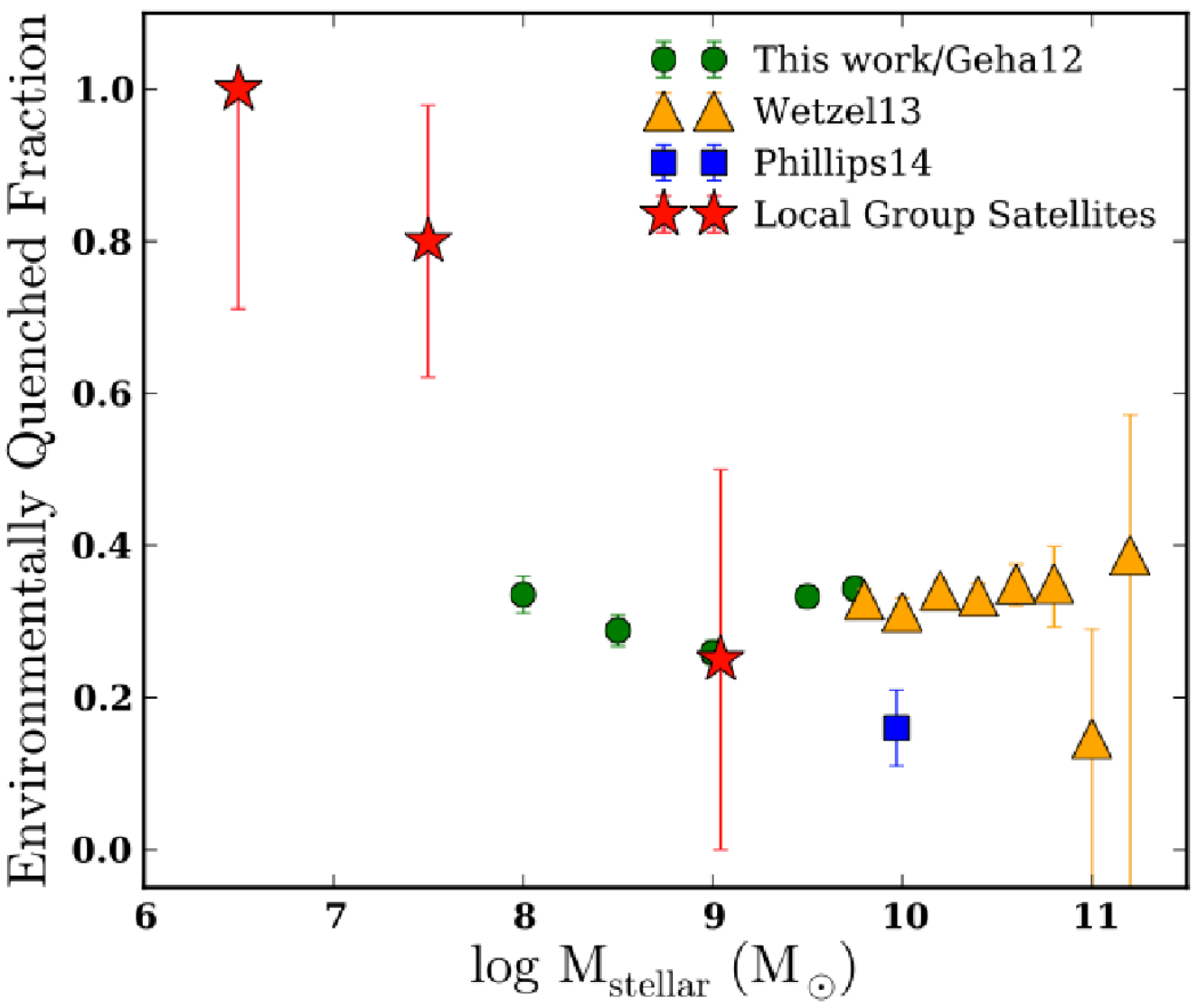

*Figure 5-16: The dependence of the environmentally quenched fraction on satellite stellar mass; that is, the fraction of satellites that are quenched in excess of that expected in the field, e.g., the fraction of satellites that are quenched because they are satellites, as a function of stellar mass. While environmental quenching seems to have an approximately constant efficiency of ~30% at stellar masses from $10^8$ to $10^{11}$ $M_\odot$, there is a dramatic upturn in quenching at lower stellar masses (if the Local Group is typical). TMT will allow spectroscopic studies to probe this regime beyond the Local Group, measuring SFRs and kinematics for large populations of low-mass systems. Figure from Wheeler et al (2014).*

Beyond the apparent breakdown of abundance matching, the limitations in our understanding of galaxy evolution at low masses are further illustrated by the inability of models to reproduce the number of passive or quenched satellites in the local universe, such that models overpredict the quenched satellite population locally (e.g., Weinmann 2006, 2010; Kimm et al. 2009; Phillips et al. 2014). Recent observations of the quenching efficiency for satellite galaxies find that only about 25–30% of infalling satellites are quenched in local groups and Milky Way-like systems. Moreover, this quenching efficiency is relatively independent of satellite stellar mass, such that lower mass systems (> $10^8$ $M_\odot$) are equally likely to quench as their more massive counterparts (Wetzel et al. 2013; Wheeler et al. 2014). From studies of the Local Group, however, we can conclude that this trend likely breaks down at yet lower stellar masses — below ~$10^8$ $M_\odot$, nearly all satellites in the Local Group are quenched (see figure 5-16), potentially indicating a dramatic change in the efficiency of satellite quenching at ~$10^8$ $M_\odot$.The major limiting factor in understanding the quenching of satellite galaxies, or more generally, the evolution of low-mass systems, is the limited observational information at stellar masses < $10^8$ $M_\odot$. While simulations have pushed their resolution limits down to very low masses, our observational knowledge is severely lacking. *N*-body simulations are able to resolve halo masses as small as $10^4$–$10^5$ $M_\odot$ (e.g., Springel et al. 2008), while observations of galaxies with stellar masses of $\lesssim 10^8$ $M_\odot$ are limited to the few systems found in the very local universe. TMT will provide a dramatic leap forward in our observational capabilities, pushing observations of low-mass galaxies (~$10^6$ $M_\odot$) beyond the Local Group and more importantly out to intermediate redshift ($z \sim 1$) over a broad range of environments. In particular, TMT will place invaluable constraints on the stellar mass-halo mass relation (and its evolution), while testing models of galaxy formation (including satellite quenching) in the yet-poorly-understood low-mass regime.

**Spatial distributions of stellar populations:** Stellar mass functions of quiescent and star-forming galaxies evolve with time in such a way that quiescent galaxies become progressively more dominant with time. This trend is further accelerated in high-density environments, such that clusters, even those at high *z*, are often dominated by quiescent galaxies. The detailed physics of quenching, however, is yet to be fully understood and is one of the major unresolved issues in our understanding of galaxy evolution. The superb spatial resolution of TMT may give us a handle on this long-standing question.

Deep, AO-assisted, IFU spectroscopy will allow us to study the spatial distribution of stellar populations within galaxies. By examining spectral features that are sensitive to star formation on different timescales (e.g., Hδ vs. $D_{4000}$), we can see which part of a galaxy is quenched and on what timescale. There are a number of physical processes that are expected to shut off the star-formation activities of galaxies. Their quenching efficiencies differ between the central part and outskirts of galaxies — e.g., ram-pressure stripping first occurs in the outskirts where the potential well is shallower and star formation is expected to cease progressively from the outskirts to the center. On the other hand, strangulation suppresses star formation on a galaxy-wide scale. Together with the conventional wisdom that different processes are more or less effective in different environments (e.g., ram-pressure is most effective in dense cluster cores, while galaxy-galaxy interactions may be least effective there), mapping the spatially-resolved stellar populations may help identify the physical process(es) at work. Another clue may come from the dynamical status of the outer part of the quenched galaxies. Some processes are expected to transform star-forming galaxies into S0s instead of classical, dispersion-dominated elliptical galaxies. The E/S0 classification is notoriously difficult at high $z$ even with *HST*, but they are dynamically distinct (S0 has a rotating disk, while E does not). So, measuring the dynamics of the outer parts of each galaxy would be another important clue toward solving the problem.

The spatial distribution of stellar populations is also interesting from the point of view of the size evolution of galaxies. Current theories predict that the cores of early galaxies form at high redshifts and are observed as compact, massive galaxies at $z \sim 2$. Then, low-mass galaxies accrete onto these galaxies and fill out the outer part to form an extended envelope. We can test this theory by direct observations of early-type galaxies over a range of redshifts. Low-mass galaxies are metal-poor galaxies and this model predicts the metallicity gradients of massive early-type galaxies should become stronger with time. It may be challenging to measure stellar metallicity with sufficient signal-to-noise even with TMT, but a stacking analysis would yield interesting statistical constraints.

**Synergy with Large Surveys:** TMT is not a survey telescope. Its field of view is relatively small compared to the largest panoramic telescope facilities; its power lies in spectroscopic follow-up observations and high angular resolution and sensitivity. Clusters and groups of galaxies that we have discussed in this section are rare objects and one needs to survey a large volume to cover a wide enough range of environment. There are a number of surveys such as the Hyper Suprime-Cam Survey, Dark Energy Survey, and the Legacy Survey of Space and Time that cover more than 1000 square degrees and are highly complementary to TMT. By using cluster catalogs from these surveys, we can efficiently point TMT to interesting groups and clusters to cover a wide environment parameter space. The superb spatial resolution and depth of TMT (aperture, AO, and instrument throughput) will then allow us to gain much deeper insights into galaxy properties than these wide-area surveys can achieve alone.

## 5.3 THE INTERGALACTIC MEDIUM

### 5.3.1 Background

The overwhelming majority of baryons reside in a diffuse, highly ionized plasma that lies between galaxies known as the intergalactic medium or IGM (Meiksin 2009). This IGM is a fundamental prediction of hierarchical cosmology where baryons trace the underlying dark matter distribution giving rise to the Cosmic Web (figure 5-17). Observationally, this web is manifested in the spectra of distant sources as a thicket of HI Lyα absorption-lines commonly termed the Lyα forest. Because the physics of diffuse gas can be relatively simple, the theory has told us that the neutral hydrogen (HI) optical depth, as well as the gas temperature, is controlled by the gas density and the intensity and spectral shape of the extragalactic UV background (EUVB). This simplicity makes it possible to turn a Lyα forest spectrum, as in figure 5-17, into a one-dimensional map of the density along the line of sight, as a function of redshift. These data provide information on the spectrum of initial density fluctuations without the non-linear gravitational processing that affects denser regions of the universe, and thus the Lyα forest may offer the best means of measuring the spectrum of these perturbations on small scales (McQuinn 2016).

The highest quality data on the physical properties of the IGM have been recorded for the redshift range $z$ = 1.6–3.5, for several reasons: first, Lyα must be redshifted above the atmospheric UV cutoff of 0.31μm, setting the minimum redshift; second, the Lyα forest evolves extremely rapidly with redshift, so that by $z>3.5$, it is so dense with absorption that it loses dynamic range for measuring weak spectral features; third, QSOs bright enough for echelle observations ($m < 19$) are very rare even at their peak near $z = 2.5$, and by $z = 6$ there is still only a small sample known in the entire sky. Thus, in general, our current view of the IGM is "one-dimensional", in the sense that we do not get the picture on the lower panel of figure 5-17 but only the one on the upper panel.

Indeed, experiments designed to examine the gas surrounding galaxies — their so-called circumgalactic medium or CGM — reveal a substantial reservoir of cool, enriched gas. Establishing the interplay between the IGM and galaxies is a major focus of current research and bears directly on theories of galaxy formation in modern cosmology.

Decades of research on the IGM using high-dispersion, absorption-line spectroscopy has established the basic properties of this medium: the surface density distribution of the HI gas, the PDF of its transmission with redshift, that the gas is mostly highly ionized, and temperature and thermal history constraints. These data have also revealed that portions of the IGM are enriched in heavy elements (C, Si, O). High-resolution spectra are capable of detecting trace amounts of metals, down to metallicities as low as 1/5000 of the solar abundances. In fact, such lines of sight provide a very high quality "core sample" of the intervening universe, from which we have learned that there may be no gas, anywhere in the universe, that was not polluted with the products of stellar nucleosynthesis at some early epoch. This enrichment implies an interplay between galaxies and the IGM throughout the early universe including, perhaps, the first stars. Current galaxy formation theory requires that the IGM provides galaxies with a nearly continuous supply of fuel for star formation.

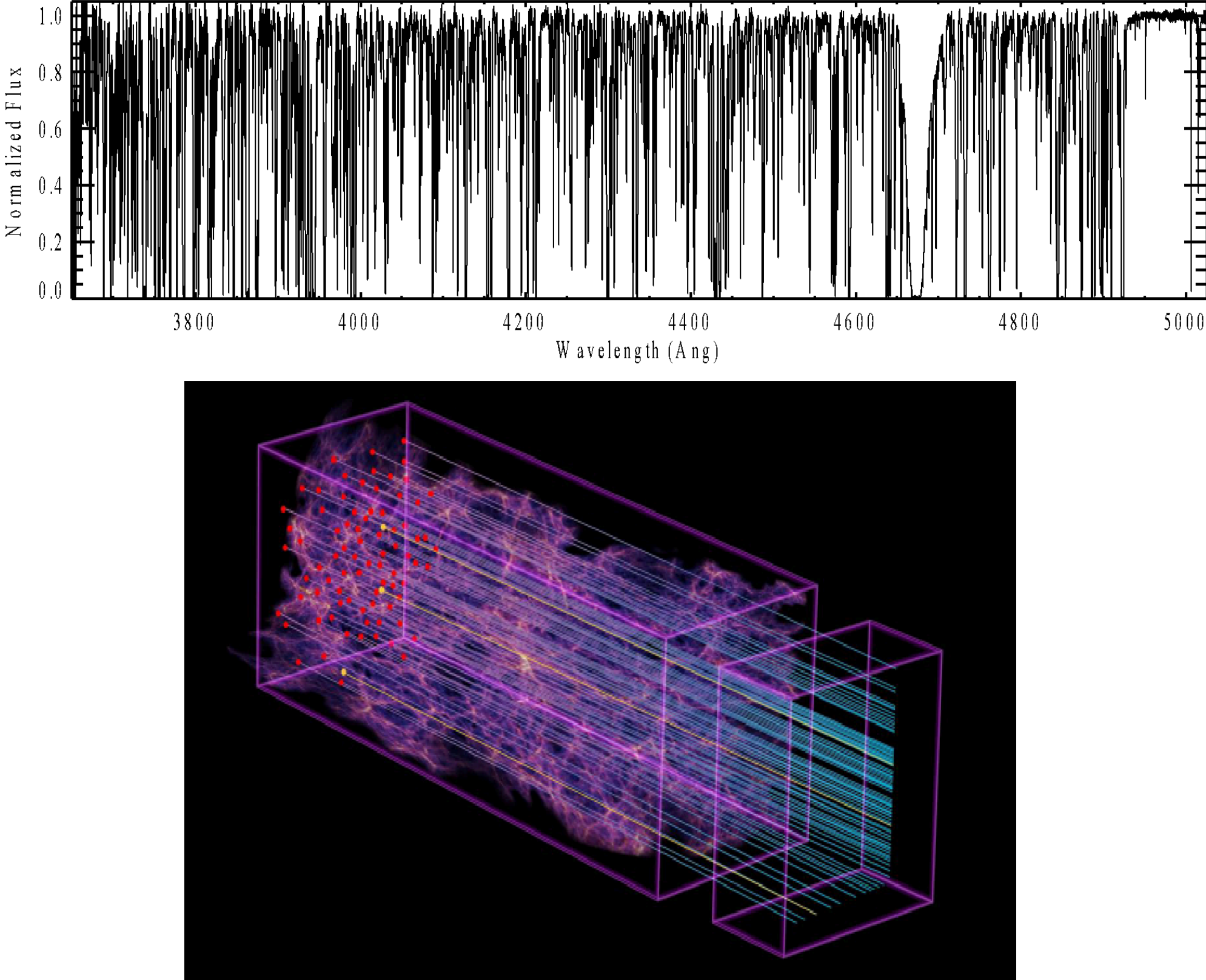

***Figure 5-17:*** *The cosmic web of the baryon distribution in a cosmological simulation. Here, HI in the IGM traces the dark matter distribution even in regions with low density contrast. A line of sight through the volume yields a one-dimensional map of both the HI and metallic species along the line of sight, as shown in the top panel. Densely sampled sightlines through a survey volume, together with detailed maps of the galaxy distribution, will provide unprecedented views of the distribution of baryons in the universe, and their relation to the sites of galaxy formation. [Illustration Credit: Zosia Rostomian, LBNL; Nic Ross, BOSS Lyman-alpha team, LBNL; Springel et al, Virgo Consortium and the Max Planck Institute for Astrophysics].*

Complementing the single-sightline, high-dispersion data, several projects have generated massive spectroscopic datasets (e.g., SDSS, BOSS) to study the IGM across large-scales and to use it as a direct probe of cosmological models. Despite the lower spectral resolution, the large sample size and sky-area-coverage offer unique analyses including measurements of the baryonic acoustic oscillations (BAO) in the IGM transmission field, cross-correlation of the HI gas to IGM metal absorption and high-z quasars and galaxies, and measurements of the mean free path to ionizing radiation. In the TMT era, there will be powerful synergy between the high-fidelity (high signal to noise and spectral and spatial resolution) data one can obtain on narrow fields with these large spectroscopic datasets spanning the full sky.

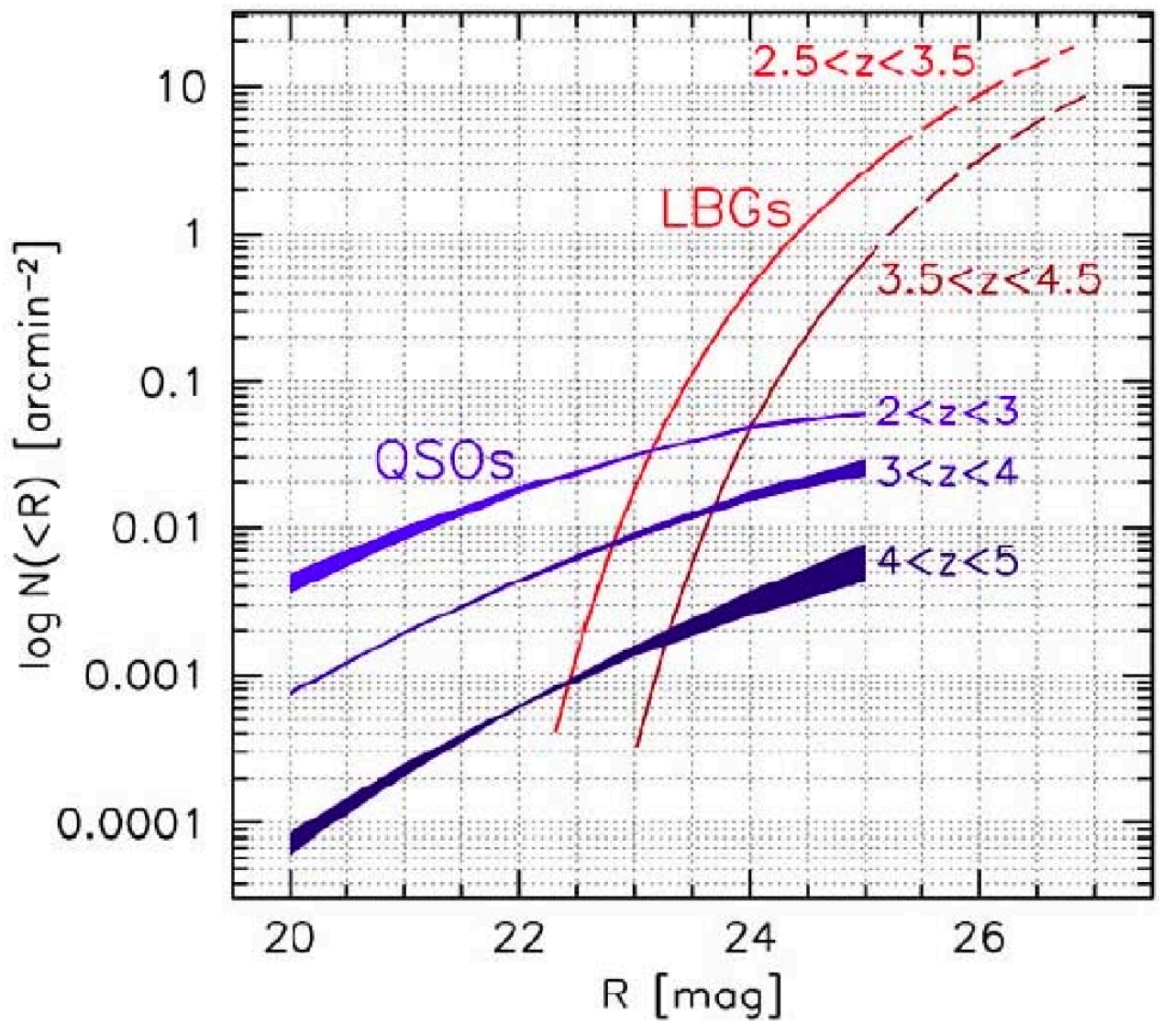

***Figure 5-18:*** *The cumulative surface density of QSOs and UV-bright galaxies as a function of R mag. The surface density of suitable IGM probes increases exponentially for R > 22 due to the very steep rest-UV luminosity function of star forming galaxies. By R ~ 24, the surface density of galaxies+QSOs exceeds 1 arcmin$^{-2}$, sufficient for tomographic mapping of the IGM on Mpc scales.*

### 5.3.2 TMT and the IGM

As emphasized above, our view of the IGM on Mpc scales is essentially one-dimensional, limited by the number density of luminous quasars on the sky. To resolve the Cosmic Web on scales of 1 Mpc we must obtain IGM spectra of a denser set of background sources. Figure 5-18 shows the space density of compact, UV-bright star forming galaxies at apparent magnitudes sufficient to yield very high-quality spectra. For galaxies with R ~ 24.5 (~0.6 μJy), approximately the apparent magnitude at which TMT/WFOS can obtain a spectrum at $R \sim 5000$ with S/N ~ 30, those in the appropriate redshift range ($z$ = 1.8–4) outnumber QSOs by more than a factor of 30.

This means that the IGM properties can be densely sampled on physical scales of < 1 Mpc, approximately the maximum sphere of influence of individual galaxies on the IGM and comparable to the expected coherence length of the undisturbed IGM. Thus, the 3D IGM can be effectively reconstructed tomographically over the range $1.6 < z < 3.5$ where the Lyman-α forest can be observed from the ground with good dynamic range.

With a sufficiently dense grid of closely-separated background sources, including faint QSOs and bright LBGs, it becomes possible to interpolate across the transverse plane to 'tomographically' reconstruct the 3D Lyα forest absorption field. Lyα tomographic maps with ~Mpc resolution will extend 3D cosmography beyond the z ~1 currently achievable by deep galaxy redshift surveys (e.g., Davis et al. 2003; Lilly et al. 2007).

In addition, as illustrated by the simulated reconstruction maps in figure 5-19, Lyα tomography is effective at mapping out overdensities of order unity and cosmic voids. Since each Lyα forest sightline probes ~ 500 $h^{-1}$ Mpc (comoving) along the line-of-sight, large cosmic volumes can be efficiently mapped. The resultant maps reveal the topology of large-scale structure in the high-z universe, constrain the power-spectrum of dark matter, and resolve extremes in the density field such as proto-clusters and voids.

Our current, one-dimensional view of the IGM has primarily been derived from several tens of echelle spectra at high S/N (>30 per pixel) recorded on 10-m class telescopes. These data pale in comparison with the S/N >> 100 per pixel one obtains for stars in our Galaxy (and its neighbors). With an echelle spectrometer on TMT one could achieve S/N>1,000 on the most luminous sources to give an unprecedented view of the low-density IGM. Such data could, for example, reveal the presence of a CIV forest, akin to the HI Lyα forest of current datasets (see figure 5-20). The spectra would establish the distribution of metals at levels below the mean density of the universe (e.g., the majority of the volume), testing models of pre-enrichment. These extremely high-quality spectra will also place high-precision constraints on the temperature of the IGM to better resolve its thermal history. The IGM temperature with redshift is a complex balance between adiabatic cooling from cosmic expansion with photoionization heating by quasars and star-forming galaxies and (possibly) heating by exotic sources (e.g., blazars) and structure formation. These data will also examine systematics that affect the large-field surveys of the IGM (e.g., continuum normalization).

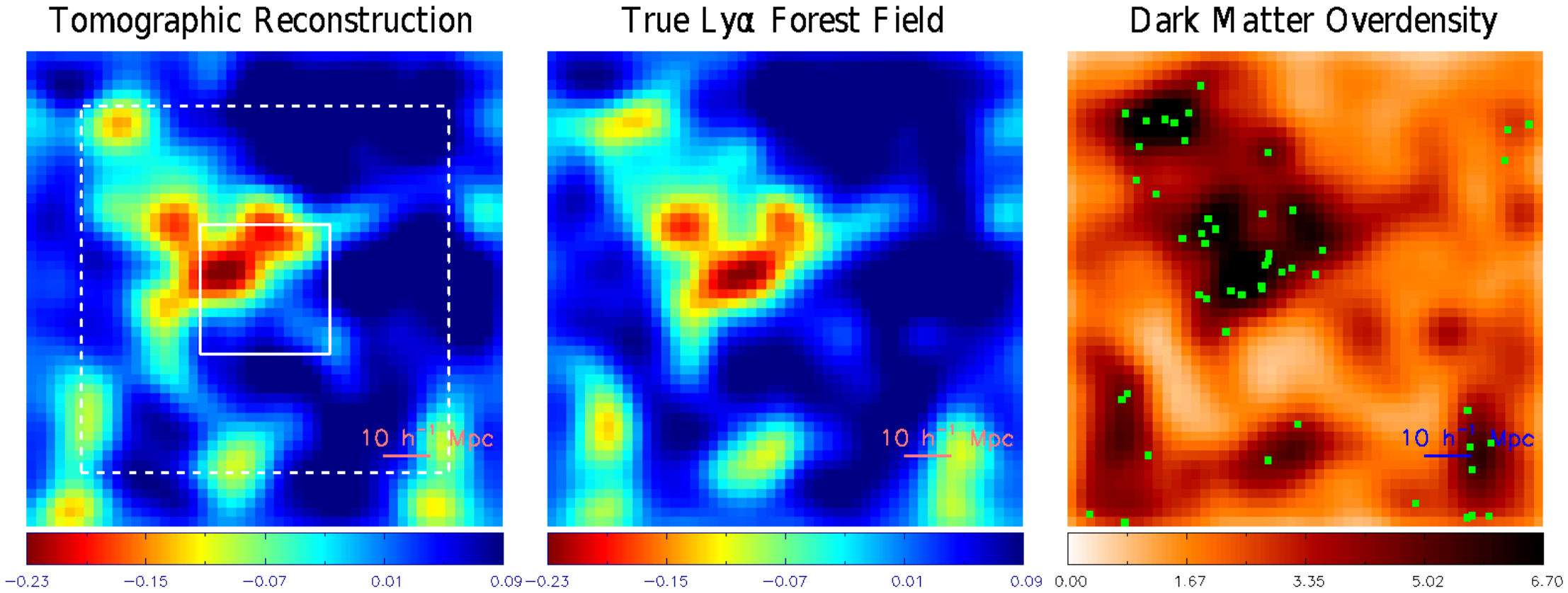

***Figure 5-19:*** *Simulation slices illustrating Lyα forest tomography with map resolutions of 3.5 $h^{-1}$ Mpc. (Left) A tomographic reconstruction from a set of simulated Lyα forest absorption sightlines from sources down to a survey depth of mAB = 24.5, which include realistic modeling of spectral S/N assuming the Reddy et al. (2008) LBG UV luminosity function. The central panel shows the true underlying 3D Lyα forest absorption field while the right panel shows the underlying dark-matter overdensity, $\Delta = \rho/\langle\rho\rangle$ (both smoothed with a 3.5 $h^{-1}$ Mpc Gaussian to match the reconstructed map). These slices have dimensions (100 $h^{-1}$ Mpc) 2 × 2 $h^{-1}$ Mpc, and the line-of-sight direction is into the plane of the page — real surveys cover ~400 $h^{-1}$ Mpc along the line-of-sight. The green dots overlaid on the DM field show positions of $R \leq 25.5$ galaxies coeval with the Lyα forest, obtained from halo abundance matching. The solid small rectangle overlaid on the left-panel shows the footprint covered by approximately 16 WFOS pointings, while the large dashed rectangle shows the area equivalent to 1 $deg^2$ on the sky. [Image adapted from Lee et al. 2014].*

Another emerging area of IGM research is to study the gas in emission. Once illuminated by ionizing photons from the EUVB or a nearby source, HI gas fluoresces with approximately 60% of the photons converted to HI Lyα (Adelberger et al 2006; Cantalupo et al 2007; Kollmeier et al 2010). Figure 5-21 shows the Slug Nebula surrounding the bright UM287 quasar at z=2.1 which extends hundreds of kpc across, e.g., far beyond the dark matter halo hosting the source. Using the boosted ionizing radiation field of the nearby quasar, gas from the surrounding IGM may be studied in emission.

Narrow band imaging and IFU spectroscopy on TMT could detect much fainter sources including optically thick gas ionized only by the EUVB. TMT's greater impact, however, would be deep optical and near-IR spectra of this extended emission to diagnose properties of the IGM in an entirely complementary fashion to absorption-line techniques.

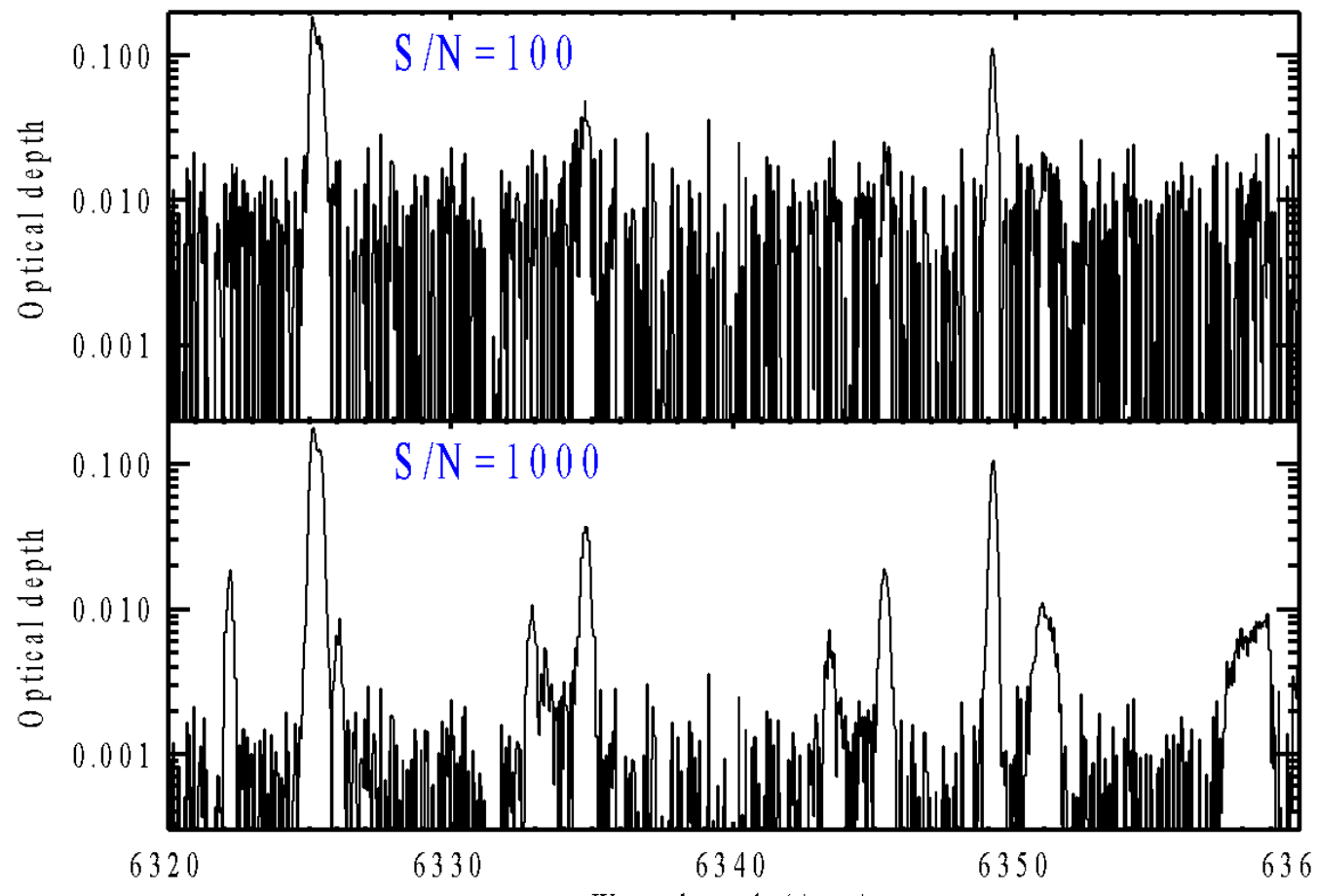

***Figure 5-20:*** *Simulated optical depth plots of CIV absorption from the IGM assuming the IGM is enriched to 1/1000 solar at the mean density. The upper panel shows the highest quality spectra that one can achieve with 10-m class telescopes. With TMT, we will reach an unprecedented S/N>1,000 spectra on the most luminous sources (including gamma-ray bursts). These spectra would directly establish whether a CIV forest exists in the z~2 universe. In turn, the data would reveal the full distribution of metals in the high-z IGM.*

The surface brightness of even the bright nebula in figure 5-21 requires several night-long integrations on a 10 m telescope to sensitively study Lyα, CIV, HeII, and Hα line-emissions. TMT observations could resolve the morphology of the gas distribution and provide information on the kinematics, enrichment, density, and ionization state. If combined with an IGM tomographic study, one would gain a complete picture of the morphology and physics of the IGM on Mpc scales.

Observations to support this science place pressure on the aperture (signal to noise and angular resolution), AO (angular resolution), wavelength range (UV to IR), multiplexing, and IFU capabilities of the TMT and its instruments (see section 12).

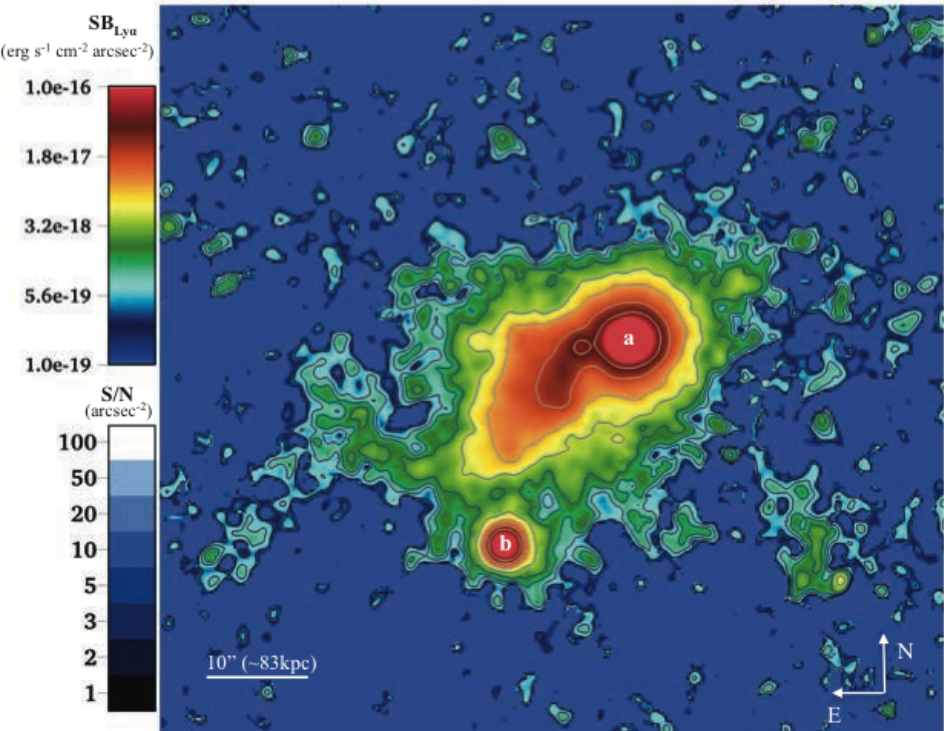

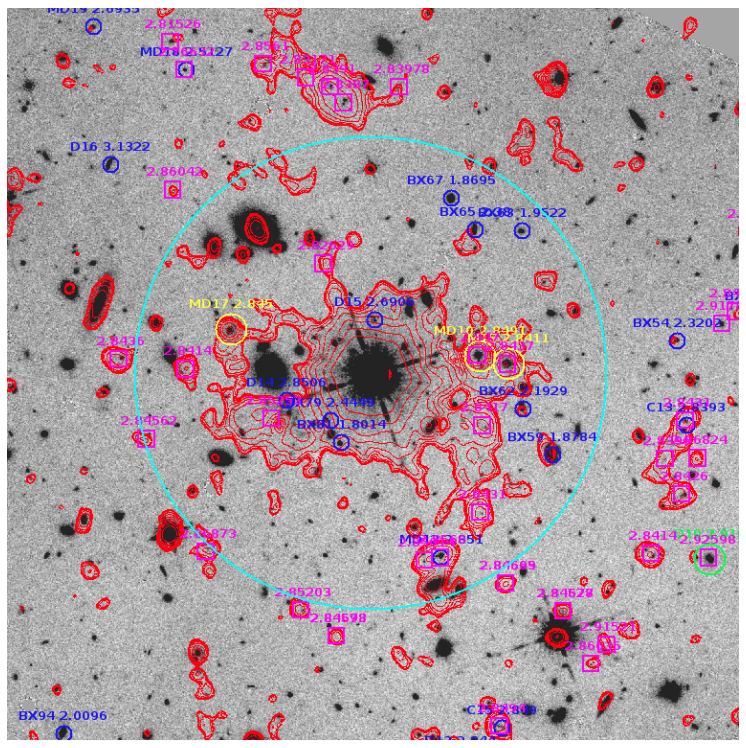

***Figure 5-21:*** *(Left) Lyα image of the Slug Nebula surrounding UM287 (labeled "a"), obtained using narrow-band imaging with Keck/LRIS-B. The color map indicates the Lyα surface brightness per arcsec$^2$. The extended emission spans a projected angular size of ≈ 55 arcsec (about 460 physical kpc), measured from ~$10^{-18}$ erg s$^{-1}$ cm$^{-2}$ arcsec$^{-2}$ contours. The nebula appears broadly filamentary and asymmetric, extending mostly on the eastern side of quasar "a" and towards south-east in the direction of a second, fainter quasar (labeled "b"). [Adapted from Cantalupo et al. 2014]. (Right) A 2' by 2' region (the cyan circle has a diameter of 1', or ~500 kpc) surrounding the z=2.84 QSO HS1549+19; the grayscale is a deep HST WFC3 F160W image (1.6µm) centered on the QSO; red contours are Lya emitting regions at the same redshift, where magenta labels mark regions with Lya spectra. (Adapted from Trainor et al 2013).*

### 5.3.3 TMT and the CGM

One of the most exciting possibilities enabled by the TMT is to combine the precision measurements of the astrophysics of the IGM using relatively bright background sources with direct observations of the luminous material — galaxies and AGN — in the same volumes of space, providing for the first time the first three-dimensional view of the distribution of baryonic material in the high redshift universe. Figure 5-22 illustrates the connection between galaxies and the IGM, with the observed nebula spanning from galactic scales (tens of kpc) to intergalactic scales (hundreds of kpc). Feedback processes from galaxies and their active galactic nuclei affect the surrounding IGM, e.g., enriching the gas, ionizing the medium, depositing heat. At the same time, gas from the IGM funnels into the galactic potential wells to fuel new stars, continuing the *baryon cycle* of feedback and accretion. Present studies of the CGM are primarily limited to examining the gas in absorption in the rare cases where a quasar or luminous galaxy lies behind a foreground galaxy, thereby obtaining single-sightline information, one galaxy at a time, and primarily at low spectral resolution with correspondingly limited constraints on the physical conditions.

The IGM tomography experiment described in the previous section may be expanded to include analysis of the CGM. Star-forming galaxies with R < 24.5 used to characterize the IGM absorption represent only the tip of the iceberg for the high redshift galaxies. The combined sensitivity of WFOS and TMT allows for spectroscopy of much fainter galaxies (to R ~ 27, albeit with lower spectral resolution than for the IGM probes) with a high degree of completeness. Galaxies at such apparent magnitudes have surface densities of >20 $arcmin^{-2}$ for each interval in redshift of $\Delta z \sim 0.5$ from $z \sim 2$ to $z \sim 3.5$ (figure 5-22). Fortuitously, this redshift range also encompasses what we believe to be the most important era for galaxy growth, encompassing star formation and massive black hole accretion. With TMT, it will be possible to simultaneously obtain a densely sampled map of the distribution of galaxies and their CGM, in 3D, providing the most complete possible census of all normal matter, and its relationship to dark matter. For the first time, the empirical picture of the high redshift universe would be of similar fidelity as those that currently exist only inside simulations. More importantly, the observations will represent the inner workings of all of the poorly understood physical processes that must be incorporated in order to understand how galaxies form — feedback (both hydrodynamic and radiative) from AGN and galaxies, gas accretion, details of the relationship between structure in diffuse gas versus that traced by galaxies.

An example of the type of exciting survey that could be mounted with TMT would be a TMT/WFOS observing program to survey both galaxies and the IGM over a volume of the $z$ = 1.8–3.5 universe that is as statistically representative as the Sloan Digital Sky Survey (SDSS) redshift survey at $z \sim 0.1$ and could be accomplished in a reasonable amount of telescope time. The relationship between angular scale on the sky and co-moving scale at the targeted cosmic epoch is vastly different between $z \sim 0.1$ (SDSS) and $z \sim 2.5$ (TMT). The SDSS was carried out with a telescope+instrument combination capable of observing over a field of view with a diameter of 2.5 degrees, or a transverse scale of ~18 Mpc (co-moving) at the median redshift of the survey. At $z \sim 2.5$, the same 18 Mpc co-moving scale is subtended by an angle of 10.6 arcmin on the sky. Thus, WFOS, with its 8.3' wide field of view, can be thought of as a wide-field spectrograph for studies of the distant universe. A survey of a representative volume of the universe (~$10^8$ co-moving $Mpc^3$) covers a solid angle of $\pi$ steradians at $z \sim 0.1$; for the proposed TMT/WFOS baryonic structure survey, the same volume is covered by ~4.5 square degrees on the sky, within which there would be ~650,000 star forming galaxies brighter than R ~ 26.5 in the redshift range $1.6 < z < 3.5$ that could be selected for spectroscopy using simple photometric criteria to within $\Delta z \sim 0.4$. The total number of targets with R < 24.5 and $1.6 < z < 3.5$ in the same volume would be ~30,000. A summary of the observing parameters, assuming a total WFOS slit length of 8.3 arcmin, is given in Table 5-2.

The WFOS survey products would also include:

- Identification in redshift space of ~1000 overdense regions that will become clusters by the present day. The physical state of potential hot gas in the proto-intracluster media can be matched against Sunyaev-Zeldovich signatures in future high resolution CMB maps, providing a complete census of baryons in all phases within the densest regions in the universe.

- Exquisite far-UV spectra of a large number of galaxies in the same redshift range for which IRIS and IRMOS can obtain rest-frame optical spectra. The far-UV spectra will provide measures of outflow kinematics, chemistry, stellar IMF, and in some cases mass outflow rate.
- 15,000 high quality sightlines through the IGM that will map intergalactic HI and metals in 3D, to be compared with the galaxy distribution in the same cosmic volumes. Even the lower resolution galaxy spectra will allow the mapping of inhomogeneities in the UV ionizing radiation field and measurement of the lifetime of bright UV sources via the transverse proximity effect (e.g., Adelberger, 2004; Trainor et al 2013).

Complementary HROS observations of the brightest background sources and IRMOS observations of the UV-faint galaxies would further add to the survey and increase our understanding of these targets.

In addition, TMT would enable a statistically meaningful sample of high-dispersion probes of the CGM. With R>10,000 spectra, one can reliably assess the column densities of multiple ions (HI, CII, CIV, etc.) to estimate the ionization state, total surface density, metallicity, and velocity fields. Presently the astrophysics of the CGM is poorly described; TMT would fully diagnose the properties of this medium

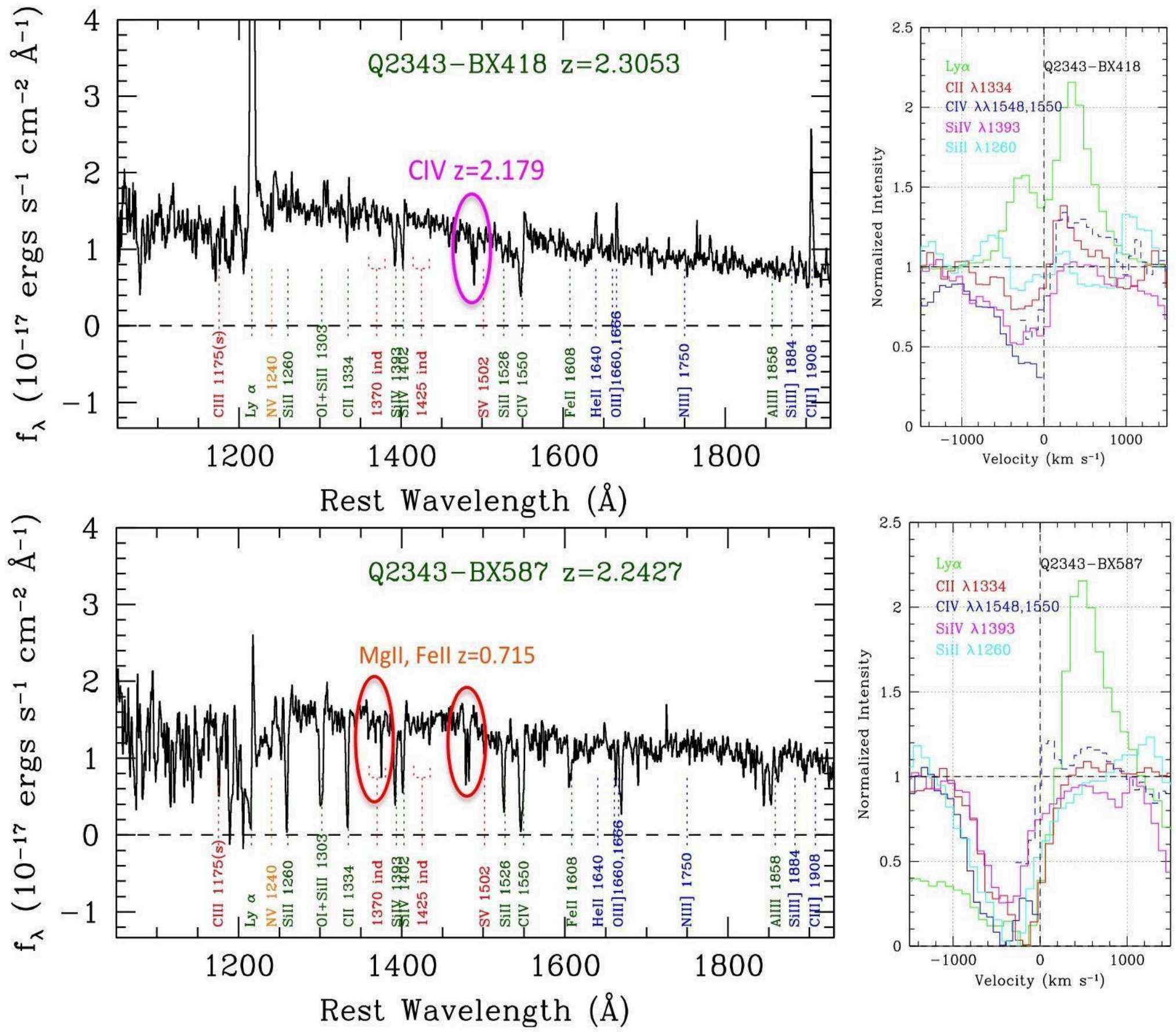

***Figure 5-22:*** *(Left) Rest-UV spectra of R = 24 galaxies observed at R=1500 obtained using Keck/LRIS, with10 hour integrations. TMT spectra of similar galaxies could be obtained with 3–4 times higher spectral resolution; spectra with similar quality will be possible to $m_{AB}$~26.5 using the R~1500 mode of WFOS. Even at this resolution, one detects metal absorption lines from intervening material (circled features), as well as measurements of outflow kinematics (right panels), showing various interstellar transitions and Lyα emission relative to the galaxy systemic redshift. Figure adapted from Steidel et al (2014).*

| Survey | Sky Area | #Targets | Exp. | λ Range | #tiles | Total |
|---|---|---|---|---|---|---|
| Galaxies: $z = 1.6 - 3.5$ | 4 x (2° x 0.56°) | 120,000 @ AB < 26.5 | 1 hour for SNR > 5 @ $R = 1000$ | 0.32 - 0.65 µm | 1000 | 1000 hours |
| IGM z = 1.6 - 3.5 | 4 x (2° x 0.56°) | 15,000 @ AB < 24.5 | 4.5 hours for SNR > 35 @ $R = 5000$ | 0.31 - 0.60 µm | 100 | 450 hours |

***Table 5-2:*** *A sample WFOS survey of baryonic structure in the high-redshift universe*

.

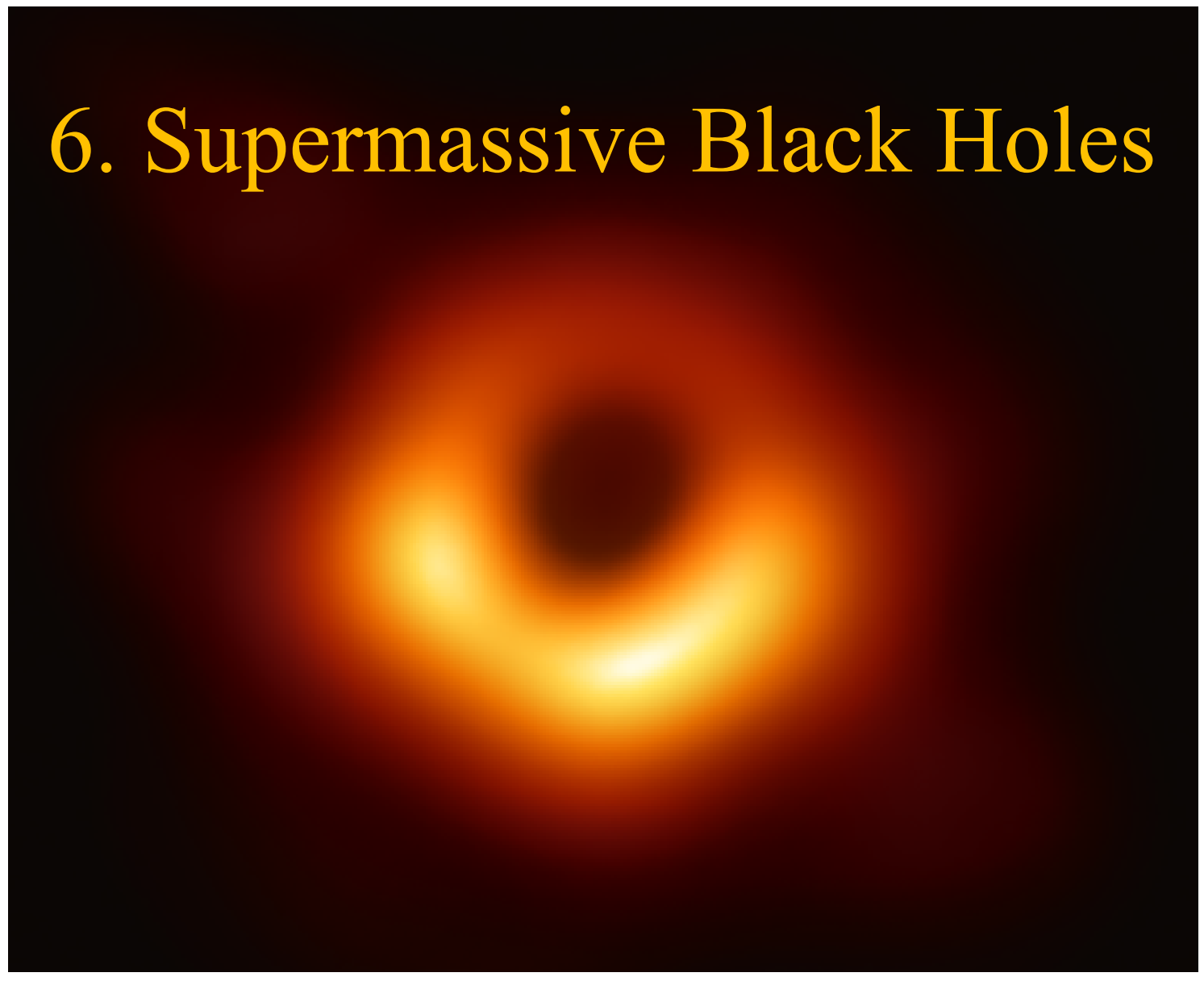

*In 2019, the image of the region of the 6.5 x*$10^9$ $M_{\odot}$ *supermassive black hole at the center of M87 was released. Named Pōwehi, at a distance of 55 million years, the SMBH gravitationally bends the light from the disk of hot gas around it and creates a dark central region as predicted by General Relativity. Image credit: Event Horizon Telescope Collaboration*

---

TMT will address the third big question in chapter 2, Q3-*What is the relationship between black holes and galaxies?,* by expanding the known census of supermassive black holes (SMBH) to much lower masses and into the regime of intermediate mass black holes (IMBH). Additionally, extending our current knowledge of SMBH masses to much earlier cosmic times will clarify the history and processes of SMBH growth. Studies of SMBHs over cosmic time also address Q1- *What is the nature and composition of the universe?* and Q2- *When did the first galaxies form and how did they evolve?*

TMT's sensitivity, diffraction limited imaging, spatially resolved spectroscopic capabilities, medium resolution spectroscopic capability are needed for these observations so that the region around the SMBH at the center of the Milky Way can be observed in exceptional detail, allowing users to probe the extreme environment in the vicinity of the SMBH, characterize the stellar population in the vicinity of the SMBH, and determine the impact of the SMBH on the region surrounding it. TMT will also be able to investigate IMBH candidates in the vicinity of the Galactic Center and the center of nearby low-mass stellar systems. In addition, these observing capabilities will enable studies of the centers of galaxies and the regions around their central SMBH across cosmic time.

The fundamental capabilities needed for these studies include TMT's spatially resolved spectroscopy and imaging at the diffraction limit. Future MCAO-fed integral field spectrographs will vastly increase the efficiency of conducting a census across SMBH masses. Time-resolved observations are necessary for AGN activity monitoring, and the AGN core dust science in particular relies on having mid-IR and polarimetric capabilities in future TMT instruments.

---

**Contributors**: Masayuki Akiyama (Tohoku University), Aaron Barth (UC Irvine), G.C. Dewangan (IUCAA), Tuan Do (UCLA), Lindsay Fuller (UT San Antonio), Andrea Ghez (UC Los Angeles), Lei Hao (Shanghai Astronomical Observatory), Fiona Harrison (Caltech), Luis Ho (KIAA Peking), Masa Imanishi (NAOJ), Mason Leist (UT San Antonio), Matt Malkan (UC Los Angeles), Claire Max (UC Santa Cruz), Leo Meyer (UC Los Angeles), Tohru Nagao (Kyoto University), Chris Packham (UT San Antonio), Michael Rich (UC Los Angeles), Yue Shen (U of Illinois), Yiping Wang (NAOC), Mansi Kasliwal (Caltech), Enrique Lopez-Rodriguez (UT San Antonio) (ISDT), Lulu Zhang (UT San Antonio)

---

# 6 SUPERMASSIVE BLACK HOLES

Supermassive black holes (SMBHs), with masses ranging from below $10^6$ to above $10^{10}$ solar masses, are now known to be present in the centers of most and perhaps all massive galaxies. The mass of the black hole (BH) is correlated with the stellar mass (the $M_{BH}$-$M_{bulge}$ relation) and velocity dispersion (the $M_{BH}$-σ relation) of the bulge of the galaxy (figure 6-1). These discoveries over the past 20 years have led to the paradigm that black holes and galaxies co-evolve and that feedback of active galactic nuclei (AGN) during the growth phases of the black hole strongly affects the gas content and star formation in the host galaxy (Ho 2004; Kormendy & Ho 2013). Understanding the formation and growth history of SMBHs, their influence on galaxy evolution, and the exotic phenomena of stellar dynamics and gas accretion in the SMBH environment has become a major theme in astronomy.

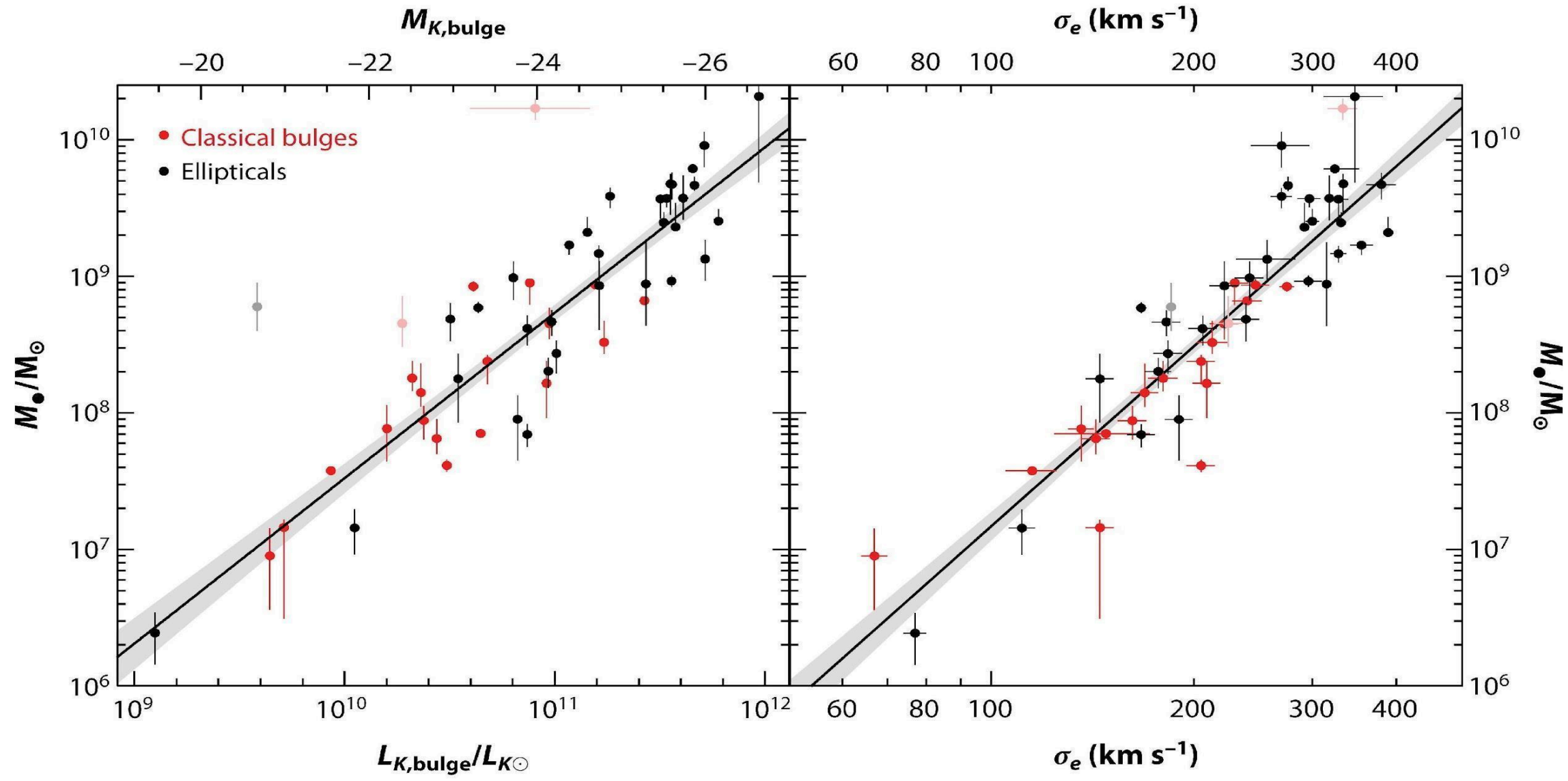

***Figure 6-1:*** *Correlation between black hole mass and bulge properties for nearby galaxies: (left) K-band luminosity, and (right) stellar velocity dispersion (Kormendy & Ho, 2013). Further studies are needed to determine the relation between black hole mass and bulge properties for all kinds of galaxies. The Paucity of measurements* $<5x10^7 M_{\odot}$ *is an opportunity for TMT.*

TMT's capabilities for high angular resolution imaging and ultra-deep spectroscopy will provide unprecedented opportunities to advance numerous areas of SMBH science. These will include precision measurements of BH masses spanning a range of more than four orders of magnitude in mass, examining the relationships between SMBHs and their host galaxy environments, understanding the physical processes of the fueling and feedback of black holes and their redshift evolution, determining the early growth history of SMBHs through observations of high-redshift quasars, and carrying out fundamental tests of General Relativity through high-precision measurements of stellar orbits around the Galactic Center. In this section, we describe a few of the most exciting science cases for understanding the fundamental properties, accretion physics, and cosmological growth history of SMBHs.

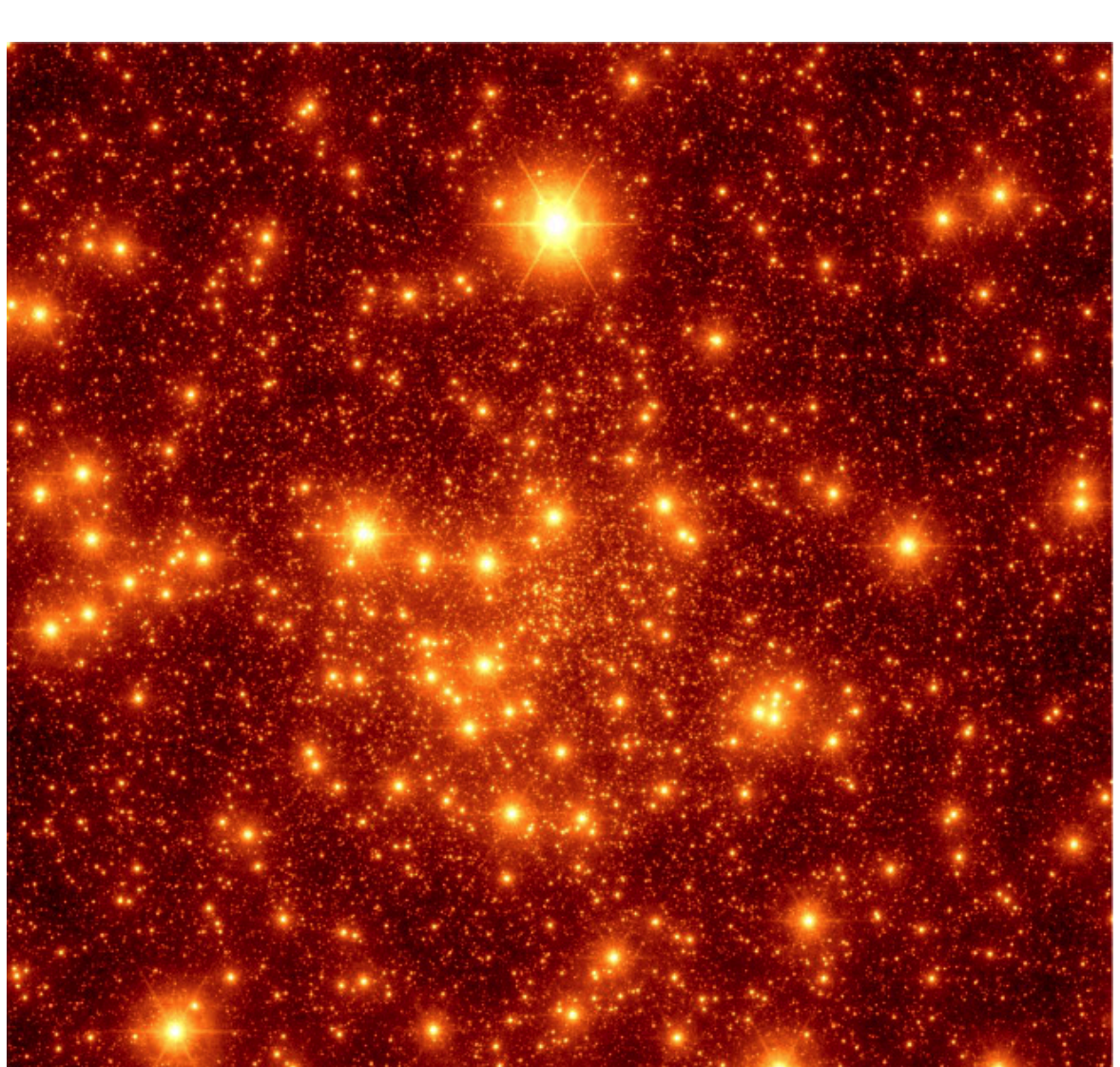

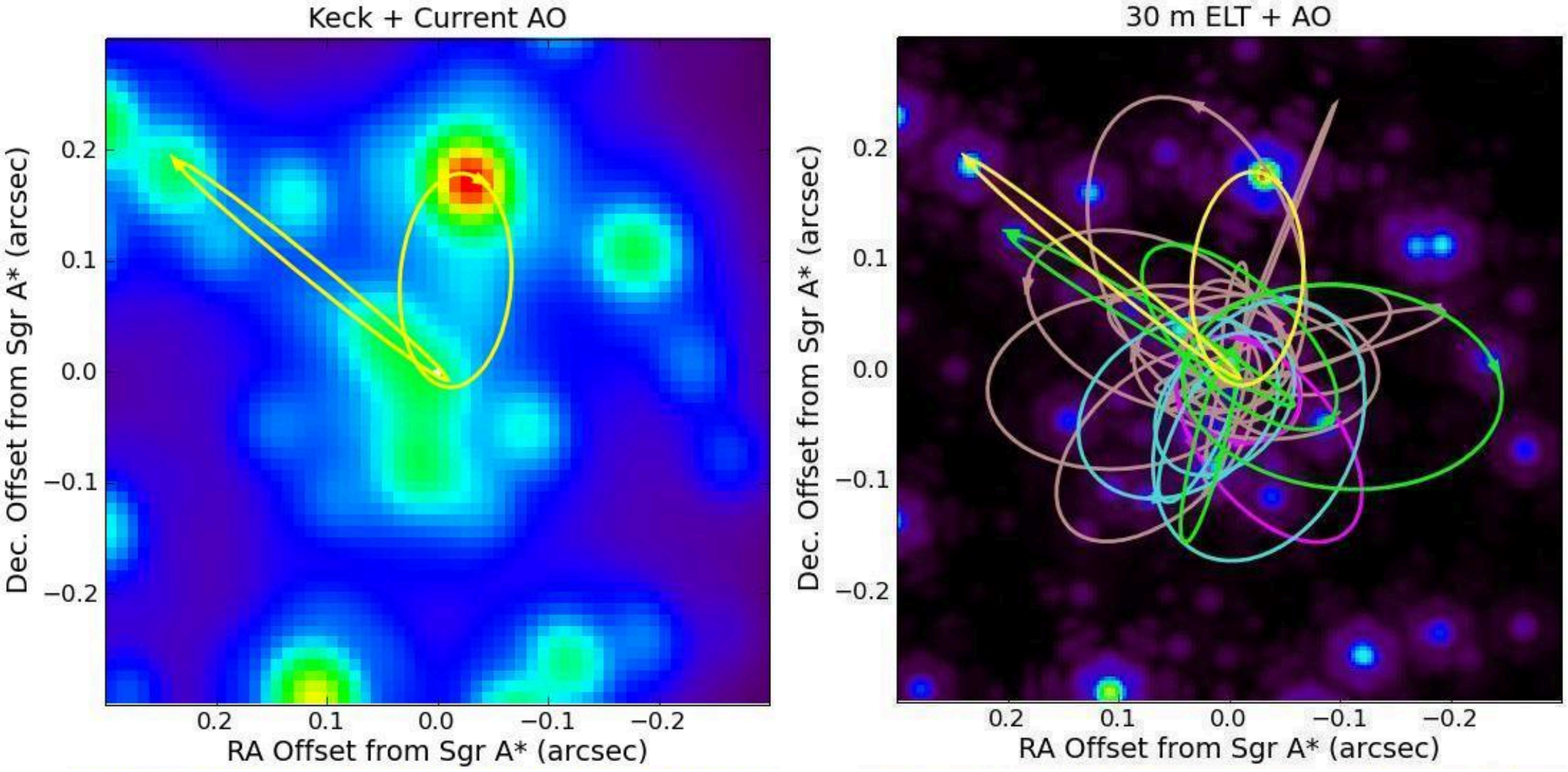

***Figure 6-2:*** *(Top) Simulated near-IR image ($t_{int}$=20s) of the central 17"x17" of the Galaxy, centered on the supermassive black hole, Sgr A* at the angular resolution of TMT. The image contains ~2×10$^5$ stars down to K ~24 mag, including ~2500 known stars and a theorized population based on the observed GC radial profile and the K-band luminosity function of the Galactic bulge. The image includes photon, background, and read noise. Credit: UCSD OIR LAB. (Bottom) Implications for short-period stars' orbits. Overlaid are all known orbits and examples of expected orbits with periods less than 23 years that are detectable both astrometrically and spectroscopically (14<K<16, yellow; K<17, green; K<18, cyan; K<19, magenta; K<20, tan). TMT/IRIS will increase the number of measurable short period orbits by an order of magnitude and also find systems that orbit the SMBH much deeper in the central potential, with orbital periods that are a factor of 5 smaller. These systems are particularly helpful for measurements of post-Newtonian effects (GR and extended mass distribution). Credit: UCLA Galactic Center Group.*

## 6.1 THE GALACTIC CENTER BLACK HOLE: OUR UNIQUE LABORATORY FOR UP-CLOSE STUDY

The Galactic Center's proximity makes it a unique laboratory for addressing issues in the fundamental physics of supermassive black holes (SMBHs) and their roles in galaxy formation and evolution. Current AO studies have transformed our understanding of the Galactic Center (GC). In the past decade, proper motion studies have determined the orbits of some individual stars moving within 0.04 pc of Sgr A* (figure 6.2). These provide the strongest evidence for the existence of a central black hole in the Galactic Center and the best mass measurement in any galactic nuclei. However, the star density in this region is so high that source confusion has limited studies to the brightest stars, and introduces biases in astrometric and spectroscopic measurements. TMT's gains in resolution and contrast will enable detection and mapping of the orbits of stars four magnitudes fainter (figure 6.3). This will allow the detection of many stars closer to the black hole than we can observe (resolve) today. The measurement of the orbits of these short period stars will be essential for tests of General Relativity in this region.

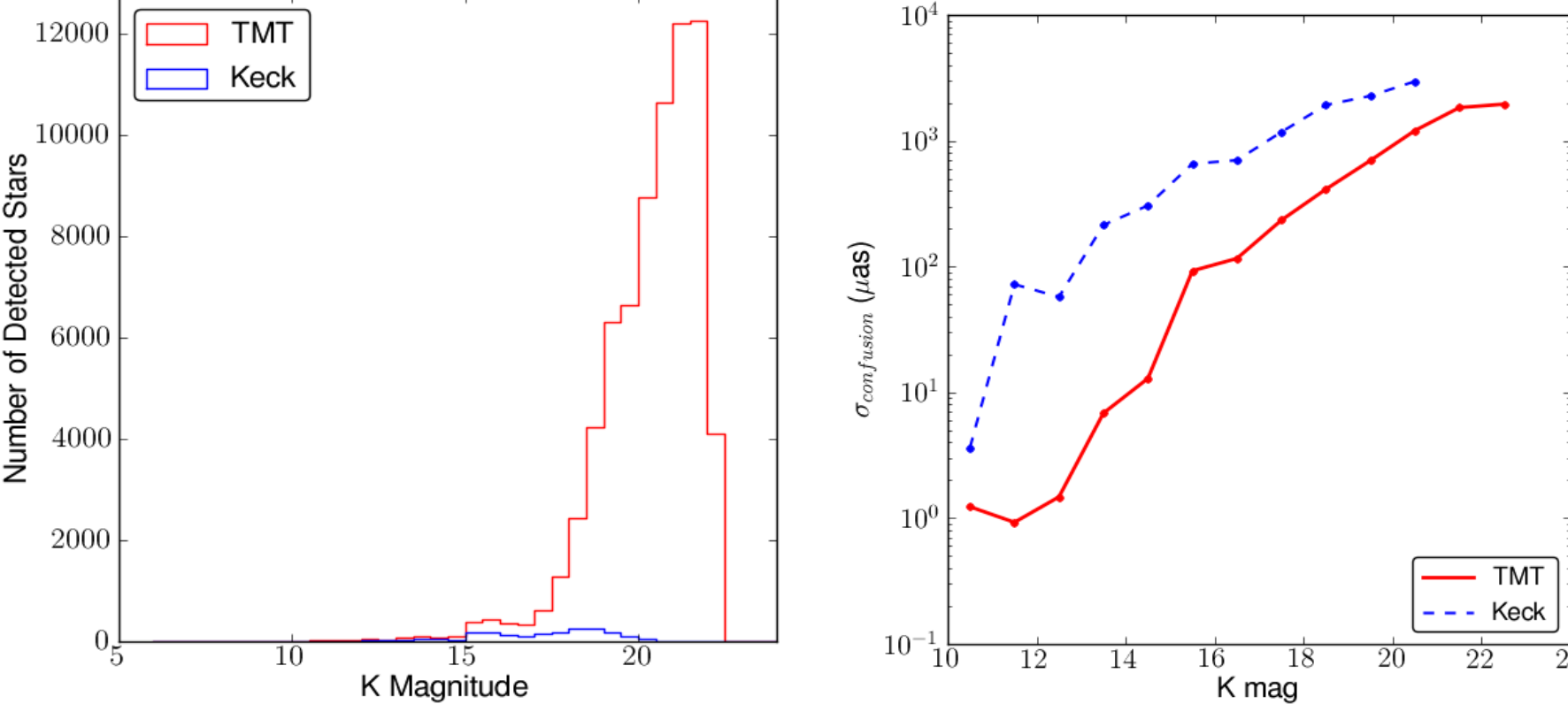

***Figure 6-3:*** *(Left) Distribution of detected stars in the simulated images for TMT (red) and Keck (blue). The number of detected stars with TMT will be two orders of magnitude greater than Keck. (Right) Astrometric error resulting from confusion plotted as a function of K-band magnitude of the planted star (red solid). Data were binned in 1-mag bins and the RMS error of the input-output positions were calculated and taken as $\sigma_{confusion}$. Confusion effects in Keck data are shown for comparison (blue dashed).*

### 6.1.1 TMT takes General Relativity tests into an unexplored regime

Measuring the orbits of short-period stars around the Galactic center black hole will probe the structure of space-time very close to a SMBH in a gravitational potential 100 times stronger and a mass scale $10^5$ times larger than any existing test. In 2017 and 2019 as star S0-2 made its closest approach to the Milky Way SMBH, AO-equipped telescopes measured the relativistic redshift of the light from the star (Do et al. 2019) and its Schwarzschild precession was measured using interferometry (Abuter et al. 2020). The measurements were consistent with predictions from general relativity and inconsistent with Newtonian gravity, however the uncertainties in the measurements still allow for small amounts of distributed mass or a compact mass in the vicinity. The general relativistic precession of the periapse of S0-2 has been measured using the VLTI-GRAVITY interferometer (Abuter et al. 2020), and while this demonstrates the effects of general relativity, it is limited to just one, relatively bright target and many more measurements are needed to better constrain the magnitude. Stars with periods shorter than S0-2 are both fainter and closer to the black hole.

TMT's increased resolution and sensitivity will increase the number of observable short-period stars by an order of magnitude. As these shorter period stars are revealed, it will be possible to measure the precession of the periapse of many of them, and therefore (1) test the specific metric form of General Relativity in an unprecedented regime and therefore distinguish between different metric theories of gravity (see section 3.2.4), (2) probe the distribution of dark stellar remnants and dark matter around the black hole and thereby test fundamental models of galaxy evolution and N-body dynamics (Rubilar & Eckart 2001, Weinberg et al. 2005, Will 2008, Merritt et al. 2010). In addition, improved measurements of stellar orbits will dramatically reduce the uncertainty in Ro, the distance from the sun to the Galactic Center. As Ro sets the scale of the Milky Way, precise measurements with TMT will constrain the Milky Way's dark matter halo to < 1% (Weinberg et al. 2005), contributing to our understanding of the origin and evolution of the Milky Way.

The spin of the black hole is the last remaining unknown properties of SgrA*. According to the no hair theorem, the spin is predicted to cause a drag on spacetime that can be measured using precise radial velocities of the stars orbiting close to SgrA* (see section 3.2.4). The precision required is <100 m/s, therefore, TMT-MODHIS has a unique role to play in furthering our understanding of this important region of our galaxy.

### 6.1.2 How the GC black hole interacts with its unusual environment

Outstanding puzzles include the origin of the GC's massive young star cluster, whose formation should have been suppressed by the SMBH, and an unexpected dearth of old red giants around the SMBH. These aspects challenge our notions of how SMBHs affect the formation and evolution of galaxies. The sensitivity of TMT will allow us to detect stars that are 100 times fainter than is possible today. Integral field spectroscopy of this expanded sample will allow us to detect the pre-main sequence population for the first time at the GC, providing an important handle on the young cluster's age and initial mass function in this extreme environment. It will also test the idea that stellar stripping can explain the deficit of old red giants. The higher resolution and sensitivity of TMT may also settle the question of whether IMBH are present in the GC environment, as the higher spatial resolution will be brought to bear on gas clouds, and the greater numbers of detected stars (along with higher resolution) may detect the binary companions of stellar-mass black holes.

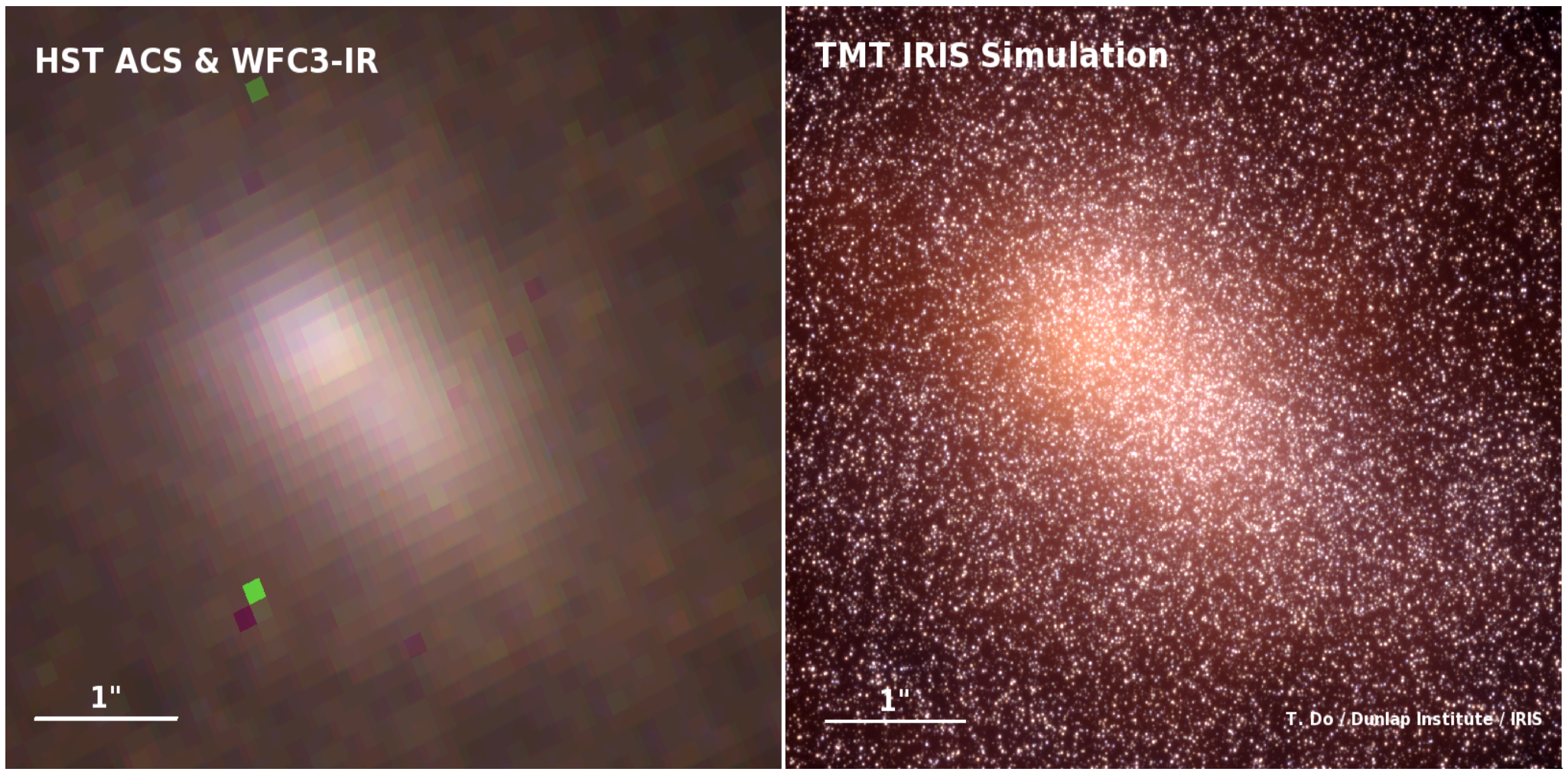

***Figure 6-4:*** *TMT will enable revolutionary studies of the nucleus of Andromeda (M31). Left: A three color image of the current capabilities using HST ACS (F814W) and WFC3 (F110W, F160W). Right: A simulated image based on our current knowledge of this region, as observed with the Z, J, and K-band using TMT's first light instrument IRIS with the NFIRAOS adaptive optics system. TMT/IRIS will provide the necessary sensitivity and spatial resolution to understand the dynamics, stellar population, and chemistry of the red giants; modeling will lead to better mass constraints for the supermassive black hole at M31's center. Images by UCSD OIR LAB.*

Meyer et al. (2006) and Nishiyama et al. (2009) reported that flares in Sgr A* are highly polarized, supporting a synchrotron origin for the NIR emission. Abuter et al. (2023), using high precision astrometry with VLTI-GRAVITY, have measured the clockwise motion of flaring material around the innermost stable circular orbit as well as the polarimetric signal with a timescale of about 60 minutes. TMT's highest-spatial resolution does not approach that of GRAVITY, but its time-domain capabilities and much higher sensitivity will enable it to perform polarimetric monitoring campaigns that will help to break model degeneracies and test relativistic effects.

The NIR (1–2.5 μm) polarimetry observing modes with TMT will be sensitive enough to monitor rapid changes on time scales of less than 5 minutes, with an accuracy of 0.1%.

### 6.1.3 Proper Motions around SMBHs in the Nearest Galaxies

The angular resolution and sensitivity of TMT will enable the measurement of individual stellar orbits around extragalactic SMBHs to measure precise black hole masses, as well as explore the dynamics and stellar populations at a level of detail only possible in the Milky Way today. For example, the center of our nearest large spiral galaxy neighbor Andromeda (M31) is about a hundred times more distant than Sgr A* (730 kpc), but compensates by harboring an SMBH roughly 35 times more massive ($1.4x10^8$ $M_\odot$). The resulting sphere of influence of the M31 black hole subtends about 6" in the sky, and the calculated apparent proper motion is about 6 percent of that in the Galactic center (see E-ELT simulations in Trippe et al. 2010). Current adaptive optics near-infrared observations with 8–10 m telescopes of M31 are strongly confusion limited. With the $D^4$ advantage of TMT in the confusion-limited case in addition to the ability to operate at Z-band, the IRIS instrument will have both the angular resolution and sensitivity to measure both the proper motion and radial velocities of individual stars allowing the possibility of reconstructing their orbits (see figure 6-4). This would be a challenging observation likely possible only in the nuclei of M31 and M32.

## 6.2 DYNAMICAL DETECTIONS AND DEMOGRAPHICS OF SMBHS

Mass is the most fundamental yet hard-to-measure property of black holes. The $M_{BH}$-$M_{bulge}$ and $M_{BH\text{-}\sigma}$ relations we observe for galaxies at $z \approx 0$ are reasonably tight for classical bulges and elliptical galaxies (intrinsic scatter 0.29 dex), but they are not for pseudo-bulges. For redshifts z>1, there are hints that both the $M_{BH}$-$M_{bulge}$ and $M_{BH}$-σ relations evolve, in the sense that the black hole tends to be over-massive with respect to the galaxy, but these findings are uncertain and controversial, depending on highly indirect estimates of $M_{BH}$ and host galaxy parameters, as well as selection biases (e.g., Lauer et al. 2007; Bennert et al. 2011; Schulze & Wisotzki 2014; Li et al. 2023).

The most reliable method to measure the mass of a black hole is from the orbital dynamics of gas or stars in the region within which the black hole dominates the gravitational potential. This region is usually defined by the radius of the sphere of influence of the hole: $r_{infl} = G\, M_{BH}/\sigma^2 = 13\ \text{pc}\ (M_{BH}/10^8\ M_\odot)^{0.5}$. TMT's IRIS instrument will provide the next great leap in observational capabilities for measurement of the kinematics of galaxy nuclei on sub-arcsecond scales. TMT's MCAO capability will enable IRIS to achieve a spatial resolution close to the diffraction limit, 8 mas (λ/μm), which surpasses the spatial resolution of the HST by almost one order of magnitude. As illustrated in figure 6-5, TMT is capable of spatially resolving the sphere of influence of a mass $M_{BH}$ at an angular distance up to $D_A = 335\ \text{Mpc}\ (\mu\text{m}/\lambda)\ (M_{BH}/10^8\ M_\odot)^{0.5}$, which corresponds to a distance of 1 Mpc for $M_{BH}= 10^3\ M_\odot$ or 3 Mpc for $M_{BH} = 10^4\ M_\odot$. At a redshift of z = 0.1, this corresponds to $M_{BH} = 10^8\ M_\odot$, and at z = 0.4, $M_{BH} = 10^9\ M_\odot$. Black holes with $M_{BH}\sim 10^{10}\ M_\odot$ can be detected at *any* redshift, provided that suitable dynamical tracers (stellar absorption lines or emission lines from ionized or molecular gas) are accessible in wavelength ranges that can be observed using IRIS.

SMBH masses can be probed using either stars or gas as dynamical tracers. The stellar-dynamical method is more general in that it can be applied to a much larger set of targets, but modeling the full distribution of stellar orbits in a galaxy is a formidable challenge. Complications include the presence of a dark matter halo, possible triaxial structure in the bulge, and stellar population gradients. These issues can be mitigated if the SMBH sphere of influence is very well resolved by the observations, and the exquisite angular resolution provided by TMT's AO system will enable major improvements in the quality and accuracy of SMBH mass measurements even for nearby galaxies in which SMBHs have already been detected using currently available capabilities. Velocity dispersion around a $10^7$ $M_\odot$ BH is ~67 km/s requiring a spectral resolution of R~9000 to properly characterize it. Measurement and modeling of the kinematics of a thin gas disk in circular rotation is far simpler, but only a small fraction of galaxies contain a circumnuclear gas disk in regular rotation that also has emission lines (from ionized or molecular gas) sufficiently bright for kinematic mapping. TMT and IRIS will have the capability to carry out both stellar-dynamical and gas-dynamical measurements of SMBH masses, and TMT's enormous collecting area and exquisite angular resolution will be an extremely powerful combination.

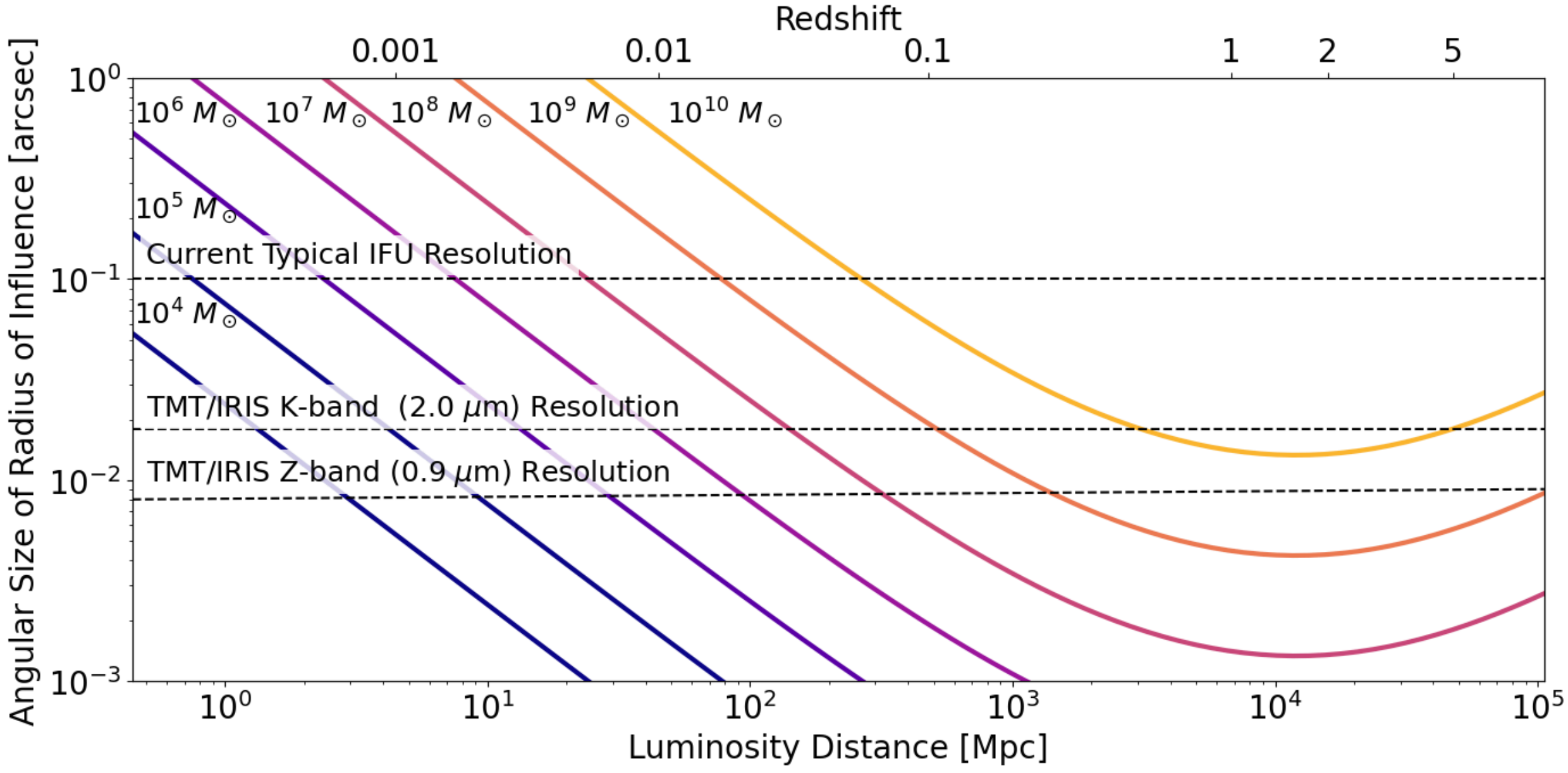

***Figure 6-5:*** *Radius of influence for various black hole masses as a function of the luminosity distance and redshift using the observed $M_{BH}$-$\boldsymbol{\sigma}$ relation. We show lines for black hole masses increasing from $10^4$ $M_\odot$ to $10^{10}$ $M_\odot$ (left to right). The increase in angular resolution will allow observations of black holes much further than possible today. Most IFS black hole mass measurements today are made at a pixel scale of 50 mas (100 mas resolution; blue, dashed), compared to TMT, which will have angular resolution of 18 mas at K band (red, solid) and 8 mas at Z band (green, solid). (Based on figure 1 from Do et al. 2014.)*

### 6.2.1 Towards a complete census of black holes in nearby galaxies

For some time after the discovery of the $M_{BH}$-σ relation, observations suggested that this correlation had very small scatter and applied uniformly to ellipticals and spirals in a single, coherent sequence. As the number of measurements increased over time however, it was recognized that this picture was a severe oversimplification. With nearly 100 dynamical measurements of SMBH masses, it is now apparent that the $M_{BH}$-σ relation shows significant scatter and clear differences between galaxy types (e.g., Kormendy & Ho 2013). In particular, the $M_{BH}$-σ relation seems to hold most strongly for classical bulges, which are believed to have formed through mergers, but not for pseudo-bulges built via secular evolution processes in spiral galaxies. However, despite years of observational effort, our understanding of SMBH demographics remains very incomplete. TMT will make major contributions across this full mass spectrum by enabling SMBH searches to be expanded to areas of this parameter space that are currently inaccessible.

At the low end of the SMBH mass range ($\lesssim 10^7$ $M_\odot$), where current stellar-dynamical observations cannot probe galaxies beyond a few Mpc, the overall shape of the $M_{BH}$-σ relation is very poorly understood, and BH detections in $H_2O$ megamaser galaxies indicate a strong departure from the $M_{BH}$-σ relation of classical bulges (Greene et al. 2010). Do et al. (2014) show that TMT with IRIS will have the ability to detect a Milky-Way equivalent SMBH ($4\times10^6$ $M_\odot$) out to the distance of the Virgo Cluster. Such observations will open up an entirely new window on SMBH demographics, enabling surveys of statistically meaningful samples of low-mass SMBHs to be carried out, and making it possible to explore the dependence of SMBH mass on a variety of host galaxy parameters such as the mass of a classical bulge or pseudo-bulge, or the presence or lack of a stellar bar.

Additionally, observations have revealed examples of galaxies with extreme and surprising properties, which demand further exploration of the full parameter space of the SMBH-host galaxy relationships. NGC 1277, for example, is a compact lenticular galaxy in the Perseus Cluster, and observations by van den Bosch et al. (2012) found evidence for an SMBH with an astoundingly large mass of $1.7\times10^{10}$ $M_\odot$. This is nearly 60% of the bulge's stellar mass, whereas in typical bulges the SMBH accounts for only ~0.1–0.2% of the stellar mass. NGC 1277 serves as a powerful reminder of the need for SMBH searches to be carried out for large and complete samples of galaxies, in order to ensure that the full complexity of the SMBH-host correlations can be explored.

### 6.2.2 The most massive black holes in brightest cluster galaxies

The heaviest black holes are expected to be hosted in the brightest cluster galaxies; the most massive galaxies known. These are likely the direct descendants of the most luminous quasars, which host the most massive black holes, in the distant universe. Using AO-assisted IFS on Gemini and Keck, McConnell et al. (2012) succeeded in detecting such ultra-massive black holes, the largest having $M_l = 2.1\times10^{10}$ $M_\odot$ in NGC 4889, the brightest galaxy in the Coma cluster. Black holes of this size have a sphere of influence that can be resolved, and thus are potentially detectable, at all redshifts by TMT. Details of the orbital dynamics need to be determined, which could show whether stellar cusps form around the SMBH, and if there is evidence of mergers. The principal limitation that places this problem beyond current telescopes and instrumentation is the low surface brightness of the centers of massive early-type galaxies. IRIS can be used to search for very massive black holes in a complete sample of the most massive galaxies between z = 0 and z = 4, straddling the peak of quasar activity at z = 2.5. To date, SMBH mass measurements have been completed for only a small handful of local BCGs, but TMT and IRIS will make it possible to achieve the S/N and angular resolution necessary to detect SMBHs in dozens of nearby BCGs (D<200 Mpc) in just a few hours of on-source integration time per target (Do et al. 2014).

### 6.2.3 Intermediate-mass black holes

Intermediate-mass black holes (IMBHs), with masses between $10^2$ and $10^6$ $M_\odot$, are a missing link between stellar-mass BHs and SMBHs (Greene, Strader, & Ho 2020). Determining the demographics of IMBHs in nearby, low-mass galaxies is of great importance since these objects are much closer to the mass scale of their original seeds, unlike high-mass SMBHs, which have essentially lost any physical memory of their original seed masses and environments. Theoretical models suggest that the occupation fraction and $M_{BH}$-σ relation of IMBHs may contain unique clues to the distribution of BH seed masses and the efficiency of their formation (Volonteri et al. 2008). IMBHs are also of great interest as sources of gravitational waves, either from binary BH mergers or from extreme mass-ratio inspiral events (in which a stellar-mass compact object inspirals into a low-mass nuclear BH). The probable hosts of these intermediate-mass objects are globular and other massive star clusters, and the nuclei of late-type bulgeless spirals and dwarf galaxies.

Some dynamical evidence has been reported for the existence of intermediate-mass black holes in globular clusters such as G1 (Gebhardt, Rich & Ho 2005, Pechetti, et al. 2022). However, the number of intermediate-mass black hole candidates and confirmed examples is still relatively small, due to the insufficient spatial resolution and sensitivity of current telescopes. Some late-type galaxies, even those that are completely bulgeless, host a central black hole with a mass as low as $M_{BH} = 10^5$ $M_\odot$ (Filippenko & Ho 2003; Greene & Ho 2007), but all such estimates are based on secondary methods of mass determination that rely on AGN broad emission lines. At the same time, HST stellar-dynamical observations have been used to set an astonishingly tight upper limit of 1500 $M_\odot$ to the mass of any BH in the nucleus of the Local Group Spiral M33 (Gebhardt et al. 2001). This observation provides the best demonstration that not all galaxies contain a central BH, but the occupation fraction of BHs as a function of galaxy mass remains almost entirely unconstrained by available data for late-type and dwarf galaxies. Recent gravitational wave (GW) observations by the Advanced LIGO/Virgo collaboration have reported a BH-BH merger event (GW190521; Abbott et al. 2020), with a remnant mass of ~142 $M_\odot$, which is in the IMBH regime. There is tantalizing evidence for IMBHs in a small number of low-mass dwarf galaxies from X-ray observations (such as the dwarf galaxy Henize 2-10; Reines et al. 2011), but the critical confirmation via spatially resolved dynamics is still lacking. At present, direct dynamical searches for IMBHs are restricted to the Local Group and its closest neighbors.

Nuclear star clusters in dwarf and late-type galaxies are the likely homes of IMBHs (if they indeed exist), and will be high-priority targets for IMBH searches in the TMT era. These clusters typically have stellar velocity dispersions of 15–30 km $s^{-1}$, and the need for simultaneously high angular resolution and high spectral resolution to resolve the kinematic structure of these objects is the primary factor limiting observational progress at present. The proposed high spectral resolution mode for IRIS (R ∼ 8,000–10,000) will be critically important for carrying out IMBH searches in nearby low-mass galaxies, and it will be a uniquely powerful capability for TMT. At a resolving power of R=8000, it becomes possible for IRIS to deliver accurate measurements of mean velocity, velocity dispersion, and higher-order moments of the line-of-sight velocity profile in individual spaxels for observations of central star clusters in galactic centers (Do et al. 2014) and for objects such as the M31 globular cluster G1. Other targets for IMBH searches with TMT will include ultra-compact dwarf galaxies (UCDs) in nearby groups and clusters; these objects may be the remnant nuclear clusters of tidally stripped low-mass galaxies and could host BHs in the range $10^5$–$10^7$ $M_\odot$ (Mieske et al. 2013). IRIS data will revolutionize the search for IMBHs, making it possible to detect IMBHs or set highly constraining limits for targets out to several Mpc distance.

### 6.2.4 Calibration of the black hole mass scale in active galactic nuclei

Reverberation mapping (Blandford & McKee 1982) utilizes the time-delay response of the emissivity of broad emission lines to continuum variations to constrain the size of the broad-line region, $R_{BLR}$. In combination with the velocity width of the emission lines, $\Delta V$, we can compute the virial product $M_{virial} = R_{BLR}\, \Delta V^2/G$, which is related to the true black hole mass through a normalization factor $f = M_{virial}/M_{BH}$ that depends on the unknown structure and kinematics of the broad-line region (Shen 2013). In practice, it is customary to calibrate an average value of $f$ empirically by assuming that AGNs obey the same M-σ relation as inactive galaxies (Onken et al. 2004, Graham et al. 2011). This is a highly uncertain and unproven procedure. There is no guarantee that AGNs follow exactly the same M-σ relation, and it is quite likely that $f$ differs from object to object (e.g., $f$ may depend on the orientation of a flattened broad-line region). Yet, this is the fundamental assumption on which most current methods for estimating black hole masses in AGNs and quasars are based.

Although recent progress in reverberation mapping studies has enabled more accurate black hole mass measurements by dynamically modeling the underlying broad-line region geometry/kinematics combined with exquisite spectroscopic monitoring data (e.g., Pancoast et al. 2014) — thus eliminating the need for the $f$ factor – the potential and limitations of such dynamical modeling analyses are yet to be fully explored. In addition, dynamical modeling on the reverberation mapping data is more resource-intensive than traditional reverberation mapping studies that only measure an average lag. Therefore, calibrating the $f$ factor across AGN host types remains a viable approach for the next few years. With IRIS, we will be able to directly calibrate $f$ in reverberation-mapped AGNs by independently deriving $M_{BH}$ through stellar-dynamical analysis. This has been attempted for a few of the nearest Seyfert 1 galaxies using AO data from 8–10 m telescopes (e.g., Davies et al. 2006; Hicks & Malkan 2008), but it is not within reach of current capabilities to expand the sample to a meaningful size. The AO capability of TMT is essential to resolve the SMBH sphere of influence, and also to reduce the size of the point-spread function of the bright central AGN point source, which otherwise completely dominates the signal from the stars.

TMT and IRIS will also make it possible to carry out a fundamental consistency check by validating stellar-dynamical SMBH measurements through direct comparison with masses measured from the dynamics of $H_2O$ megamaser disks. NGC 4258 is the prototype megamaser disk galaxy, and its BH mass has been measured to high accuracy from the rotation of the circumnuclear molecular disk on sub-parsec scales (Miyoshi et al. 1995). In recent years, maser surveys have significantly expanded the sample of such galaxies for which precise black hole mass measurements can be carried out (Kuo et al. 2011). These megamaser disk galaxies have SMBH of order $10^7$ $M_\odot$, and current capabilities are unable to resolve the stellar kinematics within the BH sphere of influence for most of these galaxies. Observations with IRIS will open up the ability to resolve the stellar kinematics within $r_{infl}$ in such targets, providing a critical cross-check between different mass measurement methods applied to the same galaxies.

## 6.3 COEVOLUTION OF SUPERMASSIVE BLACK HOLES AND GALAXIES; AGN FUELING AND FEEDBACK

Measuring the host properties of AGNs is key to understanding the co-evolution of SMBHs and galaxies. However, it is also very challenging, particularly for AGNs at higher redshift. TMT's superb spatial resolution and sensitivity will bring us a major breakthrough in separating the host signatures from the bright unresolved nucleus. This can help us to better constrain the $M_{BH}$-σ relation, particularly for AGNs at high redshift. Measuring the central velocity dispersion requires a strong spectral feature redshifted to an accessible wavelength. There are multiple absorption lines that can be used for this purpose. For example, the Ca II triplet at rest-wavelength of 0.85 µm, centering it in the JHK bands corresponds to redshifts of 0.4, 0.9, and 1.6, respectively. Given the expected spatial resolution of IRIS in each of these bands, the minimum resolved radius of influence of SMBH corresponds to 60, 100, and 150 pc at these redshifts. For comparison, note that the radii of influence for M87 ($M_{BH}$ ~ 5.4 x $10^9$ $M_\odot$) and M31 ($M_{BH}$ ~ 1.4 x $10^8$ $M_\odot$) are approximately 100 and 12 pc, respectively (Liepold, Ma & Walsh, 2023, Bender et al, 2005). Thus, it will be possible to probe the upper end of the SMBH mass function with TMT at these redshifts.

TMT will also allow us to study in much more detail the mechanisms of feeding and feedback of AGNs. It will provide observations not only for nearby AGNs in unprecedented quality, but also to galaxies reaching redshifts out to z~2, where galaxies are most active, and AGN feedback is suspected of playing a major role in shaping galaxy evolution. TMT's high-spatial-resolution spectroscopy will reveal fainter AGNs at higher redshifts, allowing us to probe the evolution of AGN luminosity function over a far wider range of luminosities.

In many ways, TMT's capabilities to explore the high-redshift AGN population and coevolution are on par with those of JWST. However, TMT has a much longer facility lifetime, better spatial resolution at the same wavelengths, and potentially higher spectral resolution, which enable unique opportunities in addressing key science questions outlined in this section.

### 6.3.1 The cosmic evolution of small and moderate-size SMBHs

Nearly all work on AGN at z>1 has been restricted to the most luminous examples, meaning the quasars that have been studied are quite rare and likely unrepresentative of the SMBH population as a whole. Current wide field surveys of AGNs have been identifying active $10^8$ to $10^9$ solar mass SMBHs during their violent growth epoch of redshift around 2. They are the most massive SMBHs in the local universe, and in order to trace the growth history of typical $10^7$ to $10^8$ solar mass SMBHs, it is necessary to identify at least one order of magnitude fainter AGNs, which are moderate- or low-luminosity AGNs at high-redshifts.

TMT's superb spatial resolution will bring significant improvement in this aspect. Spectroscopy done with narrower slits or with integral field spectrographs will be less contaminated by the light from the host, particularly for galaxies at higher redshift. This will help greatly in detecting fainter AGNs at high redshift. Galaxy redshift surveys done with TMT WFOS and IRMOS, such as the ones mentioned in section 5.1.6, can be used to push the luminosity functions of high-redshift AGNs orders of magnitudes fainter than currently possible with 8–10 meter telescopes.

Luminous high-redshift AGN are found to be heavily dust obscured (Scholtz et al. 2024), however, many moderate- or low-luminosity AGNs at high-redshifts are also heavily obscured by dust. Some of the moderate or low luminosity AGNs can be already detected in the current ultra-deep X-ray surveys as optically-faint but NIR-bright X-ray sources. However, their natures are not well known because they are fainter than current limits of optical and NIR spectroscopic observations with 8–10 m class telescopes. Thanks to its high sensitivity, TMT can unveil the physical nature of such X-ray-selected faint AGNs. Heavily obscured AGNs can be missed even with the ultra-deep X-ray surveys. One possible way to identify them is to search for AGN signatures, e.g., strong high-ionization emission lines, at the center of distant galaxies. High-spatial resolution integral-field spectroscopy with TMT should identify weak AGN activity at the center of high-redshift galaxies. By identifying a statistical number of faint X-ray AGNs and weak AGN activity at the center of galaxies, TMT can trace the accretion growth history of typical SMBHs seen in the local universe.

### 6.3.2 The first generation of accreting MBHs

Our interests in the first generation of SMBHs have led to various surveys for high-redshift quasars. The wide-field optical surveys such as SDSS, CFHT, and Pan-STARRS1 quasar searches resulted in the discovery of ~~a~~ several hundred quasars at z~6 (e.g., Fan et al. 2000, 2001, 2003, 2004, 2006; Willott et al. 2007, 2010; Banados et al. 2014, Fan et al. 2023), and wide-field near-infrared surveys such as UKIDSS and VIKING even discovered a few quasars at z~7 (Mortlock et al. 2011; Venemans et al. 2013; Wang et al. 2021). However due to the limited sensitivity of those surveys, all the discovered high-redshift quasars are already mature ones, the mass of their SMBHs being $M_{BH} \sim 10^9 M_\odot$ and the metallicity of their broad-line regions is higher than the solar metallicity (e.g., Kurk et al. 2007; Juarez et al. 2009; Mortlock et al. 2011; Venemans et al. 2013; Shen et al. 2019). For identifying the non-matured candidates for first generation quasars, we have to search for lower luminosity and/or higher redshift quasars and then examine their spectroscopic properties in detail.

Ongoing deep and wide imaging surveys such as DES and HSC have been yielding, and upcoming surveys like Rubin Observatory, and Euclid will yield many candidates of first-generation quasars at z~6-8, in the magnitude range of ~23–25 mag (that is ~2–4 mag fainter than the SDSS limit, Matsuoka et al. 2019a, 2019b, 2022). However, the efficient spectroscopic characterization of those faint candidates requires very sensitive near-infrared spectroscopic observations as recent observations are at the limit of the ability of the current 8–10 m telescopes (Onoue et al. 2019). More specifically, the determination of $M_{BH}$ (and consequently the Eddington ratio, $L/L_{Edd}$) requires the measurement of the CIV 1549 and MgII 2800 velocity profiles, while the determination of the chemical property requires the measurement of the NV 1240, CIV 1549, HeII 1640, CIII]1909, MgII 2800, and FeII multiplet fluxes. For quasars at z~6–8, those spectroscopic features are seen at $\lambda_{obs} \sim 0.8–2.5$ μm, and sensitive spectroscopic observations at this wavelength range are well-suited to IRIS on TMT. Even the discovery spectrum of a true first-generation quasar should reveal its much weaker metal lines. And quasars with low-mass SMBHs are expected to show relatively narrow (i.e., FWHM ~ 1000–2000 km/s) velocity profiles in emission lines as inferred from the properties of nearby narrow-line Seyfert 1 galaxies, and thus a moderately-high spectroscopic resolution of R~3000–4000 is needed for measurements of $M_{BH}$. Such a moderately high resolution is preferred also for avoiding the effects of the strong OH airglow emission that is crucial for sensitive near-infrared spectroscopic observations. The planned IRIS capability is perfectly matched for these kinds of observations.

### 6.3.3 Feeding and feedback of AGNs in the nearby (z<0.5) universe

AO-assisted IFU studies in the near-infrared have measured non-axisymmetric structures in the central tens of parsecs of AGN and found significant non-circular motions in the gas velocities (Davies et al. 2014). Particularly in the most highly ionized gas (traced by coronal emission lines such as [SiVI] (1.96 μm), most AGN show biconical outflows (Muller-Sanchez et al. 2011). The estimated mass flow rates are hundreds to thousands of times higher than the inferred accretion rate onto the central MBH. Although deceleration is observed, it is possible that most of the outflowing gas can escape the galaxy (Muller-Sanchez et al. 2011). In some but not all local AGN, these outflows appear strong enough to provide the AGN feedback postulated by CDM galaxy evolution theorists to restrict star formation and limit the growth of more massive galaxies. But it is still far from clear whether the observed outflows, particularly in radio-weak AGN, generate sufficient feedback to make these models viable.

The same near-IR spectroscopic maps also trace molecular gas from $H_2$ emission lines (e.g., at 2.12 μm), which often show thick, turbulent disk-like structures (Hicks et al. 2009). Comparison with a control sample of non-active galactic nuclei suggests that this concentrated molecular gas is a reservoir feeding the central engine, and in some cases obscuring it in the optical (Hicks et al. 2013). Such structures are not evident in LINERs, suggesting that Low-Luminosity AGN are not powerful enough to produce them (Muller-Sanchez et al. 2013). Although the molecular kinematics can be complex, some AGN show clear dynamical evidence of gas inflow, perhaps driven by bar-like deviations from an axisymmetric gravitational field (Muller-Sanchez 2013).

However current observations with IFUs on 8–10 meter telescopes are just able to scratch the surface of the gas dynamics in the central regions of galaxies affected by the SMBH. Observational advances will require not just much more sensitivity, but also a several-times improvement in spatial resolution, that AO on TMT will provide. With IRIS observations of the centers of nearby galactic centers of all types, it will be possible for the first time to determine how SMBHs are fueled, and how they in turn control the evolution of their host galaxies.

#### 6.3.4 AGN Feedback and its Effect on Star Formation: High spatial resolution studies over redshifts z ~ 1 to 3

In theory, during the co-evolution with their galaxies (Kormendy & Ho 2013), supermassive black holes in active phases can influence their hosts and large-scale surroundings through feedback of different modes (e.g., "quasar-mode" feedback; Di Matteo et al. 2005; Hopkins et al. 2008; "kinetic-mode" feedback; McNamara & Nulsen 2007; Weinberger et al. 2017; Davé et al. 2019). While AGN feedback has been intensively studied in many state-of-the-art cosmological simulations of galaxy evolution (e.g., Schaye et al. 2015; Pillepich et al. 2018; Davé et al. 2020, Ward et al. 2022), observations of galaxies at different redshifts on different physical scales still motivate debate on the effectiveness of this mechanism in regulating the star formation properties of galaxies (e.g., Schawinski et al. 2007; Page et al. 2012; Cheung et al. 2016; Harrison 2017; Shangguan et al. 2018; Leung et al. 2019; Shangguan & Ho 2019; Lamperti et al. 2022; Molina et al. 2022; Zhuang & Ho 2020; Xie et al. 2021; Zhang & Ho 2023). Limited by the capability of current instruments, studies of AGN feedback so far are mostly focused on the global star formation properties (e.g., Schawinski et al. 2007; Harrison 2017; Shangguan et al. 2018; Shangguan & Ho 2019; Zhuang & Ho 2020; Xie et al. 2021), or the resolved gas kinematics (e.g., Rupke & Veilleux 2011; Harrison et al. 2014; Karouzos et al. 2016; Bae et al. 2017; Husemann et al. 2019), of galaxies selected from the local universe. Although spectral observations with 8–10 meter telescopes are able to detect intermediate-redshift (z=1~3) galaxies under certain spatial resolutions of kpc scales (e.g., Leung et al. 2017; Müller-Sánchez et al. 2018; Wisnioski et al. 2019; Tiley et al. 2021), they do not have the desired spatial resolution and also wavelength coverage as TMT instruments. Through TMT observations with much-improved spatial resolution and sensitivity and wide spectral coverage, we can map the gas kinematics of galaxies of z=1~3 under 100s pc scale resolutions based on emission lines spanning a wide range of wavelength and physical conditions, thereby exploring with much more conclusive evidence the influence of AGN feedback in distant galaxies. Near-IR IFU spectroscopy will also be essential for finding evidence of outflows induced by AGN winds and jets, which are thought to provide a crucial role in preventing runaway star formation in the most massive galaxies at higher redshift (Fabian 2012). Current 8 m class telescopes are only capable of finding such outflows in rare, highly-luminous AGN at z~0.5–2 where these feedback effects are most important (e.g., Cano-Diaz et al. 2012, Liu et al., 2013, Harrison et al. 2016). TMT will be able to observe these outflows on smaller spatial scales, and in more typical, less luminous AGN. Together with ALMA observations of the corresponding molecular outflows, and sensitive VLA observations of the synchrotron emission from jets and outflows (even in traditionally radio quiet AGN), we can build up a complete picture of the nature and energetics of AGN feedback at these redshifts. Besides the gas kinematics, the canonical SED fitting based on the wide range spectroscopy of TMT spanning the optical to mid infrared bands enables a robust evaluation of star formation properties of local to distant galaxies as well. More importantly, only with the superb spatial resolution and sensitivity of TMT, we have the potential to resolve the central regions (i.e., torus) launching the feedback effects of galaxies in the local universe, which has great physical significance as detailed below. Combining TMT observations of local and distant galaxies with superb spatial resolution, we can shed more light on the physical mechanism of AGN feedback and try to understand the cosmological evolution of AGN feedback.

As forementioned, thermal-IR observations (3–14 um) on the TMT will play a key role in resolving the torus. They simulated imaging and spectroscopic observations from the TMT, as well as other ELTss, 10 m, and JWST observations by Nikutta et al. (2021). Through simulating advanced clumpy torus models at a variety of wavelengths, and partially leveraged from deconvolution techniques (i.e., Leist et al., in prep.), Nikutta et al. graphicly showed the resolved torus for the first time at these wavelengths (figure 6-6). This will potentially enable the degeneracies present in current models to be broken, helping to distinguish between competing torus models. The combination of JWST's exquisite sensitivity and TMT's fine resolution will be crucial to understand the inflow/outflow models of AGN, and their interaction to the host galaxy.

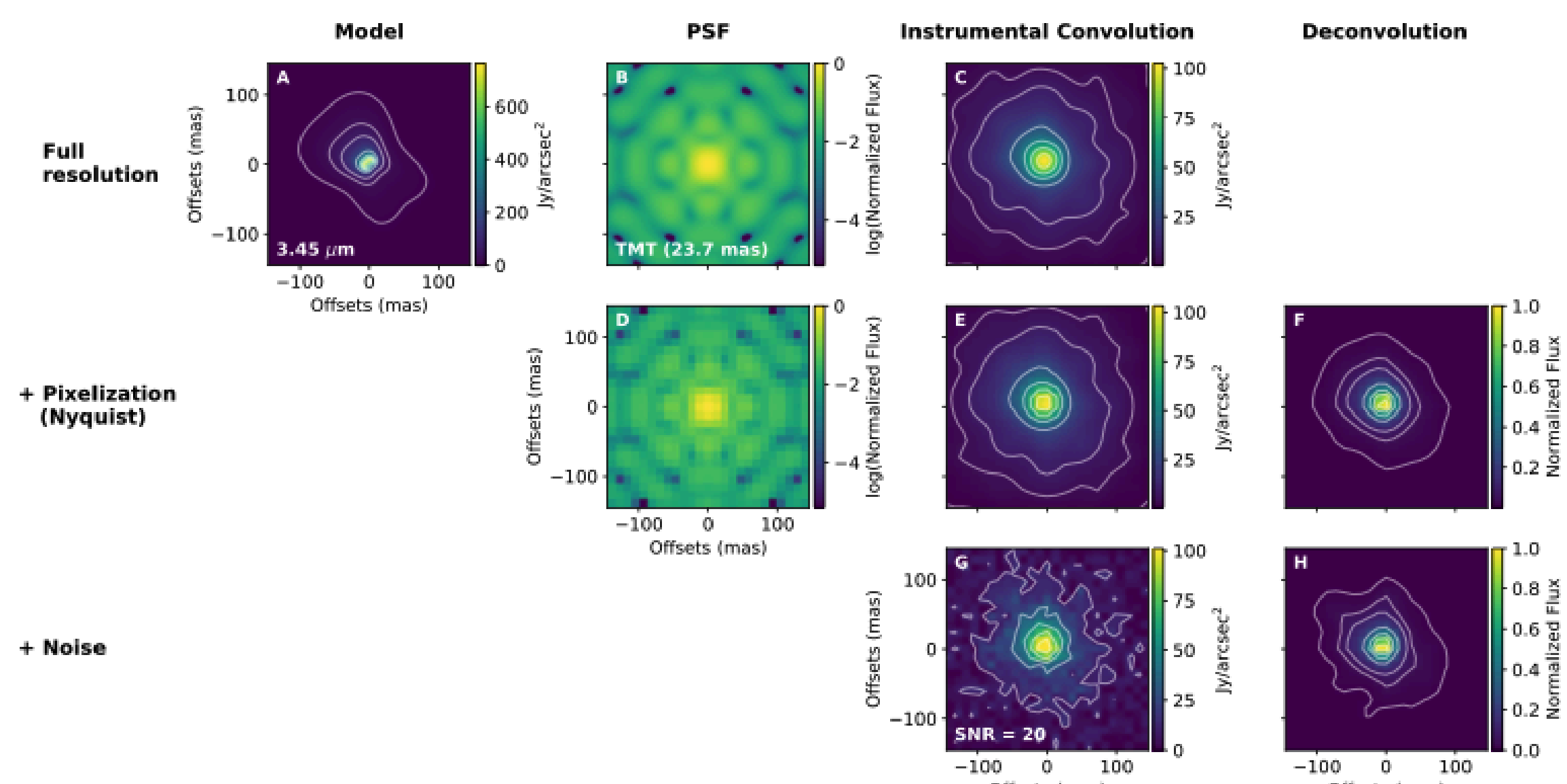

***Figure 6-6:*** *HYPERCAT step-by-step process to obtain synthetic observations from a 2D thermal emission map produced by CLUMPY models. From left to right the columns add operations on the input image by the optical system. From top to bottom the rows add detector degradations to the image (c.f. Nikutta et al. 2021).*

### 6.3.5 Distinguishing star formation in the host galaxies of AGNs

IRIS will be a powerful instrument for resolving the central bulge component of the host galaxy. A key ingredient to understanding whether a black hole and its host galaxy grew in a synchronized manner is to be able to independently constrain the growth of the black hole (accretion rate) and the host (star formation rate — see section 7.8.7). For this purpose, it is necessary to observe not only inactive galaxies but also AGNs, systems where the black hole is actively accreting. Under these circumstances it is normally very challenging to disentangle the different contributions from the AGN and star formation. The high angular resolution IFS capabilities of IRIS again will be a key new breakthrough (see figure 6-7).

For some objects, such as the Ultra luminous IR galaxies (ULIRGs), controversy still rages about the dominant IR emission process: is it a very powerful AGN or a vigorous starburst component? A highly effective methodology to distinguish between emission mechanisms is the estimation of the emission surface brightness of energy sources (Imanishi et al. 2011). The energy generation efficiency of nuclear fusion inside stars is only ~0.5% of $Mc^2$, and the maximum emission surface brightness of a starburst is found to be $\sim 10^{13} L_{\odot} kpc^{-2}$ through both observations and theory. However, the efficiency of a SMBH accreting matter is as high as ~40% of $Mc^2$. Hence, an AGN can produce a very much higher emission surface brightness than stellar fusion ($>10^{13} L_{\odot} kpc^{-2}$). Therefore, we can decisively determine the presence of buried AGN if the emission surface brightness of the energy sources of ULIRGs are $\gg 10^{13} L_{\odot} kpc^{-2}$. Due to the limited sensitivity and image size with 8 m class telescopes, this method has been applied to only a limited number of ULIRGs to date. As ULIRGs are more prevalent at higher redshift, probing the emission source from the local universe and constructing templates is crucial followup for observations using space-based telescopes. Hence, MICHI on the TMT will be able to conduct key research in this area, supporting observations with the JWST of higher-z candidates; TMT will benefit especially from higher spatial resolution. Comprehensively understanding the evolution of ULIRGs versus redshift is a powerful way to investigate the evolution of AGN and galaxies, and their effect on each other, of significant importance to cosmology.

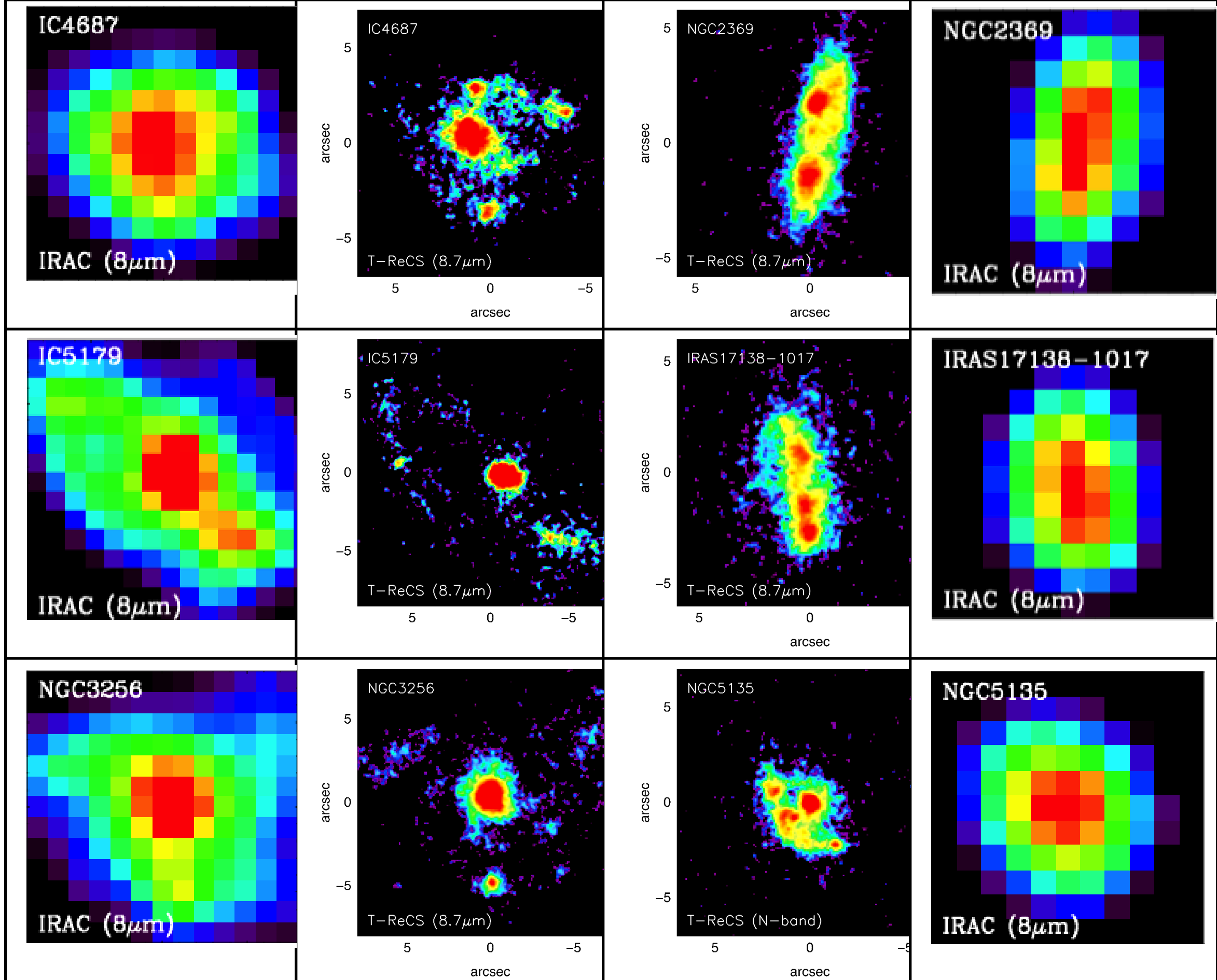

***Figure 6-7:*** *Gemini North and South observations (~0.3" spatial resolution) of active galactic nuclei and their host galaxies (center 2 columns) compared to Spitzer observations at 8 μm (outer columns) of the same objects (~2.6" spatial resolution). We can expect similar resolution gains in the era of the TMT (0.05") and JWST (0.25"). Credit: TMT ISDT.*

### 6.3.6 Binary and merging SMBH in the nearby universe

In the hierarchical paradigm of galaxy formation and evolution, galaxy mergers bring their central massive black holes together. Initially, the BHs are expected to have a separation of more than a kpc. The dynamical friction caused by stars then brings the two BHs closer on a time scale of 100 Myrs, making them a gravitationally bound binary system with a separation of about a parsec (Begelman et al. 1980).

The massive binary BHs will eventually spiral in and merge to form a single central SMBH of the new galaxy (Milosavljevic & Merritt 2003, Di Matteo et al. 2005). However, binary coalescence via emission of gravitational waves requires that the binary BH orbit must first shrink to much smaller radii (~0.001–0.01 pc) from the parsec scale separation where standard dynamical friction stops acting. Possible solutions to this "final parsec" problem include the interaction of the binary with the surrounding stars and gas e.g., slingshot ejection of approaching stars can shrink the binary orbit further. The resulting gravitational binding energy is released to the surrounding stars. N-body simulations show that this process can convert a steep power-law cusp into a shallow power-law cusp within the radius of gravitational influence of the BHs (Milosavljevic & Merritt 2001). Successive mergers will further decrease the density in the central regions and thus forming cores and resulting in mass deficit in bright elliptical galaxies (Ravindranath et al. 2002, Milosavljevic et al. 2002). The density in the central regions can become very low, thus stalling the binary orbit for a few Gyrs. Thus, a large fraction of core elliptical galaxies are likely to have stalled binary BHs. Figure 6-8 shows the stalling radii for nearby core elliptical galaxies in the Virgo cluster. At present, only the total black hole mass has been measured for these ellipticals. The current generation telescopes do not have the requisite spatial resolution to determine if the cores of these ellipticals have stalled binary BHs or a single merged BH. As shown in figure 6-8, TMT will be able to probe down to the expected stalling radii of binary BHs and reveal dual nuclei within galaxy cores if the nuclei are active. Also, stellar orbits in the innermost regions around binary BHs are expected to be different than the orbits around single BHs. Thus, studying stellar motions in the nearest core ellipticals should also provide evidence for binary BHs even in the case of non-active nuclei. In light of the recent potential Pulsar Timing Array (PTA) detection of the low-frequency gravitational wave background and its possible origin from the mergers of binary supermassive black holes (e.g., Agazie et al. 2023), the studies on binary SMBHs enabled by TMT will advance the theoretical understanding of the formation and orbital evolution of these binary SMBHs.

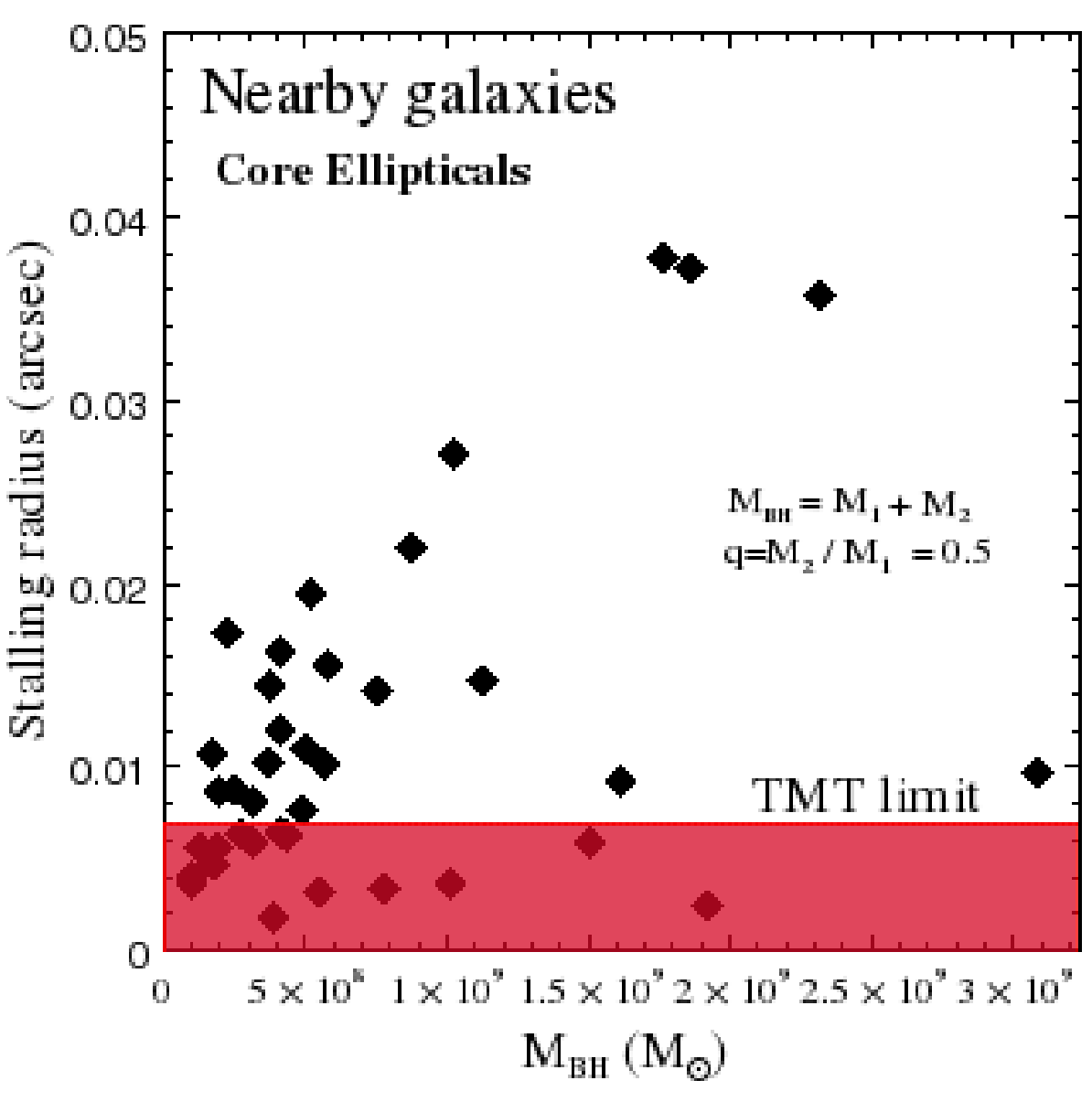

***Figure 6-8:*** *The stalling radius for merging blac in the cores of nearby galaxies in the sar Ravindranath et al. (2002). The stalling rad computed using the measured black hole mas: the prescription given in Merritt (2006). TMT op at the diffraction limit will be able to resolve black holes in a number of core ellipticals if blac are stalled. TMT will not be able to probe stallii falling in the shaded region.*

Since the presence of an active galactic nucleus is unambiguous evidence for a central massive BH, observations of double active nuclei will provide direct evidence for the presence of binary BHs. Superior angular resolution available with Chandra and Hubble Space Telescope (HST) have enabled direct observation of a pair of active nuclei separated at ~1.4 kpc in the ultra-luminous infrared galaxy NGC 6240 (Komossa et al. 2003, Max et al. 2007). Similar dual quasar systems are starting to be discovered at much higher redshift (e.g., Chen et al. 2023). TMT, with its improved spatial resolution and sensitivity, could detect and characterize pairs of active nuclei in many distant active galaxies, if they exist.

In dual AGN systems, both central engines may be buried inside the obscuring dusty structure and/or host galaxy. In such cases, NIR (1–2.5 μm) polarimetry with the TMT offers a powerful way to identify obscured dual AGN systems through the conical centrosymmetric polarization pattern centered on each AGN. For example, the polarization pattern detected in the 10” (600 pc) central regions of NGC 1068 (Capetti et al. 1995, Simpson et al. 2002) is the signature of a central point source whose radiation is polarized by dust and/or electron scattering within the ionization cones, which if rotated by 90 degrees, point directly to the illuminating source. In the case where the ionization cones can be resolved, the polarization pattern will indicate two slightly separated centers that can be identified as dual nucleus AGN.

Complementing high spatial resolution imaging searches for Massive Binary Black Holes (MBBH), polarimetric observations can also be used to detect binary SMBHs with separations of ~1 pc within an active nucleus, as their orbits are often not circular and thus generate a periodic polarization signal due to the orbital asymmetry (Dotti et al. 2022) with timescales of years due to the orbital period. Such objects are difficult to find observationally, but are critically important for correlating with gravitational wave signatures detected with the Pulsar Timing Array (PTA) and understanding the growth of black holes in galaxy centers.

## 6.4 THE STRUCTURE OF OBSCURING DUST AROUND THE CENTRAL ENGINE

Some, and perhaps most, AGNs have extremely optically thick gas, which, with dust, can produce up to tens or even a hundred magnitudes of visual extinction, radically altering our view (and classification) of the active nucleus. If, as suspected, much of this dust is close to the active nucleus, it will be warmed enough to produce the observed strong mid-IR emission. Resolving the geometry of these absorbers, possibly a torus or maybe a wind covering a large solid angle of the sky, is key to improving our understanding of AGN processes in the heavily extincted core. TMT gives us our first realistic chance to do this.

Directly observing dust in galaxy centers, and the dust which constitutes the torus, is optimally achieved at MIR wavelengths, as the characteristic black body temperature (a few 100 K) peaks at this wavelength, and contamination from stellar emission and obscuration are greatly reduced. Through observations with 8 m telescopes, the tori in AGNs were revealed to be compact (a few pc) in moderate activity AGN (e.g., Jaffe et al. 2004, Packham et al. 2005, Mason et al. 2006), or perhaps even absent in low activity AGN. The torus is suspected to be a clumpy distribution (Nenkova et al. 2002), rather than a homogeneous distribution of dust. Tentative results suggest the existence of a torus and its structure is strongly affected by the level of activity in the AGN, which in turn is related to the fueling of the central engine. However, surveys of AGN tori using 8 m class telescopes are confined to the local universe due to flux and spatial resolution limitations.

With its nearly four-fold increase in spatial resolution compared to that of 8 m telescopes, TMT can observe fainter and/or more distant objects. At z=0.5, the spatial resolution of JWST/MIRI is 1.5 kpc and AGN core observations are heavily contaminated by galactic star forming rings, etc. TMT, with a spatial resolution of 330 pc, will be able to make much cleaner direct AGN nuclei observations. Figure 6-7 illustrates a comparable improvement in spatial resolution between Spitzer and Gemini. TMT imaging and spectral observations of tori in z<~0.5 objects can produce templates that will help to calibrate and interpret complementary observations from JWST. Additionally, ALMA will probe the outermost (cold) regions of the torus structure and the combined results will reveal a much more detailed picture of AGN torus properties including the effects of and relationships between radio loudness, the host galaxy, the level of AGN activity, and redshift.

NIR polarimetric studies (Young et al. 1995; Packham et al. 1997; Lumsden et al. 1999; Simpson et al. 2002; Watanabe et al. 2003; Lopez-Rodriguez et al. 2013, 2015) of some AGN have shown that the NIR polarization can be attributed to the extinction along the line of sight by the passage of radiation of the central engine through hot (T ~1500 K) dust in the inner edge of the obscuring material surrounding the central engine. From a magnetohydrodynamic framework (Blandford & Payne 1982; Krolik & Begelman 1988; Emmering, Blandford & Shlosman 1992; Konigl & Kartje 1994; Kartje, Konigl & Elitzur 1999; Elitzur & Shlosman 2006, Wada, Scartmann & Meijerink 2016), the obscuring material is part of an outflowing wind confined and accelerated by the magnetic field generated in the accretion disk. In this scheme, the hydromagnetic wind can lift the plasma from the midplane of the accretion disk to form a geometrically thick distribution of dusty clouds surrounding the central engine. The magnetic field can induce a preferential orientation of dust grains within the dusty clouds that can give rise to a measurable degree of polarization. Thus, NIR polarimetry has the advantage of providing information about the magnetic field at locations where FIR/sub-mm emission is weak.

Lopez-Rodriguez et al. (2013, 2015) have shown that the magnetic field strength and geometry within the obscuring material of a few AGN can be determined through NIR (1–2.5 μm) polarimetry on current 8 m class telescopes. To complete the statistical analysis needed to obtain general properties and possibly recognize extraordinary objects, requires the high-angular resolution and the D^4 advantage of TMT that will 1) minimize starlight dilution within the central region, important to estimate the intrinsic polarization of the AGN; and 2) increase the number of observable objects. Such observations with the TMT will allow us to study the evolution, dynamics and morphology of the obscuring material and its interaction with the central engine and host galaxy from a magnetohydrodynamic framework. The NIR (1–2.5 μm) polarimetric capability of TMT should be able to obtain an accuracy of 0.1% in the degree of polarization for these sources.

IR (1–13 μm) imaging- and spectro-polarimetric capability in the TMT will be essential to study the dust properties in and around AGN. For example, the pioneering work by Aitken et al. (1984) and Bailey et al. (1988) on NGC 1068 using low spatial-resolution NIR imaging polarimetry and MIR spectropolarimetry observations showed a rotation in the position angle of polarization of ~70° between 4 and 5 μm. These results are consistent with the predicted angle change of aligned dust grains (Efstathiou et al. 1997). However, similar measurements were only attempted for a few objects (Packham et al. 2007). With TMT, observations and modeling of the total (Laor & Draine 1993) and polarized (Aitken et al. 2004) IR SED of AGNs in a wide wavelength range will provide crucial insights into dust grain properties such as their sizes, shapes and compositions. They will also help investigate the fraction of thermal and non-thermal polarization mechanisms in the central regions of AGNs. For bright, nearby AGN, MODHIS can detect and characterize the polarized light coming from the central region (Friedrich et al. 2010), which can be well-localized due to the exquisite spatial resolution provided by TMT+NFIRAOS, excluding unpolarized flux from the surrounding host galaxy.

## 6.5 TIME VARIABILITY, PROBING THE STRUCTURE AND PROCESSES IN THE CENTRAL ENGINE

TMT will see first light in an exciting era for time domain astronomy (see chapter 9). Particularly exciting for the study of supermassive black holes is the identification and study of Tidal Disruption Events (TDEs), which occur when a star on an orbit around SMBH makes a close approach and is ripped apart by tidal forces. Once the stellar debris rains down on the black hole, soft X-ray and UV radiation characterized by a $10^5$ K blackbody emerges from the inner accretion region. Reprocessing of this radiation in the debris results in a fraction of the emission emerging in the optical. Large numbers of TDEs are being detected by the current big transient surveys like the Zwicky Transient Factory (Villanueva & Gezari, 2021), and many more will be found in the next generation of large-area surveys such as the Rubin Observatory. Observations of TDEs are important because they in principle provide a means of measuring the masses and spins of black holes in optically normal galaxies.

TMT will allow the followup and study of TDEs up to much higher redshifts than previously possible (see section 9.8). Because of the negative K-correction (TDEs emitting primarily in the rest frame UV with a characteristic $10^5$ K black body), TDEs will be visible by TMT to redshifts of 6 or larger, enabling constraints on SMBH properties and evolution over a vast range of cosmic time.

The study of a large number of TDEs will help to anchor the observational signatures to precisely measured black hole masses. While analytic models and numerical simulations exist, there are still large uncertainties in the conversion from a measured light curve to a black hole mass estimate – the ultimate goal of TDE studies. Finding TDEs in a nearby galaxy with an independent measurement of the black hole mass, either from the M-σ relation or better yet, maser disk kinematics, will provide a better calibration of the models. With current observational capabilities the probability of finding such nearby events is low, and measuring their light curves is challenging. But, with Rubin Observatory and the greater follow-up sensitivity and angular resolution of the TMT, it should be possible to build a significant sample of low-z TDEs.

As previously mentioned in section 6.3.6, polarimetric observations of the immediate vicinity of supermassive black holes (SMBHs) will provide powerful insights into various characteristics of SMBH systems. For example, measuring time delay between polarized and unpolarized light can provide a map of the structure of the accretion disks (aka polarimetric reverberation mapping, e.g., Gaskell et al. 2012, Rojas Lobos et al. 2020). Detailed modeling of the polarization signal of blazars, which harbor highly polarized, relativistic jets, can allow for the constraint of both the the accretion disk and the magnetic properties, such as the strength and ordering of the magnetic field (e.g., Schutte et al. 2022). Detailed modeling of the accretion disks with polarimetry may also provide additional constraints on the masses of the SMBHs (e.g., Savic et al. 2018).

Variability of AGN provides a powerful means to explore and characterize their innermost regions. Time-resolved spectroscopy and spectro-polarimetry allow for highly confident identification of the excitation processes powering AGN jets. Variability studies coupled with reverberation mapping discussed in section 6.2.4 can be used to expand the present sample of a handful of AGN systems where their inner accretion disc structures and sizes can be measured up to many thousands of systems, covering the vast ranges of types and characteristics of AGN and Blazars that are observed, see also section 9.9.

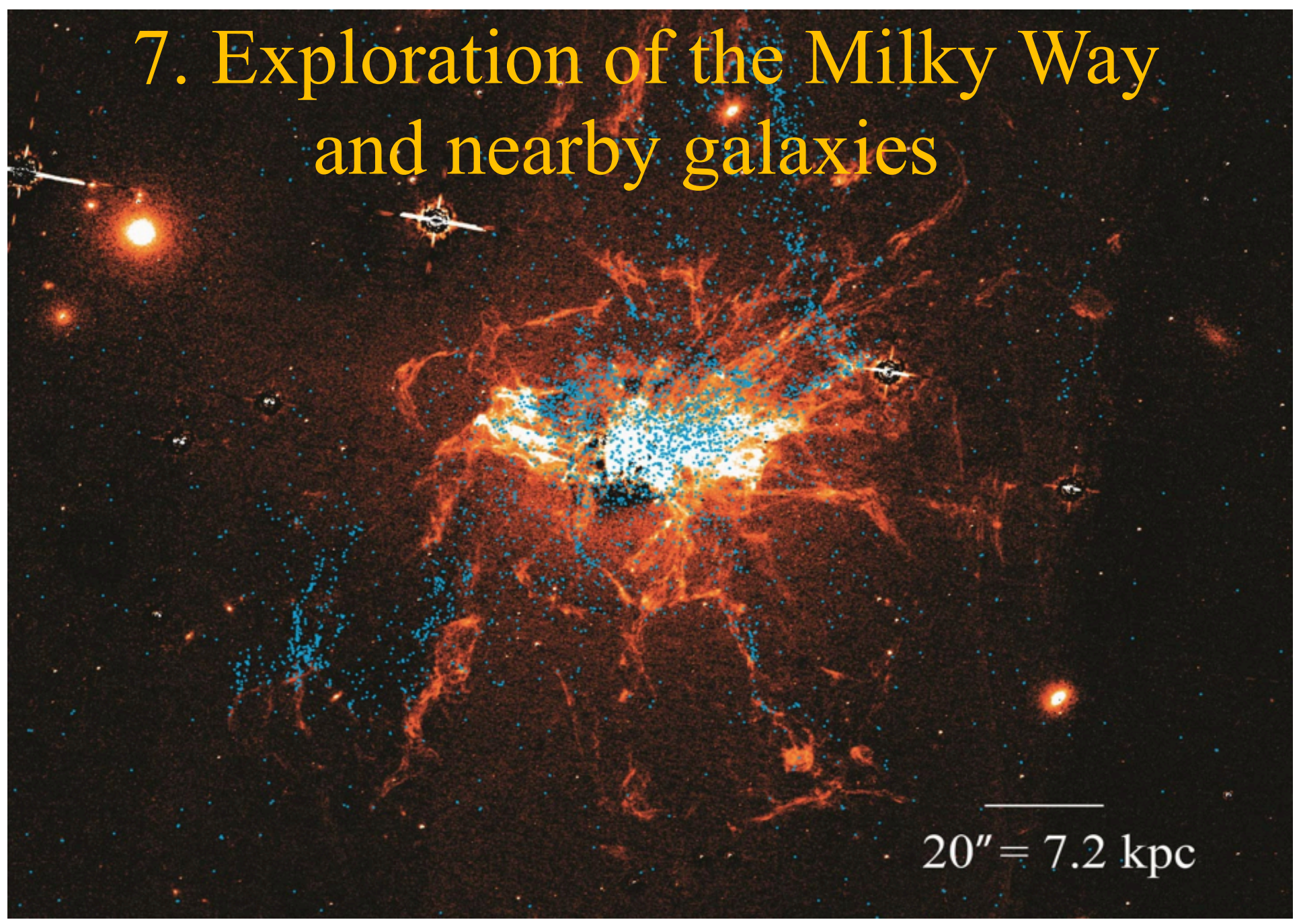

An HST image of NGC 1275 from Lim et al., 2020, the central galaxy of the Perseus Cluster, shows globular clusters forming (streams of blue dots) from filaments of cool gas (red). Once formed, they are no longer confined to those filaments and re-distribute throughout the galaxy.

---

Studying the Milky Way and the Local Group of galaxies with TMT addresses significant aspects of several of the big questions in chapter 2, Q1-*What is the nature and composition of the universe?,* Q4-*How do stars and planets form?,* and Q5-*What is the nature of extrasolar planets?* To address and answer these questions, astronomers will use TMT to study stellar populations at different levels of maturity and arising from different chemical, kinematic, and mass environments. These studies will constrain stellar evolution and star formation theories, help reveal star formation processes throughout cosmic time, uncover the star formation histories of nearby galaxies, and measure the mass-metallicity-luminosity relations and their evolution in dwarf galaxies. TMT observations to identify related collections of stars in the Milky Way and Local Group Galaxies and will provide insight into the growth and evolution processes of large galaxies. Kinematic measurements of individual stars and galactic members throughout the local group will improve our understanding of the properties of dark matter.

To advance our knowledge of our own Galaxy and the Local Group requires a significant advance in observational capabilities, requiring a significant increase in sensitivity and angular resolution that the TMT and its instruments will provide. None of the suggested new observations in this chapter could succeed without those two factors. Additionally, studying chemical abundances in many cases requires moderate to high spectral resolution spectroscopy as planned for TMT/HROS and TMT/MODHIS (high resolution) and TMT/IRIS and TMT/WFOS (moderate resolution). While most of these observations require only optical to NIR wavelength coverage, understanding the white dwarf initial-final mass relationship pushes on the TMT lower wavelength limit while the intermediate-mass black hole (IMBH) and ultra-luminous infrared galaxy (ULIRG) science greatly benefits from extending to the mid-infrared as envisioned with the MICHI instrument. IFUs and MOS modes are key to much of this science, taking advantage of the fine spatial resolution and enhancing observing efficiency. Finally, the local kinematics science, in particular, requires TMT's planned astrometric precision.

---

**Contributors**: Wako Aoki (NAOJ), Masashi Chiba (Tohoko), Mousumi Das (IIA), Richard DeGrijis (KIAA), Xuan Fang (Instituto de Astrofisica de Andalucia), Rupjyoti Gogoi (Tezpur University), Aruna Goswami (IIA), Puragra Guhathakurta (UC Santa Cruz), Hanae Inami (NOAO), K. Indulekha (Mahatma Gandhi University), Jason Kalirai (STSCI), Lucas Macri (Texas A&M), Shude Mao (NAOC), Alan McConnachie (NRC Herzberg), Stacy McGough (Case Western Reserve University), Shalima Puthiyaveettil (Indian Institute of Astrophysics), Catherine Pilachowski (Indiana University), Michael Rich (UCLA), James Schombert (University of Oregon), Matthew Taylor (University of Calgary), Sebastien Lepine (Georgia State University/AMNH), Jianrong Shi (NAOC), Annapurni Subramaniam (IIA), Judy Cohen (Caltech), Evan Kirby (Caltech)

---

# 7 EXPLORATION OF THE MILKY WAY AND NEARBY GALAXIES

The stellar contents of the Milky Way and nearby galaxies make up the fossil records that were produced during the Galaxy formation and evolution. The spatial, temporal and abundance distributions of these stars provide important clues to the underlying astrophysical processes involved and also complement the study of galaxy formation and evolution at high redshift. We discuss a number of outstanding questions we expect TMT observations will address.

## 7.1 STELLAR ASTROPHYSICS

There are still a number of areas important to stellar modeling and evolution that are usually ignored because it is unclear how to proceed — there is simply too little observational evidence to guide theoretical models in setting the relevant parameters, hence the resulting unstated uncertainties may be large. Here, we highlight two areas in which TMT can make major contributions.

### 7.1.1 Diffusion (sinking) of heavy elements in the outer parts of stars

Diffusion is believed to occur whenever the outer parts of a star are quiet without large-scale velocity fields and are then not well mixed. Gravity and temperature will preferentially concentrate the heavy elements towards the center of the star (Salaris, Groenewegen & Weiss 2000). Main sequence diffusion acts very slowly with time scales $\sim 10^9$ yr, so it is most important on the main sequence itself where stellar lifetimes are long, and is particularly important for metal-poor stars that do not have outer convective layers. Diffusion of He is important in the Sun and affects helioseismology models. Diffusion is also important in precision distance determinations based on main sequence fitting, since the abundances adopted for the model isochrone must agree with those of the stars, while the abundances deduced for the stellar atmosphere may not be those of the interior. Diffusion may also be the solution to a disagreement of a factor of ~2 between the Li abundances derived for halo turn off stars, assumed to be the primordial lithium abundance, and the (lower) value predicted by standard Big Bang nucleosynthesis models that adopt the baryonic density inferred from the current concordance cosmology of WMAP (Melendez & Ramirez 2004; Korn et al. 2006; Bonifacio et al. 2007). (We note for completeness that diffusion in some stars (e.g., white dwarfs) occurs in just a few years due to their extremely high surface gravities, informing the physics of accreting asteroids onto the star; Koester 2009; Bauer et al. 2019).

A key project to observe diffusion in action is to compare the elemental abundances for heavy elements near the Fe-peak of main sequence versus red giant and subgiant stars in metal-poor globular clusters. Stars within the lowest mass globular clusters, which may contain only a single stellar population, are sufficiently old that diffusion should have had time to act. The RGB stars have convective envelopes, and thus whatever diffusion might have occurred on the main sequence, the surface helium and heavy elements will have been restored to very close to their initial value, while the main sequence stars will be subject to the predicted larger effects of diffusion for metal-poor stars over their entire lifetime, ~13 Gyr for Galactic halo clusters. Observing these abundances requires high spectral resolution spectroscopy of main sequence stars in globular clusters with SNR difficult or impossible to achieve with existing 8–10 m telescopes. The net efficiency gain of HROS on TMT relative to current facilities will enable the required observations of these faint stars.

### 7.1.2 Evolution of massive stars with low metallicity: observational probes

Massive stars contribute a large fraction of all the heavy elements in the universe through SN explosions and the subsequent interactions between their descendants (e.g., merging neutron stars; Thielemann et al. 2017). They provide the UV ionizing flux for the ISM and their supersonic winds help shape the ISM. They are probably linked to the re-ionization of the early universe (Bromm, Kudritzki & Loeb 2001) and perhaps to the GRB phenomenon. Massive stars of very low metallicity were, until quite recently, believed not to lose very much mass through radiatively driven winds due to their weaker absorption features. This would mean that such stars might often end up as black holes, locking up their heavy element chemical inventory in perpetuity, while a solar metallicity star of similar mass would lose enough mass to end up as a neutron star. However, Meynet et al. (2007) have suggested that these stars are rapid rotators that lose up to 50% of their initial mass through a rotationally driven wind. The mass loss rates they predict for the main sequence phase of a 60 $M_\odot$ star are more than 20 times larger than if rotation were ignored. Coupled with rotationally driven instabilities that transport both angular momentum and chemical species (Zahn 1992), this means that massive low metallicity stars can significantly enrich the ISM in H- and He-burning products.

This theoretical development is very attractive as it solves a number of problems, but the observational evidence to support it is essentially non-existent at present. The low metallicity required to produce main sequence rotation rates of ~600 km/sec, close to the breakup rate at which the effective surface gravity becomes zero, is less than that seen at any place in the Milky Way where star formation is still underway. The metal depletion of the ISM in the LMC and in the SMC is also not sufficient. A population of young massive stars in a low metallicity galaxy is required, e.g., a very metal-poor star-forming dwarf galaxy. There are none close enough to obtain the required spectroscopy with existing 8–10 m telescopes. The metal-poor dwarf I Zw 18, has a distance of 18 Mpc and an oxygen abundance 1/50 that of the Sun (Skillman & Kennicutt 1993). The galaxy SBS 1415+437, with a distance of 14 Mpc, is almost as metal-poor (Aloisi et al. 2005), with oxygen below 1/20 the solar value. With WFOS on TMT, it will be possible to observe the brightest supergiants in the nearest very metal-poor dwarf galaxies to determine the mass loss rate as a function of metallicity.

### 7.1.3 Validation of theoretical scenarios for low-mass star formation at extremely low metallicity through observations

There is still no consensus for the formation of low-mass stars from metal-free clouds. Machida (2008) suggests low metallicity clouds in the early universe may preferentially produce binary stars compared to later-universe higher metallicity clouds, suggesting a multiplicity study of extremely metal poor stars may be worthwhile. Such a study would require a large sample of extremely metal-poor stars with radial velocity monitoring to obtain multiplicity statistics. Metal-poor stars are extremely rare objects, which makes finding them a great challenge (figure 7-1*)*. Large-scale systematic searches began with the HK survey by Beers et al. (1985, 1992), followed by the Hamburg/ESO Survey (Wisotzki et al. 1996, Christlieb et al. 2008) that covered ~1000 square degrees of the southern sky, collecting data of some 4 million point sources. More recent searches use medium resolution spectroscopy from large multi-object spectrographs (e.g., SDSS, SEGUE, LAMOST) or they select metal-poor candidates from large photometric survey data. The SkyMapper telescope has photometrically surveyed the southern sky in specific filter sets (e.g., Bessell et al. 2011) to allow candidate selection in a very efficient way. Follow-up high resolution spectroscopy provides candidate confirmation and detailed chemical abundance studies (Yong et al., 2021; Chiti et al., 2021). The oldest and the most Fe deficient ([Fe/H] $< -7.3$) star known so far, SMSS J031300.36–670839.3, is a discovery from this survey but the most metal-poor star with a measurement for [Fe/H] is J160540.18−144323.1 at −6.2 (Nordlander et al. 2019). Similar metal-poor objects to be found from this and other on-going surveys will provide observational constraints to test theoretical models for low-mass star formation at extremely low metallicity. TMT/WFOS would be useful not only to enhance the extreme metal-poor stars database but also for conducting detailed chemical abundance studies.

### 7.1.4 Astrophysics of rare objects

TMT equipped with a high-resolution spectrograph will provide a unique opportunity to understand the origin and evolution of rare exotic faint stellar objects through detailed chemical abundance studies. For example, observational evidence of the presence of extraordinarily strong Sr features in the hydrogen-deficient carbon (HdC) star HE 1015-2050 from medium resolution spectroscopy (Goswami et al. 2010; Goswami & Aoki 2013) brought the issue of Sr synthesis in stellar interiors to the forefront. There are now enough additional examples found that a new Sr-rich subclass of HdCs has been identified (Crawford et al. 2022), but again limited by moderate resolution.

To acquire high resolution spectra required for detailed abundance analysis to understand the origin and evolution of such faint objects (e.g., V~16.3 mag for HE 1015-2050) is extremely difficult and time consuming with the existing 8–10 m class telescopes particularly if such objects happen to undergo significant brightness variations. Likewise, the recent discovery of the unique object HE 1005-1439 with surface chemical composition that exhibits contributions from both slow (s) and intermediate (i) neutron-capture nucleosynthesis (Goswami & Goswami 2022) is quite intriguing. Finding more similar objects among faint metal-poor stars will be challenging, but extensive spectroscopic surveys offer prospects for future discoveries.

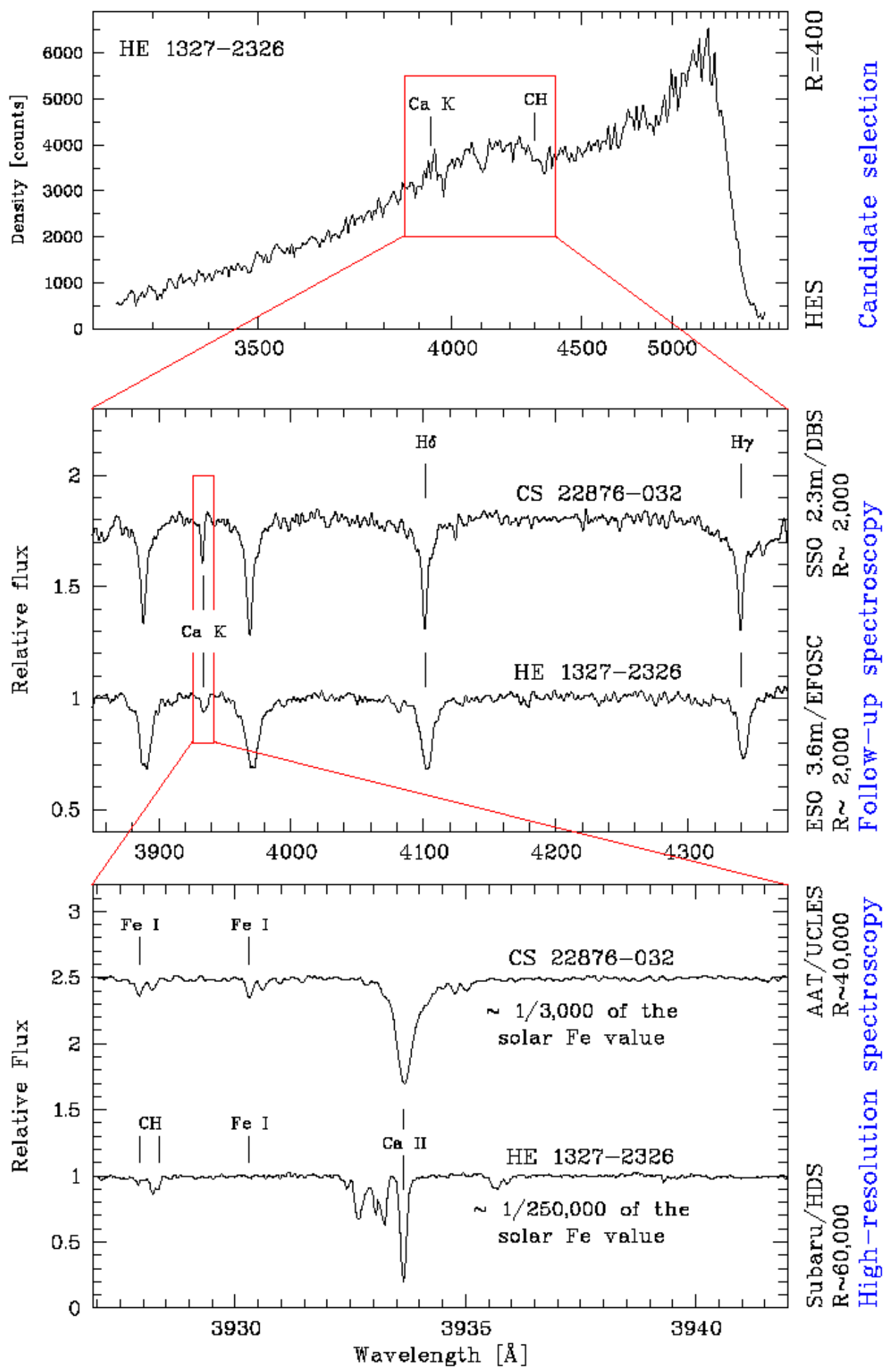

***Figure 7-1:*** *The process of finding a metal-poor star. Stars with weak Ca II K lines are identified in low-resolution spectra (top), and are selected for follow-up with medium resolution spectroscopy (middle) to get a direct measure of the Ca II K line strength. The most metal-poor stars of this sample are then selected for high resolution followup (bottom), where the abundances of other elements can be determined (Source: Jacobson & Frebel 2014).*

Similarly, the enhancement of fluorine abundance by three to four orders of magnitude relative to its likely initial abundance observed in the Extreme helium (EHe) and R Coronae Borealis (RCB) stars (Pandey et al. 2008) raises questions about F synthesis mechanisms and the prevailing conditions in the stellar interiors, although recent simulations suggest that CO-He WD mergers may play a major role (Bhowmick et al. 2020). Measurement of fluorine abundance from the HF band in the near IR in a large sample of K and M giants covering a wide range in metallicities and ages would be essential to understand the origin and evolution of fluorine. HdCs and RCBs form a rare class of supergiant stars that are thought to originate from the merger of CO/He white dwarf (WD) binary systems; the cold Galactic RCB stars are noticeably fainter than the Magellanic RCB stars (Tisserand et al. 2022). From the detection efficiency of HdCs and RCBs, Tisserand et al. (2020) estimated that including HdC stars there should be about 500 RCBs stars existing in the Milky Way. Many such potentially interesting objects will form important targets for TMT due both to the large collecting area and highly efficient spectrographs.

### 7.1.5 The Initial-Final Mass Relation – Version 2.0

Since the mid-2000s, the Keck Observatory has spearheaded an effort to measure one of the most fundamental relations of stellar astrophysics, the initial-final mass relation — IFMR v1.0. This relation connects the mass of stellar remnants, white dwarfs, to the mass of their progenitor hydrogen-burning main-sequence stars, and therefore provides a direct way to understand stellar mass loss.

The initial-final mass relation is constrained by spectroscopically measuring the masses of white dwarfs that are members of rich star clusters. The spectroscopy yields the white dwarf masses (Balmer line fitting; Bergeron et al. 1992), and the initial masses are known given the ages of the clusters. Despite the tremendous progress, two of the biggest uncertainties in the relation today are:

- **What is the upper mass limit to white dwarf formation (i.e., what is the lower mass limit to a type II supernova)?** The threshold mass separating white dwarf formation from type II supernovae is a fundamental threshold for stellar astrophysics and affects several linchpins at the heart of understanding the impact of stellar evolution on galaxy evolution. For example, the transition impacts chemical evolution models and the enrichment of the ISM (and subsequent stellar populations) by setting the balance between stellar yields from lower mass stars (e.g., He, C, N) and those from Type II supernovae (e.g., alpha elements).
- **What is the metallicity dependence of stellar mass loss?** The total amount of mass loss that a star suffers in post main-sequence evolution dictates the lifetime of the star in different phases of stellar evolution, the luminosity function of those phases, the morphology of the horizontal branch, and the structure of planetary nebulae. To accurately use the initial-final mass relation to predict the evolution of mixed stellar populations with different star formation and chemical abundance histories, we need to measure how mass loss varies with metallicity. Recent models for a subset of stellar parameters provide a framework for observational testing (Krtička et al., 2018).

The WFOS instrument on TMT offers a unique opportunity to answer these two questions. With Keck/LRIS, our current sample of target star clusters for which spectra of white dwarf members suitable for mass determination can be obtained is limited to the nearest open star clusters within ~3 kpc, and the single nearest globular cluster within ~2 kpc. Within this volume, there is a lack of very young clusters with massive turnoffs to address the first challenge, and a lack of a metallicity spread within the clusters to address the second challenge. With TMT's light collecting area, white dwarf spectroscopy can be pushed to systems at 10 kpc. This volume includes three dozen globular clusters with a wide range of metallicities, as well as numerous rich open clusters with ages from 30 to 100 Myr (i.e., present day evolving mass ranging from 5 to 10 $M_{\odot}$). Searching for white dwarfs in these clusters, and linking them to their progenitor masses, will answer these fundamental questions. The key requirements for this program that would directly impact many areas of galactic and extragalactic astrophysics are ultra high throughput at UV wavelengths down to 3600 Angstroms (covering the high order Balmer lines), MOS capability and low resolution spectroscopy with R = 2000–4000.

## 7.2 BINARY POPULATION: THE BINARY FREQUENCY OF FIELD STARS

A large fraction of stars in the Milky Way are part of binary or higher order systems, with separations < 0.04 parsecs (Larson 1995). Binaries can be identified by direct imaging or spectroscopy. The spectroscopic detection of binaries can be a laborious process that requires observations that span a long-time baseline. Two possible imaging programs are discussed here that exploit the unprecedented angular resolution of the TMT to probe the binary frequency in very different environments:

1) **A targeted study of star clusters**. A survey of targets selected according to mass, age and dynamical state will prove useful for constraining the factors that define the binary frequency. The improved angular resolution delivered by NFIRAOS will be useful for resolving multiple objects with small projected separations on the sky that might otherwise appear as a single object. Such blends will appear as erroneous points on color-magnitude diagrams (CMDs) and confuse efforts to characterize cluster properties. By measuring the brightness of the individual components of such systems as opposed to their blended light, it will be possible to obtain cleaner CMDs.

2) **A serendipitous survey of low mass binaries in the field**. The TMT has the potential to become a binary-identification factory, as binary systems will be discovered serendipitously by the NFIRAOS WFSs during routine target acquisition and set-up. The angular resolution of the NFIRAOS NGS probes in J is 0.01 arcsec FWHM, and these will guide on sources as faint as J=22. Heretofore unknown binaries (as well as extragalactic objects with modest angular extensions) could be identified by the NFARIOS RTC as objects causing problematic corrections. Binary stars could be identified visually in the WFS output display during target acquisition. If pre-imaging of the field is used to identify guide stars then proper motions can be measured to obtain distance information. Some of the selection biases due to separation and mass ratio inherent to imaging studies could be overcome, for example, by grouping systems according to proper motion. An interesting aspect of this investigation is that the stars that will be detected at the faintest magnitudes will have low intrinsic masses. With long-term observations, a binary database that spans a range of environments in the Galaxy could be sampled.

TMT opens the door to much larger samples of binary systems to study statistical dependencies between the parameters of the components, the star formation function for binaries and the question of the formation of multiple systems (Cherepashchuk 2022), and of the binarity of stellar clusters. Simulations show that binaries impact the formative phases as well as subsequent evolution of stellar clusters (Torniamenti et al. 2021). Modern machine learning methods may be leveraged for analysis of spectral survey data (section 7.1.5) on erroneous points in CMDs in conjunction with stellar spectral libraries and results from detailed spectroscopic, analytical, and simulation studies on various classes of nearby close binary systems. Gravitational wave signals from coalescing stellar mass compact object binaries are clean signals. Understanding the formation pathways of such old systems brings insights into the star formation history of the universe.

## 7.3 Star clusters: formation, evolution, disruption

### 7.3.1 Environmental dependence in star cluster formation and evolution

Stars, particularly the most massive stars, rarely form in isolation. In fact, it is now well established that the vast majority of active star formation occurs in clusters of some sort. Despite significant recent progress, the evolutionary connection between the recently formed young massive clusters (YMCs) in starbursts and old globular clusters in the nearby universe is still being developed (Forbes et al. 2018). The evolution and survivability of young clusters depend crucially on their stellar initial mass function (IMF; see also section 8.2): if the clusters are significantly depleted in low-mass stars compared to, for example, the Solar neighborhood, they will likely disperse within about a Gyr of their formation. As a simple first approach, one could construct diagnostic diagrams for individual YMCs of mass-to-light ratio (*M/L*; derived via dynamical mass estimates using high-resolution spectroscopy and the virial approximation) versus age (derived from spectral features) and compare the YMC locations in this diagram with models of simple stellar populations (SSPs) governed by a variety of IMF descriptions. However, such an approach has serious shortcomings and suffers from a number of fundamental problems like multiple generations of stars in globular clusters (GCs) and GC survivability depending also on interactions with the disk of the parent galaxy as well as with individual giant molecular clouds. Observations of a larger sample of YMCs in a variety of environments will be possible with the TMT and comparison with more realistic cluster formation and evolution scenarios can shed more light on their evolutionary connection with GCs.

The essential conditions to make a major leap forward are to obtain high-resolution spectroscopy and imaging of a significantly larger cluster sample than available at present (to distinguish between trends and coincidences), covering a much more extended age range. These observations will need to be backed up by detailed *N*-body simulations of clusters containing both a realistic number of test particles (upwards of several $x10^3$) and all relevant physical processes occurring over the clusters' lifetimes (e.g., feedback, mass loss). TMT's large aperture will allow us to obtain a statistically significant sample of YMCs and probe both the dynamics *and* the luminosity function of young and intermediate-age star clusters (and their host systems) in galaxy types spanning the entire Hubble sequence. Using this approach, the initial conditions for cluster formation (and evolution) can be modeled self-consistently. This will provide us with the firmest handle yet on potential IMF variations in external galaxies. For the expected velocity dispersions of ~7−15 km $s^{-1}$ (for masses of ~$10^5$–$10^6$ $M_{\odot}$), a spectral resolution of R > 40,000 is essential to efficiently sample the prevailing IMF conditions in a statistically significant number of YMCs. As long as the masses of the clusters are not too small ($M_{cl} > 10^5$ $M_{\odot}$, depending on the cluster size), one can extract their velocity dispersions using suitably chosen cool giant and supergiant template stars (Ho & Filippenko 1996). At R = 40,000, one can resolve velocity dispersions down to σ~6.5 km $s^{-1}$ at a wavelength of 0.85 μm, (e.g., to $M_{cl}$ ~2x$10^5$ $M_{\odot}$) for YMCs of globular cluster size. The spectral range around the CaII triplet ($\lambda_{central}$ = 0.86 μm), with its large number of metal lines, can be cross-correlated with a number of properly selected (super-)giant velocity template stars over the entire observed wavelength range to obtain velocity dispersion measurements.

GCs evince several observational features related to chemical evolution – decided by the mass of the cluster, its metallicity, and the environment of its formation (Gratton et al. 2019). They show multiple sequences in their color-magnitude diagram and multiple groups in the chromosome map pointing to multiple generations of stars. Differentials in fractional contribution of the generations to the cluster mass as well as binarity have been observed, between GCs as well as between the first and second generation of stars in GCs. TMT with ten times better sensitivity and a factor five better angular resolution than JWST, offers unprecedented scope for the detailed study of the formation and evolution of reionization and pre-reionization era stellar clusters. GCs and GC populations and their relationship with their faint, low metallicity, low-mass dwarf hosts, observed in the near universe (Weisz et al. 2023a, Weisz et al. 2023b) and from z ≳ 6 identified as lensed objects in the HST and JWST (see also §4.2) fields (Vanzella et al. 2019) can then be compared with those seen in radiation-hydrodynamic cosmological simulations (Garcia et al. 2023). Such low-mass galaxies are likely to be the more numerous type at high redshifts (see also §4.1) and their GCs are considered as significant contributors to reionization. Analysis of gravitational wave signals from compact object mergers show their numbers to be skewed towards misaligned spin binaries. GCs are considered to offer conditions conducive to the dynamical formation of misaligned spin black hole binaries. Observations with the TMT can shed light on binarity in GCs and the field of the host galaxy (see also §7.2) and provide links for constraining SFH using GW observations. Meanwhile, complementary imaging observations ideally covering at least four passbands spanning a minimum of the entire optical wavelength range will allow to independent and robustly determinations of the cluster properties (age, mass, metallicity) using the sophisticated multi-parameter AnalySED algorithm (de Grijs et al. 2003; Anders et al. 2004).

### 7.3.2 The origin and evolution of globular clusters

Research on Galactic GCs by many investigators over the last few decades has led to a new paradigm for the formation and evolution of these not-so-simple stellar populations (e.g., Bedin et al. 2004; Piotto 2008; Renzini 2008; Bekki & Mackey 2009). Photometric and spectroscopic evidence confirms the presence of multiple stellar populations in nearly all Galactic globular clusters (e.g., Marino et al. 2008; Milone et al. 2009; Zoccali et al. 2009). Both dynamical models of clusters and the total mass of first-generation stars needed to produce the second and subsequent generations suggest that globular clusters were originally 2–10 times or even more massive than they are today (e.g., Moore et al. 2006; Bekki 2011). Early results from JWST observations seem to corroborate this suggestion, as highly lensed NIRCam imaging has revealed a collection of likely GCs with formation redshifts of z~7–11, or ~0.5 Gyr after the Big Bang (Mowla et al. 2022). These early GCs may contain up to $\log_{10}(M/M_{\odot})$ ~ 8–9 providing the extra mass needed to produce second stellar generations, and requiring significant mass-loss if they are truly the progenitors of GCs we see in the local universe. It is believed that the less centrally concentrated first generation stars have mostly been lost to the population of field halo stars, whose composition the first-generation cluster stars resemble.

This new paradigm suggests that the star formation events that created GCs in the early universe were significantly more luminous and more massive than star forming regions we see in the local universe today (Moore et al. 2006; Sameie et al. 2023), and that the star formation process to produce a young GC extended over a time scale of roughly 100 million years. At the same time, we find bimodal populations of globular clusters in external galaxies, red and blue clusters with perhaps different metallicities and/or ages, and with different spatial distributions around their host galaxies (e.g., Geisler et al. 1996; Côté et al. 1998, Peng et al. 2006, Brodie & Strader 2006, D'Abrusco et al. 2015). These different cluster systems are interpreted as resulting from different formation mechanisms, either through in situ formation processes or through major or minor galaxy mergers from host galaxies that experienced different histories of star and cluster formation.

To understand the star forming regions in the early universe that may be detected with TMT, we must first learn what we can about the origin and evolution of star clusters in the local universe. While Galactic clusters can be studied with smaller telescopes up to 8 and 10 meters, understanding the implications of multiple stellar populations in globular clusters beyond the local group will require the spatial resolution and light gathering power of TMT. A study of radial gradients in colors and spectral indices that reveal the nature of the constituent stellar populations will require spatial resolution exceeding what is possible with Hubble (which can barely resolve clusters at the distance of Virgo) and even JWST. Such studies with TMT will help us constrain the properties of the original star forming events that produced GCs in different environments and allow us to connect the clusters we observe today with the observation of high redshift cluster-forming events in the early universe observed directly with TMT.

### 7.3.3 Probing IMBH candidates in extragalactic globular clusters: infrared counterparts of X-ray sources and stellar velocity profiles

Black holes of masses intermediate between the supermassive black holes in AGN and stellar mass black holes are highly challenging to identify (Greene et al. 2020; Su, et al., 2022). X-ray sources in globular clusters (Tiengo et al. 2022) are ideal candidates for such intermediate mass black holes (IMBHs): some of them are ultra-luminous, known as ULX, and have X-ray luminosities greater than $10^{39}$ erg $s^{-1}$. AGNs are much more luminous and emit equally at all wavelengths with a strong correlation between the X-ray and IR luminosities from a dust torus (e.g., Krabbe, Böker & Maiolino 2001; Lutz et al. 2004; Asmus et al. 2011; Ichikawa et al. 2012). A study of NGC1399 using archival data from Spitzer IRAC and Chandra showed that there are two categories of bright X-ray sources, one in which the mid-IR luminosity correlates with the X-ray flux as in AGNs, and the other where the IR and X-ray luminosities are uncorrelated (see figures 6 and 7 of Shalima et al. 2013), indicating that one population has a dusty environment, and the other is dust deficient. However, a major source of uncertainty in this result is possible contamination from background AGN. With high spatial resolution mid-infrared of imaging from TMT/MICHI together with spectral information, it will be possible to eliminate the background contamination and concentrate on the real IMBH candidates and the dust contents of their environments. Stellar velocity profiles can also serve to indicate the presence of globular cluster IMBHs, although to date such measurements are difficult even in Milky Way clusters (e.g., Vitral et al 2023, Göttgens 2021). Adaptive-optics assisted integral-field spectroscopy from TMT/IRMOS will enable integrated light velocity profiles to be constructed for extragalactic clusters and aid in the search for IMBH signatures.

## 7.4 THE FIRST STARS

### 7.4.1 Probing the oldest stars in the Milky Way

The first, so-called Population III stars are believed to have been formed from gas unpolluted by heavy elements, and have significant impact on the subsequent star and galaxy formation through their radiative and chemical feedback (Klessen & Glover 2023). Hence the most metal-poor stars in the Milky Way and other galaxies are the oldest stars. In their atmospheres, these old objects preserve details of the chemical composition of their birth gas clouds. These very old stars are hidden amongst a vast number of stars formed later in time. Once found, understanding the chemical composition of the oldest stars provides a direct probe of the initial conditions of star formation as well as the details of chemical evolution and nuclear astrophysics in the early universe. Recent photometric and spectroscopic surveys (e.g., Skymapper, Pristine, LAMOST) have been providing large samples of metal-poor star candidates, for which kinematic information is also available by Gaia. Whereas large efforts to identify most metal-poor stars and to determine their chemical compositions by high-resolution spectroscopy using 8–10 m class telescopes have been continued (e.g., Lardo et al. 2021; Mardini et al. 2022; Aguado et al. 2023), more sensitive high-resolution spectrometers are required to significantly expand the sample with accurate chemical composition covering wide coverage of the Milky Way structures (halo, disc bulge) and surrounding dwarf galaxies.

Since the relevant absorption lines in metal poor stars appear quite weak (figure 7.2), chemical composition studies require high spectral resolution (R ~50,000 or greater), high SNR (~100 per spectral resolution element) observations. These requirements lead to extremely long integration times on the largest telescopes currently available. For example, an 18-hour integration with UVES on the VLT was obtained in an attempt to determine the oxygen abundance in HE1327-2326, the most extreme metal-poor turn-off star known, but even with such a long exposure, the UV-OH band was not detected (Frebel et al. 2006).

The number of stars accessible to TMT will be much larger. Relative to VLT/UVES, Keck/HIRES, or Subaru/HDS, TMT/HROS will be a factor of 15–20 more efficient. For example, a 4-hour integration with HROS will enable $R$ = 40,000 spectroscopy with SNR = 100 per spectral resolution element for stars as faint as V ~ 21. The UV sensitivity that is possible at the high altitude is essential to determine N and O abundances from NH and OH molecular lines, which are key elements to constrain the mass and nucleosynthesis processes of first generations of massive stars.

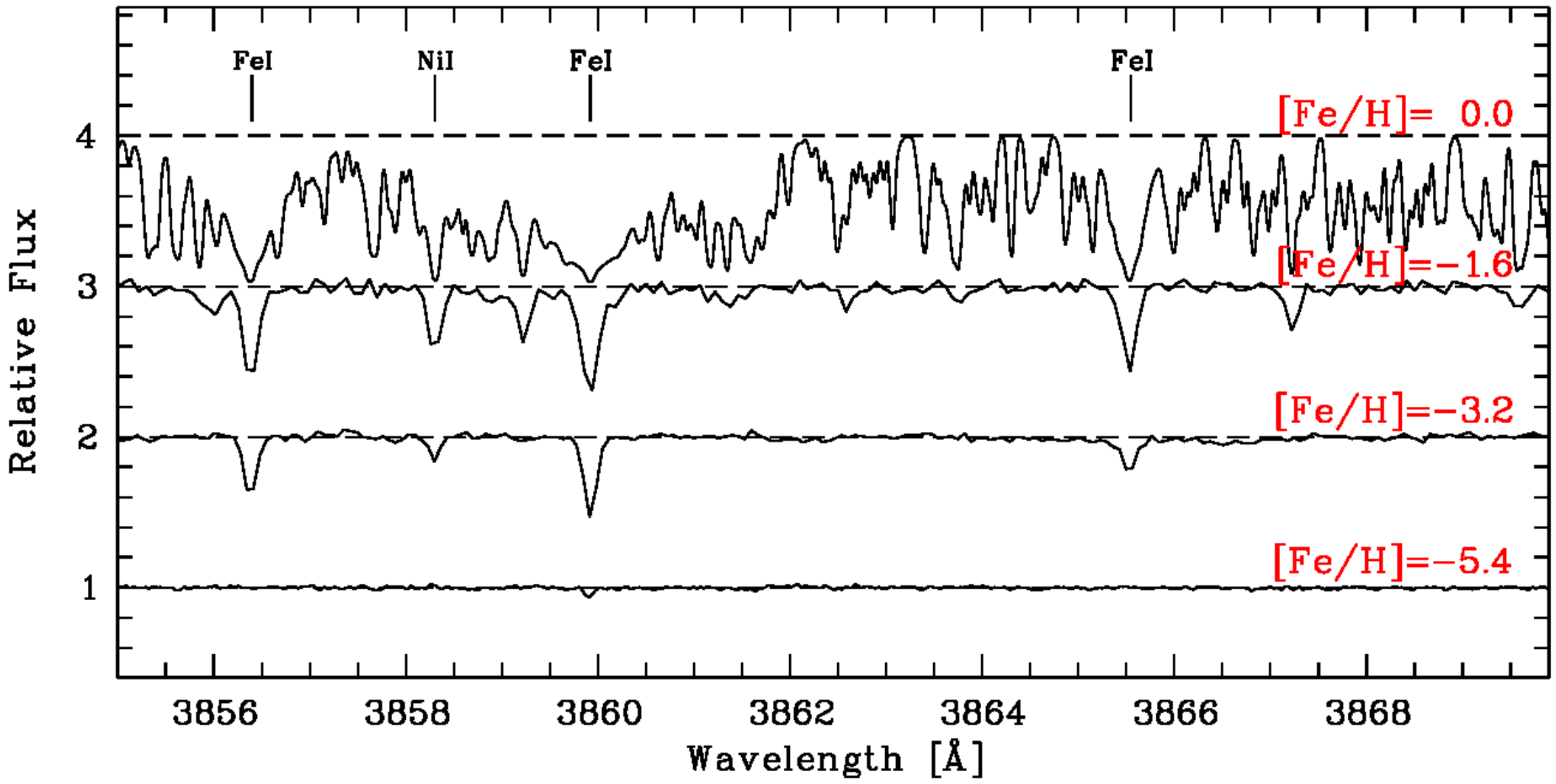

***Figure 7-2:*** *Spectral comparison of stars in the main-sequence turn-off region with different metallicities. Several atomic absorption lines are marked. The variations in line strength reflect the different metallicities. From top to bottom: Sun with [Fe/H] = 0.0, G66-30 with [Fe/H] = −1.6 (Norris et al. 1997), G64-12 with [Fe/H] = −3.2 (Aoki et al. 2006), and HE 1327-2326 with [Fe/H] = −5.4 (Source: Frebel, 2010).*

### 7.4.2 Mass distribution of the first generation of stars

Measurements of detailed chemical abundances for most metal-poor ([Fe/H]<−4) stars in the Milky Way halo in the past two decades suggest that the progenitors of such objects are massive stars with several tens of solar masses, which would be typical masses of the first stars in the universe (e.g., Klessen & Glover 2023). Detailed measurements of their chemical compositions require high-quality, high-resolution spectra like those to come from TMT/HROS. For instance, only an upper-limit of Fe abundance ([Fe/H] <−7.3) has been inferred by the current telescopes for the most iron-deficient star currently known (Keller et al. 2014). High sensitivity of the instrument is also required to identify metal-poor stars that could record the yields of very massive stars (>100 $M_{\odot}$)and Pair Instability SupeNovae (PISNe) determining their detailed chemical abundances (Aoki et al. 2014; Xing et al. 2023), as such stars are extremely rare in the current Milky Way and need to be found in a sample of faint objects. As new first generation star candidates are found from current and future surveys, TMT/HROS will be well-suited for the required confirmation follow-up observations.

## 7.5 THE STRUCTURE OF THE MILKY WAY AND NEARBY GALAXIES

### 7.5.1 Dissecting the Galactic halo: the ages and metallicities of old, nearby low-mass stars and white dwarfs

Recent estimates suggest that close to 70% of the stars in the local Galactic field population are M dwarfs, and about 6% are white dwarfs. The local population of old stars from the Galactic halo is expected to hold a larger fraction of M dwarfs and white dwarfs. These stars however have low luminosities, and are not amenable to detailed spectroscopic analysis with current telescopes unless they lie within about 100–200 parsecs of the Sun.

Low-mass stars and white dwarfs are highly useful to map out the halo population for three reasons: 1) they are by far the dominant stellar populations of the halo; they *are* the old population, holding the majority of the baryonic mass, 2) their 3D velocity components can be determined to much higher accuracy, because of their large proper motions, 3) the physical properties of low-mass stars, and in particular their metallicities, can be constrained using low to medium resolution spectroscopy, because their spectra are dominated by broad molecular bands, and with suitable spectrum synthesis techniques, can now be modeled in detail.

Deep, wide-field imaging surveys such as Pan-STARRS and those to come from Rubin Observatory will identify millions of low-mass stars through proper motion analyses. These surveys effectively provide a statistically complete census of the halo population out to several kiloparsecs from the Sun. Spectroscopic data will be required to measure and constrain the temperatures and metallicities of the M dwarfs, and the masses and ages of the white dwarfs. As the photometric distances of M dwarfs and white dwarfs are generally unreliable, Gaia and spectroscopic data can constrain their luminosities (Kim et al. 2022). Thus a better understanding of the spatial distribution and kinematics of these objects once discovered by these photometric surveys depends on follow-up observations.

The sensitivity of the TMT along with the multi-object capabilities provided by WFOS will make possible a large-scale spectroscopic follow-up of low-mass stars and white dwarfs in the Galactic halo. These observations will enable:

- Spectral classification of the candidates to confirm their Galactic halo membership, and for those too faint for Gaia, the use of existing spectroscopic distance calibrations to determine their distances to calculate their transverse motions.
- Measurements of metallicity [Fe/H] and relative abundances of critical elements [alpha/Fe] from the relative intensity of molecular band heads.
- Measurements of radial velocities to km/s precision to calculate their full 3D motions.

Along with proper motion and photometric data from large imaging surveys, the spectroscopic data from TMT will produce a detailed census of the nearby halo population, drawing their detailed distribution in velocity space. The combination of metallicity information from the M dwarfs and age information from the white dwarfs will identify possible substructure in the velocity space distribution. This will provide additional constraints on the shape of the halo, and also on the possible existence of accretion events, which would show up as streams in the local volume.

### 7.5.2 Planetary nebulae as tracers of substructures in nearby galaxies

Planetary nebulae (PNe) are descendants of low- and intermediate-mass stars. Being bright and widely distributed, they are excellent tracers of the chemistry, kinematics, and stellar contents of substructures in nearby galaxies. As the nearest large spiral system to the Milky Way, the Andromeda galaxy (M31) is an ideal laboratory in which to assess the theory of hierarchical cosmology. PNe are easily detectable at the distance of M31 (~780 kpc; McConnachie et al. 2005) and can be used to investigate the properties of the intriguing substructures such as the Northern Spur and the Giant Stream in the outskirts of M31 (figure 7-3). High-quality spectra of PNe in these regions are still conspicuously lacking. 2D maps of nebular abundances and properties of the PN progenitors would trace the different substructures.

Out of nearly 3,000 PNe in M31 revealed by Merrett et al. (2006), reliable elemental abundances have been derived for only a few dozen M31 PNe (e.g., Jacoby et al. 1999; Corradi et al., 2015). LAMOST has discovered PNe in the outskirts of M31, including the most distant one, 3.6° from the center (figure 7.3; Yuan et al. 2010). This PN is both spatially and kinematically related to the Giant Stream, and is the first one discovered in the outer streams of M31.

The α-element abundances of PNe reflect those in the interstellar medium (ISM) at the time when the progenitor stars formed. We can derive properties (luminosities, effective temperatures) of the central stars of planetary nebulae (CSPNe's) based on the nebular spectrum using models (e.g., CLOUDY; Ferland et al. 1998), given a known PN distance. Using evolution models of post-asymptotic giant branch stars (Vassiliadis & Wood 1994) and the initial-final mass relationship of white dwarfs (Catalán et al. 2008; Cummings et al. 2018), one can estimate the masses and ages of the progenitors. If the substructures are the debris of dwarf galaxies of M31, the stellar populations therein could be different from those in the disk.

Spectroscopy of PNe has seldom been used to study the substructures in M31. In order to investigate the origin of substructures, deep spectroscopy of PNe in the substructures is required, covering 0.3–0.6 μm and 0.55–1.0 μm in the blue and red channels of the proposed WFOS instrument. At a resolution $R$ ~5000, 15 target PNe can be observed simultaneously with full wavelength coverage.

The instrument setup will enable detection of the [O II] λλ3726, 3729 doublet and [S II] λλ6716, 6731 lines, which are the traditional electron density diagnostics. The temperature-sensitive [O III] λ4363 and [N II] λ5755 auroral lines will also be efficiently detected and lines such as [O II] λλ7320, 7330 and [S III] λ6312, λλ9532,9069 will also be observed, enabling determination of electron temperatures for different excitations and consequently accurate elemental abundances for N, O, Ne, S, and Ar. Given the excellent observing condition at Mauna Kea, the 30m aperture of TMT will achieve an optical spectrum with a signal-to-noise (S/N) ratio of about 10 for a faint ($m$5007 ~22) PN in M31 with only a 15 min exposure.

With a 30 min exposure for each observing field of WFOS (~15 PNe targeted simultaneously), one can cover about 12–14 fields, or ~200 PNe, in a single night. Thus, a one-week survey will produce more than 1000 high-quality spectra for the M31 PNe in a homogeneous manner. These spectra, together with the accurate luminosity information of the PNe (given that distances of all PNe are known), could be used as inputs for the nebular code CLOUDY to determine the luminosities and effective temperatures of the CSPNe. The masses and ages of the progenitor stars can then be estimated using stellar evolution models.

The elemental abundances and the properties of PN progenitors in the substructures will be compared with those on the disk to investigate the origin of substructures. The survey will allow a determination of properties of a significant number of PNe (> 1,000) across M31 and the creation of the very first 2D maps of nebular abundances and the ages and masses of progenitor stars will allow the tracing of different substructures within M31.

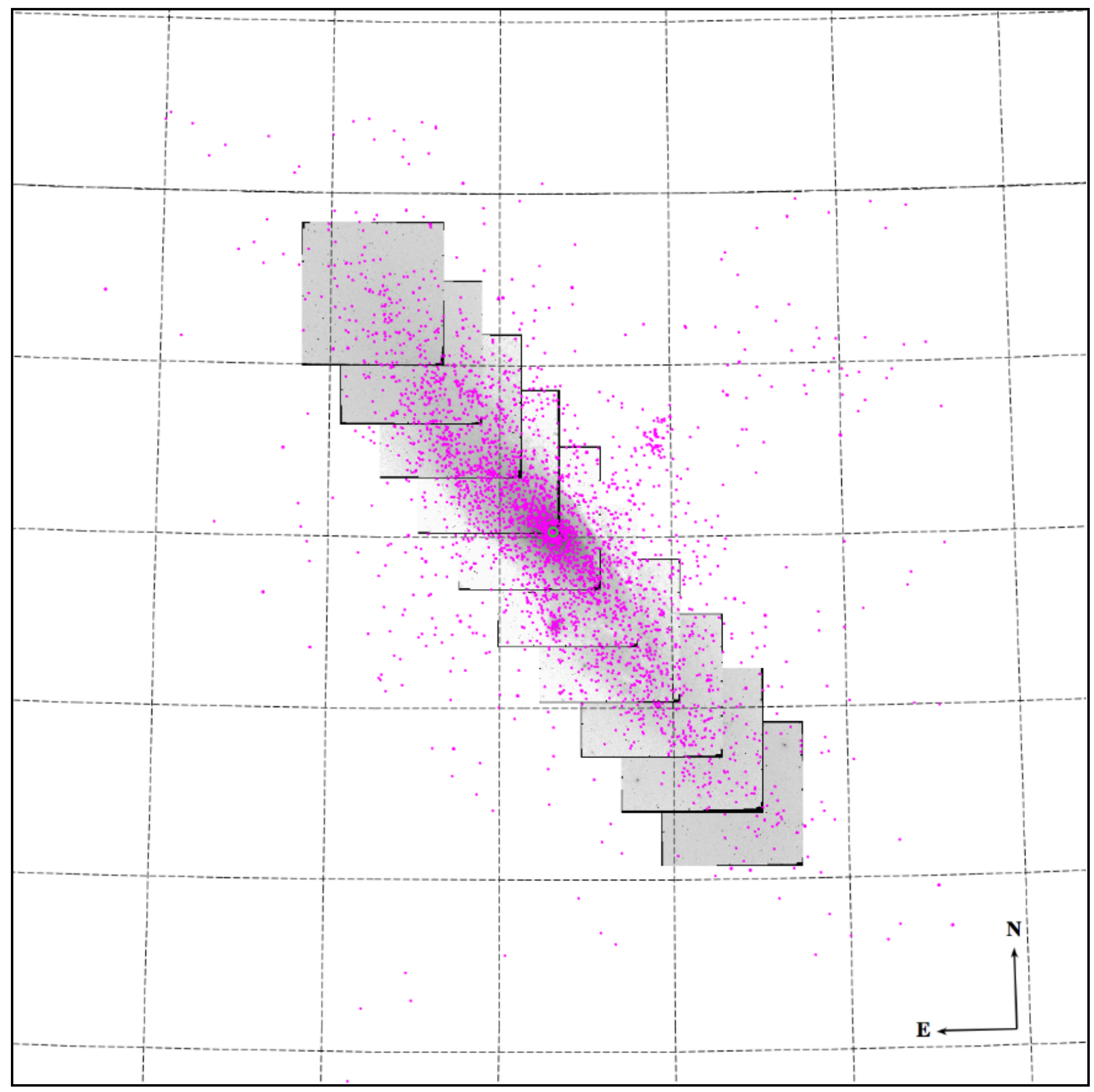

***Figure 7-3:*** *Positions of planetary nebulae (PNe; the magenta dots) overlying on the ten mosaic fields of M31 that were targeted in the Local Group Survey by Massey et al. (2006). The mosaic images are in V band (in the logarithm scale, inverted color) and each field is roughly 36' × 36' in size. The PNe sample (in total ~3000) includes those observed by Merrett et al. (2006) and LAMOST (Yuan et al. 2010; Yuan et al. 2015). Horizontal divisions of the dashed grid are 0.75°, and vertical divisions are 1°. (Source: Yuan et al. 2010).*

### 7.5.3 Chemical tagging of individual stars and Galactic Sub-structures

A systematic study of stellar motions and composition is essential to reveal the origin and evolution of the Galaxy and its present structure. The Galaxy is broadly composed of three major structural components: the disk, the bulge, and the halo. Stars can be grouped into these components based on their kinematic properties and their chemical composition (Eggen, Lynden-Bell & Sandage 1962). In addition to the above three components, the Milky Way galaxy is found to have sub-structures which are in agreement with the theoretical predictions for a hierarchical formation of the Galaxy via mergers. Understanding the different structures of the Milky Way — their origins, and dynamics — is one of the fundamental issues in astronomy.

For a decomposition of the Milky Way, measuring accurate parallaxes and proper motions for a large number of stars is essential. The Hipparcos Space Mission in 1998 provided accurate astrometry for around one hundred thousand stars in the solar neighborhood (d ~ 200 pc). Combining the ground based high resolution spectroscopy and the Hipparcos astrometry, the Galactic disk is decomposed into thin and thick disks (Reddy et al. 2003). Stars in the thick disk are old (~8–10 Gyrs), metal-poor ([Fe/H] ~ −0.6), and have distinct kinematic motions and chemical composition from those of the thin disk population. Abundances of elements with different nucleosynthesis histories suggest that the thick disk stars formed mainly from SN II ejecta, the formation of the thick disk was rapid (<2

Gyrs), and there is no current star formation. The thick disk population is believed to be the result of a major merger of a metal-poor dwarf galaxy when the Milky Way was just 1–2 Gyrs old. There are many unanswered questions regarding the thick disk and the composition of the Milky Way galaxy: Is the thick disk really a frozen entity? What is the metal-poor end of the thick disk? Are there any smaller components in the disk? What is the structure of the bulge and the halo?

To answer these questions one needs astrometry for a large number of stars beyond the Solar neighborhood and high resolution spectra to chemically tag individual stars. The space observatory Gaia has measured astrometry for nearly two billion stars up to mv~20. Gaia also measured radial velocities for nearly as many stars. Together with Gaia astrometry and radial velocities, it is possible to construct a 3D view of our Galaxy and identify the distinct components based on kinematic motion. TMT/WFOS chemical tagging of individual stars of different kinematic groups would help understand the chronological history and evolution of the Galaxy.

## 7.6 Kinematics and dynamics of the Milky Way and nearby galaxies

### 7.6.1 Kinematics of the local group

The Local Group has been the benchmark for testing and calibrating many aspects of cosmology and galaxy formation and yet, the formation and evolution of the local group of galaxies is not clearly understood. We do not have reliable estimates of the space velocities of the galaxies, for example. The major hindrance in the estimation of space velocity is measuring the very small proper motions of the distant galaxies.

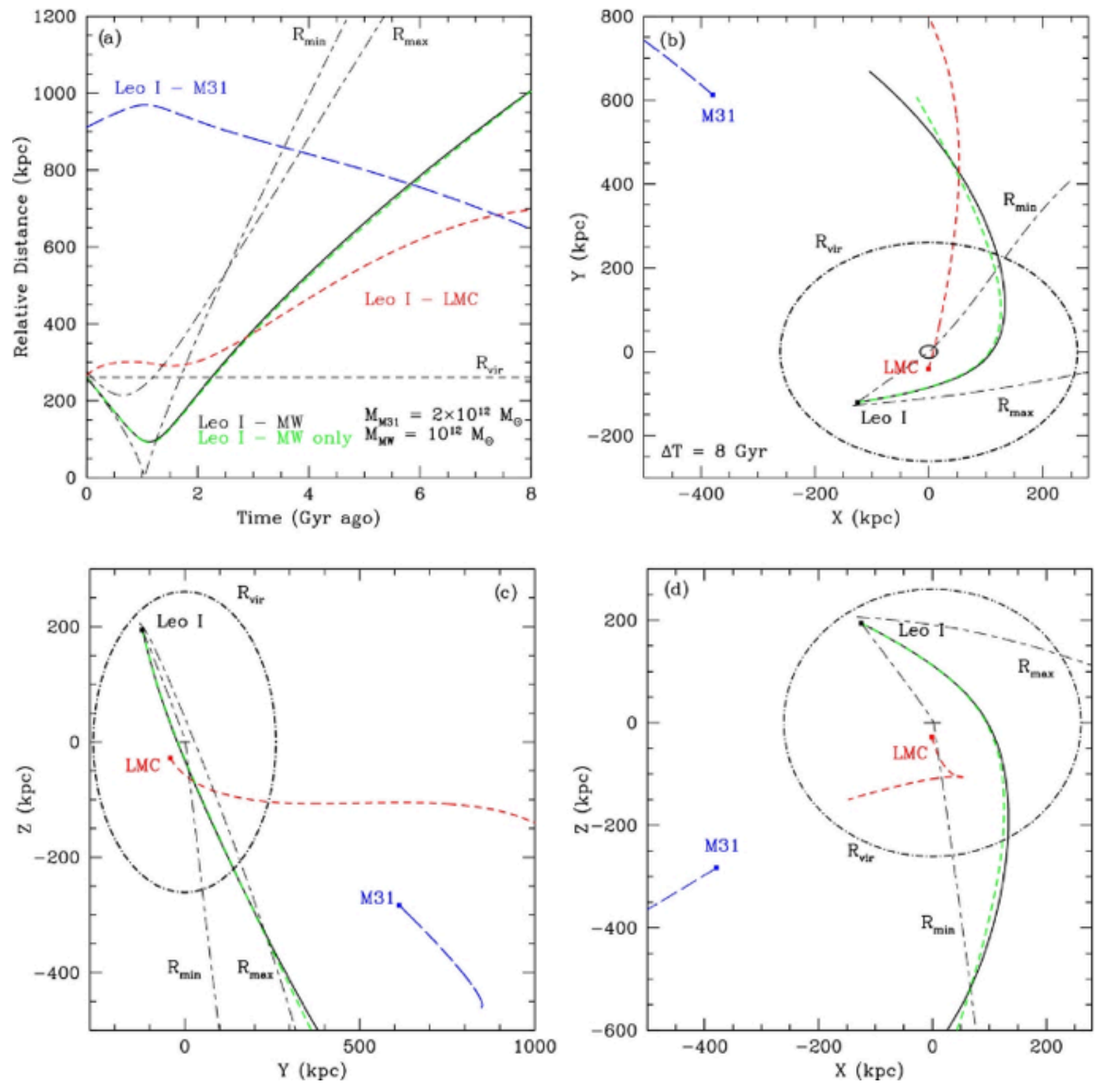

***Figure 7-4:*** *The mean orbital history of Leo I from Sohn et al. (2013)*

Recent estimations found that the Magellanic Clouds, once thought to orbit around the Milky Way, are actually on their first passage, as is the case for many of the Milky Way's dwarf satellites (Hammer et al. 2021) as determined from Gaia observations. Sohn et al. (2013), on the other hand, found that for most plausible Milky Way masses, the observed velocity implies that Leo I is bound to the Milky Way, based on the proper motion estimates from the HST. Sohn et al. (2012) obtained the first ever proper motion measurement for M31. Their proper motion is consistent with a head-on collision orbit for M31 toward the Milky Way (van der Marel & Guhathakurta 2008). The required precision needed to estimate the proper motion for the nearby galaxies is about 50 microarcsec/yr.

More generally, phase space correlations of local group satellite galaxies provide a powerful test of the ΛCDM cosmological model (Pawlowski 2021), as they are not strongly affected by uncertainties in baryonic physics. A number of surprising phase space correlations have been noted in recent years, including the vast polar structure of largely corotating Milky Way satellites, and similar structures about other galaxies. In addition, there are surprisingly many close pairs of apparently gravitationally unbound satellite galaxies. More exotic correlations have also been proposed, such as an anticorrelation between satellite dark matter density profiles and pericentric distance (Cardona-Barrero et al. 2023). However, these do not yet pose a sharp challenge to ΛCDM as they may result from once bound systems of dwarf galaxies which were accreted by their host galaxies. The accretion histories remain uncertain as a result of uncertainties in the proper motions of these satellites. Figure 7-4 shows the mean orbital history of Leo I (Sohn et al. 2013).

However, the extraordinary angular resolution of multi epoch observations using TMT with IRIS, separated by five to ten years, can reduce these uncertainties by more than an order of magnitude, to 5 km/s or less in the case of many Milky Way satellites and less than 50 km/s for many of M31's satellites, allowing a far more complete extraction of the histories of these systems. Alternatively, as in Riley et al. (2020), some of these surprising apparent correlations may disappear under the microscope of high angular resolution measurements.

The refined orbital calculations of many members of the local group will allow us to understand the kinematical history of interactions in the local volume and predict the destiny of the Local Group. Combined with the proposed studies of chemical composition and star formation history, we will be able to decipher the formation, evolution, and future of the Local Group.

### 7.6.2 Internal dynamics of dwarf-spheroidal galaxies: density profiles of dark matter halos

High-resolution spectroscopy of resolved member stars in dwarf-spheroidal galaxies (dSphs) has revealed that the measured line-of-sight velocities show much larger velocity dispersions, $\sigma_{los}$, than expected from the stellar system alone. Indeed, dynamical analysis of $\sigma_{los}$ and its spatial dependence inside dSphs has provided density profiles of dominant dark matter halos in dSphs, which can be compared with the predictions of ΛCDM theory (see section 3.1.1 for a detailed discussion).

Previous studies of $\sigma_{los}$ in dSphs have revealed several discrepancies with ΛCDM theory, which include:

- Dark halos in some dSphs show cored broader, shallower central densities in contrast to cuspy narrower, deeper ones as suggested from Navarro–Frenk-White density profiles;
- The average densities of dark halos in bright dSphs are systematically small compared to those of massive sub-halos in Milky Way-sized halos.

These results are based on the analysis of $\sigma_{los}$ alone, with assumptions for the anisotropy of the velocity dispersions of stars, such as a constant isotropy/anisotropy along the projected distance from the galaxy center. However, detailed information for velocity anisotropy and its spatial dependence for stars over the two-dimensional projected area of dSphs is actually required to obtain tighter and thus more realistic limits on the density profiles and global shapes of their dark halos (e.g., Hayashi and Chiba 2012).

TMT/IRIS can provide the needed measurements of the proper motions of member stars and their spatial distribution in the Milky Way dSphs. Provided typical internal velocities of stars are ~10 km/s, one can expect proper motions of ~0.03 mas/yr in dSphs at distance of 70 kpc from the Sun, which TMT/IRIS can measure with a few-years of coordinated observations of stellar positions. The available information of velocity dispersions perpendicular to the line of sight will allow us to construct more realistic mass models of dSphs and to set important constraints on the density profiles of dark matter halos in comparison with ΛCDM theory (Battaglia et al. 2023). For example, a single epoch of TMT observations, combined with legacy Gaia data, can exclude a cored profile for the Sculptor dwarf galaxy's dark matter halo with 4σ of confidence (Evslin 2015) and similar forecasts have been made for other dwarf satellites of the Milky Way (de Martino, et al 2022). Two epochs of observations can extend this result to some ultra faint dwarfs, which ΛCDM simulations robustly predict are cusped (Chan et al. 2015). Similarly, two epochs of TMT/IRIS observations discover, at 5σ, the ellipticity of the underlying dark matter distribution (Evslin 2016). The distribution of these ellipticities is also robustly predicted by ΛCDM (Hayashi et al. 2015). Thus TMT/IRIS may, given two epochs of observations separated by at least 5 years, robustly exclude all particulate cold dark matter models in which no non gravitational long range forces are present. Recent simulations (Martínez-García et al. 2023) also find that ΛCDM predicts a slight rotation for dwarf satellite galaxies, which is large enough to be tested by TMT/IRIS.

### 7.6.3 The mass of the Milky Way

Despite several efforts to measure it, we still do not know the total mass of Milky Way to better than about 50% due to the lack of understanding of the dark matter dominated halo, the isotropy of the orbits, and the shape of the potential (spherical or not) in the outer halo (Watkins et al., 2019). The total mass of the Galaxy is usually determined through the motion of tracers (stars, dwarf galaxies, etc.) in the outer part of the galaxy beyond 50 kpc, making assumptions regarding the isotropy of orbits and the radial density distribution of the tracers.

In an effort to determine the total mass of the Milky Way, Deason et al. (2014) considered blue horizontal branch (BHB) stars beyond 80 kpc as tracers of the outer halos. However, this sample, selected from the SDSS stripe 82 region, was found to contain mostly blue stragglers and a few QSOs, with distance uncertainties exceeding 30% in many cases. The sample contained only a few outer halo Galactic satellite dSph galaxies with reliable distances. This study reported that the number density of their outer halo stars falls very rapidly ($\rho \sim r^{-6}$ at $r > 50$ kpc) and that the radial velocity dispersion of their sample is quite cold in the outer halo; the estimated total mass for the Milky Way is rather small ~ $5 \times 10^{11}$ $M_\odot$.

Cohen et al. (2017) explored the outer halo of the Milky Way considering RR Lyrae variables found by the Palomar Transit Facility (PTF). They selected 1257 RR Lyrae variables with distances beyond 50 kpc in the Milky Way halo, specifically to probe the density distribution as well as the total mass of our Galaxy. The distance of this sample could be determined to an accuracy of 5%, and the contamination by QSOs is expected to be minimal. From a detailed analysis they found a density law, $\rho \sim r^{-3.4}$, between 50 and 85 kpc, which is close to the results found by Watkins et al. (2009) and Sesar (2010) for the inner Milky Way halo.

For stars to be used as tracers of the outer halo, one needs to know their radial velocities and the proper motions. With TMT it will be possible not only to measure the radial velocities of PTF RR Lyr stars out to 100 kpc but also the radial velocities of the RR Lyr stars to be found with Rubin Observatory out to a distance of 200 kpc and beyond, leading to an improved understanding of the total mass of the Milky Way.

### 7.6.4 Luminosity-metallicity and mass-metallicity relations for dwarf galaxies beyond the Local Group

The average metal content of a galaxy correlates with its mass; massive galaxies are more metal-rich than less massive galaxies, giving rise to the mass-metallicity relation (MZR; Tremonti et al. 2004). The MZR can be explained by the retention of metals in the galaxies' gravitational potential wells (Dekel & Silk 1986). High-mass galaxies have deep potential wells that can resist some of the expulsion of gas and metals by supernova winds, stellar winds, and galaxy-scale feedback whereas low-mass galaxies lack the gravity to resist these feedback mechanisms (e.g., Brooks et al. 2006).

Kirby et al. (2013a) compiled the metallicity of individual stars in seven gas-rich dwarf irregular galaxies (dIrrs) and dwarf spheroidal galaxies in the Local Group and produced a stellar MZR for the Local Group dwarf galaxies (figure 7-5). Further, they have combined this with the MZR for more massive galaxies (figure 7-6) observed by the Sloan Digital Sky Survey (Gallazzi et al. 2005). This relation can be used to estimate the mass of the merging satellites. The mass-metallicity relation of Kirby et al. (2013a) depends on the mean metallicity of a dwarf galaxy but does not account for potential metallicity gradients within each one, which could provide additional information about the mass distribution and star formation history of the satellite. Such metallicity gradients can be steepened during merger events that trigger an enhanced cold gas density in galaxy outskirts (Buck et al. 2023). Recent work using integral-field spectroscopy in fact reveals that the MZR across eight orders of magnitude of stellar mass is in fact non-linear, with enhanced [α/Fe] ratios observed for both the lowest ($10^{4-6} M_\odot$) and highest ($> 10^{10} M_\odot$) mass galaxies (Romero-Goméz et al. 2023). These revelations suggest that rapid bursts of star formation as a low-mass galaxy falls deeply into a massive cluster potential can significantly alter the spatially resolved stellar populations in such systems so that star-to-star chemical abundance variations are crucial to consider when interpreting their overall properties. TMT observations will make it possible to derive metallicity from individual stellar spectra for the resolved stellar population without having to use coadded spectra for metallicity determination, as was done for DEIMOS M31 satellite spectra in Kirby et al. (2013a). Improved metallicity estimates will lead to tighter relations, extending them to much fainter limits and beyond the Local group.

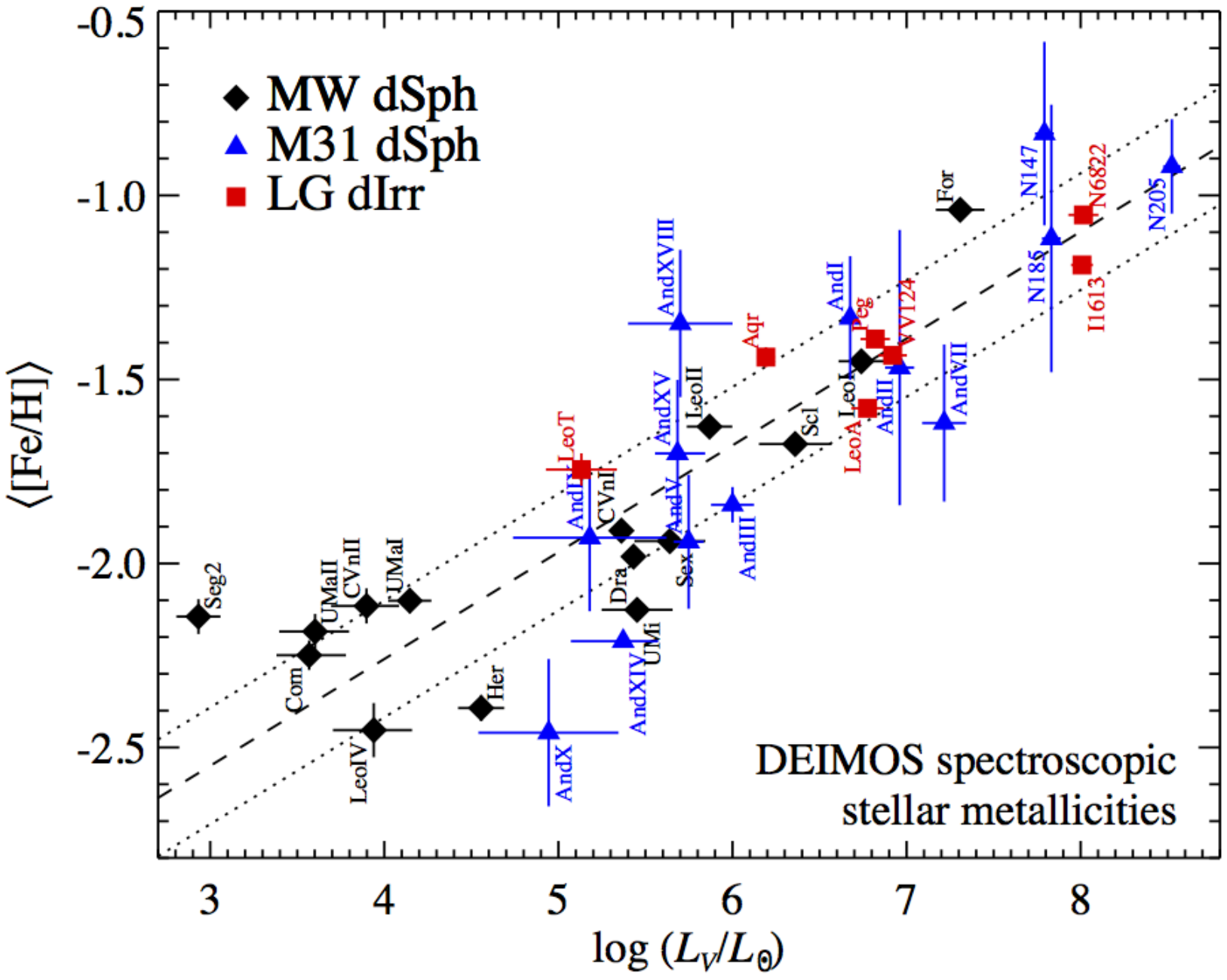

***Figure 7-5:*** *Luminosity-stellar metallicity relation for Local Group dwarf galaxies. The black diamonds (Milky Way dSphs) and red squares (dIrrs) are the average stellar iron abundances from spectroscopy of individual stars. The blue triangles (M31 dSphs) are the average stellar iron abundances from coadded spectroscopy of groups of similar stars within each dwarf galaxy. The dashed line shows the least squares fit, excluding the M31 data points and Segue 2. The dotted lines show rms about the best fit. (Kirby et al., 2013a).*

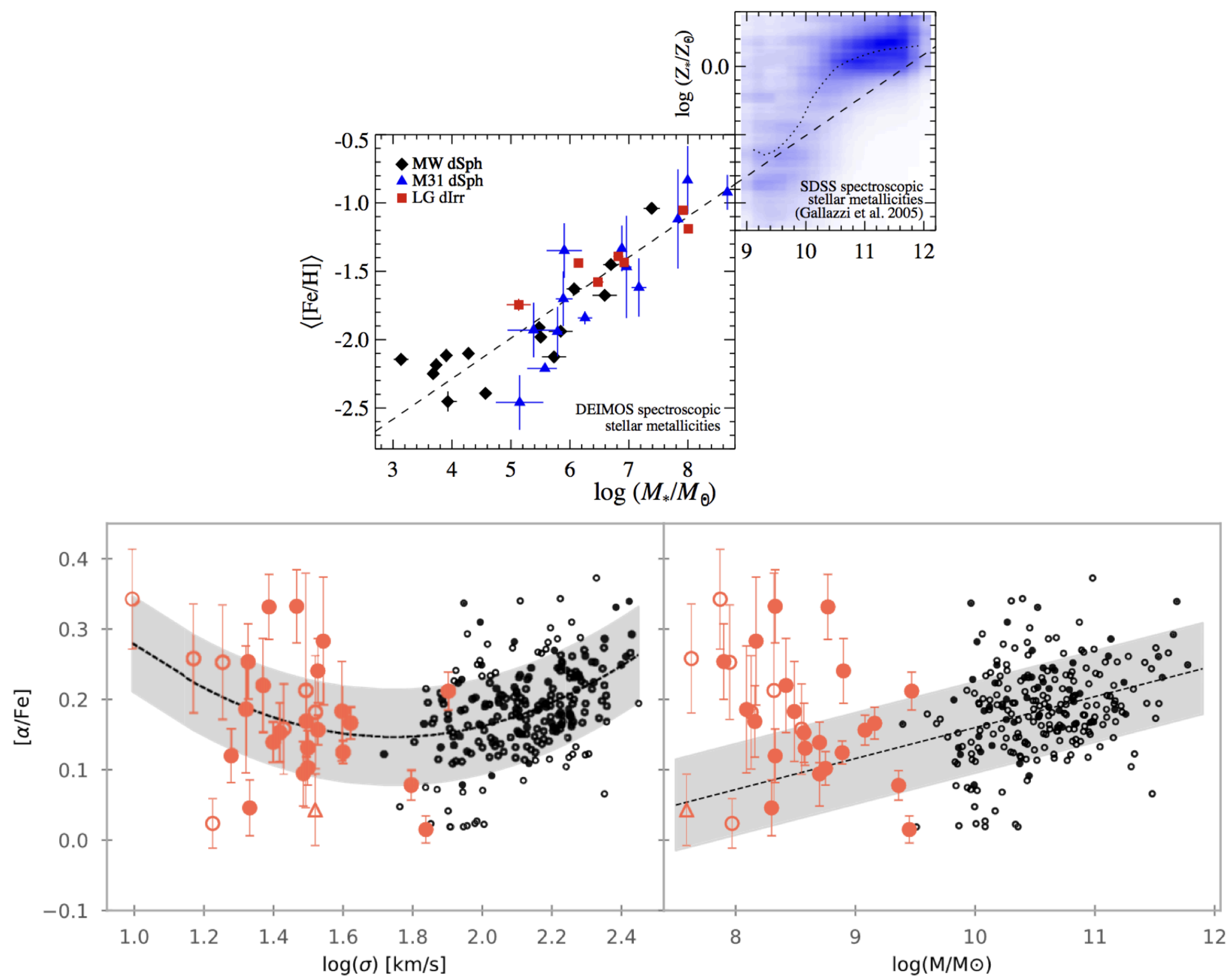

***Figure 7-6:*** *(Top) Stellar mass-stellar metallicity relation for Local Group dwarf galaxies (left) and more massive SDSS galaxies. The local group metallicities (<[Fe/H>) were measured from iron lines, and the SDSS metallicities (log Z.) were measured from a combination of absorption lines, mostly Mg and Fe. The conversion between <[Fe/H> and log Z. depends on [Mg/Fe]. The Local Group data are the same as in figure 7-5 but is plotted here vs. stellar mass rather than luminosity. The dashed line is the least-squares fit to the Local group galaxies, and the dotted line in the right panel is the moving median for the SDSS galaxies. Although the techniques used to measure both mass and metallicity differ between the two studies, the mass-metallicity relation is roughly continuous over nine orders of magnitude in stellar mass. (Kirby et al. 2013a). (Bottom) Abundance ratio [α/Fe], obtained from FSF with pPXF and the MILES library, as a function of the logarithmic velocity dispersion and the stellar mass, in the left-hand and right-hand panels, respectively. The symbols for our dwarfs are the same as in Fig. 1. For ATLAS$^{3D}$, black-filled circles are used for cluster galaxies and non-filled black circles for the rest. The black dashed line and gray shaded area on the left represent a second-degree polynomial fit to all the data visible in the figure, making a U-shape. In the right-hand panel, however, we present the linear fit* $[\alpha - Fe] - log(\frac{M}{M_\odot})$ *relation (Romero-Goméz et al. 2023).*

### 7.6.5 Milky Way satellites and dark matter distribution

The presence of low-mass galaxies swarming around the Milky Way had long been predicted. The inability to detect them was a major issue until the discovery of Segue 2 in the SEGUE survey by Belokurov et al. (2009) (figure 7-7; also called Aries ultra-faint dwarf). Many more such galaxies are currently at our ability to detect and recent years have yielded the discoveries of many of these predicted dwarfs in the form of currently disrupting systems like the Tucana III stream (Li et al. 2018) and dozens of enigmatic ultra-faint dwarf galaxies (Bechtol et al. 2015;

Drlica-Wagner et al. 2015; Kim et al. 2015a; Kim & Jerjen 2015b; Koposov et al. 2015a; Cerny et al. 2021a,b; Cerny et al. 2023; Smith et al. 2023).

Segue 2 is among the least massive galaxies in the known universe, consisting of about 1000 stars with dark matter holding them together. Keck/DEIMOS spectroscopy of 25 members of Segue 2 (Kirby et al. 2013b) revealed that the stars' [α/Fe] ratios decline with increasing [Fe/H], indicating that Segue 2 retained Type Ia supernovae ejecta despite its presently small mass and that star formation lasted at least 100 Myr. The mean metallicity ([Fe/H] ~ –2) was found to be higher than expected from the luminosity- metallicity relation defined by more luminous dwarf galaxy satellites of the Milky Way. The dynamical and chemical characteristics of Segue 2 are suggestive of two possible scenarios for its formation: Segue 2 may be the barest remnant of a tidally stripped Ursa Minor-sized galaxy that came to be ultra-faint through tidal stripping via gravitational interaction with the Milky Way leaving only the dense center of the galaxy, or it may have been born in a very low mass dark matter subhalo. Further studies and simulations are needed to test these alternatives.

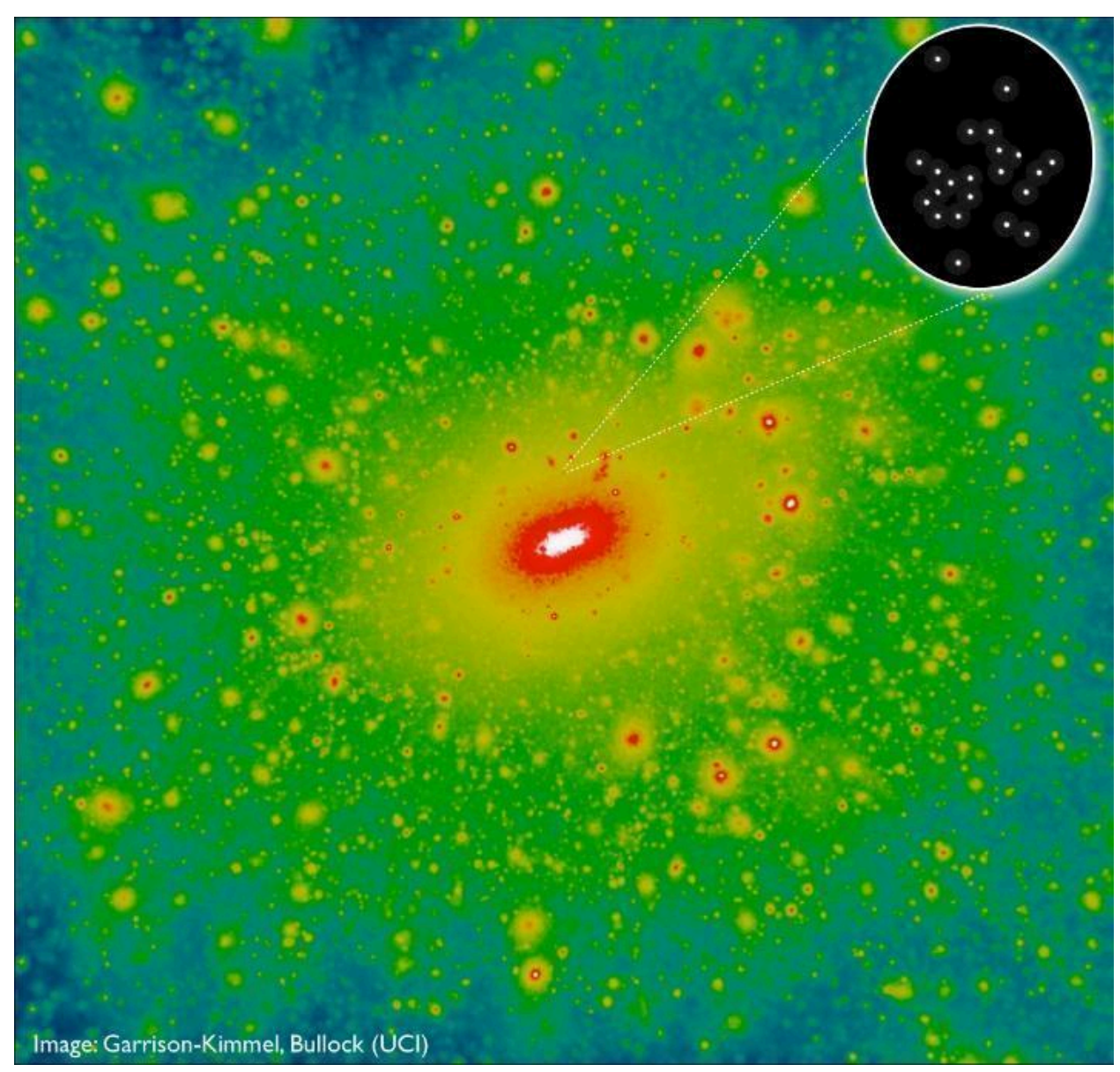

***Figure 7-7:*** *The image shows a standard prediction for the dark matter distribution within about 1 million light years of the Milky Way Galaxy, which is expected to be swarming with thousands of small dark matter clumps called halos. The scale of the image is such that the disk of the Milky Way would reside within the white region at the center. Until now, there was no observational evidence that dark matter actually clumps this way, raising concerns that our understanding of the cosmos was flawed in a fundamental way. Observations of the ultra-faint dwarf galaxy Segue 2 (zoomed image) have revealed that it must reside within such a tiny dark matter halo, providing possibly the first observational evidence that dark matter may be as clumpy as has long been predicted (Source: Garrison-Kimmel, Bullock (UCI)).*

Reliable radial velocity measurements could not be made for many of the spectroscopic targets used in the Kirby et al. (2013b) study due to the low S/N of their Keck/DEIMOS data. High resolution spectrographs like Keck/HIRES and Subaru/HDS could possibly measure the velocity distribution of ~10 stars in Segue 2 over three to four nights. As additional new dwarf satellites continue to be discovered with modern wide-field surveys (e.g., DES, Gaia, UNIONS, SkyMapper, etc.) including the imminent Rubin Observatory, multi-object spectrographs like TMT/WFOS will provide the essential combination of light collecting area and spectral resolution required for measurements of the velocities of their fainter, more common stars to measure their dynamical masses as in Segue 2; the newly discovered plethora of low-mass galaxies; as well as in yet undiscovered galaxies.

### 7.6.6 Velocity anisotropy of the distant Milky Way halo: evidence of an accretion event

The oldest and most metal-poor stars in our galaxy reside in the stellar halo, a diffuse envelope of stars extending out to a distance of ~100 kpc from the Galactic center. The orbital timescales of these halo stars are very long compared to the age of the Galaxy, thus the significant phase-space sub-structure of the stellar halo is intimately linked to its accretion history (Cunningham et al. 2019, Qiu et al. 2021, Pandey 2022). The extreme radial extent of halo stars, well beyond the baryonic center of the Galaxy, makes them excellent tracers of the dark matter halo.

Based on long baseline (5–7 years) multi-epoch HST/ACS photometry, Deason et al. (2013) measured proper motions (PMs) of 13 main-sequence Milky Way halo stars at an average distance of ~24 kpc from the Galactic center with a root-mean-square spread of 6 kpc and median proper motion accuracy of ~5 km/s at this distance. These observations point to a fairly complex velocity anisotropy profile at large distances, which is likely affected by substructure in the stellar halo. Their results suggest that the stellar halo velocity anisotropy out to a distance of ~30 kpc is less radially biased than solar neighborhood measurements — opposite to what is expected from violent relaxation and possibly indicating the presence of a shell-type structure at a distance of ~24kpc (figure 7-8). Rather than a constant or continuous decline, the velocity anisotropy is found to have a dip at 20 kpc, the location of which is coincident with a break in the stellar halo density. Although the origin of the break radius is still uncertain, Deason et al. suggest that the break radius in the Milky Way may be due to a shell-type structure built up from the aggregation of accreted stars at apocenter.

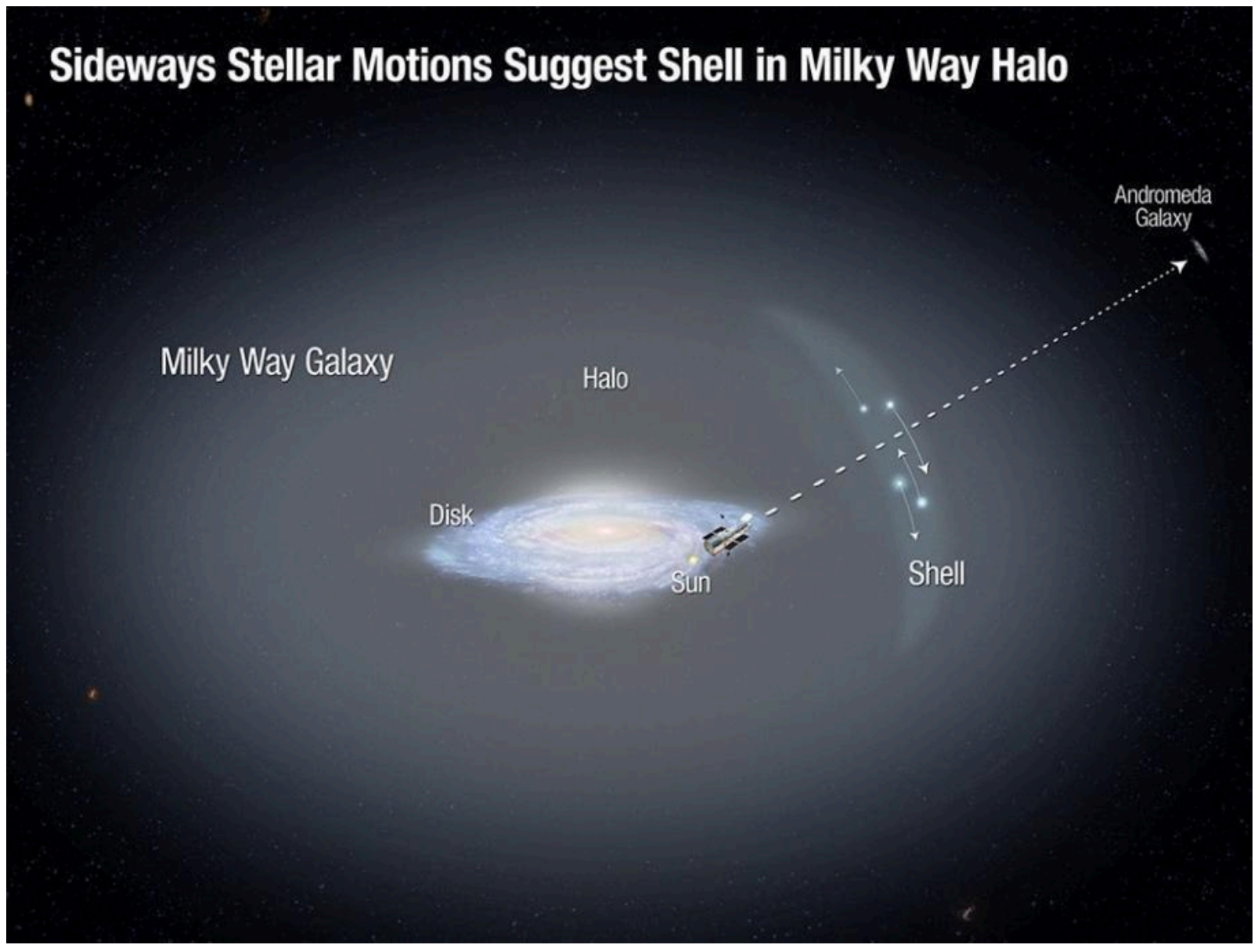

***Figure 7-8:*** *The disk of our Milky Way Galaxy, surrounded by a faint, extended halo of old stars. Deason et al. (2013) used the HST to observe the Andromeda galaxy and serendipitously identified a dozen foreground stars in the Milky Way halo. They measured the first sideways motions (represented by the arrows) for such distant halo stars. The motions indicate the possible presence of a shell in the halo, which may have formed from the accretion of a dwarf galaxy. This observation supports the view that the Milky Way has undergone continuing growth and evolution over its lifetime by consuming smaller galaxies (Credit: NASA, ESA, and A. Field (STScI).*

An independent measurement of velocity anisotropy is vital to derive the mass profile of our Galaxy. A measure of the tangential motion of distant halo stars will allow us to address whether or not this cold radial velocity dispersion is due to a shift in pressure from radial to tangential components. The Gaia mission has revolutionized our understanding of the inner stellar halo. However, even with bright halo tracers (e.g., Blue Horizontal Branch or carbon stars), its PM accuracy is unable to accurately constrain the tangential motion of very distant halo stars.

Recent progress has been made in measuring proper motions of significant stellar populations in the halo (Qiu et al. 2021), confirming the rich sub-structure hinted at by results like Deason et al. (2013). Moreover, measurements of the Milky Way ellipsoid parameters demonstrate that the velocity anisotropy is a function of field position (Cunningham et al. 2019). From the ground, the state-of-the-art requires enormous resources on 8–10 m class telescopes and/or survey baselines of over a decade. In the ideal case, we would like to directly measure the tangential motion of the halo stars along many lines-of-sight through the Milky Way halo. At large distances in the halo, 10–100 kpc, a tangential velocity of 100 km/s would correspond to a proper motion on the sky of 2–0.2 mas/year. To make significant progress on sub-decadal timescales this requires an astrometric precision that is unfeasible for current PM surveys. TMT's increased sensitivity and astrometric accuracy will enable such studies.

### 7.6.7 The Milky Way halo streams and the Galaxy's gravitational potential

Stellar tidal streams are believed to be the remnants of accreted Milky Way satellites that were disrupted by tidal forces and stretched into filaments as they orbited in the Galaxy's potential. The orbits of stars in these streams are sensitive to the properties of the potential and thus allow us to constrain the potential over the range of distances spanned by the streams, helping us understand the evolution and current dynamics of the Galaxy.

Stellar streams such as the GD-I stream (Grillmair & Dionatos 2006), the Sagittarius tidal streams (Majewski et al. 2003), and the Orphan stream (Grillmair 2006) have been used to constrain the circular velocity at the Sun's radius (Koposov et al. 2010), the total mass within 60 kpc (Newberg et al. 2010), and the shape of the dark matter halo potential (Bovy et al. 2016).

Sesar et al. (2013) studied the metallicity and spatial extent of the Orphan stream based on 30 RRab stars. These stars in the Orphan stream have a wide range of metallicity from −1.5 to −2.7 dex. The average metallicity, ~ −2.1 dex, is similar to that found for blue horizontal branch stars by Newberg et al. (2010). There exists a metallicity gradient along the stream length — the distant parts (40–50 kpc from the Sun) are about 0.3 dex more metal-poor than the parts within ~30 kpc. Comparing the distances of Orphan stream RRab stars with the best fit orbits of Newberg et al. (2010), Sesar et al. found that the best fit to distances of Orphan stream RRab stars and to the local circular velocity is provided by potentials where the total mass of the Galaxy within 60 kpc is ~ $2.7 \times 10^{11}$ $M_{\odot}$.

Many more streams likely to be found by deep imaging with PANSTARRS and Rubin Observatory will be accessible with TMT. Detailed studies of the resolved stellar populations in these streams with improved precision of stellar distances would provide great insight into the total mass of the Galaxy and the Galaxy's gravitational potential.

### 7.6.8 The Galactic halo formation: is the Milky Way halo formed by disrupting accreted dwarf galaxies?

Whether or not the Milky Way halo is formed by disrupting accreted dwarf galaxies (Searle & Zinn 1978) remains an open question. Evidence for this mechanism exists in the form of debris structures observed in space and velocity around the Galaxy (Majewski et al. 2003, Belokurov et al. 2006, Schlaufman et al. 2009). However, failures to find the observed chemically primitive stellar populations in the relatively luminous ($L > 10^5 L_{\odot}$) classical dwarf galaxies suggest a statistically significant absence of extremely metal-poor stars with [Fe/H] < −3, prompting Helmi et al. (2006) to suggest that the Galactic building blocks must have been different from the surviving dwarf galaxies. Later studies by Kirby et al. (2009) found an extremely iron-poor star, S1020549, in the Sculptor dwarf spheroidal galaxy. Follow-up high-resolution spectral analysis of this object by Frebel et al. (2010) confirmed its Sculptor membership and measured its metallicity to be [Fe/H] = −3.81. Derived chemical abundances obtained for other elements suggest that the elemental abundance ratios of this star have a chemical pattern nearly identical to that of similarly metal-poor halo stars (figure 7-9). The agreement of the abundance ratios of S1020549 with those of the halo suggests that the classical dwarf spheroidal galaxies experienced very similar chemical enrichment in the earliest phases to the Milky Way halo and ultra-faint dwarfs.

As discussed by Frebel et al. (2010), this result provides evidence that the early chemical evolution of galaxies spanning a factor of more than 1 million in luminosity is dominated by the same type of stars, and possibly the same mass function, namely massive core-collapse supernovae as indicated by the alpha-element enhancement found in all of the most metal-poor stars. This universality would also characterize any dwarf galaxies that were accreted at early times to build the Milky Way and its stellar halo. Those accreted systems are therefore unlikely to have been significantly different from the progenitors of the surviving dwarfs. In that case, the oldest, most metal-poor stars observed in present-day dwarf galaxies should be representative of the stars found in the Galactic building blocks before their destruction. These results thus support the idea that mergers and accretion of small, generally metal-poor systems, as predicted by ΛCDM models, can in principle explain the metal-poor stellar content of the Galactic halo.

The picture that is emerging from the abundance agreement suggests that the old, low-metallicity tail of the outer halo could have been populated with metal-poor stars deposited by small dwarf galaxies that were destroyed long ago. The question remains, whether enough dwarf galaxies accreted to account for all of the metal-poor halo stars. The surviving ultra-faint dwarfs are the least luminous and most dark matter-dominated galaxies (Simon et al. 2007) and they possess very few stars despite containing some extremely metal-poor stars. It is thus unclear whether the accretion of even large numbers of analogues to such systems can provide enough stellar mass to account for the entire population of low-metallicity field stars.

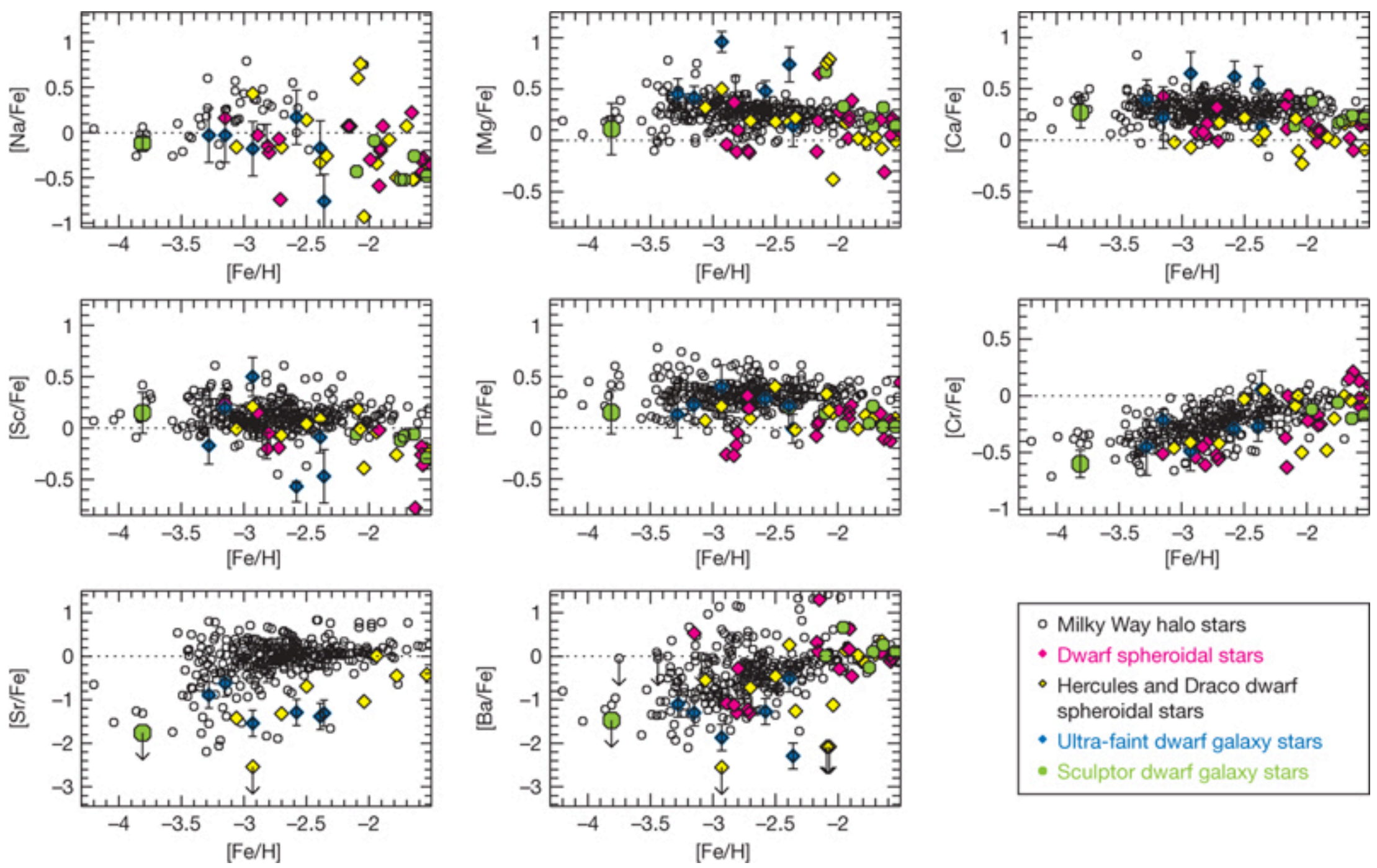

***Figure 7-9:*** *Abundance ratios as function of iron abundance in S1020549 and other metal-poor stars from the literature. In eight elements, S 1020549 (Big green filled circles, at [Fe/H] = −3.8) is compared with halo stars (black circles), ultra-faint dwarf galaxy stars (blue diamond) and the brighter dwarf galaxy stars (pink and yellow diamond). Smaller green circles indicate higher metallicity Sculptor targets (Frebel et al. 2010). Elemental abundance ratios show that this star has a chemical pattern nearly identical to that of similarly metal-poor halo stars.*

The presence of S1020549 in a relatively modest survey indicates that future searches are likely to discover more such objects in Sculptor and other dwarf galaxies. Acquiring high resolution spectra for such distant faint member stars in any of the dwarf Milky Way satellites with the existing 8–10 m telescopes will be quite challenging. TMT/HROS will open up a new window for thoroughly studying early galaxy assembly through stellar chemistry, providing great insight into the Galactic halo formation history.

Although the Tri And stellar clouds were originally viewed as the remnants of an accreted and now disrupted satellite galaxy, the suggestion by Price-Whelan et al (2015) that these stellar clouds may be a population that was kicked out of the Galactic disk is very intriguing. They carried out a critical examination of the possible mode of formation of this system in the halo by comparing the ratio of RR Lyraes to M giants with other structures in the Galaxy. They found that unlike any of the known satellites of the Milky Way, this ratio for Tri And gives a value more like the population of stars born in the much deeper potential well inhibited by the Galactic disk. Recent chemodynamical results based on high-resolution spectra obtained with 8 m class telescopes appear to support this claim, as in this space the majority of TriAnd stars are consistent with the outer thin-disk population (Abuchaim et al. 2023). N-body simulations of a Milky Way like galaxy perturbed by an impact of a dwarf galaxy demonstrate that, in the right circumstances, concentric rings propagating outwards from that of the Galactic disk can plausibly produce similar over-densities. These results provide support for the proposal by Xu et al. (2015) that, rather than stars accreted from other galaxies, the Tri And clouds represent stars kicked-out from the Galactic disk. If so, they would represent the first population of disk stars to be found in the Galactic halo.

Yet another formation scenario suggests that stars might form *in situ* from gas located in the halo itself (Eggen et al. 1962). While such populations have been seen in hydrodynamic simulations of galaxy formation (Abadi et al. 2006, Tissera et al. 2014), the observational evidence for the existence of this population remains controversial (Carollo et al. 2008). Meanwhile, with larger telescopes, Battaglia et al. (2017) have shown that some metallicity discrepancies can be seen in lower metallicity stars that suggests formation in the deeper gravitational wells of Magellanic-like galaxies rather than smaller dwarf galaxies. Bolstering this notion are dynamical simulations and observations that link halo substructure to disrupted satellites and/or merger debris. For example, the kinematically cold S2 stream - with stars consistent with formation within a disrupted primitive dwarf galaxy - is currently plunging through the Galactic disk (Aguado et al. 2021), broadly consistent with recent modeling results linking debris streams to known halo overdensities (Balbinot & Helmi 2021).

The above studies indicate that the Galactic halo is rather complex, with more than one population, each of a different origin. The outer halo, beyond 15–20 kpc from the Galactic center, has an average metallicity a factor of four lower than that of the inner halo (Carollo et al. 2007). More observations of dwarf and irregular galaxies are needed to clarify this situation, and TMT/HROS can play a significant role in unraveling the Galactic halo formation history.

## 7.7 COSMIC CHEMISTRY AND STELLAR NUCLEOSYNTHESIS

Big Bang nucleosynthesis in the first 20 minutes of the universe is believed to have created $^{1}H$, deuterium, the two isotopes of He ($^{3}He$ and $^{4}He$) and a very small amount of lithium. Almost all other elements in the periodic table are synthesized in stellar interiors and envelopes during hydrostatic and explosive burning.

Each stellar nucleosynthetic path has a different timescale and produces characteristic elemental abundance patterns. Chemical evolution in differing stellar populations traces the star formation history and age and provides insight into the chemical evolution of galaxies and their interstellar matter. For accurate abundance measurements, high spectral resolution (R >20,000) with S/N >50 ratio are required to detect and analyze faint spectral features. At this, and even higher resolutions, many important absorption lines become available for study (figure 7-10). At resolution R >90,000, high signal to noise estimates of isotope abundances become possible. TMT equipped with a high-resolution spectrograph will have a dramatic impact in nucleosynthesis studies as it will aid detection of faint spectral features and also open up currently unreachable classes of stars for study.

While elements are formed inside stars, some light elements such as D, Li, Be are also destroyed inside stars under certain conditions. The detection and abundance estimate of these elements is important to understand the stellar structure, evolution and mixing process in stars and the amount of stellar processing. We discuss a few outstanding questions related to stellar nucleosynthesis, answers to which can be sought from future TMT observations.

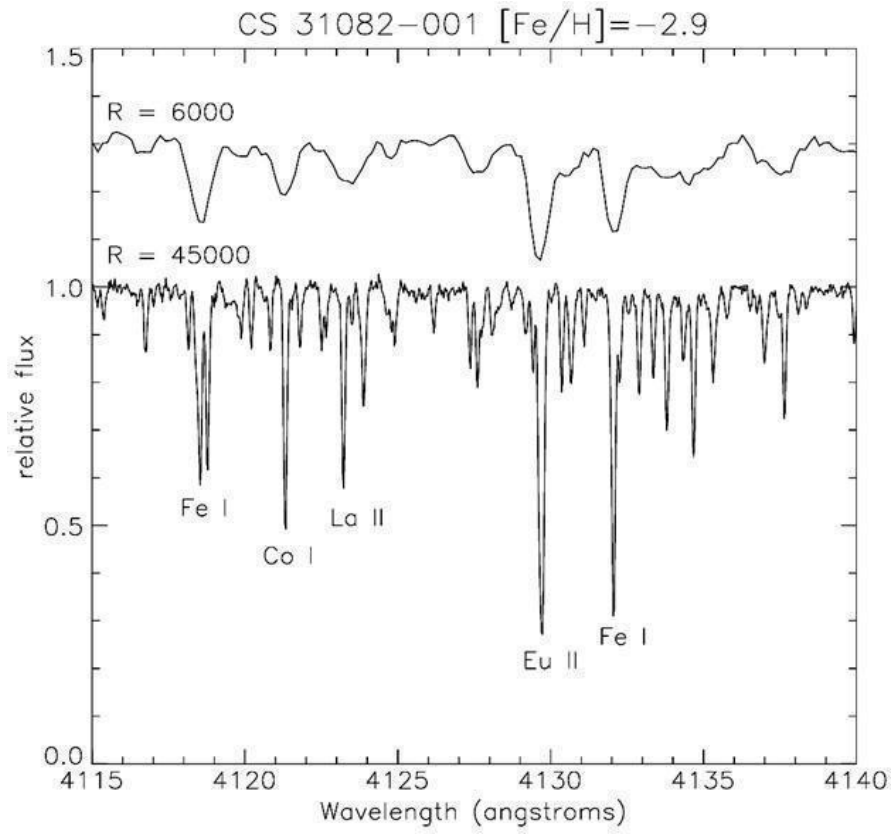

***Figure 7-10:*** *Comparison of moderate resolution (R=6000; curve) and high resolution (R=45000; lower curve) spectra metal-poor star. At higher resolution, many important absor lines become available for study. (R. Guhathakurta, UCSC).*

### 7.7.1 Li Production

Li is a unique element that traces many aspects of chemical evolution of the universe including the Big Bang, explosive events, cosmic-ray reactions, and stellar internal processes. WMAP determination of the cosmic baryon density, combined with the Big Bang Nucleosynthesis theory, tightly predicts a Li abundance value that differs from the value of the Spite Plateau (Spite & Spite 1982) by a factor of two or more. Several possible solutions, such as stellar destruction and astration, nuclear uncertainties, atomic diffusion and new physics have been proposed to explain this discrepancy.

Determinations of precise elemental abundances as a function of stellar evolutionary status are key to constrain such stellar internal processes (e.g., Korn et al. 2006), for which very high-quality spectra are required. For the significant source of Li after the Big Bang, growing evidence is obtained for classical novae that produce $^{7}$Li through synthesis of $^{7}$Be and electron capture reaction (Tajitsu et al. 2015, Izzo et al. 2015, Arai et al. 2021). $^{7}$Be is measurable from the near UV resonance lines at 313 nm in a limited phase of nova events. TMT/WFOS and TMT/HROS will have uniquely high blue sensitivity to verify these results for many more novae to ascertain that this process is sufficient to account for Li in the Galaxy.

### 7.7.2 C, N, O elements and the Be puzzle

C, N, and O are produced in all types of stars during hydrogen and helium burning. These elements play an important role in star formation by cooling the clouds through several fine structure lines. The observational evidence of the increase of C abundance ratio with respect to Fe with a decrease in metallicity may be evidence for a gradual change in the characteristic mass of the IMF, which is likely to be mediated by CMB temperatures acting as a minimum temperature (Tumlinson 2007). With the TMT it will be possible to look for such correlations in the local dwarf galaxies and test theoretical predictions.

Be abundances in the Galaxy and satellites of the Milky Way provide constraints on the pre-Galactic cosmic ray fluxes and cosmic magnetic fields. If Be and B are produced as secondary elements, the Be abundance should decrease quadratically with respect to metallicity. However, the observation of a small sample of metal-poor stars shows that there is a linear decrease in Be abundance with metallicity (Primas et al. 2000), indicating a primary production of C, N, O, and Be. To investigate the relation, further observations of very metal-poor stars discovered by the past two decades for which the current telescopes are not able to measure Be from the 313 nm line. The TMT will enable similar studies to fainter limits, extending the sample beyond the Milky Way to the local group dwarfs to help us understand the true nature of this apparent problem.

### 7.7.3 Isotopic abundance ratios and the origin of heavy elements

Very metal-poor stars ($-3<[\mathrm{Fe/H}]<-2$) show large star-to-star scatter in abundances of elements heavier than Fe; a fraction of them show large enhancements of heavy elements (e.g., [Ba/Fe], [Eu/Fe]) caused by the r- or s-process, whereas abundances of heavy elements are generally low, (i.e., $[\mathrm{Sr,Ba/Fe}] <0$) particularly in stars with $[\mathrm{Fe/H}] < -3.5$, suggesting that r- and s-processes had little contributions to the bulk of stars formed in the very early Galaxy. The heavy elements Sr and Ba are detected in almost all stars with such low metallicity, even though the abundances are quite low ($[\mathrm{Sr,Ba/Fe}] \sim -1.5$). Detailed measurements for heavy elements in extremely metal-poor stars from very weak absorption features will provide a new constraint on the r- and s-processes in the very early Galaxy.

Any neutron-capture element with multiple isotopes that are produced in different amounts by the s- and r-processes can be used to assess the relative s- and r-process contributions to the stellar composition. The isotopic abundance of these elements are fundamental indicators of neutron-capture nucleosynthesis and can be directly compared to r-process and s-process predictions without the smearing effect of multiple isotopes (Mashonkina et al. 2003, 2006). The combination of Ba, Nd, Sm, and Eu isotopic fractions can provide more complete knowledge of the n-capture nucleosynthesis, constrain the conditions (e.g., temperature, neutron density, etc.) for the r-process, and determine the actual r-process path (Roederer et al. 2008; Cescutti et al. 2014). To reconstruct the evolutionary history of neutron-rich elements in the Galaxy, it is thus important to extend our study to the isotopic level. TMT/HROS will enable high-precision measurements of the isotopic ratios of Ba, Eu, and other heavy elements, along with light elements like Li, C and Mg, throughout the Milky Way, its globular cluster system, and its halo. High spectral resolution (R~100,000) and high sensitivity are essential for such studies.

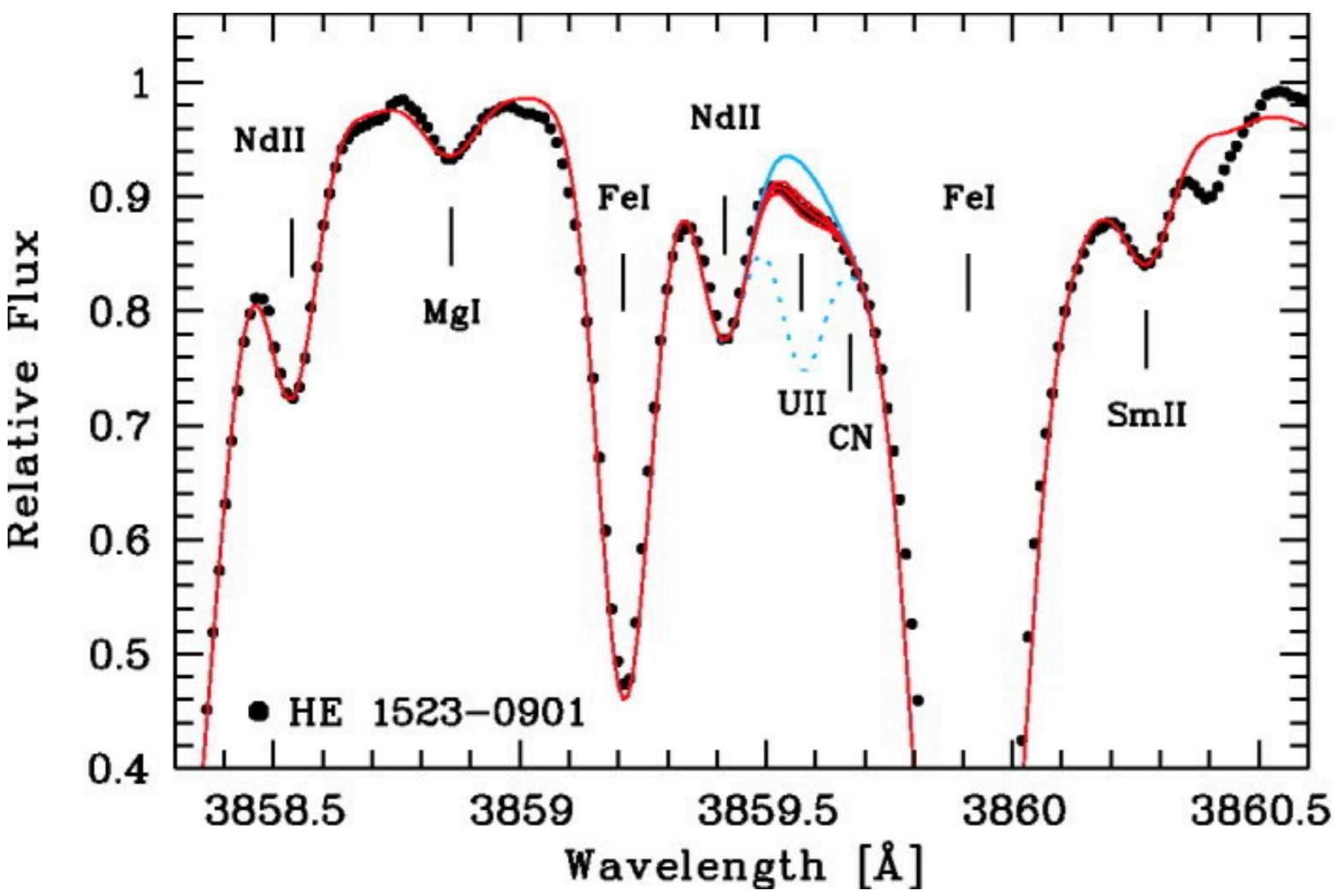

***Figure 7-11:*** *A. Spectral region around the U II line in HE 1523-0901 (filled dots). Overplotted are synthetic spectra with different U abundances. The dotted line corresponds to a scaled solar r-process U abundance present in the star if U were stable and did not decay (Source: Frebel et al. 2007).*

### 7.7.4 Cosmo-chronometry

The r-process elements thorium and uranium are radioactive and have long-lived isotopes, $^{232}$Th and $^{238}$U, with half-lives of 14 Gyr and 4.5 Gyr respectively. The spectral region around a U II feature is shown in figure 7-11. The half-lives of $^{232}$Th and $^{238}$U cover cosmic timescales that makes these elements suitable for age measurements.

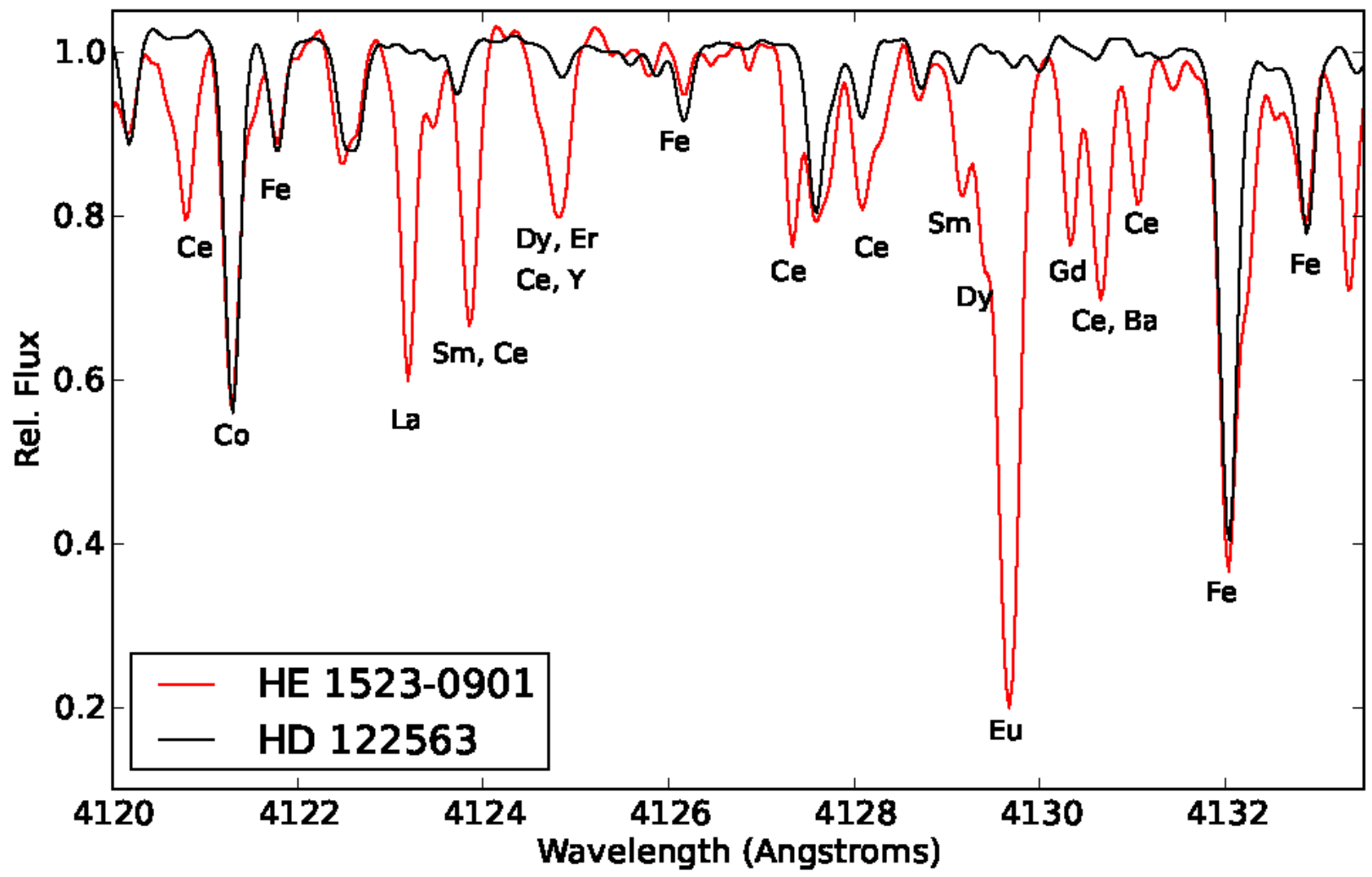

***Figure 7-12:*** *The spectra of the r-process-element-rich HE 1523-0901 (red) and the r-process-element-deficient star HD 122563 (black). These two stars have similar [Fe/H] values and atmospheric parameters (Source: Jacobson & Frebel 2014).*

Chronometers such as Th/Eu, U/Os, U/Eu, U/Th can be used to measure stellar ages. Knowing the abundances of Th, U, and Pb provides a self-consistency test for r-process calculations. These three abundances are intimately coupled to the conditions and environment of the r-process. Hence constraints on the different models yielding different abundance distributions can be obtained by explaining the stellar triumvirate of the Th, U, and Pb abundances. Such constraints lead to a better understanding of how and where r-process nucleosynthesis can occur.

Improved r-abundance calculations are crucial for reliably predicting the initial production ratios of Th/r, U/r and Th/U, which are implicitly necessary for more accurate age-dating of r-process enhanced stars. Based on seven chronometer abundance ratios, the age of HE 1523-001, the metal-poor star with the strongest enhancement in neutron-capture elements associated with the r-process that has been found thus far (figure 7-12), was found to be ~13 Gyr (Febel 2010).

The Th/U is relatively model independent and the best indicator of age. The measurement of U is, however, limited to most r-process-rich stars (Shah et al. 2023). The Pb abundance is usually measured using the Pb I line at 4057.81 Å. High resolution > 90,000 is required to separate U II lines and the Pb I line from other spectral features. With the TMT, lines due to these elements could be measured more easily in fainter old stars than can be done with existing 8–10 m telescopes, thus reducing the uncertainties in the age estimate.

## 7.8 Chemical evolution: The Milky Way, Local Group and nearby galaxies

### 7.8.1 Probing AGB stars' contribution to Galactic chemical enrichment

Low- to intermediate-mass stars (0.8–8 $M_\odot$) evolve through the asymptotic giant branch (AGB) phase; this phase although very short, is a major site of nucleosynthesis for production of carbon, nitrogen, fluorine and heavy elements such as barium and lead. Recurrent mixing episodes bring the freshly synthesized material from the core to the envelope and strong stellar winds expel these materials into the interstellar medium. Thus, the knowledge of AGB evolution and nucleosynthesis is vital to obtain an estimate of the contribution of low and intermediate mass stars to the chemical evolution of galaxies.

Measurement of isotopic ratios provides a completely new window into nucleosynthesis, galactic chemical evolution, mixing within stars, and stellar evolution. Such data, when available, significantly improve our understanding of nuclear processes in various astrophysical sites. However, as the isotopic shift is very small at optical wavelengths, very high spectral resolution and S/N (R ~90,000, S/N ~200) are required to measure isotopic ratios in stellar spectra.

There are three stable isotopes of magnesium (Mg). $^{24}$Mg is produced via routine burning of carbon in the interior of massive stars during their normal evolution and is subsequently injected into the ISM by SNII. The two heavier isotopes, $^{25}$Mg and $^{26}$Mg are secondary isotopes, believed to be produced primarily in intermediate mass AGB stars. In the young Galaxy, there was not enough time for AGB stars to contribute to element enrichment, while core collapse SNe started exploding shortly after massive stars first formed. One can thus use the ratios among the Mg isotopes to explore when the AGB stars began to contribute to the Galactic chemical inventory (Goswami & Prantzos 2000). Using R~100,000, high signal-to-noise Keck spectra of M71 giants, Melendez & Cohen (2009) demonstrated that M71 has two populations, one having weak C, N, normal O, Na, Mg, Al and a low ratio of $^{26}$Mg/Mg (~4%) consistent with models of galactic chemical evolution with no contribution from AGB stars. The Galactic halo could have been formed from the dissolution of globular clusters prior to their intermediate-mass stars reaching AGB. The second population has enhanced Na and Al accompanied by lower O and higher $^{26}$Mg/Mg (~8%), consistent with models that incorporate ejecta from AGB stars via normal stellar winds (figure 7-13; Fenner et al. 2003). Carlos et al. (2018) have derived a formation timescale of the Galactic Halo ~ 0.3 Gyr, from Mg isotopes in metal-poor dwarf stars considering only AGB star enrichment. When supernova enrichment is also considered, the upper limit for the timescale formation is found to be ~1.5 Gyr. Such studies can be significantly extended due to the expanding set of faint and old stars that can be observed with TMT/HROS, leading to an improved understanding of the contribution of the first AGBs to the Galactic chemical enrichment.

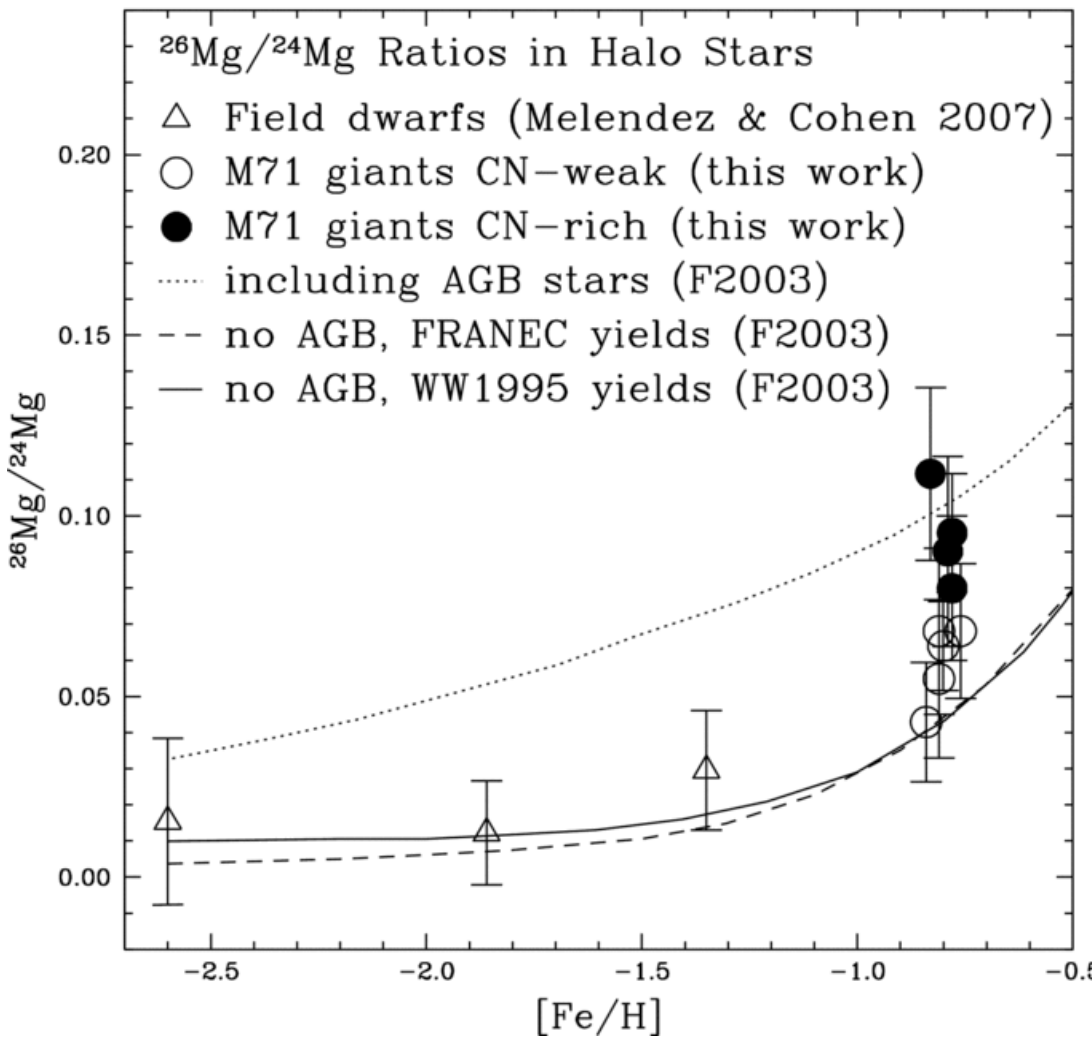

***Figure 7-13:*** *$^{26}Mg/^{24}Mg$ ratios in both field d (triangles; Melendez & Cohen 2007) and M71 g (circles) as a function of [Fe/H]. Chemical evo models by Fenner et al. (2003) including (dotted and excluding (solid and dashed lines) AGB sta shown. At the metallicity of M71 ([Fe/H] = -0. the isotopic ratios in the CN-weak stars circles) are explained by massive stars, bu CN-strong stars (filled circles) may have polluted by intermediate-mass AGB stars. (Sc Melendez & Cohen 2009).*

AGB stars are difficult to study in detail, with their cool extended dynamic atmospheres dominated by molecules. Binary mass transfer with an AGB star can lead to a variety of stellar types including CEMP-s stars, CH and barium stars, R-type stars, and R Coronae Borealis stars. The surface chemical composition of these stars can therefore be used as tracers of AGB nucleosynthesis. While the progeny of AGB stars, i.e., the post-AGB stars are rare, the warm stellar photosphere makes it possible to quantify photospheric abundances for a very wide range of elements from CNO up to some of the heaviest s-process elements including Pb that are produced during the AGB phase. Therefore, post-AGB stars can also provide direct and stringent constraints on the parameters governing stellar evolution and nucleosynthesis, especially during the uncertain AGB phase. To derive a clearer picture of element production in low- and intermediate-mass stars requires a systematic and comprehensive chemical abundance study of large samples of evolved stars in diverse metallicity environments, covering a wide range of luminosities (or initial masses). TMT will allow for an efficient exploitation of the Galactic and extragalactic sources.

The Carbon-Enhanced Metal-Poor (CEMP) stars with enhanced carbon abundance and no signatures of heavy neutron-capture elements, the so-called CEMP-no stars are the most chemically primitive objects presently known (Norris & Yong 2019; Yoon et al. 2020). In a recent study, Arentsen et al. (2019) suggested that the binary mass-transfer from extremely metal-poor AGB companion stars in the past might explain at least a few of the confirmed binary CEMP-no stars. If the binary-mass-transfer scenario holds true for the origin of CEMP-no stars, this will help to understand extremely metal-poor AGB stars and provide robust observational constraints to the AGB nucleosynthesis and mass-transfer models for metallicity [Fe/H]<–2.

### 7.8.2 Probing chemical evolution in Local Group dwarf galaxies

CDM simulations of the growth of galaxy structure suggest that halos of the galaxies like the Milky Way have accreted (and subsequently destroyed) 10s to 100s of small dwarf galaxies in the past 10 Gyr. However, detailed stellar atmosphere analyses of individual red giant branch (RGB) stars in current day Local Group dwarf galaxies show very little in common with the chemistry of stars in the Milky Way halo, disk, bulge, and moving groups (Venn et al. 2004; Navarro et al. 2004). Furthermore, the number of surviving dwarf galaxies in the immediate neighborhood of the Milky Way is far lower than predicted.

To resolve these discrepancies, it is necessary to investigate the environmental differences between the various dwarf galaxies with a detailed examination of the kinematics, metallicities, and abundance ratios of stars in dwarf galaxies beyond the Milky Way halo. With TMT/HROS, it will be possible to carry out detailed abundance analyses of stars at or just above the tip of the red giant branch throughout the Local Group.

For example, current generation high-resolution spectrographs on 8–10 m telescopes have been able to study tens of stars in Local Group dwarf spheroidal galaxies at V = 17–18 with exposure times up to 14 hours. Those systems lie as far away as 250 kpc. TMT/HROS will be able to obtain similar results for stars at V ~ 20 in more distant (~ 400–500 kpc) and hence more isolated systems. These optical measurements can be supported by K-band measurements from TMT/MODHIS, which can easily reach below the tip of the red giant branch throughout the Local Group (see Table 7-1). Such K-band measurements will be invaluable for CNO abundance measurements.

***Table 7-1:*** *Estimated limiting distances for spectroscopic observations of point sources*

| | $M_V$ (mag) | WFOS (Mpc) | HROS (Mpc) | MODHIS (Mpc) |
|---|---|---|---|---|
| Blue supergiant | −6.7 | 10.0 | 7.0 | 1.4 |
| Red supergiant | −5.9 | 7.0 | 3.5 | 3.0 |
| RGB Tip | −2.7 | 1.5 | 0.5 | 0.7 |
| NOTES: assumes 4 hour integrations at $\lambda/\Delta\lambda$ = 5000, 50 000, and 25 000 for WFOS, HROS, and MODHIS, respectively and $V - J = 2.45$ for the red stars and 0 for the blue. The corresponding central wavelengths are 0.5 μm, 0.5 μm, and 1.2 μm. The final SNR per spectral resolution element is 100, 100, and 60, respectively.<br>Original courtesy: J. Cohen (CIT), updated for MODHIS. | | | | |

With such data, the chemistry of the old stellar population in dwarf irregular galaxies (which tend to be more distant and more isolated satellites of the Milky Way) can be compared directly to the old populations of the dwarf spheroidals and Magellanic Clouds. These latter systems are within the dark matter halo of the Milky Way and are thought to have been interacting with the Galaxy over most of their lifetimes, likely affecting star formation histories and chemical evolution. A comparison with cleaner, more isolated dwarf irregulars will allow us to establish and characterize, for the first time, the effects of the environment on chemical evolution through the chemical similarities and differences in old populations. To compare dwarf irregular galaxies to the Galactic halo and other Local Group dwarf galaxies and test their impact on merging hypotheses of galaxy formation, detailed chemical analyses of their old RGB stars is required.

Analysis of spectra of young giant stars in nearby active star forming galaxies in the Local Group has become possible in the past decade due to a combination of observational data from 8–10 m telescopes and advances in model atmosphere techniques. The latter include the development of non-LTE hydrodynamic 3D extended stellar atmospheres, necessary to understand absorption features influenced by supersonic outflowing stellar winds. As shown in Table 7-1, observations with TMT will extend these studies to very large distances allowing the investigation of a variety of star formation environments and galaxy morphologies.

Sophisticated models of galactic chemical evolution are based on nucleosynthetic yields of various elements from each of the possible sources together with the amount of mass ejected from each of these sources, all as a function of stellar initial mass (and to a lesser extent stellar initial metal content). These are folded in with assumptions regarding gas flows within the galaxy and possible accretion of primordial material in addition to formation rates and an initial mass function for stars, both as a function of time. Prediction of trends of abundance ratios within the galaxy as a function of location requires knowledge of any dependencies of all of these inputs on position as well as an assumption regarding migration of stars within the galaxy.

***Figure 7-14:*** *The element ratio [Ca/Fe] as a function of [Fe/H] for a large sample of C-normal extremely-metal-poor halo stars (Cohen et al. 2013); colors of the filled points indicate the* $T_{eff}$ *of the star. The disk stars from Reddy et al. (2003) are indicated as small black dots concentrated near solar metallicity, large black dots indicate ratios for samples of stars within selected Milky Way globular clusters. The curves denote the predictions for Galactic chemical evolution by Prantzos (2007, 2014) adopting nucleosynthesis yields from various sources including Nomoto et al. (2006) or Woosley & Weaver (1995) for massive stars, Karakas (2010) for low mass stars, and Greggio (2005) for SNIa. (Source: J. Cohen).*

Figure 7-14 shows the trend for the alpha-element Ca based on detailed abundance analyses of large samples of stars in the disk and halo of the Milky Way. Figure 7-15 shows the same for the iron-peak element Co. The solid lines are predictions for Galactic chemical evolution by Prantzos (2007, 2014). The gentle rise towards lower Fe-metallicity is generally ascribed to the difference in characteristic timescale for SNII, the major production site for the alpha-elements, as compared to SNIa, where most Fe is produced. The former is much shorter than the latter, which may be as long as ~1 Gyr after the initiation of star formation.

Although creating a model of galactic chemical evolution requires many assumptions, it does work most of the time. In other words, such models can reproduce in a general way most observed trends of abundance ratios with time since the formation of the Milky Way. These models, with different inputs justified by observational data, also appear to work for the nearby galaxies, although our knowledge of their stellar populations and their chemical inventory is much more limited than is that for our own galaxy. TMT will make a major contribution to improve our knowledge of the chemical evolution of nearby galaxies. High dispersion spectroscopy using TMT/HROS and TMT/MODHIS will enable detailed abundance analyses with high accuracy for individual stars in the outer parts of the nearest groups of galaxies, including the M81 group, and for the brightest globular clusters well beyond the Virgo cluster. We will then finally obtain a full picture of the chemical evolution of the nearest galaxy groups out to the Virgo cluster and be able to compare it to that of the Milky Way.

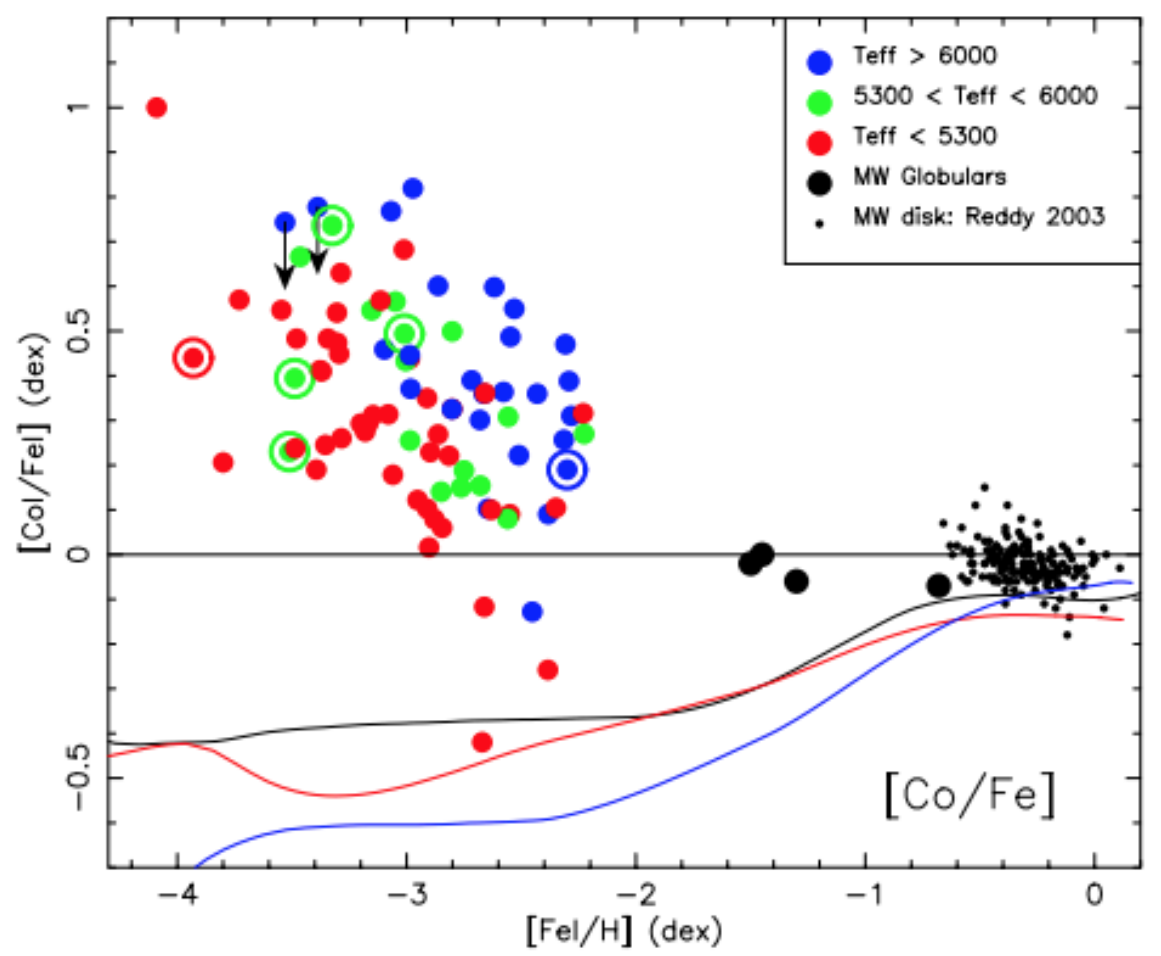

***Figure 7-15:*** *The same as figure 7-14 but for c*
*Note that in this case all the predicted models j*
*reproduce the trends seen among Milky Way halo*
*There must be a problem either in the nuclear yie*
*in the stellar abundance analysis. (Source: J. Cohe*

### 7.8.3 Abundance anomalies in ultra-faint dwarf galaxies

Ultra-faint dwarf galaxies (UFDs) are among the faintest ($L<10^5\ L_{\odot}$;; Bullock & Boylan-Kolchin 2017, Simon 2019) stellar systems in the Local Group, and hold the potential to link low-metallicity stars in currently bound systems to those populating the Milky Way halo (e.g., Kirby et al. 2008, Koposov et al. 2015b, Walker et al. 2016, Aguado et al. 2021, Chiti et al. 2023). TMT has the potential to revolutionize study of the chemical evolution of Local Group systems, including the unique cases of the UFDs in our own galaxy, and those of M31/M33.

The handful of members of UFDs in the Milky Way subjected to high dispersion abundance analysis shows striking anomalous enhancements in individual elements; they are often referred to as chemical oddballs (figure 7-16; Fulbright et al. 2004, Koch et al. 2013, Hansen et al. 2020, Tarumi et al. 2021). TMT/HROS will be able to obtain spectra of stars with r=20–21 at S/N~50 and R=15,000 or greater. WFOS will have multi-slit spectroscopy of an 8x3 arcmin field, sufficient to cover the size of UFDs in the Milky Way and dwarf galaxies in M31. For the TMT, the Andromeda/M33 system becomes available for study in much the same way that the Magellanic Clouds have been unveiled for 8 m class telescopes.

One of the striking accomplishments of our current era of 8–10 m class telescopes is that moderate to high resolution spectroscopy of stars as faint as V=18.5 has become routine. The multi-object fiber-fed FLAMES instrument at VLT has made high resolution investigation of hundreds of stars in the bulge and globular clusters a routine program. Keck with HIRES, and MIKE at Magellan have delivered high dispersion spectra of metal poor stars in the halo and dwarf spheroidals. Experiments that push the limits of high dispersion spectroscopy are underway to get spectra of thousands of giants in dwarfs and also to derive compositions from the integrated light spectra of globular clusters in M31 and beyond.

The TMT, with its high resolution spectrographs, places us on the threshold of investigating chemical evolution throughout the Local Group; and using globular cluster integrated light, throughout the Local Volume. Such work can address the causes for the multiple subclasses of metal poor stars (e.g., carbon enhanced) and chemical oddball stars, and we can determine whether the picture of chemical evolution seen in our Milky Way halo extends to the Local Group, see figure 7-16. While recent years have produced tantalizing results for UFDs orbiting within the Milky Way (Walker et al. 2016, Hansen et al. 2020, Aguado et al. 2021, Tarumi et al. 2021, Chiti et al. 2023), HROS and WFOS have the potential to finally push beyond our own halo and gather properties of these important systems located throughout the Local Group.

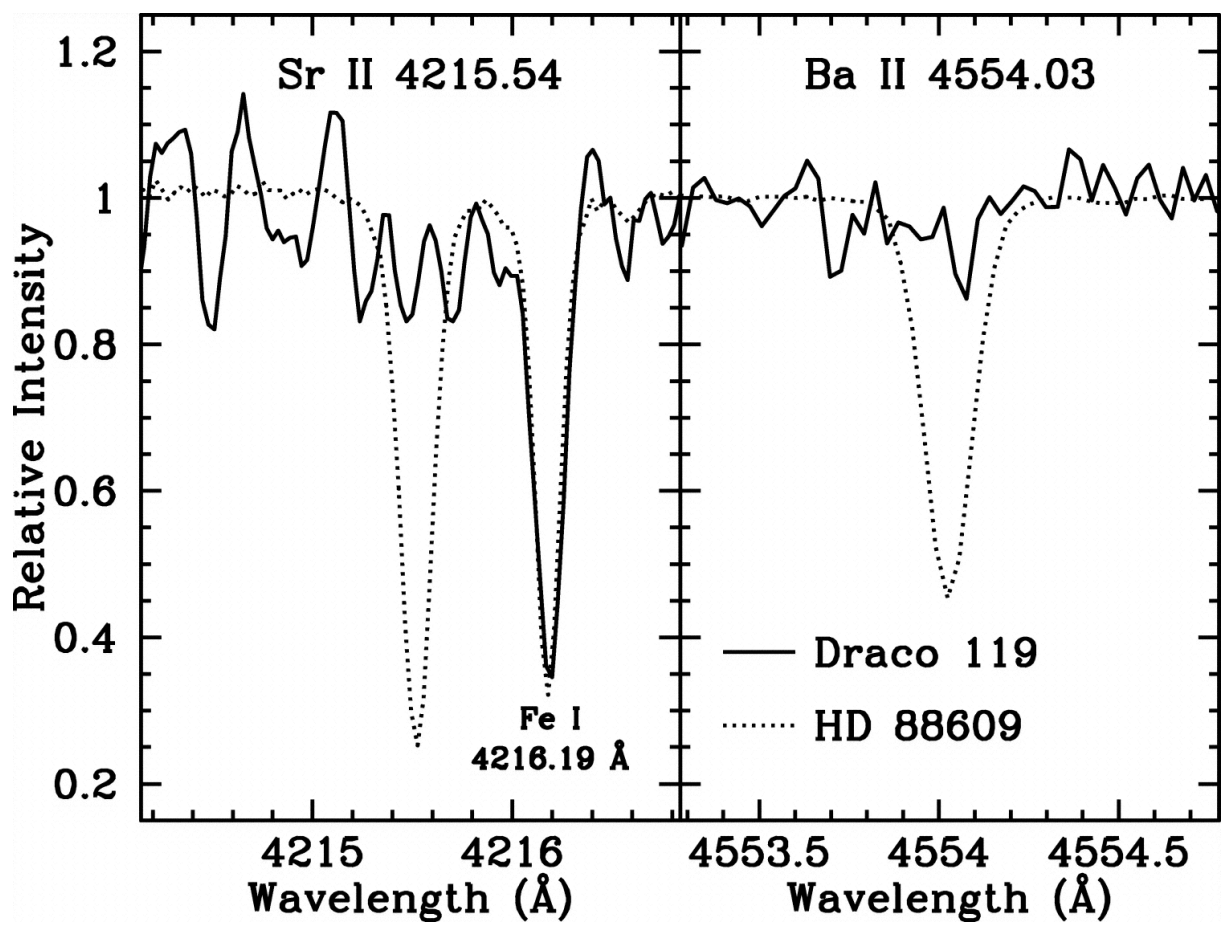

***Figure 7-16:*** *Neutron-capture elements are dej in a red giant (D119) in the Draco Dwarf Sphe. galaxy (Fe/H = −2.95) in comparison with HD 8 which has similar metallicity (Fulbright, Ric Castro 2004). The Hercules dwarf spheroidal g shows similar striking deficiencies (Koch et al. 2 TMT/WFOS will be capable of making measure on giants similar to these even as distant a Andromeda galaxy.*

Some cherished ideas are under challenge. Is the knee in the [alpha/Fe] vs [Fe/H] trend, where elemental trends move toward Solar scaled composition, due to the onset of iron produced in Type I SNe? Or is the knee arising from the natural metallicity dependent yields of massive star SNe? Is our now roughly 40 year old depiction of chemical evolution therefore in need of major restatement? What is the full range of primordial chemical enhancements that are observable in various stellar systems? Although the broad brush description of chemical evolution and nucleosynthesis is well known, it is clear that many aspects of nucleosynthesis can be improved in detail, and our best hope for advancing theoretical work is through additional observations. Large improvements in our modeling of stellar atmospheres continue to occur via non-LTE 3D models over large grids. TMT will provide the observations to test these models by revealing the contributions of the *s*- and *r*-processes to the production of *n*-capture elements (e.g., Sr, Ba), and the presence of the [alpha/Fe]−[Fe/H] knee in their abundance patterns.

### 7.8.4 Resolved stellar populations as tracers of galaxy evolution

Ground-based resolved stellar population studies of galaxies necessarily favor the Local Group and its immediate vicinity. While a range of morphological types are present, the statistics are poor and some notable galactic types are entirely absent (in particular, massive elliptical galaxies). The resolving power of the Hubble Space Telescope pushes such studies out to distances of several megaparsecs, improving statistics and sampling a broader range of galaxy type. But even here such studies are limited, and importantly, the galaxies still sample the rather benign environment of our local cosmic neighborhood. Only with the collecting power and exquisite spatial resolution of TMT can resolve stellar population studies come of age and reveal the evolutionary histories of galaxies across a very wide range of luminosities and morphologies, spanning a broad set of environments, from isolated systems, through loose groups, to groups and clusters.

Studies of the integrated light of galaxies necessarily rely on stellar population synthesis to reproduce the broad-band spectral energy distribution of the galaxy. The galaxy's light represents the sum-total of the baryonic evolution of the system over cosmic time and is contributed to by stars of varying mass and of varying age, the proportions of which depend upon the star formation history specific to that galaxy. In this respect, the detailed star formation history is one of the most significant uncertainties in these studies.

By sampling the resolved stellar populations of a galaxy and decomposing the color-magnitude diagram into its constituent stellar types, we can obtain temporally resolved information on the history of star formation, with additional information such as metallicity available for certain stellar types and/or under certain assumptions, e.g., using the red clump or red giant branch color. Figure 7-17 illustrates how the stellar color-magnitude diagram (CMD) of a complex stellar population can encode essential physical information (see Gallart et al., 2005 and Tolstoy et al., 2009, for reviews).

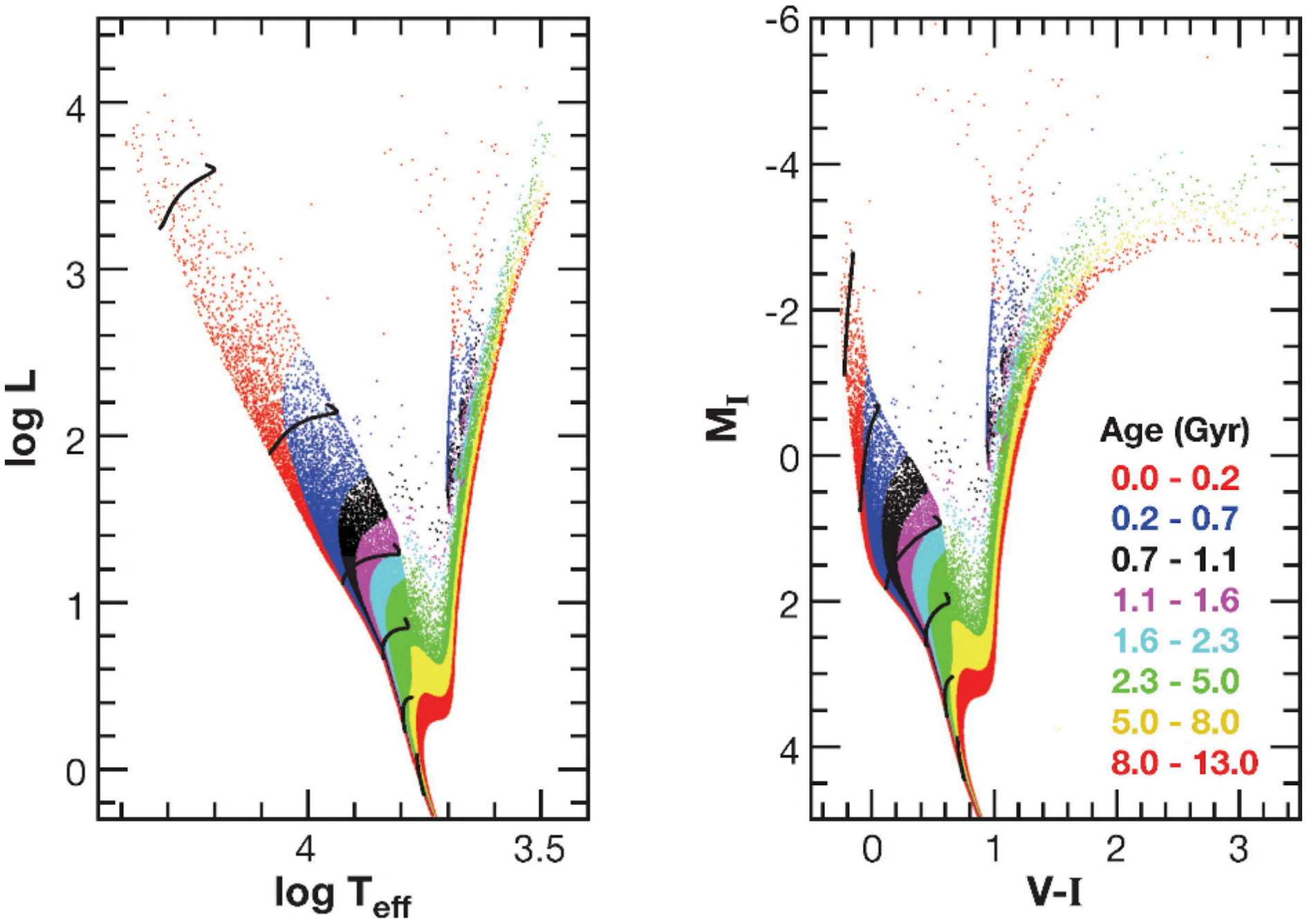

***Figure 7-17:*** *Figure from Gallart et al. (2005), Hertzsprung-Russell diagram (left) and color-magnitude diagram (right) of a synthetic complex stellar population, illustrating how different regions of the CMD are occupied by different ages of stars, the precise distribution of which allows a (partial) reconstruction of the overall star formation history of the system.*

Figure 7-18 shows cumulative star formation histories of individual galaxies for various morphological types. Such analyses can be conducted for any subset of galaxies split by any physical parameter, and it is here that TMT can enable a revolution with its increased sensitivity and angular resolution compared to current facilities.

### 7.8.5 Reconstructing the star formation histories of nearby galaxies

Reconstructing the star formation history for a given stellar system by analyzing its color-magnitude diagram is a fundamental tool for understanding the system's age and chemical composition that has been calibrated by decades of observational and theoretical work. However, progress with obtaining good CMDs within more massive galaxies (or their denser sub-components) has been limited by stellar crowding. Recently, higher spatial resolution observations have become possible using near-IR adaptive optics systems on ground-based 8–10 m class telescopes. By scaling up to a 30 m aperture and implementing AO systems with improved performance, TMT will enable another giant step forward.

JWST is starting to provide deep CMDs for galaxies within the Local Group (Wiesz et al. 2023, McQuinn et al. 2024), but these are primarily focussed on low-mass galaxies with low stellar surface densities. Figure 7-19 shows a crowding limits comparison for $\Sigma_K$ = 19 mag arcsec$^{-2}$ observed with near-IR AO-corrected 8 m and 30 m telescopes, as well as HST working in the optical at $\Sigma_V$ = 22 mag arcsec$^{-2}$. The calculations demonstrate that TMT will be able to resolve individual stars in regions with $\Sigma_K$ = 19 mag arcsec$^{-2}$ in galaxies as far away as 15 Mpc (e.g., the Virgo cluster). Of course, as distance increases, the required photometric accuracy becomes more difficult and only feasible for the intrinsically brightest stars.

A long sought goal is to construct a deep CMD for a normal elliptical galaxy. Several candidate targets lie within 10–15 Mpc, including (e.g.) NGC 3379 (d ~ 11 Mpc). While JWST will provide the sensitivity for such work, the limited resolution compared to AO-assisted 30-m class telescopes will be insufficient to overcome challenges caused by crowding. Table 7-2 provides crowding limits at three galactocentric radii for NGC 3379 observed with TMT. The crowding limit is defined to be the magnitude at which photometric errors due to crowding reach 20%, corresponding to 50% completeness. At 1 $R_e$, only the brightest RGB and AGB stars will be accessible before the crowding limit is reached. However, at 3 $R_e$ (perhaps the lowest practical surface brightness given required exposure times), it will be possible to reach the horizontal branch before reaching the crowding limit. This can be compared to actual HST/NICMOS observations at 3 $R_e$ that only reach ~ 1 mag below the tip of the red giant branch before significant crowding limits photometric accuracy (Gregg et al. 2002), although JWST can clearly do better than HST and we look forward to upcoming results.

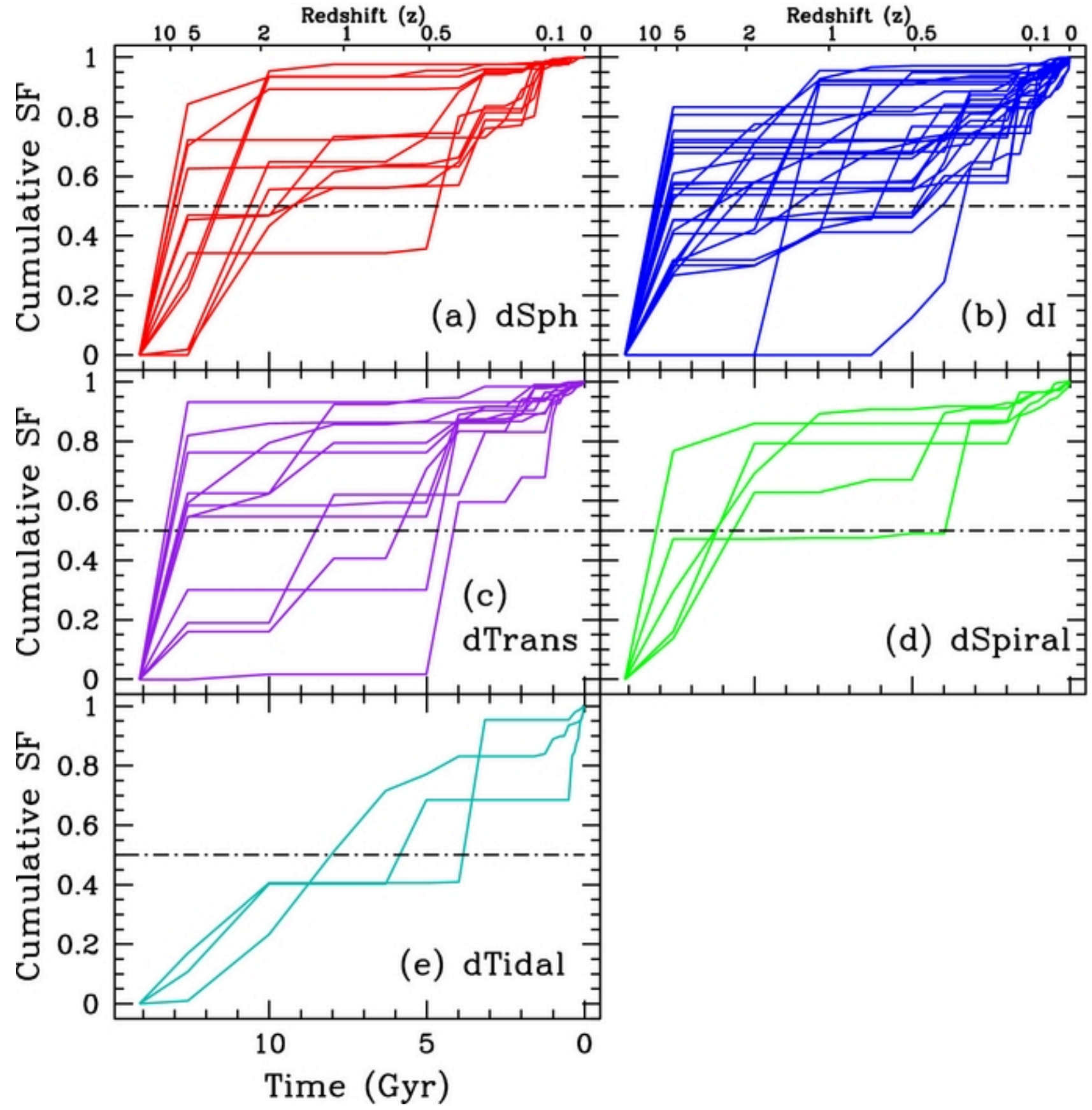

***Figure 7-18:*** *Cumulative star formation histories of 60 dwarf galaxies, split by morphological types, as derived by Weisz et al (2011). Statistical comparison of the star formation histories for each type reveals key similarities and differences, and the epochs at which they occur.*

***Table 7-2:*** *Crowding limits for NGC 3379 imaged at the K band diffraction limit of TMT. $R_c$ is the radius of the galaxy core and $R_{tot}$ is the radius to the outermost stars.*

| Name | r (arcsec) | $\Sigma_K$ (mag arcsec$^{-2}$) | $K_{lim}$ | Time (secs) |
|---|---|---|---|---|
| $R_e$ | 30 | 17.0 | 25.7 | 282 |
| $3R_e$ | 90 | 19.3 | 28.5 | 47200 |
| $R_{tot}$ | 190 | 22.5 | 31.6 | ∞ |
| K (1 hour) | — | — | 27.9 | 3600 |
| Notes: the second column (r) is galactocentric distance, the third ($\Sigma_K$) is assumed surface brightness at that distance, the fourth ($K_{lim}$) is the point source magnitude at the crowding limit (as defined in the text), and the last column (t) is the exposure time to reach these limits. For reference, the last row shows the point source limiting magnitude for t = 3600 secs in the absence of crowding | | | | |

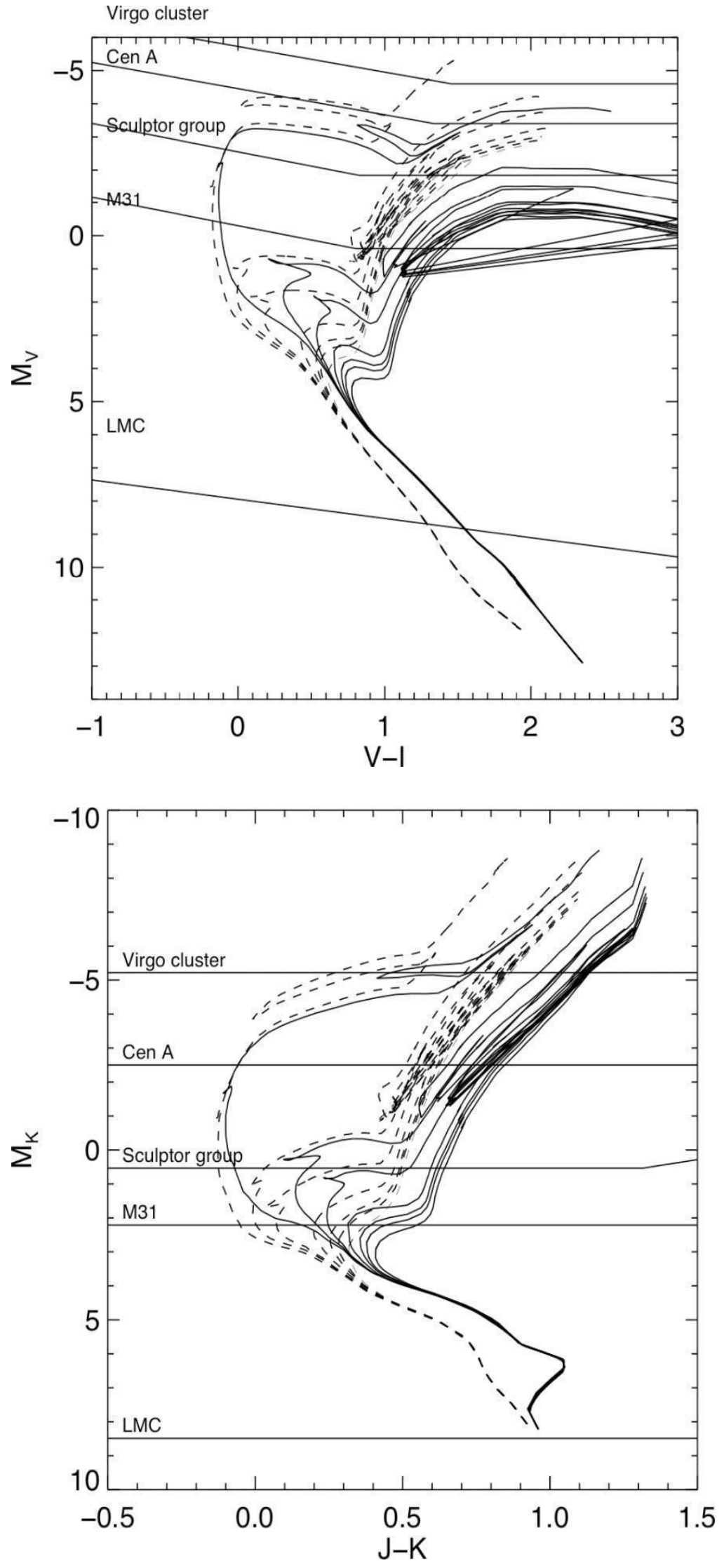

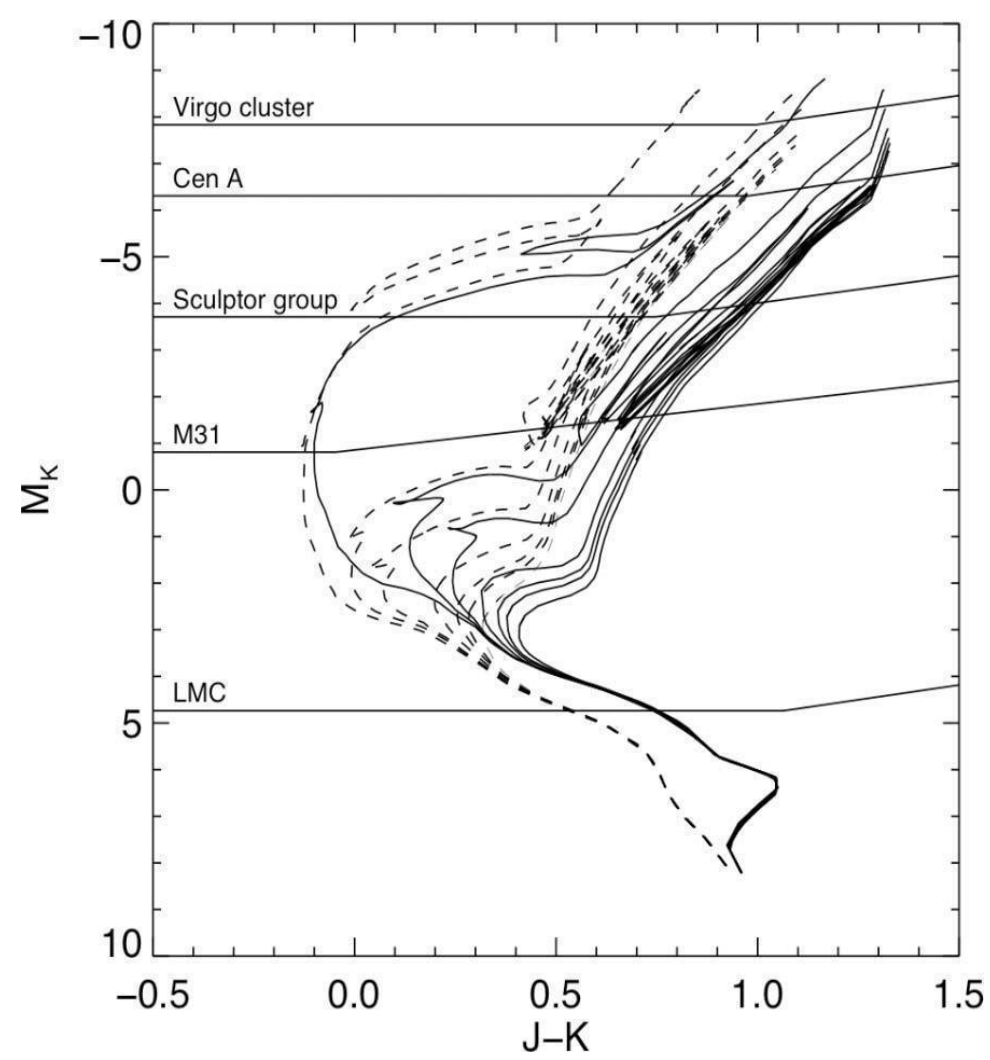

***Figure 7-19:*** *Magnitudes at which crowding limits photo to 10% accuracy in regions with $\Sigma_K$ = 19 and $\Sigma_V$ = 22 arcsec$^{-2}$at the distances indicated are shown as horizonta on top of Girardi et al. (2000) isochrones of various age metallicities. Top left: HST in the optical. Top right: G North + NIRI/Altair. Bottom left: TMT in the near-IR.*

These crowding limits were calculated analytically by comparing the contribution of surface brightness fluctuations on the scale of a telescope resolution element to that of a star of a given magnitude (Olsen, Blum, & Rigaut 2003). The ratio depends on the distance, surface brightness, and luminosity function of the stellar population as well as the photometric accuracy required for a given stellar magnitude

### 7.8.6 Probing LSB and BCD galaxies: star formation, chemical evolution, and dark matter

The class of blue compact dwarf (BCD) galaxies (Thuan & Martin 1981) exhibits recent star formation in the current epoch and appears to have survived merging or some other perturbation that would have led to an active past star-formation history. These galaxies appear to have produced stars intermittently over the Hubble time. The BCD population is somewhat diverse and so is their evolution. The amount of dark matter in BCD galaxies is of interest in testing various hypotheses regarding the stability of these HI-rich systems against perturbations that can trigger star formation. Studying the dynamics of BCD galaxies through high spatial resolution spectroscopy of individual stars and HII regions will be invaluable in this context. Integral field spectroscopy to date (e.g., James et al. 2010, Monreal-Ibero et al 2023) is limited to a handful of nearby BCDs, painting a picture of complex spatial and velocity structures attributable to both merging and disruption. Since BCDs by definition are compact objects (scale lengths < 4 kpc in most cases), the high spatial resolution offered by adaptive optics is crucial to derive their velocity fields, and the sensitivity of TMT to reach larger distances.

BCDs have complex morphologies and some show low surface brightness features (Cairós et al. 2001). The related class of low surface brightness (LSB) galaxies, have gaseous disks that are very stable against perturbations, and show very sparse star formation at the current epoch (Kaviraj 2020). It appears that some of them could be BCD galaxies in their quiescent phase, but others are comparable to normal galaxies in size and mass (O'Neil et al. 2023). More detailed study of LSBs and BCDs is required to place them in proper perspective in the picture of galaxy evolution. Large imaging surveys are yielding samples of both galaxy types (e.g., Tanoglidis et al., 2021); measuring these galaxies' distances through stellar population methods will require the sensitivity of TMT.

Some of the most metal poor galaxies that we know are dwarf or irregular LSB galaxies. The low metal content of these systems suggests that they are ideal targets to search for traces of population III stars. However, LSB disks are very faint and stellar population analyses of their integrated light require high signal to noise spectra that is not possible to obtain with current facilities. The high sensitivity of TMT will help us observe, model, and understand the chemical evolution of these low luminosity disks for comparison with models such as those of Cao et al. (2023). TMT IRMOS-IFU integral-field spectroscopy will be able to map SFR to a much higher spatial resolution than current Hα imaging. Although LSB galaxies are gas rich, their dark halos lead to low star formation rates and hence overall lower luminosities. The low luminosity and low stellar surface density makes the disks difficult to study, especially the galaxies at larger distances and beyond our local universe. The high sensitivity of the TMT will enable the detection of the disks of LSB galaxies to determine their properties, as well as sensitive searches to pin down their poorly determined incidence of supermassive black holes (Hodges-Kluck et al. 2020).

### 7.8.7 Resolving extreme star formation environments in luminous infrared galaxies at low redshift

Luminous infrared galaxies (LIRGs; LIR > $10^{11}$ $L_\odot$) and ultra luminous infrared galaxies (ULIRGs; LIR > $10^{12}$ $L_\odot$) are some of the most important targets for revealing the star formation history in the universe. They dominated the star formation rate density (e.g., Magnelli et al., 2013) at 1 < z < 3, when most galaxy growth took place in dusty environments. The spatial extent of star formation in high-z (U)LIRGs more resembles local LIRGs than local ULIRGs. With diffraction limited resolution (λ/D = 0.008″ at 1.2 μm), TMT/IRIS will reveal the morphology and gas conditions in low-z LIRGs with a physical scale of ~15 pc at 1.2 μm at the luminosity distance ($d_L$) of ~400 Mpc (or z = 0.088, the largest distance for the IRAS Revised Bright Galaxy Sample; Sanders et al., 2003). This scale is near the minimum size of typical HII regions measured in nearby galaxies with Paα emission (Liu et al., 2013). We will be able to resolve individual sites of current star formation, reveal gas properties associated with the star formation (ionization, metallicity), and explore feedback processes (outflow velocity, outflow rates) that affect conditions of the interstellar medium. TMT high spectral resolution observations of individual HII regions in each galaxy will allow us to explore the building blocks of recent massive star formation to better understand star formation in extreme environments. In addition, where an AGN exists, the high resolving power of TMT will be able to distinguish the emission from the AGN and nuclear star formation. With a planned spectral resolution of R = 4000–8000 for IRIS, we will be able to resolve diagnostic emission lines, and together with a spatial resolution that is better than HST or JWST, we will be able to disentangle heating from young stars, shocks and AGN in hundreds of starburst galaxies.

High resolution TMT observations of molecular hydrogen, $H_2$, together with atomic hydrogen emission lines will provide new insights into the molecular gas temperature and density structure of photo-dissociation regions, the primary heating and cooling processes, and the mechanism of $H_2$ formation (e.g., Sugai et al., 1997) on a scale of each individual HII region in the local universe. In the future, TMT/MICHI ground-based mid-infrared observations will provide synergy with space-based infrared observatories. TMT/MICHI yields 15 times higher sensitivity (~0.1 mJy with 5σ detection in a 1 hour integration in the 10 μm (N-band) imaging) and 4.5 times better spatial resolution (0.07″at 10 μm, corresponding to 33 pc at $d_L$=100 Mpc) than the current ground-based 8 m class telescopes. Although the sensitivity is not as high as that of JWST, the 4.5 times better spatial resolution of TMT/MICHI will allow us to investigate physical processes in deeply dust-embedded nuclei of low-z LIRGS. Using the 7–25 μm wavelength coverage of MICHI, a suite of mid-infrared features can be accessed to diagnose the physical and chemical conditions of the gas and dust in galaxies. From the ground, we expect to readily detect at least the $Ar^{+2}$, $S^{+3}$, $Ne^{+}$, and $S^{+2}$ atomic fine structure emission lines, the $H_2$ molecular emission lines from molecular gas with a few hundred K, the polycyclic aromatic hydrocarbon emission bands, and the 9.7 μm silicate features for local objects. Although all these emission lines were and will be accessible from Spitzer/IRS, JWST/MIRI, and SPICA/MCS, in the local universe, the 0.07″ spatial resolution at 10μm of TMT/MICHI will resolve the obscured dense clouds in the innermost regions of nuclei and star-forming regions where star formation is most intensive.

Spatially resolved line emission obtained via MICHI will provide detailed gas properties on each of the dusty star-forming sites. The line flux ratios of [SIV]10.51 μm/[SIII]18.71 μm and [SIV]10.51 μm/[ArIII]8.99 μm are powerful at setting constraints on the hardness of the ionizing radiation, and the [SIII]18.71 μm/[NeII]12.81 μm ratio is sensitive to the electron density (Snijders et al. 2007). Because the hardness of the radiation field is sensitive to the stellar population of galaxies, we will be able to determine the obscured stellar population in dusty galaxies. A combination of two sets of emission line ratios can be an even more powerful tool for diagnosing galaxy properties (e.g., Groves et al., 2008; Hao et al., 2009; Inami et al., 2013). With the emission lines that can be observed from the ground, comparisons of the emission line ratio diagram [SIV]10.51 μm/[ArIII]8.99 μm vs. [SIII]18.71 μm/[NeII]12.81 μm to those predicted from models can constrain age, metallicity, density, and ionization parameter with model assumptions of star formation history, IMF, stellar atmosphere models, and stellar evolutionary tracks (see figures 11 and 15 of Snijders et al., 2007). MICHI with TMT will elicit different views of the physical properties and mechanisms of star formation in regions deeply embedded in dust, which we have not yet been able to examine with optical and near-infrared facilities due to the very high extinction.

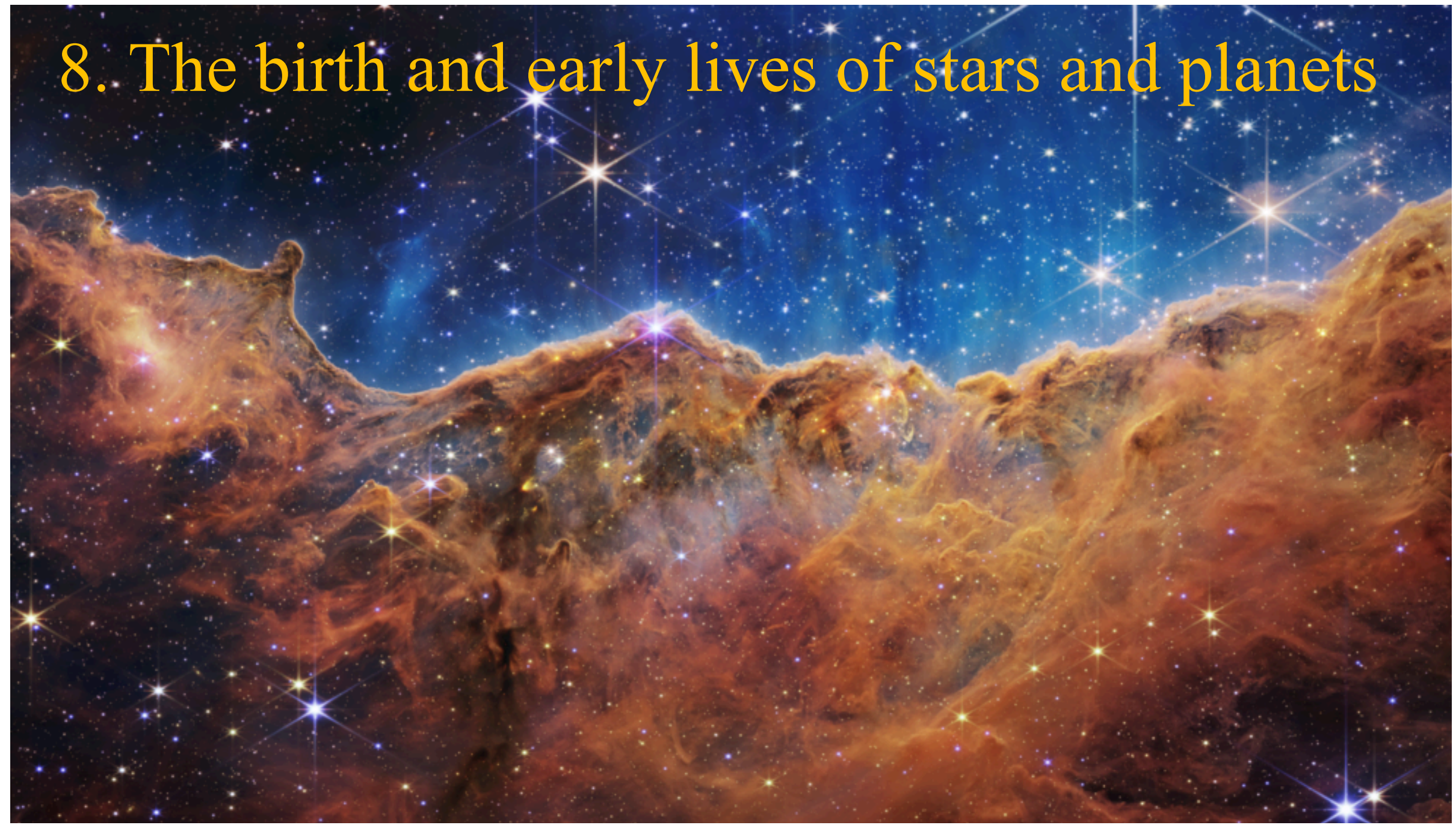

*This landscape of "mountains" and "valleys" speckled with glittering stars reveals the edge of a nearby, young, star-forming region called NGC 3324 in the Carina Nebula. Captured in infrared light by NASA's new James Webb Space Telescope, this image reveals for the first time previously invisible areas of star birth. Called the Cosmic Cliffs, Webb's seemingly three-dimensional picture looks like craggy mountains on a moonlit evening. In reality, it is the edge of the giant, gaseous cavity within NGC 3324, and the tallest "peaks" in this image are about 7 light-years high. The cavernous area has been carved from the nebula by the intense ultraviolet radiation and stellar winds from extremely massive, hot, young stars located in the center of the bubble, above the area shown in this image. Image credits: NASA, ESA, CSA, and STScI.*

---

TMT studies of the birth and early lives of stars and planets will help us answer a number of chapter 2's big questions starting from Q4-*How do stars and planets form?,* extending to Q5-*What is the nature of extrasolar planets?,* Q6-*Is there life elsewhere in the universe?,* and Q2-*When did the first galaxies form and how did they evolve?* By determining how star formation processes depend on the environment and metallicity, future TMT observations will improve our understanding of stellar physics and galactic and stellar evolution. JWST is observing characteristics of some protoplanetary disks, TMT's angular and spectral resolution are needed to understand the physical processes going on within these disks.The science cases outlined here all rely on TMT's extreme sensitivity and its AO corrected angular resolution. These fields greatly benefit from MIR observations and require the spectral resolutions R~10,000–100,000, well met by IRIS, MODHIS, and MICHI. Understanding the environments of planet formation will benefit from coronagraphic capability, like the high dispersion coronagraphic capability of MODHIS and that to come from PSI.

---

**Contributors:** Babar Ali (IPAC), Adam J. Burgasser (UCSD), Richard de Grijs (KIAA Peking University), Ruobing Dong (University of Victoria), Misato Fukagawa (NAOJ), Carol, A. Grady (Eureka Scientific, NASA GSFC), Priya Hasan (Maulana Azad National Urdu University), Gregory J. Herczeg (KIAA Peking University), Mitsuhiko Honda (Okayama University of Science), Quinn M. Konopacky (Dunlap Institute), Di Li (NAOC), Jessica R. Lu (University of Hawaii IfA), Takayuki Muto (Kogakuin University), Joan R. Najita (NOAO), Devendra K. Ojha (TIFR), Deborah L. Padgett (JPL), Klaus M. Pontoppidan (STScI), Matthew J. Richter (UC Davis), Jonathan C. Tan (University of Virginia), Chikako Yasui (NAOJ)

---

# 8 The birth and early lives of stars and planets

Star formation is the principal driver of galaxy evolution and chemical enrichment in the universe, which in turn affects the process of star birth. Additionally, it also provides the sites of planet formation and the development of life. Therefore, understanding *how stars and planets form* has long been one of the central problems in astrophysics. In fact, significant progress has been seen especially in the last decades, but the detailed processes are still quite elusive. ALMA, JWST, and advanced instruments such as SPHERE, MUSE, and SCExAOon ground-based telescopes are now yielding further scientific advances prior to the TMT's first light. However, TMT is needed to provide unprecedented angular resolution with very high sensitivity in near- to mid-infrared wavelengths to answer key current questions that limit our understanding of these processes in the innermost disk, where solar systems such as our own would form. The rest of this chapter discusses those questions and the ways in which TMT will be central to answering them.

## 8.1 Star formation

Star formation is a fundamental astrophysical process, yet we still lack a quantitative and predictive theory for how stars and clusters form (see reviews by Chevance et al. 2022 and Hacar et al. 2022).. Among the key ingredients for building such a theory are observations of young star clusters over a wide range of environments that can be used as benchmarks to test star formation models. Young clusters serve as laboratories with controlled conditions (same age, metallicity, distance, initial conditions) and many stars to sample the distributions of stellar masses, kinematics, and multiplicities that arise during the star formation process, with major advances in star formation in our local neighborhood revealed by Gaia (see reviews by Zucker et al. 2022 and Wright et al. 2022). The requisite range of initial conditions in metallicity, external pressure, cloud mass/density, and environment can only be found outside of our own galaxy, where studying clusters in detail has been extremely challenging. Fortunately, TMT's spatial resolution and sensitivity will allow individual young stars to be spatially resolved, even in the largest star clusters, within a wide range of environments throughout the Local Group galaxies and beyond. TMT observations of young clusters, which are the output of the star formation process, will complement ALMA studies of the gas clumps and cores that are the input to the star-formation process.

In the section below, we highlight TMT science cases for measuring the initial mass function (IMF), internal kinematics, and multiplicity in young clusters, followed by the sections focusing on high- and low-mass ends of the IMF. These observations will help constrain star formation theories such that reliable predictions can be incorporated into models of planet formation, stellar evolution, galaxy formation and evolution, and cosmology.

## 8.2 Developing a predictive theory of star formation

### 8.2.1 IMF vs. environment

The relationship between the interstellar medium (ISM) and the resulting stars is, arguably, the most fundamental objective in star-formation research. The Kennicutt-Schmidt (K-S) law establishes a correlation between the integrated rate of star formation and the surface density of dense gas (Schmidt 1959; Kennicutt 1998) in extragalactic sources. Heiderman et al. (2010), among others, investigated this relationship for clouds in the Milky Way and found significant (by a factor ~2) discrepancies in their measured K-S slopes compared to those of external galaxies. Other smaller-scale ($\lesssim$100 pc) measurements also tend to show deviations from the K-S law or larger scatter, which may be explained by high angular resolution observations distinguishing local, inhomogeneous distributions of dense gas and stars. The problem is that the current spatial resolution is not small enough to explore the cause of the observed behavior. Resolving stellar populations with even higher spatial resolution thereby provide a more accurate yield of star formation and knowledge of the *local* star-forming environment (e.g, metallicity, radiation field, cluster density, gas density distribution) will lead to an understanding of the origin of the K-S law, and ultimately, how physical processes in the ISM control star formation.

The TMT's combination of high spatial resolution and sensitivity will offer unprecedented capabilities for studying the IMF in distant star-forming regions not limited to our galaxy but also in the Local Group galaxies (see also chapter 7). This is the area where TMT with AO has a real niche in the field of star formation; HST and its successors including JWST are less competitive due to their significantly smaller mirror diameters.

***Table 8-1:*** *TMT limiting K-band magnitudes and corresponding lower mass limits in Arches-like clusters*

| | **Limiting *K* magnitude** | | **Limiting Mass ($M_\odot$)** | |
|---|---|---|---|---|
| **Radius (*R*e)** | **M33** | **M82** | **M33** | **M82** |
| 0.5 | 17 | <19.8 | … | … |
| 1.0 | 18.9 | <19.8 | 65 | … |
| 2.0 | 22.3 | 20 | 3 | … |
| 5.0 | 27.5 | 23.9 | 1.1 | 32 |

*Notes:* for illustrative purposes, photometric crowding and photon statistic limits have been computed for target environments in the M33 and M82 using the radial profiles of the Arches and R136 clusters, coupled with the crowding limit algorithm given by Olsen et al. (2003). The K-band magnitudes are for crowding limited photometry to 10% accuracy. The input luminosity function used for these calculations is a hybrid based on measurements in the Arches cluster (Blum et al. 2001) for the high-mass stars ($\geq 2\ M_\odot$) and measurements in the Trapezium by Hillenbrand & Carpenter (2000) for the low-mass stars ($\leq 3\ M_\odot$). The Arches radial profile is a re-fit to HST data by Blum et al. (2001).

Diffraction-limited TMT observations of rich, dense clusters in the distant Milky Way and local universe in the near-infrared will enable determination of the shape of the IMF over the entire range of masses from ~100 $M_\odot$ to below 1 $M_\odot$. TMT will allow us to probe the brown-dwarf, and even the planetary-mass regime in young clusters at distances of several kpc. It will also enable investigations of the low-mass ($\lesssim 1\ M_\odot$) stellar regime in a representative slice of the universe, including the nearest large spiral galaxies (M31 and M33; Table 8-1) and the very low-metallicity environments provided by Local Group galaxies such as NGC 6822 and IC 10. results for the low-mass IMF in the Magellanic Clouds suggest that metallicity-related differences that could affect the shape of the stellar mass distribution require significantly lower metallicities to become apparent (e.g., da Rio et al. 2009). The stellar density in a given cluster, which affects the feedback through winds, radiation field, and supernova, can be measured from such resolved observations, enabling us to establish the relationship with emerging stellar mass.

Based on the expected performance of the IRIS IFU, it will also be possible to obtain spectra of stars with masses significantly smaller than 1 $M_\odot$ at the Galactic Center, while the mass limit lies at ~10 $M_\odot$ with current 10-m class telescopes (Lu et al. 2013). With TMT, the similar detection limit as currently obtained in the center of the Milky Way can be achieved in the Local Group (figure 8-1).

ALMA, on the other hand, will provide superb spatial resolution at complementary, dust-penetrating sub-millimeter wavelengths. This is a fortunate situation, as it will allow us to probe right into the core of the most active, dust-enshrouded star-forming regions, covering the entire useful wavelength range from the infrared to millimeter waves. Multi-wavelength observations will help solve the key outstanding issue of how star formation occurs, proceeds, and is triggered, as well as the importance of the interaction between the newly-born stars and their environments. We will be able to study the early evolution and the transformation from the youngest star-forming cluster-like regions to more mature, partially-virialized systems.

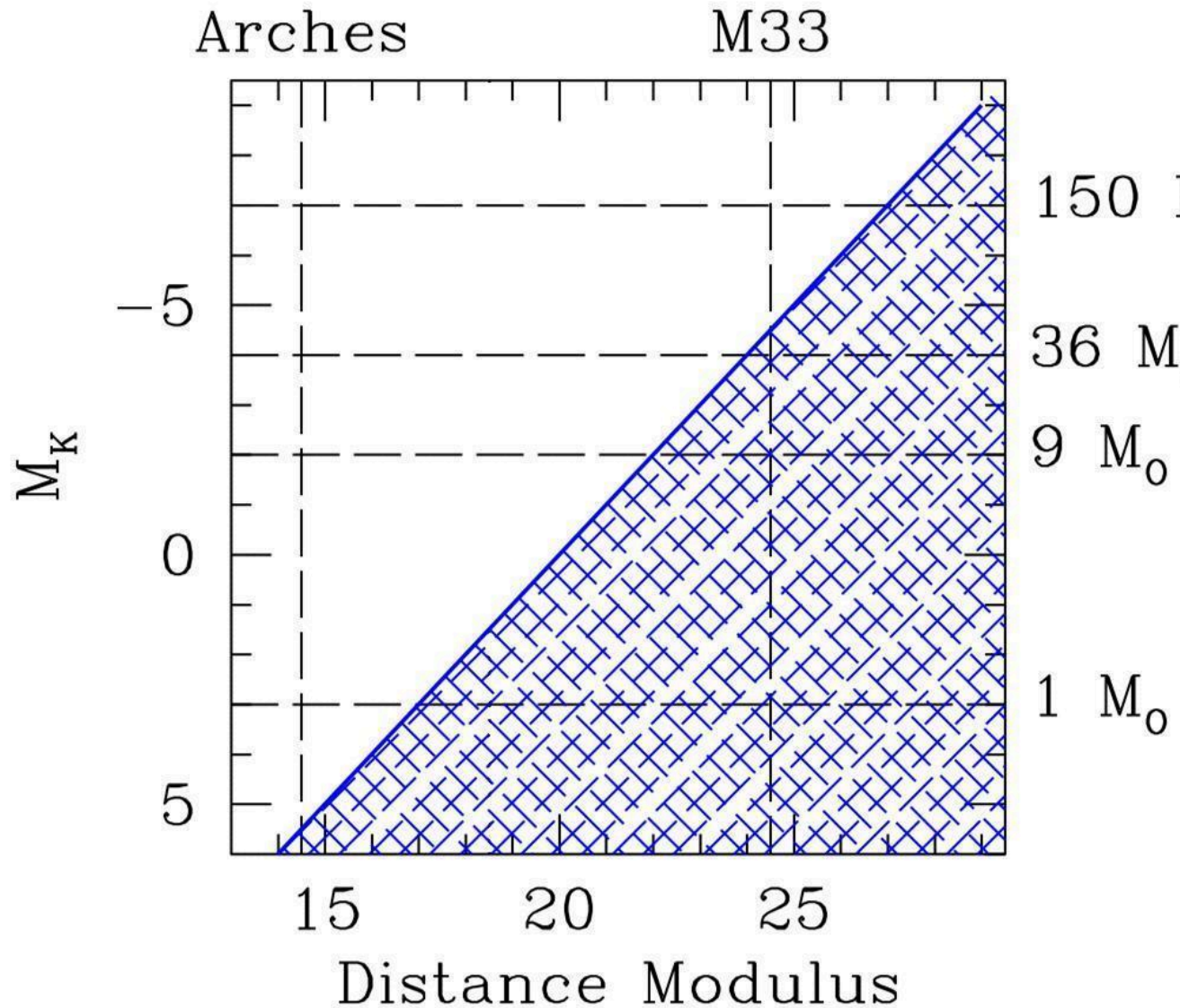

***Figure 8-1:*** *$M_K$ limits for signal-to-noise ratio (Sl 50 at R = 4,000 with IRIS in IFU mode during a 3 total exposure time. These numbers are valid fc uncrowded outer regions of clusters. At a given dis modulus, stars within the blue area are eithe observable or only observable with lower SNR.*

### 8.2.2 Kinematic evolution

Young stars within a star-forming region interact with each other in complex ways. Simulations of star cluster evolution generally assume that the stars are initially smoothly distributed and in dynamical equilibrium. However, both observations and theory tell us that this may not be how clusters form. Star formation in molecular cloud cores follows a spatial distribution imprinted by the properties of the prevailing turbulence, which can be conveniently approximated by a fractal (clumpy) distribution (e.g., Bastian et al. 2009). Numerical simulations of such initial stellar distributions suggest rapid subsequent formation of dense cluster cores occurs by the collapse of protoclusters (e.g., Allison et al. 2009). Dynamical mass segregation is predicted for stellar masses down to a few $M_{\odot}$ on timescales of a few Myr during the collapse.

To answer the question of the nature of early dynamical mass segregation, precise mass, position and velocity measurements are required. While Gaia has revealed dynamical processes in our stellar neighborhood, these processes are only measurable in high-mass star formation regions beyond 1 kpc and beyond the range where Gaia has sufficient sensitivity. It is also critical to systematically study the most likely initial conditions for cluster formation that lead to the observed configurations. The radial density profiles of young and embedded clusters can shed light on the structure of those clusters. Colors can help classify types of young stellar objects to trace the progress of star formation and help distinguish between different star formation models. Analytical techniques like the nearest-neighbor method for infrared-excess stars can be used to identify structures in star-forming regions. These data can show if clusters are expanding or contracting. The TMT instruments, IRIS, WFOS, MODHIS, and in particular MICHI will play a very important role in this field with the very high angular resolution (~10 AU at 140 pc at 10 μm) and the capability of high-dispersion (R~100,000) spectroscopy.

### 8.2.3 Multiplicity

The multiplicity properties of young stars offer a powerful diagnostic of the star formation process (see review by Offner et al. 2022). The binary fraction has been shown to be a function of primary mass, with nearly 100% of the highest mass stars having companions, down to a mere ~20% for brown dwarfs (e.g., Raghavan et al. 2010). The multiplicity fraction also seems to be a function of stellar age and environment, with younger stars exhibiting a generally higher multiplicity fraction (e.g., Duchêne 1999). It has also been demonstrated that the mass ratio distribution is roughly constant except amongst the lowest mass objects (e.g., Duchêne & Kraus 2013). These multiplicity statistics inform our understanding of the overall star formation process, as different formation theories often make different predictions for the multiplicity fraction as a function of mass and environment (e.g., Bate 2012).

Although recent technological advances allowed for the charting of new realms of parameter space, significant uncertainties still exist in the distribution of all types of binaries at all separations. In particular, the range of period/separation space typically left unprobed by current spatially resolved imaging or radial velocity monitoring is $\log(P) \geq 3$ and/or $\rho \leq 0.05$ arcsec (Duchêne & Kraus 2013). This regime is easily observable with any diffraction-limited imager on TMT, such as IRIS. Furthermore, TMT's unparalleled spatial resolution will allow us to resolve known spectroscopic binaries, previously only available to long baseline interferometry, to provide strong constraints on stellar evolution. The combination of astrometry and spectroscopic orbital monitoring provides a means of measuring component masses, which are essential for calibrating theoretical evolutionary models (e.g., David et al. 2019). In addition, possible future instrument TMT/PSI will discover high contrast ratio binaries, such as brown dwarfs around O or B type stars, at separations unachievable with current generation high contrast imagers. This will fill in currently unknown regions of mass ratio parameter space.

There is tantalizing evidence of differing multiplicity fractions for stellar types as a function of environment. It has been suggested that the density of a star-forming environment will ultimately drive the final multiplicity fractions. Dense, high mass environments may trigger dynamical interactions that ionize binaries (e.g., Bonnell et al. 2003) or disrupt disks before they have a chance to collapse and form low mass companions (e.g., Zinnecker & Yorke 2007). The resolution and sensitivity of TMT will also allow us to push multiplicity surveys to more distant, and often more massive, star-forming region such that the statistics match those of nearby low-mass star-forming regions such as Taurus and Upper Scorpius (~150 pc, e.g., Kraus et al. 2011). Current AO-fed instruments on 8–10 m class telescopes can achieve physical separation limits of ~7 AU for these regions. With TMT, the equivalent will be achievable for high-mass star-forming regions as distant as ~2 kpc, encompassing clusters such as M8, M16, and M20. Furthermore, instruments such as HROS and MODHIS will vastly improve our identification of spectroscopic binaries in star forming regions, particularly for the lowest mass objects that are too faint to target with current instrumentation. By achieving new statistics on more star forming regions, TMT with IRIS, HROS, MODHIS, and PSI will offer a new window into the complicated dynamical processes during star formation that ultimately drives the multiplicity fraction we observe amongst field objects.

An additional contribution of TMT will be its ability to probe the multiplicity of Class I protostars with high resolution using instruments like MICHI. Observing binary systems shortly after their formation has the potential to revolutionize our understanding of the binary formation process. Thus far, the multiplicity statistics of this class of objects are poorly constrained over a limited set of regions and separations (~50–2000 AU, see review by Offner et al. 2022) or biased to multiplicity seen in dust (e.g., Tobin et al. 2016) that may miss some population of stars that are dust-free at early ages. Probing closer separations has thus far been limited in spatial resolution or spectral sensitivity. With MICHI, binaries with separations <20 AU will be resolvable visually in nearby star forming regions. With the high-resolution spectroscopy mode of MICHI and the sensitivity of TMT, radial velocity variables will also be identifiable, providing important statistics on the tight binary frequency in this regime. For this science case, TMT offers a unique complement to JWST and ALMA. TMT will have superior spatial resolution for visual binary identification. ALMA will achieve similar statistics for even more embedded Class 0 protostars, providing an important comparison sample for Class I objects.

### 8.2.4 Binary Brown Dwarfs

Theoretical models for the evolution and spectra of substellar objects are ubiquitously employed to interpret discoveries of brown dwarfs and gas-giant exoplanets. For instance, planets discovered by direct imaging necessarily have their masses inferred by combining the observed bolometric luminosities and age estimates with models. However, the luminosities predicted by the models as a function of mass and age have not been rigorously validated.

Study of binary brown dwarfs is a key pathway to understanding the physical properties of substellar objects. Brown dwarfs continually cool as they age, unlike stars which occupy a fixed position on the color-magnitude diagram during the main-sequence phase. Therefore, there are degeneracies in determining the masses and ages of ordinary field brown dwarfs. Binary systems circumvent these difficulties by providing systems with common ages and metallicities.

More fundamentally, continuous astrometric monitoring of binaries can directly determine dynamical masses, which are sorely need to test models. The few precise measurements of masses and temperatures for brown dwarfs so far from HST and Keck LGS AO reveal that the theoretical models may have serious, previously unrecognized, systematic errors in their predictions of the cooling rates, perhaps related to their early formation history or the modulation of their luminosity by magnetic activity. Also, dynamical mass measurements provide a clear empirical demarcation for the lower mass limit for the star formation process.

The combination of large samples of brown dwarfs from wide-field sky surveys over the next decade (Pan-STARRS, WISE, and LSST) and diffraction-limited followup with TMT/IRIS will provide a powerful combination for advancing our theoretical understanding of substellar objects.

TMT/IRIS astrometric monitoring of binaries in the field, in open clusters of known age (e.g., Hyades and Pleiades), and in the nearest star-forming regions (e.g., Taurus and Ophiuchus) will yield a much broader sample of dynamical masses to test models, both in the raw number of objects but also spanning a wide range of ages (1 Myr to 10 Gyr). A much larger sample is possible with TMT, since tighter binaries (with shorter orbital periods and/or larger distances) can be resolved cleanly and also monitoring of wider binaries will yield orbits and masses more quickly given the great astrometric gains from TMT over Keck, VLT, or HST.

Discovery of the coldest free-floating brown dwarfs has followed a well-established pattern over the past two decades, namely such objects have been found as companions to hotter, higher mass low-mass objects. WISE, Pan-STARRS, and LSST will discover the very nearest, coldest brown dwarfs. Deep follow-up of these objects will be the unique provenance of TMT/IRIS, enabling discovery of even colder objects as binary companions.

The combination of TMT's angular resolution and sensitivity is needed for such important discoveries, since substellar binaries have very small separations (<~0.1") and the very cold (<~500 K) temperatures means the objects will be exceptionally faint, even at near-IR wavelengths. A number of distinct spectral changes have been predicted between the coolest known brown dwarfs (~600 K) and the planet Jupiter (~150 K), including the appearance of NH3 and the disappearance of alkali lines. TMT/IRIS discovery and characterization of these extremely cold objects will complete the mapping of the physical sequence between brown dwarfs and gas-giant planets.

## 8.3 STAR FORMATION AT THE EXTREME ENDS OF THE MASS FUNCTION

### 8.3.1 Formation of high-mass stars

Massive stars impact many astrophysical systems, including forming galaxies, the interstellar medium, young star clusters, and protoplanetary disks therein. Several theories for massive star formation are debated: core accretion, competitive accretion, and stellar collisions (Tan et al. 2014). These vary in their assumptions about how material is accreted to the protostar and thus make different predictions for the structure of the gas envelope, the accretion disk, outflows, and binary properties and (low-mass) protostellar crowding around the forming massive protostar.

TMT can make several important contributions to our understanding of massive protostellar environments. First, high resolution NIR and mid-infrared (MIR) imaging will reveal the structures of heated dust around massive protostars, which can arise from the accretion envelope, accretion disk (including spiral structure and binary companions forming in the disk), outflow cavities, and other protostellar companions forming as part of a fragmenting protocluster. Typical Galactic massive protostars are at ~3 kpc (the closest is in Orion KL at ~400 pc), so structures on ~1000 AU scales extend over ~0.3 arcsec. IRIS and MICHI will enable dramatic improvements over current observations of NIR (e.g., Preibisch et al. 2011) and MIR (e.g., de Wit et al. 2009) dust continuum from massive protostars, first by having a factor of three better angular resolution compared with current 8–10 m class telescopes, and second by being sensitive to fainter emission features (coronagraphic capabilities would be useful to block compact emission from the target protostar to help reveal disk, envelope and outflow structures). These images will constrain radiative transfer simulations of massive protostars (e.g., Zhang et al. 2014). In addition, imaging at wavelengths of 24 μm or longer may be feasible (potentially differentiating TMT from other ELTs), increasing the wavelength range over which radiative transfer models can be constrained. Galactic massive star-forming regions often have relatively extended MIR morphologies, requiring relatively large chop throw angles of ≳30 arcsec.

Spectroscopy with the TMT will open up additional new windows on the composition and kinematics of high-mass star-forming regions. Some probes are common to those used in studies of low-mass star formation, including $H_2$(1-0) S(1) (e.g., Hsieh et al. 2021), CO fundamental (1-0) and overtone (2-0) band emission/absorption (e.g., Ilee et al. 2018) and the shape of the 10 and 18 μm silicate absorption features. The Brγ (2.17 μm), He I (2.06 μm) recombination lines and [Ne II] line at 12.8 μm are especially relevant for massive protostars that are beginning to ionize their surroundings, with expected line widths at least ~10 km/s and typically much greater. Zhu et al. (2008) studied [Ne II] emission from ultra-compact HII regions: the small thermal width of the [Ne II] line helps reveal intrinsic kinematic structure having velocities <10 km/s. JWST will revolutionize mid-IR observations of [Ne II] but without spectral resolution; the suite of instruments on TMT will facilitate experiments where the physics is measured through the dynamics, which requires high spectral resolution, with higher angular resolution. IRIS IFU and/or MIR studies of Galactic ultra-compact HII regions have the potential to reveal their kinematic structure in unprecedented detail, and this can help to discriminate between different formation and feedback models: e.g., Is the ionized gas associated with outflows, accretion flows or both? How ordered/symmetric are the accretion and outflow structures around massive protostars? Such questions could begin to be addressed by a survey of the nearest ~30 massive protostars that sample a range of luminosities/masses and star-forming environments, ideally with a velocity resolution reaching or exceeding ~10 km/s (e.g., R~30,000).

### 8.3.2 Formation of brown dwarfs and planetary-mass objects

Stellar objects with masses less than ~75 Jupiter mass ($M_{Jup}$) are unable to sustain nuclear fusion in their cores and thus represent a separate category, referred to as brown dwarfs (BDs), which bridge the mass gap between the planets and stars. While the distinction between stars and brown dwarfs is clear enough by taking the hydrogen-burning limit as the boundary, the mass boundary between BDs and planets is still vague. Various works have adopted the criteria of free-floating objects above the deuterium-burning limit (~13 $M_{Jup}$) to distinguish brown dwarfs from planets, though the discovery of free-floating planets seems to have further muddied the water. While there is an overall consensus on the formation of stars (molecular cloud collapse) and planets orbiting parent stars (accretion of planetesimals in the circumstellar disk), the formation mechanisms of BDs and free-floating planets are still puzzling. The main sticking point of the BD formation appears to be the fact that a much larger molecular gas density than observed is required for the low Jeans mass of BDs. Dense core ejection and turbulent fragmentation are possible explanations. Free-floating, planetary-mass objects may represent an extension of BDs sharing the same formation process, but it is also possible to attribute them to dynamical ejection from planetary systems forming around stars via mutual gravitational interaction between multiple planets.

Direct, high sensitivity, high spatial resolution observations of nearby stellar clusters will help in reaching up to much lower mass limits (figure 8-2). State of the art searches for the lowest mass free-floating objects in a cluster reach about 5 Jupiter masses (e.g., Jose et al. 2020; Luhman & Hapich 2020). Meanwhile, ALMA is revealing small condensations and compact fragments (e.g., Tobin et al. 2018, Paneque-Carreno et al. 2021), yet the characterization of any proto-planet or proto-brown dwarf being formed needs spectroscopic follow-up in the infrared. TMT will provide statistically significant samples of candidate BDs and giant planets and examine and compare a number of properties (such as SED slopes) to distinguish between BDs, free-floating planets, and bound planets (see also chapter 10). This could help in solving the long-standing confusion over whether a source is to be considered a BD or a giant planet.

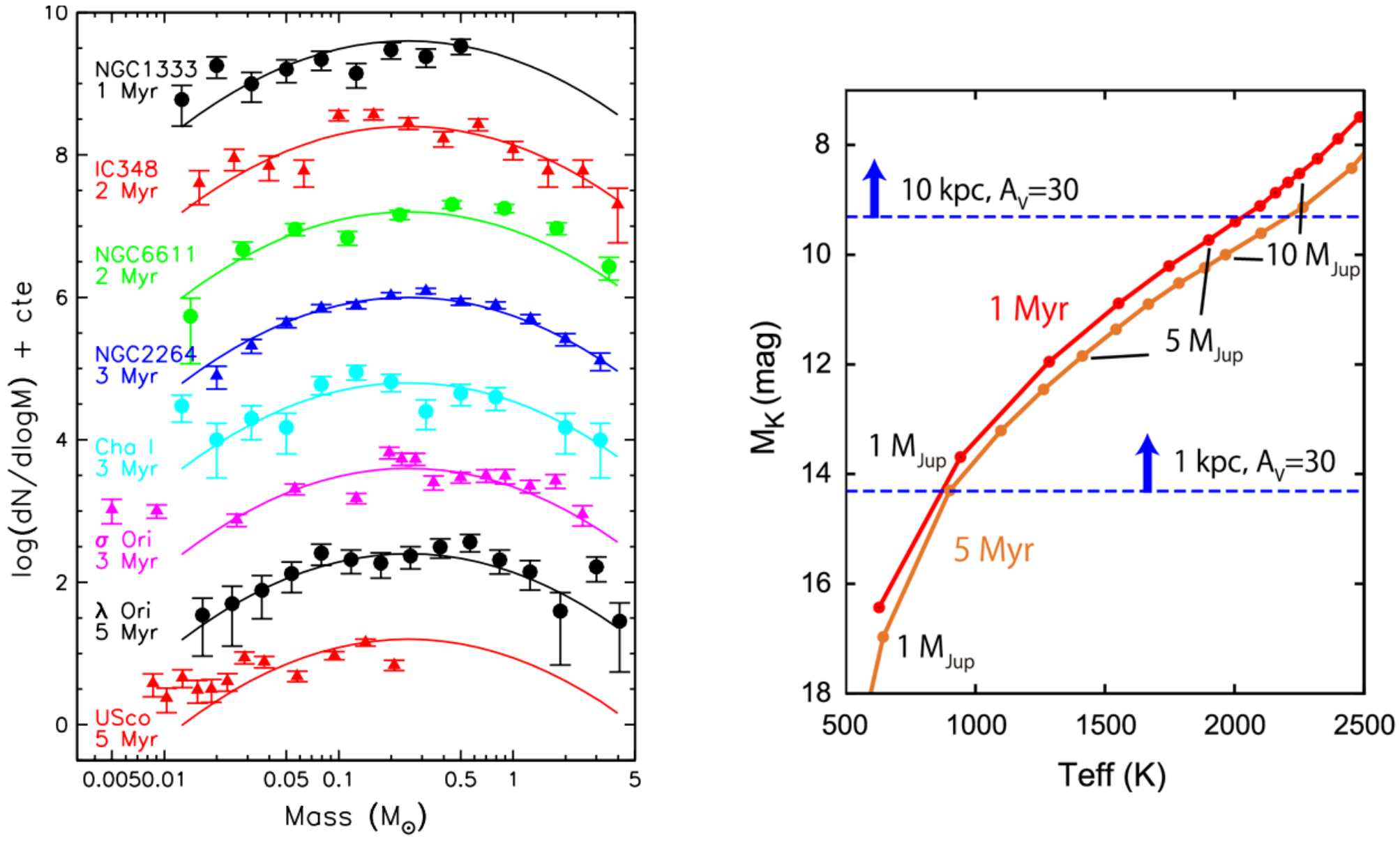

***Figure 8-2:*** *Left: IMFs for 8 star-forming regions (Offner et al. 2014 and references therein). The solid lines show the log-normal distribution by Chabrier (2005) with the normalization to roughly match the data. Right: $M_K$ for young (1 and 5 million years) Jupiter-mass (1 $M_{Jup}$=$10^{-3}$ $M_\odot$) objects predicted in the evolutionary model by Baraffe et al. (2003). IRIS photometry can detect ~1 $M_{Jup}$ objects at 1 kpc with $A_V$=30, at SNR=5 with a total exposure time of 30 minutes.*

The bottom of the IMF has long been a significant topic of analysis. Though there is wider consensus on the high mass end, the sub-stellar end still needs significant observations and proper analysis (figure 8-2). Spectroscopy is important to better constrain the IMF as it provides a more reliable estimate of temperature, thus mass and age, compared to using photometric color information. Alves de Oliveira et al. (2012, 2013) have carried out such a study in the ρ Ophiuchi and the IC 348 regions. They have achieved the detection limit of the order of ~10 $M_{Jup}$ using various telescopes. Much deeper samples can certainly be obtained with TMT, especially in close-by stellar clusters (e. g. Scholz et al. 2012). It should also be noted that constraining the multiplicity could play a key role in IMF determinations (section 8.2.3).

In addition, spectroscopy of BD candidates enables us to detect signatures from accretion disks/jets/outflows as well as their chemical properties, helping us understand if they form in any manner similar to stars. Thus far, it has been difficult to do this on a statistically significant basis due to the lack of BD candidates. With TMT, BDs of ~1 Myr with 13 $M_{Jup}$ at the distance of a few kpc can be observed in NIR spectroscopy, depending on the amount of visual extinction, which will greatly enlarge the current BD sample.

MODHIS has the potential to revolutionize our understanding of the detailed properties of the coldest stars and brown dwarfs, the late-M, L, T, and Y ultracool dwarfs (Kirkpatick 2005). PRV measurements will enable the first detection of gravitational redshift from ultracool dwarfs, whose distinct mass/radius dependence compared to surface gravity provides a measurement of the substellar, semi-degenerate mass/radius relationship (Burgasser et al. 2019). MODHIS provides a unique opportunity for late-M/L plus T dwarf spectral binaries for which both primary and secondary RVs can be captured simultaneously from different parts of the spectrum (Burgasser et al. 2010; Bardalez Gagliuffi et al. 2014). MODHIS spectroscopy will enable rotational velocity studies of low-mass stars and brown dwarfs in the field and in clusters of various ages, providing new insights into angular momentum evolution as a function of mass and age in the regime of fully convective interiors and neutral atmospheres (Irwin et al. 2011).

Using Zeeman line broadening or splitting of magnetic-sensitive spectral features we can test fully-convective dynamo models (Browning 2008; Brown et al. 2020) by examining correlations between magnetic field strength and rotation (Reiners & Basri 2006; Terrien et al. 2022), and potentially map magnetic structures (Zeeman Doppler Imaging; Donati & Brown 1997; Morin et al. 2010. Spectral monitoring of ultracool stars and brown dwarfs will also provide insights into the 3D structures and dynamics of condensate clouds, which form in both ultracool dwarf and giant exoplanet atmospheres (Charnay et al. 2018). By monitoring atomic and molecular features in velocity space it is possible to generate a Doppler imaging map of cloud structures (Crossfield 2014). With concurrent monitoring of regions of continuum absorption and scattering and of high and low atmospheric opacity, we can build a combined vertical and azimuthal map of cloud structures and compare that to 3D atmosphere models (Buenzli et al. 2012; Apai et al. 2013; Marley & Robinson 2015).

## 8.4 YOUNG PLANET-FORMING DISKS

Protoplanetary disks are flattened rotating structures of gas and dust, formed as a natural outcome of the star formation process. They are the birthplaces of planets and hence provide the initial conditions and ingredients for planet building. Disks also interact with newly-born planets, resulting in the redistribution of gas and dust and possible orbital migration of planets. After most of the gas dissipates, the evolution of a planetary system further proceeds via orbital re-configuration and growth of rocky bodies in debris disks around young main-sequence stars (see section 10.1). Therefore, observational understanding of young disks is essential to address the questions of when, where, and how planets form and evolve.

What is most critical for measuring dust in protoplanetary disks is high angular resolution and, for ExAO fed coronagraphic instruments, polarimetric imaging. It is required to spatially resolve the local physical and chemical conditions in a disk to finally constrain planet formation theories. In this respect, there is no doubt that TMT will play a major role in this science field, providing a resolution of ~1 AU at 1 μm for disks in the nearest star-forming regions at 140 pc. The high sensitivity of TMT allows us to significantly increase the sample size for resolved observations and to discuss any dependence in planet formation as a function of stellar properties. The inner (≲10 AU) regions of protoplanetary disks are particularly interesting for forming the broad diversity of other system architectures revealed by Kepler and TESS (see section 10.2). Given the abundance of exoplanets, we know that protoplanets must also be common in the 0.5–10 AU range (Zhang et al. 2018), while current direct detection surveys show that they are rare beyond ~30 AU (see review by Currie et al. 2022). These inner regions also intersect the habitable zone of their parent stars. Such inner disks can be resolved either directly, or kinematically where the (nearly) Keplerian rotation of disks can be used to separate disk regions in velocity (and hence radius). Spectro-astrometry is another advanced, promising technique to probe spatial scales much smaller than the nominal spatial resolution. Given the expected temperatures of these disks, atomic and molecular lines of interest will lie between 1 and 25 μm. The classical infrared fingerprint region contains transitions from many molecular species, including important bulk tracers, such as water, CO and $CO_2$, as well as a potentially long list of organic species (HCN, $C_2H_2$, $CH_4$, etc.) (figure 8-3).

Disk observations, including detecting the footprints of planets, can be extensively done from the early phase of TMT without coronagraphy or polarimetry. As high contrast second generation instruments become available on TMT, lots of detections of exoplanets will be expected. This may be the most compelling and exciting period since the connection between disks and planets can be established, which represents substantial progress toward the ultimate goal of understanding planet formation.

TMT will be highly complementary to JWST and ALMA. TMT and JWST will probe the hotter inner disk region while ALMA more typically traces the cooler dust and gas in the outer disk (figure 8-3). Because of its higher spatial and spectral resolution, TMT will be much superior to JWST in constraining the spatial origin of disk emission features and their dynamics.

In the following, we describe some intriguing science topics where a 30m class telescope can uniquely contribute: conditions of planet formation, disk-planet dynamical interaction including growth of gaseous planets and circumplanetary disks, and distribution of ice, water, and organic material in young plant-forming disks.

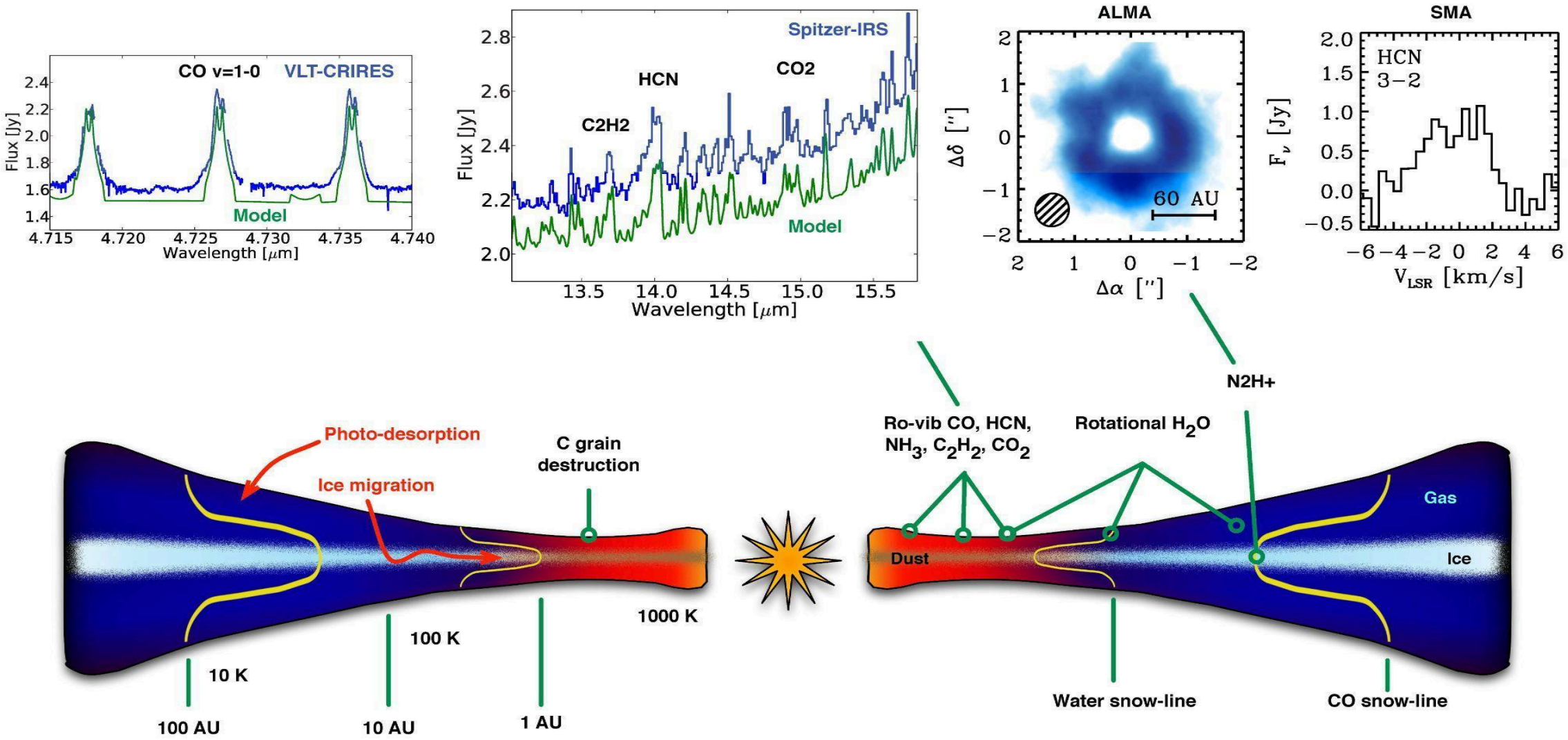

***Figure 8-3:*** *Sketch of different probes of molecules in disks, highlighting the synergistic nature of exoplanet formation studies. Millimeter and submillimeter facilities detect signatures from the outer parts of the disk. The mid-infrared predominantly traces molecular lines in the inner few AU, corresponding to the terrestrial planet-forming region.*

## 8.5 CONDITIONS FOR PLANET FORMATION

### 8.5.1 Planet formation vs. host star properties

The diversity of exoplanets and circumstellar disks seen thus far is thought to be related to properties of the stellar hosts. Metallicity is strongly correlated with the presence of giant planets but not necessarily small planets (see review by Zhu & Dong 2021). Warm debris disks, indicative of collisions between planetesimals near the star in asteroid-like belts, are much more common around intermediate mass stars than solar mass stars, with significant biases from sensitivity (e.g., Su et al. 2006, Marino 2022). Debris disks are also surprisingly frequent for binary star systems (Trilling et al. 2007), and the Kepler survey confirms that planets around and within multiple star systems are present, although the statistics are not yet robust (Armstrong et al. 2014). All of these observations indicate that properties of the host star strongly influence protoplanetary disks and their evolutionary paths towards new planetary systems. For distant star-forming regions, TMT will obtain metallicities and spectral types for stars with disks, as well as resolving their multiplicity.

With an inner working angle of approximately 70 mas using high contrast coronagraphy, TMT will be able to detect young planets in K and L bands within disks found by Spitzer, WISE, and Herschel in nearby star-forming regions and young associations out to the distance of Orion, greatly expanding the range of stellar properties that can be directly related to characteristics of disks and planets. In some cases these planets may be detectable spectroscopically with TMT, so atmospheric composition of the planet may be compared with the abundances of the host star. Although TMT may not have the mid-infrared power of JWST or cryogenic space surveys for detecting disks, with high quality AO correction it can resolve the sources of near and mid-infrared excess emission from young stars in crowded fields and tight multiples.

TMT will also have the ability to resolve sharp structures within disks and will be especially powerful for study of the gaps in transitional disks. Disk structures seen with ALMA and with VLT/SPHERE point to sites of likely planets (see review by Benisty et al. 2022), in some cases with specific predictions on planet location (Xie et al. 2021) but without the sensitivity to detect the predicted planet.

The question *How do planets form?* is best answered by characterizing them near the epoch of their formation. This can be achieved on solar system scales (1–100 AU) with MODHIS, which combines high angular resolution, high-contrast, and high-resolution spectroscopy (Snellen et al., 2014, 2015; Hoeijmakers et al., 2018; Bowler et al., 2019), a technique also known as high dispersion coronagraphy (Wang et al., 2017; Mawet et al., 2017). The AO system separates the light from the star and the companion, whose signals are then individually fed to the spectrograph via single mode fibers (SMFs) for further spectral differentiation. Furthermore, using high dispersion coronagraphy as a pathway to habitable worlds with the ELTs is one of the four key capabilities laid out by Astro2020 (RD4, see their §1.1.1). MODHIS will be able to survey directly imaged giant planets beyond 0.1 AU (figure 8-4), enabling the spectroscopic characterization of ~100 giant planets.

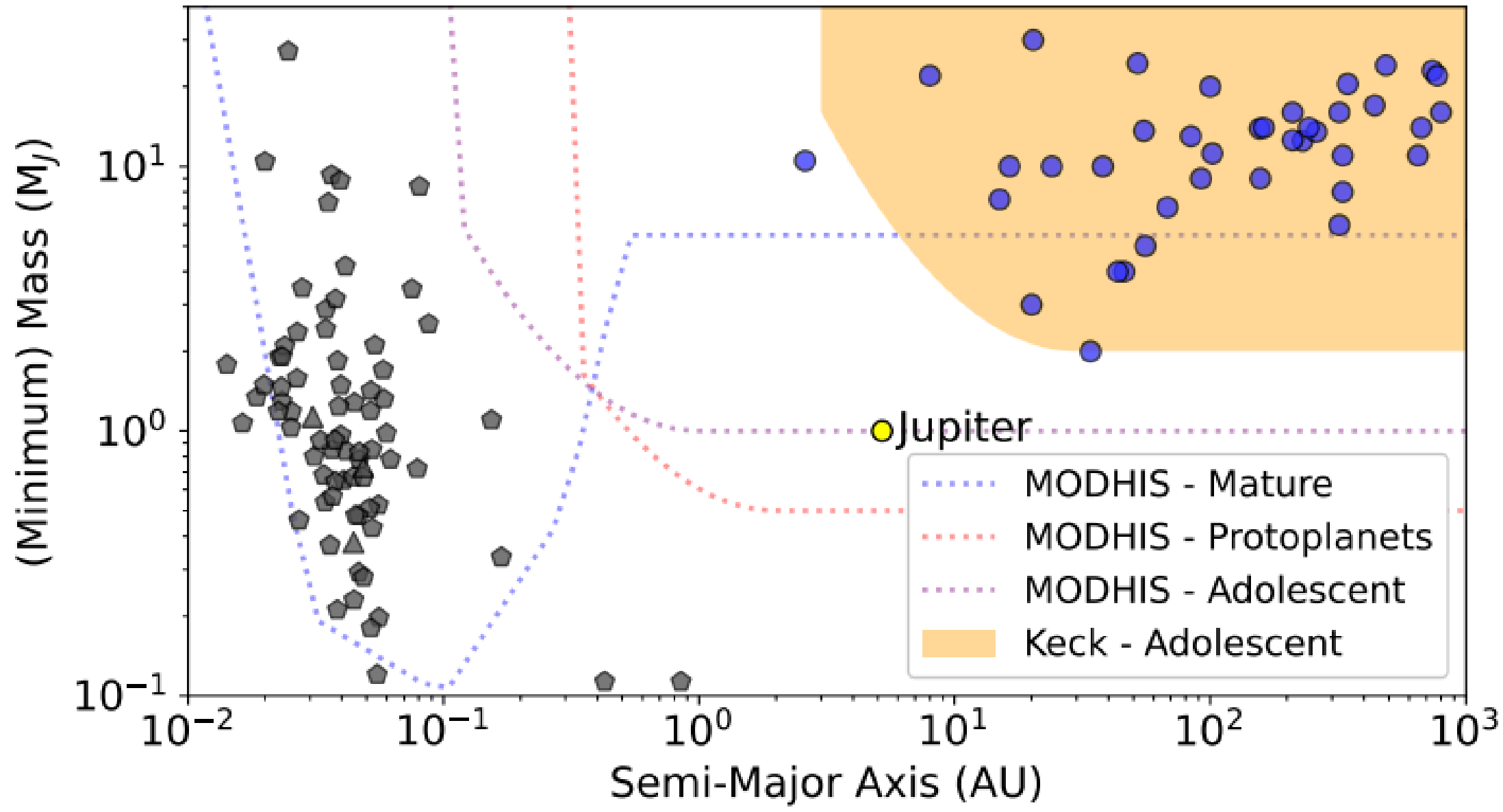

***Figure 8-4:*** *Sensitivity of MODHIS for direct spectroscopy of exoplanets for three different classes of planets: protoplanets (1 Myr old) residing in the closest (140 pc) star-forming regions, adolescent (40 Myr) planets residing in nearby (30 pc) young moving groups, and mature (1 Gyr) planets residing in the Solar neighborhood (10 pc). Sensitivity limits were computed with the MODHIS exposure time calculator. For protoplanets and adolescent planets, the planets are dominated by the glow from their heat of formation. For mature planets close-in (< 1 AU), thermal heating and reflected light become dominant contributors, allowing for increased sensitivity close-in.*

MODHIS's diffraction-limited angular resolution will uniquely enable the study of transiting planets in close binaries (<0.5″) which are typically ignored by seeing-limited instruments (Feinstein et al. 2019). Knowing which star of a pair hosts the planet makes a dramatic difference to the derived planet radius and density (Ciardi et al., 2015) and a demographic study will inform how planet formation proceeds in binary systems, a topic of great theoretical interest (Winn and Fabrycky, 2015; Kraus et al., 2016; Moe and Kratter, 2021).

### 8.5.2 Gas dissipation timescale of protoplanetary disks

The gaseous component of protoplanetary disks is expected to dissipate by processes such as accretion onto the star, winds launched by MHD and photoevaporation processes, and giant planet formation. The timescale for gas dissipation provides a valuable upper limit on disk evolution and the timescale for giant planet formation, as giant planets must accrete their gaseous envelopes on a shorter timescale than this (see reviews by Manara et al. 2022 and Pascucci et al. 2023). The dissipation of the gaseous disk also limits the time available to circularize terrestrial planet orbits through interactions with the gas disk. For example, if the late stages of terrestrial planet formation occur during an epoch when the disk has a surface density ~0.01–0.1% that of the minimum mass solar nebula, planets with the masses and eccentricities of Earth, Venus, Mars, and Mercury can be produced (Kominami & Ida 2002).

While stellar accretion rates decline with age (e.g., Fig 2 of Sicilia-Aguilar et al. 2010), measure the amount of disk gas that reaches the star, it is difficult to be certain how much gas remains in the disk without an understanding of the mechanisms that drive disk accretion. Hence what is needed is *in situ* measurements of the gaseous reservoir. As one example diagnostic, the 12.8 μm [Ne II] line is appealing because Ne is expected to remain in the gas phase; [Ne II] is expected to be ionized and heated by stellar X-rays, which are long lived compared to the expected gas dissipation timescale, and the 12.8 μm line is expected to robustly probe small column densities of gas ($N_H \sim 10^{19}$–$10^{20}$ cm$^{-2}$; Glassgold et al. 2007). While JWST will provide a robust sample of disks with [Ne II] emission, the emission will be spatially and spectrally unresolved; the high resolution follow-up with MICHI on TMT will be required to understand the origin and dynamics of this emission.

The 4.7 μm CO fundamental rovibrational lines are also powerful diagnostics of the gaseous disk at larger column densities. The high abundance of CO in disks and the modest A-values of the fundamental transitions have made them a reliable and often used tracer of the inner regions of disks surrounding T Tauri stars and Herbig Ae/Be stars. A versatile diagnostic, the formation of CO is robust in disks even in the absence of grains (Glassgold et al. 2004; Bruderer 2013), and the transitions can be excited into emission by UV fluorescence (e.g., Brittain et al. 2009), IR pumping, as well as thermally.

In addition to measuring the gas dissipation timescale, these diagnostics may also probe the dissipation process itself. Theoretical arguments (Gorti et al. 2015, Picogna et al. 2019) and blue-shifted line profiles (e.g., Sacco et al. 2012) suggest that photo-evaporative winds can be probed by [Ne II] (see also Wang et al. 2019). The CO ro-vibrational lines can also show line profiles suggestive of a wind origin, suggesting its potential role in studying the gas dissipation process (Pontoppidan et al. 2011).

The measurement of line profiles as well as spectro-astrometric signatures of these diagnostics can be used to determine where the emitting gas arises (disk or wind, and at what radii) and to constrain the gas reservoir in the disk as a function of stellar age. When combined with theoretical model predictions, these measurements could also be used to measure the rate of gas removal from the disk (e.g., the disk photo-evaporation rate).

Giant planet formation itself is another disk dissipation pathway and locating CO emission within a disk can be used to look for radial gaps created by planets (see next section). Constraints from the lack of UV fluorescence emission are particularly useful in showing that there are gaps in the gas (e.g., Brittain et al. 2013).

## 8.6 PLANET-DISK INTERACTION - DISK STRUCTURE BY PLANETS

Planets can dynamically interact with their parent disk, leaving footprints that in most cases, are more readily detected than the planets themselves (see the review by Paardekooper et al. 2023). Rings, cavities, spirals, and azimuthal asymmetries are commonly identified in ALMA surveys of cold dust (figure 8-5) and SPHERE observations (figure 8-6) of disk-surface dust. Each of these substructures may be caused by the presence of planets that already exist or may be caused by other physical processes (see reviews by Andrews 2020, Bae et al. 2022). In some cases, the gas structures pinpoint the location of a planet as the likely cause of these substructures (see review by Pinte et al. 2022).

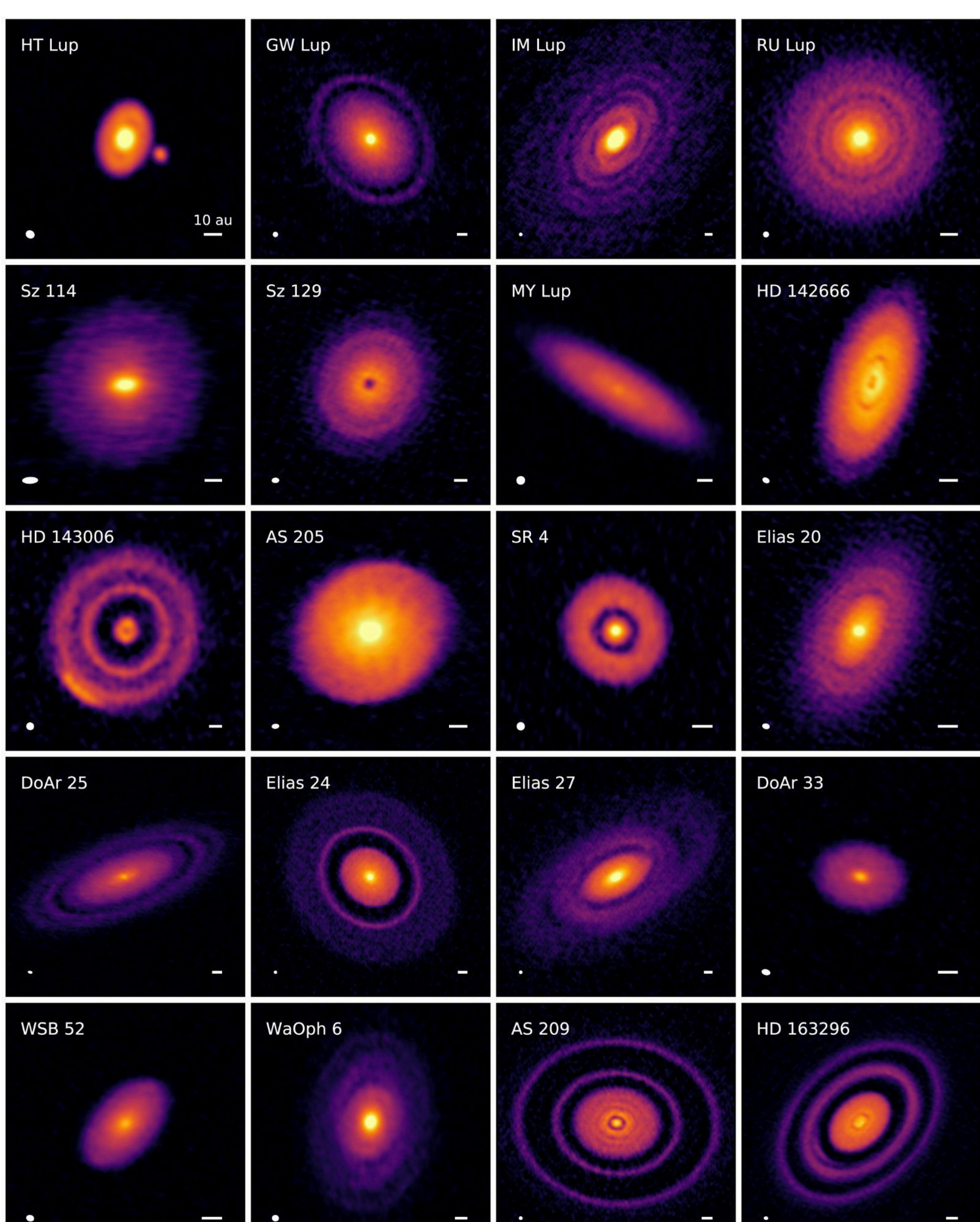

***Figure 8-5:*** *Andrews, et al. 2018 showing 240 GHz (1.25 mm) continuum emission images for the disks in the DSHARP sample*

Once formed, the gap can excite instabilities due to the pressure gradient at its outer edge, resulting in dust traps that can become favorable locations of planetesimal formation and growth, providing the formation site of additional members of a young planetary system (e.g., Raettig et al. 2021; Jiang & Ormel 2022). Dynamical processes can also excite small-scale structures such as spiral waves in association with instabilities or companions, even for those insufficiently massive to carve a gap. In fact, direct imaging with 8–10 m class telescopes has uncovered gaps in transitional disks of T Tauri and Herbig Fe/Ae/Be stars, while spirals were detected preferentially toward warmer disks around earlier type stars in scattered light (e.g., Muto et al. 2012, Garufi et al. 2021). Yet, with the current instruments, the primary focus is limited to outer disk regions of several tens of AU.

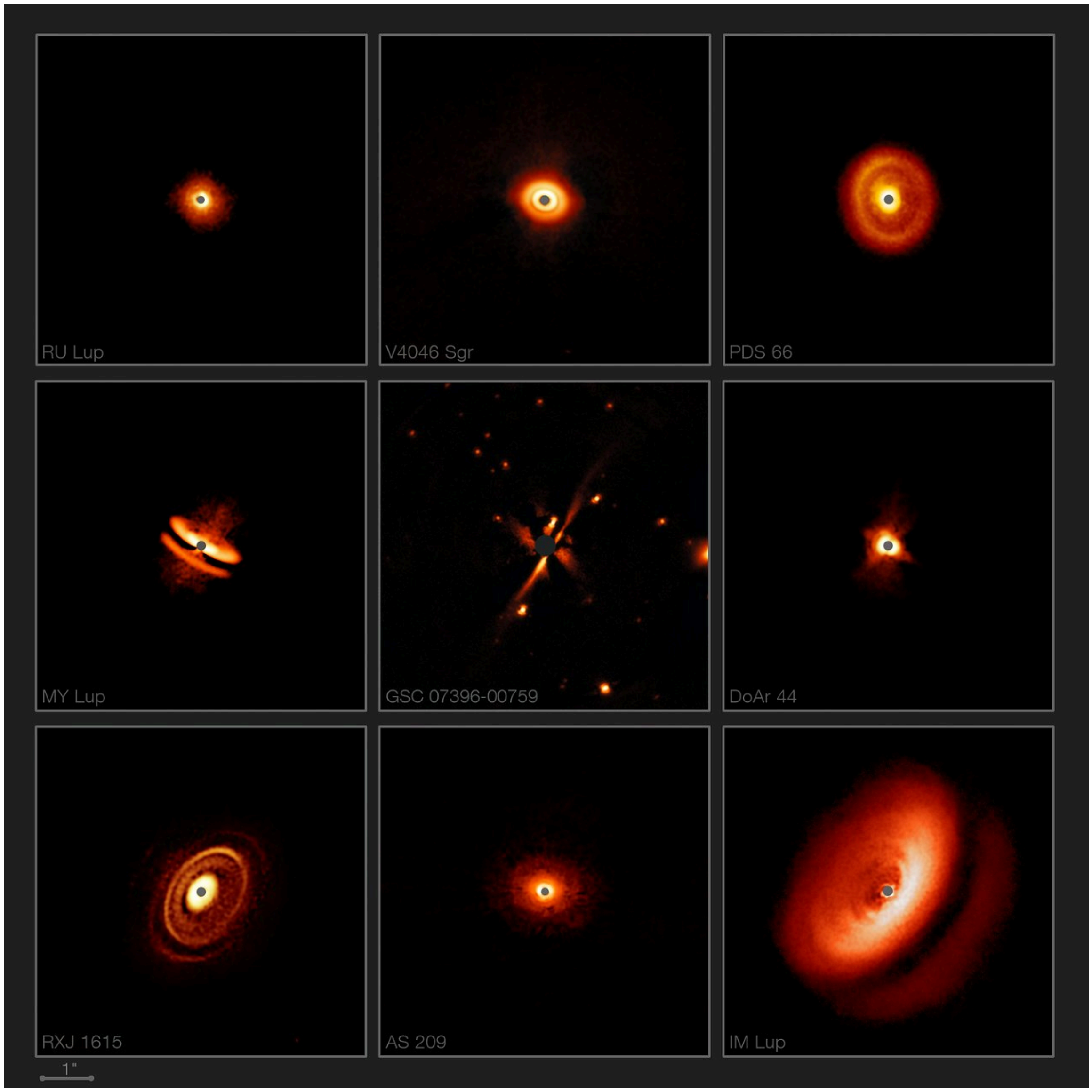

***Figure 8-6:*** *Images from the SPHERE instrument on ESO's Very Large Telescope (Avenhaus et al.2018) are revealing the dusty discs surrounding nearby young stars in greater detail than previously achieved. They show a bizarre variety of shapes, sizes and structures, including the likely effects of planets still in the process of forming.*

Higher angular resolution and contrast, afforded by TMT with its advanced AO system, are the keys to detect planetary signatures in the inner disk region and to resolve smaller-scale structures. For instance, the typical requirement for the detection of spirals is to resolve the spatial scale comparable to the disk vertical thickness. In the case of nearby star-forming regions at 140 pc, a resolution of about 0.01 arcsec is needed to distinguish tightly-wound spirals at 30 AU in a T Tauri disk colder than Herbig systems, with the aspect ratio (the ratio of the scale height to the radius) of ~0.1. A gap and spirals by a planet at 10 AU can be detected in the mode of direct imaging, preferably with polarimetry. Structures within 10 AU can be studied with the smaller inner working angle provided by PSI. Figure 8-7 shows synthetic H-band scattered light images of disk-planet models for TMT, highlighting its capability to detect structures produced by true solar system gas and ice giant planet analogs in nearby protoplanetary disks at 140 pc, thanks to its superb angular resolution and inner working angle. Furthermore, the innermost region lying beneath the coronagraphic mask or the bright stellar halo in NIR can be explored in thermal emission in MIR. The diffraction-limited resolution in the H-band is ~0.011 arcsec for TMT, which allows us to study distributions of warm dust with a spatial scale of about 0.02 arcsec, corresponding to 3 AU at 140 pc.

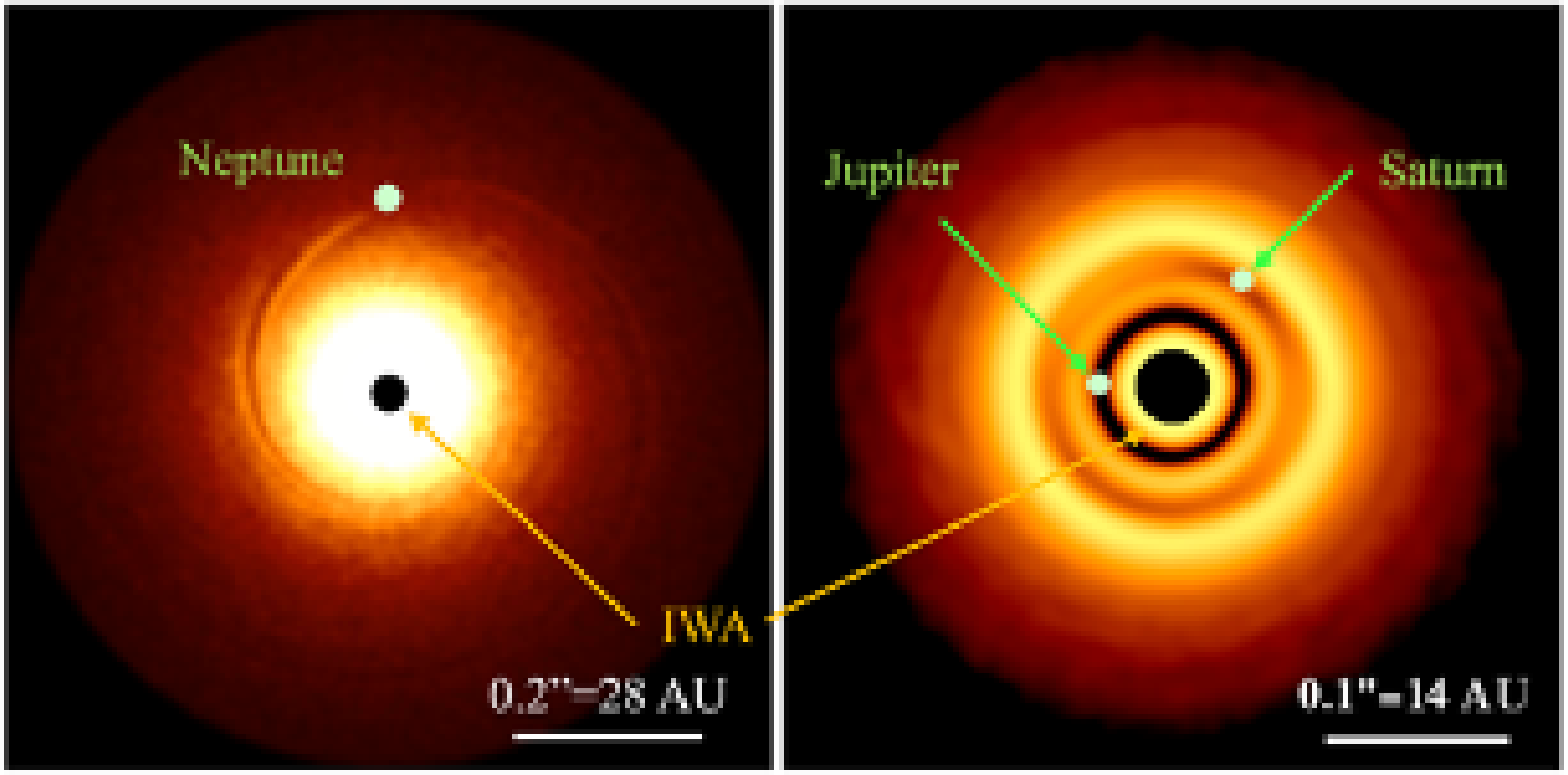

***Figure 8-7:*** *Model predictions for TMT in H-band scattered light imaging, showing spiral arms excited by Neptune (left) and gaps opened by Saturn and Jupiter (right), with the planets being on their current orbits in disks at 140 pc (Image credit: Ruobing Dong (see Dong & Fung 2017a,b) for method). Aided by extreme AO, TMT's superb angular resolution and inner working angle (IWA) in theory will enable the detection and characterization of disk features produced by true solar system giant planet analogs in nearby systems.*

It is worth noting that multi-epoch observations are quite useful to put a strong constraint on planet location through disk dynamics. For instance, a spiral arm caused by a planet in a disk is expected to co-rotate with the planet, resulting in the pattern rotation velocity different from the local Keplerian velocity, as now seen for wide-separation spirals (Ren, et al. 2020; Xie et al. 2021). With the spatial resolution of TMT, the rotation of the spiral will be better resolved at small physical distances from the star, and can be detected by observations several years apart if a planet is orbiting at 10 AU around a solar-mass star.

Near- and mid-infrared observations are sensitive to the distribution of small, micron-sized grains at the surface of an optically thick protoplanetary disk, while (sub-)mm observations probe large, mm-sized grains in the mid-plane in the continuum as well as the gaseous component at various heights through line emission. The synergy of ALMA with TMT is thus of great importance to understand the 3D distribution of different components that make up the disk. Such comprehensive understanding will have a strong impact on theory of planet formation and is critical to understanding the origin of the observed structures as a possible dust trap or planet (Bae et al., 2022).

## 8.7 GROWTH OF PLANETS

Giant planets form during the gas-rich protoplanetary disk phase within the first few million years after the formation of the central star. Beyond a certain mass, the forming protoplanet will interact with its natal disk in multiple ways that may lead to direct or indirect detections of the planet itself. A giant planet may open a gap in the gas disk, inducing local non-Keplerian velocity fields with characteristic structure and time variability. This mechanism has been proposed to explain observed spectroscopic signatures in the 4.7 μm rovibrational CO lines (e.g., Regály et al. 2014). The planet may form a relatively massive, moon-forming, circumplanetary disk (Lubow et al. 1999; Machida et al. 2008; Brittain et al. 2013, Isella et al. 2018), and it may even produce strong emission lines from accretion flows (Haffert et al. 2018). In two cases, PDS 70 and AB Aur, growing planets have been detected within the protoplanetary disk (Keppler et al. 2018; Currie et al. 2022). All of these tracers will be accessible to TMT. Indeed, they are already being pursued with existing 8–10 m class telescopes as well as with ALMA, and initial results are promising.

Therefore, while TMT may detect protoplanets through broad-band imaging of their thermal continuum emission, there is also the prospect of using kinematic gas tracers to detect the *growth* of giant protoplanets. TMT is able to detect gas tracers in the optical and infrared at very high spatial resolution; 40 mas for the four main CO isotopologues around 4.7 μm that trace the velocity field of the protoplanetary material, as well as a host of tracers of accretion onto the planet itself, including hydrogen recombination lines such as Brγ at 2.16 μm (Betti et al. 2022). Spectrally resolved line profiles with MODHIS will reveal how these planets accrete (Thanathibodee et al. 2019, Aoyama & Masahiro 2019), with implications for the radius evolution of the planet. With the velocity resolution of several km/s (i.e., R=100,000), the motion of gas moving close to the escape velocity of a Jupiter mass planet (~60 km/s) can be captured.

It should be emphasized that the detection of a gap-opening, accreting planet will have a strong impact on the theory of planet formation and migration. Measurements of the gas accretion rate onto a planet along with the depth and width of the gap are critical to calculate the angular momentum transfer rate through the gap, and therefore the final mass and type-II migration of the planet (Tanaka et al. 2002; Crida et al. 2006; Fung et al. 2014).

Planet and/or moon formation can be caught in action even after the gas dissipation (see also section 10.2.3). There have been detections of warm dust belts in a number of debris disks (e.g., Morales et al. 2011) and more recently, optical and infrared variation that may indicate planetary collision around a ~300 Myr star is reported (Kenworthy et al. 2023). This is the dust originating in the inner region at ~1–10 AU where the habitable zone resides, and can be interpreted as evidence for ongoing, oligarchic growth of terrestrial bodies. The high spatial resolution of TMT will be critical in characterizing this terrestrial material to gain new unique insight into the building blocks and mechanisms of terrestrial planet formation. In nearby debris disks, imaging from K to N band will be capable of resolving the 1 AU region. The precise structure such as azimuthal asymmetries and sharp boundaries of the disk edges can inform us of the presence of planets and help establish the connection between the dust belt and planet formation/evolution activity.

Similarly to the detection of planets with radial velocity (RV) surveys of stars, measuring the radial velocity of planets and brown-dwarfs is a promising technique to search for binary planets and exomoons (Vanderburg et al. 2018). The mass ratio of moons relative to their planet is expected to be around $10^{-4}$ (Canup & Ward 2006), which is consistent with Ganymede around Jupiter for example. This small mass ratio could explain the lack of current detections.

Batygin & Morbidelli (2020) suggests that the moon mass roughly scales with the planet mass to the 3/2 power, making larger planets harbor comparatively larger moons. Therefore, scaling the Galilean moons to the mass of HR 8799 c, similar moons would be detectable with TMT/MODHIS in only a few nights of observations. Additionally, other moon formation mechanisms are likely to generate larger objects such as collision or capture, which would be more easily detectable.

Forming gas giant planets are expected to have an accretion disk similar to the circumstellar disk from which they form. Because much can be learned about the processes that form both gas giants and their moons by exploring the disks from which they feed and form, detecting and characterizing circumplanetary disks is an exciting area of study for TMT. Information about the structure, temperatures, compositions, and grain sizes in the disk can be gleaned from spectral energy distributions. To break SED fitting degeneracies, information about the scattering properties of the grains via polarization measurements is needed (e.g., Esposito et al. 2020). Spectropolarimetry can be used to map the rotational structure of the disk and determine whether or not an inner hole is present (e.g., Vink et al. 2005).

Recent work (Szulágyi & Garufi 2021) explored the potential detectability of circumplanetary disks in J band polarized light. They found that directly imaged circumplanetary disks are detectable in polarized light with signal ranging from 1.5–3.2% for a number of configurations, and distinguishable from the circumstellar disk, particularly around high mass planets.

An ideal program for MODHIS would be deep observations in spectropolarimetry mode of known young gas giants to fully characterize the nature of their disks and probe exomoon formation.

## 8.8 MAPPING THE PREBIOTIC LANDSCAPE IN PROTO-PLANETARY DISKS

### 8.8.1 Snow line and beyond

Ices of various compositions are present in the various environments of star and planet formation such as molecular clouds, protostars, and protoplanetary disks. One example is the infrared spectra shown by the Ice Age early release science program with JWST, showing the signature of ices associated with stars at different stages of formation (see figure 8-8).

The snow line is a condensation/sublimation front of (water) ice in a protoplanetary disk; water is present in the form of vapor (gas) within the snow line and as ice (solid) beyond it. The formation of giant planet cores is believed to be enhanced because the solid surface density increases and the inward radial drift of solids may be slowed-down across the snow line (Brauer et al. 2008). It helps to explain why Jupiter-like gas giants are not the closest to the Sun in our solar system. While the snow line is located at a radial distance of ~3 AU in the current solar system, it may have been as close as ~0.7 AU in the early, optically thick phase (Garaud & Lin 2007), and therefore abundant water ice may have been present in the terrestrial planet region, which is not consistent with the current water abundance of the terrestrial planets in our Solar system. Thus understanding the location of the snow line at various evolutionary stages is important given its relation to both planet formation and the origin of the water in terrestrial planets. Recent ALMA studies constrained the snow line radius toward bursting protostellar disks (e.g. V883 Ori; Tobin et al. 2023). However, no strong constraint is obtained to the snow line radius of class II protoplanetary disks so far.

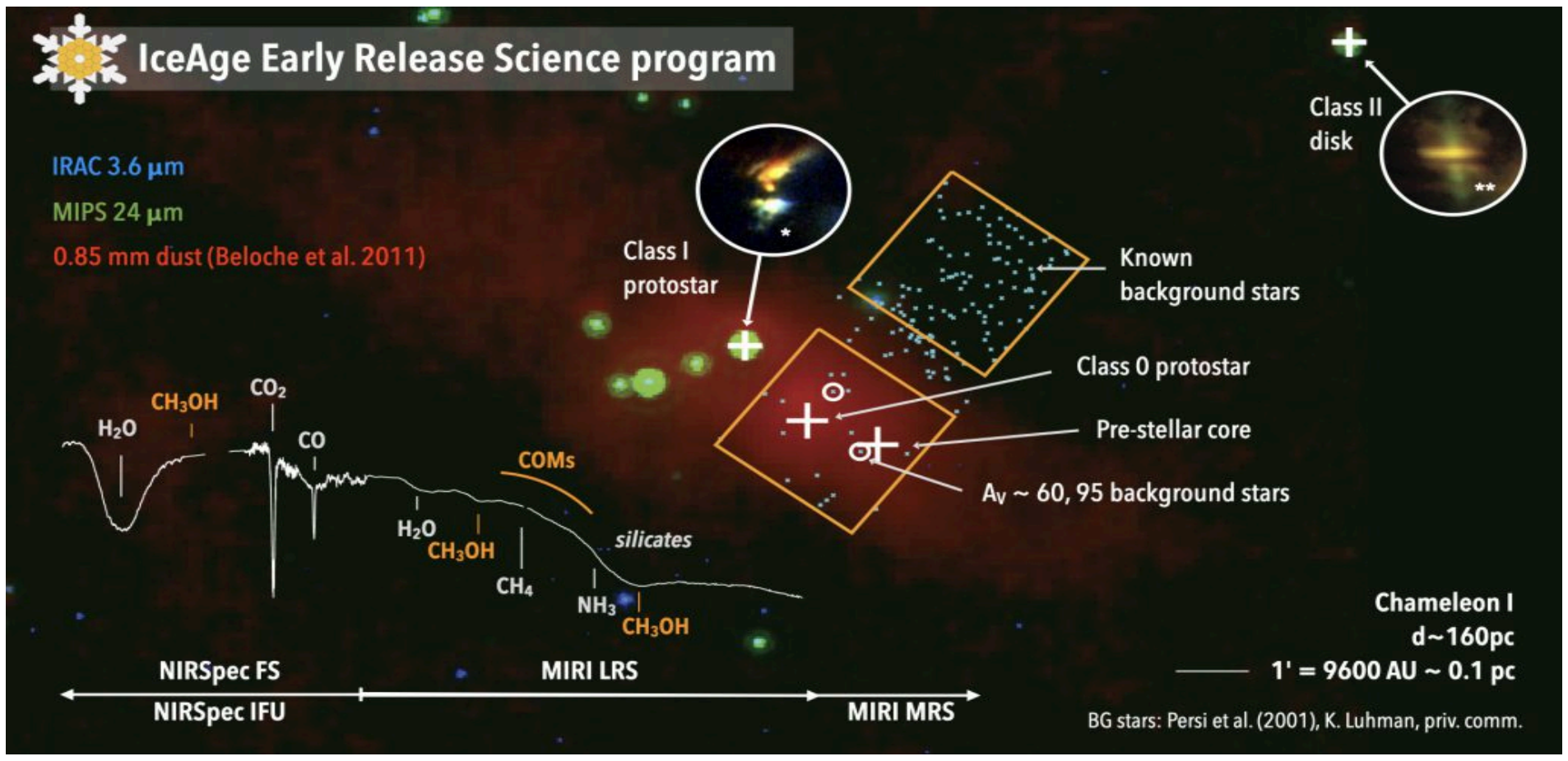

***Figure 8-8:*** *The field at the center of the Chameleon I molecular cloud that contains the pre-stellar targets of the Ice Age JWST early release program to search for signatures of ices in pre-stellar systems at different stages of formation. Image credit: M.K. McClure.*

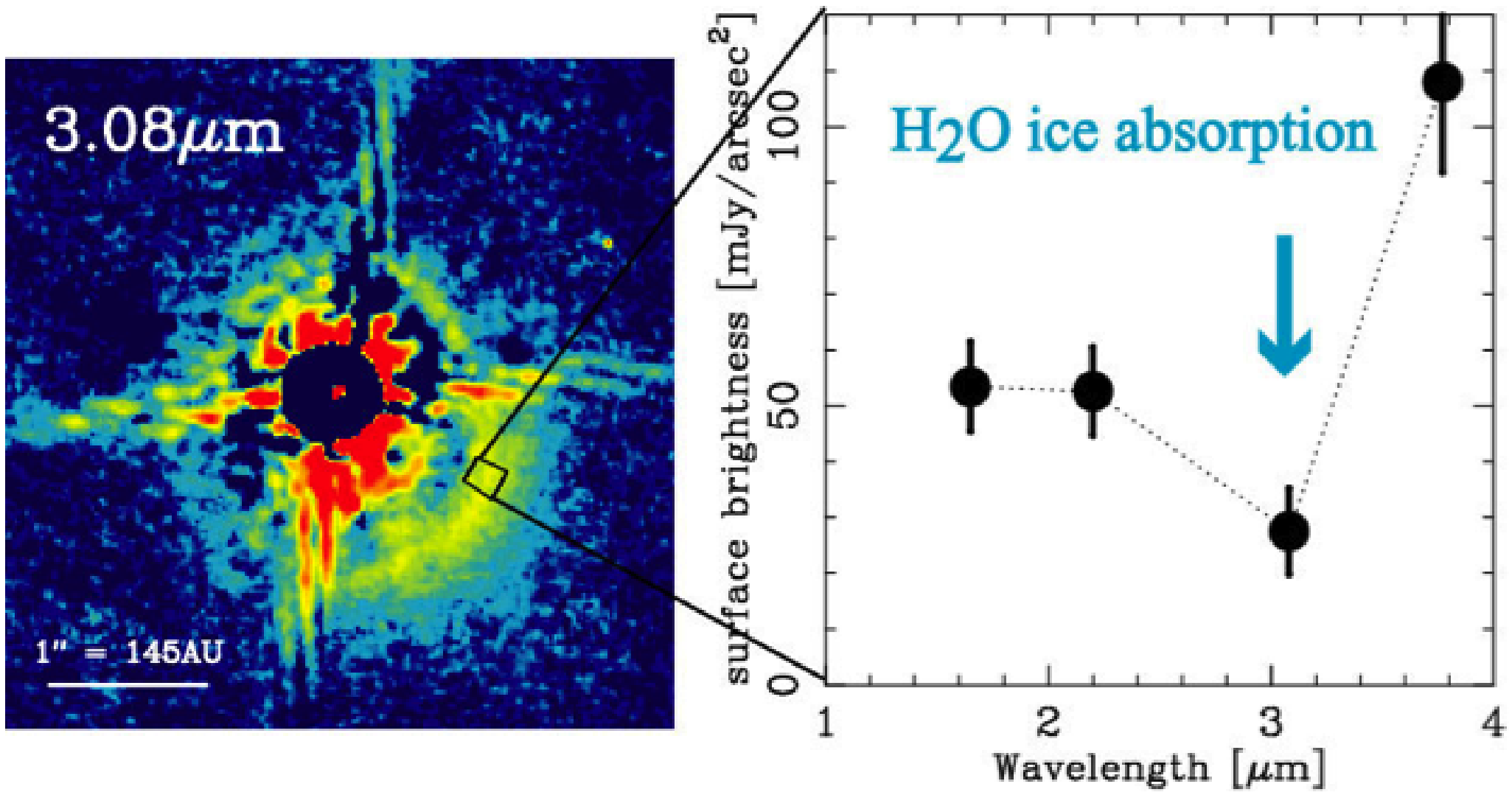

***Figure 8-9:*** *Scattered light image of the disk around the Herbig Fe star HD 142527 at 3.08 μm (left) and the scattered light spectra (right). The ~3.1 μm absorption feature, probably due to the scattering by water ice grains, is clearly seen in the spectra. As we go closer to the central star inside of the snow line, this absorption band will disappear due to the absence of water ice, which will directly show the snow line location at the disk surface. (Honda et al. 2009).*

The snow line in protoplanetary disks can be predicted through observations of water vapor. It is an indirect method in which interpreting the spectral emission features are used to extract the physical and chemical properties. Water emission lines across range of wavelengths from NIR to MIR can be used to measure how the line fluxes vary with different excitation energies, and molecular distribution, e.g., the ice-line can be retrieved via comparison with disk model (e.g., Zhang et al. 2013). This approach was successful in measuring the snow line location of ~4 AU in the disk surrounding the nearby young star, TW Hya (Zhang et al. 2013). In another approach, the radial extent of water vapor in the warm disk atmosphere can be estimated by observing the molecular Doppler shifts (e.g., Pontoppidan et al. 2010a; Salyk et al. 2019). However, all these approaches need high spectral resolution, with MICHI (R~100,000), the number of sources studied in this way can dramatically increase, enabling us to investigate the evolution of the snow line as a function of stellar age and mass. Additionally, combined with spatially resolved disk images, the spectral emission observations are helpful in unveiling interconnection between dust and gas evolution, e.g, pebble drift, inner disk depletion, and composition of planet-forming material (Banzatti et al. 2020).

The distribution of water ice can also be traced via spatially resolved spectroscopy of disk scattered light (figure 8-9). Water ice has a strong absorption band at ~3.1 μm that should be imprinted in the scattered light spectra (Honda et al. 2009). Disk models predict that the surface snow line in protoplanetary disks around intermediate-mass young stars ($L_{star} = 10\ L_{\odot}$) is located at ~20 AU (Oka et al. 2012), which corresponds to ~0.14 arcsec at 140 pc. This is expected to be within reach of TMT-IRIS without coronagraphy, and the observations will surely be feasible with NFIRAOS-MODHIS and PFI owing to the much better inner working angle combined with coronagraphic modes.

The outer disk region well beyond the snow line can be the formation site of icy planets. Comets can also form in such cold regions and could be ingredients of icy planets. Although comets are known to contain pristine frozen material, they also have incompatible, high-temperature products such as crystallized silicate grains, which infers dynamical mixing between the hot inner and the cold outer regions. It is still a matter of debate how the mixing occurs and how the high-temperature material is incorporated into comets. A key to understanding the process is obtaining the spatial distribution of various kinds of grains in young planet-forming disks. With the high-spatial resolution of TMT, MICHI will offer a unique opportunity to uncover both mineralogical evolution of solids and transport of material within a disk by spatially-resolved MIR spectroscopy (e.g., Okamoto et al. 2004).

### 8.8.2 Formation and evolution of prebiotic molecules

Planets in the habitable zone are generally expected to be depleted in the elements required to support life as we know it; carbon, hydrogen, nitrogen, oxygen, phosphorus, and sulfur (CHNOPS). However, some amounts of these elements and the molecules that carry them do end up on terrestrial planetary surfaces, and the processes that deliver them are intimately linked to protoplanetary disks (see reviews by Pontoppidan et al. 2014; van Dishoeck et al. 2014). The amount of CHNOPS molecules delivered to potentially habitable planets ultimately provides the basis for the formation and evolution of terrestrial atmospheres. Complex organic species may also be synthesized in the warm irradiated environments of disks. Indeed, the warm, energetic environments of disks within a few AU of the central star are thought to be highly chemically rich, with short timescales for the formation of a wide range of organic species (e.g., Henning & Semenov 2013; Miotello et al. 2021).

Spitzer observations indicated that strong emission from gas-phase species comes from disks around solar-mass young stars (e.g., Carr & Najita 2008, 2011; Pontoppidan et al. 2010b). Favorable geometries can also lead to deep absorption lines from disks (Lahuis et al. 2006), enabling high dispersion observations of much rarer species (Knez et al. 2009). We now know that planet-forming regions are chemically rich, highly heterogeneous objects, with strong chemical differences between demographic groups. Herbig stars have relatively simple surfaces, dominated by harsh, destructive UV fields (e.g., Fedele et al. 2011), while inner disks around solar-mass stars are rich in both oxygen and carbon-dominated species (Salyk et al. 2011a; Salyk et al. 2011b), and inner disks around low-mass stars are dominated by organics (Pascucci et al. 2013).

Critically, it was also demonstrated that many of these lines are accessible to ground-based facilities (e.g., Pontoppidan et al. 2010a; Mandell et al. 2012). JWST observations are changing many of our previous ideas of the processes in protoplanetary disks. They are providing information that is helping us understand the inward flux and eventual sublimation of icy pebbles (Banzatti et al. 2023), and the effect of irradiation on the molecular-rich inner 10 AU of protoplanetary disks (Ramirez-Tannus et al. 2023). Even though JWST is improving the characterization of these species and the disks in which they reside, the high angular and spectroscopic resolution of TMT is needed to identify the location of the emission to measure the physical processes in the disk.

There are several aspects of prebiotic chemistry that can be uniquely explored using molecular emission lines from warm gas. 1) The warm gas represents the most volatile reservoir that is lost to planetesimal and terrestrial planet-formation. Observations of it therefore test whether the CHNOPS depletions observed in the Earth and the inner solar system are universal. For instance, the Earth's carbon is orders of magnitude more depleted than what would be expected given the relatively large fraction of refractory organic carbon in the interstellar medium and in comets (Lee et al. 2010). Observations of volatile carbon carriers with the TMT can constrain models for carbon grain destruction. 2) Icy bodies in protoplanetary disks are generally affected by aerodynamics. In particular, they migrate against pressure gradients, leading to a net flow of ice inwards, across the snow-line (Ciesla & Cuzzi 2006).

Thus, the observed warm molecular gas may be recently sublimated, comet-like material, and the observable infrared lines trace chemical and photo-chemical processing of pre-biotic material. The action of such a process has been suggested as an explanation for the observed oxygen isotopic fractionation in solar system meteorites (Lyons & Young, 2005). It may also affect the local carbon/oxygen ratio, thereby producing a sharp transition from an oxygen to a carbon-dominated chemistry, with potentially profound consequences for prebiotic chemistry (Najita et al. 2013). 3) Finally, observations of some exoplanetary atmospheres have compositions that suggest the influence of large-scale element redistribution in their natal disks (Öberg et al. 2011). Observations of planet-forming CHNOPS molecules with the TMT will therefore provide observable links to the growing population of exoplanets.

Medium (R~10,000 for maximizing the line-to-continuum ratio) and/or high-resolution (R~100,000, for kinematic information) spectroscopy in the thermal infrared (3–14μm) on the TMT is poised to deliver transformative data on prebiotic chemistry in planet forming disks. The increase in sensitivity over the previous generation of mid-infrared spectrometers on 8–10 m class telescopes, both due to the larger aperture and improvements in detector technology, leads to a tremendous increase in the numbers and types of protoplanetary disks for which molecular lines can be observed. Specifically, while ground-based high resolution N-band spectroscopy can only reach the few brightest protoplanetary disks, the TMT can select targets from a reservoir of 100–1000 protoplanetary disks across the stellar mass range (see figure 8-10). JWST will be more sensitive, but has too low spectral resolution to resolve lines or to separate weak lines of rare species from the forest of bright water lines that is typical for protoplanetary disks.

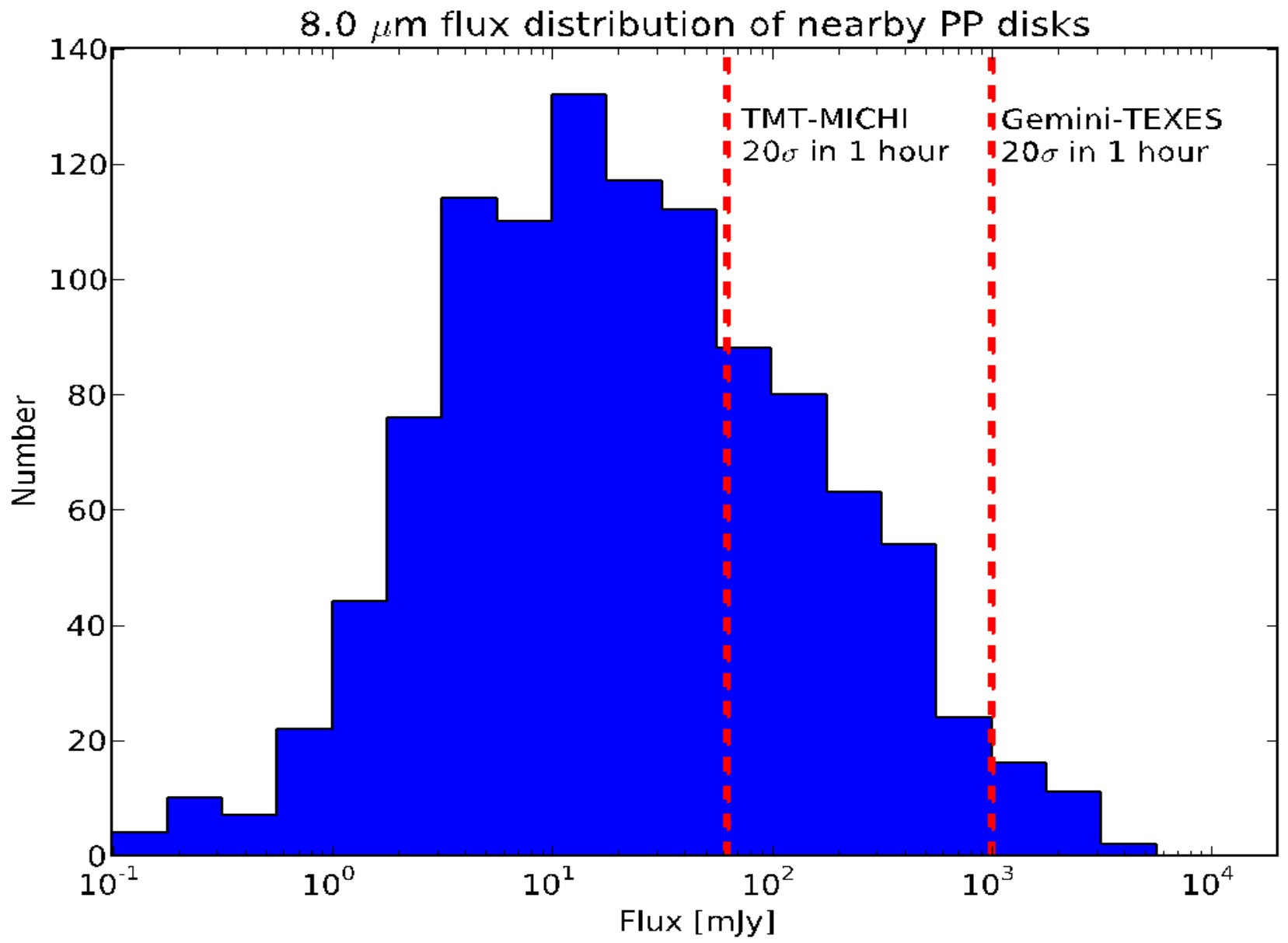

***Figure 8-10:*** *The 8 μm brightness distribution of protoplanetary disks and young stars in nearby star-forming clouds as observed with the Spitzer Cores to Planet-Forming Disks legacy survey (Evans et al. 2009). It shows how many more protoplanetary disks can be reached with the TMT relative to current high-resolution mid-infrared spectrometers.*

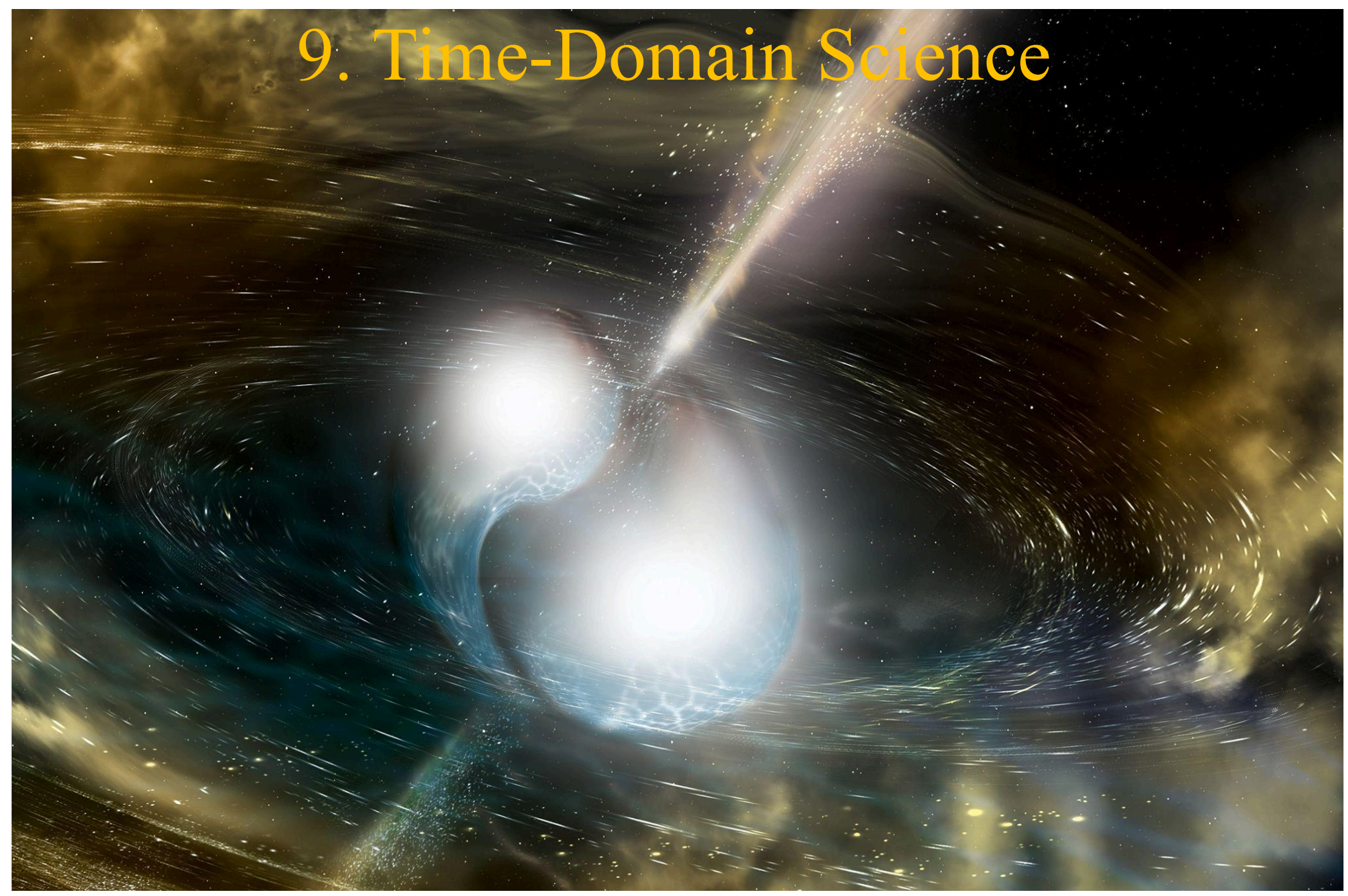

*Merging neutron stars, as shown in this artist rendition, play the key role in the generation of heavy elements in the universe. Observing the explosions with the Thirty Meter Telescope, will advance multi-messenger astronomy by allowing the detailed outcomes of gravitational wave events to be studied, with important ramifications for our understanding of the evolution of heavy elements in the universe. (National Science Foundation/LIGO/Sonoma State University/A. Simonnet).*

---

Time-domain Science with TMT directly addresses big questions Q2-*When did the first galaxies form and how did they evolve?* and Q1-*What is the nature and composition of the universe?)* TMT observations of transient, explosive targets that contribute to the chemical enrichment of the universe, at early times and beyond, will help us unravel the mysteries of galactic evolution. TMT observations of Type Ia supernovae at larger redshifts than is currently possible will extend their use as part of the cosmic distance ladder, further increasing our understanding of the universe and its evolution. As advanced LIGO and other gravitational wave detectors come on line, TMT will be able to rapidly follow up on sources to explore these as part of multi-messenger campaigns, learning about the exotic physics of binary neutron stars and the origin of heavy elements in the universe. TMT observations of tidal disruption events in black holes in other galaxies will help us address Q3**-***What is the relationship between black holes and galaxies?* Capabilities required for time domain astronomy, namely the ability to schedule time critical observations and conduct sequences of time-resolved observations, will also support studies of periodically varying systems such as exoplanets, RR Lyrae and Cepheid variable stars, addressing big questions Q1 and Q5-*What is the nature of extrasolar planets?*

*Responding to Gamma Ray Burst discoveries exemplifies the science cases that drive the rapid response capabilities of the entire observatory.* As chapter 13 summarizes, rapid ToO response times from ~1 hr to days, and sample times as low as milliseconds are essential for most of the observations needed to support the science discussed here. In all cases, TMT's high sensitivity and angular resolution allow new classes of observations compared to what is possible now. In many cases, wavelength coverage extending to the ultraviolet and the mid-infrared are essential and spectropolarimetric and coronagraphic capabilities would add additional value. In the optical and near-infrared, IRIS and WFOS are well-suited to these observations.

---

**Contributors**: G.C. Anupama (IIA), Manjari Bagchi (Tata Institute of Fundamental Research), Varun Bhalerao (IUCAA), Sarah Gallagher (Western University), U.S. Kamath (IIA), Shigeo Kimura (Tohoku University), Lucas Macri (NSF NOIRLab), Keiichi Maeda (Kyoto University), Ashish Mahabal (Caltech), Stan Metchev (Western University), Amitesh Omar (IITK), Shashi Bhushan Pandey (ARIES), Enrico Ramirez-Ruiz (UCSC), Warren Skidmore (TMT), Guy Stringfellow (University of Colorado, Boulder), Masaomi Tanaka (Tohoku University), Nozomu Tominaga (NAOJ), Lingzhi Wang (NAOC), Xiaofeng Wang (Tsinghua), Chao Wu (NAOC), Xufeng Wu (Purple Mountain Observatory), Alan Marscher (Boston University), Vishal Kasliwal (Drexel University), Enrique Lopez-Rodriguez (UT San Antonio)

---

# 9 TIME-DOMAIN SCIENCE

Time domain astronomy as discussed here is the study of transient and variable sources, see figure 9-1. A few transients arise from extrinsic effects at optical wavelengths, like binaries, microlensed objects and transiting planets. However, most transients are intrinsically variable, e.g., Cepheid variables, stellar flares, Be stars, Luminous Blue Variable stars, cataclysmic variables, young stellar objects, AGN. Accretion has a short dynamical timescale, and so it can result in variability not just because of accretion events, but because of instabilities in the disks, for example. This chapter discusses mostly energetic intrinsic transients. Intrinsically variable transients often result from some kind of explosion or collision that leads to a change in the physical character of the source (e.g., supernovae, gamma-ray bursts, neutron star mergers), or a result of accretion of matter (nova and dwarf nova outbursts, tidal disruption events, AGN flares, and flares on young or forming stellar objects).

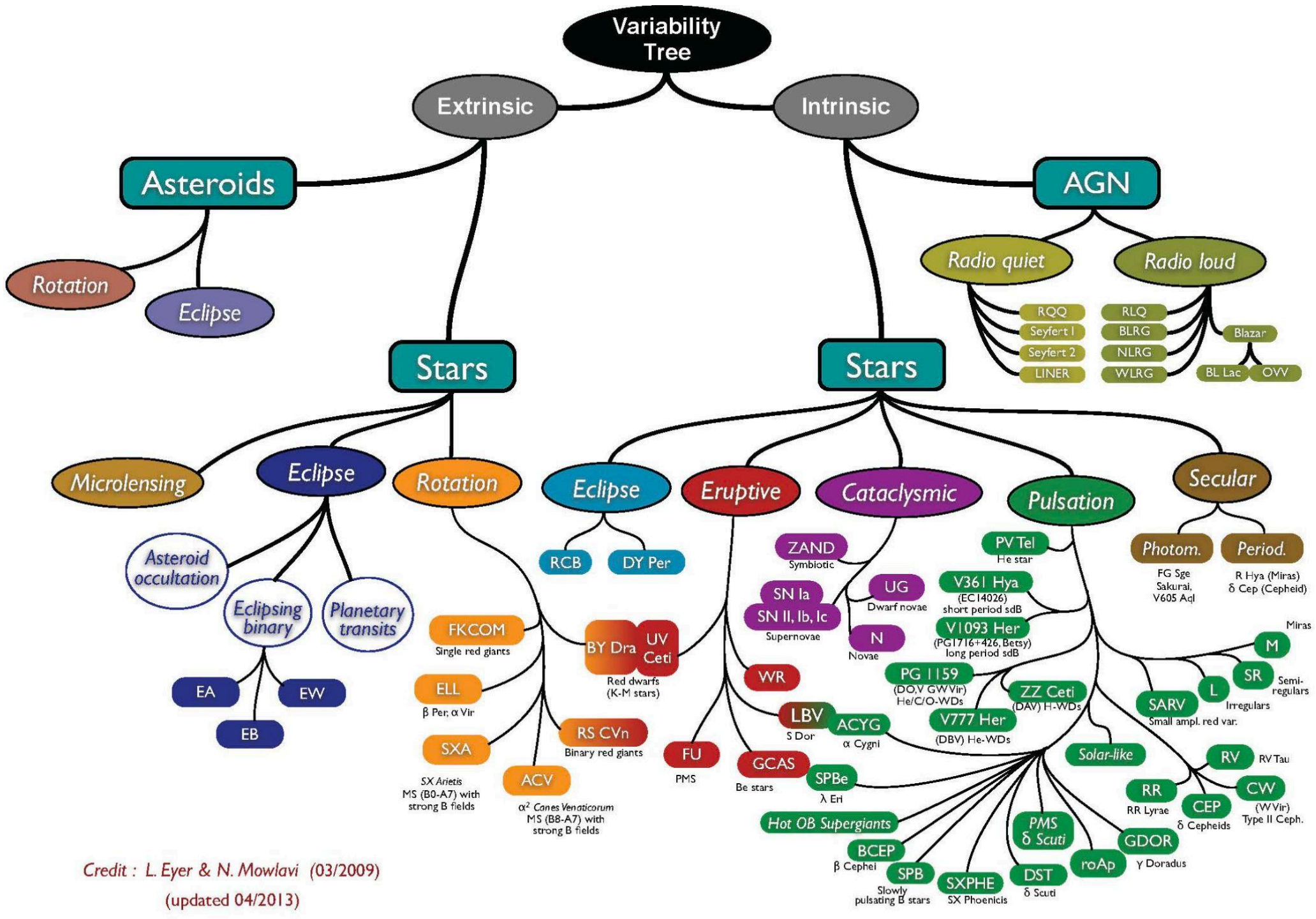

***Figure 9-1:*** *A variability tree illustrating the vast range of variable phenomena in the universe. This figure is not comprehensive and within one type of variable there can be complex combinations of different physical processes than conveyed here. In this chapter we describe a subset of examples that have high scientific impact and require more stringent observing capabilities and operational modes. All of the examples in this chapter are energetic intrinsic transients, of which many occur at cosmological distances.*

These events are unpredictable, show a temporal evolution of the physical conditions, and often fall below the detection threshold when faint. Such Target of Opportunity (ToO) transient events generally require a rapid response to a trigger of observations. Variability observations can be time dependent or time critical, both of which require observations that are time-resolved.

Time-domain astronomy will be an exciting area of research in the TMT era, in part thanks to the synergy with various surveys from up-coming facilities such as the Rubin Observatory and its Legacy Survey of Space and Time, Roman, and Euclid, and the search for gravitational wave sources (section 9.1). In past decades, the fast localization and generation of alerts for Gamma Ray Bursts (GRB) has set the most aggressive requirements on rapid followup by ground based observatories. More demands are expected for so-called "multi-messenger" astronomy to realize real-time synergetic observations with gravitational waves (GWs) and neutrinos. At the present time, large strides are being made in the application of machine learning to confidently refine and characterize targets based on variability and external contextual information out of the vast number of variable and transient candidates found in astronomical surveys. Alerts with high confidence for the correct classification of many forms of explosive or rapidly evolving transient phenomena can be generated very quickly. In the era of TMT, fast generation of alerts where confidence in the classification is high and the need for rapid followup in the early stages of transient events is going to be commonplace.

In this chapter, we discuss a few interesting time-domain science cases that benefit from the unique capabilities of TMT, notably rapid response, blue sensitivity and adaptive queue scheduling. Important science areas that are not discussed in this chapter but that could definitely benefit from the rapid response, adaptive scheduling and time resolved observing capabilities of TMT include, amongst other things, followup of microlensing events to characterize the lens and lensed objects (sections 3.3 and 10.4), Young Stellar Object variability, exoplanet transits and eclipses to identify atmospheric components, exoplanet growth processes and improve the sensitivity of masking effects from host star activity (section 10.3.4), various processes on solar system objects (sections 11.2.2 and 11.3.2), and astrometric studies of stellar motions to study black holes, star and galaxy formation, and dark matter (sections 3.1, 7.6, 6.1.3). The observing requirements of science programs discussed in detail here are summarized in figure 9-15, figure 9-16 and Table 9-1.

## 9.1 MULTI-MESSENGER OBSERVATIONS OF GRAVITATIONAL-WAVE SOURCES

Multiple gravitational-wave (GW) detectors, such as advanced LIGO, advanced Virgo, KAGRA, and LIGO India, will be operating in the TMT era. The first detection of GWs from a binary neutron star (BNS) merger in 2017 (GW170817) and the subsequent observation of the electromagnetic (EM) counterpart (figure 9-2) have shown the great potential for these events both as a means to study the origin of r-process elements, and as the distance can be determined from the GW signal itself, as a method to measure the Hubble constant at the location of the merger that is independent from the traditional cosmological distance ladder. In the literature, the terms *kilonovae* and *BNS merger* are used interchangeably, usually but not always depending on whether the EM (kilonova) or GW (BNS merger) signatures are being described. TMT will greatly contribute to the multi-messenger observations of BNS mergers.

The optical and near-infrared EM emission powered by radioactive decay energy following a BNS merger is of great interest as these are the so called kilonovae (Li & Paczynski 1998; Kulkarni 2005; Metzger et al. 2010) and are associated with short-GRBs (Gompertz, et al., 2018), as discussed in section 9.6.The observations of kilonova AT2017gfo, the identified EM counterpart to the BNS merger event GW170817, clearly demonstrated for the first time that BNS mergers are a viable site of r-process nucleosynthesis. Through analysis of the multi-band light curves of the kilonova it has been concluded that heavy elements including lanthanides (elements with an atomic number 57–71) were produced in the BNS merger (e.g., Kasen et al. 2017; Tanaka et al. 2017). Specifically, the presence of heavy r-processed elements such as Sr (Z=38) and Ce (Z=58) were identified (figure 9-2) (e.g., Watson et al. 2019; Domoto et al. 2022). These demonstrate the power of multi-messenger observations to directly study the site of heavy nucleosynthesis, which had never been possible in the history of astronomy.

However, as of mid-2024, there is only one multi-messenger observation of a BNS merger event. Therefore, an entire picture of r-process nucleosynthesis in BNS mergers still remains unclear. In the universe, BNS merger events can occur with different NS masses or mass ratios, and thus, a variety of nucleosynthetic patterns is expected. In fact, the second BNS merger event detected with GWs (GW190425 - without a detected EM counterpart) has very different component masses from GW170817. Therefore, multi-messenger observations of a variety of BNS mergers with different component masses are essential to fully understand the origin of heavy elements in the universe.

In the TMT era, the sensitivity of GW detectors will exceed the horizon distance of 200 Mpc. The number of BNS merger detections will be $\gtrsim$10 per year (e.g., Abadie et al. 2010), and the first meaningful statistical samples will be available. With the GW detection alone, the position of the sources can only be moderately determined with a localization of about 10–100 deg$^2$ (e.g., Abadie et al. 2012; Nissanke et al. 2013). Transient search observations with wide-field 8 m class telescopes in the optical and wide-field space telescopes in NIR (e.g., Roman) are crucial to detect the electromagnetic counterparts of GW sources. Such deep transient surveys will discover a large number of SNe with different spectroscopic signatures than BNS events within the GW localization area. Therefore, rapid ToO response follow-up spectroscopy (R~500) within hours with either TMT WFOS or IRIS will be critical to identify the detected transient as a true GW source. In fact, Chornock et al. (2019) estimate that ELTs will be sensitive to kilonovae out to z~0.2. For events occurring within 200 Mpc, TMT can provide the optical and NIR spectral evolution of kilonovae, which enables us to study the heavy element production in a variety of BNS mergers.

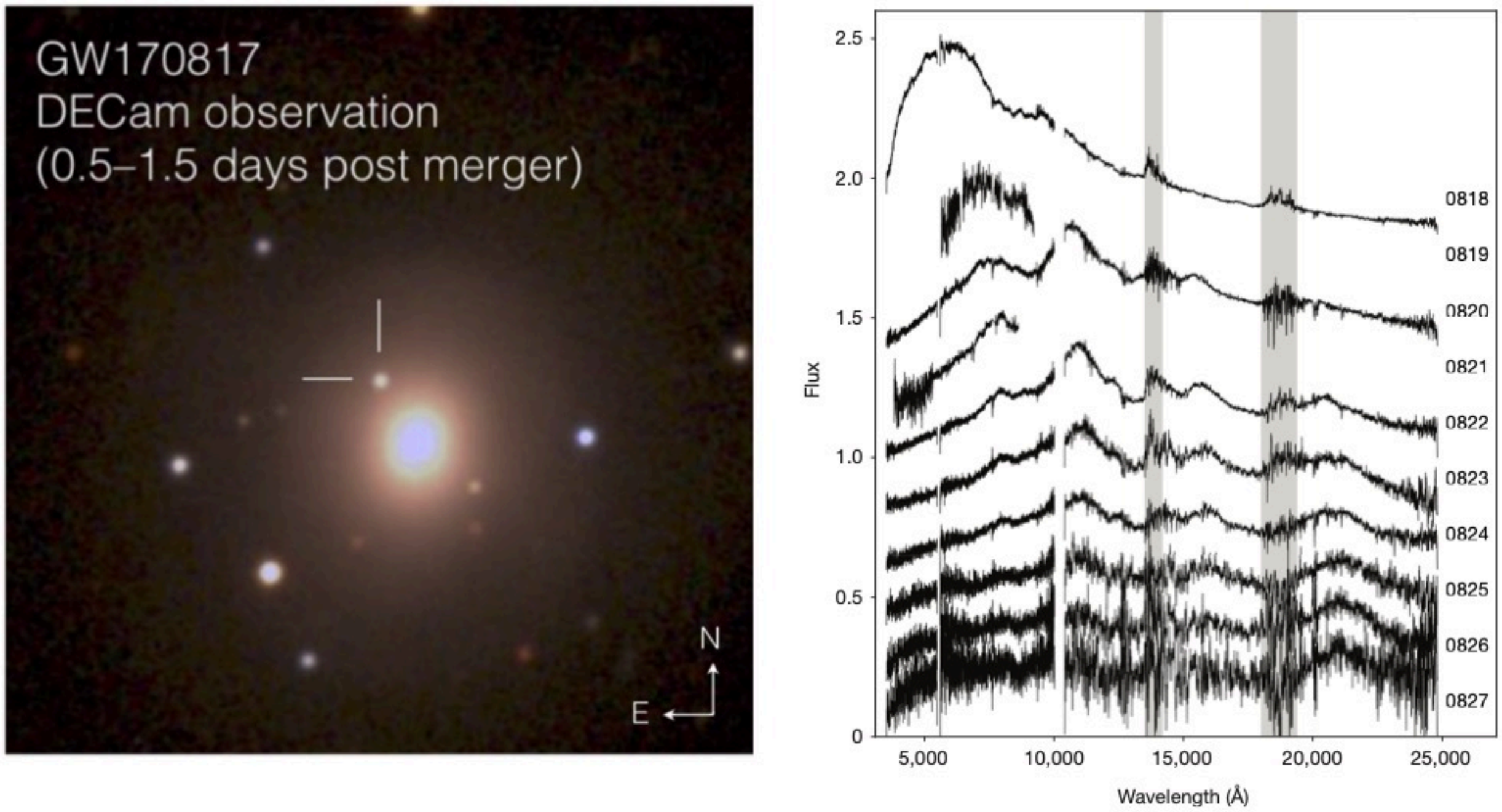

***Figure 9-2:*** *(Left) Optical image of the EM counterpart of BNS merger GW170817 (Soares-Santos et al. 2017). (Right) Spectral evolution of the EM counterpart (Pian et al. 2017) with dates indicated on the right. Note the rapid EM spectral evolution beginning from day one. Rapid response to transients is crucial to secure the necessary diagnostic data that enables classification and elucidation of the physics.*

## 9.2 MULTI-MESSENGER OBSERVATIONS OF HIGH-ENERGY NEUTRINO SOURCES

Our universe is filled with high-energy charged particles (cosmic rays), whose origin has been unknown for more than a century. Since high-energy cosmic rays produce neutrinos via interactions with ambient protons or photons, high-energy neutrinos are thought to be a smoking-gun signature to identify cosmic-ray sources. IceCube, a Giga-ton volume neutrino detector in Antarctica Ice, reported the evidence of extraterrestrial high-energy neutrinos in 2013, which opened a new window of the universe (Aartsen et al. 2013). IceCube has been detecting high-energy neutrinos for more than 10 years and established the existence of the cosmic high-energy neutrino background in TeV-PeV energies (Aartsen 2020). The origin of the cosmic high-energy neutrino background has been actively debated since its discovery, but it is still veiled in mystery. To identify the neutrino-emitting objects, IceCube constructed an IceCube alert system (Aartsen et al. 2017), which promptly circulates information of cosmic neutrino candidates and enables multi-wavelength follow-up observations. This strategy led to identification of the first neutrino emitting object, a flaring blazar TXS 0506+056 (IceCube Collaboration et al. 2018, figure 9-3). In addition, two tidal disruption events, AT2019dsg and AT2019fdr are reported to be associated with high-energy neutrino events (Stein et al. 2021; Reusch et al. 2022).

Nevertheless, these associations are statistically not enough to conclude what are the dominant sources of the cosmic high-energy neutrino background. We need to identify more counterparts to cosmic neutrino signals. The difficulty to identify neutrino sources stems from numerous unrelated transients appearing in neutrino error regions. The error regions by IceCube events are typically a few deg$^2$, and the expected redshift to neutrino sources are $z \sim 1$ (Yoshida et al. 2022). Then, we would find ~100 unrelated supernovae in the error regions if we perform deep photometric observations. Subclasses of core-collapse supernovae, such as interacting supernovae and jet-driven supernovae are proposed as neutrino sources, but currently, it is highly challenging to identify core-collapse supernovae as a neutrino source.

In the TMT era, we will have next-generation neutrino detectors in both glacial Ice (IceCube-Gen2: Aartsen et al. 2021) and in deep sea (KM3NeT: Adrián-Martínez et al. 2016; TRIDENT: Ye et al. 2023), whose angular resolution could be smaller than 10' for neutrino events of PeV energies. After a neutrino detection, 8 m class or space telescopes will be able to identify an optical transient, including core-collapse supernovae, at a cosmological distance and alert followup facilities of the source location. Next, TMT will be able to perform spectroscopic observations of faint transients, which would enable us to identify any type of transients, including interacting supernovae and jet-driven supernovae, as neutrino counterparts in the 2030s.

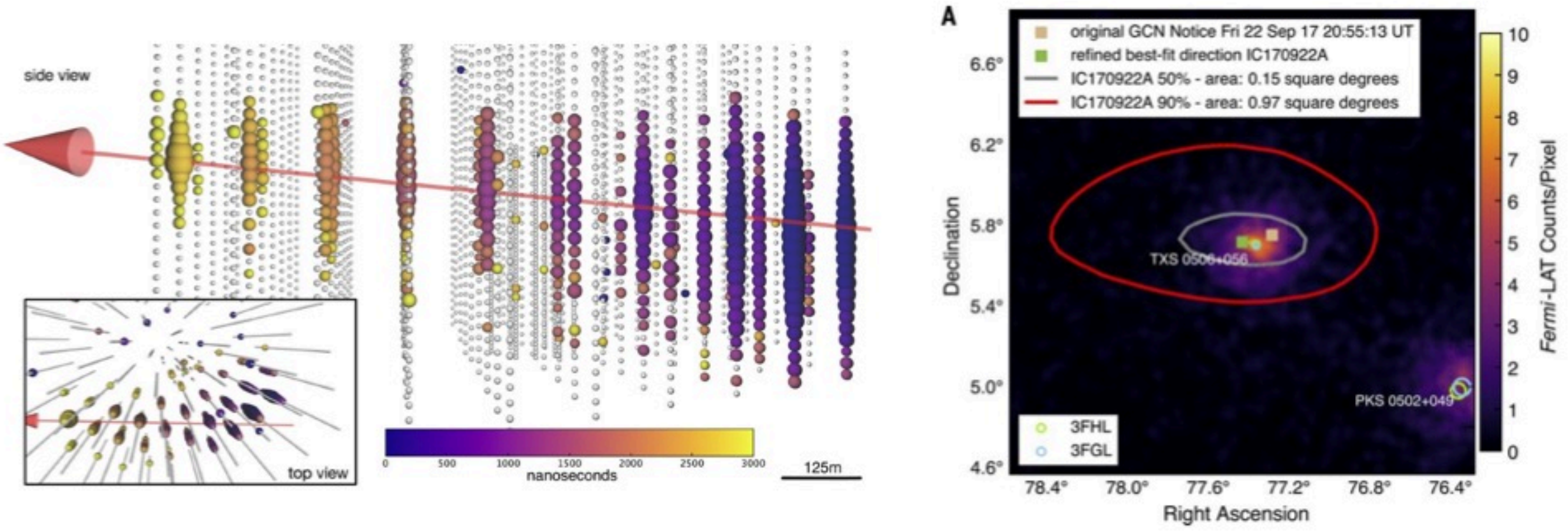

***Figure 9-3:*** *(Left) Neutrino event IceCube-170922A. (Right) Gamma-ray image of the EM counterpart TXS 0506+056. Image from IceCube Collaboration (2018).*

## 9.3 UNDERSTANDING THE NATURE OF TYPE IA SUPERNOVAE

Over the past decades, Type Ia supernovae (SNe Ia) have been used as one of the best distance indicators to measure extragalactic distances. Observations of such events at higher redshifts directly led to the discovery of the accelerating universe (Riess et al. 1998; Perlmutter et al. 1999). Of the existing methods to measure the dark energy equation-of-state (EOS) parameter $w$, the distances to SNe Ia provide the best single constraint to date. This inspired many wide-field supernova surveys such as the Palomar Transient Factory (PTF), the La-Silla Quest Supernova Survey (LQSS), and the Dark Energy Survey (DES) to search for and discover thousands of SNe Ia that exploded in the local and distant universe. With planned future wide field surveys, this number will increase dramatically, especially at high redshifts .

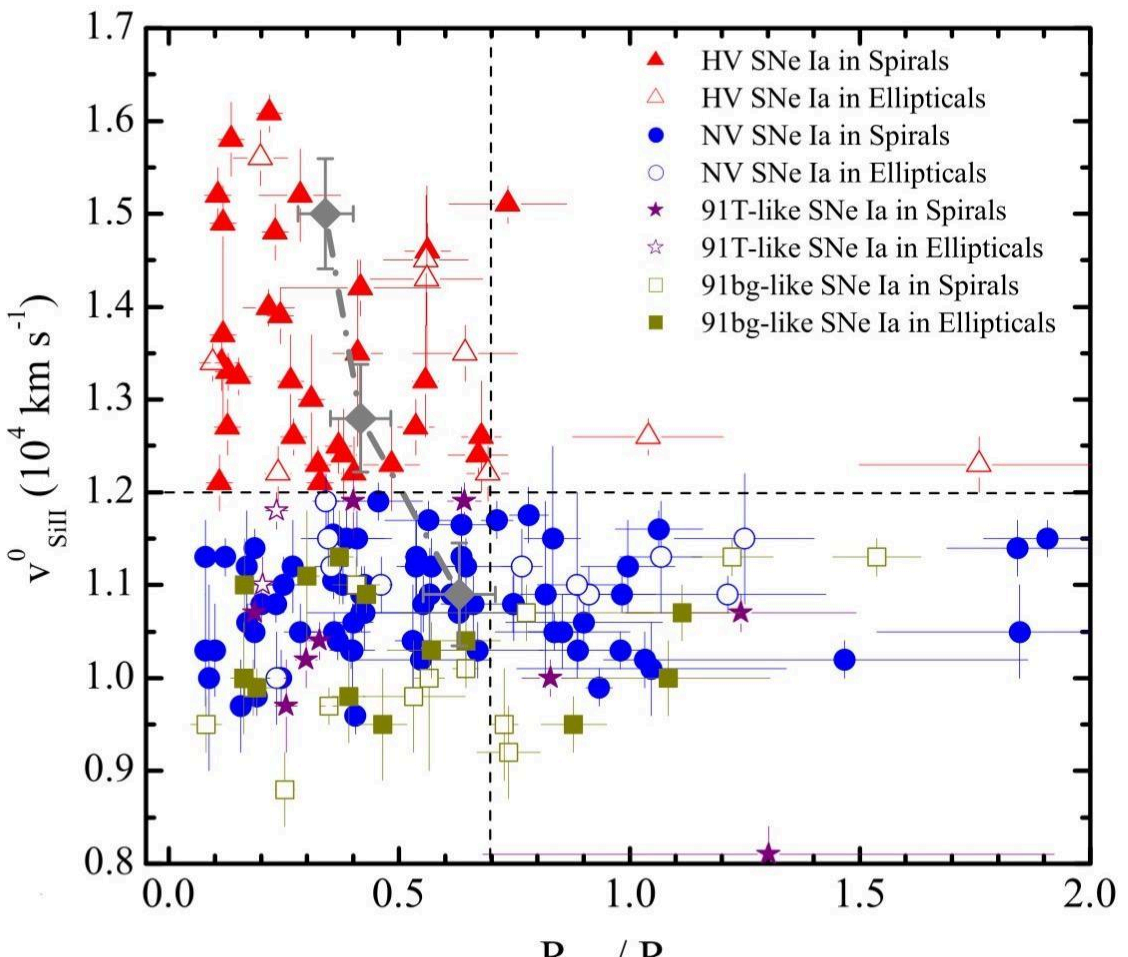

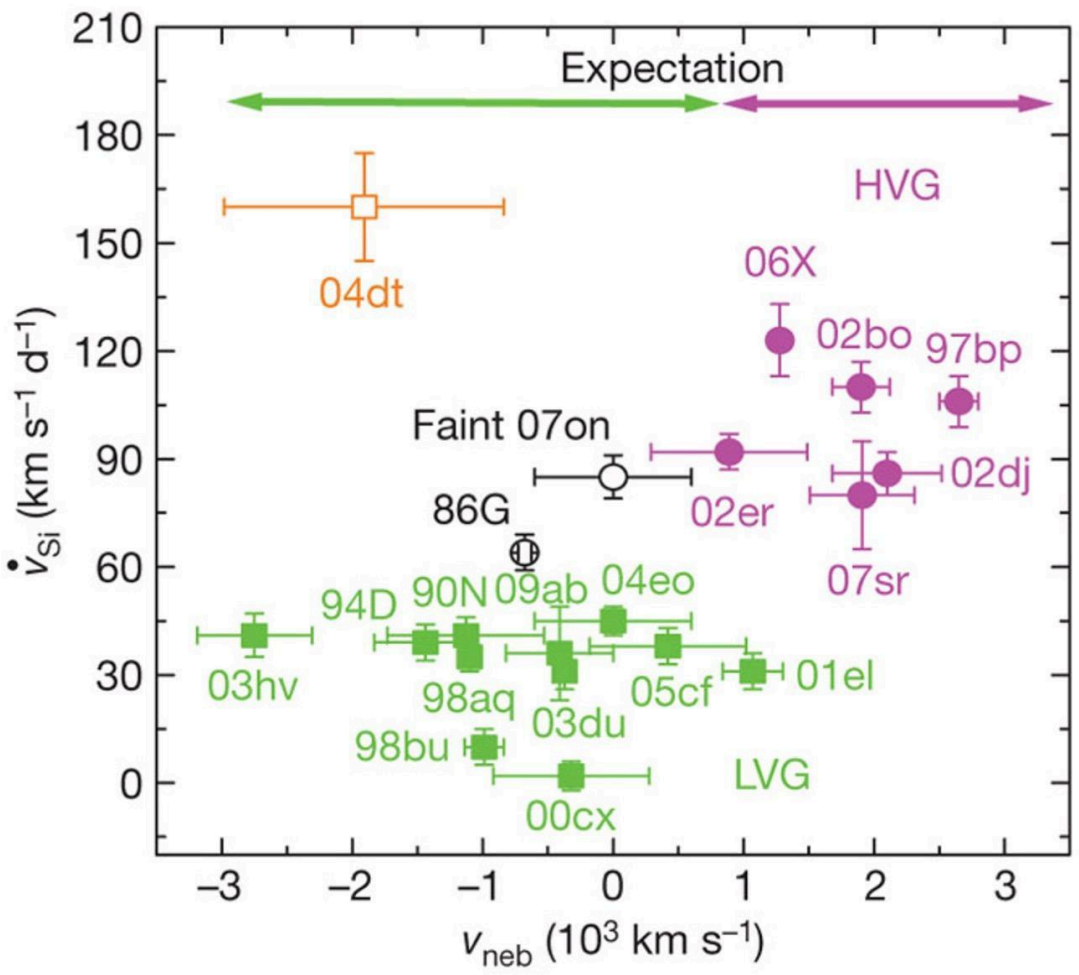

***Figure 9-4:*** *Observational indications for the origin of the observational diversity in SNe Ia. (A) High-velocity (HV) SNe Ia (triangles) are more concentrated in the inner and brighter regions of their host galaxies than normal-velocity (NV) SNe Ia (circles; figure from Wang et al. 2013), indicating that velocity is related to progenitor metallicity. (B) HV (magenta) and NV (green) SNe Ia have systematically different velocity shifts in their nebular emission lines but both are consistent with the expectations for the velocity gradient relation with velocity shift predicted by a hydrodynamic explosion models viewed from different angles and without any relation to the progenitor metallicity (Figure from Maeda et al. 2010).*

An increased sample of SNe Ia will reduce the allowed ranges of various cosmological parameters due to improved statistics (Pandey, 2013). However, accurate determination of the dark energy EOS parameter still relies on a better control of the systematic effects in determining the distance to each SNe Ia. The larger systematic uncertainties include dust absorption, environmental dependencies of the observables, and possible evolution with redshift (e.g., Howell 2011; Wang et al. 2013). TMT will make important contributions in unveiling the nature of SNe Ia due to its large light collecting area, IR capability, and powerful AO system, thereby greatly improving the precision of which these cosmological techniques and resulting derived parameters can be determined.

### 9.3.1 Characterizing high-z Type Ia Supernovae: Towards a Better Standard Candle

Recent reports suggest that SNe Ia with high-velocity (HV) ejecta (e.g., V(SiII) > 12,000 km/s) tend to inhabit larger and more-luminous hosts and are substantially more concentrated in the inner and brighter regions of their host galaxies than the normal-velocity (NV) sample (V(Si II) ~ 10,000 km/s; figure 9-4A). The HV and NV populations have different colors around maximum brightness (Wang et al. 2009). These results suggest that HV SNe Ia are more likely to originate from more metal-rich progenitors than NV SNe Ia, and are restricted to galaxies with substantial chemical evolution. As an alternative to the probable distinction between the HV and NV SNe Ia as different populations, Maeda et al. (2010), based on the ejecta kinematics in the deepest layer as seen in the optically thin, late-phases, suggest that viewing angle is another factor to distinguishes the HV and NV appearances (figure 9-4B). The dependence of SNe Ia ejecta velocity on the progenitor environment could be relevant when using SNe Ia as cosmological yardsticks, as this may lead to a change in the ratio of the HV and NV population with redshift. Thus, an important task in precision cosmology is reconciling the observed indications of a relation with metallicity and environment against the ability of a single model to explain the velocity differences based on viewing angle. The observed relative fraction of the HV to NV population might become smaller at great distances due to a real decrease in the HV SNe Ia rate in low-metallicity environments and/or may appear to be smaller due to the increased difficulty of observationally spectroscopically classifying fainter apparent magnitude SNe in the central regions of more distant galaxies. Observationally testing for evolution in the low-z and high-z samples is important for precision cosmology, but can be only explored with future high-redshift SNe Ia samples having better spectroscopic classifications, as the spectral quality and wavelength coverage of current high-z samples are not good enough for this kind of analysis.

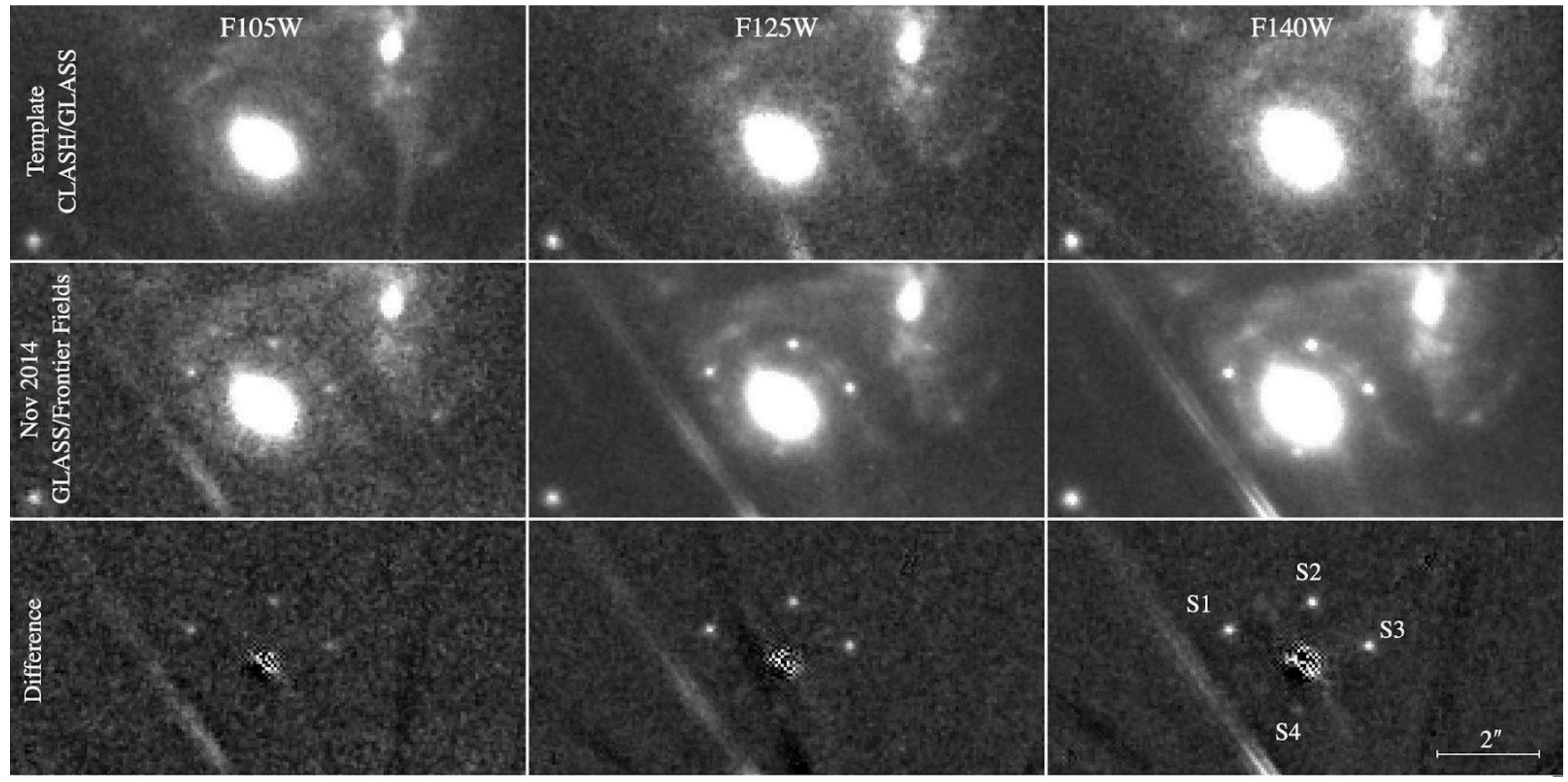

***Figure 9-5:*** *HST WFC3 images of a supernova gravitationally lensed by an early type galaxy cluster (Kelly et al. 2015)*

The Rubin Observatory will discover thousands of high-redshift SNe Ia ($z \sim 1.0$) within 1 to 2 days after explosion, and spectroscopic follow-up for these SNe will be needed beginning within a few hours after their discovery, along with repeat observations starting with 1 day cadence to understand the progenitors and explosion mechanisms. Near-infrared space missions such as Euclid and Roman can discover SNe Ia at higher redshifts ($z > 1.0$). Gravitational lensed SNe can also be discovered at higher redshift ($z > 1.0$, Amanullah et al. 2011), requiring TMT for follow-up. Surveys looking for lensed SNe behind massive galaxy clusters are ongoing with 8 m class telescopes and HST, extending the redshift range beyond $z \sim 2$ (see section 3.1.6). SNe can also be strongly lensed by a relatively faint galaxy, as seen by Quimby et al. (2013, 2014) and Kelly et al. (2015) and shown in figure 9-5. Discovery of such systems could be accelerated by new selection techniques. Present ground-based facilities cannot obtain high-quality spectra for SNe Ia beyond $z \sim 1.5$.

IRIS on TMT can easily obtain spectra of SNe up to $z \sim 4$ with better wavelength coverage, enabling accurate spectroscopic classification for future samples of high-z SNe Ia and IRIS will provide important information on the host galaxy such as the SFR. Combined with the low-z SNe Ia, the high-quality spectra obtained with TMT for the high-z SNe Ia discovered with Rubin, Roman and/or Euclid could investigate possible evolution of SNe Ia populations with redshift. It should become possible to use a specific subclass of SNe Ia that are more uniform to constrain the nature of the dark energy EOS parameter and its dynamic behavior.

### 9.3.2 Unveiling The Explosion Mechanism of Type Ia Supernovae and Characterizing the Circumstellar environment around Supernovae - Clues to the Identity of the Progenitor Systems

The explosion mechanism of SNe Ia is yet to be clarified. The explosion mechanisms are dependent on the progenitor systems, e.g., single degenerate vs. double degenerate, but also there are variants within the same progenitor model. Since the explosion mechanism sets the ejecta structure and $^{56}$Ni yields, which determine the optical luminosity and light curve, understanding the explosion mechanism is imperative toward developing precision SNe Ia cosmology. Understanding the physical causes of peculiar SNe Ia, such as SN 2005hk (Sahu et al. 2008, Foley et al. 2013) and 2005gj (Hughes et al. 2007; Prieto et al. 2007), will help to constrain models for normal SNe Ia and illuminate the underlying reason for the diversity of normal SNe Ia. Understanding the diversity of SNe Ia has been dramatically evolved over the last decade (e.g., Taubenberger 2017); the emerging diversity requires multiple-channels, probably both in the progenitor systems and explosion modes (e.g., Maeda & Terada 2016). It is becoming feasible to map different observed subclasses of SNe Ia with theoretically proposed scenarios, with a few subclasses robustly linked to specific scenarios (e.g., Jiang et al. 2017). The increasing number of high-quality time-resolved light curves of SNe starting just after explosion has been the key in model-data comparison (Maeda et al. 2018), complemented by the power of still-rare very-early spectra (Maeda et al. 2023).

Early-phase spectra, so far restricted to the local universe, are highly dependent on the explosion mechanism. Early-phase spectra reveal the outermost ejecta composition, which is highly dependent on the mode of the explosion. The carbon abundance, especially, is quite sensitive to the progenitor evolution and thermonuclear flame propagation mode. Recent results suggest that unburned carbon frequently remains in the outermost layer of SNe Ia, more commonly for the NV SNe (Parrent et al. 2011; Folatelli et al. 2012). Given the environmental and explosion mechanism details between the HV and NV SNe, investigating the carbon signature in the high-z sample is an important step. In addition to carbon, the near-IR magnesium feature should be present (Marion et al. 2009), and measuring this oxygen-burning product is key to determining the extent of thermonuclear burning.

An early response follow-up of a series of observations with a few minute cadence spanning a few hours within 2 days of the start of explosion with IRIS or WFOS, for a large sample of intermediate-to-high-redshift SNe Ia will render important progress to understanding the details of the explosion mechanism and resulting nucleosynthesis of the ejecta. A large systematic data set spanning the evolution of the spectrum over a broad wavelength range from soon after the explosion for several days as the explosion proceeds will facilitate the mapping between theory and observation using both statistical analyses (e.g., Ogawa et al. 2023) and deep-learning techniques (O'Brien et al. 2021). Early-time spectra also provide powerful diagnostics on the nature of a companion star and/or dense CSM in the very vicinity of the progenitor system through their possible observable signatures in the first few days before they are overwhelmed by the direct SN emission (e.g., Maeda et al. 2023).

The very late-time spectra in the evolution of SNe provides a strong diagnostic of the ejecta kinematics and thermonuclear products at the explosion trigger, thus being a direct method to constrain the explosion mechanism (Maeda et al. 2010). The sensitivity the IRIS and WFOS spectrographs when fed by TMT will result in a formidable tool for analyzing the explosion mechanism and ejecta in unison by opening new time-domain windows for exploration: (1) NIR spectroscopy at ∼1 year and (2) a very late-time (at >1 year after the explosion) optical spectroscopy. Sensitivity limits imposed by current ~10 m class facilities limit optical spectroscopic observations to roughly 1 year after explosion. Current capabilities in the near-IR, where few observations exist for ~10-m class facilities, show unique and fundamental signatures that provide complimentary analysis in tandem with optical spectral analysis (Motohara et al. 2006), as can be seen for example in GW-BNS explosions (figure 9-2).

Very late-time spectroscopy is essentially unexplored due to the faintness of the SNe at late evolutionary times, along with the required spatial resolution to resolve the SNe from the blended background light of the nearby extragalactic environment in which they reside and are fading into. Such observations to constrain the details of the explosion mechanism will greatly expand our understanding of processes in supernovae, but the observations are impossible with 8–10 m class observatories and require the sensitivity and spatial resolution that the IRIS-IFU will provide. During late times, the dominant energy source is no longer γ-rays, but positrons from the $^{56}$Co decay and other minor radioactive species (Seitenzahl et al. 2013). Late-time spectroscopy will further allow detection and analyses of newly-formed dust that might take place in some SNe Ia subclasses, once combined with mid-IR data (as demonstrated by the first clear detection of the dust emission by JWST and some ground-based 8 m class telescopes; Siebert et al. 2024), which provides a new direction in identifying the progenitor systems. TMT's light gathering power and instrument sensitivity will allow these types of observations.

Detection of circumstellar material (CSM) around SNe could provide an alternative way of discriminating different evolution scenarios. If SNe Ia arise from single white dwarfs in binary systems, one generally expects that there should also be considerable H-rich (or possibly He-rich) CSM within tens of AU around the progenitor system. Claims for detection of such CSM around a few nearby SNe Ia have been made based on the discovery of time-variable Na I absorptions in several cases (Patat et al. 2007; Simon et al. 2009, Dilday et al. 2012; figure 9-6). On the other hand, there is also evidence for non-detections of variable Na I lines in SNe Ia (Simon et al. 2007; Patat et al. 2013). This apparent dichotomy might be related to differences in the progenitor environments, as suggested by the correlation of the ejecta velocity $V_{SiII}$ with the location in their host galaxies (section 9.3.1). Nevertheless, the origin of the absorbing clouds is still controversial as such absorptions may arise either intrinsically in a double degenerate scenario or from the interstellar medium along the line of sight (Chugai 2008; Shen et al. 2013). To address this issue, one needs a larger sample of SNe Ia with multi-epoch, high-resolution spectra.

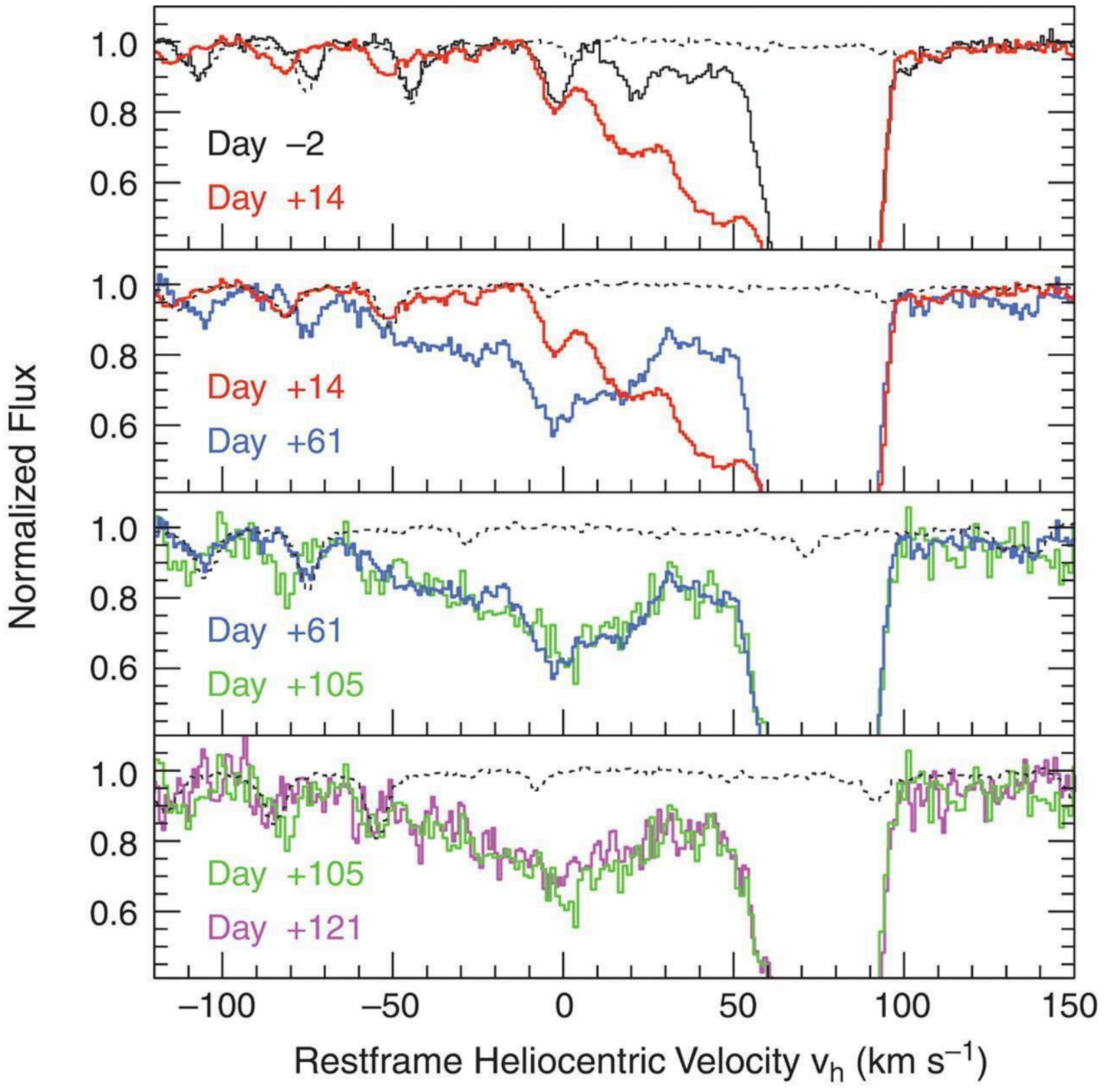

***Figure 9-6:*** *Time series of high-resolution spectra of the SNe Ia SN 2006X in the Virgo cluster galaxy NGC 4321(Patat et al. 2007). Time variability of Na I absorption indicates that absorbing material is located in the CSM around SNe Ia. TMT with WFOS will be able to gather similar observations at distances approaching 60–70 Mpc.*

At the distance of the Virgo cluster (~16 Mpc), the typical R-band magnitude for a SNe Ia within 1 day after explosion is about 19–20, with maximum brightness occurring 10 to 20 days later. TMT/WFOS can provide the highest sensitivity with complete wavelength coverage (0.33–1.0 μm) in a slit-viewing target acquisition mode. A resolution of R~7500 enables measurements of the overall evolution of Na I and Ca II absorption, allowing important parameters of the absorbing material which may be associated with the progenitor system to be determined. This includes the distance to the absorbing material, recombination timescale of Na ions, and outflow velocity, parameters important for understanding the origin of the absorbing cloud and hence the nature of the companion stars.

This will be further enabled with the multiple-mode and multiple-wavelength observations to be available in the TMT era. The CSM is expected to manifest itself in radio emission, as has recently been reported for the first time for a specific SN Ia-CSM with a massive CSM surrounding the progenitor system (Kool et al. 2023). With new radio facilities in the TMT era (e.g., ngVLA, SKA, ALMA2), the substantial increase in the number of detected cases is anticipated. A polarization capability of TMT would provide an exciting new opportunity — rapid spectral-polarization observations of a number of local SNe Ia within a day of the explosion. This would trace both the structure of the outermost layer and CSM, the methodology as highlighted recently by such activity for an extremely nearby core-collapse SN 2023ixf (Vasylyev et al. 2023).

## 9.4 IDENTIFYING THE SHOCK BREAKOUT OF CORE-COLLAPSE SUPERNOVAE

Stars with masses between 8 and ~50 solar masses will end their lives as core-collapse SNe with neutron star remnants. As their nucleosynthetic fuel expires, the interior collapses inward, eventually halted by neutron-neutron interactions. This results in the infalling interior rebounding and expanding, producing a shock wave that propagates outward. Eventually shock breakout occurs at the surface of the supernova. The signature of shock breakout (see figure 9-7) is manifested as a bright fireball radiating initially at X-ray and ultraviolet wavelengths that subsequently brightens at redder wavelengths as the overall bolometric flux falls (Tominaga et al. 2009). The resulting supernova light curve peaks and subsequently shifts to longer UV and optical wavelengths as the breakout material cools.

How core-collapse SNe appear observationally after shock breakout depends on the prior mass loss history of the progenitor, which depends in part on the initial stellar mass. When and how the outer layers of a supernova progenitor are shed can result in the local supernova environment either being relatively void of circumstellar material or instead having a hydrogen shell surrounding it that was previously ejected from the progenitor. As the explosion propagates through any local circumstellar environment, observational signatures reveal the nature of that material and the progenitor itself. The supernova ejecta, whose properties depend again on stellar mass, itself will expand and cool. In the case where the outer hydrogen shell has been stripped, the ejecta expand and cool producing a linear decline in the light curve after breakout - Type II-L SNe. In the case when the hydrogen shell is retained, it functions as an opacity source that the ejecta first ionizes as it passes through (the plateau phase) and then subsequently cools and recombines, resulting in the ejecta becoming transparent — Type II-P SNe.

Recent high-cadence transient surveys are discovering Type II supernovae just after the explosion. Analysis of the light curves indicate that the rise time to peak light is shorter than theoretical expectations and that spectra during the rise exhibit narrow emission lines of highly ionized ions (Khazov et al. 2016; Yaron et al. 2017, figure 9-8). These observations reveal that most progenitors of Type II supernovae are surrounded by a dense and confined circumstellar medium (CSM) believed to have formed through enhanced mass loss of the progenitor prior to undergoing core collapse and explosion (Forster et al. 2018; Moriya et al. 2018; Morozova et al. 2017). However, there are cases of Type II-P SNe where shock breakout has been detected (Gezari et al. 2008; Schawinski et al. 2008) suggesting a less dense CSM environment. Additional observations are needed to answer the questions: what fraction of progenitors produce dense and confined CSM; how are high mass loss rates achieved near core collapse when they are not predicted with current theories; what is the thickness and geometry of the CSM; what is the dependency on metallicity for all these scenarios. That current theories do not predict enhanced mass loss close to the time core collapse ensues is at odds with the limited observations at present, though theoretical scenarios where this might occur have been suggested (pulsations of red supergiants, Yoon & Cantiello 2010; core neutrino emission, Moriya 2014; near-surface energy deposition, Quataert et al. 2016; nuclear flash, Woosley & Heger 2015).

In order to answer the questions about progenitor type and mass loss prior to core collapse, rapid time-series imaging and spectroscopic observations are needed. The first role of TMT for studies of shock breakouts is spectroscopic identification. This is achieved only with a rapid (< 1 night) target of opportunity (ToO) observation. As shock breakout arises first at blue wavelengths with a predominantly featureless smooth spectra, the most suitable instrument is WFOS with $R \lesssim 500$. Spectroscopic data taken with 1–2 hr sampling would give unique information of the temperature evolution on the first day of SNe. The second role is a continuous set of a few observations per night for ~10 days following the first ToO observation to reveal the spectral evolution of supernovae during the very early epochs in which metal lines gradually become prominent (see figure 9-8). Coordination with other observatories to gather observations that are more equally separated in time will be a powerful enhancement. The evolution will provide clues to the properties of both the shock breakouts and the supernovae themselves, e.g., pre-supernova radius, CSM structure, and mass loss at the last stage of pre-SN evolution.

If there is the shock breakout at the stellar surface, it has UV-peaked spectra and it will be most effectively detected by optical facilities such as Subaru/Hyper Suprime-Cam (HSC) and the Rubin Observatory, with deep and wide-field capability (Tominaga et al. 2011). The redshift range of the detection extends to $z \geq 2.5$ with the limiting magnitude in the g' band of 27.5 mag (figure 9-7). TMT's rapid response capabilities and sensitivity in the blue with WFOS will be crucial for following up these detections and gathering the observations described above.

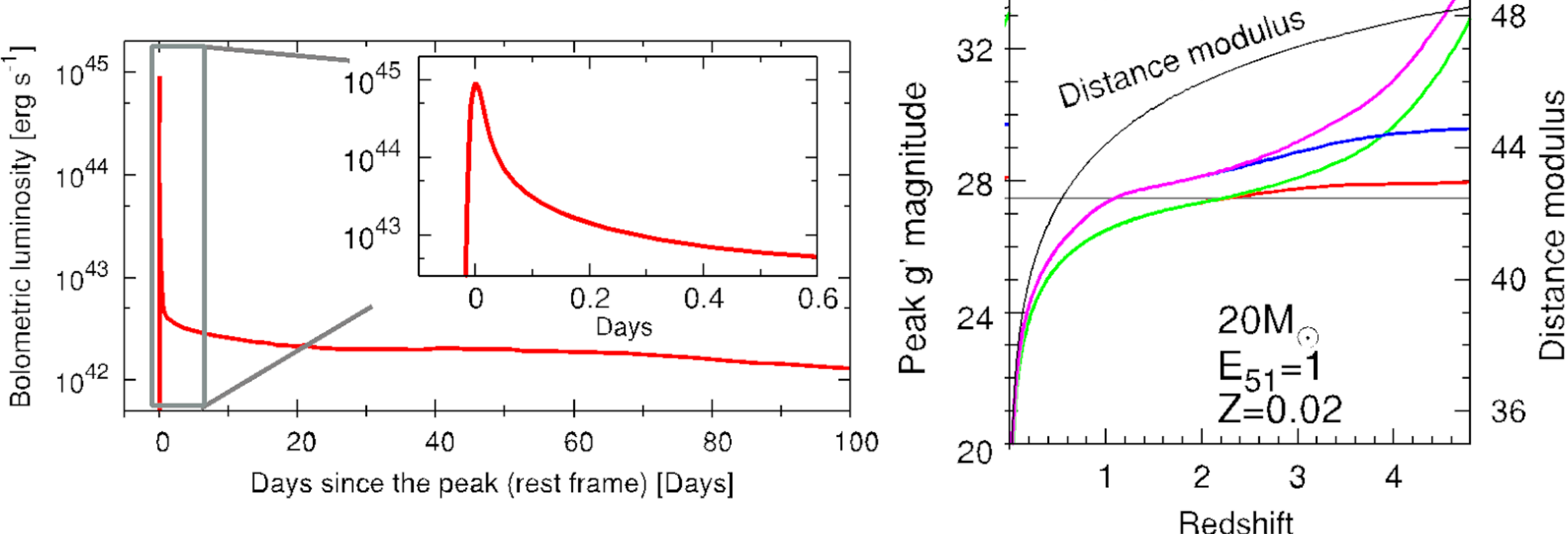

*Figure 9-7:* *(Left) Light curve of SN shock breakout (from Tominaga et al. 2009). Right: Expected peak magnitude of SN shock breakout as a function of redshift. Different colors show models with different assumptions for the reddening (from Tominaga et al. 2011). The horizontal black line shows the 1 hr limiting magnitude for Subaru/Suprime-Cam ($m_{g',lim}$ = 27.5 mag).*

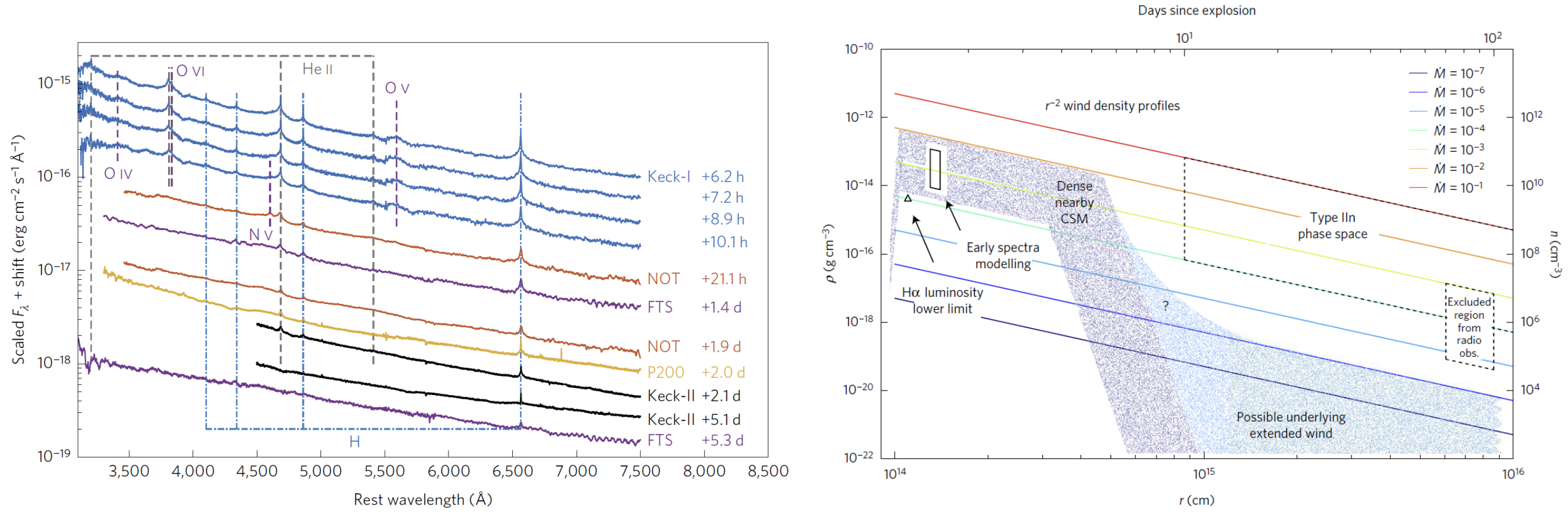

*Figure 9-8:* *(Left) A time-series of spectral observations of the core collapse supernovae iPTF 13dgy showing the spectral evolution over several days. High-ionization emission lines are visible just after the explosion. (Right) The derived density structure of CSM around iPTF 13dgy due to mass loss from the progenitor prior to the explosion (Figures from Yaron et al. 2017). TMT observations of a similar form as those shown here of core collapse supernovae would yield information about the CSM using the constraints and methods indicated. TMT would extend the redshift range and sample size significantly from the limited number of low redshift examples accessible with current facilities. Images from Yaron et al. (2017).*

## 9.5 TRACING THE HIGH-Z UNIVERSE WITH SUPERNOVAE

Core-collapse SNe are the fates of massive stars with short lifetimes and thus can be used to trace the star formation history of the universe. Although such studies are difficult at high redshift due to object faintness, a combination of planned transient surveys with 8–10 m class telescopes and spectroscopic identification by TMT will dramatically change this situation. Since SNe are detectable even in diffuse galaxies, the derived cosmic star formation history is independent/complementary to that derived from local galaxy studies.

Very bright subclasses of SNe are one of the most promising targets. With optical surveys using 8–10 m class telescopes, bright Type IIn SNe (SNe with narrow emission lines, previously discussed in section 9.4 above) and superluminous SNe can be detectable even at z > 6. Spectroscopic observations with TMT will be critical to firmly identify such high-z events, as current 8–10 m telescopes do not have the spectral sensitivity needed to identify SNe at z > 3. These classes of SNe tend to show blue spectra, thus TMT/WFOS will be the best to obtain the rest-frame far-UV emission. A response time of ~10 days is enough as these classes of SNe show relatively long-lived light curves that are further slowed due to time dilation at high redshift, but as discussed previously time-series spectra starting early and continuing until the explosion passes through the entire CSM and subsequent decline after the plateau phase will enable mapping of the extent of the CSM and investigation of the effects and implications of progenitor mass loss prior to the explosion. An interesting application of identification of high-z SNe is probing the IMF in the early universe as the supernovae rate is directly related to the star formation rate (Lazar & Bromm 2022). At high redshifts, we may also be able to detect pair instability SNe (Scannapieco et al. 2005). Since superluminous SNe and pair instability SNe are thought to arise from very massive progenitor stars, 25–90 $M_{\odot}$ and 130–250 $M_{\odot}$ respectively, a number ratio of these types of SNe and normal SNe will provide information about the high mass end of the initial mass function.

To study the high-z universe with SNe, a better understanding of the SN rate in the low-z universe is also critical. Currently, SNe searches and follow-up in the most active star-forming regions are rather limited. While such regions can provide a large fraction of the star formation in the local universe and even a dominant fraction toward high redshifts, interstellar extinction is a major limiting observational factor (Mattila et al. 2012). The main problem is that the spectroscopic follow-up is currently nearly impossible due to their faintness, even at near-infrared wavelengths, thus even the SN typing has not been secured — at z ~ 0.03, a typical SN would have a peak magnitude of J ~ 22–23 if J band extinction toward the SN is ~ 5 mag. TMT will make spectroscopic follow-up for these obscured SNe possible, clarifying for the first time the SN populations and stellar evolution in active star-forming galaxies.

## 9.6 Understanding Progenitors of Gamma-ray Bursts: Connection to Supernovae and Kilonovae

Gamma-ray bursts (GRBs) are observed through the whole electromagnetic spectrum, from GHz radio waves to 10 MeV gamma-rays. Although each is unique, the bursts fall into two rough categories. Bursts lasting less than two seconds are classified as short, and those that last longer, the observed majority, as long. All of the confirmed long GRB host galaxies are actively forming stars. The total stellar mass in these host galaxies is low and their colors are bluer than present-day spiral galaxies. The ages implied for the progenitors of long GRBs are estimated to be < 0.2 Gyr, which is significantly younger than the minimum ages derived for the early-type galaxies found to be associated with short GRBs. These results suggest that long GRBs arise from young massive stars undergoing a core collapse SNe (e.g., Woosley & Bloom 2006), and short GRBs are the results of neutron star mergers or kilonovae (e.g., Berger 2013). Apparently there is more to this than this simple picture as recent multi-band observations of several GRBs (GRB 200826A; Ahumada et al. 2021, GRB 211211A; Troja et al. 2022, and GRB 230307A; Levan et al. 2024) suggest rethinking our understanding of the progenitors and the associated ambient medium of these energetic stellar explosions.

Observations of long GRBs at low redshifts clearly show that SNe arise from the GRBs themselves. However, clear spectroscopic confirmation of SNe is limited to low-redshift GRBs because at z>0.5–1.0, the SN component is too faint and most of the important features are redshifted to NIR wavelengths. TMT/IRIS will enable spectroscopic identification of SNe in long GRBs at z>0.5. For this purpose, intensive spectroscopic monitoring from ~1 day to 30 days after the GRB is important. Even at low redshifts, for some GRBs, no SN component was discovered down to a flux limit at least hundreds of times fainter than the expected SN flux (Gehrels et al. 2006; Della Valle et al. 2006; Gal-Yam et al. 2006). Spectroscopy with TMT/WFOS will be important to search for possible faint SN components in low-redshift GRBs.

In contrast to long GRBs, any supernova-like event accompanying short GRBs is currently limited by observation to being over 50 times fainter than normal Type Ia SNe or Type Ic hypernovae (Kann et al. 2011). In 2013, a NIR excess associated with a short GRB (Tanvir et al. 2013; Berger et al. 2013) was found to be broadly consistent with the expectation of an r-process powered kilonova (see section 9.1 for further discussion of kilonovae, short-GRBs and GW sources). More recently, kilonovae were also observed in association with apparently long GRBs 211211A (Troja et al. 2022) and 230307A (Levan et al. 2024), which complicates ideas about progenitor scenarios. Current observational data are still limited to multi-band photometry or spectra only at a few epochs. To fully understand the progenitor of short GRBs, TMT/IRIS will be the ideal instrument to perform ToO NIR spectroscopic observations of kilonova events associated with short GRBs, starting at a timescale of <~ 1 day.

## 9.7 Using Gamma-ray Bursts to Probe the Structure and Formation History of High-z Galaxies

While interesting on their own, long GRBs are powerful tools to study the high-redshift universe and galaxy evolution due to their apparent association with massive star formation and brilliant luminosities (e.g. Pandey, 2013, Bolmer et al., 2019, Rossi et al., 2022). There are three basic ways of investigating the evolution of luminous matter and gas in the universe: 1) direct detection of host galaxies in emission (in the UV/optical/NIR for the unobscured components, in the FIR/sub-mm/radio for the obscured component), 2) the detection of galaxies selected in absorption along the lines of sight to luminous background sources, traditionally quasars, and 3) exploring diffuse extragalactic backgrounds. Studies of GRB hosts and afterglows can contribute to all three of these methodological approaches, bringing in new, independent constraints for models of galaxy evolution and of the history of star formation in the universe.

Absorption spectroscopy of GRB afterglows is now becoming a powerful new probe of the ISM in evolving galaxies, complementary to traditional studies of background-quasar absorption line systems (Bolmer et al., 2019). GRBs probe lines of sight in the dense, star-forming regions of their host galaxies. In contrast, the background-quasar absorption systems are selected by the gas cross section, and favor large impact parameters, mostly probing the gaseous halos of intervening or foreground field galaxies, where the physical conditions are very different to the dense central regions where GRBs occur. The growing body of data on GRB absorption systems (Cucchiara et al., 2015; Heintz et al., 2023; Schady et al., 2024 and references therein) shows exceptionally high column densities of gas (Damped Lyman-alpha systems, DLA), when compared to the typical quasar absorption systems. This opens the interesting prospect of using GRB absorbers as a new probe of the chemical enrichment history in galaxies in a more direct manner than with the quasar absorbers, where there may be very complex dynamics of gas ejection, infall, and mixing at play.

Possibly the most interesting use of GRBs in cosmology is as probes of the early phases of star and galaxy formation (e.g., Robertson & Ellis 2012; Tanvir et al. 2012; Trenti et al. 2012; McGuire et al. 2016), and the resulting re-ionization of the universe. With the adaptive optics enhanced sensitivity of TMT-IRIS, GRBs are bright enough to be detectable out to much larger distances than those of the most luminous quasars or galaxies detected at present. Within the first minutes to hours after the burst, the optical light from afterglows is known to have a range of visual magnitudes far brighter than quasars, albeit for a short time only. The observed variation of fine-structure transitions lines (e.g. SiII*) over minute to hour timescales (e.g. De Cia et al 2011; Fryer et al 2022; Hartoog et al. 2013) demonstrated the impact of the GRB afterglow on the surrounding interstellar medium. GRBs are unique probes of the process involving the end of massive stellar bodies during the epoch of reionization and beyond and a critical observational test for PopIII stars models.

To identify GRBs as high-z events, prompt NIR spectroscopic observations of GRBs are essential. Required response time is ~5 minutes from the discovery (discovery is essentially instantaneous) and brightness changes of up to a factor 10 per minute can occur requiring rapid repeat spectroscopy that yields sufficient signal-to-noise for subsequent structural analysis.

## 9.8 Studying Tidal Disruption Events and Supermassive Black Holes

The nuclei of some galaxies undergo violent activity, with quasars being the most extreme instances of this phenomenon. Such activity is short-lived compared to galactic lifetimes, and was more prevalent when the universe was only about one-fifth of its present age (Schmidt and Green 1983). Dead quasars — massive black holes now starved of fuel, and therefore quiescent — should be more common than active quasars and are now being discovered in nearby galaxies (Gebhardt et al. 2000). The presence of these black holes is not surprising. We expect to find a black hole in most galaxies on the basis of the number density of quasars and their typical lifetimes (Soltan 1982; Chokshi and Turner 1992). But we must ask a further question: can a black hole lurk in these quiescent galaxies without showing other evidence for its presence? Could a black hole be so completely starved of fuel that it does not reveal its presence? The search for switched off, dim or dead engines, starving black holes, has therefore become one of the hottest topics in extragalactic astronomy.

Each star near a massive black hole traces out a complicated orbit under the combined influence of all other stars and the black hole itself. Due to the cumulative effect of encounters with other stars, the orbits gradually diffuse. If a star wanders too close to the black hole it is violently ripped apart by the black hole's tidal field — a tidal disruption event (TDE) (Rees 1988). About half of the debris of tidal disruption eventually falls back and accretes onto the black hole. This accretion powers a flare, which is a definitive sign of the presence of an otherwise quiescent supermassive black hole (SMBH) and a powerful diagnostic of its properties (see also section 6.5).

Tidal disruptions offer a unique opportunity to study a single black hole under a set of conditions that change over a range of timescales. Tidal disruption events offer the firmest hope of studying the evolution of their accretion disks for a wide range of mass accretion rates and feeding timescales.

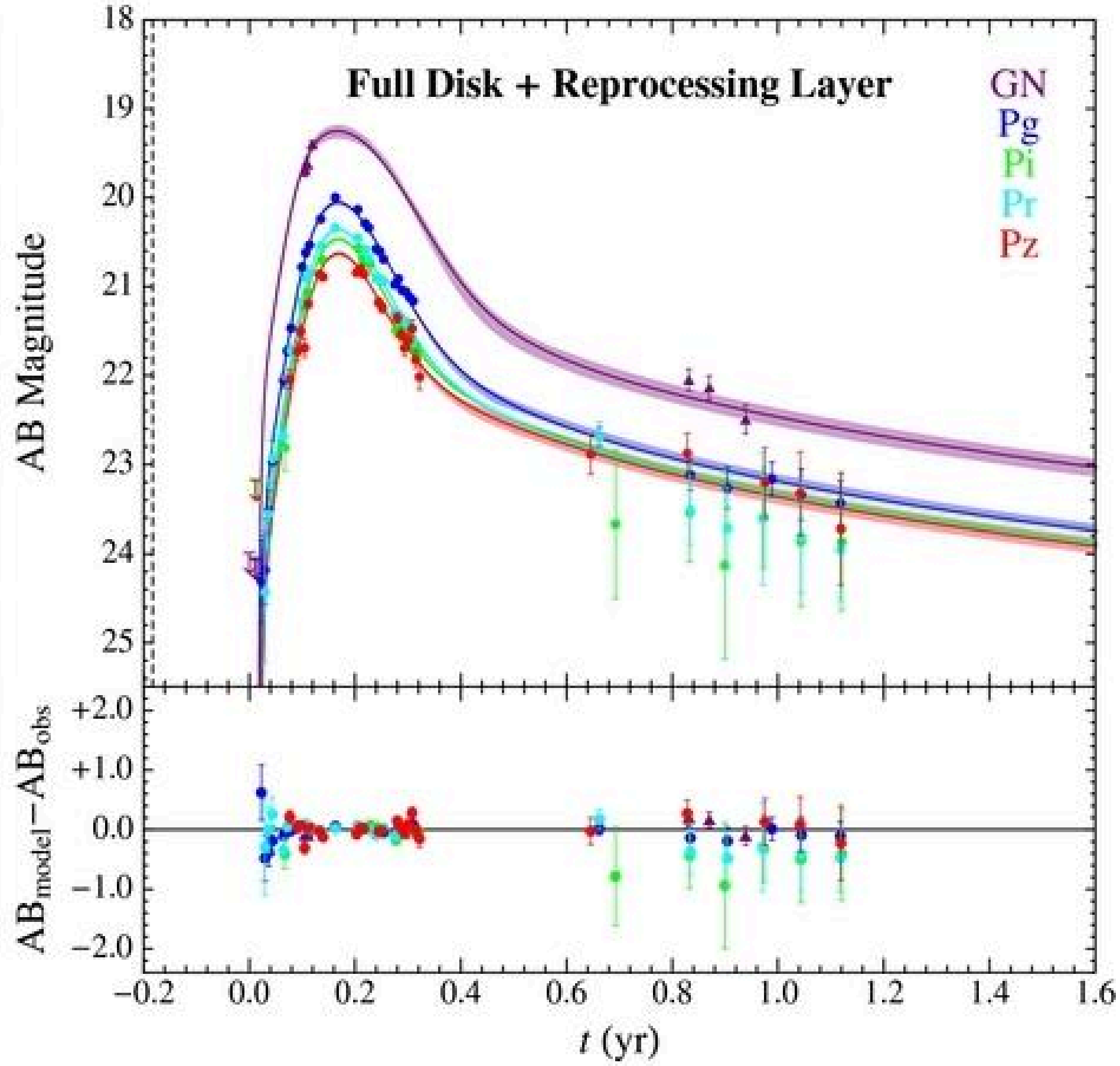

***Figure 9-9:*** *Observed and modeled light curves of TDE PS1-10jh (Guillochon et al. 2014). PS1-10jh was at a distance of 815 Mpc (Gezari et al. 2012), with TMT, such events can be studied out to much greater distances to review quiescent supermassive black holes beyond the local universe.*

Only two observations of claimed TDEs capture the rise, peak and decay of the flare (Gezari et al. 2012 and Chornock et al. 2014; figure 9-9). Capturing all three phases photometrically and with additional spectroscopic information, these impressive data sets have already shown several interesting behaviors indicating that TDEs are not as simple as one might first guess. The light curve is consistent with the bolometric luminosity closely following the rate of mass fallback (Guillochon et al. 2014), suggesting that the accretion disk viscous time is significantly shorter than the fallback time, and that the returning material must circularize by the first epoch of observation. The spectra resemble a single blackbody, with a temperature that evolves weakly in time, and a radius 10 times larger than the tidal disruption radius. As the light from the accretion disk itself should appear as a superposition of hotter blackbodies, this hints at the presence of a spatially extended reprocessing region surrounding the disk. The fact that broad HeII emission lines are seen, but hydrogen lines are missing, suggests that material may be highly ionized and outflowing at velocities ~10,000 km/s, reminiscent of the broad-line regions (BLRs) found around some steadily accreting AGN.

Ongoing and future transient surveys and radio surveys are predicted to discover 10–100 such events per year (van Velzen et al. 2011; Lin, Jiang, Kong 2022; Bricman, Gomboc, 2020; Bower 2011). Depending on the black hole mass, these events are expected to last for a few days to a few months. High resolution imaging with TMT's IRIS would precisely establish the position of the source. Optical spectroscopic follow-up observations provide diagnostics which help elucidate the demography of massive black holes in the local universe. Data should be taken in the rise, peak, and decay phases with frequent cadence. Each of these phases of a TDE contains vital information about the disruption, and can be used to constrain the properties of the host black hole and the object that was disrupted.

The modeling of tidal disruption will also be significantly improved once a large sample of TDEs has been collected. For a disruption with a well-resolved light curve, models permit a significant reduction of the number of potential combinations of star and black hole properties, enabling a better characterization of the massive black hole and the dense stellar clusters that surround them. TMT will allow the detection and study of TDEs up to much higher redshifts than previously possible. Because of the negative K-correction (TDEs emitting primarily in the rest frame UV with a characteristic $10^5$ K black body), TDEs will be visible by TMT to redshifts of 6 or larger, enabling constraints on SMBH properties and evolution over a vast range of cosmic time.

The recent discovery of the tidal disruption event AT2019DSG that coincides with IceCube high-energy neutrino event IC191001A (Stein et al. 2021, see also section 9.2) also opened up a new window of multi-messenger study to understand the nature of tidal disruption events (Winter & Lunardini 2021), the jet launching mechanism (Lee et al. 2020), and the intervening supermassive black holes (Liodakis et al. 2023).

## 9.9 TIME DOMAIN STUDIES OF AGN AND BLAZAR VARIABILITY

Blazars, the most extreme variety of AGN and the most luminous long-lived individual objects in the universe, are powered by accretion of gas onto supermassive (~$10^6$–$10^{10} M_\odot$) black holes (BHs). A pair of oppositely directed jets of magnetized, high-energy plasma continuously flow outward at speeds up to ~0.998c, presumably along the rotational poles of the BH system. Since the radiation observed from blazars is non-thermal and the luminosities are so high ($10^{11}$–$10^{15}$ $L_\odot$) yet extremely time-variable, electrons (including any positrons) in the jet must be accelerated with high efficiency to energies exceeding ~$10^4$ to $10^5$ $mc^2$ as blazars are seen to emit TeV gamma-rays.

Variability of AGN in the optical is typically about 10% over all timescales from less than an hour to years. There is a wide range of stochastic behavior between different AGN. Fully characterizing the stochastic properties of spectro-photometric measurements is vital to refining the model for the emission process and discerning between the possible AGN emission mechanisms. TMT will provide new capabilities for studying the structure of AGN. Moderate resolution optical and mid-IR spectra will make it possible to observe rapid changes in the spectral features of nearby AGN on timescales as short as ~1 hour, sampling every few minutes, allowing us to map out the structure of the accretion disk and broad-line regions with great accuracy and providing new insights into the structure of the jet region.

Shock waves, turbulence, and magnetic reconnections have all been proposed as the main means of particle acceleration, and all three might be operating inside relativistic jets. The efficiency of particle acceleration by these processes is strongly dependent on the magnetic field geometry. In the case of shocks, charged particles follow the magnetic field lines back and forth across the shock front. The efficiency of acceleration depends on the angle that the magnetic field subtends to the shock. In MHD turbulence, the energization occurs statistically as particles bounce off randomly moving regions of stronger-than-average magnetic fields. In reconnections, particles become trapped in shrinking magnetic flux tubes or in converging, oppositely directed field lines created by current sheets. Each of these processes has a different signature of time-variable linear polarization: in shocks, a favored polarization direction parallel to the shock front during peaks in flux; in turbulence, low, rapidly fluctuating polarization vectors with higher amplitudes of variations on longer time-scales; and in reconnections, polarization in a direction that remains stable during a flare but changes from one flare to the next.

Discerning among these possibilities requires precise spectro-polarimetric monitoring on time scales as short as a few minutes with TMT in order to measure fluctuations in polarization and flux. Such observations would determine the power spectrum of the fluctuations down to small size scales, testing emerging models of particle acceleration and blazar variability (e.g., Sironi & Spitkovsky 2014, Marscher 2014). With TMT, it will be particularly important to follow the changes in optical/near-IR polarization during the night of a blazar being observed by the future Cherenkov Telescope Array in order to relate acceleration of the highest-energy electrons to the magnetic field direction. As mentioned in section 9.2, further major advances in our understanding of the processes in the extreme central regions of Blazars and AGN will definitely benefit from synergistic observations from large neutrino detectors (Cerruti 2020) and time-resolved observations from TMT.

Figure 9-10 shows light curves collected with Kepler of two AGN, the power spectra show that neither lightcurve is consistent with the simple stochastic random walk model for variations (Mushotzky et al. 2011; Carini & Ryle 2012; Kasliwal et al. 2015). To move from an empirical random walk model to a physical model requires much more than simple lightcurves, but the lack of faster sampled spectro-photometric measurements constrain further understanding and the Kepler data are insufficient to study AGN variability as a function of radial position within the accretion disk of the AGN.

TMT's instruments will also be able to probe the structure of the accretion disk region by performing rapid optical and mid-IR observations ($t_{samp}$~10 mins) to carry out reverberation mapping. Observational evidence (Bhatta, et al., 2013) has led to the development of a model for microvariability that is based on stochastic synchrotron pulses and predicts variations on timescales of less than 30 seconds, these predictions are supported by Webb, et al. (2014) but the full investigation is severely limited by the need for high S/N optical and NIR low-resolution wide wavelength coverage spectro-photometry with $t_{samp}$~1 s to 3 s.

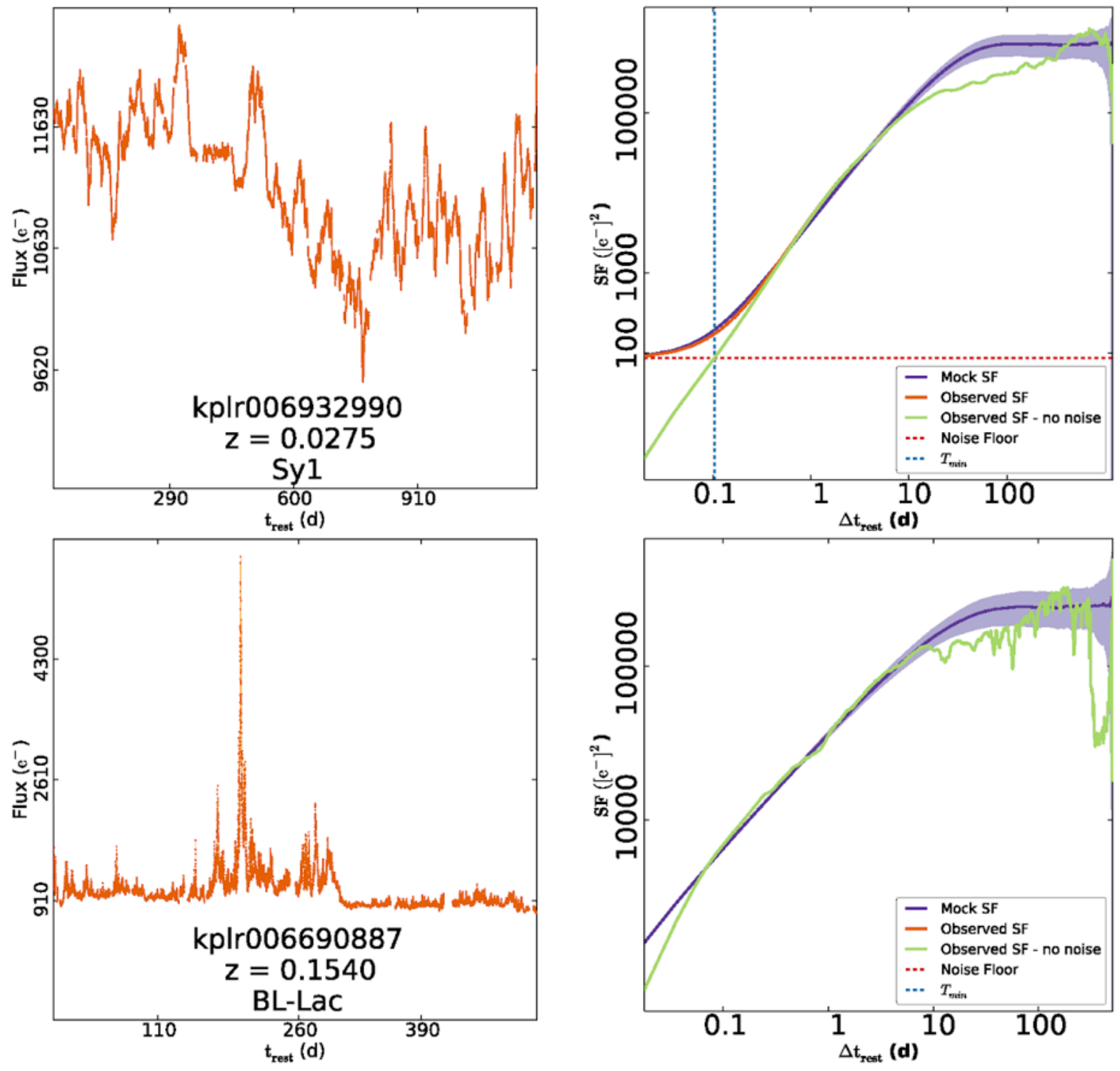

***Figure 9-10:*** *Top left: Kepler lightcurve of Seyfert Zw 229-15 and, Right: The power spectrum indicating that the fastest variations are on a timescale of ~2hr. Bottom left: The Kepler lightcurve of BL Lac kplr006690887 showing fast flares and, Right: The power spectrum indicating that the fastest variations are faster than the 26 min rest frame probed by Kepler. TMT's sensitivity and range of instrumentation will enable time-resolved spectro-photometric studies from the optical to the mid-IR that can sample the fastest timescales of variability arising in the region around accreting SMBH. Credit: TMT ISDT.*

Much of the physics underlying the processes occurring in AGN is applicable to the much smaller microquasars or X-ray binary systems (XRBs), though the timescales of variability are much shorter in the microquasar systems.

## 9.10 Cataclysmic Variables

Cataclysmic Variables (CVs) are interacting binary systems with an accreting white dwarf primary and a main-sequence mass-losing M star secondary. The binary orbital periods in these systems typically range from tens of minutes to a few days. CVs display a wide range of variability, be it accretion disk related phenomena or nova outbursts due to thermonuclear runaway in the accreted hydrogen-rich material.

Low-mass X-ray binaries (LMXBs) are very similar systems to CVs except the primary component is a neutron star or stellar mass black hole. LMXBs are less numerous than CVs and are generally much fainter in optical and infrared bands, while being very bright in X-rays. Those LMXBs that exhibit micro-quasar behavior provide a means to study the physics of relativistic jets and the fundamental properties of black holes using the same high time resolution techniques needed to study CVs that are mentioned here.

### 9.10.1 Investigating the Dissipative Process in Cataclysmic Variable Accretion Discs and Disc Evolution During Outburst Cycles

Accretion is a very common process in astrophysical systems, occurring over a huge range of scales from forming stars to supermassive black holes, but is very poorly understood. CV accretion discs provide an opportunity for studying the angular momentum dissipation mechanism and the relations between dissipation and disc density, temperature, magnetic field and velocity field using high sensitivity, time-resolved spectroscopy. CVs display rapid variability related to the accretion process that has timescales from sub-second to hours due to a number of reasons: interactions of the disc and mass transfer stream, processes in the accretion disc itself (as modeled by Ribeiro & Diaz, 2008, Scepi et al., 2018), and accretion onto the compact primary, which may be influenced by magnetic fields. The location of the source of the variability can be isolated by studying the temporal variations in the velocity of the emission and in some systems, by eclipse mapping.

These studies will also significantly enhance our understanding of angular momentum dissipation mechanisms in stable accretion disks (Ishioka et al. 2004). TMT will be transformative in this field as such observations require light gathering power unavailable with 8–10 m class observatories. Combining the velocity and temporal information from time-resolved spectroscopy with the method of eclipse mapping and monitoring disc changes on a weekly basis will transform our understanding of the processes occurring within CV accretion discs, particularly in systems whose discs display changes in state or have regular outbursts on timescales of months.

For example, in rapid (72 ms) spectroscopic observations of a bright ($M_v$~11) dwarf nova, short (2 to 3 minute) flares were found whose temporal and spectroscopic behavior were consistent with a flares modeled as arising at discrete locations within the accretion disc (figure 9-11) and could be described with a model of small (<<disk) fireballs (Pearson et al. 2005). From broadband photometry, Baptista & Bortoletto (2008) and Baptista et al. (2011) respectively derived the spatial distribution of flickering in the discs of the nova-like UU Aqr and the systematic changes in the flickering distribution across the disc through an outburst of the dwarf nova HT Cas when the disc collapsed.

Time-resolved optical spectroscopy with WFOS configured for R=4000 and several local standard stars for spectrophotometric correction, with a duration about 20 minutes centered on mid-eclipse and $t_{samp}$=50 ms will give S/N~1000 in each resolution element for an Mv~15 system. At this magnitude there are an order of 100 known eclipsing CVs (http://www.mpa-garching.mpg.de/RKcat/cbcat) that show a range of different accretion disc behaviors. Developing high fidelity maps of the physical characteristics of different accretion disks (temperature, density, chemistry, emission, even local turbulence, eddies and shear) will allow the development of a comprehensive understanding of accretion disc physics and the processes occurring within disks that is impossible to obtain with existing facilities.

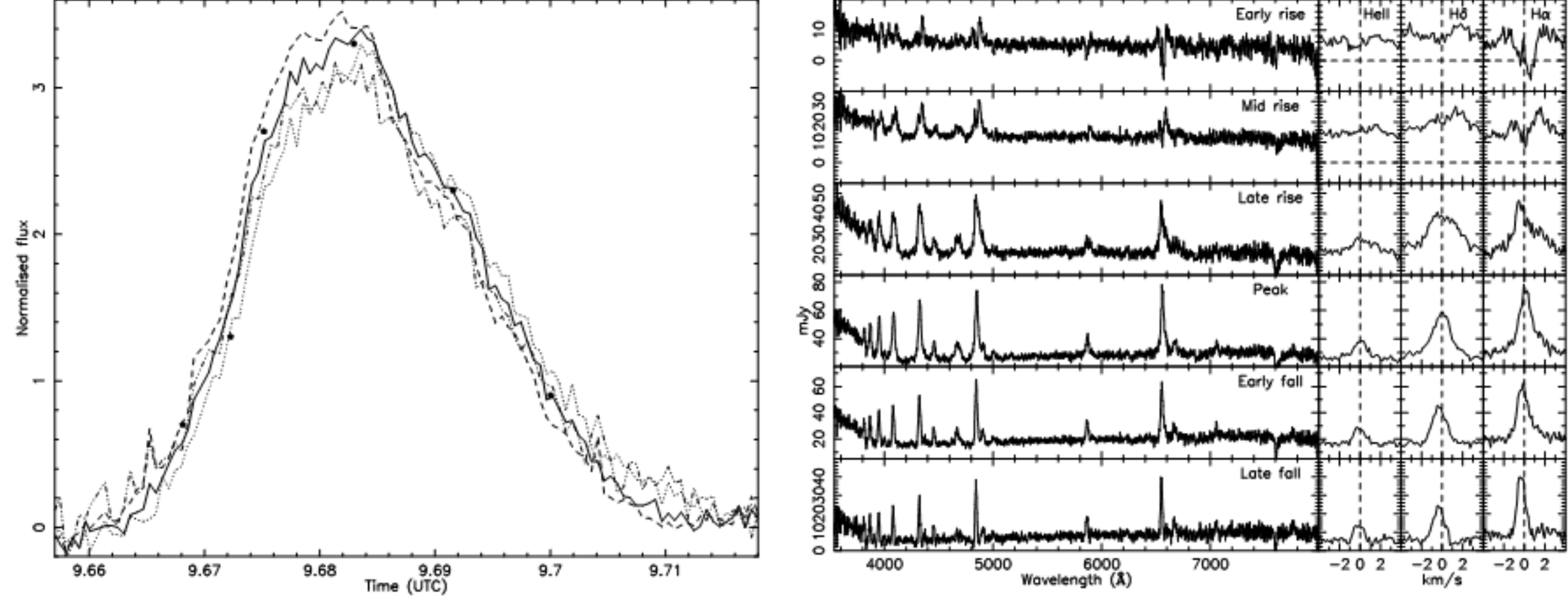

***Figure 9-11:*** *Left: Continuum lightcurves of a small flare seen in the Dwarf Nova SS Cyg. Observations were made with Keck/ LRIS. Lightcurves are 3615 Å (dot-dashed), 4225 Å (dashed), 5500 Å (solid) and 7320 Å (dotted). Right: Spectra at times (filled circles in left panel) throughout the flare. (Skidmore et al. 2004).*

### 9.10.2 Revealing the Geometry and Populations of Classical Novae

The high luminosities of novae, coupled with a rate of ~ 35 /yr in a galaxy like our own (Shafter 1997), make novae ideal for probing the properties of close binary stars in varied (extragalactic) stellar populations. Nova systems serve as valuable astrophysical laboratories in the studies of the physics of accretion onto compact, evolved objects; thermonuclear runaways on semi-degenerate surfaces that give insights into nuclear reaction networks; and line formation and transfer processes in moving atmospheres. Nova outbursts are seen in all wavelengths from γ-ray to radio and can reach $M_v = -10$ mag, placing them among the brightest transient sources known. The outburst intervals range from decades for recurrent novae to thousands of years for classical novae. Novae can rise in luminosity by >6 mags in a day, with a rise time of about 1 to 2 days in the very fast novae, with the slowest novae having a rise time of the order of weeks.

The outbursts generally cause the system to increase in brightness by about 10 mags at the peak. The decline rates can be between ~1 mag/day to ~0.03 mag/day, depending on the speed class of the nova, the rate is believed to be related to the mass accretion rate, mass of the hydrogen-rich envelope and the mass loss rate due to wind (Hachisu, et al. 2020). The nova outburst characteristics are particularly sensitive to the mass of the accreting white dwarf (e.g., Starrfield et al. 2012).

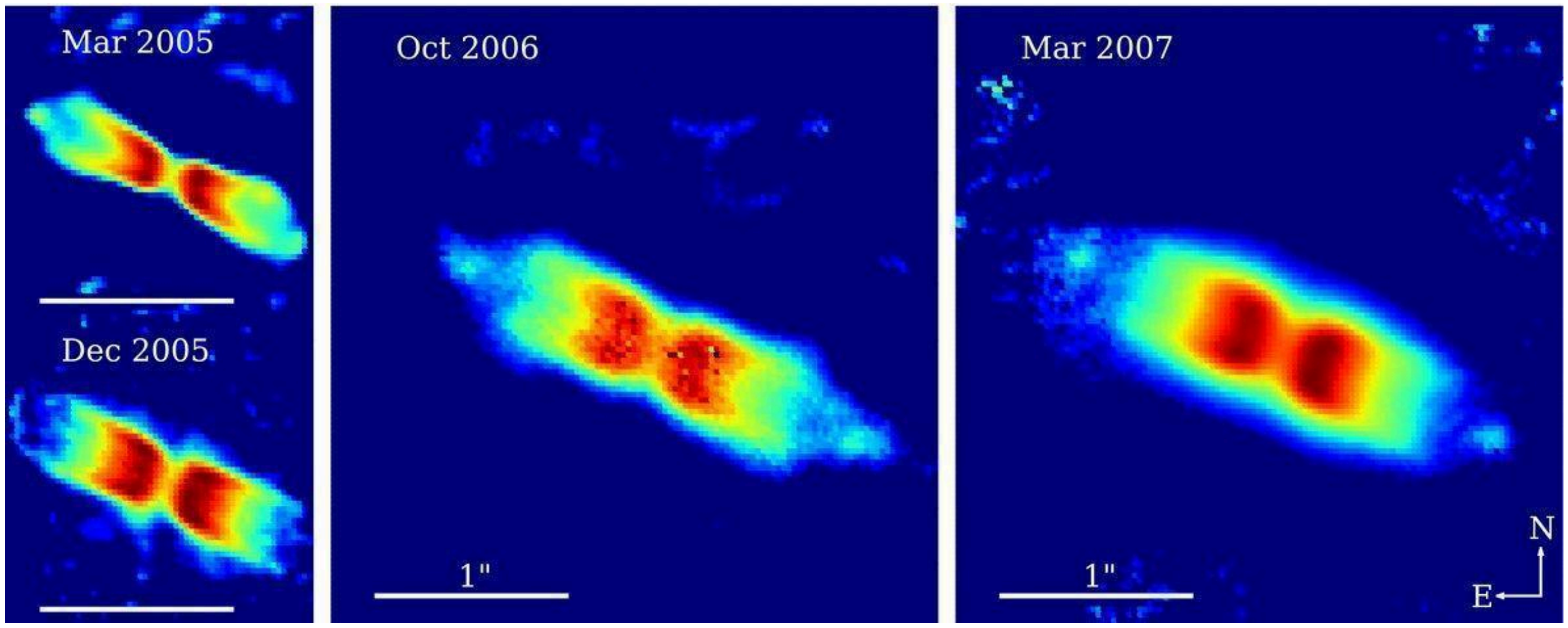

***Figure 9-12:*** *The expanding bipolar shell of the helium nova V445 Pup (2000) imaged in the near-infrared Ks band using the NAOS/CONICA adaptive optics (AO) system on the Very Large Telescope (VLT), Woudt et al. 2009. TMT+IRIS would be able to spatially resolve the expanding shell in just a couple of days for the closest, fastest novae and track the evolution of species in the shell and the shell characteristics using spatially resolved spectroscopy.*

The geometry of novae explosions is important for understanding the explosive process on the surface of the compact star. Nova shells have revealed polar-blob, equatorial-ring geometry and/or multiple blob-like condensations in the shell. There is also increasing evidence that nova phenomenology is best explained by non-spherical ejecta (e.g., Woudt et al. 2009; Ribeiro et al. 2011; figure 9-12). One leading scenario to explain such observations as well as the spectrophotometric evolution of novae is bipolar (although not jet-like) mass ejection (Shore 2013). Alternatively, the ejecta may consist of blobs of very high density gas that expand to create within each clump a wide range of densities, ionization, and velocities, while the circumbinary envelope is more homogeneous (Williams 2013).

The geometry of the post-outburst structure of novae will be further constrained by coordinated spectroscopic follow up of novae over all wavelength regimes. Since important aspects of the ejecta activity occur at the time of outburst, spectral studies are best conducted as early as possible. The timely discovery of outbursts before maximum light will be an outcome of the large time-domain surveys, such as Rubin Observatory. Alert systems will inform TMT to trigger and secure early observations. Spectrophotometry and spectropolarimetry (R ~ 5000–30,000) are needed to establish the presence of the two distinct components and blobs, and to follow the evolution of the spectrum as the fireball progresses. Spectropolarimetry, in particular, holds the key as polarimetric measurements will help quantify the degree of non-spherical symmetry in the early expanding nova shell.

The geometry of the expanding fireball can also be studied with high-spatial resolution imaging. A nova at 5 kpc (novae have been observed at distances of 1–10 kpc) that is ejecting mass at 1000 km/s (ejection velocity ranges from 300 km/s for slow novae to 10000 km/s for the very fast novae), as seen from the Earth, will have an ejected shell of size 10 mas (0.01"; ~2 times the pixel sampling of IRIS) in around 87 days. For a very fast nova, the time taken would be around 9 days to reach a similar size. If the nova is at 1 kpc and ejects mass at 1000 km/s, it will have a size of 10 milliarcseconds in 17.4 days. Thus, outbursting novae can be observed as extended objects after the first week of outburst. A nova situated even 10 kpc away would be very bright (e.g., M = −9 corresponding to m = 11 even with 5 magnitudes of extinction) for TMT.

A solution to this would be to use AO combined with coronographic techniques. For studying the early fireball phase of novae, the occulting mask needs to be 10 mas or slightly bigger. The geometry and evolution of the expanding fireball have in recent years been studied using NIR interferometric imaging (e.g., Chesneau & Banerjee 2012 & Schaefer, et al. 2014) but studies are limited to the very rare cases where the system rises to $M<6$. Direct imaging with TMT+AO will open up several orders of magnitude more targets to studies with IFU type instruments.

In addition to Galactic novae, novae have been observed in more than a dozen galaxies, some as distant as the Coma Cluster (Shafter et al. 2011, 2014). Photometric and spectroscopic observations of novae in the Local Group suggest that galaxies dominated by a lower metallicity, younger stellar population (M33 and the LMC) generally tend to host novae with a faster photometric evolution, but theoretical models predict that lower metallicity populations should give rise to slower novae (Starrfield, Schwarz, Turan & Sparks. 2000). The question that arises is; are there really two distinct populations of novae that are dependent on the metallicity and star formation history of their host galaxies? Observing the spectral evolution of a larger number of extragalactic novae, along with pushing to larger distances within the local group, would provide an important dataset in addressing this question. So far, 8–10 m class telescopes have been efficiently used to photometrically detect novae in the Fornax cluster of galaxies at 20 Mpc, but spectroscopic observations have been difficult. With TMT, one could easily do optical spectroscopy of novae at this distance.

### 9.10.3 The properties of Brown Dwarfs in interacting binaries

Stellar population synthesis models predict that there should be many highly evolved CVs whose mass-losing secondary stars have reduced in mass past the point where they can burn hydrogen. These systems with brown dwarf mass-losing secondaries are evolving toward longer orbital periods having passed the period minimum when the mass-losing secondary transitions from hydrogen burning to degenerate, the so-called 'Period Bouncers'. The number of robust candidates for such systems is very low, only a few tens, and the number of systems where the brown dwarf character of the secondary is spectroscopically confirmed is even fewer (less than ten; Neustroev & Mäntyen, 2023).

This discrepancy between observation and theory of the numbers and the properties of Period Bouncers is a long running issue and while implying that during the final phase of CV evolution, either the mass donor secondary star is destroyed or the systems detach and mass transfer ceases (Inight et al., 2023), solving this discrepancy is a powerful constraint on population synthesis models that has broad ramifications. There is limited observational evidence that sub-stellar mass brown dwarfs in period bouncers appear similar to individual brown dwarfs (Hernández Santisteban et al., 2016), but the connection is tenuous.

Great progress could be made if the properties of the mass-losing secondary star in bonafide Period Bouncers (those with measured mass ratios confirming the secondary mass of $<0.06\ M_\odot$) could be measured and compared against those of field brown dwarfs. This would identify differences in the properties that may arise due to the different formation processes between brown dwarf secondaries in period bouncers and brown dwarfs formed directly during stellar formation processes.

Such observations are presently too challenging because the spectrum of the very faint secondary star must be captured during the short eclipse time in the few low mass transfer rate Period Bouncers known. A low mass transfer rate is important to avoid additional accretion related flux from overwhelming the faint secondary star flux. High sensitivity, near-infrared spectroscopy over J, H and K bands with R~3500 that temporally resolves the eclipse will allow the donor star spectrum on the unirradiated side to be isolated and compared to regular brown dwarfs. Complementary time-resolved optical spectroscopy, both outside and during the eclipse, with R~1500 over UBVRI bands will allow the white dwarf and accretion spectrum to be characterized and extrapolated so that any residual flux present during eclipse in the infrared can be accounted for. Short exposure times are needed with resulting high signal-to-noise spectra and thus requires an ELT.

### 9.10.4 Micronovae in Magnetic Accreting White Dwarf Systems

Micronovae are a long theorized phenomenon caused by thermonuclear bursts from localized regions on the surface of magnetic accreting white dwarfs, where the stochastic flow of accreted material is strongly constrained by the magnetic field (Shara, 1982), figure 9-13. Recent time-resolved photometric observations of magnetic accreting white dwarf systems show flares lasting several hours that are consistent with localized thermonuclear runaway events, figure 9-14 (Scaringi, et al., 2022a; Scaringi, et al., 2022b). However, interpretation of the flares is controversial and other explanations are possible (Schaefer, Pagnotta & Zoppelt, 2022).

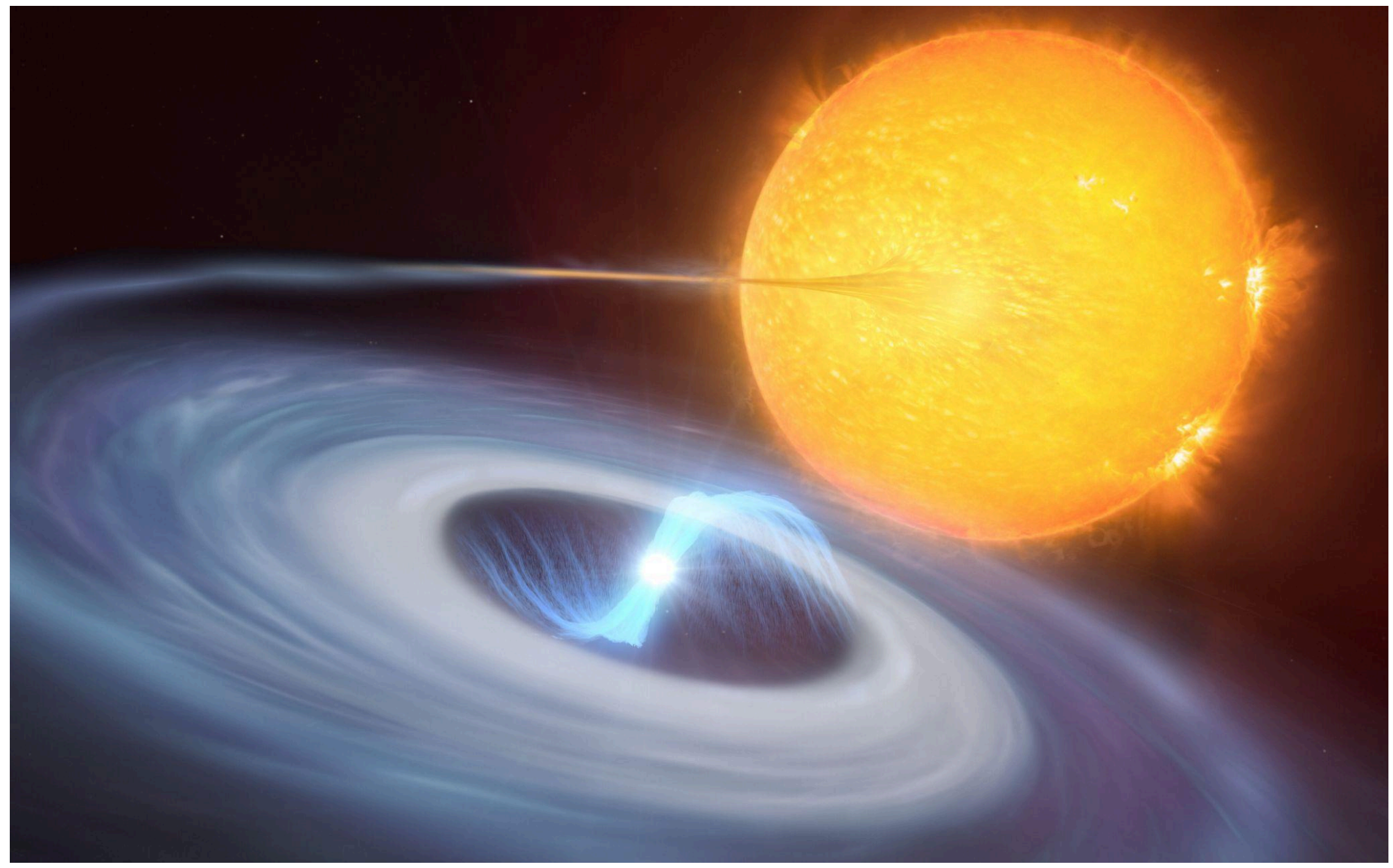

***Figure 9-13:*** *An artist's impression of a magnetic accreting white dwarf system where micronovae may occur at the region of impact of the magnetically constrained accretion flow onto the white dwarf. Image credit: ESO, M. Kornmesser, L. Calçada. Full movie of micronova simulation at https://www.youtube.com/watch?v=_zIMVjqj6VY.*

***Figure 9-14:*** *Figure 1 of Scaringi, et al., 2022a. Top: A TESS light curve of TV Col (with 20 s sampling) spanning almost 60 days. Bottom, left to right: 16.8 h light curves focussed on three detected rapid bursts showing the rise with a timescale of about 30 minutes, and the decay with a timescale of about 12 hours.*

Confirming that the mechanism of the flares in magnetic accreting white dwarf systems is indeed thermonuclear runaway, and improving our understanding of localized thermonuclear runaway processes can improve our understanding of nuclear fusion in general. Localized thermonuclear reactions may provide explanations for asymmetry in nova shells.

To characterize the properties of the flares attributed to localized thermonuclear runaway events requires time-resolved spectroscopy with sampling frequency sufficient to fully sample the white dwarf spin period (typically 12 min to 20 min), covering U to I band in the optical with spectral resolution of ~1500, sufficient to identify white dwarf absorption features and ejecta velocities of 300 km/s or more.

Near infrared spectroscopy with spectral resolutions of ~3000 to identify high ionization lines and their temporal evolution will provide information about the source of acceleration and driving of winds by the epicenter of the thermonuclear runaway region, high spectral resolution near infrared observations will provide detailed information about the wind and shock structures.

These are ToO programs and the response time needs to be much less than the 30 minute rise timescale from the time the alert is received.

## 9.11 FAST RADIO BURSTS (FRBS) AND LUMINOUS FAST BLUE OPTICAL TRANSIENTS (LFBOTS)

Recent high cadence surveys have discovered classes of intriguing and energetic objects that have as yet defied physical understanding, including fast radio bursts (FRBs) and luminous fast blue optical transients (LFBOTS). Fast radio bursts (The high sensitivity of TMT with a fast response time to triggers offers the promise to reveal their underlying cause.

FRBs) are bright millisecond radio transients whose origin is unknown (e.g., e.g., Cordes & Chatterjee 2019). Hundreds of FRBs are detected per year and they are transient phenomena mostly seen at cosmological distances but with some now found within the Milky Way. To understand the origin of FRBs, multi-wavelength follow-up observations are needed, ideally simultaneous multi-wavelength observations, potentially through coordination with other facilities. High time resolution is important to enhance the sensitivity for possible optical-NIR transients on sub-second timescales. A tiling of observations may be needed to cover a set of candidate FRB counterparts.

High slewing and rRapid slewing and target acquisition are key required capabilities. High sensitivity, high spatial resolution observations will show the environment in which FRBs appear (Dong et al. 2024).

Luminous Fast Blue Optical Transients (LFBOTs) are enigmatic events that cannot all be explained by any single identified mechanism. Possible mechanisms involve various forms of post-common envelope mergers between stars and compact companions and forms of supernovae and failed-supernovae (Metzger 2022). Due to the faint apparent magnitude and flux changes by a factor 2 within a day, high sensitivity and queue scheduling are required to track the evolution of the spectrum over time.

So far only photometric measurements have been obtained and these show a smooth broadband spectrum, but the short term temporal behavior and investigation of subtle spectral signatures require TMT's sensitivity.

## 9.12 IMPROVING THE HUBBLE CONSTANT AND MEASURING EXTRAGALACTIC DISTANCES

Despite sustained effort, the increasing number of high quality distance ladder studies are not converging on the value of the Hubble constant. In fact higher precision measurements have highlighted the Hubble Tension (see section 3.2.2). The Hubble Tension arises due to systematic, not statistical, uncertainties dominating the discrepancies between two independent methods of measuring $H_0$, one using the traditional distance ladder, the other using measurements of the Cosmic Microwave Background.

The high resolution and high sensitivity of TMT will allow for validation of the methods for distance determination that utilize variable stars and standard candles (e.g. SNe Ia).

Reducing the uncertainty of the Hubble constant below 1% helps to improve the determination of other cosmological parameters (Weinberg et al. 2013). By providing largely orthogonal constraints with other cosmological probes (e.g., high-redshift supernovae, cosmic microwave background anisotropies and baryon acoustic oscillations), a precise late-time determination of $H_0$ would serve to critically test the need for new physics beyond the ΛCDM model, such as time-dependent dark energy, additional relativistic species in the early universe, among many other possibilities (Abdalla et al. 2022).

Discrepancies between the determination of the Hubble Constant using the distance ladder and the CMB have given rise to the Hubble Tension, see section 3.2. This is likely to be a critical area of study for many years to come.

Cepheid variables are the primary distance indicators for galaxies with recent star formation (Riess et al., 2022). They are very luminous at NIR wavelengths, with $M_{abs}$ from −6 to −8.5 for periods of 20 to 100 days, respectively. The intrinsic dispersion of the Cepheid Period-Luminosity relation (also known as the “Leavitt Law”, Leavitt 1912) is ~0.08 mag in the NIR (Riess et al. 2019), making it possible to obtain Cepheid-based distances with statistical uncertainties of a few percent for modest sample sizes. In the absence of recent star formation, RR Lyrae variables provide a suitable alternative, although their range is much more limited given their considerably fainter luminosities ($M_K = -0.55$ at P = 0.5 d, Benedict et al. 2011).

Gaia measurements of the parallaxes and light curves for thousands of Milky Way Cepheids and RR Lyrae stars are significantly improving the Period-Luminosity relations (Clementini et al. 2023). TMT can utilize these improved relations to measure distance much more accurately, extending to galaxies actually in the Hubble flow. We can readily envision two applications of TMT for extragalactic variables, both relying on diffraction-limited imaging with IRIS: 1) improving the determination of the Hubble constant ($H_0$) and 2) measuring extragalactic distances out to the Coma cluster (D ~ 100 Mpc).

Beyond trigonometric distance of Milky Way Cepheids and RR Lyrae variables, we can also establish extragalactic distance anchors using eclipsing binaries and constrain the Leavitt Law in nearby galaxies. For example, Vilardell et al. (2010) were able to determine the distance of M31 to 4% level with two eclipsing binary systems composed of early type stars. Pietrzynski et al. (2019) were able to refine the distance of the LMC to 1% level with 20 late-type eclipsing binary systems.

While there are hundreds of eclipsing binaries in external galaxies reported by various surveys, such as the DIRECT project (Macri et al. 2001), the PAndromeda project (Lee et al. 2013), only a handful of them are bright enough (V~20 mag) to be even spectroscopically followed up by 8–10 meter class telescopes.

However, with the extremely large aperture of TMT, we can obtain spectra of much fainter eclipsing binaries and statistically reduce the uncertainty of extragalactic distance anchors out to beyond the local group of galaxies, improving by several fold the overlap between Cepheid Variables and eclipsing binaries in the Cosmic Distance Ladder. Uncertainties in the calibration of the first steps of the Cosmic Distance ladder will reduce to about 60% of their present levels.

Besides Cepheids and RR Lyrae, some long-period variables (LPVs) such as Miras have also been shown to follow tight period-luminosity relations in the infrared (Huang et al. 2018). Miras are comparable to Cepheids in terms of absolute magnitude, but they are much more numerous and present in all types of galaxies, which might allow distance estimates using significantly smaller survey areas compared to Cepheids, ideal for the field-of-view of TMT.

In addition to the distance ladder method, gravitational lensing time-delay provides an alternative one-step method to estimate the Hubble constant, which also shows a discrepancy with Hubble constant estimates from the cosmic microwave background. Currently the gravitational lensing time-delay method relies on multiply lensed quasars (H0LiCOW, Suyu et al. 2017; Wong et al. 2020), which is susceptible to mass-sheet degeneracy (i.e. different mass distribution in the lensing cluster that can result in the same time delays; Yıldırım et al. 2023).

To break such degeneracy, it is imperative to use lensed sources with well-characterized intrinsic luminosity, such as SNIa. The first multiply lensed SNIa, iPTF16geu (Goobar et al. 2017) has image separation less than 0.5” and time-delay on the order of hours (More et al. 2017). The sharp image quality and fast-response time delivered by TMT will be key to spatially resolve these strongly lensed supernovae and provide accurate $H_0$ estimates.

Diffraction-limited imaging with TMT/IRIS will give a >10× increase in angular resolution and 25× finer pixel scale relative to HST, removing photometric biases due to crowding and yielding a 3× improvement in distance uncertainties. S/N = 50 photometry can be obtained in less than 120s for any object of interest (the faintest being a 20-day Cepheid at 50 Mpc, with K = 26.8 mag). Several neighboring fields within 2’ of each other containing Cepheids of the same host galaxy could be observed in quick succession.

JWST is undertaking several programs that are further confirming the Hubble Tension (Freedman & Madore 2023) through improvements in the distance ladder, but the limited spatial resolution and near-IR sensitivity of JWST compared to TMT, the finite lifetime of JWST, the timescale before the next generation of large space assets begin operation, mean that TMT will play a unique role in extragalactic distance work for several decades by undertaking the programs here. Precise extragalactic distances have many other useful applications; e.g., the accurate calibration of the bulge luminosity and black hole mass relation, which cannot be performed using samples in the Hubble flow. NIR AO imaging with TMT/IRIS will deliver extragalactic distances beyond 10 Mpc, using either RR Lyrae ($P_{cyc}$~0.5 day) out to D ~1 Mpc (requiring ten ~1 hr exposures) or Cepheids out to the Coma cluster (D = 100 Mpc, requiring 1.5 hours of telescope time per integration and 10 observations over several months) for late-type systems.

## 9.13 SUMMARY OF CAPABILITIES

Table 9-1 summarizes the key parameters for the telescope and instruments to realize the science cases described in this chapter. In order to maximize scientific outcome from time-domain astronomy, response time for ToO observations and sampling time for time-resolved observations are especially important; these requirements are summarized in figure 9-15 and figure 9-16.

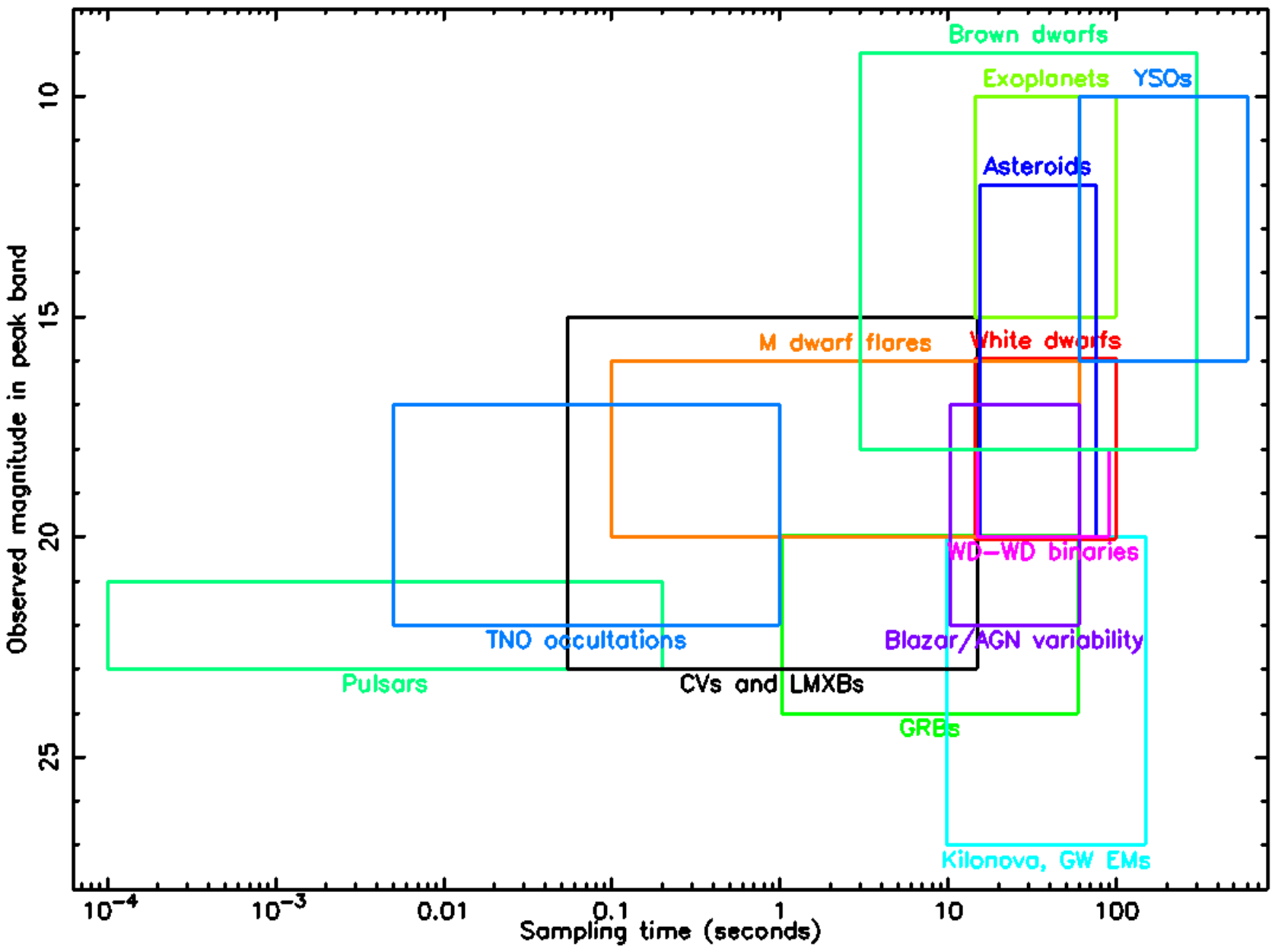

***Figure 9-15:*** *The ranges of sampling times required for observations of variable objects. In some cases the objects are continuously variable, and observations can be scheduled at any time. In other cases rapid observations are needed to follow a transient object after a rapid response ToO program is initiated (see figure 9-16). Credit: TMT ISDT.*

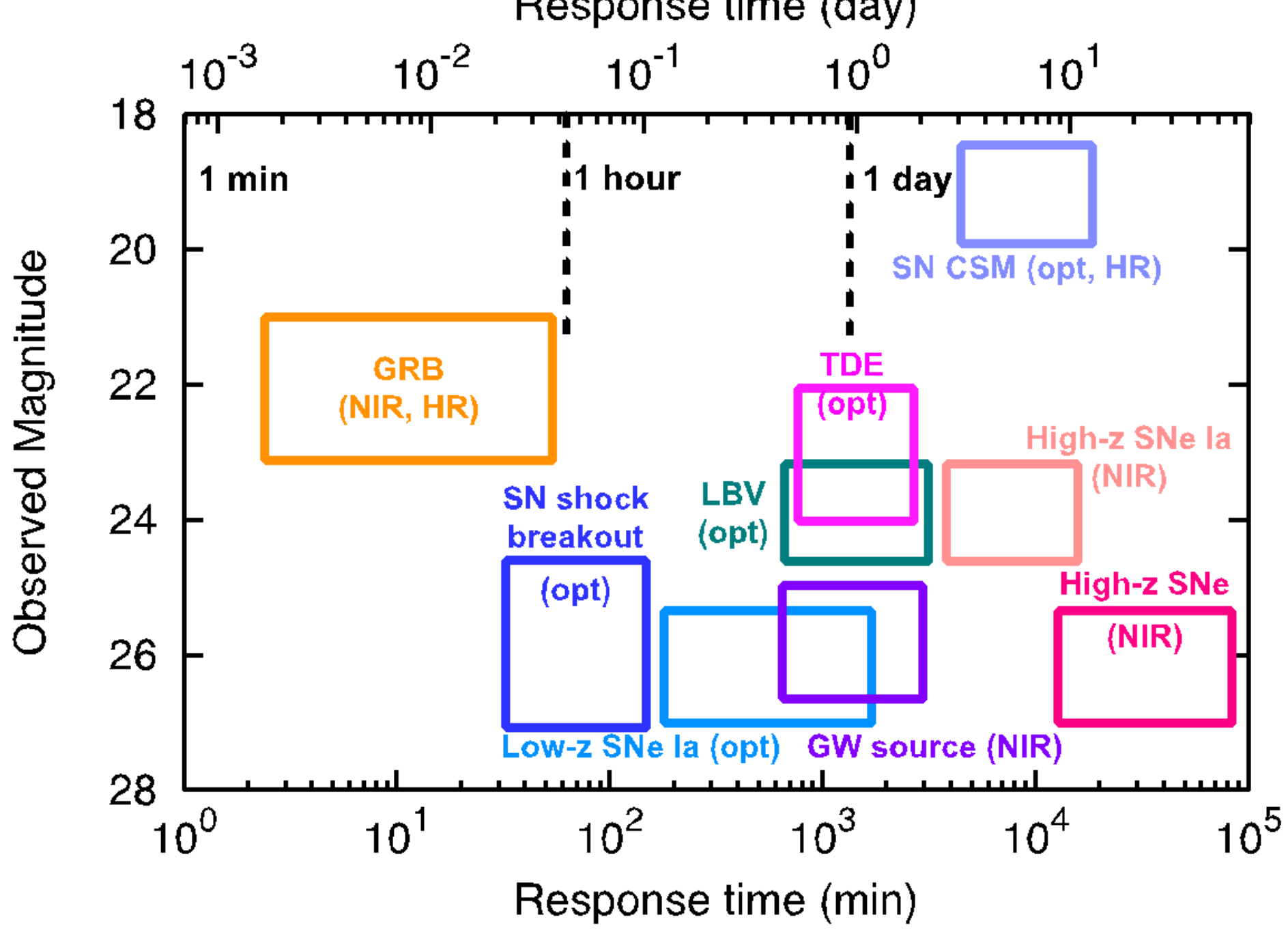

***Figure 9-16:*** *Required ToO response time for different science cases. Credit: TMT ISDT.*

***Table 9-1:*** *Summary of key parameters for Time-Domain Science. Capabilities which are not planned for the first-light instruments are written in Italic font. The absolute accuracy on the timestamp depends on the particular science program but a general rule of thumb is $t_{error}$<0.1 $t_{int}$ where $t_{error}$ is the uncertainty on the mid-point of the time when photons are actually hitting the detector.*

| Science Case | Response time | Sampling time | Cadence | Brightness (mag) | Observing mode | Notes |
|---|---|---|---|---|---|---|
| SNe Ia evolution (9.3.1) | 3-10 days | - | 3–10 days | 24 (NIR) | R = 500 NIR spectroscopy | Synergy with surveys |
| SNe Ia explosion mechanism (9.3.2) | 1 hr - 1 day | - | Initially 1hr<br>After 30 days weekly for 2 years | 26 (opt.)<br>23 (NIR) | R = 500 – 3000 opt./NIR spectroscopy | Synergy with surveys<br>Benefits from *spectro-polarimetry* |
| CCSNe shock breakout (9.4) | 1 hr | - | 1hr – 10 days | 26 (optical) | R = 500 opt. spectroscopy | Synergy with surveys |
| Luminous high-z SNe (9.5) | 10 days | - | 10 days | 26 (opt., NIR) | R = 500 opt./NIR spectroscopy | Synergy with surveys |
| NIR search Low-z SNe (9.4) | 3 days | - | 10 days | 25 (NIR) | R = 500 NIR spectroscopy | |
| SNe CSM (9.3.2) | 1 day | - | 1 day | 19-20 (optical) | *R~30000 opt. spectroscopy* | |
| GW sources (9.1) | 1 day | - | 1 day | 26 (opt)<br>24 (NIR) | R = 500 opt./NIR spectroscopy | Synergy with GW observatories |
| High-energy neutrino sources (9.2) | 3 days | - | 10 days | 26 (opt., NIR) | R = 500 opt./NIR spectroscopy | Synergy with neutrino observatories |
| GRB-SN/kilonova (9.6) | 1 day | 1 hr<br>(For short GRBs) | 1 day | 28 (opt)<br>26 (NIR) | Opt/NIR spectroscopy (R = 500) | Synergy with X-ray/γ-ray surveys |
| High-z GRBs (9.7) | < 1hr | 1 min initially | 1hr – 1 day after 1–2 hours | >22 | NIR high-resolution spectroscopy | Synergy with X-ray/γ-ray surveys |

| Science Case | Response time | Sampling time | Cadence | Brightness (mag) | Observing mode | Notes |
|---|---|---|---|---|---|---|
| TDEs (9.8) | 1 day | - | 1–10 day | >22 | Optical/NIR spectroscopy | Synergy with surveys |
| Accretion disk in CVs (9.11.1) | <1 hr from discovery with Rubin | 50 ms for up to a few hours | daily to weekly | 15 (opt) | time-resolved spectroscopy (R = 4000) | *Spectropolarimetry for magnetic CVs* |
| Classical novae - Early evolution (9.11.2) | 1hr -1 day | - | 1 hr | 10–25 | *R ~30,000 Opt./*NIR spectroscopy/*spectro-polarimetry* | Synergy with wide field surveys |
| Classical novae - geometry (9.11.2) | ~10 days | - | None | 10–19 | NIR AO imaging with occulting mask | |
| Extragalactic distance (9.12) | - | - | 3–5 days Cepheids, 1h RR Lyr | 27 (NIR) | NIR imaging | Synergy with Rubin, HST, and JWST |
| AGN reverb mapping (9.10) | - | >10 mins for a few hours | - | 17 to 22 | R=1k to 4k Opt./NIR spectroscopy | |
| Blazar jet μ-variability (9.10) | - | 1 s to 3 s | - | 17 to 22 | R=500-3000 Opt./NIR spectroscopy | |
| Blazar jet Polarimetry (9.10) | - | ~30 s to 2 minutes | - | 17 to 22 | *R=500–3000 Opt./NIR spectro-polarimetry* | |

# 10. Exoplanets

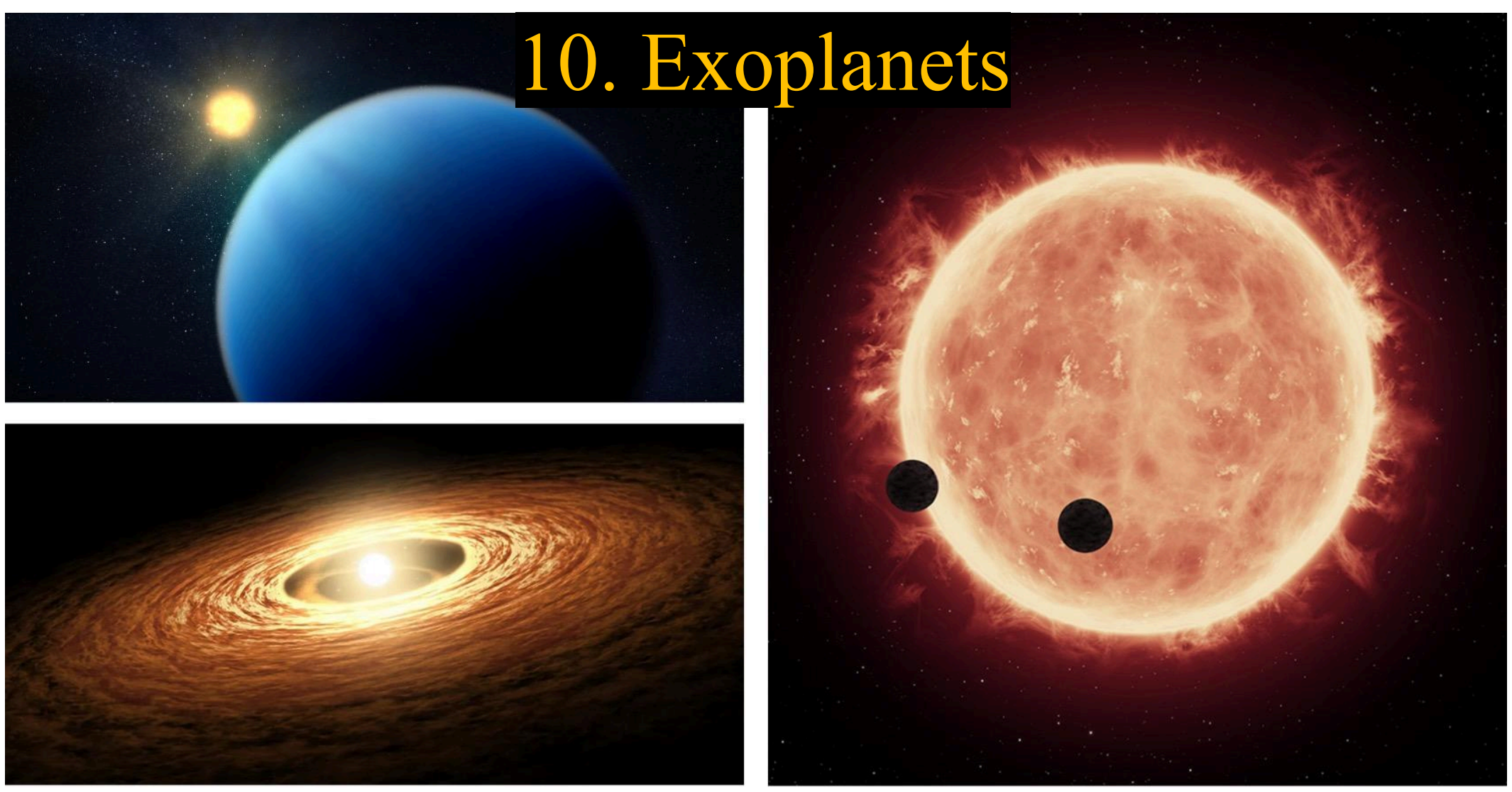

*The characterization of exoplanets and studies of exoplanet formation with TMT will include direct imaging and high resolution spectroscopy of reflected light or flux from self-luminous exoplanets (upper left), transit studies of exoplanets that pass in front of their host star (right), and direct imaging of protoplanetary disks (lower left). Image credits: NASA, NASA-JPL, ESA, CSA and STSCI.*

---

In early 2022, the NASA Exoplanet Archive logged its 5000th confirmed exoplanet and a few years later, that number is still climbing. In just 30 years, astronomers have developed a large collection of exoplanets ranging from sub-Earths to super-Jupiters that need more precise study to understand their true nature and the context in which they exist. Working together with GAIA, JWST, TESS, PLATO, Roman Space Telescope, and the Habitable Worlds Observatory, TMT will study exoplanet systems toward answering two big questions Q5-*What is the nature of extrasolar planets?* and Q6-*Is there life elsewhere in the universe?* In addition to answering these questions directly by the characterization of each exoplanet's atmosphere, specifically looking for biosignatures, TMT will investigate the stars associated with exoplanets to improve our understanding of the effects of stellar activity on our measurements. The science cases presented here comprise a suite of programs that will enable users to characterize exoplanets with lower masses, temperatures, and ages than is now possible, moving us toward identifying the *census* of exoplanets, and possibly the *specific* exoplanets where life may reside.

The laser guide star MCAO mode of TMT's first light AO system (NFIRAOS) provides the very high sky coverage and available field of view with the diffraction-limited angular resolution needed for microlens observations and direct imaging of young gas giant planets around low mass host stars. NFIRAOS will also provide natural guide star AO modes that support coronagraphic high spectral resolution observations that will be used to characterize many gas giant and small planets. High resolution spectroscopy is particularly important for the possible detection of biosignature gasses in exoplanet atmospheres and for characterizing microlensing systems. Mid-infrared capabilities, to be provided by MICHI, will enable additional science particularly for the direct planet imaging and biosignature gas detection cases. TMT's planned future extreme adaptive optics and coronagraphic capabilities (PSI) will significantly enhance the obtainable science from direct imaging of exoplanets and the links between exoplanet properties and the nature of their host stars.

---

**Contributors**: Ian Crossfield (University of Kansas), Thayne Currie (University of Texas-San Antonio), Christian Marois (NRC Herzberg), Carl Melis (UC San Diego), William Welsh (San Diego State University), Michael C. Liu (University of Hawaii), Bruce Macintosh (Stanford University & University of California Observatories), Norio Narita (NAOJ & University of Tokyo), Angelle Tanner (Mississippi State University), Subo Dong (KIAA), Eric Gaidos (University of Hawaii), Taro Matsuo (Kyoto University), Stephen Kane (SFSU)

---

# 10 EXOPLANETS

In just 30 years, astronomers have revealed thousands of extrasolar planets. Some exoplanets have similar properties (e.g., mass, radius, age, orbital separation) to those in our own solar system. However, short-period exoplanets detected primarily by Doppler radial velocimetry and transit photometry, and distant exoplanets discovered by microlensing and direct imaging, have revealed entirely new classes of planets such as hot Jupiters, super-Earths, and wide-separation super-jovian planets.

All these worlds require precise, detailed study to understand their true nature and the context in which they exist. Future studies of exoplanet systems with TMT will be aimed toward answering the big-picture question, *What is the nature of extrasolar planets?,* and taking the next steps to answer the question, *Is there life elsewhere in the universe?*.

In addition to answering these questions directly via the spectroscopic characterization of exoplanetary atmospheres, TMT will measure precise system architectures via imaging, astrometry, and radial velocity spectroscopy, as well as investigate the stars associated with exoplanets.

The science cases presented here comprise a suite of programs that will enable users to characterize exoplanets with lower masses, temperatures, and ages than is now possible, as well as permitting far deeper characterization of planets than is presently feasible.

Although extrasolar planets discovered thus far have provided important clues about the context within which the Earth and other solar system planets fit, our knowledge of the overall census remains incomplete. We have good constraints on the (high) frequency of planets larger than the Earth with orbital periods less than 100 days and on (the dearth of) the widest-separation, most massive planets (e.g., Burke et al. 2015; Vigan et al. 2021).

However, other classes of planetary systems are almost entirely unexplored — e.g., Jupiter-like planets beyond ~5 AU, and planets with radii and masses less than Earth's (see figure 10-1), especially around sun-like stars. Furthermore, our knowledge about the physical properties of all these exoplanets (e.g., atmospheric composition, chemistry, dynamics, and thermal structure) is weaker and is restricted to a smaller number of favorable cases.

While both rocky planets and planets with Earth-like orbits are known (e.g., Howard et al. 2013), we have barely scratched the surface of true (potentially habitable) Earth twins.

The Thirty Meter Telescope will provide an enormous advance in our ability to identify and characterize extrasolar planets. Historically, new technological advances have driven a large number of exoplanet discoveries – including high-precision Doppler measurements, high-precision space-based photometry, advanced adaptive optics, and precise *Gaia* astrometry.

TMT's key advantages in exoplanet science are its high angular resolution (permitting direct characterization of previously inaccessible planets, including habitable-zone planets around the nearest stars) and high spectral resolution (facilitating atmospheric spectroscopy and Doppler observations impossible for today's facilities).

TMT will generate an incredible number of additional discoveries, drastically expand the kinds of planets we can detect, provide a rich understanding of these planets' physical properties, and potentially yield detections of habitable rocky planets, possibly even accompanied by the detection of $O_2$ and/or other biosignature gasses in their atmospheres.

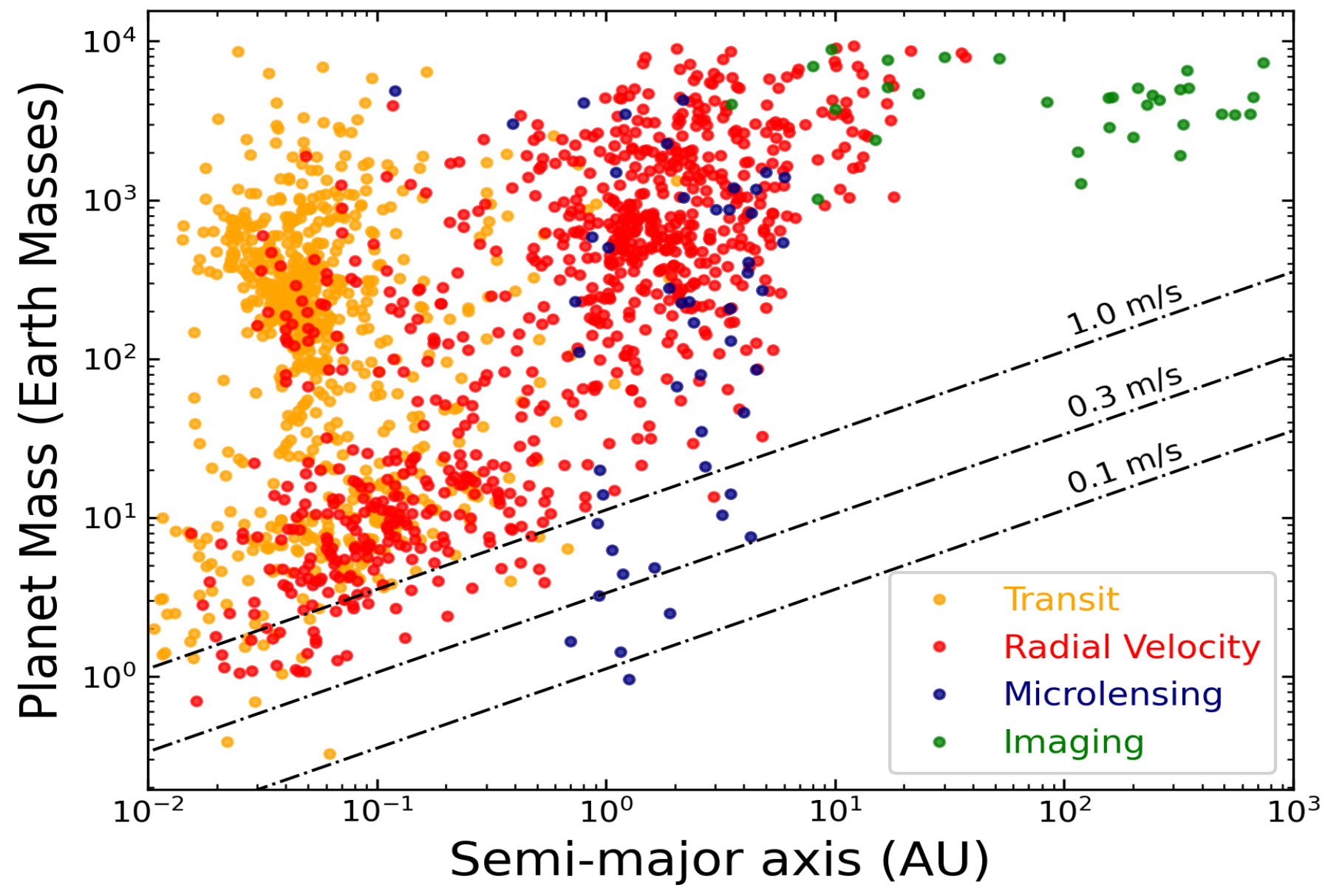

***Figure 10-1:*** *Known extrasolar planets with well-measured masses (points). Dashed lines show the typical detectable mass for a given level of Doppler precision, assuming edge-on circular orbits around a solar-type star.*

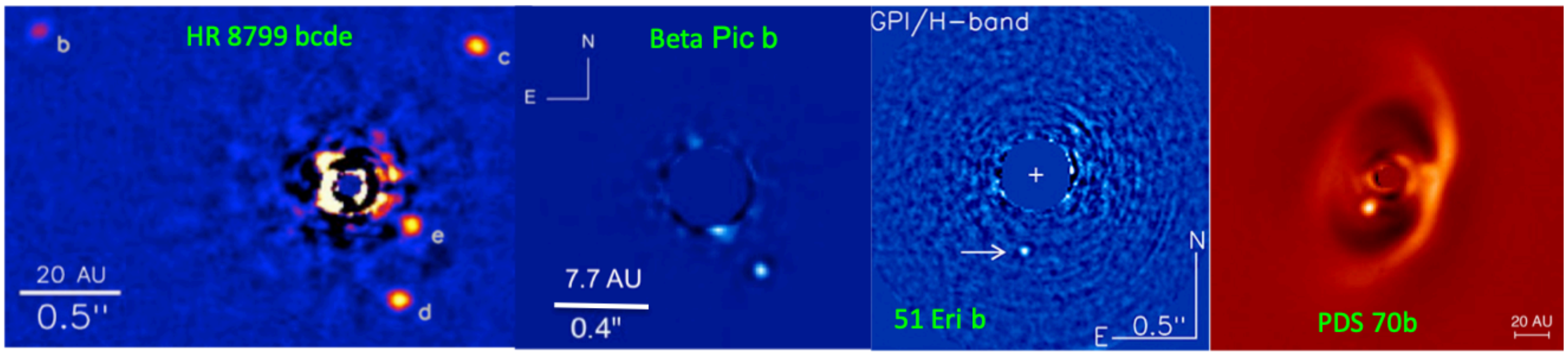

***Figure 10-2:*** *A gallery of emblematic directly-imaged planets detected with AO on 8–10 m class telescopes: fully-formed planets HR 8799 bcde (5–7 Mj; 40 Myr), beta Pic b (12 Mj; 23 Myr), and 51 Eri b (2 Mj; 23 Myr), and the PDS 70 b protoplanet (Marois et al. 2008, 2010; Lagrange et al. 2010; Macintosh et al. 2014, 2015; Keppler et al. 2018; Mueller et al. 2018). First-light TMT direct imaging will more fully characterize planets like these; dedicated planet-imaging instruments with TMT will planets lower in mass and more solar system like, including rocky planets.*

## 10.1 Direct imaging and spectroscopy of exoplanets

*Direct imaging* is a means — possibly *the* means — by which we will eventually confirm and characterize a habitable, rocky planet around a nearby star and is a key focus outlined for the US-ELT Program in the Astro 2020 Decadal Survey. Compared to indirect methods like RV and transits, direct imaging is currently best suited for detecting planets on wider, ~10–100 AU orbits and around stars significantly younger than the Sun (~1–200 Myr), when jovian planets are self luminous and detectable by infrared thermal emission (not reflected light) with adaptive optics (AO) systems on ground-based 8–10 m class telescopes. Of the ~5,000 extrasolar planets known thus far, just over 20 have been directly imaged (Currie et al. 2023a), starting with the discovery of the HR 8799 planets over 15 years ago (Marois et al. 2008).

Companions imaged thus far comprise the extremes of planet formation, with typical orbital separations of 30 AU or greater and typical masses of ~5–15 $M_{Jup,}$, and have predominantly been found around stars more massive than the Sun (e.g. Carson et al. 2013; Chauvin et al. 2017; Nielsen et al. 2019). Planet-to-star contrasts for these companions range between $10^{-4}$ to $10^{-6}$. Most directly imaged planets have estimated temperatures between ~1000 K and ~2000 K (e.g. Barman et al. 2011; Bonnefoy et al. 2018), overlapping with temperatures for brown dwarfs but significantly hotter than the gas giant planets in our own solar system. A few directly-imaged multi-planet systems have been discovered; at least two systems host imaged *protoplanets* (i.e. planets in the process of forming) (e.g. Marois et al. 2010; Lagrange et al. 2010; Nowak et al. 2010; Haffert et al. 2019; Currie et al. 2022). Several directly-imaged planets have also been detected with indirect methods RV and/or (recently) with precise astrometry (Snellen et al. 2018; Currie et al. 2023b; Franson et al. 2023; Mesa et al. 2023; De Rosa et al. 2023).

Direct imaging targets are particularly amenable to atmospheric characterization. During the timeframe probed by direct imaging with current ground-based facilities — the first ~200 Myr — jovian exoplanets are expected to undergo substantial evolution in their luminosities and atmospheric properties (e.g. temperature, surface gravity, and clouds), the exact trends for which may depend on their formation conditions (Spiegel and Burrows 2012; Marley et al. 2012). The first photometric and spectroscopic measurements of imaged planets provided key insights into the atmospheric properties of young gas-giant planets (figures 10-2 and 10-3). Near-infrared photometry indicates that young planets with temperatures of ~1000–2000 K and spectra characteristic of L-type dwarfs or (especially) L/T transition objects are dustier/cloudier than their older, field brown dwarf counterparts (e.g. Currie et al. 2011); near-IR low-resolution spectra reveal evidence for low surface gravities for many imaged exoplanets (e.g. Barman et al. 2011). Such low gravities lead to thicker condensate clouds and enhanced non-equilibrium carbon chemistry in young planetary photospheres compared to field brown dwarfs (e.g., Marley et al 2012; Skemer et al 2012, 2014). Near-IR spectra for the first, cooler (~600-800 K) T-dwarf planets discovered by direct imaging (e.g. 51 Eri b; Macintosh et al. 2015) show evidence for methane absorption and modest differences with field substellar objects that may be due to thin clouds (e.g. Rajan et al. 2017).

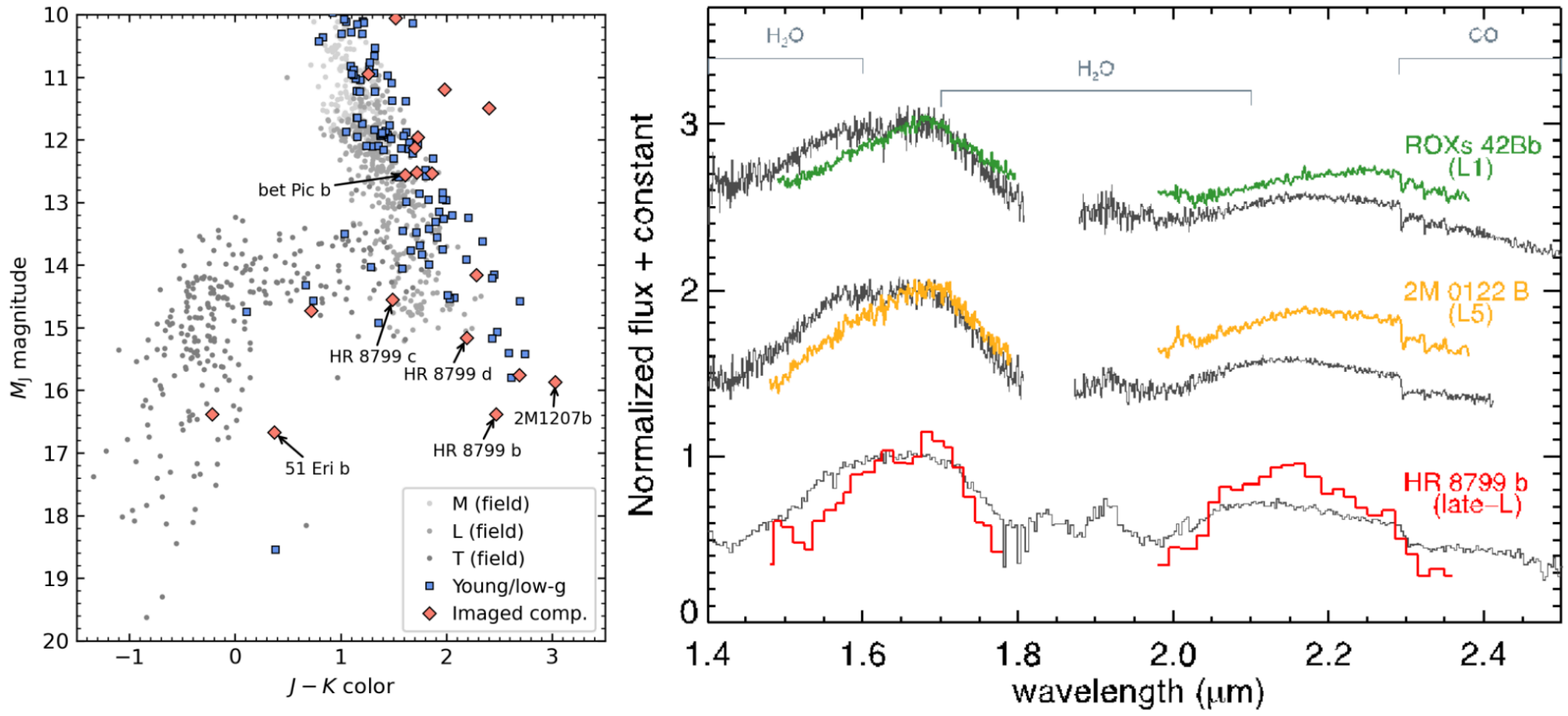

***Figure 10-3:*** *Atmospheres of directly imaged planets. Left: Near-IR color-magnitude diagram comparing field brown dwarfs (gray symbols) and young substellar companions (colored points; from Currie et al. 2023a). The young objects form a distinct sequence in the diagram from the field objects, most of them being redder and some extending to fainter J-band magnitudes. Right: Near-IR spectra of the young companions ROXs 42Bb (~9 $M_{Jup}$, ~3 Myr, Currie et al. 2014; spectra are from Bowler et al. 2014), 2MASS 0122-24B (~13 $M_{Jup}$, 125 Myr; Bowler et al. 2013), and HR 8799b (~5 $M_{Jup}$, 30 Myr; Barman et al. 2011), arranged from the hottest object at the top to the coolest object at the bottom. For each companion, a field object of comparable*

*spectral type is also plotted, normalized to the peak H-band flux. The differing colors and continuum shapes reflect the lower surface gravities (e.g., younger ages) of the companions.*

Near-IR high-resolution spectropolarimetry of directly imaged young Jovian exoplanets will provide a means to study cloud characteristics such as grain composition and cloud heights and structure (Marley & Sengupta 2011). The relative polarized signal from thermal emission from the planet itself reflected off clouds is much smaller than the potential signal from host-star reflected light, opening the possibility of spectropolarimetry enhancing the ability to study smaller, cooler, more mature exoplanets compared to spectroscopy alone (Bailey et al. 2018, Chakrabarty & Sengupta 2021, Sanghavi et al. 2021).

Thermal infrared (3–5 micron) spectra of directly-imaged super-jovian exoplanets diagnose carbon chemistry — e.g. methane and carbon monoxide at 3.3 and 4.7 microns, respectively — and is also shaped by clouds (e.g. Miles et al. 2018, 2020). Early L-type exoplanets (e.g. kappa And b) have thermal IR spectra matching those of young, low-gravity brown dwarfs (Stone et al. 2020); modeling of thermal IR data for cooler exoplanets like HR 8799 b reveal additional evidence for non-equilibrium carbon chemistry (Skemer et al. 2012, 2014). Medium and high-resolution spectra (R ~ 1,000-100,000) of imaged exoplanets obtained in the near-IR data measure rotation periods and radial velocities, resolve molecular lines, and may yield clues to planetary formation histories by quantifying the relative abundances of carbon, oxygen, and other elements and isotopologues (Snellen et al. 2014; Konopacky et al. 2013, Barman et al. 2015; Moillere et al. 2020; Ruffio et al. 2021).

Recent major direct imaging surveys with leading extreme AO systems — e.g. SPHERE/SHINE and Gemini/GPIES — detect super-jovian planets and very low-mass brown dwarfs around ~several percent of nearby, young stars between 10 and 100 AU (Nielsen et al. 2019; Vigan et al. 2021). Nearly all of these planets now have low-resolution near-IR spectra. But these surveys are largely unable to image planets with masses comparable to or lower than Jupiter or those at sub-Saturn orbital separations. However, RV surveys indicate that the jovian planet frequency peaks at ~3 AU and sub-jovian-mass planets are more common than super-jovian ones (e.g. Fulton et al 2021). Furthermore, only a subset of known imaged planets – predominately the brightest, most massive, and/or most widely separated – are amenable to medium-to-low resolution spectroscopy with current facilities.

Exoplanet direct imaging science capabilities are heavily driven by technology development, in particular AO and coronagraphy. Facility (conventional) AO systems on 8–10 m telescopes equipped with simple coronagraphs (or no coronagraphs at all) were used for the first direct imaging searches and yielded the first discoveries and earliest atmospheric characterization studies. In the near IR, imaging detections with conventional AO systems on these telescopes are generally limited to massive, bright, and very young exoplanets at wide angular separations (~0.5” or greater; typically 30–50 AU or greater). In the thermal IR, the same AO systems have generally been more successful with imaging detections because young, self-luminous jovian planets are intrinsically brighter in the mid-IR while their host stars are fainter (e.g. Marois et al. 2010; Skemer et al. 2014).

The past 10 years have seen the rapid development and deployment of *extreme* AO systems on 8–10 m telescopes, with spatially-filtered and more sensitive wavefront sensors, more wavefront control with two-level woofer-tweeter pairs, advanced coronagraph designs, and an IFS operating in the near-IR (e.g. Beuzit et al. 2008; Chilcote et al. 2020; Groff et al. 2016). These enhanced technical capabilities provide far better image quality (e.g. Strehl ratios of ~0.9 in the near-IR) and contrasts of $10^{-6}$ at <0.5” separations, a factor of 10–100 improvement over conventional AO systems (Macintosh et al. 2014). This improved performance results in more precise spectra of known planets and discoveries of planets with lower masses and located at smaller angular separations (e.g. Macintosh et al. 2015; Zurlo et al. 2016; Currie et al. 2023b; De Rosa et al. 2023). Examples of extreme AO systems include the Gemini Planet Imager, Spectro-Polarimetric High-contrast Exoplanet REsearch instrument (SPHERE), and the Subaru Coronagraphic Extreme Adaptive Optics project (SCExAO) (Macintosh et al. 2014; Buezit et al. 2019; Jovanovic et al. 2015). Some current and upcoming thermal IR IFS instruments are fed light by AO systems providing a high-fidelity correction of wavefront errors, such as LBT/ALES and Keck/SCALES (Skemer et al. 2018; Skemer et al. 2022).

Planet-to-star contrasts regularly achieved with the best current extreme AO systems after post-processing at 0.5" ($10^{-6}$) are still a factor of 10 to 100 times brighter than those needed to image the brightest and most easily-detectable gas giant planets in reflected light ($10^{-7}$–$10^{-8}$) or rocky, habitable zone planets ($10^{-8}$) around the nearest low-mass stars at much smaller angular separations (Currie et al. 2023a). However, the key technical barriers to achieving deeper contrasts are now better understood and measures to overcome them are being actively developed (see next subsection). Raw contrast limits, especially those within several λ/D, are primarily set by servo lag error (i.e. the speed of an AO system correction vs. timescale for the atmosphere to change) and WFS measurement error and secondarily by chromatic wavefront errors and other non-common path aberrations (e.g. Guyon 2005; Guyon et al. 2018). With precise calibration of these wavefront error sources and a high-performance coronagraph, raw contrasts below $10^{-5}$ (approaching $10^{-7}$ after post-processing) near the diffraction limit are possible with 8–10 m telescopes. On TMT, the same level of wavefront error mitigation results in a factor of 10–100 deeper raw contrasts at smaller angular separations of ~0.03–0.05" (Guyon 2005).

Thus, our current knowledge of exoplanets by direct imaging with 8–10 m class telescopes represents only a glimpse of the full range of planet demographics across planet masses and orbital separations and an even coarser understanding of detailed planet atmospheric characterization. High-contrast imaging technology honed over the next decade on 8–10 m telescopes and deployed on TMT opens up a vast new discovery space, including mature gas giant planets and habitable-zone rocky planets.

The Thirty Meter Telescope has the potential to fundamentally reshape our understanding of extrasolar planets detectable by direct imaging (see figure 10-4). It will deepen our understanding of the formation, orbits, and atmospheric properties of gas giant planets that are or will be detectable with current 8–10 m class telescopes, expand the discovery space to mature and more solar system-like planets both in reflected light and thermal emission, and provide a first look at small exoplanets within the habitable zones of the nearest stars.

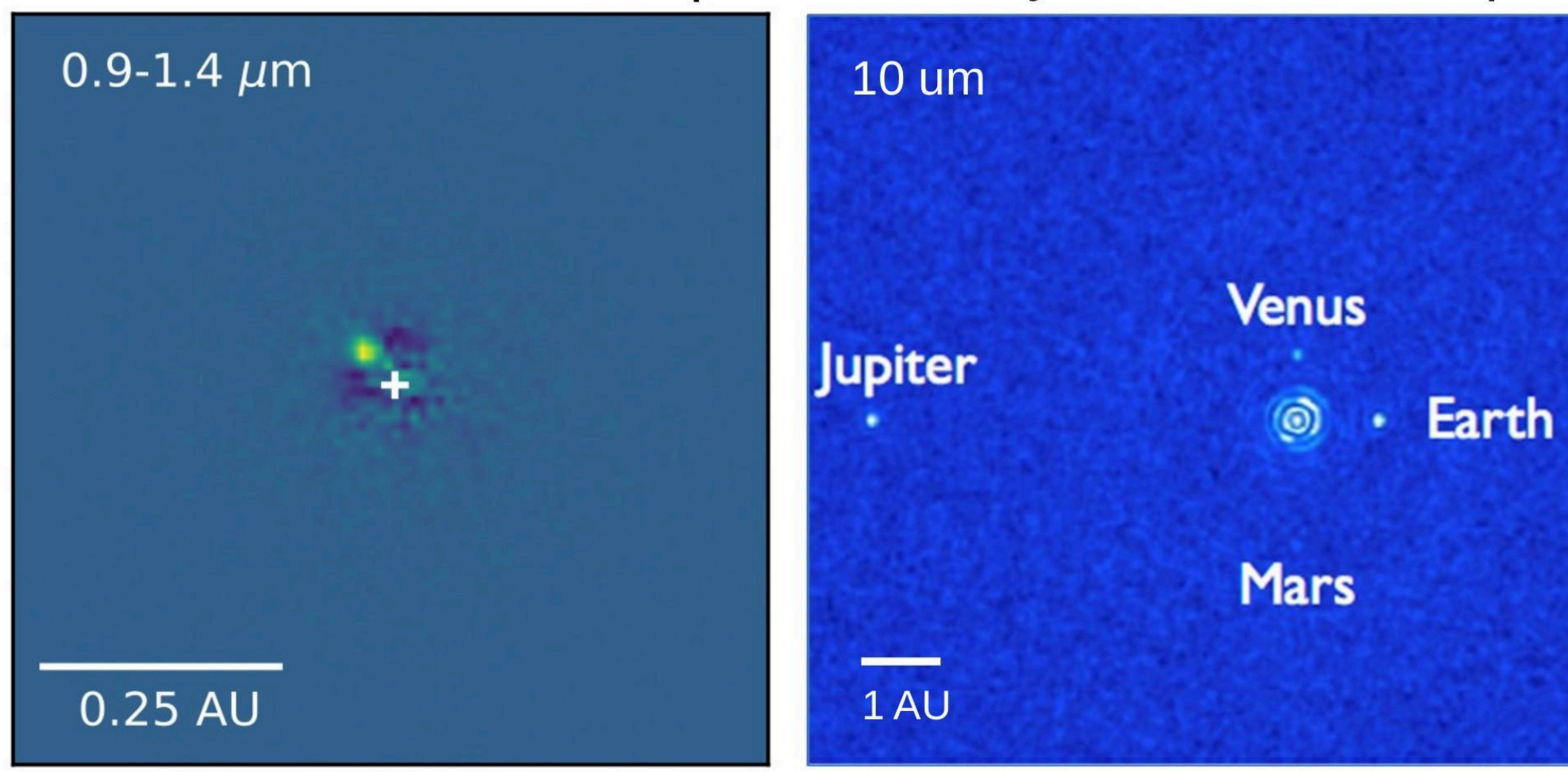

***Figure 10-4:*** *Simulated TMT observations of a habitable-zone terrestrial planet at 3.5 pc around an M dwarf (left, collapsed PSI-Blue datacube for a 30 hours observation; Sallum et al. 2022) and a nearby G dwarf (right, MICHI image with integration time of several hours). TMT will revolutionize our understanding of exoplanets beyond what is possible today: probing habitable-zone worlds around the nearest stars via direct characterization and in transit, measuring atmospheric chemistry, composition, and dynamics in systems young and old and from rocky planets to gas giants; and via precise radial velocities and high-resolution imaging.*

TMT's chief science focuses with direct imaging include both those achieved first-light instrumentation and dedicated direct imaging instruments that will come online later. First-light TMT observations will provide unique and powerful follow-up characterization observations of gas-giant planets discovered by current-generation, 8–10 m class telescopes and new discovery potential for other gas giant planets:

- NFIRAOS+IRIS will yield high spectral-resolution spectroscopy that will establish chemical abundances for a large sample of known exoplanets. High-precision astrometry will enable monitoring of planetary orbits, providing a unique window into the planets' dynamical histories. High-dispersion coronagraphy with MODHIS will measure the compositions and spins of a large number of self-luminous planets and possibly mature planets at AU scales, along with radial velocities to measure the planets' orbital velocities and perhaps even detect massive satellites orbiting these targets (see Radial Velocity section).
- NFIRAOS+IRIS will also target optically faint stars that are inaccessible to most current AO systems and identify these planets at smaller angular separations. Low-mass M dwarfs and brown dwarfs may be the most common type of planetary host (Currie et al. 2023a), and NFIRAOS+IRIS will be able to detect sub-Jupiter mass planets around these stars. Similarly, high-contrast imaging of young stars in the nearest star-forming regions will observe planet formation in action during the critical ~1–10 Myr epoch when gas-giant planets are assembling.

Building upon two decade's worth of technological development with extreme adaptive optics and coronagraphy on 8–10 m telescopes within the TMT community, TMT's dedicated, second-generation planet imaging instruments will explore currently uncharted discovery space in exoplanet properties and provide direct images of rocky exoplanets in the habitable zones of the nearest stars:

- TMT's Planetary Systems Imager (PSI) will detect planets on AU orbits, separations previously only accessible to long temporal baseline Doppler monitoring. Combining direct imaging data with RV or precision astrometry will directly constrain the planet masses and orbits simultaneously, providing a mapping between a planet's atmospheric properties and its mass. Fed by a mid-IR AO system, MICHI will provide imaging and high resolution spectroscopy capabilities.
- While current direct imaging detections are limited to young, self-luminous planets, PSI will image mature, temperate Neptune to Jupiter-sized planets in reflected light on AU scales.
- PSI will be capable of imaging rocky planets in reflected light in the habitable zones around the nearest low-mass stars (<0.5 $M_\odot$) and perhaps provide the first probes of biosignatures.
- With PSI and/or the Mid-Infrared Camera High-Disperser & IFU-Spectrograph (MICHI), TMT may be capable of detecting large rocky, habitable zone planets at 10 microns around a dozen of the nearest stars.
- Around young nearby stars where the final assembly phases of terrestrial planet formation is underway, PSI will be sensitive to molten rocky planets and could identify signposts of active planet assembly, providing a unique window into the early history of the solar system's terrestrial planets.

### 10.1.1 Exoplanet imaging: landscape in 2030

Direct imaging with the Thirty Meter Telescope will build upon technology matured with current and upcoming exoplanet imaging facilities. Additionally, TMT's exoplanet imaging capabilities will complement upcoming space missions. Figure 10-5 displays a timeline for the development of these capabilities, summarizing TMT's unique science focus and the other platforms that will be contemporaneous with TMT.

The next 5 years (2024–2029) will see the development of second-generation extreme AO systems — the upgraded Subaru/SCExAO, SPHERE+, and GPI-2.0 — which will yield contrasts below $10^{-6}$ at several diffraction beamwidths, a factor of 10 gain over current performances. Furthermore, these systems prefigure high-contrast imaging architectures with TMT and prototype technologies needed to image planets in reflected light with TMT (e.g. Lozi et al. 2022; Chilcote et al. 2020). For example, at Subaru AO-3K will replace Subaru's faculty AO system (188-actuator DM and optical wavefront sensor) with a 3,000 actuator DM and a near-IR Pyramid WFS, leaving SCExAO to calibrate residual wavefront errors and/or perform focal-plane wavefront sensing (Lozi et al. 2022).

These systems will gain improved performance via transformative advances in wavefront control, calibration, and starlight suppression. Capabilities include sensor fusion, combining measurements from multiple wavefront sensors to improve wavefront calibration, and predictive control, which optimizes how wavefront sensing data are used to estimate wavefront errors (e.g. Guyon et al. 2020; Males and Guyon 2018). Focal-plane wavefront sensing techniques, speckle nulling, electric field conjugation, and linear dark field control, being honed on these systems will enhance raw contrasts (e.g. Miller and Guyon 2017). Such systems, together with further mature aggressive coronagraph designs, will maximize starlight suppression at small angles (e.g. Nguyen et al. 2022).

In parallel with new technological developments, exoplanet direct imaging programs are substantially maturing. Unbiased or blind direct imaging surveys carried out with extreme AO and GPI/SPHERE have constrained the frequency of gas giant planets beyond 10 AU but have resulted in a low rate of detections. Many surveys carried out now instead target only stars showing evidence for the dynamical pull of a companion, usually from calibrated Hipparcos and Gaia precision astrometric data. These surveys, while in their infancy, have shown exceptional success in yielding a higher frequency of imaged exoplanets and low-mass brown dwarfs (Currie et al. 2023b; Franson et al. 2023; De Rosa and Nielsen 2023; Mesa et al. 2023). Additionally, the combination of direct imaging with astrometry far better constrains the planets' orbits and yields direct dynamical mass measurements instead of highly uncertain inferred masses gained through evolutionary models (Brandt et al. 2021). The next Gaia data release (DR4) will identify astrometric evidence for lower mass planets than identifiable with current data, which can be imaged with 2nd-generation extreme AO systems. As a result, the TMT era will begin with a far larger sample of imaged planets than known now, with measured masses and orbits, whose atmospheres can be constrained in more detail with follow up data.

Exoplanet imaging with the recently launched *JWST* has demonstrated the power of the thermal IR for exoplanet atmospheric characterization (Miles et al. 2023; Petrus et al. 2023) and the need for such capabilities with TMT. While *JWST's* sensitivity at very wide, background-limited separations (>2") is and will be unmatched, current 8–10 m class telescopes outperform *JWST* in raw planet detection capability within 0.5", and TMT will have an even greater advantage. As a consequence, *JWST* does not obviate the need for TMT but provides a complementary probe of planets by direct imaging, well suited for exploring extremely cold and wide separation planets vs. TMT's focus on planet detection on solar system scales.

The Coronagraphic Instrument (CGI) on the *Roman Space Telescope* will provide the first demonstration of wavefront control in space, a key stepping stone towards a future NASA flagship mission focused on imaging exo-Earths around Sun-like stars in reflected light. CGI will focus on exoplanet imaging at ~0.15–0.5" separations and will be capable of imaging small numbers of self-luminous planets and at least some mature planets in reflected light. CGI will have an inner working angle of 0.15", or roughly 15 diffraction-beamwidths for TMT at 1.6 microns, but may be capable of achieving contrasts at ~0.15"–0.5" nearly a factor of 10 deeper than possible with even PSI on TMT without additional fundamental advances in ground-based AO (e.g. Bailey, et al. 2023). Thus, TMT and *Roman* CGI likewise probe a complementary phase space. For some nearby stars, *Roman* and TMT could both search for and characterize planets in different regions: Neptunes to Jupiter-sized planets near the ice line for *Roman* and super-Earths to Neptune-sized planets at smaller separations for TMT.

Longer term (2040s), NASA plans to fly the Habitable Worlds Observatory (HWO), a flagship coronagraphic mission focused on imaging and spectrally characterizing Earth-like planets in the habitable zones around the nearest ~100 stars in reflected light. With current and reasonably foreseeable technology, this science case is best carried out in space, as the contrasts required to image a true Earth twin around a Sun-like star in reflected light are 2 orders of magnitude deeper than the $10^{-7}$ requirement for the *Roman* CGI (the HWO requirement is $10^{-10}$). However, here too, TMT and HWO are highly complementary. HWO's planned 6 m primary is too small to access the habitable zones of the nearest low-mass stars that can be resolved with TMT/PSI. Some planets detected in thermal emission at 10 microns with TMT could be detected in reflected light with HWO.

A reflected light and thermal emission detection of the same habitable zone planet constrains the planet's energy budget, temperature, and radius better than a single detection alone (e.g. Wang et al. 2019). Working in concert, TMT and space-based missions like HWO will transform our knowledge of the diversity of nearby exoplanetary systems and take significant steps toward being able to image and characterize a habitable Earth-like planet. Figure 10-6 shows the synergies and separate capabilities of ground based ELT and spaced based coronagraph systems.

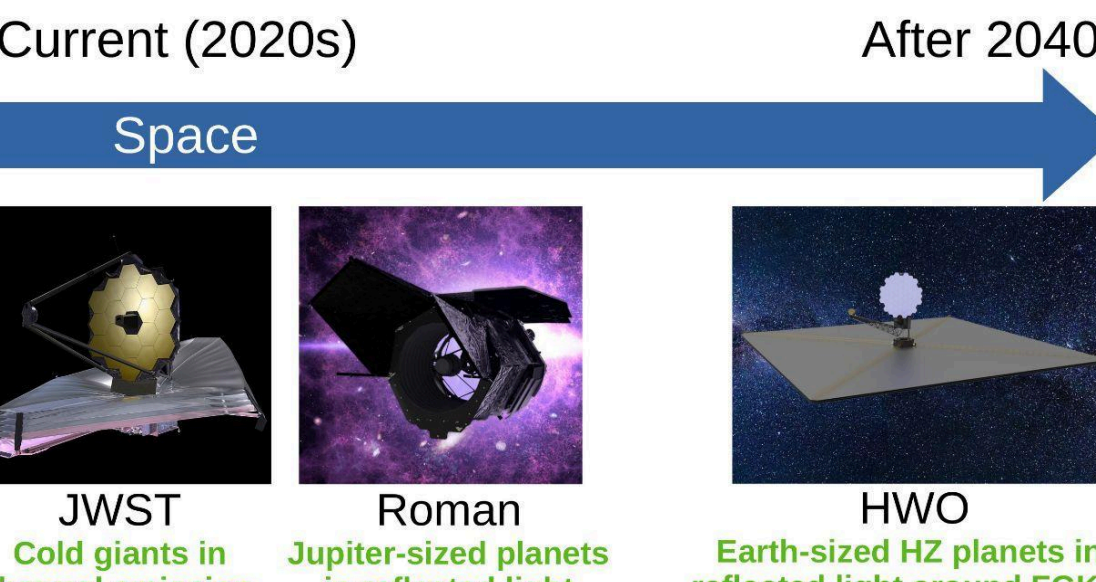

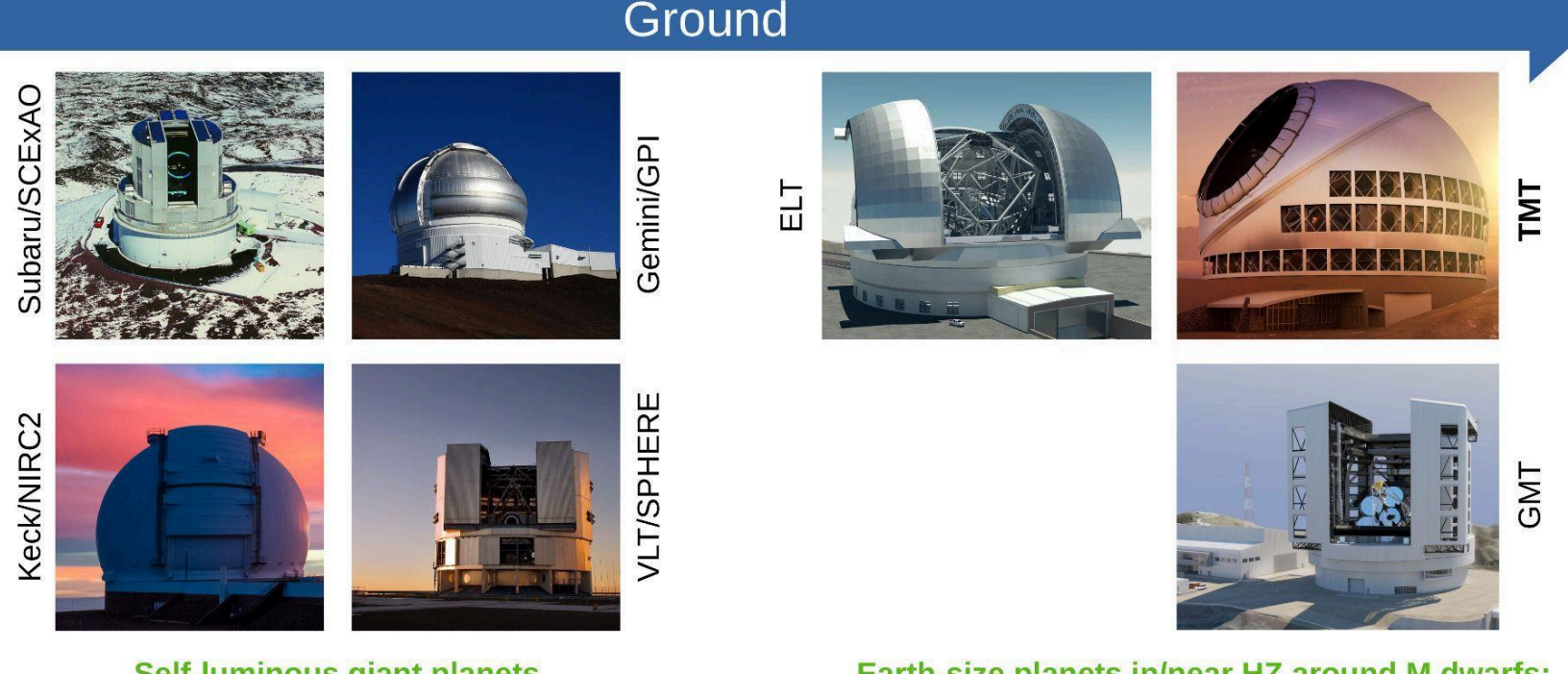

***Figure 10-5:*** *Timeline of direct imaging-capable space-borne missions and ground-based facilities from the present (2024) to TMT first light (and beyond) and their unique phase space*

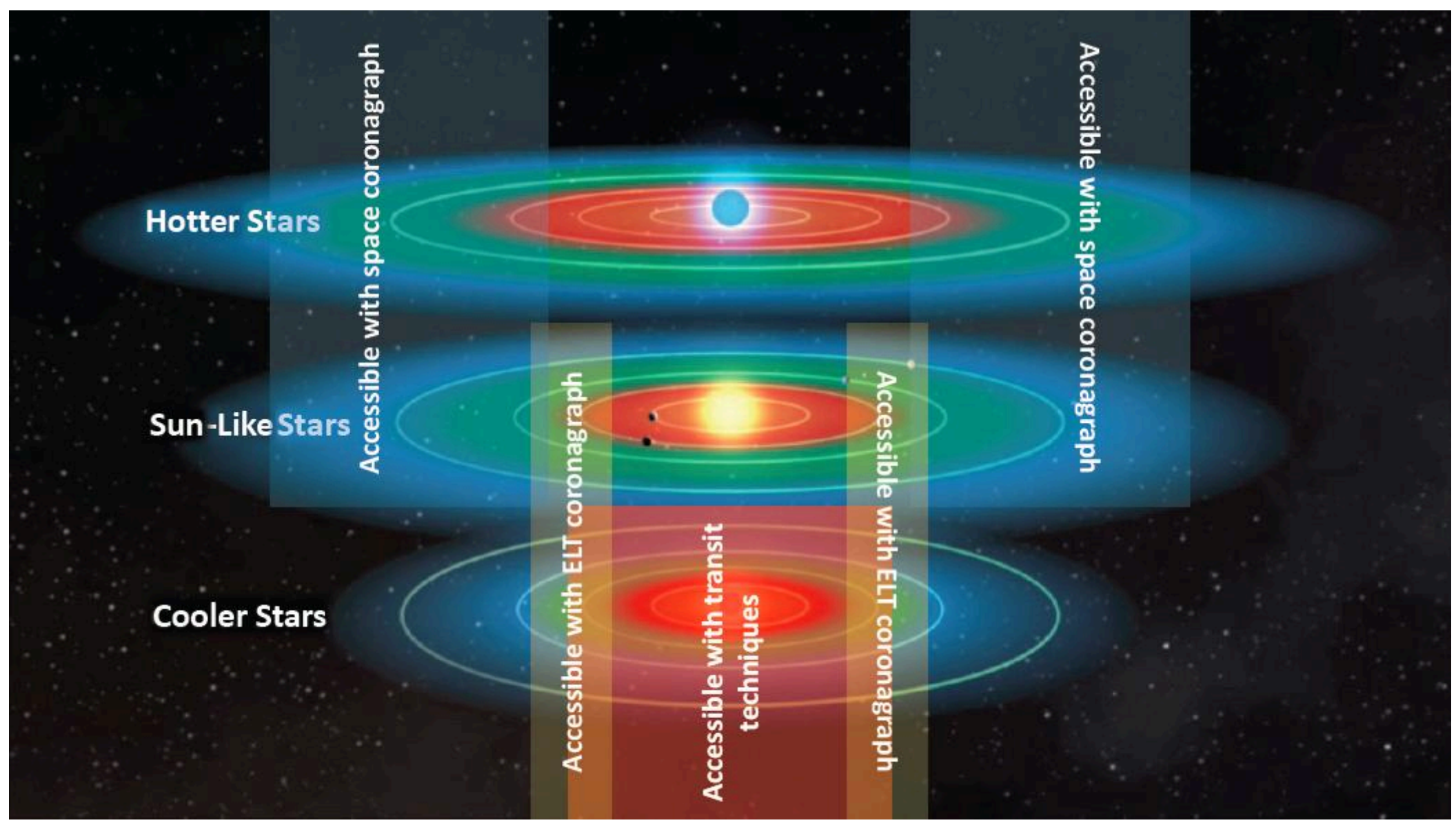

***Figure 10-6:*** *The habitable zones around stars of varying mass where liquid water can exist and the space based and ground based telescopes needed to probe these regions using coronagraphic and transit techniques (figure 2.1.1 of ASTRO2020 Decadal Review)*

### 10.1.2 First-Light Spectroscopy and Imaging of young gas giants with NFIRAOS, IRIS and MODHIS

The AO-fed TMT/IRIS, as well as high-dispersion coronagraphy with MODHIS, will provide a powerful exoplanet direct imaging and spectroscopy capability at TMT's first light (figure 10-7). IRIS will detect planets at factors of 3–4 smaller angular separation than possible with 2nd-generation extreme AO systems (e.g. upgraded SCExAO; GPI-2.0; SPHERE+), owing to TMT's larger aperture. It will also re-detect at medium spectral resolution planets discovered with 8–10 m class telescopes. MODHIS will harness the power of recent developments in coronagraphic spectroscopy to characterize these (and other, new) planets in detail. Spectra from both IRIS and MODHIS will therefore provide a wealth of information about atmospheric composition and chemistry for a large sample of gas giant planets that is inaccessible from photometry or low resolution spectroscopy. Their data will yield resolved spectral lines (e.g. carbon monoxide, water, methane) and rotation rates (e.g. Konopacky et al. 2013). Modeling these spectra will constrain chemical abundances by measuring multiple elemental abundances (e.g. C, N, O, Fe, along with other elemental and isotopic ratios), and spin rates via $v \sin i$. Such measurements may ultimately help to probe the planets' formation environments (e.g. Oberg et al. 2011).

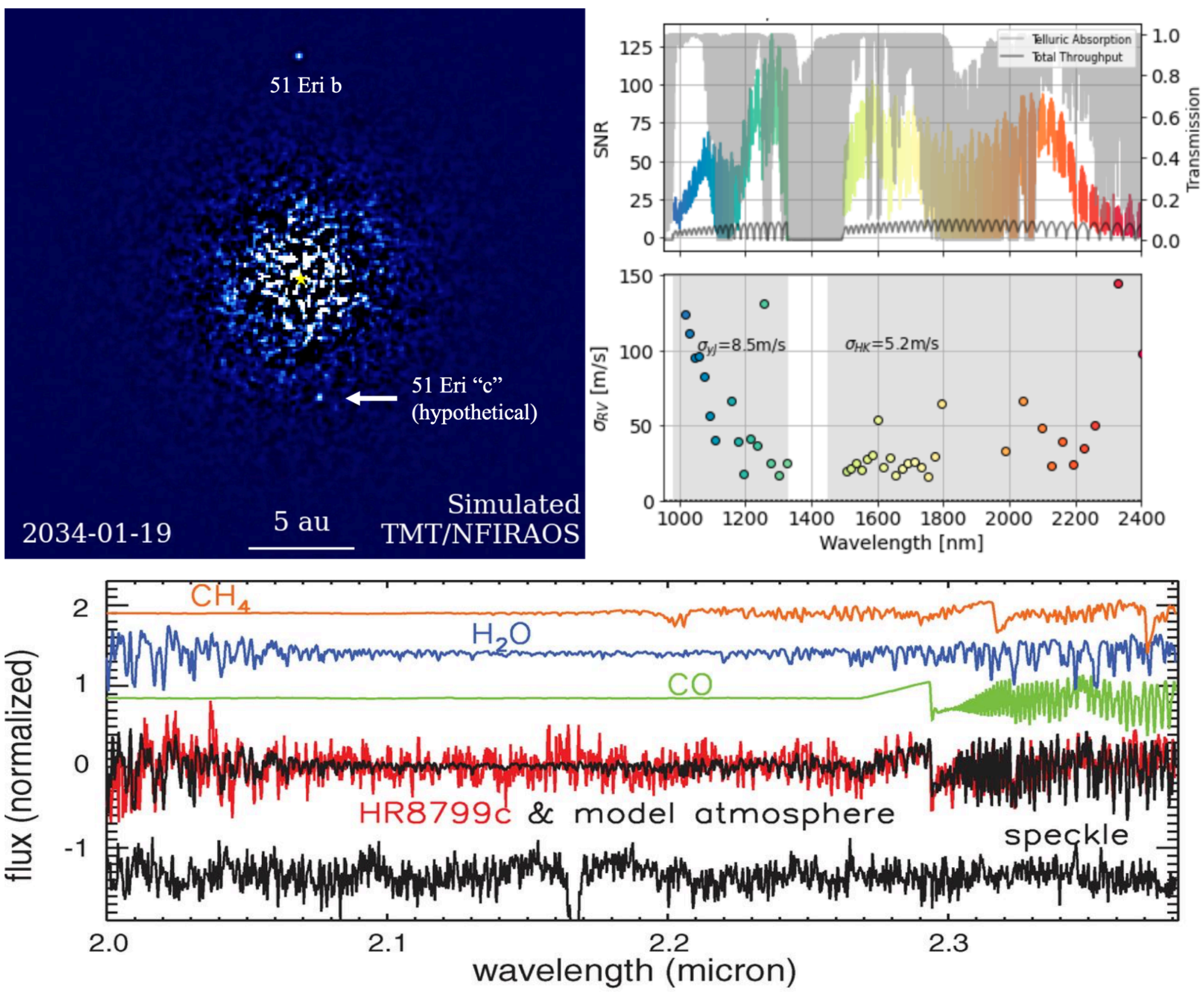

***Figure 10-7:*** *Exoplanet imaging science with TMT at first light. Top-left: Simulated NFIRAOS/IRIS image of the 51 Eri planetary system courtesy of J. Wang, showing a high SNR detection of 51 Er b (2 Jupiter masses) and the detection of a hypothetical second jovian planet with the same mass (51 Eri "c") at a Jupiter-like separation made possible by TMT's improved angular resolution. (Top-right): MODHIS high-resolution spectroscopy of the HR 8799 c planet. The panels show the expected SNR as a function of wavelength and limits on radial-velocity precision obtained by analyzing HR 8799 c's resolved spectral lines. (Bottom) Medium-resolution spectrum of HR 8799 c obtained from Keck/OSIRIS (Konopacky et al. 2013) showing multiple water and CO lines but little evidence for methane. IRIS spectra will provide similar probes of atmospheric abundances for a far larger sample of younger gas giant planets.*

TMT's superior angular resolution over 8–10 m class telescopes will deliver far more precise astrometry for known planets detected with NFIRAOS/IRIS (e.g. ~3 mas precision for most planets detected at a high SNR). Combined with new astrometric data from *Gaia* and/or precision RV obtained over the next decade, NFIRAOS/IRIS will provide exceptionally precise limits on the orbital parameters and dynamical masses of a large sample of young gas giant planets probing a range of ages. Orbital constraints will investigate spin-orbit alignments between the planets and their host stars, planet eccentricity distributions, mass functions, and a mapping between a planet's atmospheric properties and its dynamical mass. A NFIRAOS/IRIS program characterizing known exoplanets will then provide robust investigations of how the atmospheres of planets of different masses evolve with time.

High dispersion coronagraphy with MODHIS can also be used to conduct direct spectroscopy of gas giant planets, constraining their compositions, radial velocities, and spins. For the most favorable exoplanets (as well as for large numbers of brown dwarfs), Doppler Imaging techniques will permit two-dimensional, global maps via Doppler Imaging (figure 10-8). MODHIS will utilize a vortex fiber nuller (Ruane et al. 2018), already demonstrated at Keck (Echeverri et al. 2023), which combines nulling interferometry with high resolution spectroscopy and enables exoplanet detection at or even interior to one diffraction beamwidth. MODHIS will thus be able to characterize the atmospheres of planets on small orbits at various key stages in their formation and evolution: 1) newborn (1 Myr) protoplanets residing in nearby (~140 pc) star-forming regions to witness planet formation in action; 2) adolescent (~30 Myr) planets that recently finished forming and are still radiating away the signatures of their formation process, and 3) some mature (> 1 Gyr) planets. These measurements will provide key insights into how gas giant planets evolve. Planetary orbits evolve through gravitational interactions and the amount orbits change informs us how dynamically hot planetary systems are. Atmospheric composition changes due to accretion of solids and gasses from the circumstellar disk. If magnetic breaking with the protoplanetary disk is indeed responsible for setting planetary spin rates, protoplanets likely have a different spin distribution than older planets for which angular momentum is conserved.

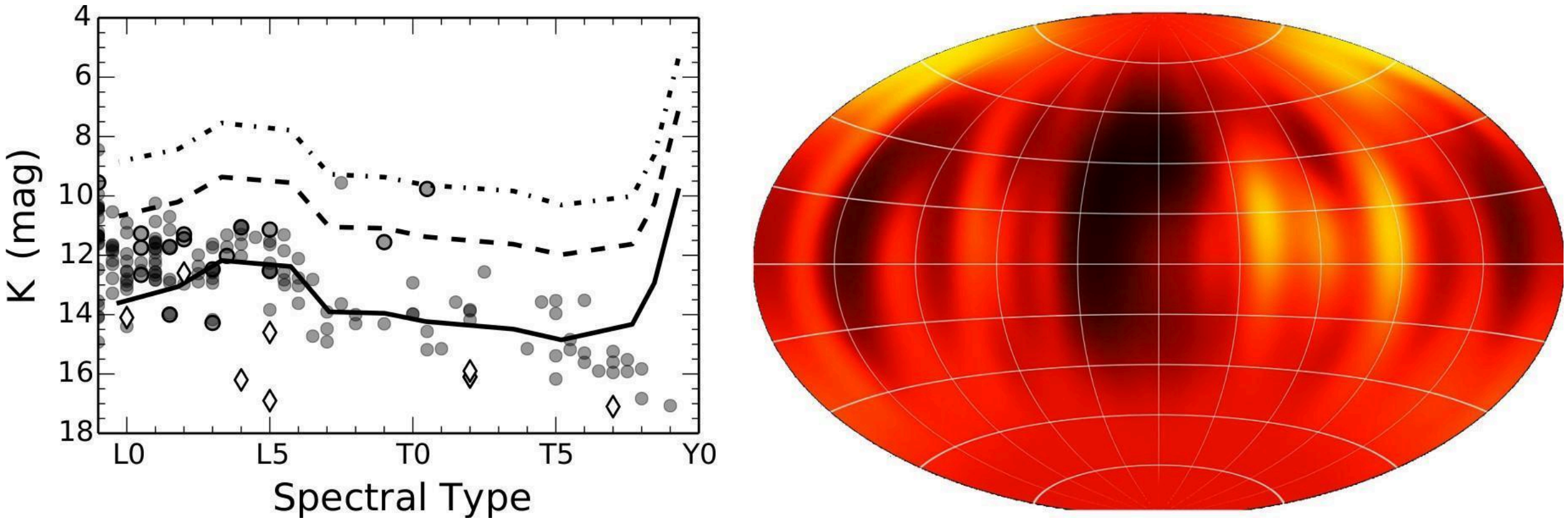

***Figure 10-8:*** *Left: The solid line shows the estimated sensitivity for K-band Doppler Imaging with TMT/MODHIS compared to known brown dwarfs (circles) and directly imaged planets (diamonds), from Crossfield (2014). TMT will facilitate global weather mapping of several dozens of substellar objects and several directly-imaged planets and dozens of brown dwarfs, as shown at Right for brown dwarf WISE 1049B (Crossfield et al. 2014).*

Much like standard Doppler imaging, direct spectropolarimetry of rotating gas giant exoplanets can be used to create Zeeman Doppler images of exoplanets' magnetic fields from variation in the line profiles in polarized light. A goal of Zeeman Doppler imaging with MODHIS will be to create the first magnetic field maps for gas giant exoplanets. Bright, massive planets like Beta Pictoris b or HR 8799 bcde are amenable to high SNR spectropolarimetric observations, but are inaccessible with current facilities because of the need for AO and high-contrast techniques. MODHIS naturally fills this niche with its fundamental design as a high-contrast instrument fed by AO (e.g., Plummer & Wang 2023).

Magnetic fields are found in the gas giants in our solar system such as Jupiter, but have not been definitively measured on any other exoplanets. Beyond simple detection of magnetic activity, MODHIS will be able to explore the structure of the magnetic field and therefore provide hints about the interiors of these planets. This will also shed light on possible interactions between the magnetic field of the planet and orbiting moons, such as those observed between Jupiter and Io that give rise to aurora on Jupiter (e.g., Khurana et al. 2004).

One unique and powerful capability offered by TMT will be laser guide star (LGS) AO high-contrast imaging. Low-mass stars (<0.5 $M_{\odot}$) are the most common types of stars in the stellar neighborhood and they represent the most common hosts of planetary systems. However, many nearby low-mass stars are too faint for ExAO systems on current telescopes doing wavefront sensing at optical wavelengths (which are limited to I = 8–9 mag) or even the near-infrared (H < 9–10).

Stars in embedded clusters probing the earliest stages of planet formation likewise are often too faint in the optical or even the near infrared for high-quality wavefront sensing. Identifying and characterizing imaged planets on Jupiter-like separations in these systems, complete a picture of planetary system evolution and diversity that would otherwise be missing. Planets on such orbits are too distant for most precision RV surveys. *JWST* lacks the angular resolution to probe these regions around all but the nearest young stars.

Finally, NFIRAOS+IRIS LGSAO high-contrast observations will provide unprecedented sensitivity to planet detections around nearby low-mass stars and embedded, newly-born stars. Since low-mass stars are much less luminous than more massive stars, NFIRAOS+IRIS data analyzed with advanced post-processing techniques (e.g. algorithms utilizing angular/spectral differential imaging) will be able to detect young Jupiter planetary companions at the same contrast limit. These companions will have atmospheres that are cooler than more massive planets that have been imaged around young stars like HR 8799 bcde and beta Pic b and will help us probe the diversity of exoplanet atmospheres. IRIS detections of planets around the youngest stars will probe the earliest stages of gas giant planet formation on Jupiter-like scales.

### 10.1.3 Opportunities at First Light to Explore the Nearest Terrestrial Worlds

One of the top goals of exoplanet science is finding other habitable worlds. By combining the high angular resolution and high spectral resolution capabilities together, MODHIS offers us the first attempt to directly measure the spectra of exoplanets residing in the habitable zone around other nearby stars. In particular, MODHIS operating behind the NFIRAOS AO system is capable of reaching the necessary starlight suppression to detect and characterize systems that are like Proxima b, the closest terrestrial planet that resides in the habitable zone of its host star (Wang et al. 2017).

Although Proxima Centauri itself is too far south for TMT, there are a handful of other nearby low-mass stars that MODHIS is capable of detecting habitable zone exo-Earths around (Ruane et al. 2019). For these stars, there is incomplete knowledge whether they harbor terrestrial planets in their habitable zones, but upcoming precise infrared spectrographs like MODHIS would be capable of detecting these planets indirectly before performing any direct spectroscopy observations. While it is only a handful of systems and the habitability of planets around low-mass stars is still uncertain, this would likely be our first attempt to perform direct spectroscopy of an exo-Earth candidate, and should not be missed.

### 10.1.4 Probing Clouds in Exoplanet Atmospheres

The measurement of a large number of featureless transmission spectra of smaller Neptune-like planets has been attributed to the presence of hazes or aerosols in the planet's atmosphere (e.g., Sing et al. 2016). With current spectral datasets, clouds are extremely challenging to characterize and are inherently complex three-dimensional structures that tend to be approximated as one dimensional, uniform sources of opacity. Part of the complexity stems from the uncertainty in the composition and the distribution of sizes of particles in the clouds.

Size and composition can be explored with spectropolarimetry of directly imaged exoplanets to probe different atmospheric depths, enabling modeling of the cloud structure as a function of height and providing information about grain composition. Signal strengths of up to 1.5% in the J band can result from thermal emission from the planet being scattered off the clouds (Marley & Sengupta 2011).

In addition, reflected host-star light from a variety of exoplanets, scattered from particles in the clouds, will offer constraints of cloud properties (e.g., Bailey et al. 2018, Chakrabarty & Sengupta 2021, Sanghavi et al. 2021). Predicted signal depends strongly on orbital phase and inclination of the orbital plane, ranging from zeros to upwards of 60% polarization fraction in favorable geometries, see figure 10-9.

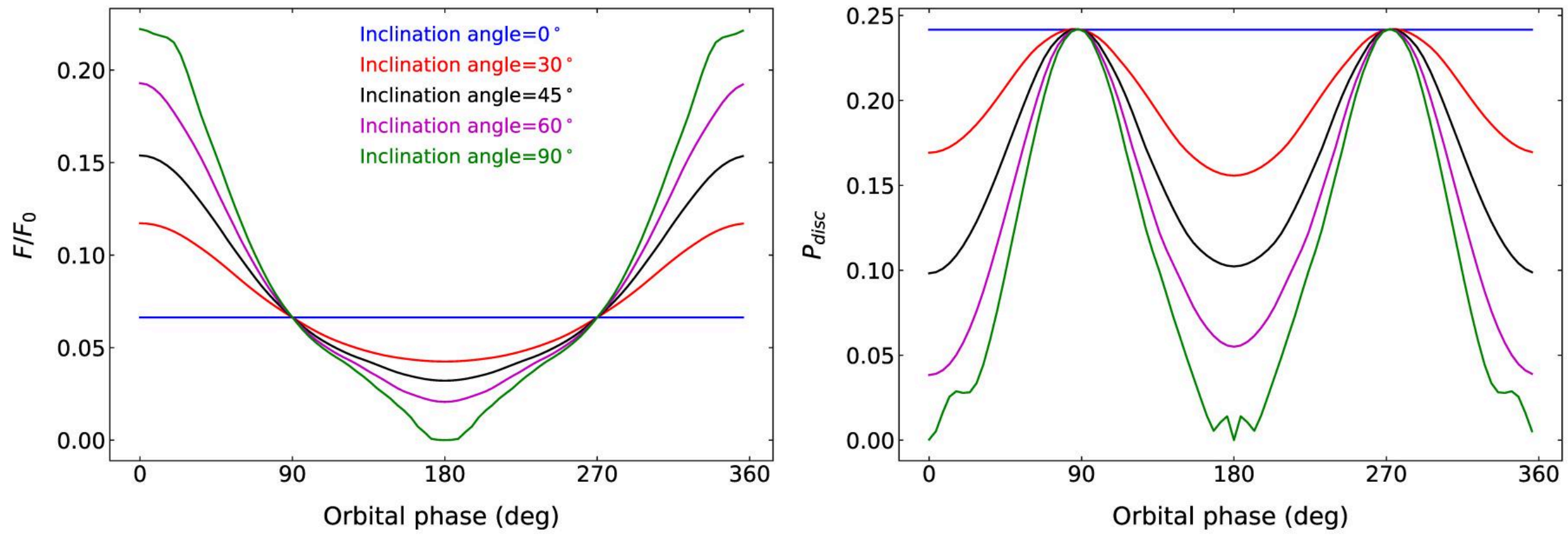

***Figure 10-9:*** *The albedo (left) and total expected polarization (right) at 0.5 µm as a function of orbital phase (Chakrabarty & Sengupta 2021) for different orbital inclinations for a cloudy planet with multiple scattering from cloud particulates expected for Hot Jupiters (primarily forsterite, Mg2SiO4). In all cases, the planet reaches 25% polarization. At the wavelengths of MODHIS the signal will likely be smaller than at 0.5 µm, but still detectable.*

### 10.1.5 Exoplanet imaging and spectroscopy with TMT ExAO: Giant Planets, Rocky Planets, and Planetesimals

Dedicated exoplanet imaging instrumentation harnessing the power of extreme adaptive optics and advanced coronagraphy will enhance TMT's capabilities well beyond those available at first light. Proposed/planned TMT exoplanet imaging instruments, the Planetary Systems Imager (PSI) and MICHI, cover a wide wavelength range, from the red optical to greater than 10 microns, to image and characterize exoplanets in reflected light and thermal emission (Fitzgerald et al. 2019; Packham et al. 2018).

The PSI design builds upon the architecture honed by upcoming state-of-the-art 8–10 m class extreme AO systems like the upgraded SCExAO, GPI-2.0, and SPHERE+ and thermal IR-focused IFS instruments like SCALES on Keck (Skemer et al. 2022). It consists of a first-stage, high-order common path "woofer" correction with wavefront sensing performed in the near-IR. Sharpened starlight redward of 2 microns is then sent to the PSI-Red and PSI-10 channels for 2–5 micron and 10 micron high-contrast imaging. Shorter-wavelength light (0.6–1.8 microns) is sent to the PSI-Blue channel, where a second-order visible-light AO loop provides an additional correction for deeper contrast. Starlight that is sharpened and further suppressed by coronagraphs in all three channels is then fed to imagers and integral field spectrographs yielding imaging detections and planet spectra. PSI-Blue and PSI-Red can feed light into MODHIS through fibers.

MICHI also builds upon recently-honed high-contrast imaging detector technology advances. Its 10 micron imaging detection capabilities are enhanced by using a Geosnap array to substantially reduce or even eliminate excess low-frequency noise that limits sensitivity (Leisenring et al. 2023). Its design includes vector vortex coronagraphs, which have demonstrated success in starlight suppression in the 2–5 micron thermal IR window and at 10 microns (Serabyn et al. 2017; Wagner et al. 2021).

Detecting rocky, habitable zone exoplanets around the nearest stars and identifying biomarkers in their atmospheres is likely the highest-impact objective for TMT's direct imaging science case. The performance needed to achieve this goal, e.g. $10^{-8}$ contrast at several times the diffraction limit for reflected light detections of Earths around low-mass stars, requires honing of new technologies and extensive instrument calibration after commissioning.

Thus, extreme AO direct imaging with TMT will follow a gated approach to development and science milestones (Sallum et al. 2022) that maps onto different unique scientific breakthroughs that it can provide: detection/characterization of temperate gas and ice giant planets in thermal emission, reflected-light detections of Neptune-to-Jupiter sized planets, and finally detecting habitable-zone rocky planets in reflected light and thermal emission. PSI will also provide key direct insights into the formation and evolution of rocky planets.

***Detecting and Characterizing Neptune to Jupiter-Sized Giant Planets***

At commissioning, the red channel of PSI provides an enormous opportunity to vastly increase our sample of imaged exoplanets, characterize their masses, orbits, and atmospheres in detail. Extreme AO systems on 8–10 m class telescopes over the next 10 years should image a few dozen superjovian (3–20 Mj) planets responsible for astrometric accelerations seen with *Gaia*. However, these systems will typically lack the angular resolution to find companions interior to ~5–10 AU, where the jovian planet frequency peaks. PSI-Red will greatly expand this discovery space and improve our understanding of planet atmospheric evolution by detecting planets with even lower masses at smaller, 1–3 AU scale orbits around stars within 50 pc and around older stars where planets are cold and undetectable with 8–10 m telescopes in the near IR (Crossfield 2013, Quanz et al. 2015).

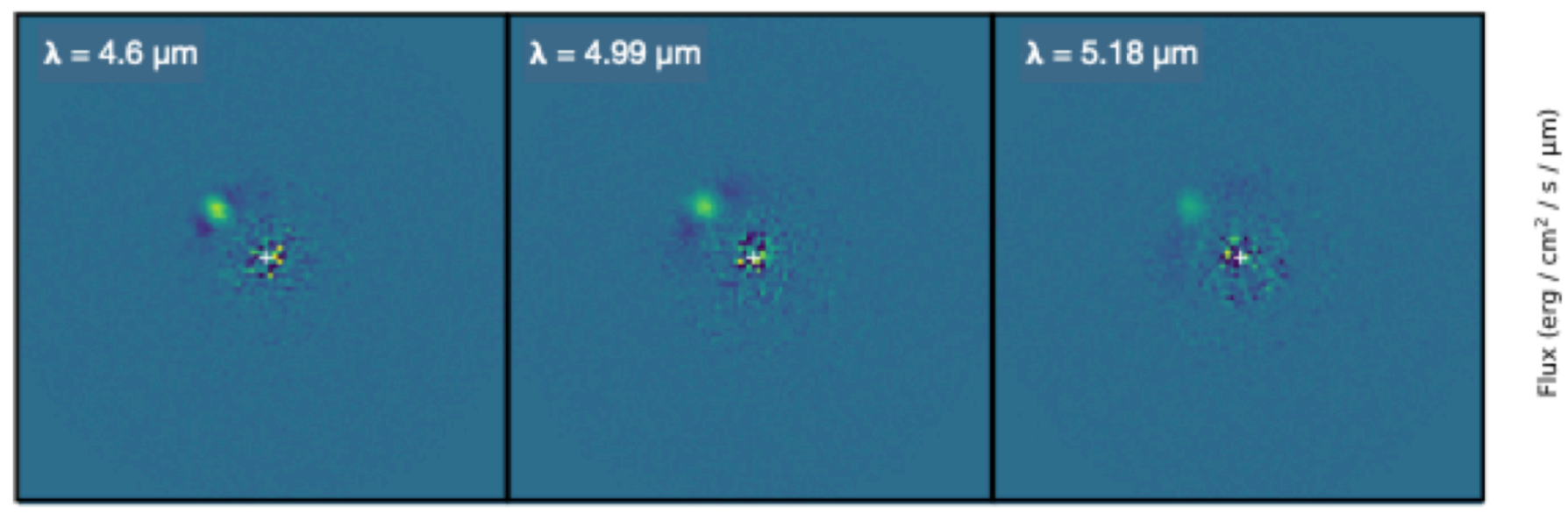

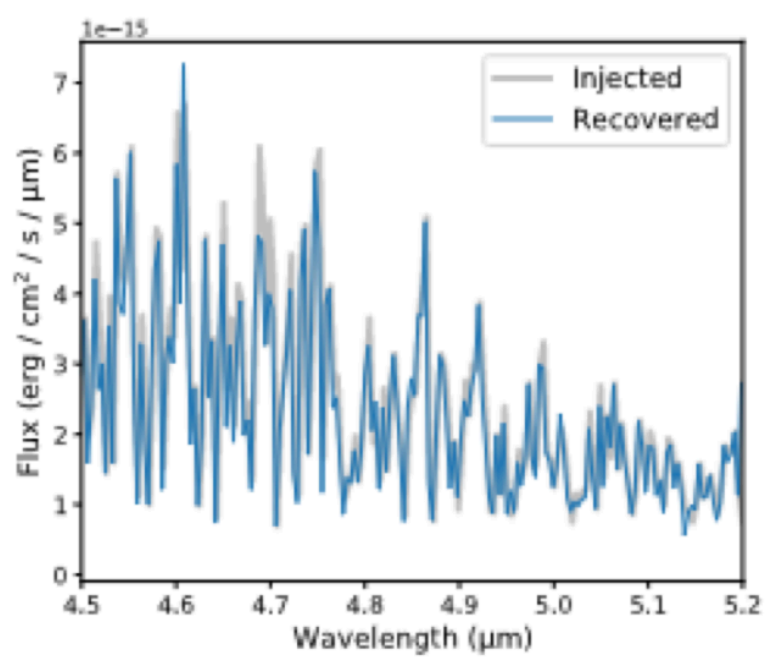

***Figure 10-10:*** *Exoplanet characterization with PSI-Red integral field spectroscopy (from Sallum et al. 2022). (left three panels) Simulated IFS wavelength slices from the PSI-Red IFS for a temperature sub-Neptune planet orbiting 1 AU from a Sun-like star at 10 pc (20 hour cumulative exposure; R~140 spectral resolution). (right) Simulation showing of PSI-Red's ability to yield high-fidelity spectra in M band for this planet. PSI-Red will achieve recon-level detections in far shorter exposures.*

Once calibrated, PSI-Red will be sensitive to detecting temperate planets with masses comparable to or even below that of the ice giants in our solar system (figure 10-10). Thermal IR spectra with PSI-Red will constrain the abundances of key molecules such as methane and water to within ~1 dex precision (Sallum et al. 2022). At medium spectral resolution, PSI-Red will be able to constrain molecular abundances in the atmospheres of more massive planets to an even greater precision (~0.1 dex).

In the blue channel, PSI will provide reflected-light detections of a large sample of mature gas giant planets previously only seen with precision RV. Astrometry for these planets, when combined with RV, will constrain orbits and yield dynamical masses. Low-resolution reflected-light spectra for these planets will provide key empirical comparisons with gas giant planets in our own solar system, illuminating how the atmospheres of these planets are sensitive to insolation and mass. For the nearest of these planets, adding a thermal emission detection at 3–10 microns with PSI and/or MICHI will help constrain planet radii, phase angles, and equilibrium temperatures (Wang et al. 2019).

***Detecting Habitable-Zone Rocky Planets***

The habitable zone defines orbital separations where a terrestrial planet could maintain liquid surface water. However, these planets are located at small angular separations (e.g., tens of milliarcseconds) inaccessible by 8–10 m ground-based telescopes. The last development step with PSI will be achieving contrasts necessary to detect Earth-sized planets in the habitable zones of the nearest stars. Guyon et al. (2018) model sources of wavefront errors whose presence sets the contrast floor for PSI-Blue and a roadmap for mitigating them. If mitigation is successful, raw contrasts of $10^{-6}$ near the optical axis should be achievable. Typical post-processing gains with ~one-to-two hour extreme AO observations range between factors of 10 and 100 (e.g. Bailey et al. 2016): these gains scaled to TMT suggest a final contrast floor approaching $10^{-8}$ .

Figure 10-11 (left panel) shows the discovery potential for habitable zone rocky planets with PSI-Blue with different levels of success in mitigating wavefront errors. Even in the pessimistic case where residual errors limit contrasts to ~$10^{-7}$ at 3–5 lambda/D, PSI-Blue may be capable of detecting rocky habitable-zone planets around a few nearby M stars if the planets are somewhat larger than the Earth (e.g. 1.5–2 Earth radii). Simulations of PSI's performance in the case of effective mitigation of wavefront errors and post-processing speckle suppression (baseline, optimistic cases shown) suggest that PSI could be capable of detecting Earth-sized habitable-zone planets around nearly one to two dozen nearby low-mass (M) stars. Mini-Neptune planets in the habitable zone roughly 2–3 times the size of the Earth may be detectable around hotter (K) stars.

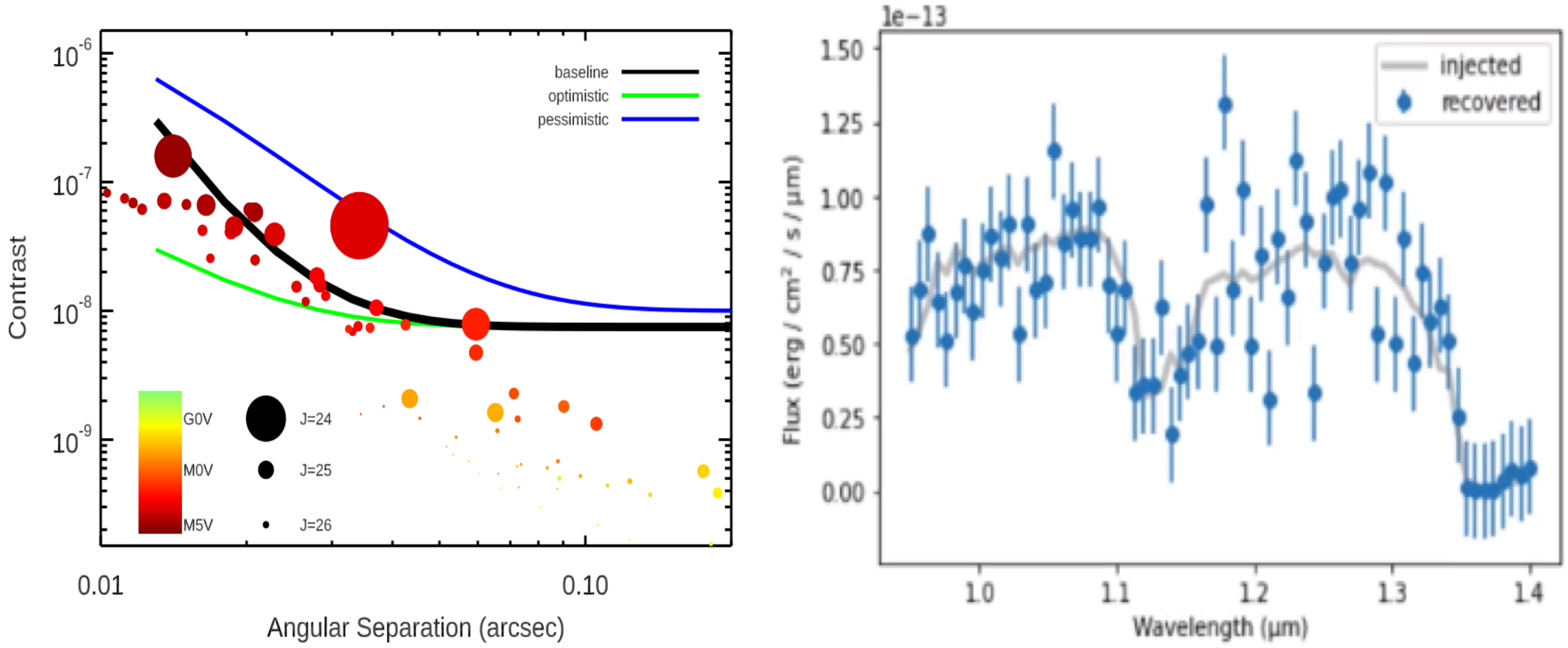

***Figure 10-11:*** *Rocky exoplanet detection and characterization potential with PSI-Blue. (left) Contrasts for an Earth-sized planet with an Earth-like insolation around nearby (<~7 pc) stars targetable from TMT's primary site compared to predicted contrast curves. Apparent exoplanet magnitude is indicated by circle size and host star type is indicated by color. The contrast curves draw from PSI team simulations of a sixth-magnitude star assuming a simple AO model within 5 diffraction beamwidths (blue curve; `simple integrator) or a more advanced linear predictor (black and green curves). The simulated contrast is then adjusted by a factor of 100 at wider separations (representing post-processing) but degraded in contrast near the optical axis due to self-subtraction and imperfect wavefront error mitigation. (Updated from Fig. 4 of Rodler 2018). (right) Injection and recovery of the spectrum of a simulated Earth-sized planet at 2.5 pc over 20 hours of integration time at a contrast of ~$10^{-7}$ (from Fitzgerald et al. 2022).*

If it imaged the Earth itself from the vantage of a very nearby star, it would be sensitive to diatomic oxygen produced by photosynthesis due to plant life (Owen 1980; Angel et al. 1986; Léger et al. 1993, 2011; Selsis et al. 2002; Kaltenegger et al. 2010). Equipped with an IFS, PSI-Blue will be able to yield planet spectra, providing our first direct identifications of biomarkers in the atmospheres of habitable zone rocky planets. In the PSI-Blue IFS wavelength coverage, $O_2$ has absorption features at 0.68 microns, 0.72 microns, and 1.27 microns (Lopez-Morales et al. 2019; Kawahara et al. 2012). The latter feature may be a particularly strong focus for PSI-Blue given the higher Strehl ratio imaging available in the near IR than in the red optical. Water has multiple prominent absorption features that can be covered by PSI-Blue: 0.94, 1.13, and 1.4 microns.

While detections of rocky, habitable zone planets in reflected light may be possible with a few hours of integration time, spectra with a fidelity needed to decisively identify biomarkers may require multi-night campaigns for individual targets with PSI-Blue (figure 10-10). Raw AO performance within several diffraction beamwidths is critical to achieving the goal of detecting a habitable rocky planet with PSI-Blue. For a contrast floor of $10^{-8}$, Kawahara et al. (2012) estimate that biomarkers could be detectable in less than 2 hours of observing. For poorer performance (e.g. driven by degraded AO quality due to faint primary star signals), 20–30 hours may be required to yield detections that can constrain biomarkers.

Rocky habitable zone planets may also be detectable at 10 microns with either MICHI or PSI-10 around about a dozen of the nearest stars. These wavelengths probe the peak of the blackbody thermal emission from Earth-temperature planets. For the hypothetical case depicted in the chapter title page – a solar system analogue at an Alpha Centauri-like distance – an Earth twin would be detectable at a high SNR in a modest exposure time (~an hour). A SNR ~ 8 detection is predicted for a 290 K planet around the closest near-solar mass star accessible with TMT (epsilon Eridani) in several hours (3.12 pc; Lopez-Morales et al. 2019). For most other real systems accessible by TMT, exposure times will be tens of hours, significantly longer due to high thermal background and greater system distances.

Biosignatures may also be probed at 10 microns with the MICHI or PSI-10 instrument channel for the nearest systems in large multi-night campaigns. Here, precipitable water vapor is a key factor determining sensitivity and detectability. Ozone has a prominent, wide absorption feature near 10 microns. Methane and carbon dioxide have more subtle features at 8 microns and 14 microns, respectively, which may be challenging to detect in small exoplanet atmospheres due to low transmission through the Earth's atmosphere.

With a large sample of rocky planets with PSI spectra and 10 microns imaging, we can begin to study the diversity and evolution of terrestrial planet atmospheres directly. The Earth's reflectance spectrum is predicted to have changed substantially from the Archean era and Proterozoic era to the present day, tied to the evolution of oxygen in our atmosphere. Analyzing spectra for many rocky worlds will allow us to compare their atmospheres to that in various stages of Earth's history.

***Witnessing Rocky Planet Assembly*** – In addition to detecting fully-formed rocky planets, the TMT will provide crucial information about their formation history. During the terrestrial planet formation process, a typical rocky world like the Earth will likely experience on the order of 10 giant impact events with Mars-sized bodies. These events impart an immense amount of energy, melting and vaporizing the planet surface. As the energy from impact-heating fades over $10^5$ years, the planet is significantly brighter. Such planets will be very close to their parent star, requiring an ELT to spatially resolve them. A dedicated TMT ExAO instrument may be capable of imaging molten terrestrial planets figure 10-12).

For Sun-like stars, giant impact-type events related to terrestrial planet formation are only observed when the host star is between 10–100 Myr of age, consistent with expectations from Solar system formation simulations (Melis et al. 2010). Lupu et al. (2014) estimate that a survey of 15–40 stars in this age range would yield one post-impact planet. Such young stars are numerous in the Solar neighborhood, but the expected orbital semi-major axis of these terrestrial worlds combined with the inner working angle of TMT's AO-fed instruments sets a tight distance constraint. For a nominal inner working angle of 50 mas for PSI to achieve $10^{-8}$ contrast, we would need stars within 50 pc to probe separations of 2.5 AU or closer (corresponding to separations within the snow-line where terrestrial planets are thought to form). Stars in the β Pictoris (10–20 Myr), AB Doradus (~100 Myr old), and Argus (~40 Myr old) moving groups are at the right distance. *Gaia* will also greatly expand the sample of such stars.

***Mapping Planetary Debris Systems*** – Planetesimal belts are more than just signposts for larger planets. They are crucial components of a complete planetary system, capable of influencing the final configuration of the matured planets. Dynamical friction between these small rocky bodies and planets can drive orbital migration of the planets and circularize planet orbits. In our solar system, such processes are envisioned to have occurred through interactions with the Kuiper Belt that drove the giant planets into new orbits (the so-called Nice model). Thus, to have a complete understanding of any planetary system and its current configuration and develop robust formation and evolution models, it is necessary to characterize any belts of planetesimals that may also be orbiting the same host star.

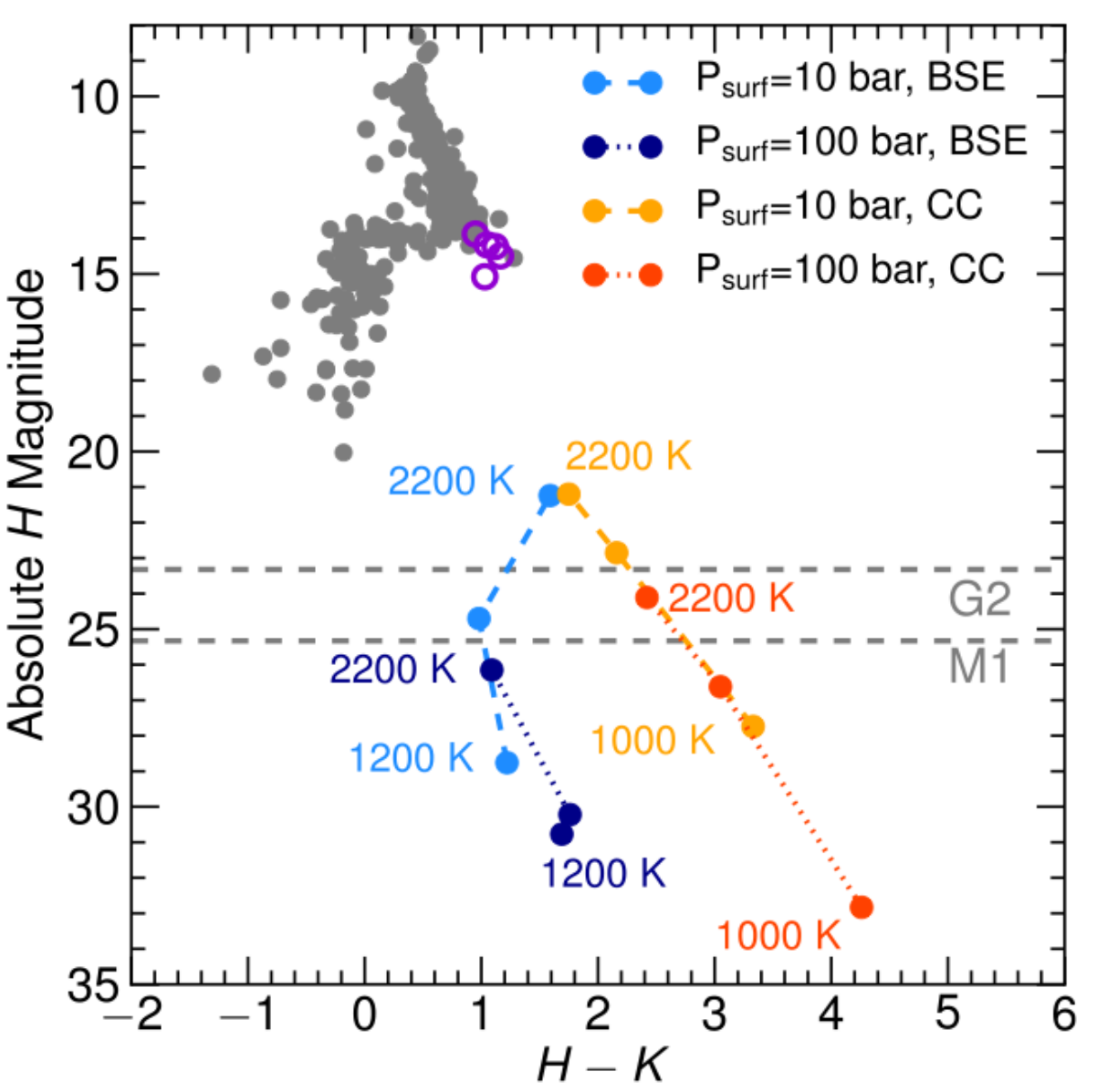

***Figure 10-12:*** *Detecting post giant-impact emission from forming terrestrial planets, from Lupu et al. (2014). The figure shows the color-magnitude diagram of worlds that recently experienced an energetic impact with a planetary embryo as a function of composition (BSE = bulk silicate Earth, CC = continental crust), atmospheric pressure, and surface temperature. For comparison, observed brown dwarfs are shown as gray circles and directly imaged planets (4 HR 8799 planets and 2M1207b) are shown as purple open circles. The dashed horizontal lines show detection limits for a TMT ExAO instrument assuming a contrast of $10^{-8}$ and a Sun-like (G2) or low-mass (M1) star.*

TMT near-IR ExAO instruments and the mid-IR instrument MICHI will be capable of detecting and characterizing planetesimal belts through the thermal emission from small dust grains that are generated during collision events. This probe of the innermost regions of nearby debris disks will complement the work done by ALMA and JWST at wider separations. Near-IR ExAO imaging will detect light scattered and polarized by small dust grains, while MICHI will resolve thermal emission from the debris (see section 8.6.1). According to the models of Roberge et al. (2012), debris-producing planetesimals located at 1 AU from their host stars could be identified out to 10 pc, while asteroid belt-like separations could be detected around stars out to 30 pc. If these planetesimal belts are in a steady-state collisional cascade, then around Sun-like stars we could detect collections of small rocky bodies with a total mass of roughly ten times the mass of our own asteroid belt and roughly a tenth of the mass of our Kuiper Belt.

***Polarization Signatures of Molecules, Earthshine and Biosignatures in directly imaged Exoplanets*** – Atmospheric molecules such as water, surface features such as liquid oceans, and even organisms such as grass and trees, polarize light via Rayleigh scattering. Polarization from atmospheric molecules (as opposed to condensates) has been explored extensively both for gas giants and for smaller planets (e.g., Madhusudhan & Burrows 2012, Karalidi et al. 2013, Chakrabarty & Sengupta 2021, Rossi et al. 2022). Spectropolarimetry measurements of the earthshine reflection off of the Moon (Miles-Paez et al. 2014) show that species such $O_2$ might be more easily found via spectropolarimetry than with traditional absorption line spectroscopy. PSI will have the capability to carry out direct spectropolarimetry of Earth-like worlds to look for polarized signals in reflected light, though such observations will be very challenging, with only a small number of possible targets.

## 10.2 TRANSITING EXOPLANETS

A planet whose orbit intersects our line of sight to its host star repeatedly transits the stellar disk, producing a small and (in general) periodic attenuation of the light. Transiting planetary systems can usually be better characterized than systems detected by radial velocity, astrometry, or microlensing because the radius of the planet and the inclination of the orbit are established. Observations during transit can probe a planet's atmospheric composition, orbital obliquity, and enable one of TMT's most compelling science cases: the potential for detection of $O_2$ or other biosignature gasses in rocky exoplanets via transmission spectroscopy. In addition to using direct spectropolarimetry to study the atmospheres and surface features of directly imaged exoplanets, MODHIS can be used to search for polarimetric signatures from transiting Earth-like exoplanets.

### 10.2.1 Exoplanet transits: landscape in 2030s

With its unprecedented aperture and access to three-quarters of the sky (and as the only ELT in the northern hemisphere), TMT will be well poised to carry out observations of transiting systems detected by other observatories such as *TESS*, *Plato*, and *Roman*. TMT's observations of transiting planetary systems will involve spectroscopy to obtain more information about the properties of the planet or its host star, as well as high-spatial resolution (AO) imaging to identify stellar companions to stars or unassociated background objects. The primary transit science cases for TMT are: 1) atmospheric studies of planets from hot Jupiters to habitable-zone rocky worlds via high-resolution optical/infrared spectroscopy; 2) measurements of system architectures via orbital obliquity RV measurements; and 3) characterization and validation of faint host stars of transiting (candidate) planets using high-resolution spectroscopy and high angular-resolution imaging.

The NASA *Kepler* mission (2009–2013) dramatically expanded the scope of transit studies to Earth-size (and smaller) planets and to populations of thousands of planets on which robust statistical methods can be applied. *Kepler* surveyed ~1/400th of the sky for over 3 years; its deep, narrow survey has since been complemented by the nearly all-sky wide, shallow *TESS* survey (in roughly 27-day fields) and will be further complemented by the intermediate-depth and -duration *Plato* transit survey mission. The *Roman* Galactic Bulge Survey will also discover thousands of transiting planets that are fainter and more distant than the bulk of even the *Kepler* sample (figure 10-13; Wilson et al. 2023). The host stars of many of these transiting systems are relatively distant (>1 kpc) and faint (V>14), making follow-up observations such as Doppler radial velocity measurements or transit spectroscopy difficult or impossible even with 8–10 m class telescopes. For example, *Kepler* detected 23 candidate Earth-size planets in the HZ; because these worlds constrain eta-Earth, they are arguably some of the most interesting planets we know of. Yet none of these have been confirmed: the RV signature is too small to measure with current facilities.

Furthermore, many transiting planet candidates at such distances are probably unresolved binaries hosting a planet, while others turn out to be eclipsing binaries and some are indeed simple one-star, one-planet systems. Observations must establish *which* star a planet transits to accurately determine its mass and rule out the possibility of a false positive (see section 10.3.4): the K-band diffraction limit of a 10 meter telescope at 1 kpc is 45 AU and spectroscopic detection of single-line systems with separations of tens of AU requires high-SNR, high-resolution spectra that are difficult to obtain for these faint stars.

Transit surveys have discovered many new planets since *Kepler's* launch opened the floodgates: as of Nov 2023, roughly 3700 planets have been confirmed from *Kepler, K2,* and *TESS* with over 7500 planets candidates from these missions still waiting to be confirmed. As of Nov 2023, *TESS* in particular continues to reveal roughly 500 new candidates per year, including Earth- to super-Earth-size planets around the closest, brightest stars (figure 10-12). ESA's PLAnetary Transits and Oscillations of stars (*Plato*) mission will launch in 2026 and is expected to discover hundreds of Earth- and super-Earth-size planets (as well as thousands of larger ones) around 4th–11th magnitude stars. Stars in this magnitude range offer the opportunity for high-cadence, high-spectral resolution observations of transits to characterize the planets' atmospheres, orbital alignments, and to measure physical properties of the stellar disk. Furthermore, the *Roman* Galactic Bulge Survey will reveal tens of thousands of transiting planets around distant, fainter stars whose follow-up observations will require TMT's large aperture (figure 10-12; Wilson et al. 2023). By first light, TMT will therefore have access to thousands of transiting planets. *JWST* is characterizing roughly 50 transiting planets per year; ESA's *Ariel* mission, due to be launched in 2029, will also conduct low spectral resolution transit and eclipse spectroscopy and will identify many targets with potentially interesting atmospheres for more detailed follow-up with TMT.

MODHIS transit observations of close-in exoplanets will allow us to explore the compositional diversity of planets and their atmospheres by determining the bulk elemental compositions of giant planet envelopes. Transit spectroscopy explores the nature of clouds and hazes on these planets, and can be used to search for atmospheres of sub-Neptunes and terrestrial exoplanets — plausibly even with temperate conditions. MODHIS will also be able to conduct high-resolution spectroscopic observations of close-in exoplanets to measure atmospheric abundances, dynamics, and temperature structure, and can do so for both transiting and non-transiting exoplanets. Excitingly, the large aperture of TMT and the high sensitivity of MODHIS mean that we will be able to obtain high-resolution spectra of exo-planetary atmospheres to directly constrain their temperature structures, measure specific elemental abundances, determine dominant chemical processes, and probe global atmospheric dynamics.

The precision of ground-based transit spectroscopy can be improved with the simultaneous observation of multiple reference stars to reject common-mode temporal variations from telluric or instrumental sources. At visible wavelengths, WFOS (field of view of 25 sq. arc minutes) could be used to observe some of the faintest Kepler transiting systems along with comparison stars and make higher spectral resolution measurements during transit than ESA's Ariel mission.

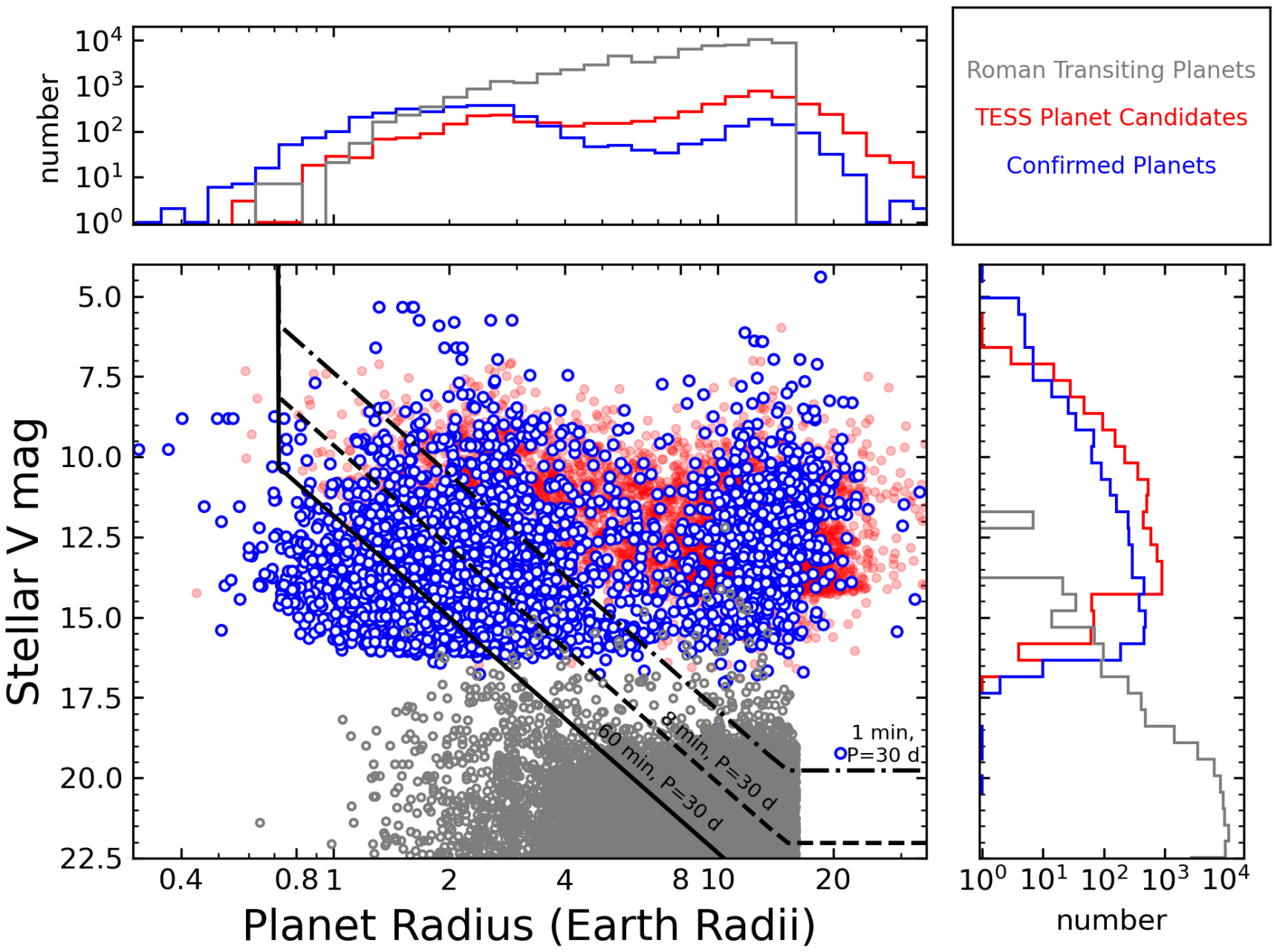

*Figure 10-13: Known planets (blue), unconfirmed TESS planet candidates (as of Nov 2023, red), and transiting planets predicted from the Roman Galactic Bulge Time Domain Survey (Wilson et al. 2023). The black curves show TMT's approximate photon-limited Doppler uncertainties for a 30-day period assuming exposure times of 1, 8, or 60 min.*

### 10.2.2 Orbital Obliquity Measurements

The orbital architecture and atmosphere of a planet are two complementary fossils of its formation and migration history. For example, when planets smoothly migrate through their natal disk the systems may remain flat and dynamically 'cool,' while systems that evolved by planet-planet scattering may become dynamically 'hot' – measuring the spin-orbit alignment (obliquity) constrains these pathways (Winn & Fabrycky, 2015).

To date, most Jovian-mass planets have aligned orbits for $a/R_* < 10$ around cooler stars, with more misaligned orbits around hotter stars. For the smallest, coolest stars most Jovians have aligned orbits but the sparse sampling for smaller planets on wider orbits is largely unexplored. Obliquities are most frequently measured via the Rossiter-McLaughlin effect, which reveals the changes in the system's RV signature during a transit. These obliquity measurements also provide simultaneous constraints on the transiting planet's atmospheric composition (see next section).

Measurements of exoplanet orbital obliquity could be the quickest and most powerful route to high-impact exoplanet RV observations for TMT. Such observations give a high science per hour because the planet targets are already known beforehand and each observation lasts only a few hours. For example, a night's observation with TMT could confirm a candidate transiting Earth-size planet orbiting in the habitable zone of a Sun-like star, since the amplitude of the R-M effect is much larger than the Doppler RV amplitude.

With the excellent instantaneous sensitivity provided by an efficient instrument on a 30m telescope, TMT could measure the obliquities of over 1300 planets known as of Nov 2023, and many more pending planetary candidates from *TESS, Plato*, and *Roman*. In particular, precise TMT RVs could measure obliquities of >100 planets ranging in size from super-Earths to sub-Neptunes orbiting M dwarfs. Only one such system currently has a published obliquity measured with a precision better than 10 degrees (AU Mic b; Hirano et al., 2020). The statistics of the population for young and mature stars will provide important constraints to theories of planet formation and evolution.

### 10.2.3 High-dispersion spectroscopy and biosignature gasses

Transit observations at high spectral resolution can determine the atmospheric composition and structure of both transiting and non-transiting planets (Brogi et al. 2012), measure atmospheric wind speeds and circulation patterns (Showman et al. 2013), and potentially detect molecular oxygen or other biosignature gasses (Web & Wormleaton 2001, Snellen et al. 2013, Rodler & Lopez-Morales 2014, Currie et al. 2023a). By peering between telluric lines, these measurements do not require simultaneous reference stars, and regularly reach photon-limited precision (Brogi et al. 2014), making them an especially exciting growth opportunity for TMT. MODHIS will offer high spectral resolution at first-light while later instruments HROS and MICHI will further enhance this capability, supporting a range of transit observations TMT can make to advance this field. These high-resolution measurements are a unique science mode for TMT, as they are highly complementary to medium- and low-resolution spectroscopy obtained with *JWST* and *Ariel.*

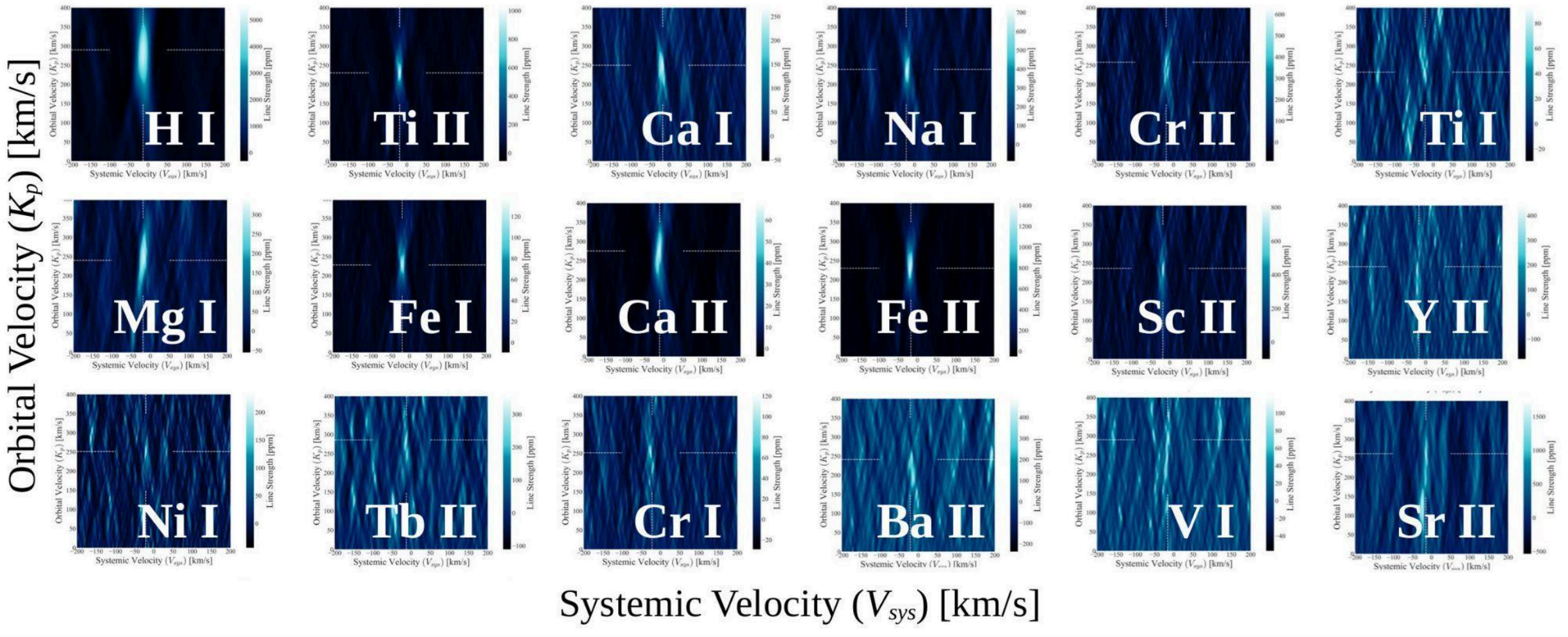

***Figure 10-14:*** *High spectral resolution detection of eighteen atomic and ionic species in the ultra-hot Jupiter KELT-9b using spectra from HARPS-N and CARMENES on 3.5m telescopes (Borsato et al. 2023). TMT will extend such studies to an even wider array of elements, molecules, and planets to determine atmospheric composition while also measuring atmospheric dynamics and temperature structure.*

In the infrared, spectroscopy at high resolution of planets with $H_2$-dominated atmospheres will extend current detections of atoms, ions, and molecules (figure 10-14) to more exotic species – hydrocarbons such as HCN and $C_2H_2$ which have not yet been observed in exoplanet atmospheres (de Kok et al. 2014), as well as rarer species. For the hottest gas giant planets, wind speeds of up to 1–2 km/s will be directly measured via the mean Doppler shift of the planet's absorption spectrum (Kempton et al. 2012). Molecular detections and measurements of thermal structure will be measured as a function of rotational phase for a large sample of both transiting and non-transiting hot Jupiters and hot Neptunes, complementing *JWST's* and *Ariel's* observations of these planets.

The most exciting possibility TMT provides is the ability to search for biosignature gasses in the atmospheres of nearby planets. The most well-studied spectral feature is from the molecular oxygen band heads at 0.77 and 1.27 micron, which could be detected with a set of dedicated HROS and MODHIS observations of a rocky planet transiting a nearby M dwarf. The power of conducting these observations at high spectral resolution is that the exoplanet's atmosphere would be distinguished from the Earth's via the Doppler shifts resulting from the two planets' relative orbital motions; detection would come not in a single night but would require observations spread over many years (figure 10-14; Rodler & Lopez-Morales 2014, Currie et al. 2023a). Although any detection would be challenging and no single gas is an unambiguous indicator of life nor uniquely associated with it (Tian et al. 2014), the detection (or significant nondetection) of such molecules in a nearby rocky planet's atmosphere would be a landmark result for the study of planetary atmospheres, habitability, and astrobiology.

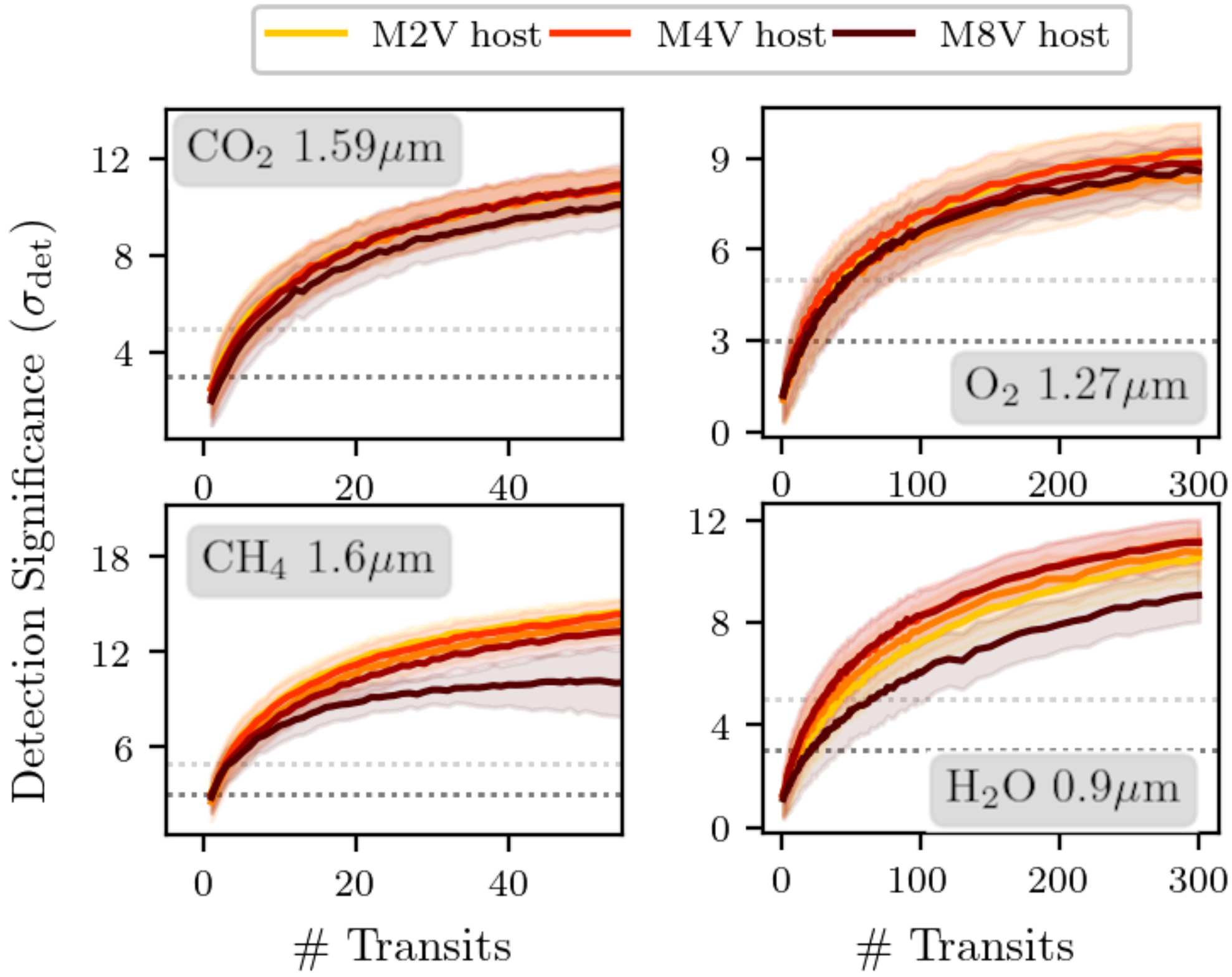

*Figure 10-15: The number of high-dispersion TMT transit observations required to detect $CO_2$, $CH_4$, $O_2$, and $H_2O$ in the atmosphere of an Earth analogue transiting an M dwarf at 5 pc (Currie et al. 2023c). ~$1\sigma_{det}$ is considered a non-detection, typical detection thresholds are 3 or $5\sigma_{det}$. $CO_2$ and $CH_4$ should be detectable in $<10$ transits, while $O_2$ and $H_2O$ can be detected with several dozen transits. Only a TMT-like facility will be sensitive to potential biosignature gasses in the atmospheres of habitable-zone, rocky planets transiting nearby cool dwarfs.*

Bailey et al. (2018) modeled the polarimetric signatures for transiting Jupiter-sized exoplanets due to Rayleigh scattering from small atoms and molecules in the atmosphere and scattering from clouds. Measuring the polarimetric signatures of larger exoplanets should be possible with TMT-NFIRAOS-MODHIS for a number of targets. Gordon et al. (2023) explored the wavelength-dependent polarimetric signal from surface features on Earth-like planets, including ocean, sand, and trees. Oceans strongly polarize light, with up to 70% polarization fraction around ~1–1.2 µm. Ice, forests, sand, and grass are also strongly polarizing, often in similar wavelength regimes, and combinations of surface features can be distinguished by comparing the ratio of various Stokes parameters. With MODHIS having the sensitivity to detect these types of planets in transit with NFIRAOS in unpolarized light, the capability of spectropolarimetry becomes critical for the goal of characterizing Earth-like worlds. MODHIS has a key role to play in the era of characterization of Earth-like planets and the search for life signs on these worlds.

### 10.2.4 Characterization of transit planet hosts

A number of effects can mimic the periodic dip in the light curve of a transiting planet in a star's light curve, thus stars with transit-like signals remain candidate systems until false positives (e.g., grazing transits or larger bodies, or a diluted eclipsing binary) can be ruled out to an acceptable level. Moreover, the presence of unresolved stellar companions can make interpretation of the signal ambiguous (it is necessary to know which star is being transited) as well as requiring correction for the fact that the transit signal is being diluted by the star that is not being transited. Finally, planet occurrence rates and the properties around single vs. multiple stellar systems provide tests of theories of planet formation and evolution (Quintana et al. 2007).

Most *Kepler* systems are too faint and the planets too small to have detectable Doppler radial velocity signals (c.f. figure 10-12). Ten-meter class telescopes can rule out large Doppler signals from stellar companions, but these measurements are observationally expensive and not practical for the hundreds or even thousands of systems of interest, but TMT/MODHIS could rapidly screen these systems with well-timed observations at quadrature points in the orbit of the candidate planet where the maximum change in radial velocity is expected. High-resolution AO imaging using TMT could also allow rapid screening of binary systems, with detections at closer inner working angles than is possible with current smaller telescopes (Law et al. 2014). Stellar companions of any sort will be detectable with TMT AO imaging down to angular separations of ~30 mas, corresponding to 30 AU at 1 kpc — nicely complementing the sensitivity to shorter-period companions provided by *Gaia* and spectroscopic monitoring.

## 10.3 DOPPLER DETECTION OF PLANETARY SYSTEMS

During the first decade and a half of the planet discovery era, RV surveys were the dominant method for discovering new exoplanetary systems, with over 400 systems discovered between 1995 and 2009 (http://exoplanets.eu). Over that period Doppler surveys expanded their reach and progressed steadily from measurement precisions of 3–10 m/s to <1 m/s as a consequence of improvements in instrument stability and data analysis methods. With echelle spectrometers now approaching the 0.1 m/s radial velocity (RV) measurement precision threshold needed to detect habitable Earth mass planets around Sun-like stars (Pepe et al. 2011), the continuation of Doppler surveys on state-of-the-art telescopes is imperative. Such surveys, conducted with the optical and infrared echelle spectrometers being built for the TMT, have the potential to complete the census of Earth-mass planets in our stellar neighborhood. The collection of such systems will provide a valuable sample for follow-up programs to characterize their atmospheres through direct imaging and spectroscopy.

The radial velocity method originally focused on finding exoplanets around main sequence FGK stars. These targets are ideal, both due to their similarity to our Sun, and due to the abundance of sharp atomic lines in their spectra. The gravitational velocity perturbation exhibited by the host star is directly related to the mass of the planet and is inversely related to its orbital distance. Hence, the RV detection method is most sensitive to massive, close-in planets. This ease of detection led to a substantial number of early discoveries of what we now call hot Jupiters, but as observing temporal baselines grew longer and as measurement sensitivities improved, the Doppler searches gradually expanded into a larger exoplanet phase space. As more planets were discovered, key physical trends began to emerge. Notable examples include a correlation between planet occurrence and stellar metallicity (Fischer & Valenti 2005; Johnson et al. 2010), a peak in the distribution of planets with orbits of ~3 days (Cumming et al. 2008), a gradual increase in the number of planets with increasing period and decreasing mass (Cumming et al. 2008, Howard et al. 2010) and a peak in the semimajor axis distribution at ~3 AU (Fernandes et al. 2019, Fulton et al. 2021). RV planet detection programs are naturally limited to planets with periods that are shorter than the length of the survey, which currently amounts to somewhat more than 10 AU for the longest-running projects (Jupiter, for example has a 11.86 year orbital period and induces a 12 m/s radial velocity half amplitude).

In the *TESS* and *Kepler* eras, characterization of new transiting planets has also become a major focus of RV observations; the joint detection of a planet via radial velocimetry and transit photometry is exceptionally powerful, allowing measurements of planetary densities.

With multiple ongoing RV surveys focused on nearby stars, and improvements in both the hardware and RV data analysis techniques, observations from several facilities now regularly reach on-sky precision of <50 cm/s on the quietest stars (Zhao et al. 2021, Trifonov et al. 2021, Pepe et al. 2021). Doppler precision measured in mere centimeters per second is accomplished with instruments that have very high spectral resolution (R>100,000) and fiber-scrambling capabilities to aid with spectral stabilization. Using such techniques, modern instruments have reached the point where stellar noise is the limiting factor on RV measurements of all but the most quiescent of stellar targets. Achieving this precision also requires very high signal-to-noise and hence either a bright star with long exposures or a telescope of TMT's aperture. What TMT really offers is the ability to extend the current state-of-the-art to much fainter stars: intrinsically faint very-low-mass stars, as well as more typical FGK targets at greater distances.

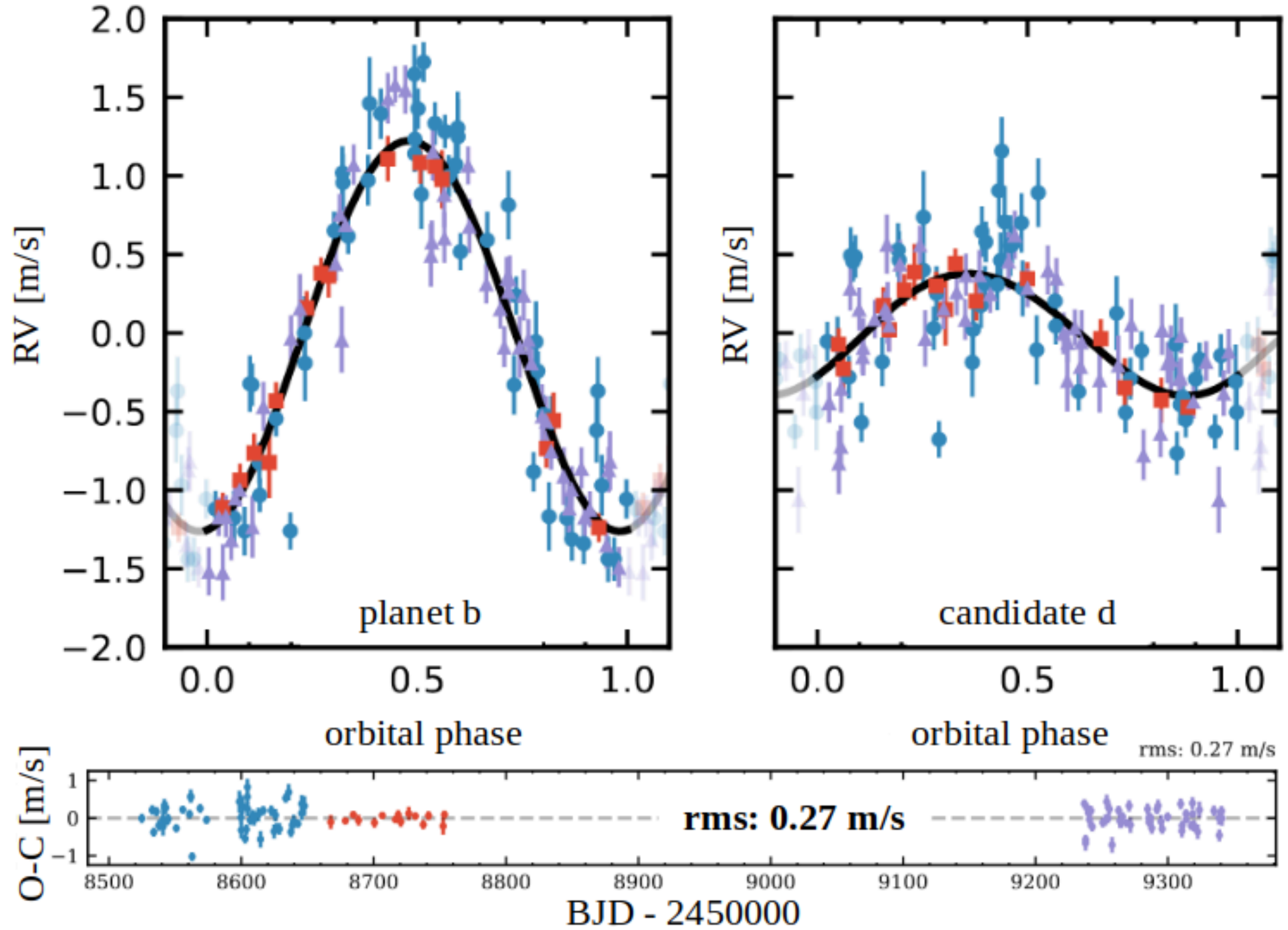

***Figure 10-16:*** *ESPRESSO measurements of radial velocities for Proxima Centauri, showing the separate phased components for the habitable-zone planet b (left) and planet candidate d (right). The residuals to the fit (bottom) have an RMS of just 27 cm/s (Faria et al. 2022). TMT observations will meet or exceed this precision for larger numbers of more distant, fainter exoplanetary systems.*

RV techniques exhibit a powerful synergy with transit observations. Only for planets with both transit radius measurements and RV mass measurements can we determine density, and by extension, clues to the physical composition of the planets. With *Kepler*'s and *TESS'*s demonstration that small (1–4 $R_E$) planets are common, the next step in understanding the frequency and nature of Earth-size planets was the determination (via RV masses) that small planets exist in at least three classes: rocky super-Earths, $H_2O$-rich water worlds, and $H_2$-rich mini-Neptunes (Luque & Palle 2022). The time needed for such RV observations are a significant bottleneck in completing the scientific legacy of large-scale transit missions. The *PLATO* and *Roman* surveys will (like *TESS* and *Kepler* before them) have a similar need for RV follow-up, requiring high precision and often around fainter stars (see figure 10–13; Rauer et al. 2014, Wilson et al. 2023). A high-precision RV capability on TMT (particularly one optimized for late-type stars) would be a powerful component of an integrated exoplanet roadmap, and would forge a near-complete understanding of nearby planetary systems.

MODHIS offers extreme sensitivity for precise radial velocity measurements for cool stars and close binaries requiring high angular resolution. MODHIS will be able to detect and characterize rocky Earths and super Earths orbiting in the Habitable Zones (HZ) of the over 50 dwarfs later than M5 that lie within 25 pc. Many planetary systems transiting late-type stars have been discovered by *TESS, Roman,* and *Plato*; the IR-sensitive MODHIS will be optimized to determine RV masses for these new small, temperate, planets orbiting the coolest stars. Furthermore, MODHIS's diffraction-limited angular resolution will also enable the study of transiting planets in close binaries (<0.5″) (Feinstein et al. 2019). Knowing which star of a pair hosts the planet makes a dramatic difference to the derived planet radius and density (Ciardi et al., 2015). Demographic studies will also inform how planet formation proceeds in binary systems.

Beyond the study of planets, TMT will have the precision to detect and characterize satellites ("exomoons") around nearby directly imaged planets. Such satellites are an important missing piece of our understanding of exoplanet formation as they can strongly influence the atmospheres of giant exoplanets and icy moons around gas giants could develop life (Reynolds et al. 1983), but no such bodies are currently known beyond the solar system. Scaling the Galilean moons to the mass of HR 8799 c, their reflex motion would be detectable with TMT/MODHIS in only a few nights of RV observations, and given favorable orbital alignments confirmation would be swift via the Rossiter-McLaughlin effect (Ruffio et al. 2023).

### 10.3.1 TMT's role in Doppler studies of exoplanets

The TMT's primary advantage for both optical and infrared radial velocity studies will be an increased sensitivity due to its larger collecting area as well as the suite of state-of-the-art echelle spectrometers that will be designed with RV exoplanet detection in mind. With current optical RV surveys limited to stars with V~14 or brighter, the larger collection area of the TMT will allow us to complete an RV survey of much fainter targets than is currently accessible. In particular, TMT will offer a powerful synergy with transiting planets found in the *Roman* Galactic Bulge survey (Wilson et al. 2023): these systems will be so faint as to be almost wholly inaccessible to current RV facilities (figure 10-12). The advantage of TMT's large aperture will be maximized with an echelle tailored for precise RV measurements. MODHIS is an infrared, first light instrument with precise RV capabilities, and the High Resolution Optical Spectrometer (HROS) is an optical-wavelength PRV instrument contemplated for the first decade of the TMT science program. Both these instruments will have spectral resolutions ~100,000 to the current echelle spectrometers responsible for hundreds of RV discovered planets, and will be designed to reach instrumental precision comparable or superior to current world-class instruments such as VLT/ESPRESSO (Pepe et al. 2021).

In the infrared, MODHIS will significantly increase the number of M dwarfs that can be surveyed for terrestrial, habitable-zone planets. The benefits of this investment would be substantial, and could significantly extend the surveys being completed with infrared PRV spectrographs such as SPIRou and NIRPS (Donati et al. 2020). Together with instrumental advances to push infrared RV precision down from the 1-2 m/s level achieved by SPIRou and NIRPS to below 1 m/s would allow a complete survey down to sub-Earth-mass planets across the entire range of the habitable zones for the more massive M3–M4 dwarfs that are the most common star in the Solar neighborhood. TMT's larger aperture will extend our ability to detect planets around the latest M dwarf spectral types that are too faint to be efficiently surveyed with the RV precision necessary to detect habitable-zone, Earth-mass planets. By combining the results from both optical and infrared surveys, the TMT will enable the definitive census of potentially habitable planets around our nearest neighbors.

RV surveys of young stars have yet to produce a sufficient sample of planetary systems within which we can search for changes in the architecture of planetary systems with age. Such trends would provide insight into planet formation and evolution models. At distances of a hundred parsecs and more, only the most massive stars in nearby star formation regions like Taurus, the Hyades, Orion and Sco Cen are bright enough for RV planet search programs (e.g., Paulson et al. 2004; Quinn et al. 2014). TMT will be able to push detection limits to solar-mass stars and below as well as allow for thorough surveys of hundreds of stars in much less time than what is currently possible with 8–10 m telescopes. Direct imaging techniques (section 10.2) will also favor young stars; the combination of Doppler monitoring and direct characterization could completely sample the giant planets around nearby young stars; extending discoveries such as the RV-discovered beta Pic c (20 Myr old; Lagrange et al. 2019) to many more populous, but more distant, star forming regions. Section 10.1.3 discusses how we might overcome the photospheric noise inherent to young stars.

Depending on its final levels of stellar and instrumental noise sources, TMT's RV capabilities may augment, rather than replace, large-scale programs on 8–10 m telescopes for blind RV searches for earth-mass planets, mass determination of transiting planets, and RV discoveries to known direct-imaging and/or transiting planet hosts. TMT may be most important for follow-up of fainter (more distant and/or ultra-cool) stars hosting highly unusual or interesting planet candidates (e.g., *Plato* and *Roman* discoveries or low-mass stars imaged with direct techniques), high cadence observations (e.g., Rossiter-McLaughlin effect; see section 10.3), and of course for the mass and atmospheric characterization of the transiting planets themselves (section 10.3).

### 10.3.2 Radial Velocities: landscape in the 2030s

In addition to completing the survey of nearby habitable-zone rocky planets, the TMT will provide follow-up observations for numerous space-based NASA missions that have already launched or are expected to launch prior to TMT first light. Both MODHIS and HROS will measure the masses of small transiting planets discovered by *TESS* (Ricker et al. 2014) and *Plato* (Rauer et al. 2014) in specific high-value cases where it is imperative to obtain spectra with the highest possible resolution and S/N. TMT will also be able to collect supporting observations of the stars hosting planets detected with J*WST, Gaia* and the *Roman* Space Telescope. By the time TMT will be on-sky, some of these planets will have been known for many years, however, the S/N needed to reach the RV precisions for planet confirmation will not be readily achieved without the TMT. TMT will deliver sufficient precision to determine whether the planetary mass-radius diagram for large terrestrial-mass planets differs for M dwarfs as compared to more massive solar type stars. This will directly probe the differences in planet formation, evolution, and migration as a function of stellar mass. Given the expected plethora of newly discovered and potentially complex planetary systems from these missions, the shorter total on-sky time needed to collect a statistically robust sample of RV measurements at RV precisions of < 1 m/s will be key to the efficient characterization of exoplanetary systems.

The early evolution of young, still-contracting planets is poorly understood (Fortney et al. 2008; Spiegel and Burrows 2013). Many exoplanets have been found transiting young host stars (David et al. 2018; Mann et al. (2017; Ciardi et al. 2018). Taking advantage of the demonstrated 2–4× reduction in the effects of stellar activity relative to visible wavelengths for these young stars (e.g., Johns-Krull et al. 2016; Carleo et al. 2018; Klein et al. 2021), MODHIS will be able to determine the masses of young transiting planets and address the initial density of exoplanets, an area of intense observational and theoretical interest which will require extensive observations for its resolution. The accurate determination of the masses and radii of young planets will be critical to understanding the mechanism(s) responsible for the Fulton Gap in the Planet Radius-Orbital Period plane (David et al. 2021; Pascucci et al. 2019). We can also learn about the occurrence rate of habitable zone Earth-mass planets that future space missions will specifically be targeting by understanding whether or not the population of dense super-Earths found in transit surveys initially formed as rocky bodies or started out as larger, gas rich bodies but lost their $H_2$ rich atmospheres to end up as rocky cores (Pascucci et al 2019).

### 10.3.3 Exomoons

As part of the formation process of exoplanets, exomoons are an important missing piece to better understand planetary systems and can strongly influence the atmospheres of giant exoplanets (e.g. interaction between Io and Jupiter). Icy moons around gas giants are also interesting targets as they could develop life (Reynolds et al. 1983). Several exomoon candidates have been proposed as of late 2023 (e.g. Kipping et al. 2022, Lazzoni et al. 2020), but none have been confirmed. Exomoon mass upper limits rule out Jupiter-mass moons orbiting the 7 Jupiter masses planet HR 8799 c in periods shorter than 1 day (Vanderburg & Rodriguez 2021).

As mentioned in section 8.7, the moon mass is suggested to roughly scale with the planet mass to the 3/2 power (Batygin & Morbidelli 2020), meaning larger planets will harbor larger moons. By scaling the Galilean moons to the mass of HR 8799 c, TMT/MODHIS (with an RV precision of 5 m/s in 10-min exposures at R=100,000) will be able to detect moons around HR 8799 c in only a few nights of observations. Other moon formation mechanisms, such as collision or capture, are likely to generate more easily detectable larger objects.

NIR spectropolarimetry will allow exomoon detection through variations or asymmetry in the polarimetry signal during the transit of an exomoon across a rotating giant cloudy planet, an effect somewhat analogous to the well-known Rossiter-McLaughlin effect (Sengupta & Marley 2016). The transiting body blocks part of the light of its host, yielding a change in the integrated signal from that host.

### 10.3.4 Limits to Doppler studies due to stellar activity

Even when spectrographs are able to regularly reach instrument precision of ~0.1 m/s, it is not yet clear whether we will be able to reach this precision for all nearby stars due to their intrinsic photospheric RV noise from starspots, granulation, and p-modes. Such stellar jitter has led to the misidentification of a few extrasolar planets including the one around the young star TW Hydra (Setiawan et al. 2008) that was quickly shown to be a false-alarm produced by starspot activity (Huélamo et al. 2008). Future RV surveys of young stars are being designed to collect measurements at both optical and infrared wavelengths, where the starspots produce a reduced RV perturbation (e.g., Crockett et al. 2012). Similarly, the combination of RV measurements from the TMTs optical and near-IR echelle spectrometers could be used to discern between low-mass planets and the stellar activity that plagues both M dwarfs and young stars.

The discovery of planets around Proxima Cen with semi-amplitudes of 50-100 cm/s (figure 10-15; Faria et al. 2022) demonstrate that high cadence observations can overcome the limitations of stellar jitter. Since jitter is best measured and removed when instrumental sensitivity is high, TMT should have a clear advantage over current facilities in measuring the smallest-amplitude signals – especially in the presence of stellar RV noise. Furthermore, all-sky, high-cadence *TESS* photometry will have provided the data to estimate photometric variability (flicker, Bastien et al. 2014) and use it to estimate RV variability for most desired targets. With all this in mind, future TMT RV programs can and will be able to detect rocky planets around nearby stars with improved instrumental sensitivity, broad wavelength coverage, careful target vetting, and high cadence observations.

## 10.4 GRAVITATIONAL MICROLENSING

A Galactic gravitational microlensing event originates from the rare occurrence of an extremely close alignment (within ~1 mas) of two stars on the sky. A planet accompanying the foreground lens star reveals its presence by producing a short-duration deviation on the otherwise smooth and symmetric amplification pattern of the companion-less background source star (Mao & Paczynski 1991). The planet detection sensitivity of microlensing peaks at intermediate orbital separations (~1–10 AU; e.g., Gould & Loeb 1992; Bennett & Rhie 1996; Dong et al. 2006), which bridge the close-in planet discoveries by RV/transit detection surveys and the wide-separation regime probed by direct imaging programs. Unlike all other planet detection methods, microlensing is sensitive to the mass of the host star and the planet rather than the light, so it is capable of discovering planets around stars across the stellar mass function.

Thousands of microlensing events are discovered per year by ground-based monitoring of hundreds of millions of giant stars in the dense stellar fields toward the Galactic bulge (see, e.g., Gaudi 2012). The *Roman* Galactic Bulge survey will carry out a dedicated space-based microlensing survey, currently planned as six campaigns of 72 days each (Penny et al. 2019, Johnson et al. 2022). *Roman* will measure planet occurrence via the discovery of ~1400 microlensing planets, including ~200 terrestrial planets (Penny et al. 2019). However, the science gains from microlensing planet discoveries are limited without knowledge of the physical properties of the host stars. TMT high-resolution and deep follow-up observations will allow unprecedented accurate and detailed characterizations of planet hosts, significantly enhancing the information provided by these microlensed planetary systems.

### 10.4.1 Microlensing: landscape in the 2030s

The rate of microlensing detections is steadily increasing. Between 2004 and 2021, 36 microlensing planet discoveries have been published, with most of them coming from a two-stage survey/follow-up strategy (Gould & Loeb 1992; Griest & Safizadeh 1998). These discoveries have offered us some early glimpses into the distribution of planets more massive than ~3 $M_{Earth}$ near or beyond the nebular snow line (Gould et al., 2006; Gould et al., 2010; Cassan et al., 2012). Modern surveys such as KMTNet are now discovering ~25 microlensing planets per year (Gould 2023) via three dedicated 1.6 m, 4 $deg^2$ FOV telescopes completely covering the longitudes to observe the bulge continuously with ~10 minute cadence (Kim et al. 2016). The planned launch of the wide-field *Roman* Space Telescope at the end of the decade will bring a further order-of-magnitude leap in planet detection sensitivity (Penny et al. 2019).

### 10.4.2 TMT Microlensing Science

TMT AO imaging with IRIS and PSI will play a significant role in characterizing the host stars of current (KMTNet) and future (*Roman*) microlensing candidates. For the majority of microlensing planet detections, even though the precise planet-to-star mass ratios and angular planet-to-star separation can be directly inferred from the microlensing event itself, the masses and distances of planet host stars cannot be separately constrained from the microlensing light curves alone. An effective way to break the degeneracy that doesn't involve a new space mission (Gould & Horne 2013; Dong et al., 2007) is to directly measure the flux of the host star via late-epoch AO imaging. Ground-based microlensing targets are typically blended with significant flux from unrelated stars due to crowding in the bulge field. High-resolution follow-up observations from the Hubble Space Telescope (HST) and with AO systems on ground-based 8–10 m class telescopes have resolved the unrelated stars in some cases (figure 10-17; e.g., Alcock et al. 2001, Bennett et al., 2006, Dong et al., 2009, Janczak et al. 2010, Batista et al., 2014). The typical relative proper motion between the source and lens is ~6 mas/yr for a lens in the disk and ~4 mas/yr for a bulge lens, so it generally takes several decades for the lens to be separately resolved from the background star even with high-resolution images (Han & Chang 2003). With current AO facilities, constructing a full microlensing census (after resolving all blending and degeneracies) may take until 2050 (Gould 2023).

With a 30m telescope, we can immediately produce a full microlensing census at first light. At TMT commissioning the lens star would be separated from the source by ~2 λ/D in H-band in ~5 years after the peak of the detection of the event. Therefore, at the time of commissioning, TMT/IRIS will be able to resolve essentially all of the KMTNet planet hosts (see, e.g Henderson et al. 2015, Gould 2023). An average source would be at H ~ 18, and for a typical ~0.3 $M_\odot$ M dwarf lens at ~4 kpc, the required contrast ratio is about 10:1, easily observable by IRIS LGS. At ~2–3 λ/D, the achievable contrast ratio is ~1000:1, allowing the detection of the lens population down to the bottom of the stellar mass function for the majority of events (excepting only those with the brightest giant sources, a few percent of all KMTNet detections; Henderson et al. 2015).

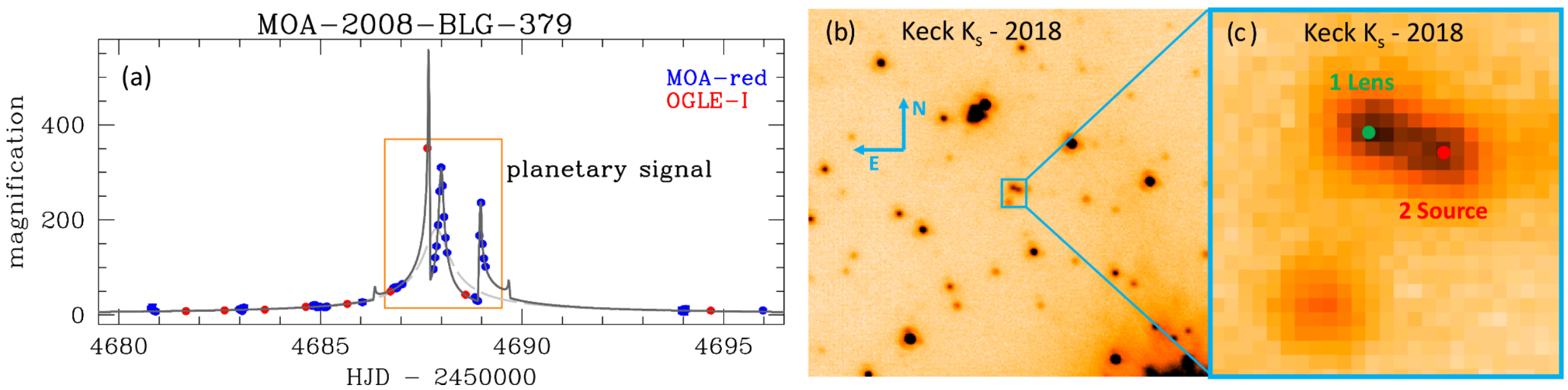

***Figure 10-17:*** *High-resolution imaging of a planetary microlensing signal (panel a) by Keck AO imaging (panels b & c) clearly resolves the lens and source stars 10 years after the microlensing detection (Bennett et al. 2024). This analysis revealed that the planetary lens was a 2.4 Jupiter-mass planet at 2.7 AU around a mid-M dwarf 3.4 kpc away. TMT will be essential for conducting rapid, timely observations to disambiguate microlensing signals detected by KMTNet and the Roman Galactic Bulge survey.*

During the *Roman* survey, the lens and source are typically not sufficiently separated to be completely resolved. However, thanks to a well-calibrated PSF it may be possible to measure lens flux via image modeling even when the lens and source are separated by a fraction of the PSF (Bennett et al, 2007; Spergel et al., 2013). However, roughly 1/3 of *Roman* microlensing detections will require AO imaging to disentangle lens and host stars, as with the KMTNet sample. The wait time for the lens and host star to be sufficiently separated is roughly ~5 yr (and inversely proportional to proper motion), so TMT/IRIS AO follow up of *Roman* targets could begin essentially immediately at TMT first light (Gould 2023). Thus TMT is essential to maximize the *Roman* science yield, and to do it on a reasonable timescale. TMT would be used to study a set of key targets that would be used to validate the analysis methods being applied to the Roman and KMTnet samples.

For both *Roman* and KMTnet followup, once the lens is resolved an exciting prospect for TMT is to obtain a medium resolution (R~4000) spectrum of the star with the IRIS IFU at its smallest plate scale of 9 mas. An approximately ten minute exposure would result in a spectrum with S/N≥10 per wavelength channel for a typical lens. The IRIS spectrum would enable identification of the planet host's stellar type and metal abundance as well as constraint system kinematics through radial velocity measurements. Prior RV surveys have demonstrated that the frequency of close-in Jovian planets increases as a function of host metallicity (Santos et al., 2004; Fischer & Valenti 2005). Recently, there is evidence from RV surveys and from the *Kepler* transiting planets that smaller planets at short-period orbits are distributed over a wide range of metallicity (Mayor et al., 2011, Buchhave et al., 2012). Spectroscopic characterization of microlensing hosts by TMT would open up a new window in studying whether the distribution of planets ranging from super Jupiters to sub-Earths at long periods depends on the environment of the star (metal rich vs. metal poor, bulge vs. disk).

Finally, for some nearby lens systems it is also possible to image brown dwarf planet hosts. In these cases, the lens identification is considerably more secure, as its separation from the source can be checked with the independent relative source-lens proper motion due to finite-source effects, which are well measured for most planetary microlensing events. The lens flux measurements would offer a mass-distance relation for the lens after including models of extinction and mass-luminosity relation. The relative lens-source proper motion and event timescale provide tight constraints on the angular Einstein radius, which offers another mass-distance relation for the lens. The mass and distance of the lens star can be constrained down to ~10% by combining these constraints.

# 11. Our Solar System

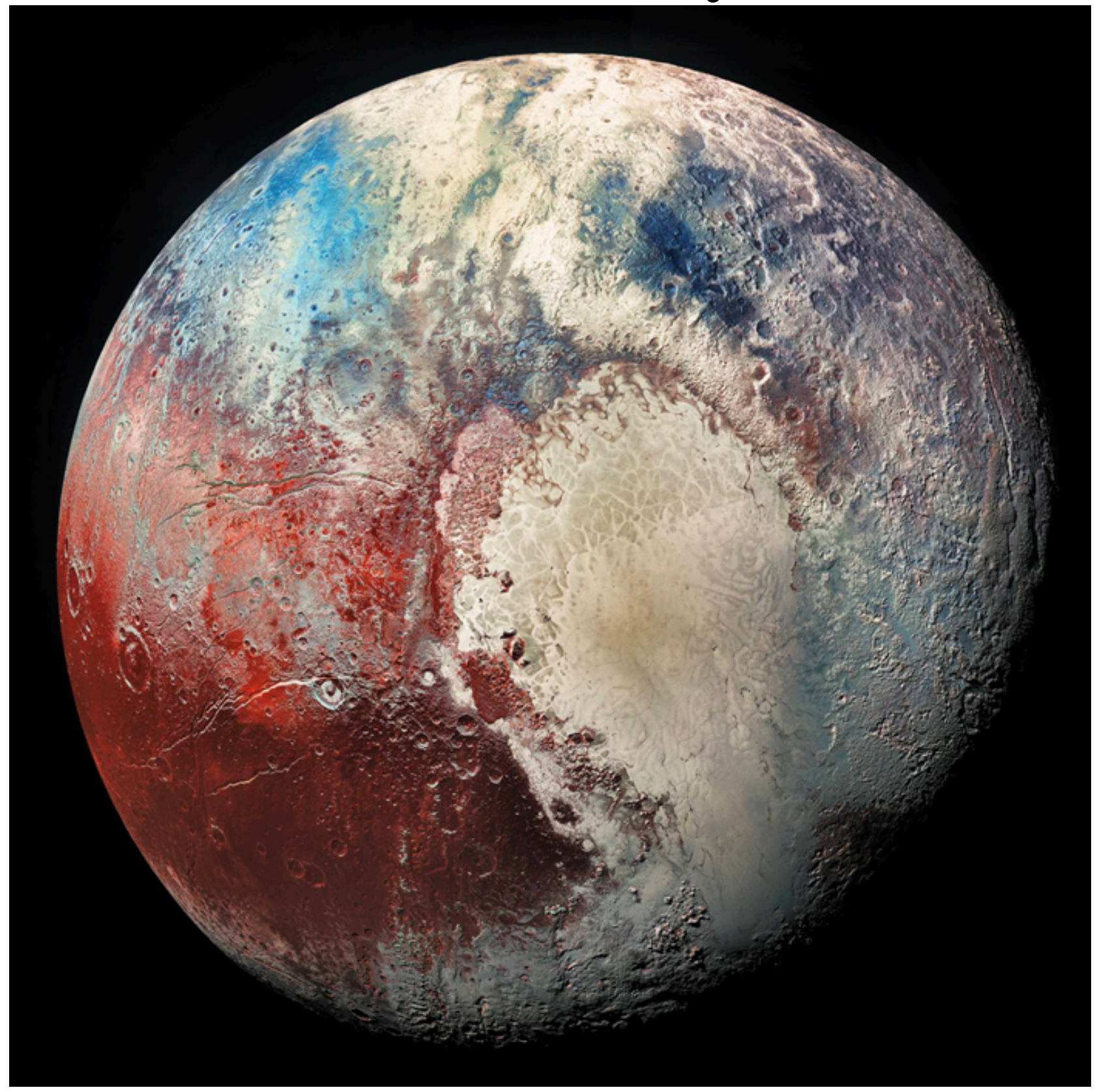

*Image of Pluto obtained by the New Horizons spacecraft as it flew past the dwarf planet in July of 2015. In this image (http://photojournal.jpl.nasa.gov/catalog/PIA19952), the most reflective white surface patches correspond to the infrared luminance channel of the Ralph/MVIC instrument on-board while the green and blue luminance channels have been removed. The color data come from the original image and were layered over the infrared channel, and further enhanced, to bring out the abundant compositional and geological variation across the surface of Pluto.*

---

Using TMT's high sensitivity and diffraction limited capabilities to study objects in our own solar system will provide powerful insights into the big question Q5-*What is the nature of extrasolar planets?* In addition, our solar system provides a test of planetary system formation models, addressing Q4-*How do stars and planets form?* Studies of solar system objects, including the recent discovery of cryovolcanism on moons of the giant planets (Enceladus, Europa), allow us to test ideas of how habitable planets are formed, how materials are delivered to those planets, and tie those answers into exoplanet formation models, addressing Q6-*Is there life elsewhere in the universe?* By answering these questions, TMT's explorations into the nature of our own solar system and the properties of the objects in our cosmic backyard will help us better understand our own place in the universe.

All the science observations discussed here require TMT's large light gathering power and fine angular resolution. Many of the solar system studies described here require TMT's LGS MCAO to further concentrate the light of faint targets. TMT's time domain ability and its responsiveness to cover targets of opportunity is going to be of use for discovery and characterization of interstellar objects and other transient or periodic events.

---

**Contributors**: Michael A'Hearn (University of Maryland), Christophe Dumas (TMT), Leigh Fletcher (University of Oxford), Thomas Greathouse (SWRI), David Jewitt (UCLA), Jianyang Li (PSI), Junjun Liu (Caltech), Franck Marchis (SETI Institute), Karen Meech (University of Hawaii), Michael Mumma (NASA GSFC), Glenn Orton (NASA JPL), Angel Otarola (TMT), Shalima Puthiyaveettil (IIA), Tomohiko Sekiguchi (Hokkaido University of Education), Feng Tian (Tsinghua University), Paul Wiegert (University of Western Ontario), Bin Yang (University of Hawaii).

---

# 11 Our Solar System

The solar system is the closest and best-studied planetary system. Observations of the planets, satellites, and small bodies in the solar system provide indispensable information about planet formation and evolution processes that remain unattainable for other planetary systems. TMT's immense capabilities will enable us to tackle long-standing questions concerning the formation of the solar system (and habitable worlds), the origin of planetary volatiles, the physics of the ice and gas giants, and unravel the complex dynamical history recorded in the Kuiper Belt. TMT's exquisite spatial resolution (see Table 11-1) will enable solar system studies at a level of detail only possible from in-situ space missions to-date.

***Table 11-1:*** *Spatial resolution achieved by diffraction-limited performance of TMT with the use of adaptive optics for several solar system targets. Distance is that from Earth during favorable viewing conditions. For unresolved objects, the integration time needed to achieve a given S/N scales as the square of the ratio of the telescope apertures. For resolved objects, the integration for a given S/N is the same for all telescopes with apertures large enough to resolve the source. However, the spatial resolution scales as the ratio of the telescope apertures.*

| **Spatial Resolution (mas)** | **7** | **14** | **34** | **55** | **83** | **138** | **193** |
|---|---|---|---|---|---|---|---|
| **Wavelength (µm)** | **1** | **2** | **5** | **8** | **12** | **20** | **28** |
| **Source Radius (km) - Distance (AU)** | **Spatial Resolution (km)** | | | | | | |
| Mars (3390) - (0.5) | 2.5 | 5 | 12.5 | 20 | 30 | 50 | 70 |
| Venus (6052) - (1) | 5 | 10 | 25 | 40 | 60 | 100 | 140 |
| Ceres (470) - (2) | 10 | 20 | 50 | 80 | 120 | 200 | 280 |
| Io (1830) - (4) | 20 | 40 | 100 | 160 | 240 | 400 | 560 |
| Titan (2575) - (9.5) | 47 | 95 | 240 | 380 | 570 | 950 | 1330 |
| Uranus (25,559) - (19.2) | 95 | 190 | 380 | 770 | 1150 | 1920 | 2680 |
| Neptune (24,764) - (30) | 150 | 300 | 750 | 1200 | 1800 | 3000 | 4200 |
| Ixion (707 mean) - (39) | 194 | 390 | 970 | 1560 | 2330 | 3890 | 5450 |
| Pluto (1151) - (39.5) | 200 | 400 | 990 | 1600 | 2400 | 3940 | 5500 |

## 11.1 Primitive bodies

Studies of small primitive bodies with TMT will address many of the questions identified in the Planetary Science and Astrobiology Decadal Survey (NASEM, 2022, chapters 4–7):

- What were the initial conditions in the solar system? Isotopic signatures of volatiles in small bodies can provide information about the conditions in the early solar system. Tracing this in small body populations through high resolution spectra of outgassed volatiles can help trace variation across the disk.
- How did the giant planets gravitationally interact with each other and smaller bodies in the outer solar system? Remote sensing of the composition and physical properties of Trans Neptunian objects, Centaurs and comets by TMT will help unveil the scattering that occurred in the early solar system and the chemical conditions during formation.
- How and when did the asteroids and inner solar system protoplanets form? This will be addressed by understanding the diversity of compositions of remnant planetesimals (asteroids, comets). Determining the contribution of outer solar system material to the terrestrial planets will be addressed through measurements of water-rich asteroids and comets.

- How have planetary bodies collisionally and dynamically evolved throughout solar system history? Determining the size distribution of small Kuiper belt objects, long period and short period comets will address the collisional environment in the outer solar system.

### 11.1.1 Asteroids

Asteroids in the main belt between Mars and Jupiter are probably remnants of high temperature accretion in the inner solar system, with additional bodies accreted from the outer solar system in an early phase of dynamical chaos. Their low total mass ($3\times10^{-4}$ Earth mass) reflects the clearing of perhaps 0.5 Earth mass of material by dynamical processes that remain to be elucidated from the dynamical imprints. While nearly a million asteroids have been identified that are larger than 1 km in size, their physical properties remain relatively poorly known.

However, as the source of meteorites, the main belt asteroids can, with care, be directly linked to samples in the laboratory, providing unique opportunities for hands-on, detailed isotopic and compositional measurements that are otherwise limited to the small samples coming from complex space missions.

### 11.1.2 Active Asteroids

A recently discovered hybrid population of asteroids also exists. The so-called active asteroids, have the orbits of asteroids but lose mass, giving them the appearance of comets (Snodgrass et al 2017). Some of these objects may develop dust coma because of collisions, or rapid spin up, but some contain water ice (Kelley et al 2023), trapped from beyond the snowline in the solar system; these are the main belt comets (MBCs). When exposed to the sun by the removal of a protective layer of debris, this ice sublimates, producing the comet-like appearance. MBCs are potentially important because some are located in a region of the main belt from which the terrestrial planets are thought to have acquired their volatiles and some biogenic precursor molecules. Study of the MBCs may help us to understand the origin of inner solar system volatiles and provide clues to the formation of habitable planets. TMT will be useful because most active asteroids are faint. TMT's large aperture will permit visible and near-IR spectroscopic studies in search of volatiles released by outgassing.

### 11.1.3 Asteroid Satellites

Over 500 asteroids and other small solar system bodies are known to have satellites and radar and photometric observations have shown that satellites are extremely common. Asteroidal satellites are thought to form via a variety of scenarios including collisions, capture through 3-body interactions, and/or disruption. Asteroids with satellites (Merline et al 1999) are the best study cases for understanding the dynamical history, impact process, and physical properties of the primary. Asteroid moons provide the most accurate way to measure the mass and the internal structure of the asteroids (Beauvalet, Marchis & Ruffio, 2013). Density, which is directly related to the composition and internal structure of the asteroid, can be estimated from their diameter, even more precisely if the shape of the primary is available. If the primary is associated with a young dynamical family, then the combination of the moon(s) and the dynamical family properties provides strong constraints on the family-forming process. TMT's unprecedented angular resolution provided by NFIRAOS and IRIS (first-light instrument), and PSI (first-decade instrument), will allow to resolve and directly image both tightly-bounded systems and systems that have high primary-to-satellite mass ratios.

### 11.1.4 Physical properties of the outer-belt asteroids

The origin of the primitive outer belt asteroids (namely the Cybeles, Hildas and Trojans) is unclear. They may consist of materials that formed the cores of Jupiter and Saturn, or they may be objects formed at larger distances (perhaps in the Kuiper Belt) and then scattered inwards to their present locations. The physical and dynamical properties of these objects enable key tests of the competing scenarios for the evolution of the early solar system.

Composition of an asteroid can be best determined through moderate resolution (R ~ 1,000) spectroscopy in the near-infrared (1–2.5 μm), where most important ices and organic materials show diagnostic features. It is expected that hundreds, possibly thousands, of km-size Cybeles, Hildas and Trojans will be observable spectroscopically by IRIS on TMT. Investigating the outer-belt objects with TMT, especially the smallest bodies, may provide answers to the outstanding issues, such as: What are the cores of the giant planets made of? Where and how did the Trojans form? What is the dynamical history of the solar system?

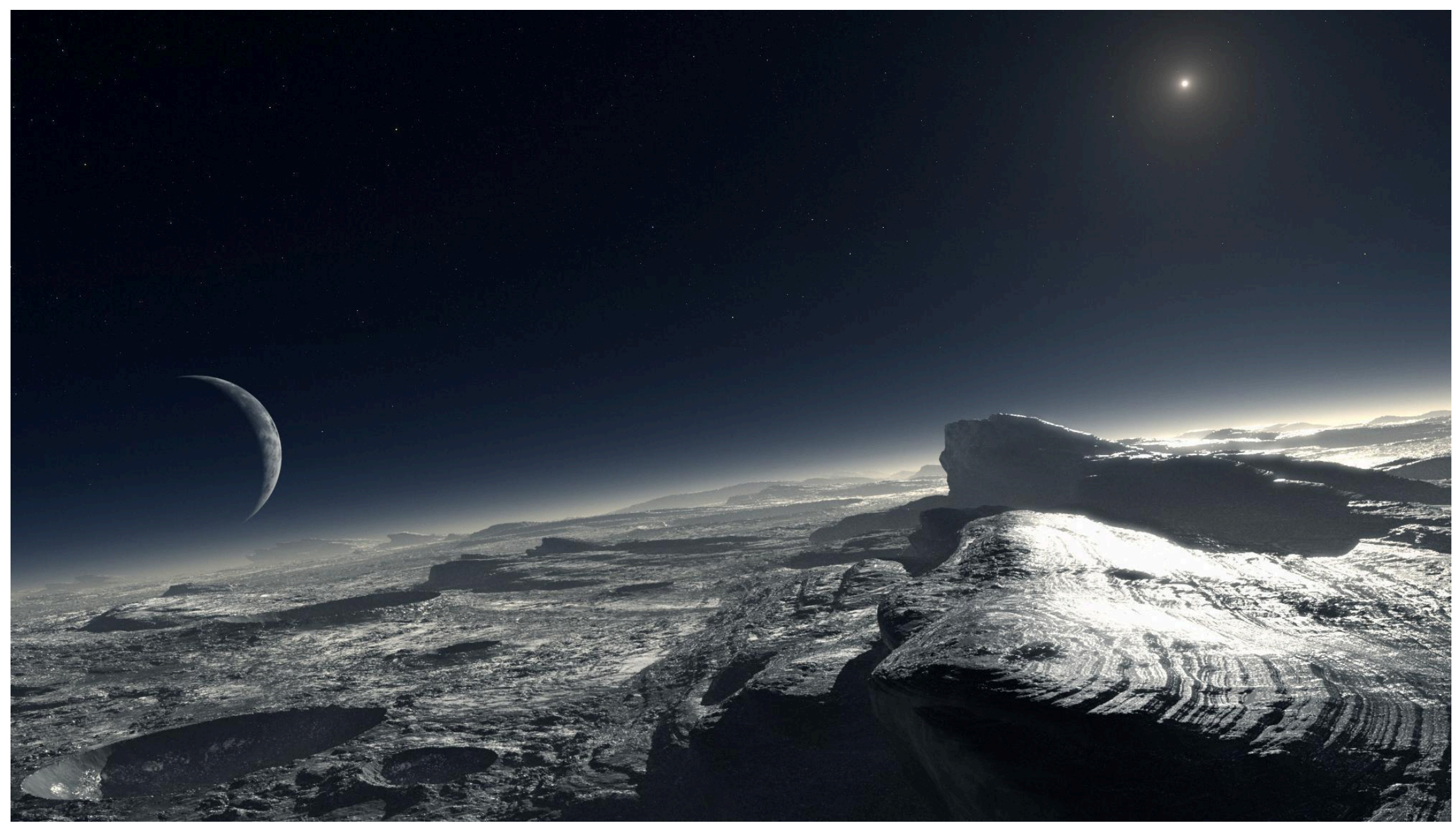

***Figure 11-1:*** *Artist's representation of an icy Kuiper belt object showing the faint Sun in the upper right. Courtesy ESO.*

### 11.1.5 Kuiper Belt objects

Beyond Neptune, a much larger population of small bodies exists in the Kuiper Belt (figure 11-1). Kuiper Belt objects (KBOs, also referred to trans-Neptunian objects, or TNOs) outnumber asteroids by 1000:1. These objects are the result of low temperature accretion in the outer regions of the solar system and they are thought to be ice-rich. JWST has shown a mixture of water, $CO_2$ ice, hydrocarbons and complex organics on some of the larger bodies (Emery et al., 2024). More than 3000 KBOs are known to date but physical properties have been established for only about 100 of them. TMT will greatly increase the sample size of KBOs with measured physical properties, and the study of collisional processes in the Kuiper Belt will inform the interpretation of the much more distant (but numerous) debris disks around main-sequence stars, about which much less is known.

The mean density of distant planetesimals is expected to be markedly different from that of native main-belt asteroids. The most promising method to estimate the shape, and subsequently infer the density, is via analyzing light curves of binary systems. For ground-based measurements, the precision of the inferred diameter is the limiting factor of the most accurate density estimates. Given the advantage of TMT's high-angular resolving power (~7 mas at 1μm), IRIS and PSI will be able to resolve a dozen large KBOs and about 50–100 Trojan asteroids, providing a good estimate of their sizes. If moons are detected, TMT will also derive high-precision densities of these objects.

Because of the distance (30–50 AU) of this region from the Sun and the Earth, even with the largest telescopes currently available, only the largest KBOs (> several hundred km in diameter) are currently observable. The Kuiper Belt is therefore the solar system region about which we know the least. Due to its low temperature and the relatively slow dynamical evolution, the Kuiper Belt can be considered a debris disk of our planetary system, comparable to those of other stars (e.g., Kalas et al., 2006; Trilling et al., 2008). The belt contains essential information about the planetary formation processes, including both the cold disk that harbors the objects that are thought to form *in situ*, and the hot/scattered disk that is the refuge of objects that are dynamically scattered during the dynamical evolution of the inner solar system. Comparisons of the Kuiper Belt with the debris disks around other stars provide important information about both the Kuiper Belt itself and the planetary environment around other stars.

TMT, with its immense light-gathering ability, will significantly contribute to these studies by characterizing many more small KBOs. Compositional information will greatly aid the study of the dynamical families, which are generally formed during collisions of their parent bodies. Finally, spectroscopic observations from the TMT may also reveal evidence of evolutionary processes in KBOs, such cryovolcanism, volatile loss and surface gardening.

### 11.1.6 Centaurs

Centaurs are recently (within 10 Myr) escaped Kuiper belt objects on their way to becoming Jupiter family comets. They are relatively nearby (5–30 AU), compared to the Kuiper belt, making them relatively easier to observe using the high resolution of TMT. They also remain pristine, enabling us to determine the composition of the early solar nebula. The IRIS instrument, with its diffraction limited imaging and spectroscopic capabilities, is ideal for studying the actual shapes and composition of these objects. TMT will be capable of resolving complex structures such as the rings around Chariklo (Braga-Ribas et al. 2014) that have a diameter of ~0.1”, and its high sensitivity will enable us to characterize several new centaurs as small as 20 km in size.

### 11.1.7 Comets

Comets represent the icy planetesimals that are leftover building blocks from the collapse of the solar nebula (figure 11-2). A majority have been stored at temperatures ranging from 10 K to 40 K in one of two reservoirs: the Kuiper Belt that extends from Neptune’s orbit at 30 AU to at least several thousand AU; and the Oort cloud, a spherical assemblage that reaches 50,000 to 100,000 AU from the Sun. These reservoirs contain, respectively, 1 billion and 100 billion comets larger than about 1 km. Comets are important as carriers of the most primitive material in the solar system. Their study allows us to probe the chemical make-up of the solar system at its origin.

The 2.9–5.0 μm region (3450–2000 $cm^{-1}$) is the single most important region for spectroscopy of simple molecules (up to 8 atoms). Virtually all simple gasses have at least one vibrational fundamental band in this spectral region. For emission spectra, resolving powers to 100,000 (even 300,000) are preferred in order to discriminate densely packed and blended lines. But this spectral window is difficult to access from the ground and some important work can still be done in the 1–2.5 μm region. With TMT/MODHIS, it will be possible to measure emission lines from the dominant primary volatile ($H_2O$) at 2.0 μm in active distant comets. It may be possible to measure hot bands of $CO_2$, the CO overtone (2.4 μm), and various other species as well. Ices can also be investigated through their solid-state absorption bands in a variety of objects from active comets to satellites and KBOs.

Manx comets are a separate type of comets. They are small (~km-sized) bodies on long period comet orbits which suggest a source region from the Oort cloud, yet they exhibit little or no activity when close to the sun – atypical of long period comet behavior. The first of these discovered was found to have a spectrum similar to inner solar system rocky material, yet it had low-level outgassing (Meech et al. 2016). The hypothesis is that this object formed near the solar system’s ice line and was ejected outwards during giant planet formation. A study of the chemical properties of Manx comets and the relative amount of inner solar system material can provide constraints on dynamical models of our solar system’s formation. Manxes are faint objects, but with TMT/WFOS, it will be possible to obtain low resolution spectra to map their surface mineralogy - particularly in the 0.4–1.0 micron region.

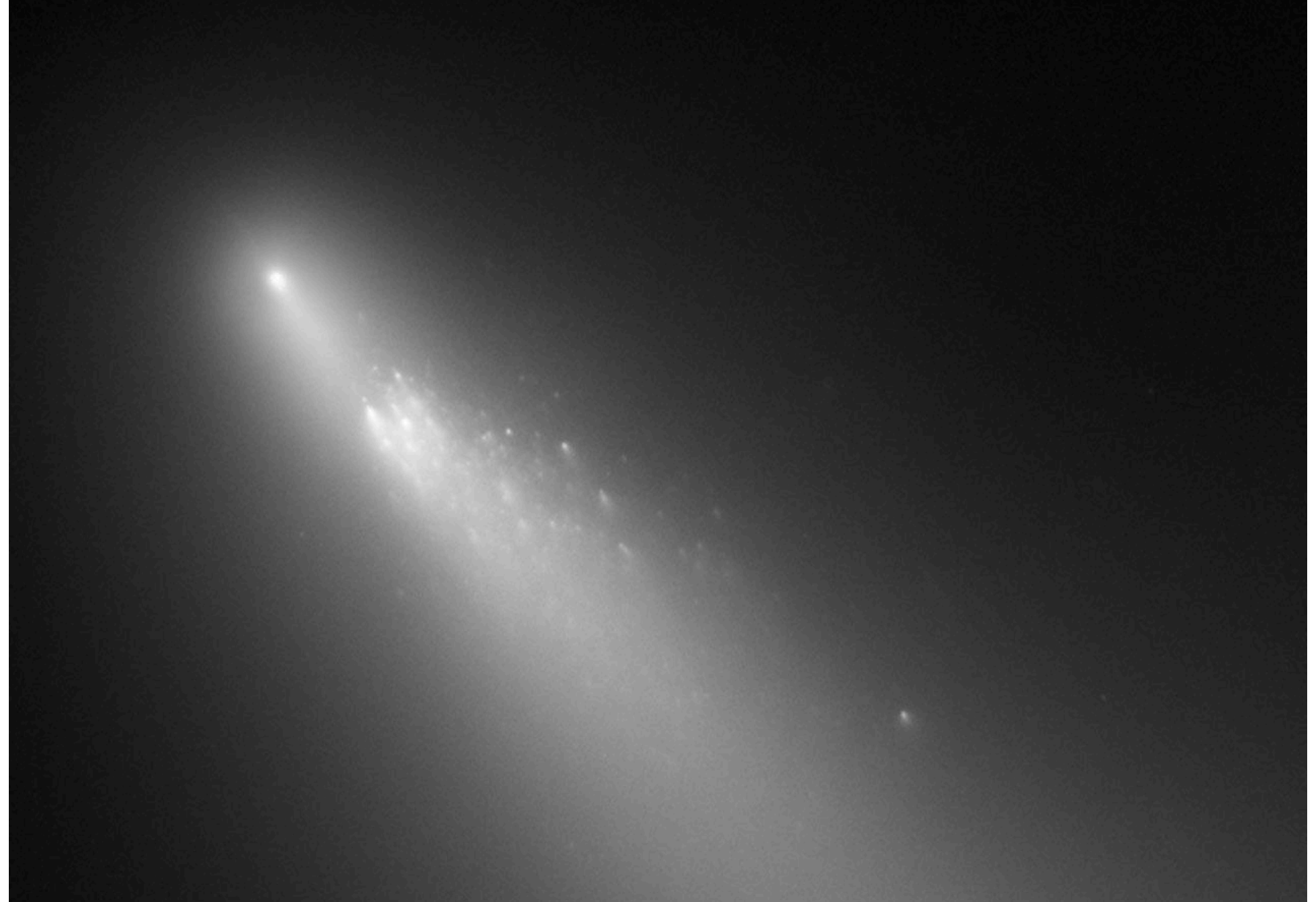

***Figure 11-2:*** *Comet Schwassmann-Wachmann 3 disintegrating. TMT will be able to explore the properties of cometary nuclei. (Courtesy STSCI – NASA/ESA).*

At optical wavelengths, high resolution spectroscopy can be used to determine nuclear spin temperatures in $NH_3$ and $H_2O$ through emission lines of their dissociation products $NH_2$ and $H_2O^+$. The nuclear spin temperature is believed to be preserved indefinitely after the formation of a molecule, and hence gives an estimate of the temperature prevailing at the time of the last condensation of the ice. Extending these measurements to fainter objects and to a larger sample will allow us to examine temperature differences that might correlate with formation location in the protoplanetary disk.

### 11.1.8 Source of Planetary Volatiles

Comets are likely originators of part of Earth's water, while outer-belt asteroids, which also contain water, are now believed to constitute a larger reservoir.

Since 2005, a new class of comets has been identified residing in the main asteroid belt, known as main belt comets (MBCs). MBCs have shown recurrent dust production near their perihelia, but only recently has outgassing water been detected on one of them (Kelley et al. 2023), reinforcing the role of MBCs are a promising reservoir of terrestrial water. Direct evidence of water sublimation has also been detected on 1 Ceres, the largest main belt asteroid (Kupers et al., 2014), using the Herschel Space Observatory. . TMT will excel in detecting low-level sublimation, which will enable identification of orders of magnitude more MBCs and likely allow direct detection of volatiles within them.

Addressing the science cases presented here relies on TMT's increased sensitivity and AO-enhanced angular resolution compared to existing facilities. High resolution spectroscopy with $R \gtrsim 100{,}000$ is a key enabler in many cases, as is the full wavelength coverage of the infrared provided by IRIS (up to 2.4μm) and MICHI (mid-infrared). section 12 further summarizes key science needs and their corresponding requirements on TMT and its instruments.

### 11.1.9 Interstellar Objects (ISOs)

The discoveries of an interstellar asteroid in 2017 (1I/2017 U1, now named 'Oumuamua), followed soon by an interstellar comet (comet 2I/Borisov) in 2018, extended our definition of 'multi-messengers' in astronomy (Fitzsimmons, A. et al. 2024). This pair of kilometer-sized bodies are the first known objects to have recently entered our solar system from interstellar space. They are also quite different from each other in their properties. 'Oumuamua is seemingly asteroid-like and gas-free but with acceleration tentatively attributed to outgassing, while Borisov appears as a more-or-less typical comet with abundant gas production.

As a new population of unstudied and heterogeneous objects of an unknown nature, almost all observations are of value. Specifically, the science case for ISOs is similar to that of other distant solar system bodies, relying on TMT's ability to characterize them at larger distances than today's facilities (e.g. compositional studies based on spectroscopy of light reflected from their surfaces, or from emission from outgassing). Many more of these objects will be discovered as large transient surveys like LSST get underway. TMT queue scheduling will be important for enabling responsive observations of unique objects with potentially brief visibility windows. Although solar system time-domain science is not discussed in chapter 9, needs are shared between time-domain astrophysics and time-domain solar system science: observations with rapid response time and specific science-driven cadences.

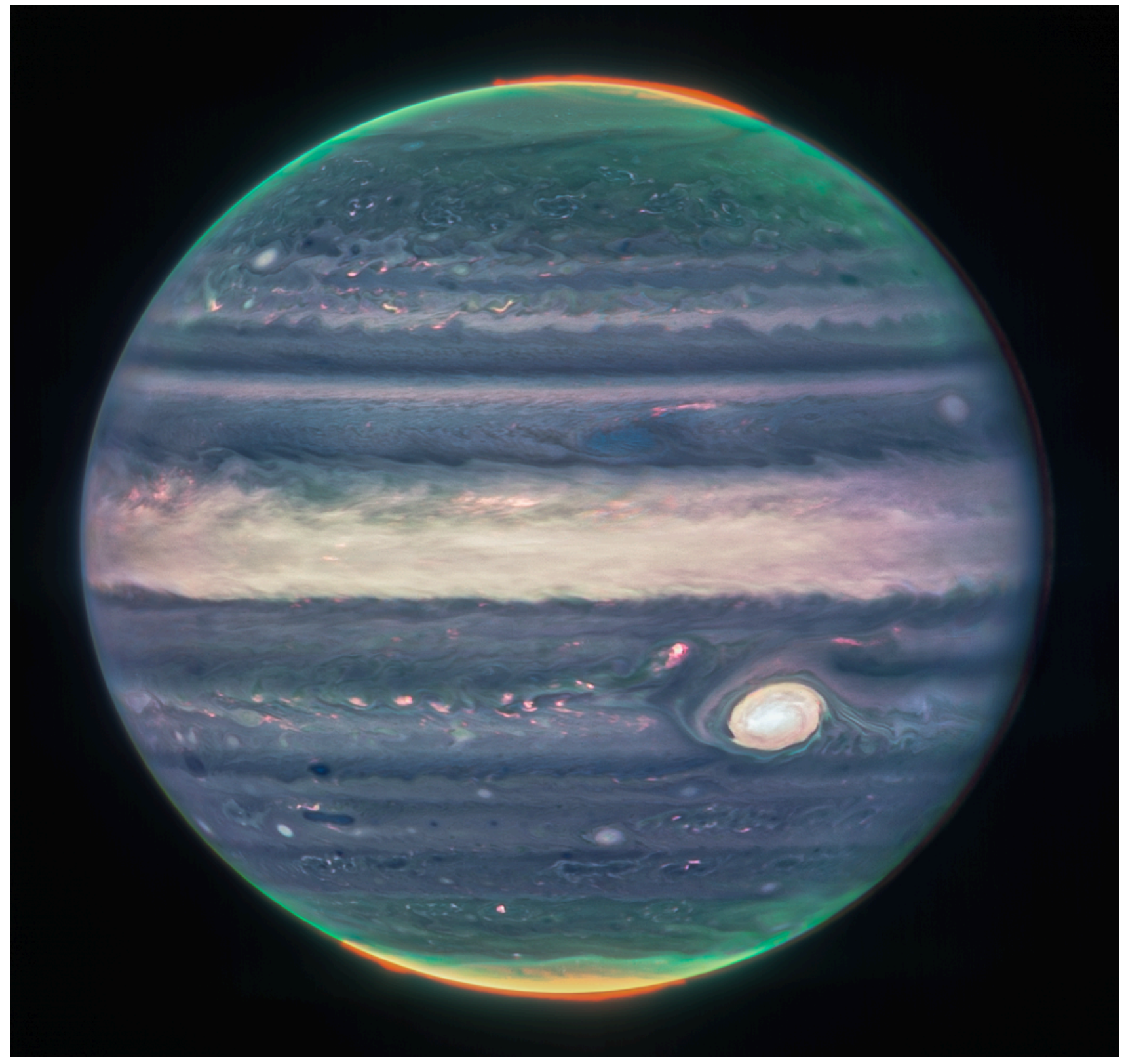

***Figure 11-3:*** *JWST infrared Jupiter image processed for high dynamic range shows a wide range of phenomena that will be probed with TMT: aurora and airglow emission from thermosphere/ionosphere, stratospheric photochemical and auroral hazes, and cloud features at different tropospheric altitudes shaped by jets, vortices, and convective storms. (Credit: NASA, ESA, CSA, Jupiter ERS Team; image processing by Judy Schmidt).*

## 11.2 GIANT PLANETS

The Planetary Decadal Survey identified five major priority science questions in relation to giant planet structure and evolution (NASEM, 2022, Ch. 10). TMT capabilities will help answer all of these five questions, and enable new discoveries along the way:

1) What are giant planets made of and how can this be Inferred from their observable properties?
   TMT spectroscopic observations will constrain chemical processes, vertical mixing, and dynamical transport in all four giant planets by simultaneously measuring multiple tracers such as temperature, condensable and disequilibrium species. TMT's angular resolution will enable new advances in our understanding of the horizontal variation of these tracers on fine spatial scales.

2) What determines the structure and dynamics deep inside giant planets and how does it affect their evolution?
   TMT high-resolution imaging will measure velocity fields, and thus the dynamics of vortices, storms, and thermal waves, at fine spatial scales. Spectroscopic measurements of disequilibrium gasses will probe mixing at unseen deep levels.

3) What governs the diversity of giant planet climates, circulation, and meteorology?
   TMT imaging and spectroscopic time series observations will address this question very effectively, deriving high-resolution wind fields at multiple altitudes, determining the spatial and temporal variability of stratospheric temperatures, trace gasses, and aerosols, and constraining storm evolution on timescales from hours to years.

4) What processes lead to the dramatically different outcomes in the structure, content, and dynamics of the outer planets' magnetospheres and ionospheres?
   Magnetospheric processes will be traced by observations of auroral emission in the infrared, stratospheric aerosols, and temperatures and densities inferred from hydrocarbon emissions.

5) How are giant planets influenced by, and how do they interact with, their environment?
   Giant-planet impact phenomena range from clouds of dark stratospheric aerosol debris that evolve over periods of days to weeks, to compositional anomalies that can last hundreds of years in the stratosphere. High resolution imaging and sensitive spectroscopy are ideal for studying these unpredictable phenomena, as well as the interactions with seasonal solar radiation and solar wind.

### 11.2.1 Vortices and thermal waves

Jupiter's red spot and other cloud features (figure 11-3) are large-scale coherent vortices whose physics is still unclear. Hubble's OPAL program (Simon, Orton, Wong, 2023) and programs with JWST (e.g. Hueso, et al., 2023) are making significant strides in observationally characterizing the atmospheric and meteorological processes of Jupiter (and the other gas giant planets), however the spatial and spectral resolution afforded by TMT is necessary to develop a deep understanding. TMT will enable time-resolved high resolution spectroscopy and imaging of these structures and provide important information about the coupling of chemistry and dynamics in the vortices. Similar structures are observed on Saturn, which TMT will also be able to explore.

Thermal waves are ubiquitous in giant planet atmospheres. The characteristics of the thermal waves (phase speed, propagation direction, and wave number) provide important clues in understanding the generation mechanism of these waves and the formation of jets in giant planet atmospheres (Li, et al., 2004). However, due to the combination of long integration times and low spatial resolution of current observations, many characteristics of these waves, such as the phase speed and propagation direction, have not been accurately measured. High spatial- and temporal-resolution observations by TMT will enable mapping thermal waves in giant planet atmospheres in much detail and reveal the thermal wave characteristics on different giant planets.

Io's calderas can be used as occultation sources to study waves propagating through Jupiter's atmosphere and probe its vertical structure. The volcanic calderas have temperatures of ~1,300 K (Spencer et al. 2007) with spatial extents of 10–50 $km^2$. These bright emission sources are just below the diffraction limit of TMT at all wavelengths, so they can serve as point-like emission sources to be observed while Io is eclipsed by Jupiter's atmosphere. Io's 42-hour revolution around Jupiter affords repeated ingress and egress observation possibilities every 4 Jovian rotations. There are often several volcanoes active on Io simultaneously. IRIS' integral field units could observe all of Io's disk simultaneously, gathering as many as 5–10 occultation light curves nearly simultaneously, yet slightly offset both in time and space.

### 11.2.2 Planetary Seismology from Impacts on giant planets

The internal structures of giant planets are much less well known than those of main-sequence stars because of uncertainties in the equation of state of degenerate gas, the composition (typically non-solar), the interaction with the magnetic field and, in the upper layers, the relative magnitudes of internal heat and energy deposited from the sun. Giant planet interiors are inaccessible to direct study from above, but oscillations excited by asteroid and comet impacts can generate atmospheric waves that are potentially observable. Such waves will propagate through the planetary interiors, allowing giant planet seismology to constrain internal structure in much the same way as done for our planet using earthquakes.

Impacts by asteroids or comets, such as Jupiter's collision with comet Shoemaker-Levy 9 in 1994, would provide the conditions for detecting planet-wide atmospheric waves, something that the technology did not allow back then. TMT will be extremely efficient to respond to such targets of opportunity events and measure the propagation direction, the propagation speed, and the energy of the atmospheric waves. Such measurements will probe the atmospheric structure and composition, providing unique information useful not just in the solar system but also in the study of Jupiter-like exoplanets, where no comparable data will be available for the foreseeable future.

### 11.2.3 Source of internal heat in Uranus & Neptune

Ice giants Uranus and Neptune have only been visited once (by Voyager 2). They remain the least understood and most mysterious planets in the solar system and yet the Kepler mission has already shown that planets of similar mass (in between that of Earth and Jupiter) are widespread outside the solar system. TMT observations, therefore, can help us understand not only Uranus and Neptune, but a large number of known exoplanets as well. High angular and high spectral resolution observations by TMT will be used to constrain the planets' bulk compositions and the temperature and gas abundance structures in their atmospheres, shedding light on how planetary atmospheres form and evolve as a function of distance from their host stars.

TMT observations can also determine the spatial variability of temperature and gas abundances on Uranus and Neptune, which is connected to the atmospheric dynamics, such as the vertical propagation of waves from the troposphere to the stratosphere.

## 11.3 ROCKY PLANETS AND MOONS

### 11.3.1 Titan

Titan maintains a high pressure (> 1 bar) nitrogen atmosphere and an active hydrological cycle driven not by water, as on Earth, but by hydrocarbons. Its atmosphere is often compared to that of the young Earth, albeit cooled to much lower temperatures (<90 K) than ever found on our planet. It offers a valuable opportunity to study a high mass atmosphere on a solid-surface planetary body and informs models of atmospheric circulation, precipitation and seasonal response. Current observations suggest that Titan has lakes mainly in the northern polar region and has tropospheric clouds mainly in the southern middle latitudes and southern polar region. Numerical simulations predict dramatic climate change over the next decades. Models suggest that the lake formation is due to the cold-trapped methane accumulated in the polar region, and predict that prominent clouds are being formed while lake levels are expected to rise over the next ~fifteen years due to the seasonally varying solar radiation on Titan. TMT will be able to observe and monitor this exciting climate change. In particular, the high spatial and spectral resolution offered by TMT will reveal the spatial distribution and temporal variation of Titan's methane clouds, and separate the high clouds generated by deep convection from the low clouds formed over the surface methane reservoirs, providing needed observations to test current general circulation models and our understanding of the hydrological cycle and seasonal variations on Titan.

### 11.3.2 Planetary atmospheres

The TMT, with its high spatial resolution, will provide ground based observers with unique observing capabilities currently possible from spacecraft only, such as planetary limb sounding. The limb sounding method offers an observer the ability to probe a planetary atmosphere by looking through the planet's atmosphere at distinct tangent points, providing remarkable vertical resolution. These types of observations can be used to look for vertical variations of the thermal structure and chemical composition of a planet's atmosphere. Many of the planetary scale atmospheric waves are expressed as thermal variations with respect to the background atmosphere and such waves would be easily observable with the TMT. Table 11-1 shows that TMT spatial resolution in the near infrared (~1–5 μm) is more than adequate to resolve the 11 km scale height of the Martian atmosphere. At 8 μm, where HDO and $H_2O_2$ abundances can be mapped, the TMT would provide a 20 km spatial resolution over the > 100 km vertical extent Martian atmosphere.

Regarding temporal variations, atmospheric features change on timescales that can be probed with TMT. Mars experiences global dust storms in addition to its annual cycle of polar atmospheric freeze-out. Giant planets have their cloud deck evolving in dramatic and unpredictable ways, as shown by the emergence of a super-storm on Saturn (Sayanagi et al 2014). Multi-wavelength imaging and spectroscopy with TMT will bring capabilities for high-resolution observations of sudden atmospheric changes that cannot be matched by telescopes in space.

### 11.3.3 Volcanism (Io)

Io is the most active volcanic world in our solar system and spacecraft imaging reveals a world littered with volcanic eruptions (figure 11-4).

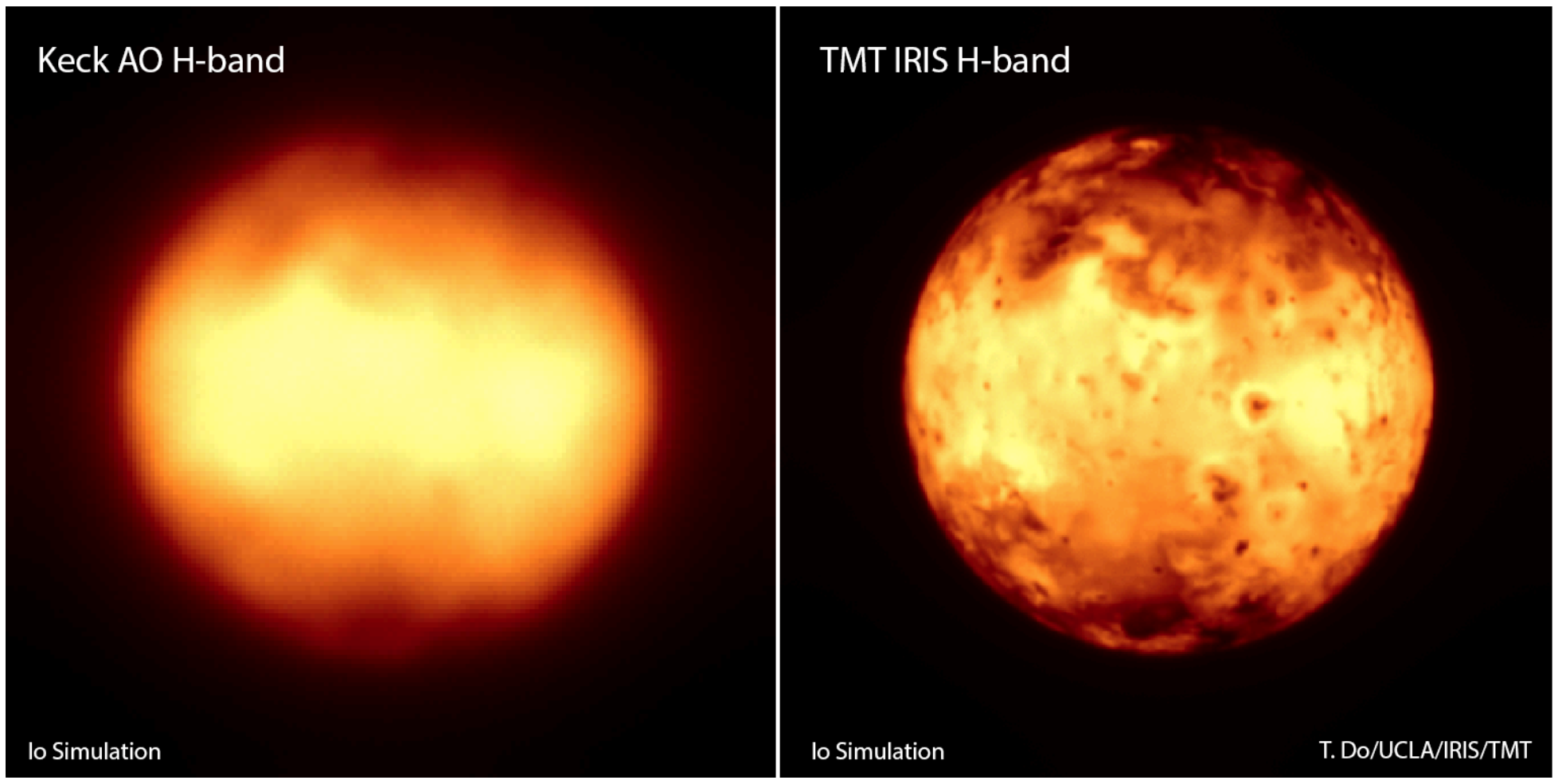

***Figure 11-4:*** *The Sulfur-covered surface of Jupiter's satellite, Io. These simulated images show a comparison between adaptive optics observations of Io surface in reflected near-infrared light (H band) made with the 10 m Keck telescope (left) and the TMT (right). Circular features are regions of ejecta fallback from active volcanoes. No impact craters are visible owing to the extreme youth of the surface. Io will fill ~600 TMT pixels at 0.7 μm, enabling long-term monitoring of its activity. The volcanic heat source is thought to be tidal flexing by the finite eccentricity of the orbit, itself a result of a Laplace resonance between the Galilean satellites.*

Current research points towards an atmosphere originating from the sublimation of $SO_2$ frost by solar heating, and composed of $SO_2$ gas together with some other gasses released during volcanic eruptions. $SO_2$ has strong molecular bands at 7, 8 and 19 µm, making its disk fully resolved by TMT (26 and 10 spatial resolution elements across the disk at 7 and 20 µm, respectively). Direct imaging of Io's surface with IRIS and MICHI will enable a spatially resolved monitoring of its activity and surface composition, and will help constrain how much of its atmosphere comes from frost sublimation *vs* volcanic eruption.

Io is in eclipse approximately every two days, when the satellite is located within Jupiter's shadow. These events, which last for a few hours, offer another way of testing atmospheric production models. Eclipse events inhibit solar heating required for frost sublimation and observing Io during eclipses provide means to measure how much of its atmosphere condenses onto the cold surface. This process cannot be observed with today's 8–10 m class telescopes but TMT's large aperture will provide the sensitivity and spatial resolution to carry out such programs.

Additionally, we can expect that individual volcanic plumes will be recognizable. During the New Horizons flyby of Jupiter, the volcano Tvashtar was imaged with a plume height varying between roughly 320 and 360 km and a full width of about 1,100 km, consistent with the diameter of the pyroclastic deposits. Thanks to TMT's high spatial resolution, we will be able to directly measure the composition of similar-sized volcanic plumes with IRIS and MICHI.

# 12 NEW REALMS

Chapters 3 to 11 describe a broad range of astronomical science cases for the TMT observatory. These science cases require a range of observing capabilities that cover large areas of the potential discovery space, spanning a wide range in wavelengths, spatial and spectral resolutions, response times and cadence, and fields of view, all the while pushing toward maximum sensitivity.

Experience with every previously constructed general purpose observatory that had a set of motivating science cases tells us that the science that is produced with such a facility rarely follows what was originally included in the science case documents. The purpose of the science cases in the previous chapters are to act as examples of what can be achieved and to guide us to an observatory design that has exceptional capabilities that can be directed toward almost any conceivable astrophysical science question.

This updated version of the 2015 TMT Detailed Science Case identifies numerous new discoveries since the 2015 science case was published. In all cases that we can envision, prospective new studies that may arise as a result of these recent discoveries can be conducted with the planned TMT instrument suite, giving us much confidence that the planned TMT Observatory is a highly capable and flexible facility that can address much more science than is captured in the Detailed Science Case.

While new knowledge always raises more questions, the ability to attempt to answer questions is limited by the technology and fundamental observing capabilities that can be brought to bear on those questions.

And so, with the large leap forward in spatial resolution and sensitivity that TMT will provide, and the ability to harness these new capabilities rapidly and apply them across a range of wavelengths, spectral resolutions and fields of view, the potential scientific discoveries defy our ability to imagine them at this time. Exciting and as yet unanticipated discoveries await the users of the TMT, the astronomical community and society as a whole.

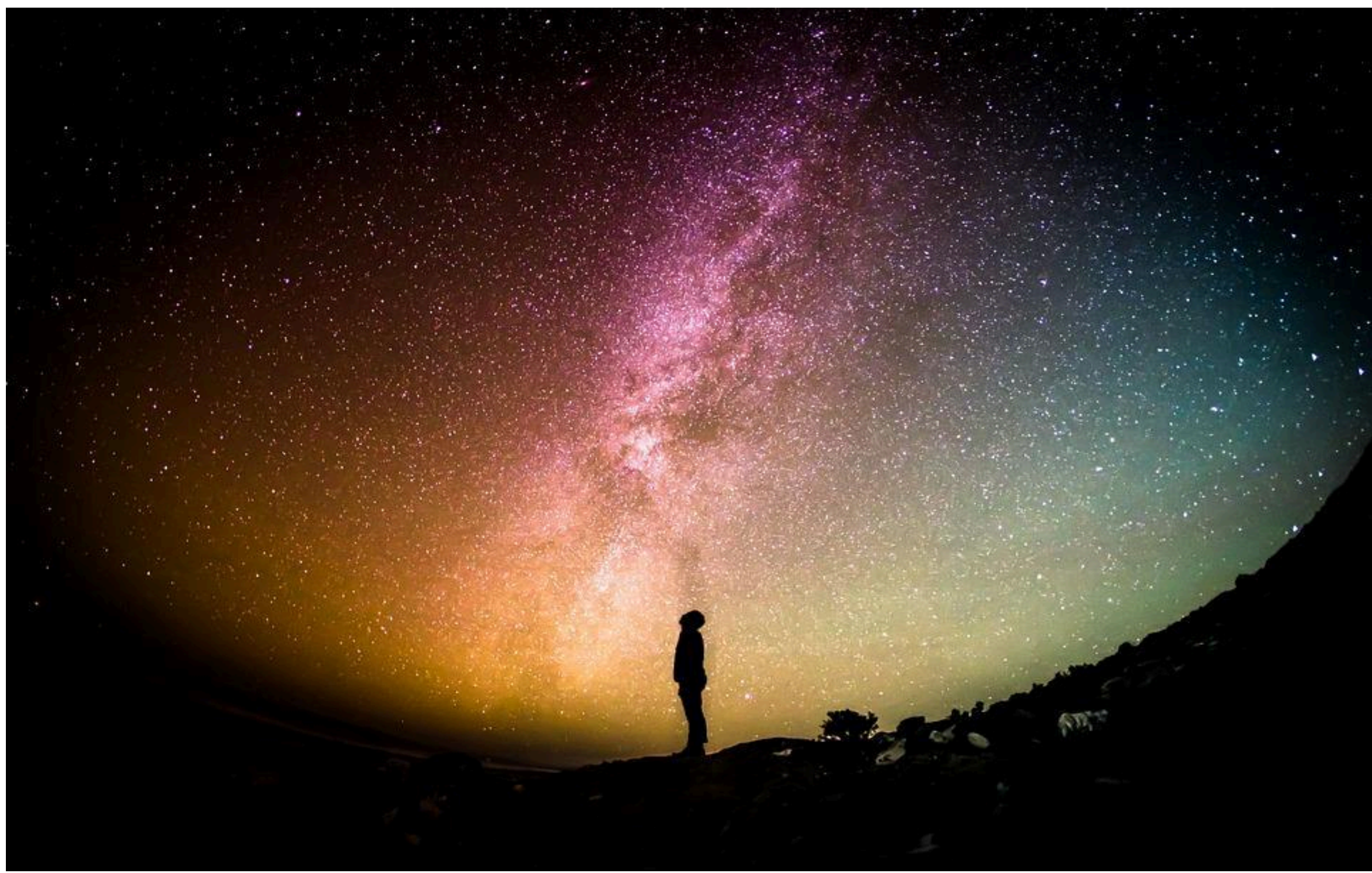

*Image credit: https://www.dreamglossary.com/*

# 13 SUMMARY OF TMT SCIENCE CASES VERSUS TMT INSTRUMENTS/AO CAPABILITIES

***Table 13-1:*** *Summary of the relations between the fundamental questions, science themes and specific science areas, and the forms of observations and instrument/AO capabilities required to support them. A particular form of Observation may support multiple Science Areas and a single Science Area may require more than one form of Observation . [SL=Seeing-Limited, LGS MCAO=Laser Guide Star Multi-Conjugate Adaptive Optics, MOAO=Multi-object Adaptive Optics, ExAO=Extreme Adaptive Optics, MIRAO=Mid-IR Adaptive Optics].*

| **Fundamental questions** | | **Science Cases** | | **Instruments/AO Facility** | | |
|---|---|---|---|---|---|---|
| **Qs ID** | **Detailed Science Case Chapter** | **Science Areas** | **Observations** | **Requirements** | **Capabilities at First light** | **First Decade** |
| Q1<br>Q2<br>Q4 | (Chapter 3)<br>Fundamental physics and cosmology | · Dark matter structure on large and small scales<br>· The effect of dark energy and dark matter on the cosmic expansion rate<br>· Neutron star equation of state<br>· Tests of General Relativity<br>· Variations of physical constants over cosmic time | · Proper motions in dwarf galaxies<br>· IFU high-SNR spectroscopy of lensed images<br>· Wide-field optical spectroscopy of R = 24.5 galaxies<br>· Transient events lasting > 30 days<br>· High resolution spectroscopy of QSOs and gamma-ray bursts (GRBs) | · $\Delta\lambda$ = 0.31–0.62, 2–2.4μm<br>· Seeing-limited FOV > 10′<br>· 4-mas/pixel K band imaging FOV > 30″<br>· R = 1,000–50,000<br>· Very efficient acquisition<br>· 0.05-mas astrometry stable over 10 years | SL/WFOS<br>LGS MCAO/IRIS imager<br>LGS MCAO/MODHIS | SL/HROS |
| Q1<br>Q2<br>Q3<br>Q4 | (Chapter 4)<br>The early universe | · Metal-free star formation in first light objects<br>· Gravitationally lensed first light objects<br>· Early galaxies and re-ionization<br>· Structure and neutral fraction of z > 7 IGM | · Faint object multiplexed, spatially-resolved spectroscopy<br>· High-resolution NIR spectroscopy<br>· Diffraction-limited NIR imaging | · $\Delta\lambda$ = 0.8–2.5 μm<br>· R = 3,000–30,000<br>· J, H, K 4-mas/pixel imaging<br>· $F = 3 x 10^{-20}$ ergs $s^{-1}cm^{-2}Å^{-1}$<br>· Exposure times > 15 ks | LGS MCAO/IRIS IFS<br>LGS MCAO/IRIS imager<br>LGS MCAO/MODHIS | MOAO/IRMOS |

| Fundamental questions | | Science Cases | | Instruments/AO Facility | |
|---|---|---|---|---|---|
| **Qs ID** | **Detailed Science Case Chapter** | **Science Areas** | **Observations** | **Requirements** | **Capabilities at First light** / **First Decade** |
| Q1<br>Q2<br>Q3<br>Q4 | (Chapter 5)<br>Galaxy formation and the intergalactic medium | · Connecting the distributions of stars and dark matter<br>· Baryon cycle during era of peak galaxy formation<br>· Evolution of star formation rates, extinction, metallicity, and kinematics of z = 5.5 to < 1.5 galaxies<br>· IGM/CGM properties on scales < 300 kpc | · Optical/NIR multiplexed and Seeing-/Diffraction-limited spatially-resolved spectroscopy of distant galaxies, AGNs and R ~ 27 high redshift objects | · $\Delta\lambda$ = 0.31–2.5 μm<br>· R = 3,000–5,000, 50,000<br>· Very efficient acquisition<br>· High multiplexing (goal > 100) | SL/WFOS<br>LGS MCAO/IRIS IFS<br>———<br>SL/HROS<br>MOAO/IRMOS |
| Q1<br>Q2<br>Q3 | (Chapter 6)<br>Super massive black holes | · Tests of general relativity (GR) with Milky Way (MW) SMBH<br>· Demographics of low-mass black holes<br>· Measuring curved space-time in other galaxies<br>· Measure AGN SMBH mass to z=0.4 with reverberation mapping<br>· Scaling relations for SMBHs out to z = 2.5 and masses at z > 6<br>· SMBH growth and host galaxy evolution | · Diffraction-limited NIR IFU spectroscopy of galaxy cores<br>· Diffraction-limited NIR imaging and IFU spectroscopy of MW center<br>· Radial velocity variations of stars at the Galactic Center | · $\Delta\lambda$ = 0.8–2.5 μm<br>· R = 3,000–5,000<br>· J, H, K 4-mas/pixel imaging FOV > 30”<br>· 0.05-mas astrometry stable over 10 years<br>· Vrad < 100 m/s | LGS MCAO/IRIS IFS<br>LGS MCAO/IRIS imager<br>LGS MCAO/MODHIS<br>———<br>MOAO/IRMOS |

| Fundamental questions | | Science Cases | | Instruments/AO Facility | | |
|---|---|---|---|---|---|---|
| **Qs ID** | **Detailed Science Case Chapter** | **Science Areas** | **Observations** | **Requirements** | **Capabilities at First light** | **First Decade** |
| Q1<br>Q4<br>Q5 | (Chapter 7)<br>Exploration of the Milky Way and nearby galaxies | · Chemical abundances of old stars in the MW<br>· Initial-Final Mass relation and stellar astrophysics<br>· Kinematics/chemical evolution in Local Group<br>· Diffusion and mass loss in stars<br>· Resolved stellar populations out to Virgo cluster | · High-resolution optical/NIR spectroscopy<br>· Crowded-field high-precision photometry<br>· Medium-resolution multi-object optical spectroscopy | · $\Delta\lambda$ = 0.33–0.9,1.4–2.4 µm<br>· R = 4,000, 40,000–90,000<br>· NIR photometric precision = 0.03 mag @ Strehl = 0.6 | LGS MCAO/IRIS imager<br>SL/WFOS<br>MCAO/MODHIS | SL/HROS<br>MIRAO/MICHI |
| Q4<br>Q5<br>Q6 | (Chapters 8 & 10)<br>The birth and early lives of stars and planets; Exoplanets | · Direct detection of reflected-light from sub-Jovians and super-Earths<br>· Characterization of exo-atmospheres /biosignatures<br>· Understanding the full star formation process<br>· Deposition of prebiotic molecules onto protoplanetary surfaces<br>· Architecture of planetary systems | · Diffraction-limited, high-resolution near- and mid-IR spectroscopy<br>· High-Strehl AO-assisted imaging and spectroscopy<br>· Crowded-field high-precision photometry<br>· High-resolution optical spectroscopy | · $\Delta\lambda$ = 0.6–14 µm<br>· R = 4,000, 30,000–100,000<br>· Low telescope emissivity<br>· Dry site (precipitable water vapor (PWV) < 5-mm median conditions)<br>· Fixed gravity vector & thermal control of instruments<br>· Very efficient acquisition<br>· First Decade contrast ratio of $10^8$ | LGS MCAO/MODHIS<br>LGS MCAO/IRIS imager | SL/HROS<br>ExAO/PSI<br>MIRAO/MICHI |

| Fundamental questions | | Science Cases | | Instruments/AO Facility | |
|---|---|---|---|---|---|
| Qs ID | Detailed Science Case Chapter | Science Areas | Observations | Requirements | Capabilities at First light<br><br>First Decade |
| Q1<br>Q2<br>Q3<br>Q5 | (Chapter 9) Time-domain science | · Gravitational wave and other transient source progenitors<br>· Nuclear processes/particle acceleration in extreme environments (neutrino and cosmic ray sources) | · Rapid response to transient sources and time resolved spectroscopy | · $\Delta\lambda$ = 0.33–2.4 μm<br>· R = 100–2,000<br>· Fast response time < 10-min<br>· Integration time < 10-s<br>· J, H, K 4-mas spatial sampling | LGS MCAO/IRIS IFS<br>LGS MCAO/IRIS imager<br>SL/WFOS |
| Q4<br>Q5<br>Q6 | (Chapter 11)<br>Our Solar System | · Composition, surface physics, structure of asteroids/asteroid satellites/Kuiper Belt Objects/comets/interstellar visitors<br>· Volcanism and tectonic activity<br>· Atmospheres/weather of terrestrial and giant planets | · Spatially-resolved spectroscopy of solar system objects<br>· Transient events (hours to years) | · $\Delta\lambda$ = 0.4–10 μm<br>· R = 1,000–100,000<br>· Non-sidereal tracking<br>· Fast response time | LGS MCAO/IRIS imager<br>LGS MCAO/IRIS IFS<br>LGS MCAO/MODHIS<br><br>SL/HROS<br>MIRAO/MICHI<br>ExAO/PSI |

# 14 Abbreviations and Acronyms

| | |
|---|---|
| ACS | Advanced Camera for Surveys (HST instrument) |
| ADI | Angular Differential Imaging |
| AGB | Asymptotic Giant Branch |
| AGN | Active Galactic Nucleus |
| ALMA | Atacama Large Millimeter Array |
| AO | Adaptive Optics |
| AU | Astronomical Unit |
| BCD | Blue Compact Dwarf galaxies |
| BCG | Brightest Cluster Galaxy |
| BH | Black Hole |
| BLR | Broad Line Region |
| CARMENES | Calar Alto high-Resolution with Near-infrared and optical Échelle Spectrographs |
| CMB | Cosmic Microwave Background |
| CMD | Color-Magnitude Diagram |
| CRIRES | Cryogenic Infrared Echelle Spectrometer (VLT) |
| CSPNe | Central Star of Planetary Nebula |
| dSphs | dwarf Spheroidal galaxies |
| DCC | Document Control Center |
| DE | Dark Energy |
| DEIMOS | Deep Imaging Multi-object Spectrograph (Keck) |
| DSC | Detailed Science Case |
| ELT | Extremely Large Telescope |
| ESO | European Southern Observatory |
| ESPRESSO | Echelle SPectrograph for Rocky Exoplanet Stable Spectroscopic Observations |
| FOV | Field Of View |
| GPI | Gemini Planet Imager |
| GRB | Gamma Ray Burst |
| HIA | Herzberg Institute of Astrophysics |
| HIRES | High Resolution Echelle Spectrometer (Keck) |
| HROS | High-Resolution Optical Spectrometer |
| HST | Hubble Space Telescope |
| IFU | Integral Field Unit |
| IGM | Inter-Galactic Medium |
| IMBH | Intermediate Mass Black Hole |
| IMF | Initial Mass Function |
| IR | Infrared |
| IRIS | InfraRed Imaging Spectrometer |
| IRMOS | InfraRed Multi-Object Spectrometer |
| ISM | Inter-Stellar Medium |
| ISDT | International Science Development Team |
| JWST | James Webb Space Telescope |
| KBO | Kuiper-Belt Object |
| KMTNet | Korea Microlensing Telescope Network |
| LINER | Low-Ionization Nuclear Emission-line Region active galactic nuclei |
| LMC | Large Magellanic Cloud |
| LOS | Line Of Sight |
| LSB | Low Surface Brightness galaxies |
| MBCs | Main Belt Comets |
| MBH | Massive Black Hole |
| MCAO | Multi-Conjugate Adaptive Optics |
| MIR | Mid-Infrared |
| MIRES | Mid-Infrared Echelle Spectrometer |
| MOAO | Multi-Object Adaptive Optics |
| MODHIS | Multi-Objective Diffraction-limited High-resolution Spectrograph |
| MOSFIRE | Multi-Object Spectrometer for Infra-Red Exploration (Keck) |
| NASA | National Aeronautics and Space Administration |
| NFW | Navarro, Frenk & White (dark matter density profile) |
| ngVLA | next generation Very Large Array |
| NIRES | NearInfraRed Echelle Spectrometer |
| NIR | Near InfraRed |
| NIRSPEC | Near Infrared (echelle) Spectrograph (Keck) |
| NFIRAOS | Narrow Field Infrared Adaptive Optics System |
| NOAO | National Optical Astronomy Observatory |
| NRC | NAS National Research Council (USA) or National Research Council of Canada |
| NSF | National Science Foundation |
| NS | Neutron Star |
| PFI | Planet Formation Instrument |
| PNe | Planetary Nebulae |
| PSF | Point Spread Function |
| QSO | Quasi-Stellar Object (AGN) |
| RGB | Red Giant Branch |
| RM | Rossiter-McLaughlin (effect) |
| RV | Radial Velocity |
| SAC | Science Advisory Committee |
| SCExAO | Subaru Coronagraphic Extreme Adaptive Optics |
| SDI | Spectral Differential Imaging |
| SDSS | Sloan Digital Sky Survey |
| SFR | Star Formation Rate |
| SKA | Square-Kilometer Array |
| SMBH | SuperMassive Black Hole |
| SNe | Super Novae |
| SNR | Signal-to-Noise Ratio |
| SPIRou | SpectroPolarimètre Infra-Rouge |
| SPHERE | Spectro-Polarimetric High-contrast Exoplanet Research (VLT) |
| STScI | Space Telescope Science Institute |
| TDE | Tidal Disruption Event |
| TEXES | Texas Echelon Cross Echelle Spectrograph |
| TMT | Thirty Meter Telescope |
| UCI | University of California at Irvine |
| UCLA | University of California at Los Angeles |
| UCSC | University of California at Santa Cruz |
| UKIDSS | United Kingdom Infrared Deep Sky Survey |
| ULIRG | Ultra Luminous Infrared Galaxy |
| URL | Universal Resource Locator |
| UV | UltraViolet |
| UVES | Ultraviolet and Visible Echelle Spectrograph (VLT) |
| VLT | Very Large Telescope |
| WD | White Dwarf |
| WFC3 | Wide Field Camera 3 (HST instrument) |
| WFOS | Wide-Field Optical Spectrometer |
| WMAP | Wilkinson Microwave Anisotropy Probe |
| YSO | Young Stellar Object |
| YMC | Young Massive Cluster |

# 15 INDEX

## 15.1 TABLE OF CONTEXT

## 15.2 TABLE OF FIGURES

## 15.3 TABLE OF TABLES

# 16 CONTRIBUTORS

## 16.1 TMT INTERNATIONAL SCIENCE DEVELOPMENT TEAMS AND OTHER CONTRIBUTORS

Many scientists from across the TMT collaboration and outside in the broader international astronomical community contributed to this document and previous version. The majority of these scientists are members of the TMT ISDTs, and each chapter was developed by members of a particular ISDT, led by a chapter editor from each ISDT, supported by the ISDT conveners. Additional input was sought by ISDT conveners or DSC editors where needed. All contributors and authors for each chapter are listed at the beginning of each chapter.

## 16.2 TMT DSC 2024

### 16.2.1 Editing Team

Warren Skidmore (Thirty Meter Telescope International Observatory, US) (Editor in-chief)

Updates were carried out by TIO staff Warren Skidmore, Scot Kleinman, Bob Kirshner, Gelys Trancho, and David Andersen. Final compilation was done by Nancy Han.

### 16.2.2 TMT Science Advisory Committee Readers

Roberto Abraham (University of Toronto, Canada)
Masayuki Akiyama (Tohoku University, Japan)
Len Cowie (University of Hawaii, US)
Ian Dell'Antonio (Brown University, US)
Christophe Dumas (TMT & UK ATC)
Mitsuhiko Honda (Okayama University of Science, Japan)
Bruce Macintosh (University of California Observatories, US)
Karen Meech (University of Hawaii, US)
Stan Metchev (Western University, Canada)
Surhud More (Inter-University Centre for Astronomy and Astrophysics, India)
Norio Narita (University of Tokyo, Japan)
Amitesh Omar (Indian Institute of Technology, Kanpur)
Eric Peng (NOIRLab, US)
Manoj Puravankara (Tata Institute of Fundamental Research, India)
Chuck Steidel (California Institute of Technology, US)
Sivarani Thirupathi (Indian Institute of Astrophysics)
Tommaso Treu (University of California Los Angeles, US)

## 16.3 TMT DSC 2022

### 16.3.1 Editing Team

Warren Skidmore (Thirty Meter Telescope Project Office, US) (Editor in-chief)

Updates were carried out by TIO staff: Warren Skidmore, Christophe Dumas, George Jacoby, Scot Kleinman, and Gelys Trancho.

#### 16.3.2 TMT Science Advisory Subcommittee Readers

Masayuki Akiyama (Tohoku University, Japan)
Ian Dell'Antonio (Brown University, US)
Tommaso Treu (University of California Los Angeles, US)
Len Cowie (University of Hawaii, US)
C. Pilachowski (Indiana University, US)
Gregory Herczeg (KIAA, Peking University, China)
Bob Kirshner (TIO)
Norio Narita (University of Tokyo, Japan)
Karen Meech (University of Hawaii, US)

### 16.4 TMT DSC 2015

#### 16.4.1 Editing Team

Warren Skidmore (Thirty Meter Telescope Project Office, US) (Editor in-chief)

Ian Dell'Antonio (Brown University, US)
Misato Fukugawa (Osaka University, Japan)
Aruna Goswami (IIA, India)
Lei Hao (Shanghai Astronomical Observatory, China)
Paul Hickson (University of British Columbia, Canada)
David Jewitt (University of California Los Angeles, US)
Greg Laughlin (University of California Observatory Lick, US)
Luc Simard (NRC Herzberg, Canada)
Matthias Schöck (Thirty Meter Telescope Project, NRC Herzberg, Canada) (editorial support)
Charles Steidel (California Institute of Technology, US)
Tommaso Treu, UC Los Angeles (editorial support, ISDT coordinator)

The editors thank Judy Cohen (Caltech) for suggestions, comments and information that greatly strengthened this document.

#### 16.4.2 TMT Science Advisory Committee Readers

Masayuki Akiyama, Tohoku University
Michael Bolte, UC Santa Cruz
Ray Carlberg, University of Toronto
Judith Cohen, Caltech
Timothy Davidge, NRC/HIA
Ian Dell'Antonio, Brown University
Mark Dickinson, NoirLab
Taotao Fang, Xiamen University
Paul Hickson, University of British Columbia
Jiangsheng Huang, Harvard-CfA
Garth Illingworth, UC Santa Cruz
Masanori Iye, NAOJ
Nobunari Kashikawa, NAOJ
Shri Kulkarni, Caltech
Xiaowei Liu, KIAA-PKU
Jennifer Lotz, STScI
Lori Lubin, UC Davis

Shude Mao, NAOC
Chris Martin, Caltech
Jerry Nelson, UC Santa Cruz (Project Scientist)
Shashi Bhushan Pandey, ARIES
V. Pankonin, NSF
C. Pilachowski, Indiana University
A.N Ramapraksh, IUCAA
B. E. Reddy, IIA
D. Silva, NOAO
A. Subramaniam, IIA
Charles Steidel, Caltech
Tommaso Treu, UC Los Angeles
Tomonori Usuda, NAOJ
Suijian Xue, NAOC

Tommaso Treu, Todd Boroson and Mark Dickinson masterminded the establishment of the ISDTs and oversaw ISDT activities during the time leading up to and including the update of the 2015 DSC.

# 17 ACKNOWLEDGEMENT

The TMT International Observatory (TIO) gratefully acknowledges the support of the TIO Members and the Gordon and Betty Moore Foundation. The Members and their respective collaborating institutions are as follows: California Institute of Technology (Caltech); the National Research Council of Canada (NRC-CNRC); the Association of Canadian Universities for Research in Astronomy (ACURA); the Department of Science and Technology, Government of India (DST); the Department of Atomic Energy of India (DAE); the National Institute of Natural Sciences of Japan (NINS); the National Astronomical Observatory of Japan (NAOJ); and the Regents of the University of California (UC).